\newif\ifnoteu
\noteutrue 
           
\documentclass{ctatdrF}
\usepackage{rotating}
\usepackage{nicefrac}
\usepackage{amssymb}
\usepackage{placeins}
\usepackage{aas_macros}
\usepackage[titletoc]{appendix}
\usepackage{geometry}
\usepackage[font=small,labelfont=bf]{caption}
\usepackage{hyperref}
\usepackage{multirow}
\usepackage{aas_macros}
\usepackage{float}
\usepackage{lineno}
\usepackage{tocloft}
\usepackage{colortbl} 
\usepackage{tabu}
\usepackage{incgraph,tikz}

\newcommand{\dg}{$^{\circ}$}
\newcommand{\degr}{$^{\circ}$~}

\renewcommand{\appendixname}{Appendix}
\def\mydots{\leavevmode\xleaders\hbox to 0.25em{\hfil.\hfil}\hfill\kern0pt}
\definecolor{amethyst}{rgb}{0.6, 0.4, 0.8}
\definecolor{green}{rgb}{0.55, 0.71, 0.0}
\definecolor{apricot}{rgb}{0.98, 0.81, 0.69}
\definecolor{auburn}{rgb}{0.43, 0.21, 0.1}
\definecolor{babyblueeyes}{rgb}{0.63, 0.79, 0.95}
\definecolor{bittersweet}{rgb}{1.0, 0.44, 0.37}

\begin{document}

\title{Science with the Cherenkov Telescope Array}
\shorttitle{Science with CTA}
\version{3.1}
\editor{Rene Ong}
\distribution{CTA}
\comment{Version for public release}

\newglossaryentry{CTA}{name={CTA},description={Cherenkov Telescope Array}}
\newglossaryentry{CTAC}{name={CTAC},description={CTA Consortium}}
\newglossaryentry{IACT}{name={IACT},description={Imaging Atmospheric Cherenkov Telescope}}
\newglossaryentry{LST}{name={LST},description={Large Sized Telescope}}
\newglossaryentry{MST}{name={MST},description={Medium Sized Telescope}}
\newglossaryentry{SST}{name={SST},description={Small Sized Telescope}}
\newglossaryentry{FoV}{name={FoV},description={Field of View}}
\newglossaryentry{PSF}{name={PSF},description={Point Spread Function}}
\newglossaryentry{WIMP}{name={WIMP},description={Weakly Interacting Massive Particle}}
\newglossaryentry{KSP}{name={KSP},description={Key Science Project}}
\newglossaryentry{VHE}{name={VHE},description={Very High Energy}}
\newglossaryentry{HE}{name={HE},description={High Energy}}
\newglossaryentry{MW}{name={MW},description={Milky Way}}
\newglossaryentry{GPS}{name={GPS},description={Galactic Plane Survey}}
\newglossaryentry{LMC}{name={LMC},description={Large Magellanic Cloud}}
\newglossaryentry{DM}{name={DM},description={Dark Matter}}
\newglossaryentry{SNR}{name={SNR},description={Supernova Remnant}}
\newglossaryentry{GRB}{name={GRB},description={Gamma-ray Burst}}
\newglossaryentry{AGN}{name={AGN},description={Active Galactic Nuclei}}
\newglossaryentry{CR}{name={CR},description={Cosmic Ray}}
\newglossaryentry{SKA}{name={SKA},description={Square Kilometre Array}}
\newglossaryentry{UHECR}{name={UHECR},description={Ultra-high Energy Cosmic Ray}}
\newglossaryentry{PWN}{name={PWN},description={Pulsar Wind Nebula}}
\newglossaryentry{UV}{name={UV},description={Ultraviolet}}
\newglossaryentry{IR}{name={IR},description={Infrared}}
\newglossaryentry{SMBH}{name={SMBH},description={Supermassive Black Hole}}
\newglossaryentry{BH}{name={BH},description={Black Hole}}
\newglossaryentry{ToO}{name={ToO},description={Target of Opportunity}}
\newglossaryentry{GC}{name={GC},description={Galactic Centre}}
\newglossaryentry{IC}{name={IC},description={Inverse Compton}}
\newglossaryentry{GW}{name={GW},description={Gravitational Wave}}
\newglossaryentry{EBL}{name={EBL},description={Extragalactic Background Light}}
\newglossaryentry{IGMF}{name={IGMF},description={Intergalactic Magnetic Field}}
\newglossaryentry{LHC}{name={LHC},description={Large Hadron Collider}}
\newglossaryentry{SM}{name={SM},description={Standard Model}}
\newglossaryentry{CDM}{name={CDM},description={Cold Dark Matter}}
\newglossaryentry{SUSY}{name={SUSY},description={Supersymmetry}}
\newglossaryentry{ALP}{name={ALP},description={Axion-like Particle}}
\newglossaryentry{LIV}{name={LIV},description={Lorentz Invariance Violation}}
\newglossaryentry{CP}{name={CP},description={Charge Parity}}
\newglossaryentry{QCD}{name={QCD},description={Quantum Chromodynamics}}
\newglossaryentry{FP7}{name={FP7},description={Framework Programme 7}}
\newglossaryentry{ALMA}{name={ALMA},description={Atacama Large Millimeter/Submillimeter Array}}
\newglossaryentry{MWL}{name={MWL},description={Multi-wavelength}}
\newglossaryentry{MM}{name={MM},description={Multi-messenger}}
\newglossaryentry{SED}{name={SED},description={Spectral Energy Distribution}}
\newglossaryentry{MoUs}{name={MoUs},description={Memoranda of Understanding}}
\newglossaryentry{VLBI}{name={VLBI},description={Very Long Baseline Interferometry}}
\newglossaryentry{eMERLIN}{name={eMERLIN},description={Multi-Element Radio Linked Interferometer Network}}
\newglossaryentry{JVLA}{name={JVLA},description={Jansky Very Large Array}}
\newglossaryentry{LOFAR}{name={LOFAR},description={Low Frequency Array}}
\newglossaryentry{MWA}{name={MWA},description={Murchison Widefield Array}}
\newglossaryentry{ASKAP}{name={ASKAP},description={Australian Square Kilometre Array Pathfinder}}
\newglossaryentry{VLITE}{name={VLITE},description={VLA Low Frequency Ionosphere Transient Experiment}}
\newglossaryentry{LOBO}{name={LOBO},description={Low Band Observatory}}
\newglossaryentry{HEAT}{name={HEAT},description={High Elevation Antarctic Teraherz telescope}}
\newglossaryentry{DATE5}{name={DATE5},description={Dome A Terahertz Explorer 5 m}}
\newglossaryentry{OIR}{name={OIR},description={Optical/Infrared}}
\newglossaryentry{E-ELT}{name={E-ELT},description={European Extremely Large Telescope}}
\newglossaryentry{TMT}{name={TMT},description={Thirty Meter Telescope}}
\newglossaryentry{GMT}{name={GMT},description={Giant Magellan Telescope}}
\newglossaryentry{HST}{name={HST},description={Hubble Space Telescope}}
\newglossaryentry{JWST}{name={JWST},description={James Webb Space Telescope}}
\newglossaryentry{iPTF}{name={iPTF},description={intermediate Palomar Transient Factory}}
\newglossaryentry{ZTF}{name={ZTF},description={Zwicky Transient Facility}}
\newglossaryentry{Pan-STARRS}{name={Pan-STARRS},description={Panoramic Survey Telescope and Rapid Response System}}
\newglossaryentry{GW}{name={GW},description={Gravitational Wave}}
\newglossaryentry{LSST}{name={LSST},description={Large Synoptic Survey Telescope}}
\newglossaryentry{TDE}{name={TDE},description={Tidal Disruption Event}}
\newglossaryentry{RTA}{name={RTA},description={Real-time Analysis}}
\newglossaryentry{UV}{name={UV},description={Ultra-violet}}
\newglossaryentry{RXTE}{name={RXTE},description={Rossi X-ray Timing Explorer}}
\newglossaryentry{NuSTAR}{name={NuSTAR},description={Nuclear Spectroscopic Telescope Array}}
\newglossaryentry{MAXI}{name={MAXI},description={Monitor of All-sky X-ray Image}}
\newglossaryentry{eROSITA}{name={eROSITA},description={extended ROentgen Survey with an Imaging Telescope Array}}
\newglossaryentry{NICER}{name={NICER},description={Neutron star Interior Composition ExploreR}}
\newglossaryentry{ISS}{name={ISS},description={International Space Station}}
\newglossaryentry{LOBSTER}{name={LOBSTER},description={Large-angle OBServaTory with Energy Resolution}}
\newglossaryentry{HXMT}{name={HXMT},description={Hard X-ray Modulation Telescope}}
\newglossaryentry{XTP}{name={XTP},description={X-ray Timing and Polarization mission}}
\newglossaryentry{SVOM}{name={SVOM},description={Space Variable Objects Monitor}}
\newglossaryentry{LOFT}{name={LOFT},description={Large Observatory for X-ray Timing}}
\newglossaryentry{SMEX}{name={SMEX},description={Small Explorer}}
\newglossaryentry{XMM}{name={XMM},description={X-ray Multi Mirror}}
\newglossaryentry{INTEGRAL}{name={INTEGRAL},description={International Gamma-Ray Astrophysics Laboratory}}
\newglossaryentry{Swift-BAT}{name={Swift-BAT},description={Swift Burst Alert Telescope}}
\newglossaryentry{Fermi-GBM}{name={Fermi-GBM},description={Fermi Gamma-ray Burst Monitor}}
\newglossaryentry{NASA}{name={NASA},description={National Aeronautics and Space Administration}}
\newglossaryentry{ESA}{name={ESA},description={European Space Agency}}
\newglossaryentry{Fermi-LAT}{name={Fermi-LAT},description={Fermi Large Area Telescope}}
\newglossaryentry{AGILE}{name={AGILE},description={Astro-rivelatore Gamma a Immagini LEggero}}
\newglossaryentry{PANGU}{name={PANGU},description={PAir-productioN Gamma-ray Unit}}
\newglossaryentry{CAS}{name={CAS},description={Chinese Academy of Sciences}}
\newglossaryentry{HAWC}{name={HAWC},description={High Altitude Water Cherenkov}}
\newglossaryentry{LHAASO}{name={LHAASO},description={Large High Altitude Air Shower Observatory}}
\newglossaryentry{H.E.S.S.}{name={H.E.S.S.},description={High Energy Stereoscopic System}}
\newglossaryentry{MAGIC}{name={MAGIC},description={Major Atmospheric Gamma-Ray Imaging Cherenkov Telescopes}}
\newglossaryentry{VERITAS}{name={VERITAS},description={Very Energetic Radiation Imaging Telescope Array System}}
\newglossaryentry{KM3Net}{name={KM3Net},description={Cubic Kilometre Neutrino Telescope}}
\newglossaryentry{ToO}{name={ToO},description={Target of Opportunity}}
\newglossaryentry{LIGO}{name={LIGO},description={Laser Interferometer Gravitational-wave Observatory}}
\newglossaryentry{BLR}{name={BLR},description={Broad Line Region}}
\newglossaryentry{RoboPol}{name={RoboPol},description={ROBOtic POLarimeter}}
\newglossaryentry{BL Lac}{name={BL Lac},description={BL-Lacertae}}
\newglossaryentry{VOEvent}{name={VOEvent},description={Virtual Observatory Event}}
\newglossaryentry{NIR}{name={NIR},description={Near Infrared}}
\newglossaryentry{WSRT}{name={WSRT},description={Westerbork Synthesis Radio Telescope}}
\newglossaryentry{pMSSM}{name={pMSSM},description={Phenomenological Minimal Supersymmetric Model}}
\newglossaryentry{NFW}{name={NFW},description={Navarro, Frenk and White}}
\newglossaryentry{LEP}{name={LEP},description={Large Electron-Positron Collider}}
\newglossaryentry{LSP}{name={LSP},description={Lightest Supersymmetric Particle}}
\newglossaryentry{WMAP}{name={WMAP},description={Wilkinson Microwave Anisotropy Probe}}
\newglossaryentry{VIB}{name={VIB},description={Virtual Internal Bremsstrahlung}}
\newglossaryentry{CMB}{name={CMB},description={Cosmic Microwave Background}}
\newglossaryentry{LOS}{name={LOS},description={Line of Sight}}
\newglossaryentry{dSph}{name={dSph},description={Dwarf Spheroidal Galaxy}}
\newglossaryentry{EAGLE}{name={EAGLE},description={Evolution and Assembly of GaLaxies and their Environments}}
\newglossaryentry{ARGOS}{name={ARGOS},description={Advanced Rayleigh Guided Ground Layer Adaptive Optics System}}
\newglossaryentry{BRAVA}{name={BRAVA},description={Bulge Radial Velocity Assay}}
\newglossaryentry{GAIA}{name={GAIA},description={Globales Astronomisches Interferometer f\"ur die Astrophysik}}
\newglossaryentry{GIBS}{name={GIBS},description={GIRAFFE Inner Bulge Survey}}
\newglossaryentry{MUSE}{name={MUSE},description={Multi Unit Spectroscopic Explorer}}
\newglossaryentry{VLT}{name={VLT},description={Very Large Telescope}}
\newglossaryentry{IRF}{name={IRF},description={Instrument Response Functions}}
\newglossaryentry{CMR}{name={CMR},description={Circum-Nuclear Ring}}
\newglossaryentry{GO}{name={GO},description={Guest Observer}}
\newglossaryentry{HEGRA}{name={HEGRA},description={High Energy Gamma Ray Astronomy}}
\newglossaryentry{ISM}{name={ISM},description={Interstellar Medium}}
\newglossaryentry{ARGO-YBJ}{name={ARGO-YBJ},description={Astrophysical Radiation with Ground-based Observatory at YangBaJing}}
\newglossaryentry{Milagro}{name={Milagro},description={Milagro Gamma Ray Observatory}}
\newglossaryentry{MeerKAT}{name={MeerKAT},description={Karoo Array Telescope}}
\newglossaryentry{STAC}{name={STAC},description={Scientific and Technical Advisory Committee}}
\newglossaryentry{STP}{name={STP},description={Short-term Program}}
\newglossaryentry{LTP}{name={LTP},description={Long-term Program}}
\newglossaryentry{FITS}{name={FITS},description={Flexible Image Transport System}}
\newglossaryentry{HEASARC}{name={HEASARC},description={High Energy Astrophysics Science Archive Research Center}}
\newglossaryentry{ATCA}{name={ATCA},description={Australia Telescope Compact Array}}
\newglossaryentry{MOPRA}{name={MOPRA},description={Mopra Radio Telescope}}
\newglossaryentry{VISTA}{name={VISTA},description={Visible and Infrared Survey Telescope for Astronomy}}
\newglossaryentry{ROSAT}{name={ROSAT},description={ROentgenSATellit}}
\newglossaryentry{PILOT}{name={PILOT},description={Polarized Instrument for Long Wavelength Observations of the Tenuous ISM}}
\newglossaryentry{CSM}{name={CSM},description={Circumstellar Medium}}
\newglossaryentry{ATNF}{name={ATNF},description={Australia Telescope National Facility}}
\newglossaryentry{IRAS}{name={IRAS},description={Infrared Astronomical Satellite}}
\newglossaryentry{FSRQ}{name={FSRQ},description={Flat Spectrum Radio Quasar}}
\newglossaryentry{LF}{name={LF},description={Luminosity Function}}
\newglossaryentry{EGRET}{name={EGRET},description={Energetic Gamma Ray Experiment Telescope}}
\newglossaryentry{OVRO}{name={OVRO},description={Owens Valley Radio Observatory}}
\newglossaryentry{RA/Dec}{name={RA/Dec},description={Right Ascension/Declination}}
\newglossaryentry{MC}{name={MC},description={Monte Carlo}}
\newglossaryentry{NS}{name={NS},description={Neutron Star}}
\newglossaryentry{BH}{name={BH},description={Black Hole}}
\newglossaryentry{TF}{name={TF},description={Transient Factory}}
\newglossaryentry{EGS}{name={EGS},description={Extragalactic Survey}}
\newglossaryentry{FRB}{name={FRB},description={Fast Radio Burst}}
\newglossaryentry{LIV}{name={LIV},description={Lorentz Invariance Violation}}
\newglossaryentry{GCN}{name={GCN},description={Gamma-ray Burst Coordinate Network}}
\newglossaryentry{IGMF}{name={IGMF},description={Intergalactic Magnetic Fields}}
\newglossaryentry{KAGRA}{name={KAGRA},description={Kamioka Gravitational Wave Detector}}
\newglossaryentry{EM}{name={EM},description={Electromagnetic}}
\newglossaryentry{GVD}{name={GVD},description={Gigaton Volume Detector}}
\newglossaryentry{AMON}{name={AMON},description={Astrophysical Multimessenger Observatory Network}}
\newglossaryentry{SMC}{name={SMC},description={Small Magellanic Cloud}}
\newglossaryentry{APEX}{name={APEX},description={Atacama Pathfinder EXperiment}}
\newglossaryentry{IRAM}{name={IRAM},description={Institut de Radioastronomie Millim\'etrique}}
\newglossaryentry{LBL}{name={LBL},description={Low-frequency peaked BL Lacertae}}
\newglossaryentry{IBL}{name={IBL},description={Intermediate-frequency peaked BL Lacertae}}
\newglossaryentry{HBL}{name={HBL},description={High-frequency peaked BL Lac}}
\newglossaryentry{UHBL}{name={UHBL},description={Ultra-high-frequency peaked BL Lacacertae}}
\newglossaryentry{EHBL}{name={EHBL},description={Extremely-high-frequency peaked BL Lacertae}}
\newglossaryentry{LLAGN}{name={LLAGN},description={Low-luminosity Active Galactic Nucleus}}
\newglossaryentry{NLSy1}{name={NLSy1},description={Narrow-Line Seyfert 1 Galaxy}}
\newglossaryentry{SSC}{name={SSC},description={Synchrotron Self Compton}}
\newglossaryentry{ATOM}{name={ATOM},description={Automated Telescope for Optical Monitoring}}
\newglossaryentry{GMRT}{name={GMRT},description={Giant Metrewave Radio Telescope}}
\newglossaryentry{TA}{name={TA},description={Telescope Array}}
\newglossaryentry{ANTARES}{name={ANTARES},description={Astronomy with a Neutrino Telescope and Abyss environmental RESearch}}
\newglossaryentry{JEM/EUSO}{name={JEM/EUSO},description={Japanese Experiment Module/Extreme Universe Space Observatory}}
\newglossaryentry{ICM}{name={ICM},description={Intra-cluster Medium}}
\newglossaryentry{WHIPPLE}{name={WHIPPLE},description={Fred Lawrence Whipple Observatory}}
\newglossaryentry{CANGAROO}{name={CANGAROO},description={Collaboration of Australia and Nippon (Japan) for a GAmma Ray Observatory in the Outback}}
\newglossaryentry{KASCADE}{name={KASCADE},description={Karlsruhe Shower Core and Array Detector}}
\newglossaryentry{CREAM}{name={CREAM},description={Cosmic Ray Energetics and Mass}}
\newglossaryentry{PAMELA}{name={PAMELA},description={Payload for Antimatter Matter Exploration and Light-nuclei Astrophysics}}
\newglossaryentry{AMS-02}{name={AMS-02},description={Alpha Magnetic Spectrometer}}
\newglossaryentry{TOTEM}{name={TOTEM},description={Total, elastic and diffractive cross-section measurement}}
\newglossaryentry{OSETI}{name={OSETI},description={OpticalSearch for Extraterrestrial Intelligence}}
\newglossaryentry{SCT}{name={SCT},description={Schwarzschild-Couder Telescope}}
\newglossaryentry{SFS}{name={SFS},description={Star-Forming System}}
\newglossaryentry{SFR}{name={SFR},description={Star-Formation Rate}}
\newglossaryentry{SN}{name={SN},description={Supernova}}
\newglossaryentry{CMZ}{name={CMZ},description={Central Molecular Zone}}
\newglossaryentry{CWB}{name={CWB},description={Colliding-Wind Binary}}
\newglossaryentry{ULIRG}{name={ULIRG},description={Ultraluminous Infrared Galaxy}}

\maketitle

\hphantom{dummy}
\vspace{22.5cm}
\noindent Cover Illustration: Galactic centre background image created by NASA, ESA,
SSC, CXC, and STScI. CTA southern hemisphere array rendering created
by Gabriel P\'{e}rez D\'{i}az, IAC.
\clearpage

\section*{Executive Summary}
\addcontentsline{toc}{section}{Executive Summary}

The Cherenkov Telescope Array, CTA, will be the major global observatory
for very high energy gamma-ray astronomy over the next decade and beyond. 
The scientific potential of
CTA is extremely broad: from understanding the role of relativistic
cosmic particles to the search for dark matter. CTA is an explorer of 
the extreme universe, probing environments from the immediate
neighbourhood of black holes to cosmic voids on the largest
scales. Covering a huge range in photon energy from 20~GeV to 300
TeV, CTA will improve on all aspects of performance with respect to
current instruments. Wider field of view and improved sensitivity 
will enable CTA to survey hundreds of times faster than previous 
TeV telescopes. The angular resolution of CTA will
approach 1 arc-minute at high energies --- the best resolution of any 
instrument operating above the X-ray band --- allowing detailed 
imaging of a large number of gamma-ray sources. A one to two order-of-magnitude collection area improvement makes CTA a powerful instrument for time-domain astrophysics, 
three orders of magnitude more sensitive on hour timescales than Fermi-LAT at 30 GeV.
The observatory will operate arrays on sites in both hemispheres to
provide full sky coverage and will hence maximize the potential for the
rarest phenomena such as very nearby supernovae, gamma-ray bursts or
gravitational wave transients.
With 99 telescopes on the southern site and 19 telescopes on the northern site, 
flexible operation will be possible, with sub-arrays available for specific tasks.

CTA will have important synergies with many of the new generation of
major astronomical and astroparticle observatories. Multi-wavelength
and multi-messenger approaches combining CTA data with those from other
instruments will lead to a deeper understanding of the
broad-band non-thermal properties of target sources, elucidating the
nature, environment, and distance of gamma-ray emitters. 
Details of synergies in each waveband are presented.

The CTA Observatory will be operated as an 
open, proposal-driven observatory, with all
data available on a public archive after a pre-defined proprietary
period (of typically one year). 
Scientists from institutions worldwide have combined together
to form the CTA Consortium. This Consortium
has prepared a proposal for a Core Programme of 
highly motivated observations.
The programme, encompassing approximately 40\% of the available observing time
over the first ten years of CTA operation, 
is made up of individual Key Science
Projects (KSPs), which are presented in the subsequent chapters. 
The science cases have been prepared over
several years by the CTA Consortium, with community input gathered via
a series of workshops connecting CTA to neighbouring communities. A
major element of the programme is the search for dark matter via
the annihilation signature of weakly interacting massive particles (WIMPs). 
The strategy for dark
matter detection presented here places the expected cross-section for
a thermal relic within reach of CTA for a wide range of WIMP masses from
$\sim$200~GeV to 20~TeV. This makes CTA extremely complementary to other
approaches, such as high-energy particle collider and direct-detection experiments. 
CTA will also conduct a census of particle acceleration over a wide
range of astrophysical objects, 
with quarter-sky extragalactic, full-plane Galactic and Large Magellanic Cloud
surveys planned. Additional KSPs are focused on transients, acceleration up to PeV energies in our own Galaxy, active galactic nuclei, 
star-forming systems on a wide range of scales, and the Perseus cluster of galaxies.
All provide high-level data products which will benefit a wide community, and together they will provide 
a long-lasting legacy for CTA.

Finally, while designed for the detection of gamma rays, CTA has considerable potential for a range of astrophysics and astroparticle physics based on charged cosmic-ray observations and the use of the CTA telescopes for optical measurements.

\clearpage

\vspace{1cm}

{\Huge \bf \ctablue Authors}

\vspace{0.2cm}

{\bf \large \ctablue The Cherenkov Telescope Array Consortium:}

\begin{center}
B.S.~Acharya$^{1}$,
I.~Agudo$^{2}$,
I.~Al Samarai$^{3}$,
R.~Alfaro$^{4}$,
J.~Alfaro$^{5}$,
C.~Alispach$^{3}$,
R.~Alves Batista$^{6}$,
J.-P.~Amans$^{7}$,
E.~Amato$^{8}$,
G.~Ambrosi$^{9}$
% ...

 \bigskip 
 \bigskip 
$^{1} \ $Tata Institute of Fundamental Research, Homi Bhabha Road, Colaba, Mumbai 400005, India
 \vspace{-0.25cm} 

$^{2} \ $Instituto de Astrof\'{i}sica de Andaluc\'{i}a-CSIC, Glorieta de la Astronom\'{i}a s/n, E-18008, Granada, Spain
 \vspace{-0.25cm} 

$^{3} \ $University of Geneva - D\'{e}partement de physique nucl\'{e}aire et corpusculaire, 24 rue du G\'{e}n\'{e}ral-Dufour, 1211 Gen\`{e}ve 4, Switzerland
 \vspace{-0.25cm} 

$^{4} \ $Universidad Nacional Aut\'{o}noma de M\'{e}xico, Delegaci\'{o}n Coyoac\'{a}n, 04510 Ciudad de M\'{e}xico, Mexico
 \vspace{-0.25cm} 

$^{5} \ $Pontificia Universidad Cat\'{o}lica de Chile, Avda. Libertador Bernardo O' Higgins No 340, borough and city of Santiago, Chile
 \vspace{-0.25cm} 

$^{6} \ $University of Oxford, Department of Physics, Denys Wilkinson Building, Keble Road, Oxford OX1 3RH, United Kingdom
 \vspace{-0.25cm} 

$^{7} \ $LUTH and GEPI, Observatoire de Paris, CNRS, PSL Research University, 5 place Jules Janssen, 92190, Meudon, France
 \vspace{-0.25cm} 

$^{8} \ $INAF - Osservatorio Astrofisico di Arcetri, Largo E. Fermi, 5 - 50125 Firenze, Italy
 \vspace{-0.25cm} 

$^{9} \ $INFN Sezione di Perugia and Universit\`{a} degli Studi di Perugia, Via A. Pascoli, 06123 Perugia, Italy
 \vspace{-0.25cm} 

% ...

\end{center}

\clearpage

\section*{Contents}

\vspace{5mm}

\noindent Chapters and corresponding authors:

\vspace{5mm}

{
\renewcommand{\arraystretch}{2}
\setlength\tabcolsep{0mm}
\begin{tabu} to \textwidth {p{155mm}r}
\ref{sec:sci_intro}. Introduction to CTA Science --- {\it \footnotesize J.A. Hinton, R.A. Ong, D. Torres} \mydots &  \pageref{sec:sci_intro} \\
\ref{sec:sci_synergies}. Synergies --- {\it \footnotesize S. Markoff, J.A. Hinton, R.A. Ong, D. Torres} \mydots &  \pageref{sec:sci_synergies} \\
\ref{sec:obs_prog}. Core Programme Overview --- {\it \footnotesize J.A. Hinton, R.A. Ong, D. Torres} \mydots &  \pageref{sec:obs_prog} \\
\ref{sec:DM_prog}. Dark Matter Programme -- {\it \footnotesize E. Moulin, J. Carr, J. Gaskins, M. Doro, C. Farnier, M. Wood, H. Zechlin} \mydots & \pageref{sec:DM_prog} \\
\ref{sec:ksp_gc}. KSP: Galactic Centre -- {\it \footnotesize C. Farnier, K. Kosack, R. Terrier} \mydots & \pageref{sec:ksp_gc} \\
\ref{sec:ksp_gps}. KSP: Galactic Plane Survey -- {\it \footnotesize R.C.G.~Chaves, R. Mukherjee, R.A. Ong} \mydots & \pageref{sec:ksp_gps} \\
\ref{sec:ksp_lmc}. KSP: LMC Survey -- {\it \footnotesize P. Martin, C.-C. Lu, H. Voelk, M. Renaud, M. Filipovic} \mydots & \pageref{sec:ksp_lmc} \\
\ref{sec:ksp_eg}. KSP: Extragalactic Survey -- {\it \footnotesize D. Mazin, L. Gerard, J.E. Ward, P. Giommi, A.M. Brown} \mydots & \pageref{sec:ksp_eg} \\
\ref{sec:ksp_trans}. KSP: Transients -- {\it \footnotesize S. Inoue, M. Rib\'{o}, E. Bernardini, V. Connaughton, J. Granot, S. Markoff, P. O{\hspace{1mm}}Brien, F. Schussler} \mydots & \mydots \pageref{sec:ksp_trans} \\
\ref{sec:ksp_acc}. KSP: Cosmic Ray PeVatrons -- {\it \footnotesize R.C.G.~Chaves, E. De O\~{n}a Wilhelmi, S. Gabici, M. Renaud} \mydots  & \mydots \pageref{sec:ksp_acc} \\
\ref{sec:ksp_sfs}. KSP: Star Forming Systems -- {\it \footnotesize S. Casanova, S. Ohm, L. Tibaldo} \mydots & \mydots \pageref{sec:ksp_sfs} \\
\ref{sec:ksp_agn}. KSP: Active Galactic Nuclei -- {\it \footnotesize A. Zech, D. Mazin, J. Biteau, M. Daniel, T. Hassan, E. Lindfors, M. Meyer} \mydots & \mydots \pageref{sec:ksp_agn} \\
\ref{sec:ksp_clust}. KSP: Clusters of Galaxies -- {\it \footnotesize F. Zandanel, M Fornasa} \mydots & \mydots \pageref{sec:ksp_clust} \\
\ref{sec:nongamma}. Capabilities beyond Gamma Rays -- {\it \footnotesize R. B\"uhler, D. Dravins, K. Egberts, 
J.A. Hinton, R.D. Parsons} \mydots  & \mydots \pageref{sec:nongamma} \\
\ref{sec:MC}. Appendix: Simulating CTA -- {\it \footnotesize G. Maier} \mydots  & \mydots \pageref{sec:MC} \\
Acknowledgments \mydots & \mydots \pageref{sec:ACK}\\
References \mydots & \mydots 193 \\
Glossary \mydots & \mydots \,209 \\
\end{tabu} 
}

\section{Introduction to CTA Science}
\label{sec:sci_intro}

Ground-based gamma-ray astronomy is a young field with enormous
scientific potential. The possibility of astrophysical measurements at
teraelectronvolt (TeV) energies was demonstrated in 1989 with the
detection of a clear signal from the Crab nebula above 1 TeV 
with the Whipple 10m imaging atmospheric Cherenkov telescope
(IACT). Since then, the instrumentation for, and techniques of,
astronomy with IACTs have evolved to the extent that a flourishing new
scientific discipline has been established, with the detection of more than 150
sources and a major impact in astrophysics and more widely in
physics.
The current major arrays of IACTs: H.E.S.S., MAGIC, and
VERITAS, have demonstrated the huge physics potential at these
energies as well as the maturity of the detection technique. Many
astrophysical source classes have been established, some with many
well-studied individual objects, but there are indications that the
known sources represent the tip of the iceberg in terms of both
individual objects and source classes. The Cherenkov Telescope Array
(CTA) will transform our understanding of the high-energy universe and will explore questions in
physics of fundamental importance.
As a key member of the suite of new and upcoming major astroparticle
physics experiments and observatories, CTA will exploit synergies with
gravitational wave and neutrino observatories 
as well as with classical photon observatories.
CTA will address a
wide range of major questions in and beyond astrophysics, which can be
grouped into three broad themes:

{\bf Theme 1: Understanding the Origin and Role of Relativistic Cosmic Particles}
\begin{itemize}
\item  What are the sites of high-energy particle acceleration in the universe?
\item  What are the mechanisms for cosmic particle acceleration? 
\item  What role do accelerated particles play in feedback on star formation and galaxy evolution? 
\end{itemize}

{\bf Theme 2: Probing Extreme Environments}
\begin{itemize}
\item  What physical processes are at work close to neutron stars and 
  black holes?
\item  What are the characteristics of relativistic jets, winds and explosions?
\item  How intense are radiation fields and magnetic fields in cosmic voids, and how do these evolve over cosmic
  time?
\end{itemize}

{\bf Theme 3: Exploring Frontiers in Physics}
\begin{itemize}
\item What is the nature of dark matter? How is it distributed?
\item Are there quantum gravitational effects on photon propagation?
\item Do axion-like particles exist?
\end{itemize}

This chapter introduces the key characteristics of the observatory
and describes
the involvement of the wider scientific community in the derivation of
the scientific requirements for CTA. The broader multi-wavelength and multi-messenger context
is presented in the following chapter and 
the scientific programme proposed for the Key Science Projects, to be carried out
by the CTA Consortium using guaranteed time, is presented in later chapters of this document.

\subsection{Key Characteristics \& Capabilities}
\label{sec:sci_capabilities}

CTA will be an observatory with arrays of IACTs on two sites, aiming to: 
\begin{itemize}
\item improve the sensitivity level of current instruments by an order of magnitude at 1 TeV, 
\item significantly boost detection area, and hence photon rate,
  providing access to the shortest timescale phenomena,
\item 	substantially improve angular resolution and field of view and hence ability to image extended sources, 
\item provide energy coverage for photons from 20~GeV to at
  least 300 TeV, to give CTA reach to high-redshifts and extreme accelerators,
\item dramatically enhance surveying capability, monitoring
  capability, and flexibility of operation, allowing for simultaneous
  observations of objects in multiple fields,
\item serve a wide user community, with provision of data products and
  tools suitable for non-expert users, and
\item provide access to the entire sky, with sites in two hemispheres. (In this document, the two
sites are referred to as CTA-South and CTA-North).
\end{itemize}

CTA will be operated as an open, proposal-driven observatory for the first time
in very-high-energy (VHE, E$>$20\,GeV) astronomy. The observatory-mode operation of CTA
is expected to significantly boost scientific output by
engaging a research community much wider than the historical ground-based
gamma-ray astronomy community.

The very wide energy range covered by the southern CTA array necessitates the use
of at least three different telescope types: referred to as Large,
Medium and Small-Sized Telescopes (LSTs, MSTs and SSTs). The LSTs provide sensitivity at the lowest energies and SSTs at the highest. 
There are multiple
strong motivations for the wide CTA energy range: the lowest energies
provide access to the whole universe (avoiding significant gamma-gamma
absorption on the extragalactic background light); the highest
energies are needed to study the extreme accelerators which we know
from direct cosmic-ray measurements are present in our galaxy; a wide
energy range maximises the chances of serendipitous detection of new
source classes with unknown spectral characteristics, for example in
the search for dark matter with an unknown WIMP mass; a wide energy
range is key for discrimination between scenarios and to identify
features. 
{\em All} objects which have been studied over a wide
energy range with good signal to noise using current IACT arrays exhibit
features in their gamma-ray spectra. 
Conversely, the narrow energy range and
lower signal to noise measurements more typical of current generation
instruments invariably result in spectra which are consistent with
power-law forms.
In the north, where the inner regions of the Galaxy are not visible,
there will be a greater emphasis on extragalactic targets.
Therefore, in the
interest of optimisation of the observatory, the northern CTA array
will be implemented with only LSTs and MSTs. 

Access to the full sky is necessary as many of the phenomena to be studied by CTA
are rare and individual objects can be very important. For example, the most
promising galaxy cluster, the brightest starburst galaxy and the only
known gravitationally-lensed TeV source are located in the north. The inner Galaxy and the
Galactic Centre are key CTA targets (see e.g.
Chapter~\ref{sec:ksp_gc}) and are located in the south. Full sky coverage
ensures that extremely rare but critically important events (for example a Galactic supernova explosion,
bright gravitational wave transient,  or nearby gamma-ray burst)
will be accessible to CTA.

Individual CTA telescopes will have Cherenkov cameras with wide field of view: $>$4.5\degr for the
LSTs, $>$7\degr for the MSTs and $>$8\degr for the SSTs. The wide camera
field serves a dual purpose: to provide contained shower images up to
large impact distance (improving collection area and resolution) for
on-axis gamma rays and to increase the gamma-ray field of view of the system as
a whole.
This characteristic of CTA is critical for the observation of very extended objects and regions of
diffuse emission, as well as for surveys. Furthermore, the
wide field provides reduced systematic errors, with a uniform response
over regions much larger than the point-spread-function size (not always the case for current generation instruments).

The large telescope number ($\sim$100 in the south) 
and individual wide telescope fields of view result in
a CTA collection area which is one or more orders of magnitude larger than
current generation instruments at essentially {\em all} energies, with
substantial benefits for imaging, spectroscopy and light-curve generation.
Multi-square-kilometre collection area is essential at the highest energies where there is essentially zero background even
in long exposures and sensitivity is limited by the collection of sufficient signal
photons. 
For very short timescale phenomena, CTA is
background free over much of its energy range and the large collection
area is the key performance driver.

For events incident in the central parts of the CTA arrays, the number
of recorded shower images will be large ($>$10) for all but the
lowest energies. These high image multiplicities, combined with the
contained nature of events and superior image information to existing
instruments, provide excellent energy and angular
resolution. A precision of 1 arc-minute on individual photons will be
obtained for the upper end of the CTA energy range, the best resolution achieved
anywhere above the X-ray domain.

The ability to rapidly respond to external alerts, and to rapidly issue
its own alerts, is built into the CTA design. In particular the LSTs,
where the energy range covered provides access to essentially the
whole universe, are optimised for rapid movement, with a goal slewing
time of 20~s (minimum requirement 50~s) to anywhere in the observable sky.
A real-time analysis pipeline will enable the identification of
significant gamma-ray activity in any part of the field of view and the issuing of alerts to other instruments within one minute.

\begin{figure}[htbp]
  \centering
\vspace{-1.0cm}
\includegraphics[width=0.95\textwidth]{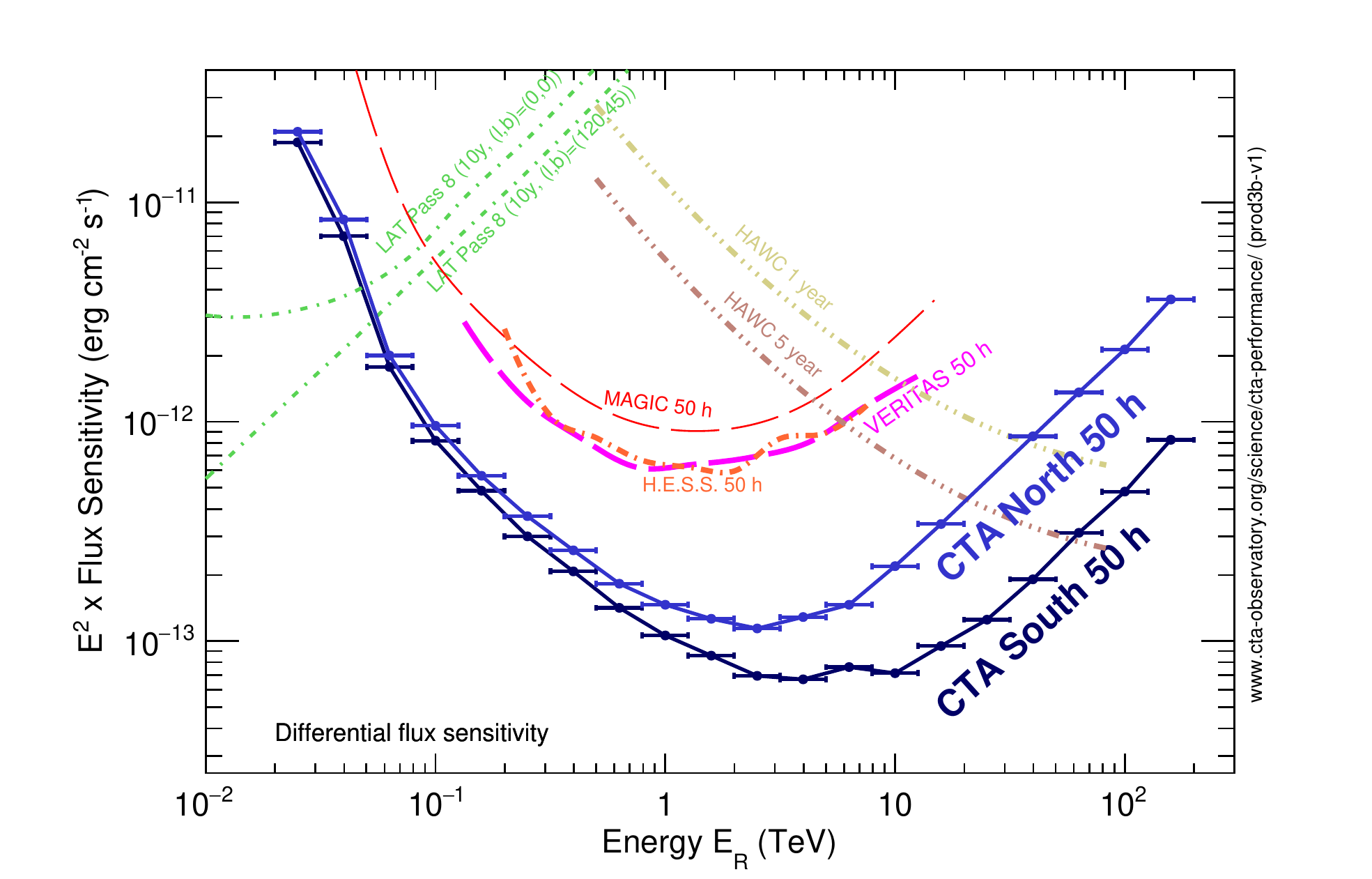}
\includegraphics[width=0.60\textwidth]{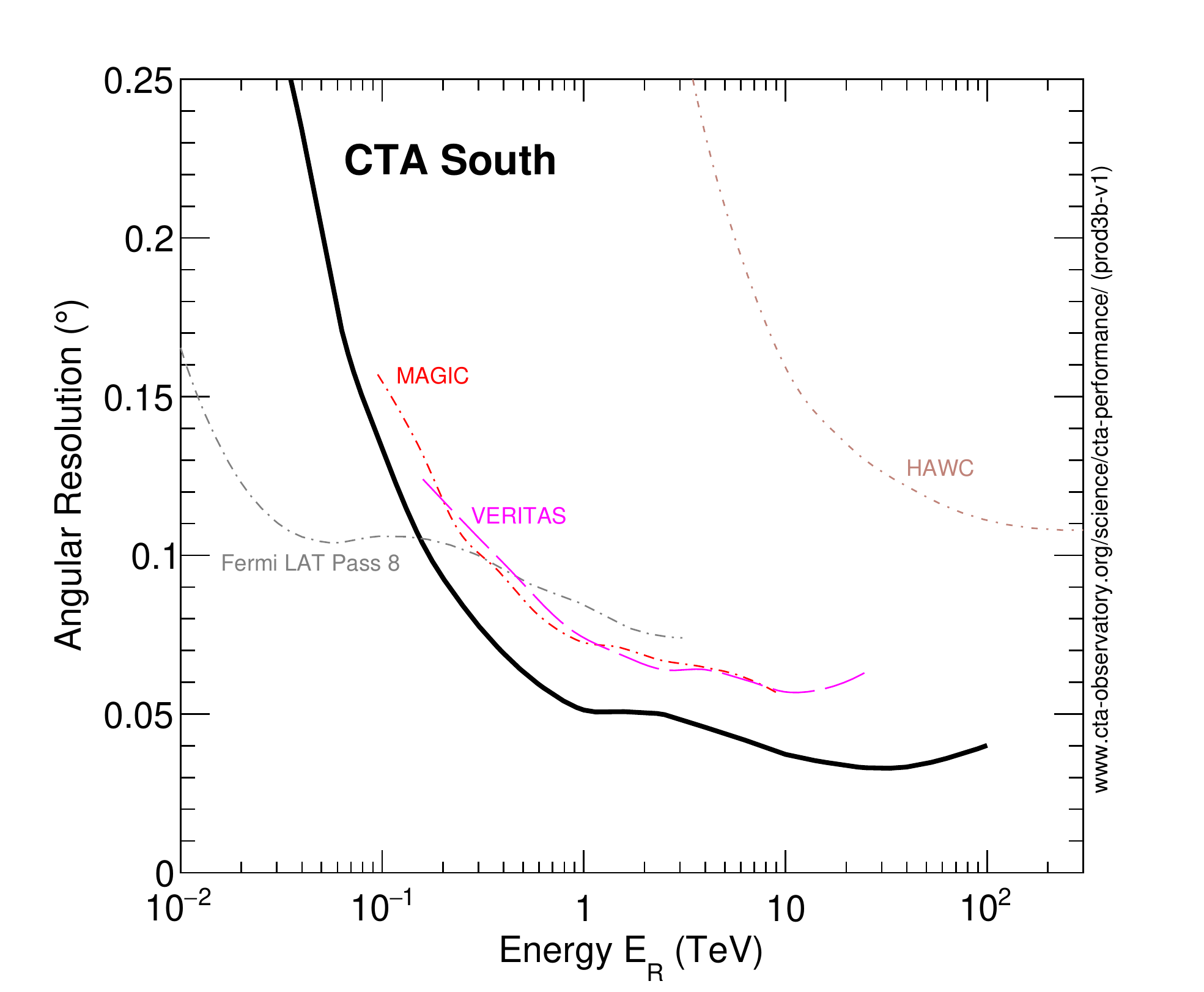}
\caption{Comparisons of the performance of CTA with selected existing gamma-ray instruments. 
Top: differential energy flux sensitivities for CTA (south and north) 
for five standard deviation detections in five independent logarithmic bins per decade in energy. For the CTA sensitivities, additional criteria are applied to require at least ten detected gamma rays per energy bin and a signal/background ratio of at least 1/20. The curves for Fermi-LAT and HAWC are scaled by a factor of 1.2 to account for the different energy binning.
The curves shown give only an indicative comparison of the sensitivity of the different instruments, as the method of calculation and the criteria applied are different. In particular,
the definition of the differential sensitivity for HAWC is rather different due to the lack of energy reconstruction for individual photons in the HAWC analysis. 
Bottom: angular resolution expressed as the 68\% containment radius of reconstructed gamma rays (the resolution for CTA-North is similar). 
The sensitivity and angular resolution curves are based on
the following references:
Fermi-LAT \cite{FERMIperf}, HAWC \cite{Abeysekara2017},
H.E.S.S. \cite{HESSperf}, MAGIC \cite{MAGICperf}, and VERITAS \cite{VERITASperf}.
The CTA curves represent the 
understanding of the performance of CTA at the time of completion of this
document; for the latest CTA performance plots, see 
https://www.cta-observatory.org/science/cta-performance.
}
\label{fig:cta_sens}
\end{figure}

The dramatic improvement in the point-source sensitivity of CTA with respect to
current instruments is a consequence of the combination of improved
background rejection power, increased collection area and improved
angular resolution. The improved background rejection power is achieved primarily through high image multiplicity and is
particularly important for the study of extended, low-surface
brightness objects and for low-flux objects where deep exposures are required. Figure~\ref{fig:cta_sens} compares the sensitivity and angular resolution of the CTA arrays to a selection of existing gamma-ray detectors.

Finally, the number of individual telescopes in the CTA arrays, and
the ability to operate multiple sub-arrays independently, provides
enormous flexibility of operation. CTA will operate with
different pointing directions for different sub-systems, for example
with the LSTs pointed to a distant active galaxy and the MSTs and SSTs observing a hard-spectrum
Galactic source. Furthermore, small groups of MSTs or SSTs may be used
to monitor up to ten variable sources simultaneously. The pointing
pattern of the CTA telescopes may also be optimised for the purpose of
surveying an extended region of arbitrary shape, for example the error box from a gravitational wave alert~\cite{Bartos14,Abbott16a}.
Preliminary CTA performance curves are available publicly at~\cite{PublicPerformanceCurves}.  
Below we highlight two key aspects 
of the unique instrumental reach of CTA.

\subsubsection{Surveying Capabilities}

CTA will conduct a census of particle acceleration in our universe by performing surveys of the sky at unprecedented sensitivity at very high energies.
Deep fields will be obtained for some key regions hosting prominent targets, while wider and shallower surveys will be conducted to build up unbiased population samples and to search for the unexpected.
The combination of the wide CTA field of view with
unprecedented sensitivity ensures that CTA can deliver surveys one to two
orders of magnitude deeper than existing surveys early in the
life of the observatory. Indeed over much of the sky and much of the energy
range of CTA, {\em no} sensitive survey exists and CTA measurements will be revolutionary.
The CTA surveys will open up discovery space in an unbiased way and generate legacy datasets
of long-lasting value.

The potential for surveys with CTA is described
in~\cite{Dubus13}. 
To ensure that essential surveys will be conducted by CTA early on in the observing programme, and to accommodate the related observing time demands, surveys will be part of CTA's Core Programme and are described in the Key Science Project (KSP) chapters following.

\begin{figure}[h!]
\centering
\includegraphics[width=0.99\textwidth]{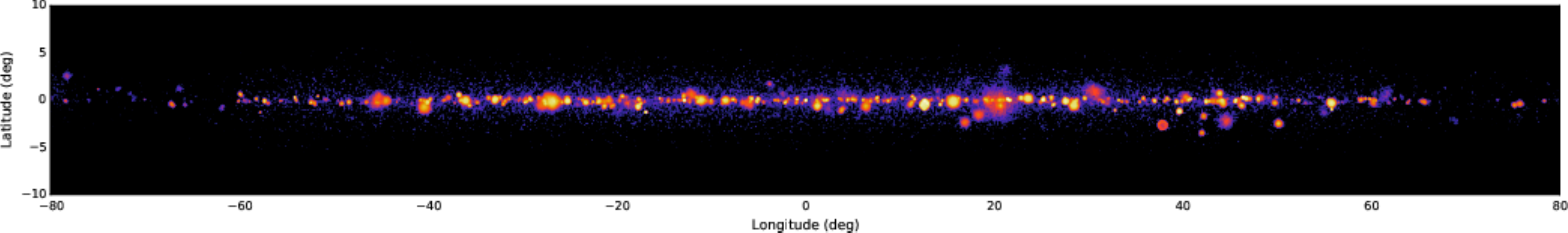}
\includegraphics[width=0.96\textwidth]{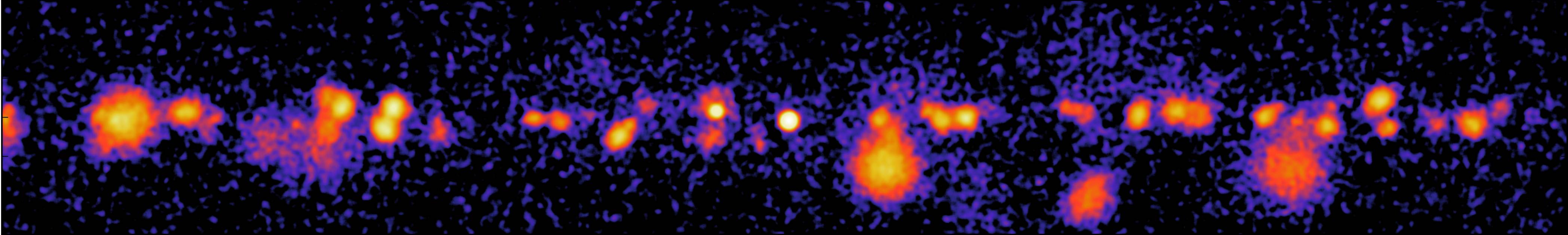}
\caption{Top: simulated CTA image of the Galactic plane for the inner region,
$-80^{\circ} < l < 80^{\circ}$, adopting the proposed GPS KSP observation
strategy and a source model incorporating both supernova remnant and 
pulsar wind nebula populations, as well
as diffuse emission. Bottom: a zoom of an example 20$^{\circ}$ region in Galactic longitude.}
\label{fig:sci_galsurvey}
\end{figure}

\begin{figure}[htbp]
  \begin{centering}
  \includegraphics[trim=0 3 2 2, clip=true, width=3.0in]{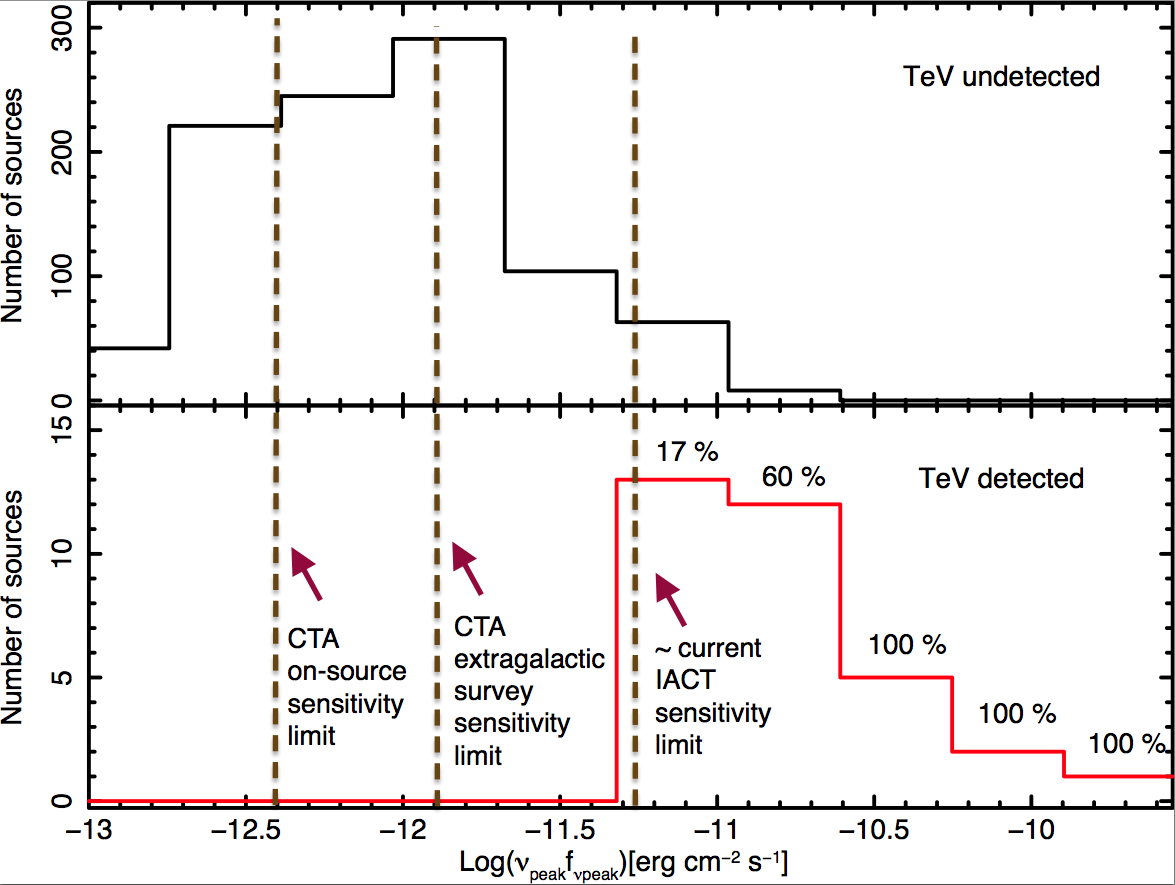}
  \hspace{0.2cm}
 \includegraphics[trim=0 2 1 0, clip=true, width=3.0in]{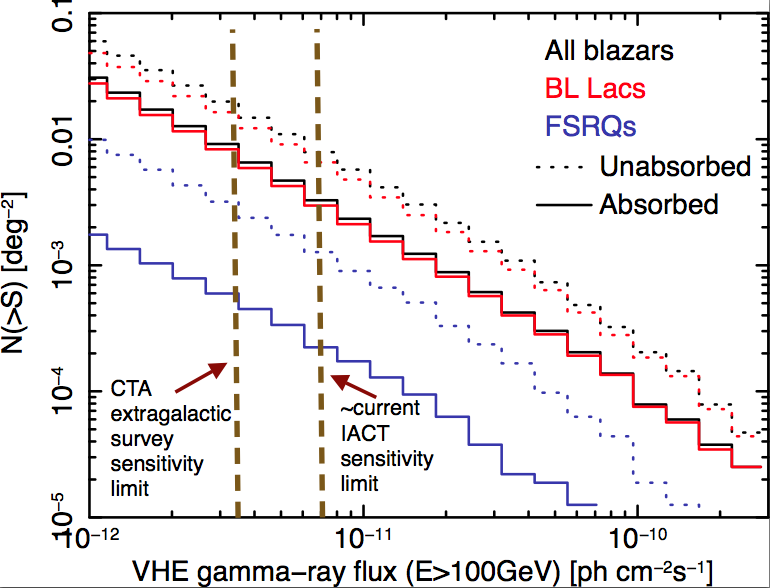}

 \caption{Predictions for the number of blazars on the sky in the GeV--TeV domain.
Left: Source counts versus peak synchrotron flux. The upper panel shows predictions by \cite{Arsioli14} together with the current and envisioned sensitivity limits of IACTs. The lower panel shows detected AGN with current instruments. Right: Expected source counts as a function of the integral gamma-ray flux above 100\,GeV, from \cite{Padovani14}.
}
  \label{fig:sci_egsurvey}
\end{centering}
\end{figure}

\begin{itemize}

\item {\bf Extragalactic Survey} (Chapter~\ref{sec:ksp_eg}) -- covering 1/4 of
  the sky to a depth of $\sim$6~mCrab. No extragalactic survey has ever been
  performed using IACTs, and the existing VHE surveys using ground-level
  particle detectors \cite{Abdo07b, Amenomori05} have 
  modest sensitivity, limited angular and energy resolution, and limited
  energy range. A 1000~hour CTA survey of such a region will reach the same
sensitivity as the decade long H.E.S.S. programme of inner Galaxy
observations and will cover a solid angle $\sim$40 times larger, providing a
snapshot of activity in an unbiased sample of active galactic nuclei (AGN)
(see Figure~\ref{fig:sci_egsurvey}).

\item {\bf Galactic Plane Survey (GPS)} (Chapter~\ref{sec:ksp_gps}) -- consisting of
  a deep survey ($\sim $2~mCrab) of
  the inner galaxy and the Cygnus region, coupled with a somewhat shallower
  survey ($\sim$4~mCrab) of the entire Galactic plane. For the typical luminosity
  of known Milky Way TeV sources of $10^{33\mathrm{-}34}$ erg/s, the
CTA GPS will provide a distance reach of $\sim20$ kpc, detecting
essentially the entire
population of such objects in our galaxy and providing a large sample of objects
one order of magnitude fainter. The excellent angular resolution of
CTA is critical here to avoid being limited by source confusion rather than flux (see Figure~\ref{fig:sci_galsurvey}).

\item {\bf Survey of the Large Magellanic Cloud (LMC)} (Chapter~\ref{sec:ksp_lmc})
  -- providing a face-on view of an entire star-forming galaxy, resolving
  regions down to 20~pc in size with sensitivity down to luminosities of
  $\sim10^{34}$ erg/s. CTA aims to map the diffuse LMC emission as well as
  individual objects, providing information on relativistic particle transport.

\end{itemize}

These surveys will establish the populations of VHE emitters in
Galactic and extragalactic space, providing large enough samples of objects to understand source evolution and/or duty
cycle. Data products from the survey KSPs include catalogues and flux
maps which will serve as valuable long term resources to a wide community.

Some other KSPs are also effectively surveys due to the wide field of view.
For example, a deep observation of the Perseus Cluster is envisaged (see Chapter~\ref{sec:ksp_clust}) 
providing a sample of low redshift
galaxies and with sensitivity to the low end of the luminosity
function of active galaxies as well as to diffuse emission associated
with accelerated hadrons or dark matter annihilation.

The search for an annihilation signature of dark matter, throwing light
on the nature of the dark matter particles, is a key part of the CTA research
programme. The prime targets are the 
Galactic Centre (Chapter~\ref{sec:ksp_gc}) and Milky Way satellite galaxies, but the surveys introduced above will probe concentrations of
dark matter in the LMC and Milky Way, providing complementary datasets. The 
strategy for dark matter detection with CTA is introduced in Chapter~\ref{sec:DM_prog}.

Surveys will in general be conducted in a mode with telescopes
co-pointed, but a {\it divergent mode} is also possible
and
under consideration for the Extragalactic Survey, offering increased
instantaneous field of view ($\sim$ 20\dg $\times$ 20\dg) and
survey depth at the expense of angular and energy resolution.

\subsubsection{Short Timescale Capabilities}

\begin{figure}[h]
\begin{center}
\includegraphics[width= 0.7\textwidth]{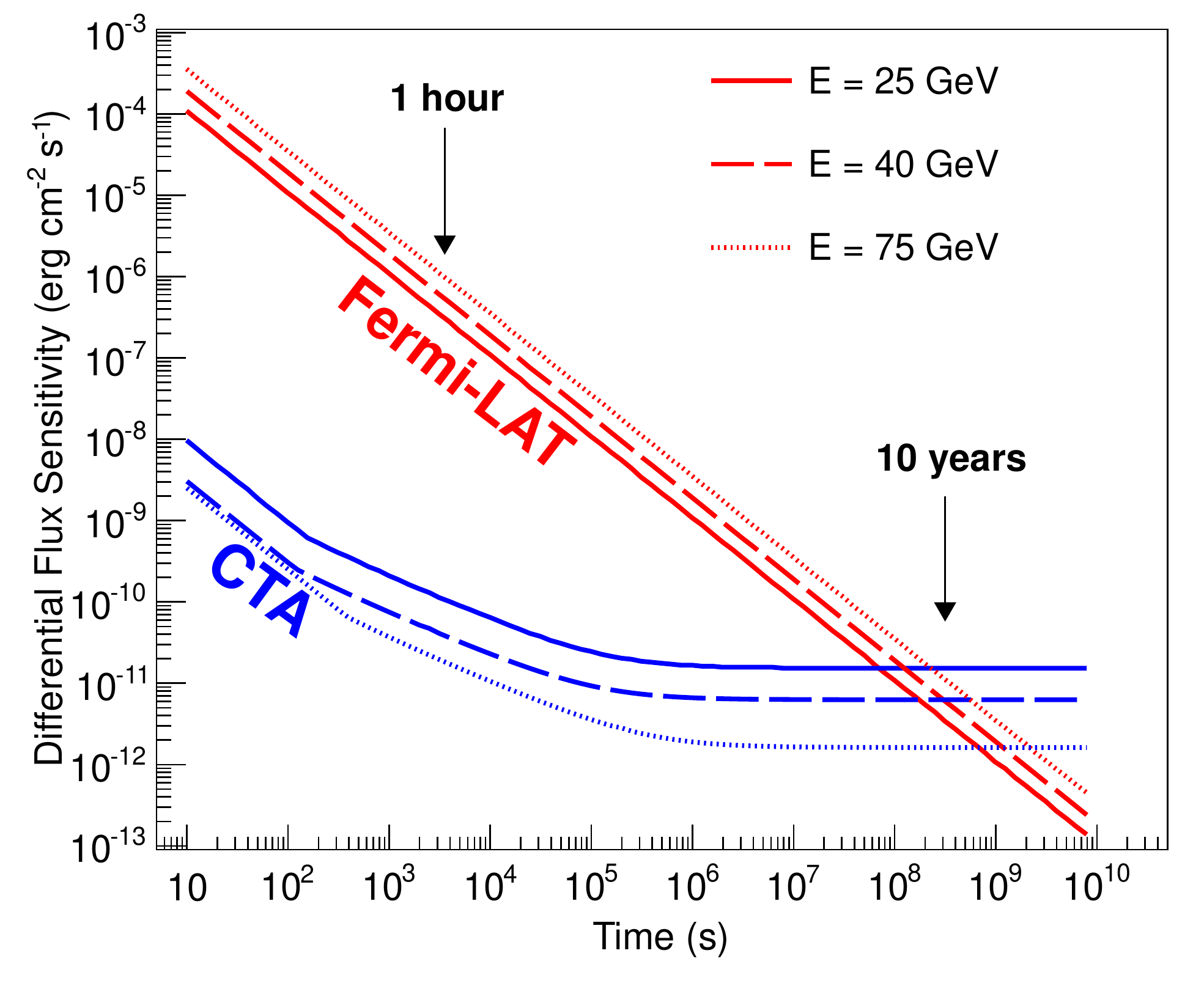}
\caption{Comparison of the sensitivities of CTA and Fermi-LAT in the energy
  range of overlap versus observation timescale. Differential flux
  sensitivities at three energies are compared. Adapted from~\cite{Funk13}.
Note that the Pass 6 sensitivity is shown for Fermi-LAT and the
 CTA sensitivity is calculated using an early model of the arrays;
thus, better sensitivities for both Fermi-LAT and CTA are now expected.}
\label{fig:sci_fermicta}
\end{center}
\end{figure}

CTA is a uniquely powerful instrument for the exploration of the
violent and variable universe. It has unprecedented potential both in
terms of energy reach and sensitivity to short timescale phenomena.
Figure~\ref{fig:sci_fermicta} compares the sensitivity of CTA to that
of Fermi-LAT as a function of observation time. 
CTA has four orders of
magnitude better sensitivity to minute timescale phenomena at energies
around 25 GeV.
Even at variability timescales of 1 month, CTA will
be a factor 100 more sensitive than Fermi-LAT. 
Only for emission
which is stable over the full mission lifetime of Fermi are the
sensitivities of the two instruments comparable in the lowest part of
the CTA energy range. Instruments such as HAWC, which have sensitivity
in the higher part of the CTA range, are also limited at short
timescales by (relative to CTA) small collection areas, as well as low
signal to noise.

\begin{figure}[h]
\begin{center}
\includegraphics[width=0.8\textwidth]{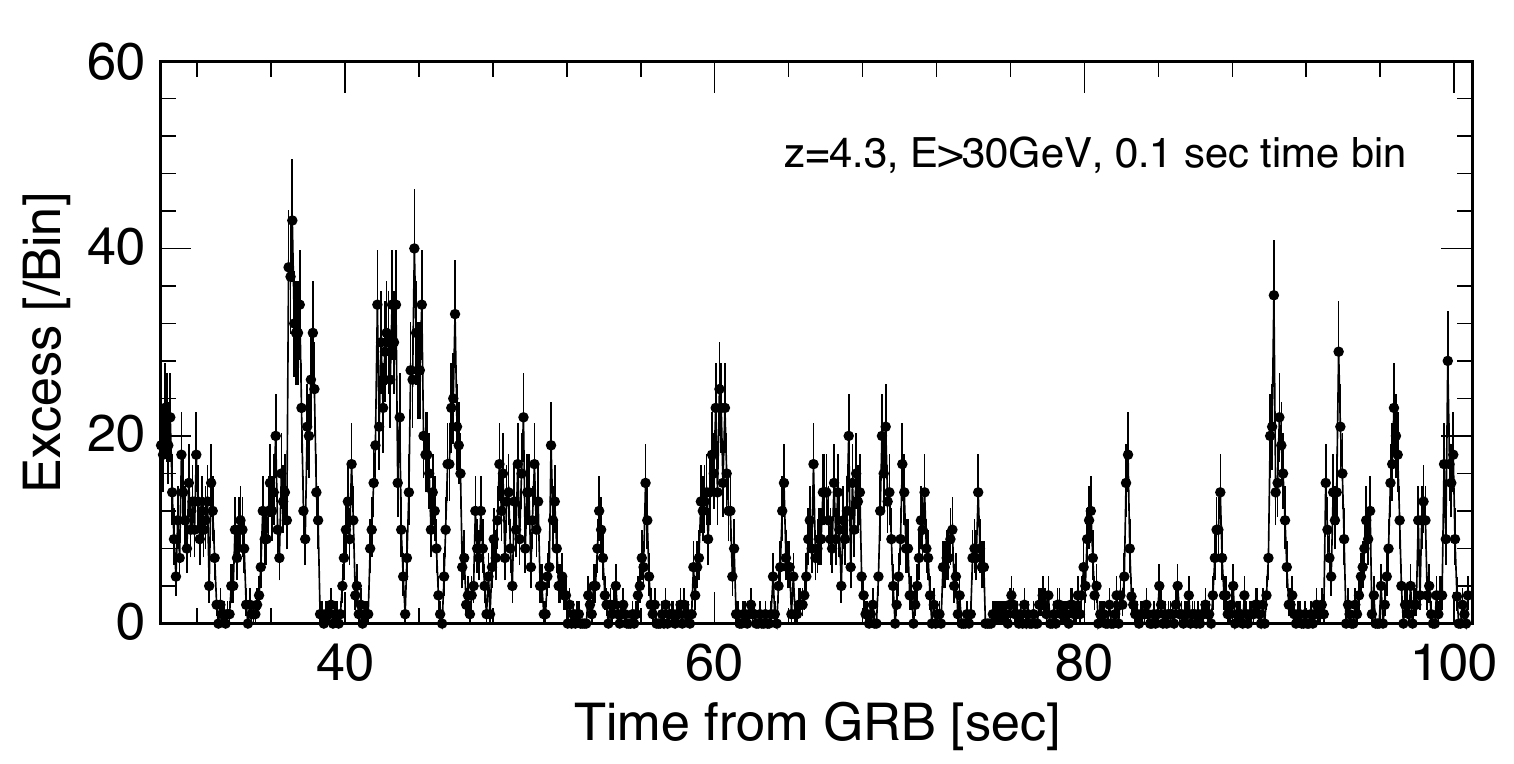}
\caption{Simulated CTA gamma-ray burst light curve, based on the
  Fermi-LAT-detected GRB~080916C at $z=$4.3. See
  Figure~9.1 for more details.}
\label{fig:sci_grblc}
\end{center}
\end{figure}

The ability to probe short timescales at the highest energies will allow CTA to explore the
connection between accretion and ejection phenomena surrounding
compact objects, study phenomena occurring in relativistic outflows, and open
up significant phase space for serendipitous discovery. Particularly
important targets for CTA are gamma-ray bursts (GRBs), 
AGN and Galactic compact object binary systems. The most
dramatic case is that of GRBs where CTA will make high-statistics
measurements for the first time above $\sim$10 GeV, probing new
spectral components which will shed light on the physical processes at
work in these systems~\cite{Inoue13} (see Figure~\ref{fig:sci_grblc}).

CTA's reach to ultra-high energies also provides access to a regime
where cooling times for electrons are extremely short and variability
is expected even for objects which are currently considered as stable
sources (for example the Crab pulsar wind
nebula~\cite{deOna13} and the supernova remnant RX\,J1713$-$3946~\cite{Uchiyama07}).

As well as having unprecedented sensitivity to emission on short
timescales, CTA will be able to respond very rapidly, both to
external alerts and in delivery of alerts to other observatories. The
absolute maximum repointing times for the CTA telescopes (to and from anywhere
in the observable sky) will be 50~s for the LSTs and 90~s for the MSTs and
SSTs, with the goal to reach shorter slewing times. This fast slewing capability is particularly important
for capturing transient phenomena such as GRBs.

The wide field of view and unprecedented sensitivity of CTA make the
serendipitous detection of transient or variable VHE emission
likely. To maximise the scientific return, CTA will be equipped with a low-latency
(effectively real-time) analysis pipeline which will monitor the field
of view for variability on a wide range of timescales. The detection
of a gamma-ray flare in the field and the issuing of an alert will be
possible within 60~seconds. CTA will itself respond to such alerts by
repositioning the telescopes, 
by modifying the observing schedule and by alerting other
observatories to allow rapid follow-up. See ~\cite{Bulgarelli13} for details.

\vspace{0.7cm}

The KSPs which rely on the short-timescale capabilities of
CTA include:

\begin{itemize}

\item {\bf Transients} (Chapter~\ref{sec:ksp_trans}) -- comprising a programme responding
  to a broad range of multi-wavelength and multi-messenger alerts,
  including gamma-ray bursts, gravitational wave transients and high-energy
  neutrino transients. Rapid feedback to the wider community on the
  VHE gamma-ray properties of transients is a key element of the KSP.
\item {\bf Active Galactic Nuclei} (Chapter~\ref{sec:ksp_agn}) -- where flaring
  activity forms a key part of the science case, with rapid
  bi-directional information flow again critical. Blazars exhibit the fastest known variability (1 minute timescales)  at
TeV energies and blazar flares can be used to search for Lorentz invariance violation, as well as cast light
on the physics of the ultra-relativistic inner jets of these systems
(see for example \cite{Begelman08}).
\item {\bf Galactic Plane Survey} (Chapter~\ref{sec:ksp_gps}) -- with multiple
  observations of every part of the Plane allowing the
  identification of objects variable on timescales from seconds to
  months, including the expected discovery of many
  new gamma-ray binaries~\cite{Paredes13}. Real-time alert
  generation from CTA will maximise the scientific return from
  short-timescale transients.
\end{itemize}

\subsubsection{Capabilities Beyond Gamma Rays}

Whilst CTA is designed as a gamma-ray observatory it will, as part of its normal operation, collect an enormous quantity of valuable information on charged cosmic rays. Of particular interest are the highest energy cosmic-ray electrons, which must be associated with nearby particle accelerators (which can therefore be studied using CTA data in both the gamma-ray and electron channels), and heavy nuclei, which can be separated using their direct Cherenkov emissions
(i.e. from the primary cosmic ray, rather than from the secondary products in an air shower). 
Cosmic-ray observations with CTA are discussed in~\cite{Picozza13} and in 
Chapter~\ref{sec:nongamma}.

Both gamma-ray and cosmic-ray observations with CTA rely on nanosecond-timescale cameras to detect Cherenkov light. Other uses for the very large optical-photon collection area of the CTA telescopes do exist. 
Longer integration time observations of {\em optical} targets with CTA could include the use of intensity interferometry, to provide unprecedented angular resolution at blue wavelengths for bright sources (see~\cite{Dravins13} and Chapter~\ref{sec:nongamma}).

\subsection{Overview of CTA Science Themes}
\label{sec:sci_themes}

Here we provide a brief overview of the main scientific questions and
topics addressed by CTA, referring forward where relevant to the KSP chapter
for details.

\subsubsection{Understanding the Origin and Role of Relativistic Cosmic Particles} 

Relativistic particles appear to play a major role in a wide
range of astrophysical systems, from pulsars and supernova remnants
to active galactic nuclei and clusters of galaxies.
Within the
interstellar medium of our own galaxy these cosmic rays are
close to pressure equilibrium with turbulent motions of gas and
magnetic fields, yet the relationship between these three components,
and the overall impact on the star-formation process and the evolution
of galaxies, is very poorly understood. CTA will provide the first
high angular resolution measurements of cosmic-ray protons and nuclei (rather
than the energetically sub-dominant electrons that produce the
non-thermal emission seen at radio and X-ray wavelengths) in
astrophysical systems, providing insights into the process(es) of
acceleration, transport and the cosmic-ray-mode feedback mechanisms in these
systems. Historically, non-thermal effects in astrophysical systems
have largely been ignored or parameterised away due to a lack of high
quality data. The insights from CTA will therefore represent a major
contribution to our deepening understanding of the processes by which
galaxies and clusters of galaxies evolve, in the era of precision
astrophysics with major instruments across all wavebands from 
radio (SKA) to VHE gamma ray (CTA).

Below we introduce the main elements of this theme, moving from the
accelerators themselves to the impact of accelerated particles.

{\bf \it Cosmic Accelerators}

The primary goal of gamma-ray astrophysics thus far has been to
establish in which cosmic sources particle acceleration takes place,
and in particular, to establish the dominant contributors to the
locally measured 'cosmic rays' which are 99\% protons and nuclei
(collectively referred to as 'hadrons'). Huge
progress has been made in the last decade in this area, with the
combination of Fermi-LAT and IACT data proving extremely effective in
identifying the brightest Galactic accelerators and providing strong
evidence of hadron acceleration in a handful of sources. However, key
questions remain unanswered: are supernova remnants (SNR) the only
major contributor to the Galactic cosmic rays? Where in our galaxy are
particles accelerated up to PeV energies? What are the sources of
high-energy cosmic electrons? What are the sources of the ultra-high
energy cosmic rays (UHECRs)?

CTA will address all of these questions and also the critical issue
of the mechanism(s) for particle acceleration at work in cosmic
sources, through two main approaches:

\begin{itemize}
\item {\bf a census of particle accelerators in the universe}, with
  Galactic and Extragalactic surveys and deep observations of key
  nearby galaxies and clusters and
\item {\bf precision measurements of archetypal sources}, where bright
  nearby sources will be targeted to obtain resolved spectroscopy or
  very high statistics light curves, to provide a deeper physical
  understanding of the processes at work in cosmic accelerators.
\end{itemize}

A general census is required to understand the populations of
accelerators and the evolution/lifecycle of these source
classes. Deep observations of individual sources are required to
acquire the very broad band spectra needed to unambiguously separate
lepton and hadron acceleration and to test acceleration to the highest
energies possible for Galactic accelerators.

While the main resources for the census are the KSP survey observations introduced above, the Guest Observer (GO) programme will provide most of the deep observations of individual sources. For example, the deep observation of the TeV-bright pulsar wind nebula (PWN) HESS\,J1825$-$137, mapping in detail its energy-dependent morphology and 
studying particle propagation and cooling in the post-shock flow~\cite{deOna13}, is anticipated as a GO proposal.

\begin{figure}[htbp]
\begin{center}
\includegraphics[width=0.9\textwidth]{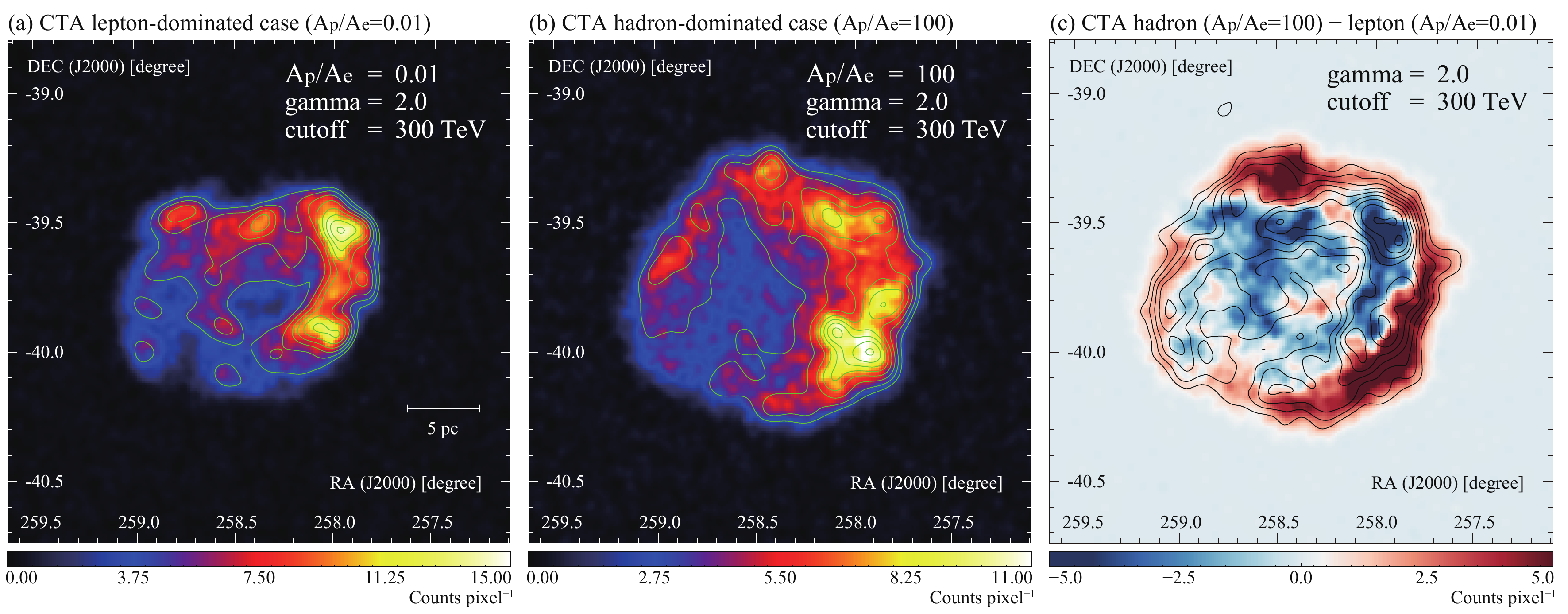}
\caption{Simulated CTA images of the TeV-bright supernova remnant
  RX\,J1713$-$3946 for different emission scenarios, showing the power of CTA to differentiate between these scenarios. Reproduced
  from~\cite{Nakamori14}.
}
\label{fig:sci_1713}
\end{center}
\end{figure}

One key object that is included in the proposed KSPs is the TeV 
bright, young supernova remnant (SNR) RX\,J1713$-$3946 (see
Figure~\ref{fig:sci_1713} and Chapter~\ref{sec:ksp_acc}), where the
dominant gamma-ray emission mechanism is unclear from current
measurements~\cite{Gabici14,Acero13}, but where CTA can resolve the
ambiguity between electron- and proton-dominated emission and resolve sub-structure within the SNR shell on scales that are important for our understanding of the acceleration process. 

Known TeV-emitting source classes where CTA data will have a
transformational impact on our understanding include PWN, gamma-ray binaries, colliding-wind binaries, massive
stellar clusters, starburst galaxies and active galaxies. There is
clearly huge potential for the discovery of new classes of
accelerators, with emission from clusters of galaxies~(see
Chapter~\ref{sec:ksp_clust}) as one of the most exciting
possibilities.

{\bf \it Propagation and Influence of accelerated particles}

Beyond the question of how and where particles are accelerated in the
universe, is the question of what role these particles play in the
evolution of their host objects and how they are transported 
out to large distances. 
On the scale of clusters of galaxies, cosmic
rays with TeV--PeV energies are thought to be confined over a Hubble
time~\cite{Volk96}. On smaller scales they typically escape from their
acceleration sites and may impact upon their environments in a number of ways:

\begin{enumerate}
\item as a dynamical constituent of the medium,
\item through generation / amplification of magnetic fields, and
\item through ionization and subsequent impact on the chemical
  evolution of, for example, dense cloud cores.
\end{enumerate}

All these effects are relevant for the interstellar medium of our own
Galaxy and are likely to be important in star-forming systems on all
scales~(see e.g. Chapters~\ref{sec:ksp_sfs} and \ref{sec:ksp_lmc}). The first aspect is also likely to be
important for the process of AGN feedback on the host galaxy cluster
and growth of massive galaxies (see e.g. Chapter~\ref{sec:ksp_agn} and~\ref{sec:ksp_clust}).

CTA will map extended emission around many gamma-ray sources and look
for energy dependent morphology associated with diffusion (in the case
of hadrons) or cooling (in the case of electrons). As the
energy dependence is expected to be opposite in the two cases, such
mapping provides another means to separate emission from these two populations.
It is CTA's unprecedented (in the gamma-ray domain) angular resolution,
energy resolution and background rejection power that make this
possible~(see e.g. Chapter~\ref{sec:ksp_acc}).

There are important synergies here with THz instruments capable of
mapping molecular material and deriving the physical conditions in the
TeV-emitting regions~(see Chapter~\ref{sec:sci_synergies}).

Again, many of the key targets for this topic are left for GO proposals, with the expectation that teams with access to key data sets at other wavelengths will propose CTA observations. For example, although there are a number of galaxy clusters and AGN with cluster-scale impact that are potentially detectable with CTA, the KSP on galaxy clusters targets only a single object (Perseus, see Chapter~\ref{sec:ksp_clust}). 

\subsubsection{Probing Extreme Environments}

Particle acceleration to very high energies is typically associated
with extreme environments, such as those close to neutron
stars and black holes, or in relativistic outflows or explosions. VHE
emission from accelerated particles can therefore act as a probe of
these environments, providing access to time and distance scales which
are inaccessible in other wavebands. VHE emission often escapes from systems
where UV and X-ray emission is absorbed, and it provides information
independent of assumptions on magnetic field strengths.  In addition,
VHE photons from distant objects can be used as a probe of the
intervening space. Gamma-gamma pair production signatures will allow us to
measure the redshift evolution of the UV-IR background, and hence the
star-formation history of the universe, to probe magnetic fields in
cosmic voids down to values many orders of magnitude below the reach
of any other technique. CTA will also establish if VHE photons heat the
gas in these under-dense regions, suppressing the formation of dwarf
satellite galaxies.

Below we consider three key areas within this theme where CTA data will
have a transformational impact, referring forward
to the Key Science Projects that will address them and to potential major 
GO observations.

{\bf \it Black holes and jets }

\begin{figure}[htbp]
\begin{center}
\includegraphics[width= 0.85\textwidth]{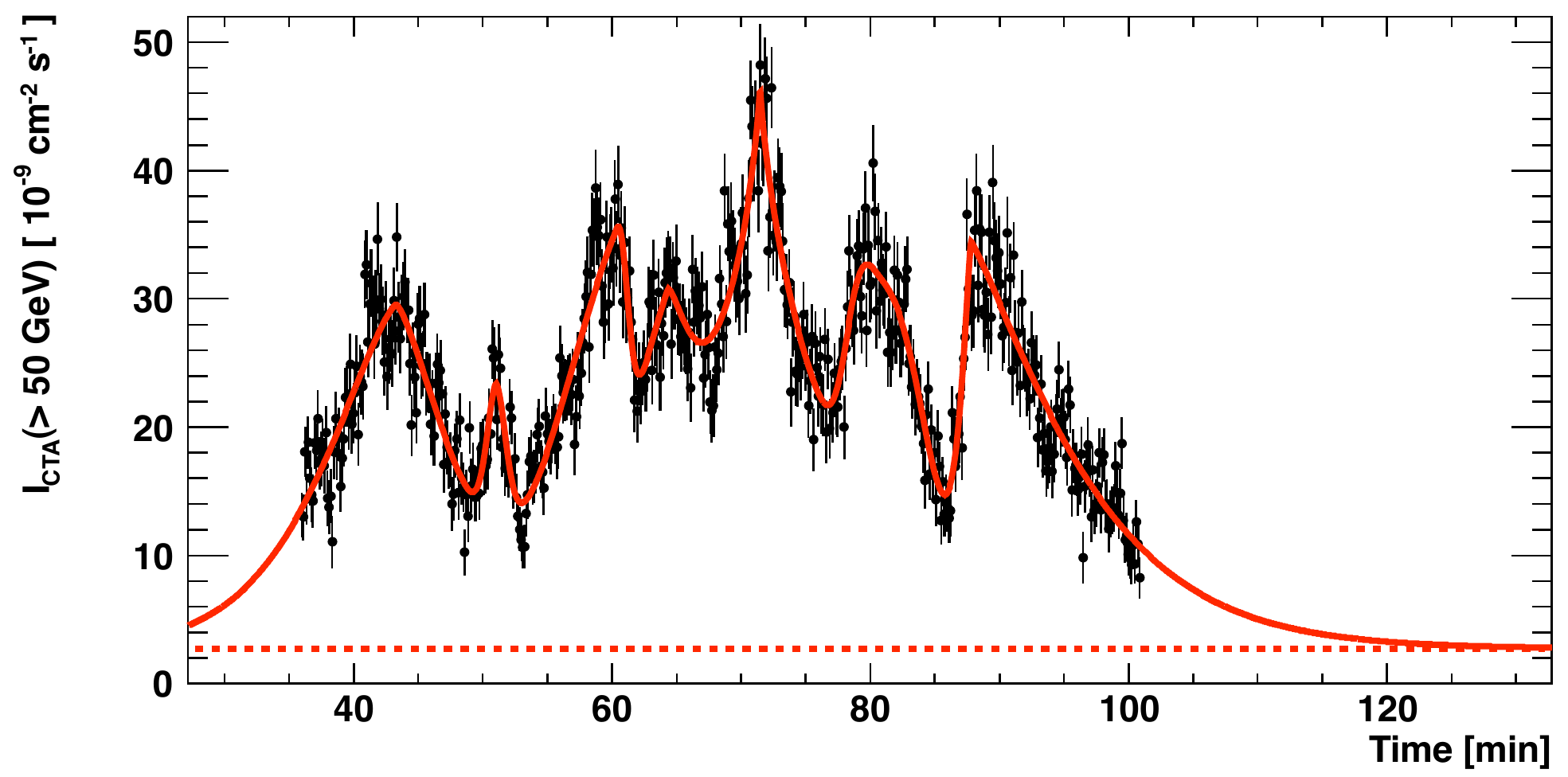}
\caption{Probing ultra-fast variability in the inner jet of an active
  galaxy: simulated CTA light curve for the 2006 flare of
  PKS\,2155$-$304~(reproduced from \cite{Sol13}). Such observations provide access to timescales much shorter than the light-crossing time of the supermassive black hole.}
\label{fig:sci_2155lc}
\end{center}
\end{figure}

Active galactic nuclei are thought to harbour supermassive black holes
(SMBHs), accreting material and producing collimated relativistic
outflows by a still poorly-understood process. Similarly, accreting stellar mass
black holes are known to produce jets, and particle acceleration seems
to be universally associated with BH-powered jets. Acceleration may
occur extremely close to the SMBH (and must do so in some systems to explain the 
remarkably short variability timescales, see Figure~\ref{fig:sci_2155lc}) or 
up to Mpc scales, where the largest AGN jets finally terminate.
Active galaxies are seen as one of the most likely sites of the
acceleration of the UHECRs, with energies up to around $10^{20}$ eV,
but so far there is no strong evidence for hadronic acceleration in AGN jets.

Simultaneous broad-band data (see Chapter~\ref{sec:sci_synergies}) is
needed to study variable jet emission in both Galactic and extragalactic
systems, with CTA data playing a key role: establishing the presence
of very high energy particles, identifying the presence of hadrons and
studying extremely short-timescale variability that provides
information on the smallest spatial scales and probes the bulk ultra-relativistic
motions of the inner jet. Time-resolved VHE spectral measurements are key to disentangling
leptonic and hadronic emission scenarios, to study jet power and
dynamics and to probe magnetic fields in this extreme environment. Whilst a significant sample of AGN will be targeted in the Active Galactic Nuclei KSP~(see Chapter~\ref{sec:ksp_agn}) and discovered in the CTA survey KSPs, it is anticipated that the majority of targeted AGN observations with CTA will occur in the open time, with many as target-of-opportunity (ToO) proposals associated with triggers from other wavelengths.

The low luminosity AGN at the heart of our own galaxy,  Sgr A$^{\star}$, is
coincident with a TeV source and has associated non-thermal emission
in the radio and X-ray bands. However, the sensitivity, resolution and
pointing precision of current gamma-ray telescopes are insufficient
to separate the emission from very nearby sources from the diffuse emission
around the Galactic Centre. CTA will map this region in
unprecedented detail (see Chapter~\ref{sec:ksp_gc}), probing the
relationship between the central source and the diffuse emission and,
on much larger scales, up to the Fermi bubbles.

There is evidence from current IACTs for TeV emission from a single
system hosting a stellar mass black hole: Cygnus X-1. The sensitivity
of CTA should allow this object to be studied in detail, 
opening the door to studies of high-energy non-thermal processes
associated with stellar mass black holes and allowing the first
comparisons to be made with SMBH systems. Whilst the Core Programme will cover the Cygnus region as part of the GPS and black-hole binary activity will be used to trigger the Transient KSP~(see
Chapters~\ref{sec:ksp_trans} and \ref{sec:ksp_gps}), the bulk of observations of such objects, including deep observations of Cygnus X-1, are expected throughout the GO programme, again likely from proposers with 
access to data from other wavelength bands.

The recent discovery of gravitational wave (GW) emission associated with the mergers of 
massive black holes~\cite{Abbott16a, Abbott16b, Abbott16aa} raises many exciting new possibilities in observational astrophysics and many questions about the evolution of high mass binary systems~\cite{Abbott16b,Abbott16e}. Clearly the possibility for jet formation and acceleration of particles to TeV energies in such systems exists, and as GW detections will, for the foreseeable future, all originate within the TeV gamma-ray horizon, such alerts will form a key target for CTA (see
Chapter~\ref{sec:ksp_trans}).

{\bf \it Neutron stars and relativistic outflows}

CTA will probe the environment around neutron stars via pulsed
gamma-ray emission from the magnetosphere of pulsars and study the
ultra-relativistic outflows of these systems via mapping
and spectral measurements of the associated synchrotron/inverse-Compton nebula and
(possibly and uniquely) the unshocked pulsar wind. Young and energetic
(in terms of available rotational energy) pulsars cluster tightly
along the Galactic plane and hence the majority of objects will be
covered by the CTA Galactic Plane Survey~(see Chapter~\ref{sec:ksp_gps}). Two key objects:
HESS\,J1825$-$137 and the Vela pulsar (and associated Vela\,X
nebula),
are very promising targets 
and are expected to be targeted by GO proposals.

Binary systems including a pulsar provide a unique opportunity to study a
relativistic outflow under changing physical conditions as the orbit
progresses, via energy-dependent light-curve measurements. Several
such systems will be covered by the CTA Galactic Plane Survey, but deep observations are again expected as GO proposals.

Merging neutron stars and other compact object mergers are the likely
counterparts of short GRBs,
and are of course the targets of the young field of gravitational wave astronomy. CTA will be able to
respond rapidly to triggers from GRB or GW instruments, and hence probe
the highest energy processes associated with such events.

{\bf \it Cosmic voids}

Much of the universe consists of extremely under-dense regions known
as cosmic voids. Very high energy photons interact within these voids 
and allow us to probe the radiation fields and magnetic fields that
they contain.
The Extragalactic Background Light (EBL) is the integrated emission
from stars and galaxies of all types throughout the evolution of the
universe. As such, it is an important tool for cosmology but it is
extremely difficult to measure directly, due to very strong
foregrounds from the Solar System and the Milky Way. However, the EBL leaves an imprint on the measured spectra of gamma-ray sources, via the
process of gamma-gamma pair production. The wide-band, high-quality
spectra measured with CTA for a large number of objects will allow
the EBL spectrum from the optical to the far infrared to be precisely measured at
redshift zero. Furthermore, with the expected large sample of blazars
up to redshift $\sim$1 detected with CTA, the evolution of the EBL
with cosmic time can be probed for the first time. See~\cite{Mazin13} and Chapter~\ref{sec:ksp_agn} for more details. 

Pair production by TeV photons interacting in voids also offers the prospect
of measuring the extremely weak magnetic fields thought to exist in
these regions. Secondary gamma rays are produced by the primary
$e^\pm$ pairs via inverse-Compton scattering on the EBL. A cascade can
then develop from further pair and inverse-Compton interactions. Depending on the
typical value of the intergalactic magnetic field (IGMF), deflections
of the secondary particles may either be small enough that secondary
components may be observable as {\em pair echoes}, which arrive with a
time delay relative to the primary emission, or as a {\em pair halo},
potentially resolvable extended emission around the primary
source. The properties of the extended emission depend on the IGMF
strength. A strong enough IGMF ($> 10^{-12}$ G) leads to full
isotropisation of the cascade emission and formation of a physical
pair halo, while weaker magnetic field leads to the appearance of an
extended emission with an IGMF-dependent size. If the IGMF strength is
in the range, $B \sim 10^{-16} - 10^{-12}$~G, the spatially-extended
emission may be detectable and resolvable by CTA by virtue of its high
sensitivity and angular resolution; e.g., for a source at a distance
of 100~Mpc, the extended emission would be on the $\sim 1^{\circ}$
scale and would be comfortably contained within the CTA
field of view. See~\cite{Sol13} and Chapter~\ref{sec:ksp_agn} for details.

If, as has been recently suggested, TeV electrons produced in
gamma-gamma interactions in the voids do not initiate cascades but
rather {\em heat} the ultra-low dense plasma~\cite{Broderick12}, CTA will allow this
hypothesis to be proven and the heating rate to be very well
constrained. As such heating could be the dominant means of heating in
low density regions after redshift $\sim$2, and could solve the problem of the missing
dwarf satellite galaxies, this measurement would be a very valuable
addition to cosmology.

\subsubsection{Exploring Frontiers in Physics} 

The reach of CTA encompasses considerable discovery space in the area of fundamental physics. CTA will reach the expected thermal relic cross-section for self-annihilating dark matter for a wide range of dark matter masses, including those inaccessible to the Large Hadron Collider (LHC).
The long travel times of gamma rays from extragalactic sources
combined with their short wavelength make them a sensitive probe for
energy-dependent variation of the speed of light due to
quantum-gravity induced fluctuations of the metric. CTA will be
sensitive to such effects on their expected characteristic scale: the
Planck scale. On their long journey, gamma rays may couple to other 
light particles such as axion-like particles (ALPs), 
under the influence of intergalactic magnetic fields. Such photon-ALP oscillations effectively make the universe more transparent to gamma rays and, 
akin to neutrino oscillations, introduce a spectral modulation. 
Each of these effects would represent a very major discovery, alone
worth the effort of constructing and operating CTA. The major step in
sensitivity and energy coverage that CTA represents brings such
effects within reach and could well  allow further issues in
fundamental physics to be addressed. 

Below we briefly consider each of these fundamental physics probes in
turn, referring forward to the Key Science Projects which will address
them in detail.

{\bf \it Dark Matter}

A major open question for modern physics is the nature of dark matter. 
On scales from kpc to Mpc, there are numerous lines of evidence
for the presence of an unknown form of gravitating matter that cannot
be accounted for by the Standard Model of particle physics. The
observation of the acoustic oscillations imprinted into the cosmic
microwave background quantifies this dark component as making up about
27\% of the total universe energy budget. Being dominant with respect
to the baryonic component which accounts for only about 5\% of the
total energy density, dark matter shaped the growth of cosmic structures
through gravitational instability. By comparing the observed
distribution in large galaxy redshift surveys with computer
simulations of structure formation, it emerges that the particles
constituting the dark matter had to be moving non-relativistically when they
stopped scattering in the early universe: hence the term cold dark matter
(CDM). The observational evidence has led to the establishment of a
concordance cosmological model, dubbed $\Lambda$CDM. Despite the fact
that the standard cosmology rests on the dark matter paradigm, we still have no
clue as to its particle nature. One of the most popular
scenarios for CDM is that of weakly interacting massive
particles (WIMPs) that comprise a large class of non-baryonic
candidates with a mass typically between few tens of GeV and a few TeV
and an annihilation cross-section of the order of the weak
interaction. Natural WIMP candidates are found in many models, for example, in 
supersymmetric extensions of the Standard Model and indeed in any model
incorporating the breaking of electroweak symmetry by the Higgs mechanism. 
The success of the WIMP scenario relies on the
observation that weak scale masses (GeV--TeV) and couplings yield a
self-annihilation cross-section that generically implies their relic
abundance to be close to the currently observed value for dark matter. This
(velocity weighted) cross-section ($\sim 3 \times 10^{-26}$ cm$^3$ s$^{-1}$) is therefore
a benchmark value that CTA aims to reach through searches for the
gamma rays arising from annihilation of dark matter in the Galaxy, as has been
attempted already by all operating IACTs.
Note that CTA is also sensitive to particle physics models in which gamma rays result
from the decay of dark matter particles.

Obtaining convincing evidence for dark matter from excesses in the measured
energy spectrum of gamma rays needs careful assessment of (uncertain)
astrophysical backgrounds as well as a good understanding of the
galactic dynamics of dark matter. There is a major complementary effort at
the LHC in attempting to create dark matter directly in the laboratory or
in detecting its virtual traces on Standard Model signals. 
Some underground direct-detection
experiments that measure the recoil energy of nuclei in a
well-shielded detector when hit by a passing dark matter particle 
have reported events in their signal region (e.g. CDMS-Si, CRESST, and
EDELWEISS-II) although these are all consistent with being residual
background and are in tension with stronger limits placed by other
experiments (e.g. XENON100, LUX, and CDMS II-Ge). Moreover, controversial
evidence has been presented of an annual modulation signal (due to our
motion around the Sun) of dark matter with mass around 10~GeV (presented
originally by DAMA/LIBRA but, more recently, also by
CoGeNT). Concerning indirect detection, CTA will have a much greater potential
for dark matter detection than the current generation of 
VHE telescopes, for a number or reasons:
1) CTA's
extended energy range will allow searches for WIMPs with lower mass,
2) the improved sensitivity in the entire energy range will improve the
probability of detection of dark matter,
3) the increased field of view with a
homogeneous sensitivity as well as the improved angular resolution
will allow for more efficient searches for extended sources and
spatial anisotropies, and 
4) the improved energy resolution will
increase the chances of detecting a possible spectral feature in the a
dark matter induced photon spectrum.

\begin{figure}[htbp]
\begin{center}
\includegraphics[width= 0.8\textwidth]{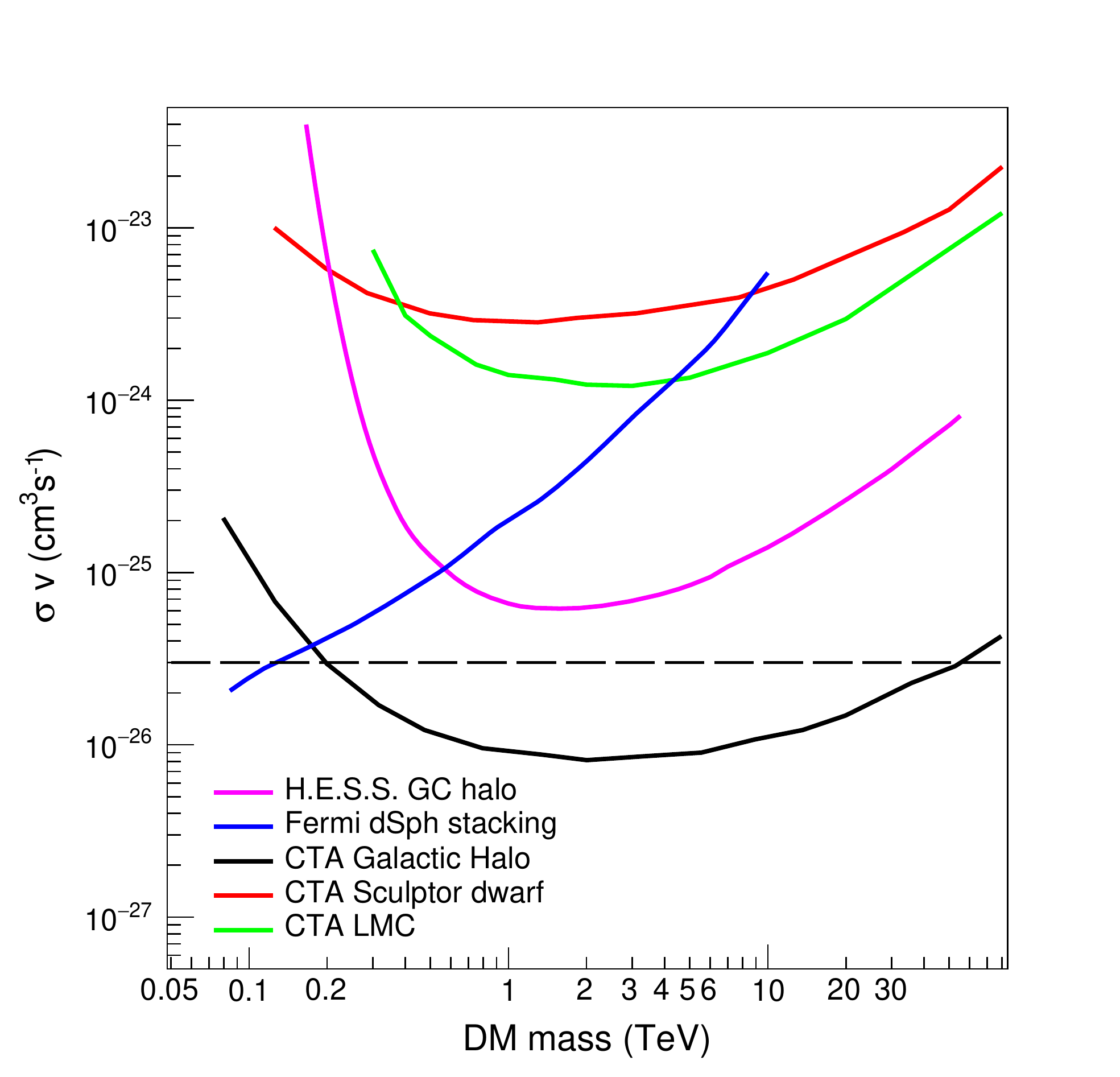}
\caption{CTA sensitivity to a WIMP annihilation signature as a
  function of WIMP mass, for nominal parameters and for the multiple
  CTA observations described in Chapter~\ref{sec:DM_prog}. The dashed
  horizontal line indicates the likely cross-section for a WIMP which
  is a thermal relic of the Big Bang. See Figure~4.1 for a discussion of the
  various sensitivities.}
\label{fig:sci_dmsens}
\end{center}
\end{figure}

By observing the region around the Galactic Centre and by adopting
dedicated observational strategies~(see Chapter~\ref{sec:ksp_gc} and
Figure~\ref{fig:sci_dmsens}), CTA will indeed reach the canonical
velocity-averaged annihilation cross-section of $\sim 3 \times
10^{-26}$ cm$^3$ s$^{-1}$ for a dark matter mass in the range $\sim$200~GeV to 20~TeV ---something
which is not possible with current instruments for any exposure
time. Together with the constraints from Fermi-LAT on dark matter lighter than
a few hundred GeV, this will seriously constrain the WIMP paradigm for
CDM in the case of non-detection. Models with a large photon yield from dark matter
annihilation will be constrained to even smaller cross-sections. In
conclusion, the WIMP paradigm, either through detection or
non-detection will be significantly impacted upon during the first
years of operation of CTA. Additional targets, including Milky Way
satellites~(see Chapters~\ref{sec:DM_prog} and ~\ref{sec:ksp_lmc}) complement the primary
GC observation, with considerable scope for Guest Observer observations.

If signatures of dark matter do appear in direct-detection experiments or at the LHC, gamma-ray observations will provide a complementary approach to identify dark matter, while the typical cutoff of the energy spectrum will allow for a precise mass determination. If such experiments do not detect dark matter, as may be the case for sufficiently heavy dark matter candidates, CTA may be the only way to look for such particles over the next decade.

{\bf \it Quantum Gravity and Axion-like Particle Search}

Photons of energy extending up to hundreds of TeV which will 
be detected by CTA from distant cosmic sources provide a powerful 
tool to search for possible new physics beyond the Standard Model. 
Apart from the search for annihilation/decay signals from dark matter, 
there is the exciting 
possibility of detecting axion-like particles (ALPs) and finding evidence 
of Lorentz invariance violation (LIV) associated with possible quantum 
gravity effects on space-time at the Planck scale. 
Blazars~(Chapter~\ref{sec:ksp_agn}) and gamma-ray bursts 
(Chapter~\ref{sec:ksp_trans}) have been identified as 
the most promising (bright and distant) target classes for both these
searches. It has been suggested that quantum gravity effects may induce time delays
between photons with different energies travelling over large
distances, corresponding to a non-trivial refractive index of the vacuum. 
High statistics measurements of GRBs and blazars over a wide energy range  
(see e.g. Figures~\ref{fig:sci_2155lc} and~\ref{fig:sci_grblc}) will allow 
CTA to probe this possibility 
significantly better than is possible with current IACTs. 
Even a negative result is rather important in this context to guide theoretical work.

Axions are a proposed solution to the strong-CP problem of quantum chromodynamics  
and also well motivated candidates to constitute a part or all of CDM.
ALPS would not have the correct properties (i.e. mass and coupling) to explain
the strong-CP problem, but they could potentially 
be an important component of the dark matter.
ALPS are expected to convert into photons (and vice versa)
as they traverse cosmic magnetic fields. In the case of a very
distant AGN, the ALP/photon coupling can result in a detectable
enhancement of the TeV photon flux (which competes with the 
absorption on the EBL), dependent on the ALP mass. The search for ALPs by CTA will
complement dedicated laboratory experiments and studies using indirect astrophysical tests and X-ray telescopes. See Chapter~\ref{sec:agn_fund_phys} for more details.

\subsection{Community Input to the Science Case}

The scientific motivations and requirements for CTA have been
developed by the CTA Consortium with the engagement of the much wider
community of scientists working in astrophysics and astroparticle physics. 
A dedicated work package (LINK) of the FP7-funded Preparatory Phase of CTA 
was created to provide this community engagement. A series of
workshops provided the main basis for reaching out to the broader
community, gathering input to refine the science goals, to
perform the scientific optimisation of the instrument, and to develop 
user interfaces and services to provide the best possible 
scientific exploitation of the observatory.

Three FP7-supported LINK workshops took place:

\begin{itemize}

\item $1^{\mathrm{st}}$:\ {\it Probing Physics Beyond the  Standard Model with CTA},
Oxford, UK, November 11-12, 2010 

\item $2^{\mathrm{nd}}$:\ {\it Links between CTA science and
 cosmic ray physics at high energies},
Buenos Aires, Argentina, November 19-21, 2012 

\item $3^{\mathrm{rd}}$:\ {\it X-raying the Gamma-ray Universe},
Hakone, Japan, November 4-6, 2013

\end{itemize}

Additional symposia and  workshops included:

\begin{itemize}

\item  {\it AGN physics in the CTA era}, Toulouse, France, 16-17 May, 2011
  
\item {\it The Highest-Energy Gamma-ray Universe},
A Joint Discussion Session of the International Astronomical Union
General Assembly, Beijing, China, August 20-21, 2012

\item {\it Extragalactic Gamma-ray astronomy with CTA},
Muonio, Finland, March 18-21, 2013

\item {\it The Gamma-ray sky in the era of Fermi \& CTA},
Symposium at the European Week of  Astronomy \& Space Science (EWASS),
Turku, Finland, July 11-12, 2013

\end{itemize}

These meetings were all targeted at different communities, where links
to CTA were apparent, but needed to be developed. In addition, a
number of national-level meetings took place, exploring links to
existing strong theoretical and experimental activities in individual
CTA member countries.

As part of these activities, a special issue of the journal
Astroparticle Physics was also produced:
\emph{Seeing the High-Energy Universe with the Cherenkov Telescope
  Array} \cite{Hinton13}, 
dedicated to CTA and with seven articles by prominent scientists outside
of the CTA Consortium, as well as detailed studies and highlight
articles written by the Consortium science teams.

\section{Synergies}
\label{sec:sci_synergies}

CTA will have important synergies with many of the new generation of
astronomical and astroparticle observatories.  As the flagship
VHE gamma-ray observatory for the coming decades, CTA plays a similar
role in the VHE waveband as the SKA in radio, ALMA at millimetre, or
E-ELT/TMT/GMT in the optical wavebands, 
providing excellent sensitivity and resolution compared to
prior facilities.  At the same time, the scientific output of CTA 
will be enhanced by the additional capabilities provided by
these instruments (and vice-versa).  Multi-wavelength (MWL) and
multi-messenger (MM) studies using CTA provide added value to the
science cases in two main ways:

\begin{description}
\item[Non-thermal emission] To understand the origin of cosmic rays
  and the extreme physical environments that produce them, it is
  necessary to study non-thermal signatures that span many orders
  of magnitude in frequency in the broad-band 
  spectral energy distribution (SED) of a given object.  In the case of
  time-variable emission, such studies require simultaneous observations
  and/or alerts and triggers between observatories.

\item[Source properties] Information on the nature of gamma-ray
  emitting sources can be provided by MWL observations,
  enabling, for example, the object class, environmental conditions or
  the distance to be established.  For this purpose, simultaneous 
  observations are in
  general not required, except for the need to characterise transient
  sources, for example in the case of gamma-ray burst redshift measurements.
\end{description}

\begin{figure}[htbp]
\begin{center}
\includegraphics[width= \textwidth]{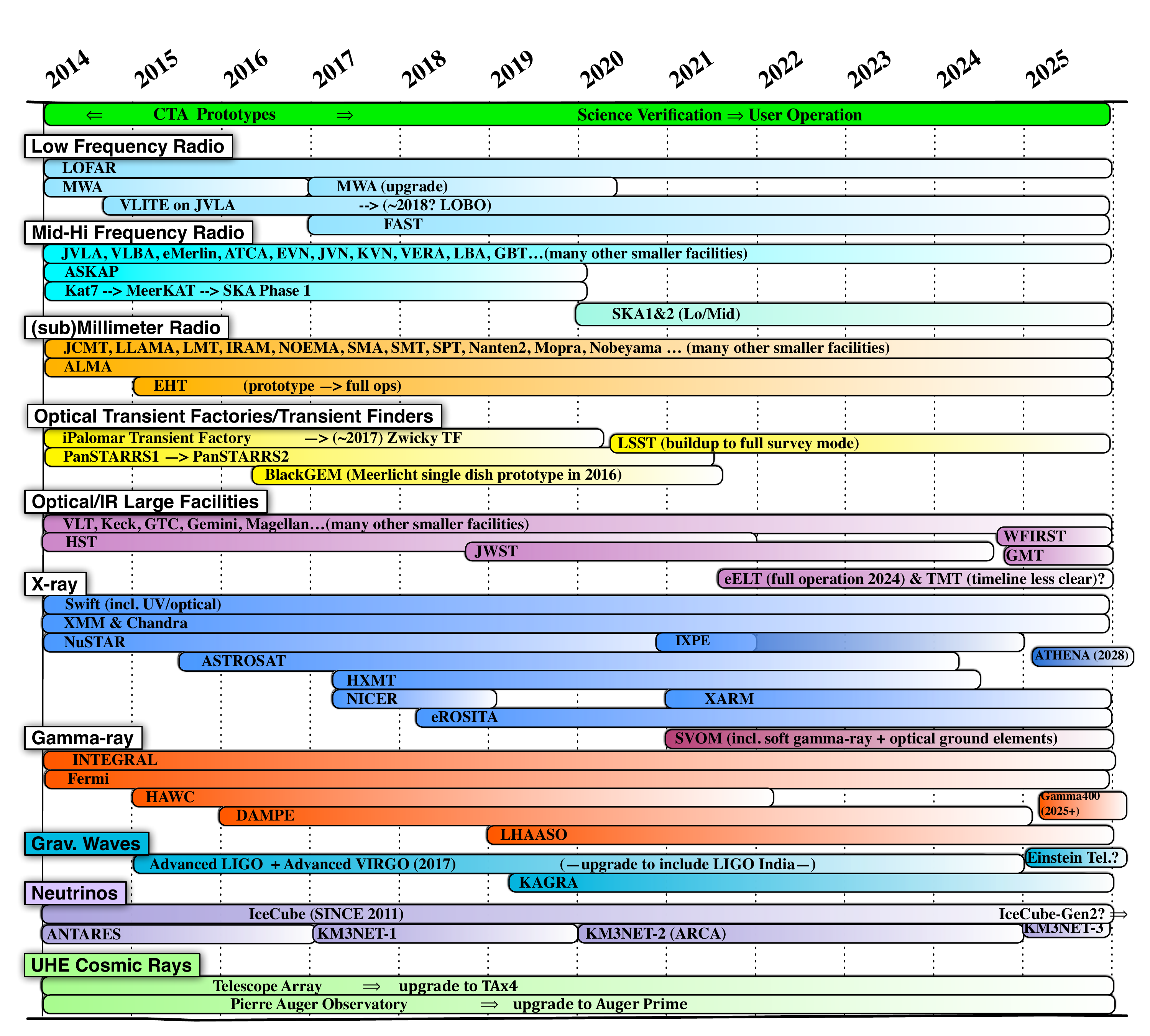}
\caption{Timeline of major multi-wavelength/multi-messenger facilities over the next
  decade.  Note that the lifetimes of many facilities are uncertain,
  contingent on performance and funding.   We indicate this
  uncertainty via the gradient, but have chosen timelines based on the
best information currently available.  Instruments still in the
proposal phase have been omitted, as have many relevant survey
instruments mentioned in the text, for the sake of space.  }
\label{fig:mwltimeline}
\end{center}
\end{figure}

The need for (simultaneous) MWL and MM observations has been
considered as a factor in the site selection process for CTA and in
the preparations for CTA science. Below we describe the main areas in
which synergies exist by waveband; see also a summary timeline of
major facilities in Figure~\ref{fig:mwltimeline}.
We also discuss cases where
agreements
between CTA (Consortium or Observatory) are desirable,
as well as cases where data can
be obtained without agreement via publicly accessible archives. 
Detailed MWL/MM plans can be found in the individual KSP chapters.  Please
note that there are many important and complementary facilities to CTA the
world over, and for the sake of space we cannot list them all,
particularly the many existing survey instruments.  We thus focus
primarily on the newest developments, and this chapter is
representative rather than exhaustive.  

\subsection{Radio to (Sub)Millimeter}

Until a few decades ago, the radio band provided our main window to
the non-thermal universe, via the cyclo-synchrotron emission of
relativistic electrons that often dominates over thermal processes below
$\sim$10~GHz.  Synchrotron emission goes hand in hand with particle
acceleration, due to the inferred presence of magnetic fields and the
presence of relativistic electrons, either directly accelerated or
produced as secondaries.   In addition, dark matter annihilation
scenarios usually lead to the production of synchrotron-emitting
secondaries along with gamma-ray emission.\footnote{See \cite{SKAScience} for many examples.}

The radio bands also have tremendous advantages for localising
acceleration zones, because of the high angular resolution (e.g. down
to 10's of microarcseconds with VLBI) and the ability to observe in
daylight.  The combination of radio measurements with those at very high energies can
provide limits on the electron density independent of assumptions about
magnetic field strengths and can help determine which of several
competing non-thermal processes dominate at the highest energies.  
Radio measurements also provide important magnetic field constraints
via Faraday rotation and provide the ephemerides of known pulsars, to guide
the search for potential gamma-ray pulsations with CTA. The success of Fermi-LAT in this regard relied on close cooperation with radio observatories~\cite{Rayetal2012}.
An exciting recent development in the radio domain is the discovery of Fast Radio Bursts \cite{Lorimer07,Thornton13}, with the possibility of high energy counterparts and potential synergies due to the wide field of view of CTA.  

After decades of incremental improvements, radio astronomy has
now again entered a rapid development phase.  Many existing
facilities have recently received major upgrades, providing much
improved bandwidth and sensitivity (e.g., JVLA, e-MERLIN). At the same time
windows to entirely new parts of the radio spectrum at both low and
high frequencies are finally being opened.  In particular, the
low-frequency bands (30--80, 120--240 MHz) are now being explored using
LOFAR, which can monitor 2/3 of the sky nightly in Radio Sky Monitor
mode and has a Transients Key Project dedicated to the detection,
triggering and cataloging of new radio transients.
In China, the Five-hundred-meter Aperture Spherical radio Telescope 
(FAST, 70 MHz--3 GHz), the largest radio telescope ever built,
had first light in 2016 and is now undergoing commissioning tests.
A key new radio project is  
the Square Kilometre Array (SKA), whose phase 1 will come online
during CTA's science verification phase, followed by a
ramp up to full operation with phase 2
by about 2024.  SKA will have unprecedented sensitivity and excellent angular
resolution, and the use of phased-array
technology allows for a very large field of view, ideal for survey studies
and transient detection (see Section~\ref{ssec:oir}). The pathfinders for SKA are very powerful instruments in their own right and will be important for early multi-frequency work involving CTA. A low-frequency pathfinder (MWA; 80--300~MHz) is well into its early science phase in
Australia with upgrades in progress, and new
projects at somewhat higher frequencies are under development in
Australia (ASKAP; 700--1800~MHz and UTMOST; 843 MHz) and South Africa (MeerKAT; eventually
3 bands between 0.6--14.5~GHz). The ThunderKAT programme for transients with 3000 hours of MeerKAT plus matching optical coverage (2017-2021) is particularly interesting for CTA. Finally, while
not strictly a pathfinder for SKA, VLITE has just
been commissioned with a wide
bandwidth 330~MHz channel and large field of view (FoV) to conduct a new low-band
survey of the sky, as well as detect new transients in real-time.
VLITE is a three year pathfinder for the proposed LOBO project, a more
extensive low-frequency radio monitoring project using the full JVLA.
Having radio facilities in both hemispheres provides important
complementarity for the two CTA sites. 

CTA's sensitivity to diffuse emission around accelerators makes
mapping of the interstellar gas over wide areas absolutely essential
to enable identification of sources in the Galactic plane and within
other large-scale surveys such as that of the LMC.  (Sub)- millimeter
wavelengths thus complement CTA science by offering a detailed
understanding of the environment into which shock waves propagate and
through which accelerated particles are transported and interact.
Most relevant to CTA are the facilities geared to degree-scale surveys
such as Mopra (Australia), APEX and Nanten2 (Chile), and Nobeyama 45~m (Japan),
whose beam sizes are well-matched to CTA's arc-minute resolution.
These telescopes measure molecular gas via a variety of molecular
lines that trace the matter density over a wide range of scales.  Of
particular interest is the missing ``dark'' molecular gas now attracting
serious attention in the ISM community, traced by THz lines, with
pathfinder telescopes in Antarctica such as HEAT (USA/Australia)
paving the way for large-scale survey instruments such as the proposed
DATE5 project led by China.

The recently completed Atacama Large Millimeter / sub-millimeter Array
(ALMA) represents a huge leap forward for (sub)-millimeter
interferometry.  With sub-arcsecond resolution and the sensitivity to
probe a very wide range of interstellar molecules, ALMA can carry out
high fidelity probes of the density, temperature and ionisation level
of material towards many CTA sources (including the LMC and nearby
starburst galaxies), helping to understand the environments in which
particles are accelerated and interact.

Furthermore, in recent years it has become clear that the
sub-millimeter range is of particular interest for studying the
particle-acceleration processes in the jets of Galactic black hole
transients (microquasars), as well as in the innermost regions of nearby
AGN.  For the former there are many new small arrays and single dish
facilities in both hemispheres available.  For the latter, the
upcoming Event Horizon Telescope will link ALMA and other
facilities in the first global VLBI array at mm/sub-mm frequencies,
offering direct imaging of the jet-launching regions of key sources
such as Sgr A${^\star}$ and M\,87.  Eventually mm/sub-mm observations together
with CTA can be used to directly study the relation between near event
horizon physics and cosmic-ray acceleration and non-thermal processes
in astrophysical jets.

At higher frequencies, the microwave all-sky survey by the Planck Satellite
(decommissioned in 2013) has produced a legacy archive that can be
searched for very extended microwave counterparts to CTA sources
within our Galaxy, complementary information on the lobes of nearby
radio galaxies and nearby clusters, and constraints on Galactic
magnetic fields.

\subsection{Infra-red/Optical through Ultra-violet and Transient Factories}
\label{ssec:oir}

Traditionally, the overlap between optical/infrared (OIR) astronomy
and gamma-ray astronomy has been considered to be fairly small.
Indeed much OIR emission has a thermal origin, such as stellar light,
heated dust, or emission from HII regions. However,
the last few years have revealed
that many compact, high-energy sources emit detectable levels of
synchrotron emission in the OIR, which can also display very fast
variability.  Some examples include blazars, microquasars and pulsar
wind nebulae, all of which are
high energy gamma-ray emitters, making OIR a
new frontier also for MWL exploration, and especially for producing
transient alerts.  Some steady gamma-ray sources also display mixed
OIR emission, such as the supernova remnant Cas A, which emits strong
thermal emission but also has IR synchrotron-emitting regions.  In
addition, OIR studies of non-radiative shocks in supernova remnants
can provide useful constraints on non-linear particle acceleration and
its influence on shock heating.  Finally, OIR observations provide an
interesting perspective in the case of gamma-ray binaries, as
properties of circumstellar discs may directly affect inter-wind
shocks and lead to light-curve evolution.
 
There are too many existing smaller facilities to list here that will
likely be useful for follow-up of CTA results, so we focus only on the
larger developments.  The high sensitivity of future telescopes will
increase the number of sources for which one can identify non-thermal
emission, or for which one can detect faint line emission from
non-radiative shocks.  By the time that CTA is operational, several
very large, ground-based facilities will also come online.  
Already quite advanced is the project to build
the European Extremely Large
Telescope (E-ELT) in Chile, which will have diameter of 39~m and is
expected to start operation by $\sim$\,2024.   A similar project on the
US side, the Thirty Meter Telescope (TMT) meant for Hawaii has run
into some uncertainty with the site, but has enough momentum that it
is likely to continue with a somewhat delayed timescale.   The Giant
Magellan Telescope (GMT) in Chile also begins commissioning in 2021,
with diameter of 24.5~m.  In space, the follow-up for the Hubble Space
Telescope (HST) is the James Webb Space Telescope (JWST), launching in
2018.  Similar to HST, the emphasis for JWST will be more on thermal
sources, however HST has made some important progress for IR
synchrotron-emitting sources and JWST's improved sensitivity will
likely prove relevant for constraining several TeV-emitting sources.
On the longer timescale, NASA has just placed highest priority on the
O/NIR WFIRST mission, a wide-field survey instrument.

Optical polarimetry, as compared for example to radio, has not historically been
of interest for MWL studies of VHE sources.  It is becoming apparent,
however, that polarisation offers an ideal way of isolating the
synchrotron/non-thermal component in cases of mixed emission.  This
technique can be employed to provide new insights in broad-band SED
correlations, for example to reveal potential low-energy signatures of
otherwise orphan VHE flares. Additionally, polarisation studies of
jets allow direct derivation of magnetic field parameters that can be
used to improve SED modeling and emission-region localisation.

\begin{figure}[htbp]
  \begin{center}
  \includegraphics[trim=18 190 18 190,clip,width= 0.68\textwidth]{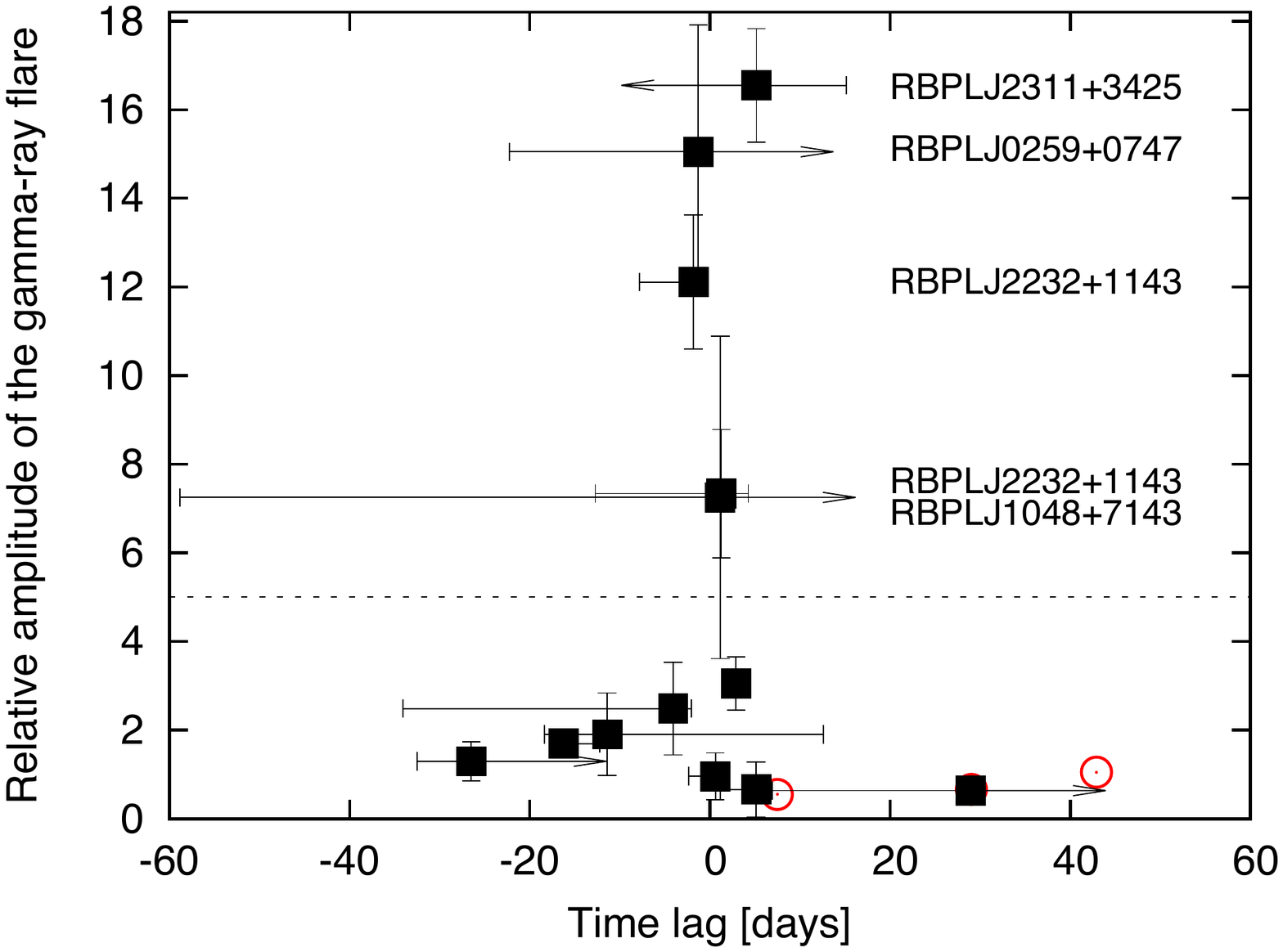}
\caption{Results from the first season of RoboPol~\cite{Pavlidou14} blazar monitoring,
  simultaneously with Fermi observations in the GeV gamma-ray range.    
  The fractional amplitude of a gamma-ray flare is plotted against the time delay
between the gamma-ray flare and observed rotation in the optical
polarisation angle.  The length of
the delay seems to be correlated with the gamma-ray flare
amplitude. Red symbols show values prior to redshift correction. Reproduced from~\cite{Blinov15}.} 
\label{fig:gammapol}
 \end{center}
\end{figure}

In general, the technical requirements for basic, but valuable, optical
studies can be met at modest cost, suggesting that the installation
of a small on-site (or nearby) optical telescope with polarimetric capability
could significantly benefit the CTA science case.
Having a dedicated facility for simultaneous, high cadence
monitoring of AGN sources, as well as to follow up transients or help
trigger CTA programmes, could be an important addition to the Observatory
capabilities. For example, several new VHE blazars were
discovered from triggers based on high optical emission states (see
e.g. \cite{Reinthal2012,Lindfors2012,Aleksic2012}) and 
optical polarisation shows interesting correlations with gamma-ray flares
(see Figure~\ref{fig:gammapol}). The addition of a dedicated optical telescope to the 
CTA baseline is under discussion within the project. 

The newest development in the CTA context are the multiple initiatives
for increasingly more sensitive, wide-field optical transient
monitoring, collectively referred to as ``Transient Factories''.
Currently in operation are two ground-based facilities with $\sim7-8$
deg$^2$ FoV, the Panoramic Survey Telescope and Rapid Response System
(Pan-STARRS) and the ``intermediate'' Palomar Transient Factory (iPTF).
The latter is itself a pathfinder for the Zwicky Transient Facility
(ZTF), that will have a very large 50 deg$^2$ FoV, a survey speed of
3750 deg$^2$/hr, and be online in 2017.  On a similar timescale, the
BlackGEM project, aimed to identify counterparts of gravitational wave
(GW) sources, will focus on transient detection, particularly before GW
sources are discovered, and will start operation in $\sim\,$2018 with
an initial deployment of three telescopes.  All of
these facilities use $<1$ m telescopes and in some sense pave the way
for the Large Synoptic Survey Telescope (LSST), an 8~metre class
telescope with a 9.6 deg$^2$ FoV that will scan the available sky every
three nights with much higher sensitivity.  By the time
CTA is starting science verification, these facilities together with SKA and
its prototypes will generate overwhelming numbers of triggers (e.g.,
thousands each of GRBs and tidal disruption events, and likely
hundreds of Galactic transients per year).  Thus in the coming years
it is key to understand the potential for VHE follow-up and define
appropriate response criteria, in order to select from the many
transient alerts that will be supplied via subscribable streams such
as VOEvent.\footnote{http://wiki.ivoa.net/twiki/bin/view/IVOA/IvoaVOEvent}
Particularly the earliest of these facilities will be proprietary in
terms of sharing their transients alerts, exactly when the response
modes of CTA need to be trained.  To that end, agreements between CTA and some of these collaborations
are likely to be beneficial and could
include also triggering of the external facilities
on CTA-detected transients.

Finally, the ultra-violet (UV) domain probes synchrotron emission of
electrons which have comparable energies to those emitting 
inverse-Compton emission
detected by CTA. As such, simultaneous UV observations of bright AGN,
blazars and other variable objects can be extremely useful, as long as
they are not too absorbed by interstellar gas.  Swift, XMM-Newton and ASTROSAT
all have UV capabilities, and other missions are being proposed, but at
the time of writing there are no definitive plans for a UV-capable
space mission on the timescale of CTA.

\subsection{X-ray}

There is an obvious synergy between gamma-ray astronomy and X-ray
astronomy. Phenomena which result in high enough temperatures for
thermal X-rays to be produced are very often associated with shock
waves, accretion or high velocity outflows, and hence with particle
acceleration and gamma-ray emission. In addition, studies of
synchrotron and inverse-Compton emission in the X-ray domain have
become increasingly important as missions capable of higher spatial
resolution and sensitivity have been launched.  In supernova remnants
for example, the X-ray emitting electrons have $\sim$100 TeV energies,
making the combination of VHE gamma-ray and X-rays extremely powerful
for constraining magnetic field strengths, the electron to proton
ratio of the accelerated particles and the particle energy
distributions.  The thermal X-ray emission from gamma-ray sources
provides valuable information about plasma properties (e.g. temperatures,
densities) and energetics (e.g. outflow/shock velocities).
Non-thermal X-ray emission also provides a natural tracer of locations
of extreme particle acceleration.

Over the last decade X-ray astronomy has been very successful thanks
to large missions such as Chandra and XMM-Newton, both
spectro-imaging missions, plus several medium missions such as
the Rossi X-ray Timing Explorer (RXTE, which pioneered many timing and
monitoring studies), Suzaku (imaging-spectroscopy), and Swift
(which continues to be extremely successful for transient detection and
follow up, including in the UV).  NuSTAR, launched in 2012 and with a
guest-observer programme started in 2015, is the first focusing telescope at hard 
X-rays and is currently making many exciting discoveries about
extreme particle accelerators in the energy range 10-80 keV.  MAXI, a
Japanese all sky monitor, is currently in operation on the
International Space Station, and the main X-ray transient detector besides Swift. More recently CALET and the UFFO pathfinder have added additional hard X-ray / soft gamma-ray transient capabilities.

Chandra, XMM-Newton and Swift will likely continue to operate
throughout the early years of CTA operation, if not beyond.  Several
new missions have recently been launched or are funded, and will be
very relevant for CTA science during its operation.
For example, the German/Russian mission eROSITA
(launch 2018; 0.3--10 keV imaging spectroscopy survey), will
overlap with CTA's early operation years and further.  eROSITA in
particular will be the first imaging all-sky survey in the 2--10 keV
range and, as such, it can be expected to provide a primary reference for
CTA source identification and multi-wavelength correspondences.  The
data however will be proprietary, with the German
side planning two data releases, one around 2021 and the other two
years later at the end of the survey, $\sim$2023.  The Russian side
will likely join in at least the final data release, but if CTA wants
earlier access to survey data, as well as first pick of transients discovered in
their offline (not real-time) pipeline analysis, a memorandum of understanding may be
necessary. The loss of the Hitomi (earlier Astro-H) mission is clearly a major blow, but 
a replacement mission (XARM) has now been approved by JAXA.

Instruments with more focused capabilities include the recently
launched (2015) Indian UV/X-ray satellite ASTROSAT, which features an
all-sky monitor that will be very valuable for triggering CTA.  An
agreement may be necessary here as well.  In June 2017, NICER was
installed on the International Space Station (ISS) for soft X-ray
timing focusing on primarily neutron stars and will be open for proposals in
the second year.  Also recently launched by China is the Hard X-ray Modulation Telescope
(HXMT), a ``super RXTE'' operating in the 20-200 keV band with a 3600 deg$^2$ FoV.

Beyond that the landscape for future X-ray missions is not yet fully
determined but looking very promising. 
SVOM is an optical/X-ray mission primarily for discovering GRBs,
planned for launch in 2021.  SVOM has similarities with Swift, triggering on bursts in both softer and harder bands, and following them in the X-ray and optical on-board.
SVOM also includes a dedicated group of
ground-based optical telescopes for wide FoV coverage before and after
transient events.

Concepts under development which could provide synergies with CTA include
the Chinese enhanced X-ray Timing and Polarimetry mission (eXTP),
which incorporates large-area soft and hard X-ray telescopes, a wide-field monitor and a
polarimeter, and
would be launched around 2025. Several concepts related to Lobster-eye wide-field X-ray optics are being explored in the US, Europe and in China. Such instruments could provide a major source of triggers for CTA transient observations.

The Athena+ project due for 2028 launch is, however, the next major
observatory class mission, with good spatial resolution ($\sim$10''),
high sensitivity and energy resolution, and excellent spectroscopic
capabilities.  This mission will be the key X-ray facility for the 2030 decade
and is designed for complementarity with radio/optical facilities
and a large scientific breadth, providing additional high-energy
constraints for CTA-detected sources.

One final development is the potential for X-ray polarimetry, a
sure-fire way to isolate X-ray synchrotron from other components and
to constrain the presence of accelerated particles. The IXPE mission has recently been selected by NASA as part of the SMEX programme.
In the M4 ESA call, the X-ray Imaging Polarimetry Explorer (XIPE) was
selected for a two-year design study as part of the Cosmic Vision
programme.  By the time of CTA's early science verification, the
future for this exciting new capability should be clear.

\subsection{Sub-VHE Gamma-ray Energies}

The hard X-ray / soft gamma-ray domain (0.1--10 MeV) represents a very
useful window on the non-thermal spectra of astrophysical sources, but
is extremely challenging experimentally. Three main instruments currently contribute here:
INTEGRAL, Swift-BAT and Fermi-GBM.
INTEGRAL was launched in 2002 and its
lifetime has been extended through to 2018 and it may be extended further into
the CTA early science period.
The Fermi-GBM and Swift-BAT
detectors are the critical current instruments for the detection of high-energy transients.
Swift was launched in 2004 with a nominal 10 year lifetime that has
already been extended.  In the 2016 NASA Senior Review it reviewed 
first out 
of six missions, with operations confirmed through 2018 and 
recommended for extension at least until 2020.
With an orbital lifetime stated as 2025 or
beyond, it will likely continue to be a resource for transient
detection during the CTA period.

At higher gamma-ray energies the synergies with CTA are stronger and
the instrumental performance better matched to CTA.  The GeV domain is
dominated by pion decay, bremsstrahlung and inverse-Compton emission,
and in combination with the TeV range can help identify the dominant
radiative mechanisms. CTA's lowest energy range overlaps with that of
the two current instruments: the Fermi Large Area Telescope
(Fermi-LAT) and AGILE.   
The Fermi mission
has been so successful that its lifetime has been extended through 2018, 
and it will likely continue operating through 2020 and potentially
beyond.  NASA does not generally decommission fully operational
missions, especially those with no clear successors, and Fermi's science
performance continues to improve while having no consumables. The 2016 NASA Senior Review panel explicitly acknowledged the added scientific value of extending Fermi's lifetime so that it overlaps with CTA operations. Thus
it seems likely that Fermi will still be able to provide triggers and
complementary coverage for a number of years of the CTA era.   Fermi 
will continue to provide the main reservoir of extragalactic targets
for CTA.

On the horizon, there are several missions upcoming or proposed to
advance the observations of gamma rays from space in diverse
domains. The Chinese Academy of Science's DAMPE (launched in 2015) and HERD
(proposed for launch in the 2020's) are going to explore the energy range
from hundreds of GeV to 10~TeV with an energy resolution approaching 1\%. A few
missions currently in the concept development phase (AMEGO in the U.S. and 
PANGU in China/Europe) aim at exploring the
energy range from 0.5 MeV to 1 GeV with much improved sensitivity, 
point spread function,
and polarization capabilities thanks to the first-time ever
combination of detection techniques based on pair production and
Compton scattering in the same instrument. 
An alternative concept
for a high-sensitivity, high-angular-resolution instrument with
polarization capabilities in the MeV to GeV domain uses
gas time projection chambers. The concept is
subtantiated in two ongoing R\&D projects, AdePT and HARPO. All of
these future (potential) missions offer new and interesting synergies
with CTA.

\subsection{Complementary VHE Gamma-ray Instruments}

Several ground-based VHE gamma-ray instruments may be operational at
the same time as CTA. None of these instruments are direct
competitors, but rather provide complementary performance. In
particular, the High Altitude Water Cherenkov (HAWC) detector is a
100\% duty cycle and very wide field of view ($\sim1$~sr) TeV range
instrument ~\cite{Tepe12}.  Seated at a high-altitude site in Mexico,
HAWC is capable of detecting the brightest known TeV sources in
$\sim$1~day and will be able to provide alerts to CTA for
active/flaring states of blazars and transients.  HAWC's modest
($\sim0.5^{\circ}$) angular resolution and somewhat limited energy
resolution gives it competitive sensitivity for
very extended emission, and by the time of CTA, it will have mapped
the northern sky to intermediate depth~(see
Chapter~\ref{sec:ksp_gps}), identifying many interesting steady
sources for CTA to investigate.

LHAASO~\cite{DiSciascio16}, is an ambitious 
multi-component project incorporating HAWC-like water
Cherenkov detectors and a very large array of scintillators at a site in China.
LHAASO will complement CTA at higher energies in a
similar way to HAWC, with modest resolution and background rejection
power offset by high duty cycle, wide field of view and large area.
A number of concepts now also exist for a ground-particle-based detector for VHE gamma rays in the southern hemisphere, strongly complementing CTA-South by providing triggers and additional information on very extended emission regions.

One or more of the current generation of IACT arrays --- H.E.S.S.-2, MAGIC
and VERITAS --- may continue operations into the CTA era. Use of these
telescopes for monitoring could be considered, under suitable
agreement between the telescope and CTA.  In particular for cases when
the CTA sites are at different longitudes than current IACTs, these can
extend monitoring of bright flaring sources to periods before and
after the CTA observations.

\subsection{VHE and UHE Neutrinos}

Essentially all mechanisms invoked for the production of high energy
neutrinos will also produce gamma rays of similar energies, and unlike
charged cosmic rays, both point back to their sources.  There is thus
strong complementarity to observations with these two messengers.
Gamma-ray telescopes, using e.g. the atmospheric Cherenkov technique,
achieve the precision pointing and sensitivity to identify and
understand populations of accelerators and to even localise
acceleration sites in nearby objects. 
Neutrino telescopes, using e.g. reconstructed muon tracks, are less able
to precisely pin-point the origin of neutrinos, but the neutrinos they detect
are the only completely unambiguous tracer of hadronic acceleration, even out to high redshifts
and beyond PeV energies, a combination that is not
possible for gamma-ray telescopes due to photon-photon absorption.

The experimental situation in the VHE neutrino domain has
recently dramatically altered, with the first strong evidence for
astrophysical neutrinos above the MeV band. The IceCube collaboration
has announced the detection of a diffuse astrophysical neutrino signal
at 0.1--1 PeV energies~\cite{Aartsen13Ta}. Individual neutrino sources
with a flux corresponding to O(1) neutrino per IceCube exposure
will be very easily detectable with CTA up to 1 PeV if they are
Galactic in nature\footnote{Unless there is very strong internal
  gamma-gamma absorption, as might be the case for some binary
  systems.} as has been suggested~\cite{Razzaque13}.

With the construction of KM3Net~\cite{Katz06} and several upgrades to
IceCube in the planning, the detection of individual neutrino sources
is a distinct possibility.  CTA is the ideal instrument to follow up
on any VHE neutrino clustering, necessary to localise and characterise
the VHE accelerators.  An important new aspect therein will be the
follow-up of neutrino-generated triggers as ToOs, in order to localize
and identify the hadronic accelerators.

\subsection{Gravitational Waves}

Now that gravitational waves from black hole-black hole merger events have
unambiguously been detected by Advanced LIGO 
\cite{Abbott16a,Abbott16b,Abbott16aa}, a new and exciting field is opening up for 
electromagnetic follow-up and identification.  Mergers of binary black holes and
neutrons stars (or mixed systems) should be detectable out to a few
hundred Mpc \cite{Aasi13}, with expectations of several to
hundreds of GW transients per year after
2018~\cite{Abadie10}. However, until the advent of
third-generation detectors such as the Einstein Telescope ~\cite{Punturo10},
the localisation errors on these transients will be relatively large
and asymmetric.  For follow-up of GW alerts, CTA has huge advantages
with respect to most other instruments and wavebands. These advantages
include the large field of view and the flexibility to map very large
and non-circular error boxes (which comes from the large telescope
number~\cite{Bartos14} and potential divergent pointing modes), the
rapid response time, and the less ambiguous nature of counterpart
identification (when compared, for example, to the optical band).

\section{Core Programme Overview}
\label{sec:KSP_strategy}
\label{sec:obs_prog}

Over the lifetime of CTA 
most of the available observation time at both of the CTA sites will be 
divided into open time, based on scientific merit and
awarded by a Time Allocation Committee to Guest Observer proposers, 
and a Core Programme consisting of a number of major legacy projects.
Smaller fractions of observation
time will be allocated to Director's Discretionary time and host reserved time.
The Core Programme, corresponding to approximately 40\% of the total observation time over the first
ten years of CTA operations, will be comprised of Key Science Projects (KSPs) to be carried out
by the CTA Consortium. 
Here we introduce the KSPs; in the following chapters they are described in more detail.

\begin{sidewaysfigure}[htbp]
\begin{center}
\includegraphics[trim=50 120 50 50,clip,width=1.04\textwidth]{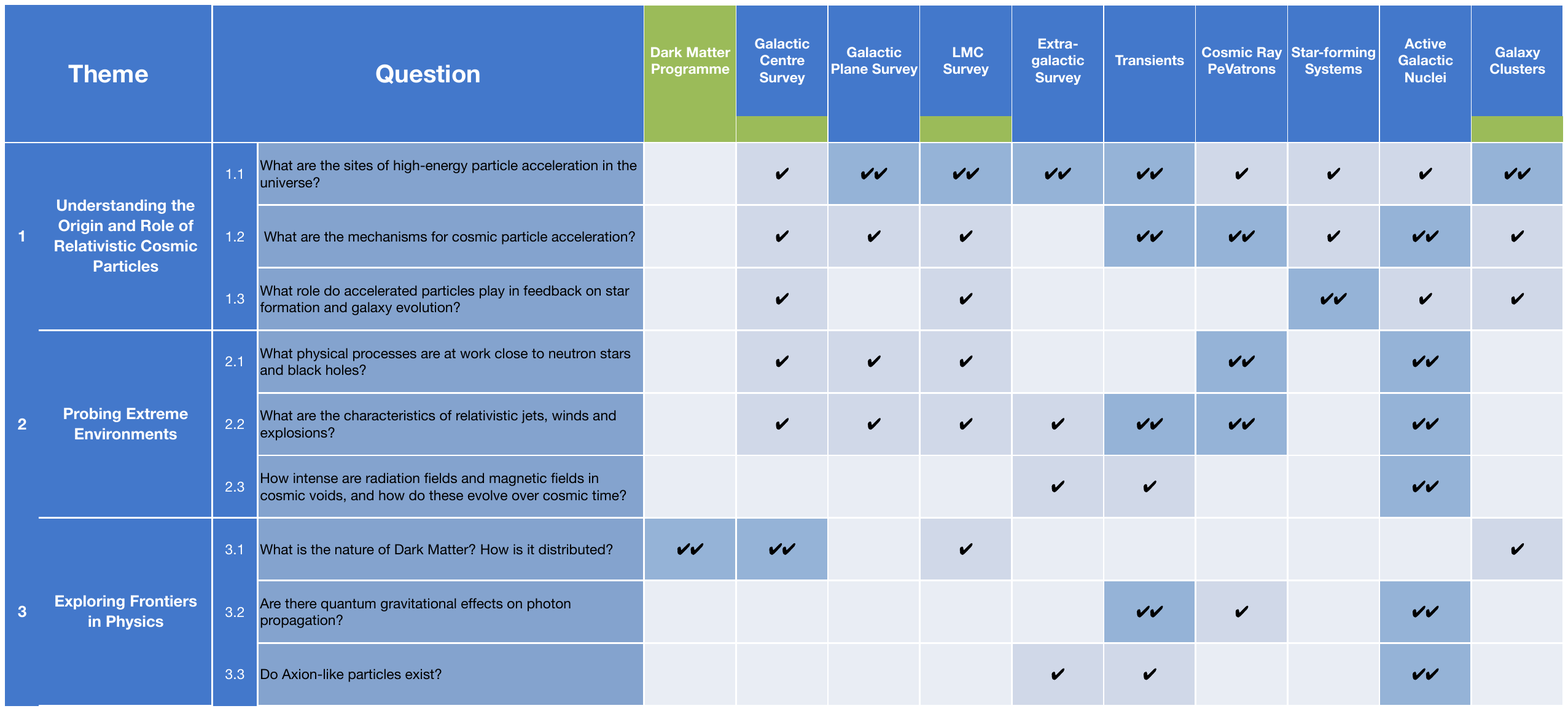}
\vspace{-1cm}
\caption{Matrix of CTA science questions and proposed Key Science Projects (KSPs). The KSPs are sets of observations addressing multiple science questions within the CTA themes. KSPs which contribute to the programme aimed at dark matter detection are indicated in green, with the exclusively dark-matter-oriented targets described entirely within the Dark Matter
Programme in Chapter~\ref{sec:DM_prog}. For KSPs simultaneously addressing dark matter and other physics/astrophysics, the motivation and context for the dark matter element is again described in Chapter~\ref{sec:DM_prog}. 
The order of the KSPs in this table starts
with dark matter due to its importance and transversal nature
and follows with surveys and more focused KSPs by increasing distance scale.
The check marks are intended to give a qualitative assessment of the impact of each KSP on a particular science question.
}
\label{fig:KSPmatrix}
\end{center}
\end{sidewaysfigure}

The CTA KSPs have been defined through a multi-year process of discussion within the CTA Consortium and in interaction with the wider community. 
They are ambitious projects with very significant scientific promise that also require considerable observation time. 
As such they are suitable for execution in the guaranteed time of the CTA Consortium. 
Figure~\ref{fig:KSPmatrix} provides a matrix of the main science questions within the 
CTA themes versus KSPs. The KSPs are multi-purpose observations designed to efficiently address the broad ranging science questions of CTA. An internal review of the KSPs was conducted in late 2014, followed by
a presentation in 2015 to the CTA Scientific and Technical Advisory Committee (STAC). 
The projects were refined and optimised accordingly.

The criteria used for selection of the KSPs were:
\begin{enumerate}
\item excellent scientific case and clear advance beyond the state of the art,
\item the production of legacy data-sets of high value to the wider community, and
\item clear added value of the project as a KSP rather than as part of the Guest Observer Programme, including:
\begin{itemize}
\item the scale of the project in terms of observing hours - very large projects will be difficult to accommodate in the open time early in the lifetime of the observatory,
\item the need of a coherent approach across multiple targets or pointings, and
\item the technical difficulty of performing the required analysis and hence 
reliance on Consortium expertise.
\end{itemize}
\end{enumerate} 

As a demonstration of feasibility, this programme 
has been scheduled in detail using the prototype software for final CTA observation scheduling, under 
the assumption of an approximate 40\% share of the observation time during the first ten years. 
The preliminary conclusions of this exercise are that the presented programme is feasible 
overall, with some adjustments needed on the total observation hours and on the relative hours among the KSPs.

The following Chapters present each KSP in turn. The Dark Matter
Programme, being of particular importance to CTA and overlapping
considerably in terms of observation fields with other science topics,
is dealt with somewhat differently, with a single Chapter
(\ref{sec:DM_prog}) describing the complete strategy.

We note that the Core Programme described below was developed over a several year period and 
was largely finalized in 2016.  It
is expected to evolve before it is implemented, due to changes in theoretical understanding and new observational insights. It is envisaged that working groups within the CTA Consortium will continuously improve and update the programme and that feedback from the broader astronomical community will be essential in refining the KSPs.

\section{Dark Matter Programme}
\label{sec:DM_prog}

The existence of dark matter as the dominant gravitational mass in the universe 
is by now well established but the detailed nature of dark matter is at present still unknown. 
Multiple hypotheses endure as to the character of dark matter, and for the most popular models discussed 
CTA has a unique chance of discovery. 
In the form of Weakly Interacting Massive Particles (WIMPs), dark matter particles can self-annihilate, converting their large rest masses into other Standard Model particles, including gamma rays.  Indirect detection from such annihilations provides a unique test of the particle nature of dark matter, {\it in situ} in the Cosmos.
Observations of the gamma rays provide the probe for the ``indirect" detection of dark matter employed by CTA. In the standard thermal history of the early universe the annihilation cross-section has a natural value, the ``thermal cross-section'', which provides the scale for the sensitivity needed to discover dark matter in this way. Particular models for WIMPs such as supersymmetry (SUSY) and theories with extra dimensions, give predictions for gamma-ray energy spectra from the annihilations which are essential ingredients towards the predictions of the sensitivity of the indirect searches. Another vital ingredient in the CTA sensitivity predictions is the distribution of dark matter in the targets observed for the search.

The priority for the CTA dark matter program is to discover the nature of dark matter with a positive detection. The publication of limits following non-detections would certainly happen but in planning the observational strategy, 
the priority of discovery drives the programme. The possibility of discovery should be considered in the light of model predictions  where the minimum goal for searches within the most widely considered models is the velocity-weighted thermal cross-section of $\rm 3\times10^{-26}\,cm^3 s^{-1}$.
The principal target for dark matter observations in CTA is the Galactic halo. These observations will be taken within several degrees of the Galactic Centre with the Galactic Centre itself and the most intense diffuse emission regions removed from the analysis. With a cuspy dark matter profile, observations of 500 hours in this region provide sensitivities below the thermal cross-section and give a significant chance of discovery in some of the most popular models for WIMPs. Since the dark matter density in the Galactic halo is far from certain, other targets are also proposed for observation. Among these secondary targets, the first to be observed will be ultra-faint dwarf galaxies with 100 hours per year proposed. 

\begin{figure}[!ht]
\begin{centering}
\includegraphics[width= 7.9cm]{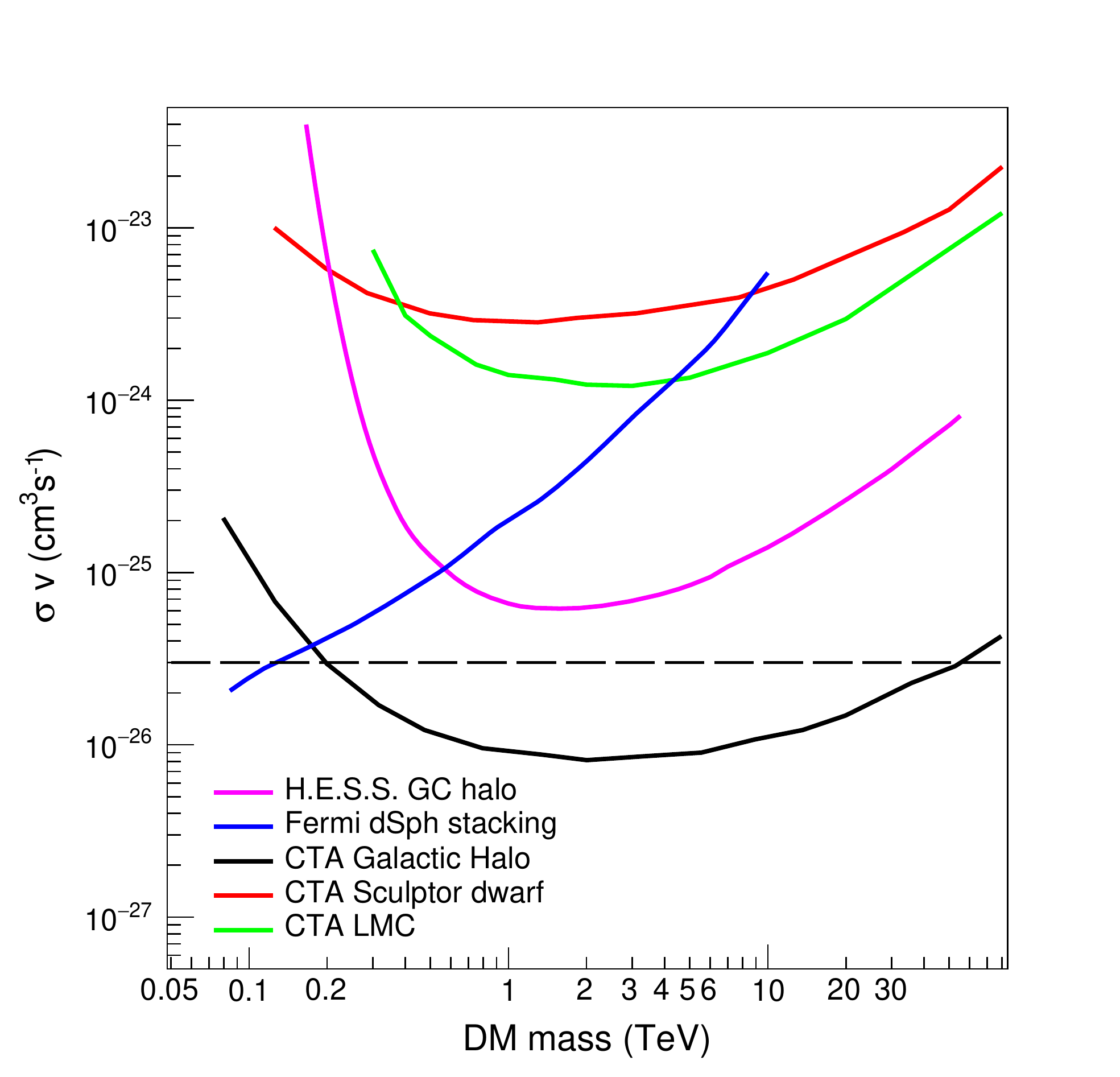}
\includegraphics[trim=140 10 140 10,clip,width= 7.3Cm]{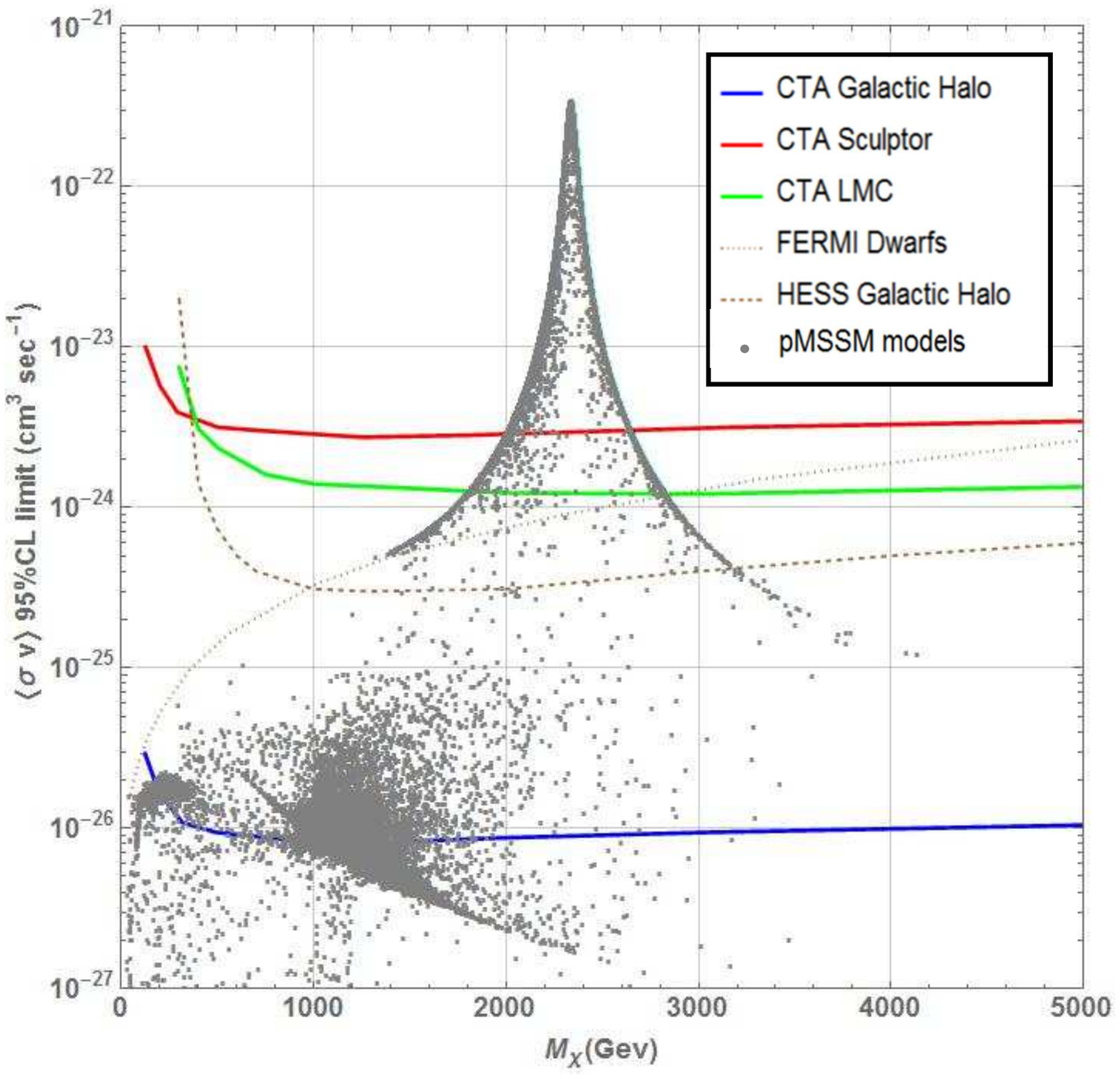}
\caption{Left: comparison of predicted sensitivities in $\langle \sigma v \rangle$ for the targets of: the Milky Way Galactic halo; the Large Magellanic Cloud (LMC) and the dwarf galaxy Sculptor. The CTA sensitivity curves use the same method and W$^+$W$^-$ annihilation modes for each target and the Einasto dark matter profile. The sensitivities for the three targets are all for 500 hours taking into account the statistical errors only; for the Milky Way and the LMC, the systematics of backgrounds must be well controlled to achieve this statistically possible sensitivity. 
The H.E.S.S. results come from the Galactic halo for the W$^+$W$^-$ channel~\cite{Abdallah16c}
and the Fermi-LAT results come from dwarf spheroidal galaxies for the W$^+$W$^-$
channel~\cite{Ackermann15c}.
The horizontal dashed line indicates the thermal cross-section at 3 $\times$ 10$^{-26}$ cm$^3s^{-1}$.
Note that the Fermi-LAT results come from a stacked analysis of many dwarf spheroidal 
galaxies and the CTA curves are for specific individual sources.
Right: the zoomed sensitivities in the TeV mass range together with model points from the 
phenomenological minimal supersymmetric model
(pMSSM),
extracted from Ref.~\cite{Roszkowski15} (see also Ref.~\cite{Roszkowski14}).
The H.E.S.S. results come from the Galactic halo for the 
Tasitiomi parametrization (close to the W$^+$W$^-$ channel 
without electroweak correction)~\cite{Abramowski11b}
and the Fermi-LAT results come from dwarf spheroidal galaxies for the W$^+$W$^-$
channel~\cite{Ackermann15c}.
}
\label{fig:summary_dm}
\end{centering}
\end{figure}

Beyond these two observational targets, alternatives will be considered closer to the actual date of CTA operations. New star surveys will extend the knowledge of possible sites of large dark matter concentrations and a detailed study of the latest data will be made to continuously select the best targets for dark matter searches in CTA. Among these new possible targets are newly discovered candidate dwarf galaxies and dark matter clumps which could be very promising sources if their locations are identified {\it a priori}. Beyond the targets proposed for observations in the present KSP, the data for the Large Magellanic Cloud (LMC) KSP will also be used to search for dark matter. Furthermore, the data from the Galactic Plane and Extragalactic surveys might give hints of gamma-ray sources which do not have counterparts in multi-wavelength data and which could be pursued as dark matter targets. The sensitivity predictions for the Galactic halo, the dwarf galaxy Sculptor and the LMC are summarized in Figure~\ref{fig:summary_dm}.  The left panel shows that the sensitivity possible with the Galactic halo observation is much better than what is possible with a single dwarf galaxy or the LMC. 

The Dark Matter Programme is very well suited to being carried out in the Core Programme by the CTA Consortium. The observations require a large amount of time with a significant chance of major discovery but with a clear risk of a null detection. For the Galactic halo the observation time used will also be of great use for astrophysics, but the time on dwarf galaxies may be of lesser use. In the phase between now and the actual operation of CTA, much will evolve in the knowledge of dark matter distributions in the various targets. Further detailed work is needed to understand the systematics in the backgrounds; however, it is very likely that to have a chance of discovery the emphasis of observation should be on the Galactic halo. To have the best opportunity to discover the nature of dark matter CTA must probe TeV scale masses with annihilation cross-sections in the range $\rm 5\times10^{-27}$ to $\rm 3\times10^{-26}\,cm^3 s^{-1}$. This seems possible if adequate observation time is allocated to the Dark Matter Programme as well as sufficient resources to develop methods to control the systematic errors.

Although there are great expectations for the search of new particles at the Large Hadron Collider (LHC), collider experiments will most probably not \emph{on their own} be able to assess if they are the whole constituent of the 
dark matter in the universe. Complementary searches, by means of direct nuclear recoil measurements and indirect searches such as the one CTA will perform, could allow an important breakthrough in the
identification of the true nature of dark matter. 
The combination of results obtained from the three major axes of dark matter searches, production at colliders, direct detection of nuclear recoil and indirect detection,
will be essential to refine our understanding of the physics beyond the Standard Model. The dark matter 
parameter space in reach for CTA is unique and will benefit the entire scientific community. 
The release of multi-dimensional likelihood functions enabling model-independent searches for new particles is foreseen.

\subsection{Science Targeted}
The nature of dark matter in the universe is one of the most compelling questions facing physics and astronomy at the present time.  There are many programmes around the world that have capabilities or are devoted to searching for non-gravitational signals of dark matter.  Among these, CTA has an almost unique capability during the next decade to explore the WIMP mass region in the few hundreds of GeV to the 10 TeV regime, above the reach of the LHC and ton-scale direct detection experiments.  To capitalize on this remarkable opportunity, it is imperative for CTA to devote substantial resources (both observing time and analysis effort) towards a comprehensive and integrated dark matter programme.  If this is done, CTA will be a cornerstone of the global, multi-faceted line of attack from different experiments probing with different methods to understand the nature of the dominant gravitational matter in the universe.

\subsubsection{Existence of Dark Matter}
\begin{figure}[h!]
\centering
\includegraphics[width=.342\textwidth]{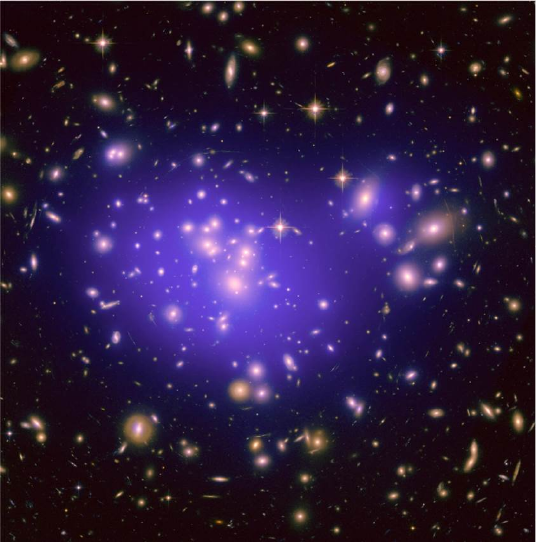}
\includegraphics[width=.475\textwidth]{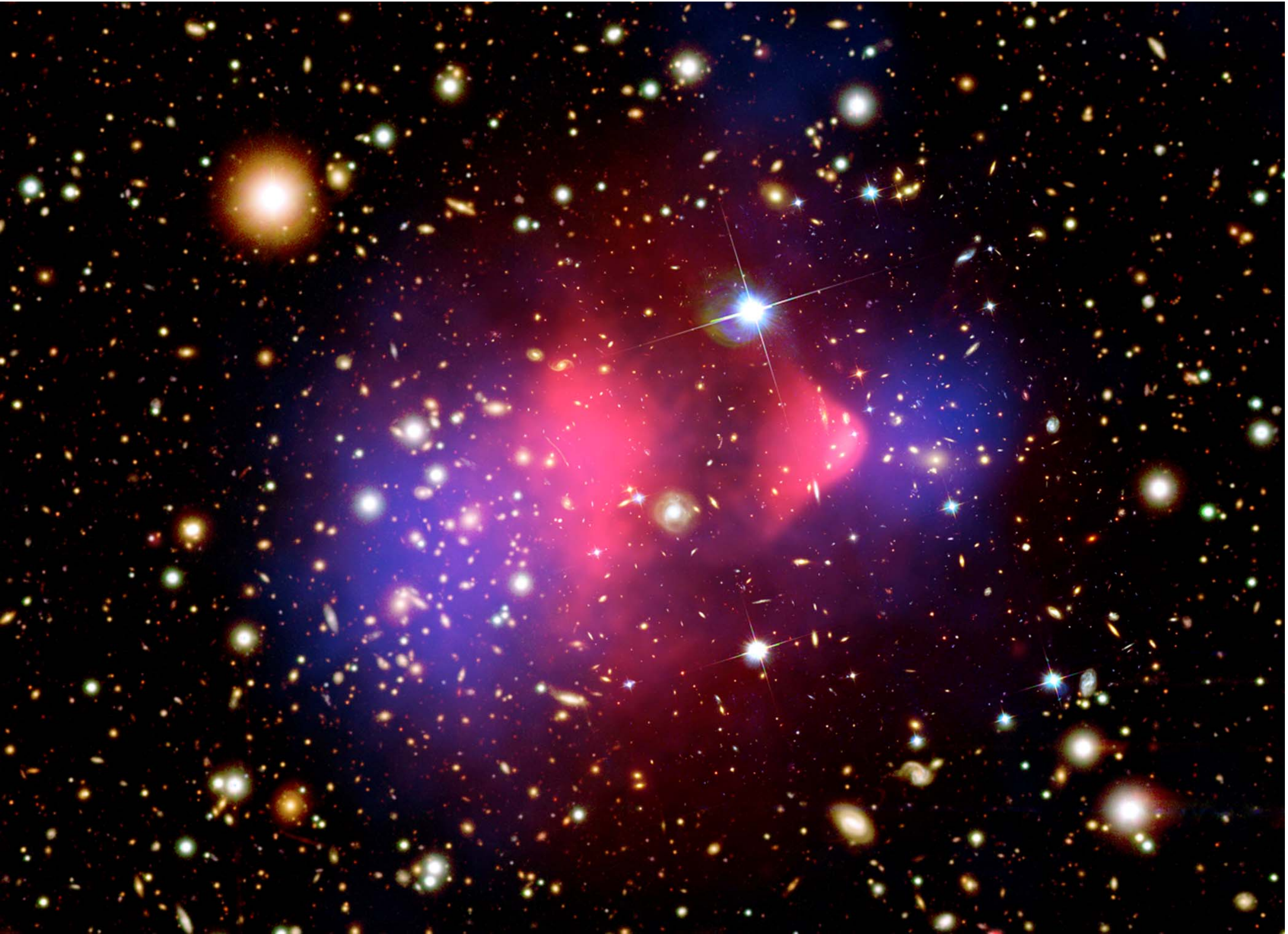}
\caption{Left: Hubble Space Telescope image of the inner region of the galaxy cluster Abell 1689. The blue overlay shows the dark matter distribution reconstructed by gravitational lensing using the multiple galaxy images seen in the telescope image. 
Credit: NASA, ESA, E. Jullo (JPL), P. Natarajan (Yale), and J.-P. Kneib (LAM, CNRS).
Right: Composite image of the galaxy cluster 1E~0657-56 (also known as the Bullet 
cluster)~\cite{Clowe06}. Hot X-ray emitting gas is shown in red and the blue hue shows the dark matter distribution in the cluster deduced from gravitational lensing. 
Credit: X-ray: NASA/CXC/CfA/M.Markevitch et al.; Optical: NASA/STScI; Magellan/U.Arizona/D.Clowe et al.; Lensing Map: NASA/STScI; ESO WFI; Magellan/U.Arizona/D.Clowe et al.
}
\label{fig:collidingclusters}
\end{figure}

The existence of dark matter in the universe was first proposed by Zwicky~\cite{Zwicky33} in the 1930's to explain the dynamics of the Coma galaxy cluster where the observed luminous matter was insufficient to provide gravitational stability. More recent studies have confirmed the presence of dark matter in galaxy clusters by gravitational lensing, for instance in the galaxy cluster Abell 1689 shown in the left panel of  Figure~\ref{fig:collidingclusters}.  Evidence for dark matter now exists on many scales. In spiral galaxies, composed of a central bulge surrounded by a luminous disk, stellar motions are dominated by rotation within the disk. The luminous component of such a galaxy decreases exponentially from the centre giving the expectation that the star rotation velocities would scale as r$^{-1/2}$ according to Kepler's laws, however, galaxy rotation curves usually remain flat far from galactic centres, typically beyond 30-40 kpc, where gas and stars are not dominant, implying a massive dark component. In elliptical galaxies, where the dynamical equilibrium is dominated
by random motion of dark matter and stars rather than the rotational motion, observations also indicate a large contribution of dark matter. Two galaxy types are particularly dominated by dark matter: low surface brightness galaxies and dwarf spheroidal galaxies. 
In galaxy clusters about 80\% of the total mass is composed of dark matter. 
Recently, gravitational lensing studies of colliding galaxy clusters have shown different behaviours of the major components of the cluster pair during collision~\cite{Clowe03,Bradac08} and provide one of the strongest pieces of evidence to date for the existence of dark matter (Figure~\ref{fig:collidingclusters}, right).

Numerous observational cosmology experiments performed over the past two decades, such as observations of the cosmic microwave background, baryonic acoustic oscillations, large scale structures and supernovae, combine to give the standard picture of the composition of matter/energy in the universe as: $\sim$68\% dark energy, $\sim$27\% dark matter, and
$\sim$5\% baryonic matter, and a small fraction of radiation in various forms.
Only $\sim$1\% of the universe is normal matter in luminous star systems.
The limited amount of normal matter in the form of baryons and luminous stars is given in independent ways from the predictions of primordial nucleosynthesis and the measurements of the spectrum of fluctuations of the cosmic microwave background. The total amount of matter, baryonic plus non-baryonic comes, again, from two independent sets of measurements: in galaxy clusters and the cosmic microwave background. It is simulations of structure formation which indicate that the non-baryonic dark matter is ``cold'' rather than ``hot'' in the sense that to explain the observed structures in the universe the dark matter must behave in a non-relativistic manner.  The baryonic dark matter could be concentrated in molecular clouds or small stellar objects with masses too low to be luminous, although searches for these latter objects have not proved fruitful~\cite{Alcock00,Tisserand06}.
With this broad set of observations it is now well established that the dominant gravitation mass in the universe is non-baryonic dark matter with a global 
density of $\Omega_{\rm DM} h^2=0.120\pm0.003$~\cite{Planck13,Planck15},
where $\Omega_{\rm DM}$ is the ratio of the dark matter density to the critical density and
$h$ is the dimensionless Hubble parameter in units of $100\,{\rm km}\,{\rm s}^{-1}\,{\rm Mpc}^{-1}$.   Independently, Big-Bang nucleosynthesis measurements indicate that most of the dark matter is non baryonic.
However despite this precise knowledge on the global amount of dark matter, its nature 
is still elusive and remains to be discovered.

\subsubsection{Distribution of Dark Matter}  
The hierarchical formation of structures in the universe is due to the gravitational amplification of primordial density fluctuations during its expansion. Due to the complexity of the physical processes that play a role in the structure formation, large-scale cosmological N-body simulations are used for modeling of the evolution from density fluctuations in a non-linear regime. These simulations give predictions for the dark matter halos that are observed to surround all systems, from galactic scale to galaxy cluster scale. Among the groups performing simulations are the Aquarius Project~\cite{Springel08} and the Via Lactea Project~\cite{Diemand08}. Until recently, simulations used only cold dark matter (CDM), included only the gravitational force, and usually predicted the dark matter density to go approximately as 1/r towards the centre of the dark matter halos.  Standard parameterizations of these simulated dark matter halos are the Navarro, Frenk and White (NFW)~\cite{Navarro95} and the Einasto~\cite{Graham06,Navarro08} profiles.  The latter one is moderately shallower on small spatial scales compared to the NFW profile. N-body simulations showed dark matter profiles that can be both steeper and shallower~\cite{Springel08,Diemand08}. Steeper profiles are usually referred to as cuspy profiles. All the dark matter simulations agree on the main halo structure at large distances but the predictive power is limited by the spatial resolution of the simulation, and the shape and density of the profile in the inner part of the halo relies on extrapolation of the simulation prediction. The existence of such a cuspy density profile is in disagreement with observations of disc and dwarf galaxies where detailed mass modeling using rotation curves suggests a flatter or cored dark matter density profile in the central region. The study of velocity dispersions of stars in dwarf galaxies suggest that they can be equally accommodated by cuspy and cored profiles~\cite{Walker09}. Recently, it has been shown that the detection of distinct stellar populations in dwarf galaxies allows for measurements of the inner slope of the dark matter profile and may allow cored and cuspy profiles to be distinguished~\cite{Walker11b}.

Incorporating baryons into the N-body simulations dramatically increases their complexity. Predictions on the dark matter and total mass distribution require a realistic treatment of the baryons and their dynamical interactions with the dark matter. Extensive work is being done to quantify the effects of 
baryons~\cite{Zeldovich80,Blumenthal85,Merritt03,Merritt06} and black holes~\cite{Gondolo99,Merritt02} in modifying the dark matter distribution. 
The centre of galaxies are complex environments and a number of astrophysical processes may likely change the initial dark matter density distribution. Because baryons dissipate energy and so collapse to smaller scales than dark matter, they constitute a sizeable fraction of the mass in the central regions. In the central regions of galaxies the gravitational potential is dominated by baryons and the dark matter distribution is expected to evolve due to interaction with these components. 
Collisionless dark matter simulations have reached maturity and much effort has been devoted recently to implement gas hydrodynamics and a description of star formation within simulations~\cite{Gnedin08,Wise10,Teyssier01,Keres11}. 
Feedback processes including supernova winds, radiation from young stars, and radiation and heat from black hole accretion play a crucial role in galaxy formation. These processes have
an impact on the scaffolding of dark matter during galaxy formation; cuspy dark matter distributions in halos may be altered and tend to produce core-like dark matter distributions, reducing the potential for a CTA discovery of DM.

\subsubsection{The Nature of Dark Matter}
Present information indicates dark matter is non-baryonic and is compatible with a  collisionless fluid of cold and weakly interacting massive particles (WIMPs). A major motivation for WIMPs is that in the standard thermal picture of the early universe a particle with annihilation cross-section and mass of the order of the weak interaction leads to the observed dark matter relic density~\cite{Planck13}. To be dark matter, WIMPs have an average annihilation cross-section (multiplied by the relative velocity of the annihilating WIMPs) of $\rm \langle \sigma v \rangle = 3 \times 10^{-26} cm^3s^{-1}$. This point is discussed further in section~\ref{subsec:sigmavWIMP}. The dark matter particle is necessarily neutral of charge and colour and must be stable on cosmological time scales. For the discussion in this document it is assumed that there is only one type of non-baryonic dark matter which makes up the full amount of the relic density $\Omega_{\rm DM} h^2$. Latest results from Fermi-LAT~\cite{Ackermann13b} and Planck~\cite{Planck15} satellites are starting to probe thermal WIMPs with masses up to $\sim$100 GeV.

No candidate exhibits the necessary properties within the Standard Model of particle physics. However, theories beyond the Standard Model, mainly built in order to solve problems inherent to particle physics, like the unification of couplings at high energy and the hierarchy and naturalness problems~\cite{Susskind82}, do have dark matter candidates. The currently most popular candidates for WIMPs with masses from a few GeV to a several tens of TeV come from the supersymmetric and extra-dimensional theories.  In many SUSY models, the lightest supersymmetric particle is the lightest neutralino.  In models with extra dimensions, the dark matter candidates include
 the first Kaluza-Klein excitation of the $B^{(1)}$ boson and the neutrino $\nu^{(1)}$.
Axion-like particles are also among the popular candidates. 
A review of particle physics candidates can be found in Ref.~\cite{Bertone04}.

\subsubsection{Search Methods for Dark Matter}
Different complementary approaches are required to establish and corroborate a dark matter signal and to extrapolate
from a discovery to understanding the properties of dark matter in the universe. The four searches to carry out for non-gravitational signatures of dark matter in the form of WIMPs are:  direct detection, indirect detection, collider experiments, and astrophysical probes sensitive to non-gravitational interactions of dark matter. 
The direct-detection method looks for interactions of dark matter particles embedded in the Milky Way's dark matter halo in Earth-based detectors (see Ref.~\cite{Cushman:2013zza} for a recent review) while the indirect-detection method looks for secondary particles emanating from dark matter annihilations or decays in our own and other galaxies. 
The two methods are complementary; positive evidence seen with distinct methods would provide convincing confirmation of the discovery of dark matter.
To elucidate the particle properties of the dark matter, collider experiments  have searched for evidence of dark matter particle candidates over the past three decades. Many searches for direct production of supersymmeteric particles (sparticles) have been made at the LEP, TeVatron and LHC colliders as well as various fixed target experiments. All such searches have led to negative results. Many parameters and branching ratios measured in accelerator experiments do, however, lead to strong constraints and indications of where the dark matter particles may lie. 
The particle properties of dark matter can be also constrained through its impact on astrophysical observables. In particular, non-gravitational interactions of dark matter can affect the densities of dark matter present in the central regions of galaxies or the amount of dark matter substructure found in galactic halos \cite{Tulin13}. Such interactions may also alter the cooling rates of stars and influence the pattern of temperature  fluctuations observed in the cosmic microwave background.

The indirect search method looks for cosmic radiation emitted from annihilations of pairs of WIMPs in regions of the surrounding universe with a high dark matter density. Different experiments search for different annihilation products and currently searches are in progress with charged cosmic rays, gamma rays and neutrinos. 
Experiments with neutral particle probes (gamma rays and neutrinos) can point directly to the annihilation sources, while charged cosmic rays, at least below energies of $\sim$10$^{19}$ eV, are deflected considerably by 
Galactic and intergalactic magnetic fields and cannot be used to trace back to any particular location. 
Indirect detection of dark matter annihilations through gamma rays has attracted much interest due to several unique properties of gamma rays. First of all, they do not scatter appreciably during their travel through
the Galaxy, but rather point back to the site where the annihilation took place. Also, absorption can generally be neglected, as the cross-section for scattering on electrons and nuclei for GeV to TeV photons is small. This means that one may use properties of the energy distribution resulting from these processes to separate a signal
from astrophysical foreground or backgrounds. And, as the electromagnetic 
cross-section of gamma rays is so much higher than the weak interaction cross-section 
for neutrinos, they are relatively easy to detect.

\begin{figure}[h!]
\begin{centering}
\includegraphics[clip=true,width=8cm,height=6.3cm]{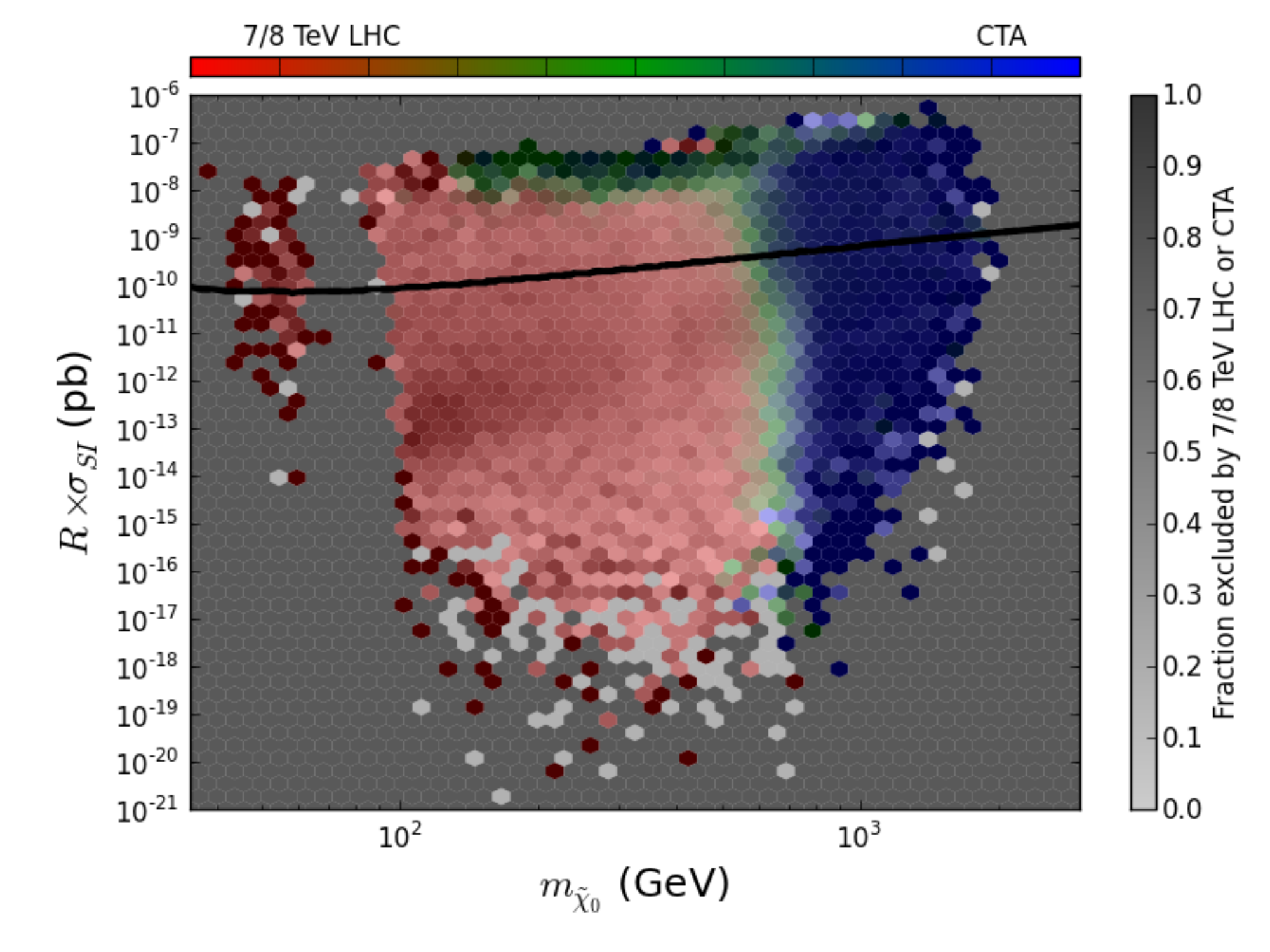}
\includegraphics[clip=true,width=7.4cm]{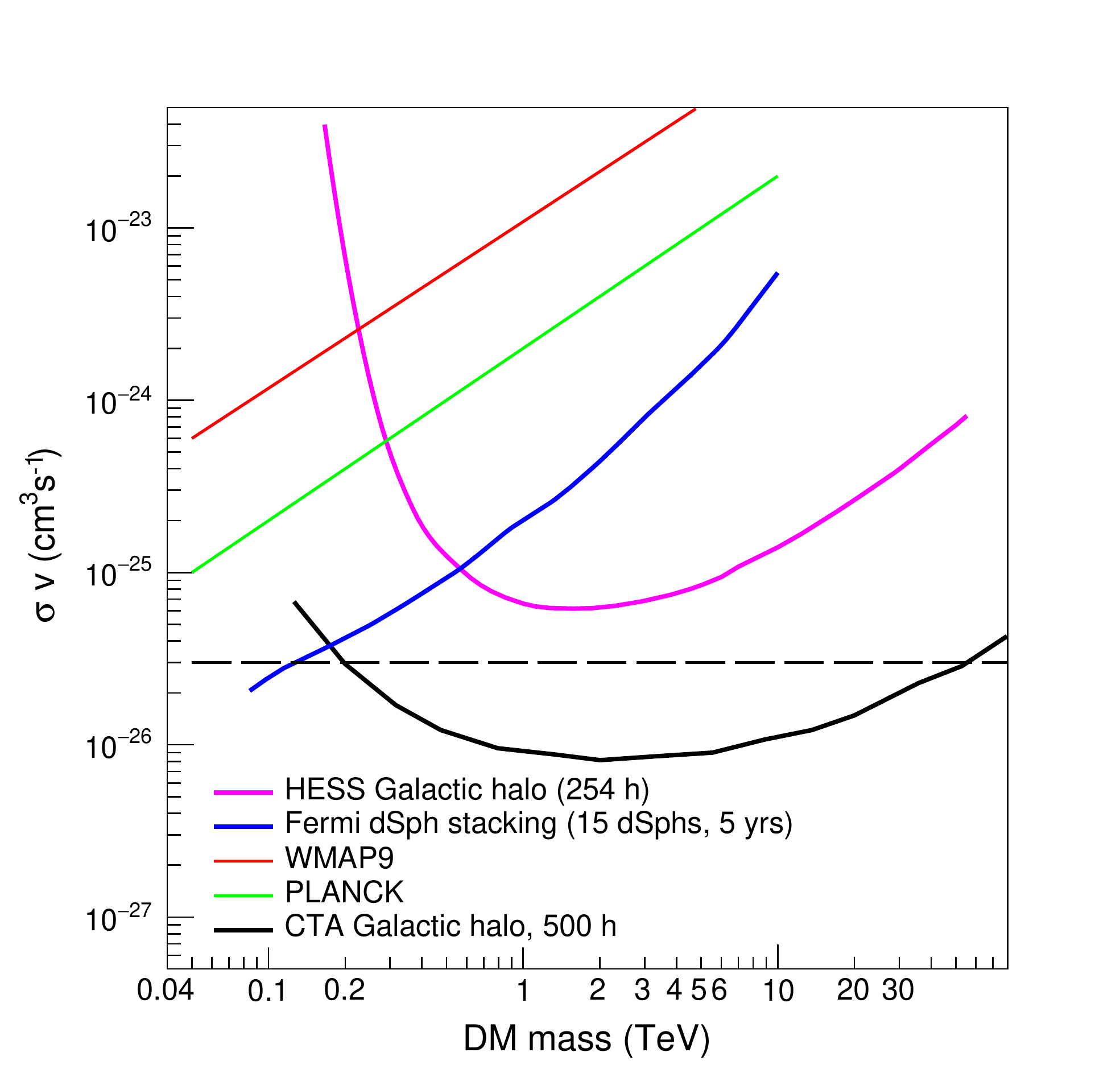}
\caption{Left: Comparisons of models from the phenomenological minimal supersymmetric model
(pMSSM) surviving or being excluded by future direct-detection, indirect-detection and collider searches in the neutralino mass-scaled spin-independent cross-section plane. The spin-independent XENON1T exclusion is shown as a solid black line. The models accessible to CTA (blue) and LHC (red) are shown.
Figure extracted from~\cite{Cahill-Rowley:2013dpa}. Right:
Current best limits on the annihilation cross-section from indirect detection 
(Fermi-LAT dwarf spheroidal galaxies stacking analysis,  W$^{+}$ W$^{-}$ channel \cite{Ackermann15c} and 
H.E.S.S. Galactic halo, W$^{+}$ W$^{-}$ channel \cite{Abdallah16c}) and cosmic 
microwave background
experiments (WMAP and Planck, $b\bar{b}$ channel \cite{Planck15}) compared to 
the projected sensitivity for CTA from observations of the Galactic halo 
for the Einasto profile and W$^{+}$ W$^{-}$ channel.
 \label{fig:complementarity}
}
\end{centering}
\end{figure}

These different search methods are sensitive to different couplings and different dark matter candidates. Also, the diverse experiments are sensitive to different dark matter particle masses. For a complete understanding of the nature of dark matter, these different techniques are complementary and essential. Figure~\ref{fig:complementarity} shows the survival and exclusion rates from the direct, indirect and LHC searches and their combinations in the plane of scaled spin-independent cross-section versus lightest supersymmetric particle (LSP) mass~\cite{Cahill-Rowley:2013dpa}. {The spin-independent cross-section is scaled to the fraction of DM provided by the WIMPs.}
The relative contributions arising from the LHC and CTA searches to the model survival/exclusion are
shown. 
It is clear that CTA dominates for large LSP masses, 
which correspond mostly to the neutral wino and Higgsino LSPs, and it also competes with the LHC throughout the band along the top of the distribution.

\subsubsection{Annihilation of Dark Matter Particles} 
For the indirect search experiments, complete rate predictions rely on calculations of the numbers and spectra of the relevant particle species in the annihilation reaction final state. To be able to self-annihilate, the dark matter candidate is most often  
a Majorana particle or a Dirac particle with no matter-antimatter asymmetry,
but a complex scalar, or even a vector-like particle, could
be a possibility. In all annihilation locations, the relative velocity of the WIMPs is low and usually annihilation rates are calculated in the null velocity limit where only
the s-wave term contributes. In this limit, the annihilation products in the leading order of perturbation theory are mostly pairs of Standard Model fermions/anti-fermions and neutral pair combinations of gauge or Higgs bosons.

\begin{figure}[h!]
\begin{centering}
\includegraphics[clip=true,width=7.5cm]{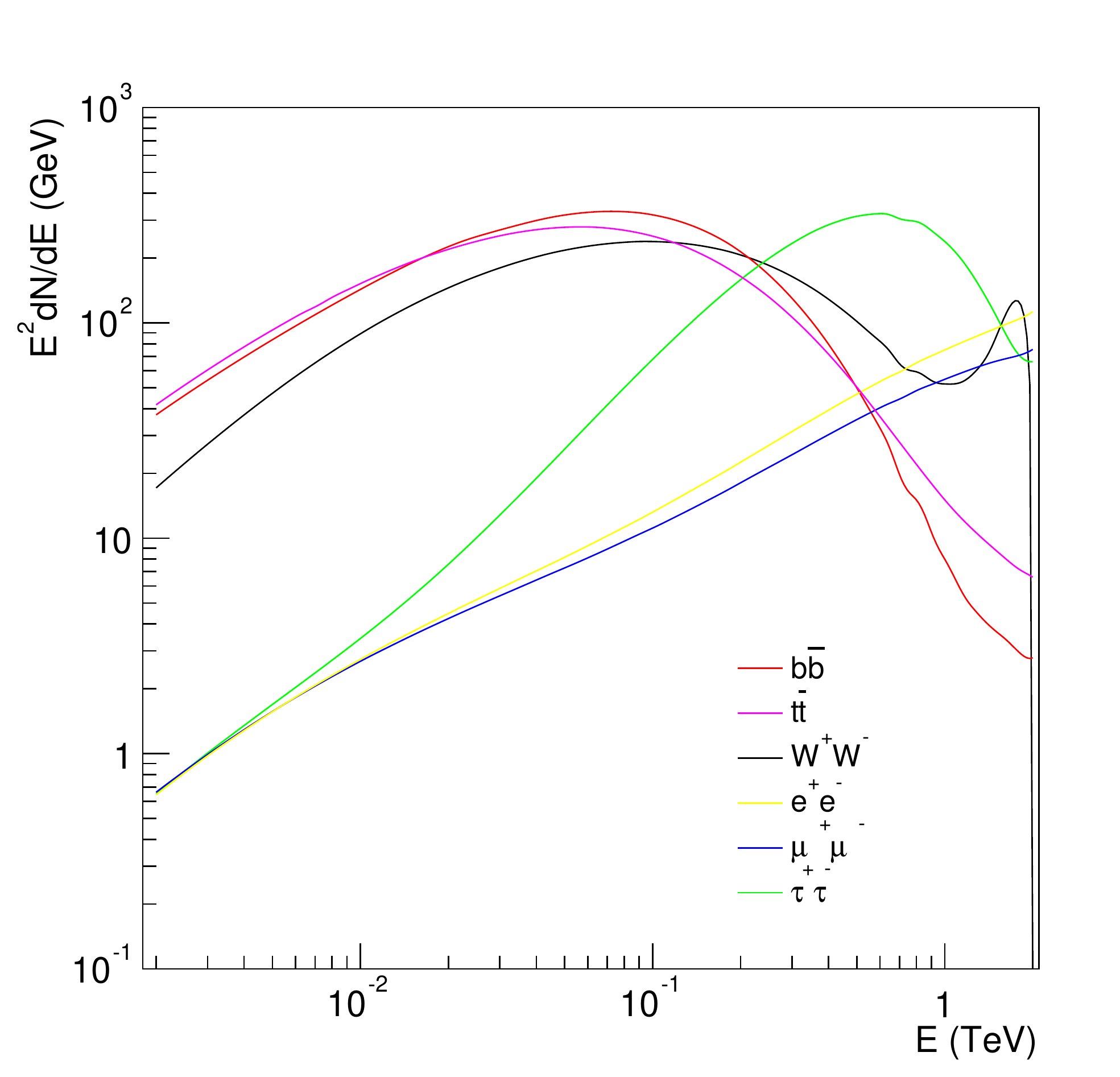}
\includegraphics[clip=true,width=7cm]{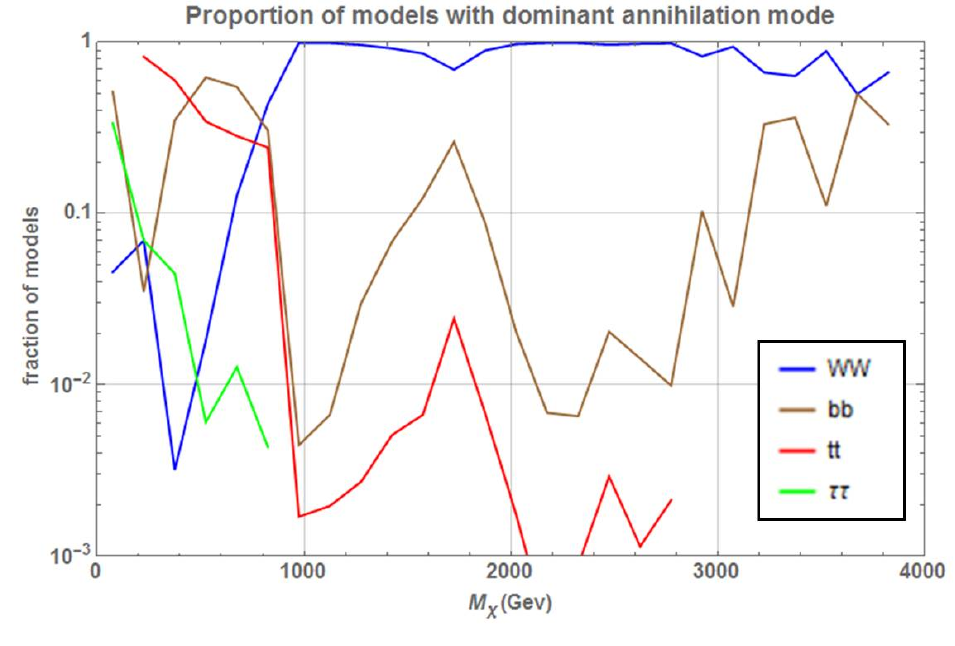}
\caption{Left: Annihilation spectra for the continuum signals from the quark, lepton and gauge boson primary channels for a 2 TeV dark matter mass. The line-like feature expected from 
the virtual internal bremsstrahlung process contribution is particularly prominent for the W$^+$W$^-$ channel.
Right: The dominant annihilation modes in the pMSSM scan of Ref.~\cite{Roszkowski14}.  As a function of neutralino mass, the plot shows the fraction of models with each of the annihilation modes as indicated. It should be noted that in general for any particular model more than one mode contributes and the dominant mode is the one with the largest branching fraction. 
 \label{fig:annihilationchannels}
}
\end{centering}
\end{figure}

Three types of dark matter annihilation spectra are expected in the final state: (i) a continuum of gamma rays up to the dark matter mass  from the decays of neutral pions produced by hadronization and/or decays of the annihilation products; (ii) a monochromatic gamma-ray signal produced by 
loop-order induced processes at E$_{\gamma}$ = m$_{\rm DM}$; and (iii) line-like features close to the dark matter mass from radiative corrections to processes with charged final states (e.g. virtual internal bremsstrahlung). These spectral features provide powerful discrimination against the more smooth spectra expected for standard astrophysical sources. The left panel of Figure~\ref{fig:annihilationchannels} shows typical spectra arising from the above-mentioned processes.
The right panel of Figure~\ref{fig:annihilationchannels} indicates the dominant annihilation modes as a function of the neutralino mass M$_{\chi}$ for the allowed models in the pMSSM scan of Ref.~\cite{Roszkowski14}. From this plot it can be seen that above 800 GeV the W$^+$W$^-$ is always the dominant annihilation mode (meaning that for the particular model, it is the mode with the largest branching fraction). Between 200-800 GeV, the $t\bar{t}$  and the $b\bar{b}$  modes dominate in different regions.  The $\tau^+\tau^-$  mode is only significant below 200 GeV.

The gamma-ray continuum from dark matter annihilation discussed in the previous section typically vastly dominates the total photon count, at least for energies E$_{\gamma}$ $\le$ 0.1 M$_{\chi}$.  However, the resulting spectrum is rather soft and does not contain any specific features that would unambiguously point to its dark matter origin. Higher-order processes, on the other hand, can add sharp spectral features to the high-end part of the spectrum, E$_{\gamma}$ $\sim$ M$_{\chi}$, and would provide potential smoking-gun signatures for the detection of particle dark matter. In fact, the detection of such features would not only help to discriminate a signal from the background~\cite{Bringmann11}, but would also provide valuable information about the particle nature of the annihilating dark matter.

The first signal considered historically of this type is the direct annihilation of dark matter pairs into $\gamma X$ (where X = $\gamma$, Z, H or some new neutral state). This process is necessarily loop-suppressed because the dark matter particles carry no charge, but it leads to the striking signature of monochromatic photons with an energy of $\rm E_{\gamma} = M_{\chi} (1-M_{X}^2/4 M_{\chi}^2)$. The discrimination of these lines is generally challenging, though annihilating Kaluza-Klein dark matter may provide a noteworthy exception in that it can lead to the fascinating signature of several equidistant lines at TeV energies. For thermally produced dark matter, one would naively expect that the process $\chi\chi \rightarrow \gamma\gamma$  happens at a rate of $\rm \alpha^2 < \sigma v>_{\rm thermal} \sim 10^{-31} cm^3s^{-1}$, where $\alpha$ is the fine structure constant. While this naive estimate falls well below the sensitivity of CTA, there are several mechanisms that can significantly enhance line signals - in particular at the high energies accessible by CTA, see, {\it e.g.}, Ref.~\cite{Arina09}.

At first order in $\alpha$, pronounced spectral features can also be generated by an additional photon in the final state whenever dark matter annihilates to charged particles. This process is known as internal bremsstrahlung and one may further distinguish between final state radiation and virtual internal bremsstrahlung, VIB~\cite{Bringmann07}. The former produces a model-independent hard spectrum with a sharp step-like cutoff at $E_{\gamma} = M_{\chi}$ \cite{Birkedal05}, like in the case of Kaluza-Klein dark matter~\cite{Bergstrom04}. VIB, on the other hand, dominates for TeV dark matter annihilating into 
W bosons~\cite{Bergstrom05} or if the tree-level annihilation of dark matter into light fermions is suppressed by a symmetry~\cite{Bergstrom89,Toma13}, such as the helicity suppression for neutralino dark matter. Given the energy resolution of CTA, VIB features are essentially indistinguishable from a monochromatic line (though in principle the exact spectral shape is highly model-dependent). At lower energies, internal bremsstrahlung of gluons or electroweak gauge bosons may also both change and significantly enhance the photon spectrum.

\subsubsection{Parameters Expected for WIMP Dark Matter}
\label{subsec:sigmavWIMP}
For a standard thermal history of the early universe, the abundance of a particle is related to the thermally-averaged annihilation cross-section times relative velocity ($\langle \sigma_{\chi\chi} v\rangle$). Initially the WIMPs were in thermal and chemical equilibrium with the hot ``soup'' of Standard Model particles. The WIMPs dropped out of thermal equilibrium 
(``freezed-out'') once the rate of interactions that change Standard Model particles into WIMPs, or vice-versa, became smaller than the Hubble expansion rate of the universe. After freeze-out, the co-moving WIMP density remained essentially constant and the dark matter relic density and the annihilation cross-section are thus 
inversely proportional
(neglecting the logarithmic dependency on the dark matter mass): $\Omega_{\rm DM} h^2 \simeq  K/\langle \sigma_{\chi\chi} v\rangle$, where the proportionality constant is $\rm K \simeq 3\times10^{-27} cm^3 s^{-1}$ and is related to the cosmic microwave background temperature and the Planck mass.  For $\rm \Omega_{\rm DM} h^2  \sim 0.1$, this gives $\rm \langle \sigma_{\chi\chi} v\rangle \sim 3\times10^{-26} cm^3 s^{-1}$, which is referred to as the thermal cross-section. The fact that this cross-section is of the order of magnitude of a weak interaction is often referred to as the  
``WIMP miracle''.

\begin{figure}[h!]
\begin{centering}
\includegraphics[clip=true]{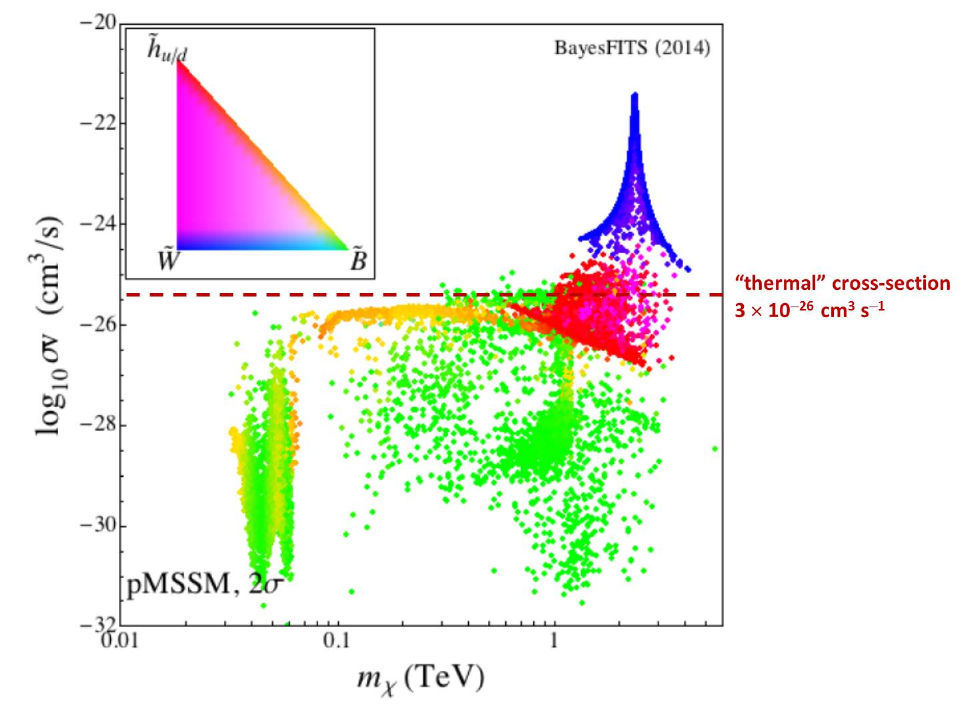}
\caption{Annihilation cross-section points from a 19-dimensional pMSSM fit from Ref.~\cite{Roszkowski15} which contains a set of basic constraints and direct search limits as explained in the reference. The colour coding identifies the composition of the lightest neutralino. Pure states are shown for the supersymmetric electroweak gauge bosons
(green for the bino and blue for the wino) and for the Higgsino (supersymmetric partner of the Higgs boson)
in red.  Admixtures are shown with intermediate colours in accordance with the legend. 
 \label{fig:annihilationxsection}
}
\end{centering}
\end{figure}

This thermal value of the cross-section is often used as a sensitivity goal for indirect searches; however, this value cannot be taken as a strict expectation, in general, since the cross-section in the early universe is not identical to the cross-section applicable to indirect searches at the present time. Detailed discussions of the differences can be found in Ref.~\cite{Roszkowski14}. In the early universe the relic density is obtained by using a momentum dependent cross-section including both annihilation of the LSP neutralino and co-annihilation with close-in-mass neutralinos and other sparticles. Due to the high temperature at freeze-out, the momentum dependence is different from present day annihilation which takes place essentially at rest. Furthermore, the present day cross-section contains no contribution from co-annihilation since the co-partners have all decayed away. Figure~\ref{fig:annihilationxsection} shows example points from a pMSSM model scan showing that many specific points are below the thermal cross-section and some are above, and hence searches should encompass a wider range of annihilation cross-sections. A strong enhancement of the annihilation cross section occurs for winos  around 2-3 TeV due to Sommerfeld enhancement.

\subsubsection{Rate of Gamma Rays in Detector}
The rate of gamma rays from dark matter is usually expressed as a separation into two terms
which characterise the astrophysical properties of the source and the particle physics contribution to the rate. The astrophysical terms are combined in a so-called J-factor term defined as :
\begin{equation}
 J_{\Delta \Omega} =  \int_{\Delta\Omega}  
   d\Omega
  \int_{\textrm{los}}  
  dl \times \rho^2[r(l)]   \,. 
\label{eq:jbar}
\end{equation}
The J-factor reflects the integral of the squared dark matter density distribution, $\rho^2$, over the line of sight (los)  and inside the observing angle $\Delta\Omega$. The dark matter density is conveniently parameterized as a function of the radial distance $r$ from the centre of the astrophysical object under consideration. Depending on the dark matter targets, the 
J-factors range from $\sim$10$^{21}$ to $\sim$10$^{24}$ GeV$^2$cm$^{-5}$.  
The number of observable events is then expressed as:
\begin{equation}
N_{\rm DM} = \frac{T_{\rm obs}\  J_{\Delta\Omega}\ \langle \sigma v \rangle}{8 \pi M_{\chi}^2} \int_{E_{\rm min}}^{E_{\rm max}} \frac{\rm d N_{\rm DM}}{\rm d E}(E) A_{\rm eff}(E) dE   \, , 
\label{eqn:expcount}
\end{equation}
where:
\begin{itemize}
\item $T_{\rm obs}$ is the live time of observation,
\item $\langle \sigma v\rangle$ is the thermally averaged velocity-weighted annihilation cross-section,
\item $M_{\chi}$ is the dark matter particle mass,
\item  $\rm d N_{\rm DM} (E)/ \rm d E $ is the energy spectrum of the gamma rays produced in the annihilation,
\item $A_{\rm eff}$  is the detector effective area, and
\item $E_{\rm min}$ and $E_{\rm max}$ are the energy limits for the measurement.
\end{itemize}
                                      
\subsection{Strategy}
\label{sec:dmstrategy}
The indirect dark matter search with CTA has several possible astrophysical targets, each with its own inherent advantages and disadvantages. The Milky Way represents a natural place to look for dark matter signatures and its centre is expected to be the brightest known source in the dark matter induced gamma-ray sky, although the exact magnitude is rather uncertain. The dark matter density profile in the Milky Way should lead to an annihilation signal  observable on large angular scales; however, astrophysical Galactic foregrounds coupled with the enormous spatial extent and the truly diffuse nature of this Galactic dark matter emission make separation between signal and background challenging. On the other hand, nearby dwarf spheroidal galaxies should provide easier separation of signal and background but yield comparatively lower signals because of both the distance and lower dark matter content compared to the Milky Way.

The concordance cosmological $\Lambda$CDM model predicts that the formation of visible structures has been guided by gravitational accretion of baryons onto previously formed dark matter over-densities. The astrophysical structures of interest result from the hierarchical formation of dark matter halos from primordial dark matter over-densities. The subsequent evolution of these dark matter halos occurred in many different ways. In particular, on galactic and 
sub-galactic scales, the process depended on halo parameters such as the halo mass and the mass density profile, the evolution history, and the conditions set by the local galactic evolution environment.
Some of the resulting halos could have been sufficiently massive to accrete enough baryons to initiate star formation and form galaxies, including the variety of satellite galaxies we actually observe in the Milky Way halo. 
In-falling dwarf galaxies ({\it e.g.}, dwarf irregular galaxies) approaching more central parts of their host halo could have evolved to dwarf spheroidal galaxies (dSphs), see {\it e.g.}~\cite{Mayer09}. These dwarf galaxies, being highly dark matter dominated and comparatively close by, form one of the primary targets for CTA observations of this programme.
However, for various reasons including the halo size, location, encounters, the baryonic content of the environment, or the presence of a central black hole, less massive dark matter clumps could have evolved differently and not into a visible dSph. They would constitute dark matter-dominated over-densities purely observable in gamma rays or cosmic rays emerging from dark matter annihilations. Such objects are commonly known as dark matter subhalos or dark matter clumps. 

It is clear that the scientific interpretation of any dark matter signal detected by CTA (or indeed a non-detection) in terms of a constraint on the properties of dark matter particles depends sensitively on our knowledge of the dark matter distribution within the targeted systems. This section considers the likely impact of observations and modeling over the next 5-10 years on our knowledge of the dark matter profiles of dSphs, the LMC and the central regions of the Milky Way.

Observations with the fully operational CTA are required, in order to maximize the sensitivity and to take advantage of the best energy and angular resolutions. In particular, the energy resolution is of utmost importance to identify possible spectral
features, {\it e.g.} bumps and the cutoff. If detected, these will provide crucial information in the identification of
the dark matter particle. In case of massive dark matter candidates, the continuous dark matter-induced gamma-ray spectra extends
down to the GeV scale. A low energy threshold is therefore mandatory to probe the largest energy range
and to be sensitive to dark matter candidates with low masses. Observations at the lowest possible zenith angle
are preferred, since they provide access to the lowest possible energies. Good weather conditions are also
required to keep an optimal energy resolution and to reduce systematic effects.

The observational strategy proposed for the CTA Dark Matter Programme is focused first on collecting a significant amount of data on the Galactic Centre. Complementary observations of a dSph galaxy will be conducted to extend the search. 
The Galactic Centre, the LMC, and galaxy clusters are valuable targets both for dark matter searches and for studies of non-thermal processes in astrophysical sources.
Data will be searched for continuum emission and line features, and strategies will be adopted according to 
the findings. Discoveries will modify any strategies defined {\it a priori}. 

Below we outline the strategy for CTA observations of the Galactic halo, dwarf galaxies, the LMC, 
and the Perseus cluster, respectively, followed by an overall summary of the targets.

\subsubsection{Milky Way}
\paragraph{Description}
The centre of the Milky Way has in the past been considered as a target for dark matter searches~\cite{Aharonian06a}. More recently, because of the rich field of VHE gamma-ray astrophysical sources in the region, the searches focus on the Galactic halo excluding the central region of Galactic latitude b$<$0.3$^{\circ}$~\cite{Abramowski11b}.
Even excluding the very central region, the total mass of dark matter in the Galactic halo together with its proximity to Earth make it the most promising source for dark matter searches with CTA. The inconvenience of this target, however, is the fact that being a diffuse source, the integration over the inner halo, while yielding a large signal, gives a very large instrumental background from misidentified charged cosmic rays. Furthermore, there are astrophysical backgrounds from various sources which must be understood, even with the very central region excluded from the analysis. It is believed that the disadvantages of the Milky Way target can be overcome with sufficient experimental effort to control systematic effects in the background subtraction and modeling. The expertise required for this analysis strengthens the case for this programme to be conducted by the CTA Consortium.

Standard astrophysical processes typically 
have steeper VHE spectra than the expected dark matter-induced gamma-ray continuum emission. Given the wealth of other high-energy emitters expected in this region, the search for a dark matter component requires a very deep exposure to enable detection and detailed spectro-morphological studies, a good understanding of the instrumental and observational systematics, and accurate measurements of other astrophysical emission in the region to be able to reduce any contamination to the dark matter signal.
A deep exposure for the Galactic Centre observation will provide the means for an in-depth study and better understanding of the  astrophysical emissions in this region. 

Stellar dynamics in the Milky Way is dominated by the gravitational potential of baryons up to the kpc scale and the dark matter density distribution in the inner kpc region can thus be accommodated by both cuspy (NFW, Einasto) profiles and cored (isothermal, Burkert) profiles. 
Important efforts are ongoing to accurately simulate the baryon impact on the dark matter distribution in the central region of galaxies.  With rapid progress being made in the field, a more comprehensive picture for
the central region of the Milky Way is expected by the time of CTA observations with reduced theoretical uncertainties on the dark matter distribution. Although the observation strategy may substantially differ for a kpc-size core profile compared to a cuspy profile, the detection of a dark matter signal and the detailed study of its morphology would help to resolve this important question. 

\paragraph{Evolution of knowledge}
The inner region of the Milky Way presents a very complex environment in which to determine the distribution of dark matter. Present kinematic and microlensing data are consistent with an inner log slope for the halo of -1.0~\cite{Iocco11}, the value predicted by dark matter only simulations of structure formation, although the observational constraints are very broad. Simulations including baryonic physics can produce steeper profiles through adiabatic contraction of gas and supermassive black hole formation. However, outflows produced by AGN activity and/or star formation have the opposite effect. The theoretically expected value is therefore almost as uncertain as the observationally determined one. 
Given the high level of activity in the field of galaxy formation, it is expected that improved simulations, {\it e.g.} EAGLE~\cite{Schaye14}, will provide further insights into theoretical expectations over the coming years. However, the interpretation of CTA observations requires tighter observational constraints. The two likely avenues for improvement on this front are from microlensing and large-scale radial velocity surveys.  Microlensing surveys can be used to probe the baryonic mass function towards the Galactic Centre and hence place limits on the baryonic contributions to the mass distribution of the inner Galaxy, {\it e.g.}~\cite{Wyrzykowski14}. Kinematic surveys such as ARGOS~\cite{Freeman12}; BRAVA~\cite{Howard08}, GAIA~\cite{gaia} and GIBS~\cite{Zoccali14} are collecting samples of thousands of radial velocities which will be used to map the dynamical structure of the bulge and bar regions. Together, the microlensing and kinematic data have the potential to provide improvements on the constraints in the mass profile of the inner Galaxy. 

\paragraph{Observational Strategy}
The Galactic Centre observations will be taken with multiple grid pointings with offsets from the 
Galactic Centre position of about $\pm$1.3$^{\circ}$  to cover the central 4$^{\circ}$ 
as uniformly as possible. The observation strategy 
defined explicitly to search for dark matter will require 525 hours to probe cuspy profile dark matter distributions. In the Galactic Centre KSP a further 300 hours are proposed for astrophysics
covering up to latitudes $\pm$10$^{\circ}$.
These data will also be included in the analysis for dark matter to improve the sensitivity for cored dark matter density profiles. 
Given the major scientific impact of a positive result, we propose that
the initial 525 h exposure be done in the first three years of CTA operation with high priority.

\begin{figure}[!ht]
\begin{centering}
\includegraphics[width= 7.4cm]{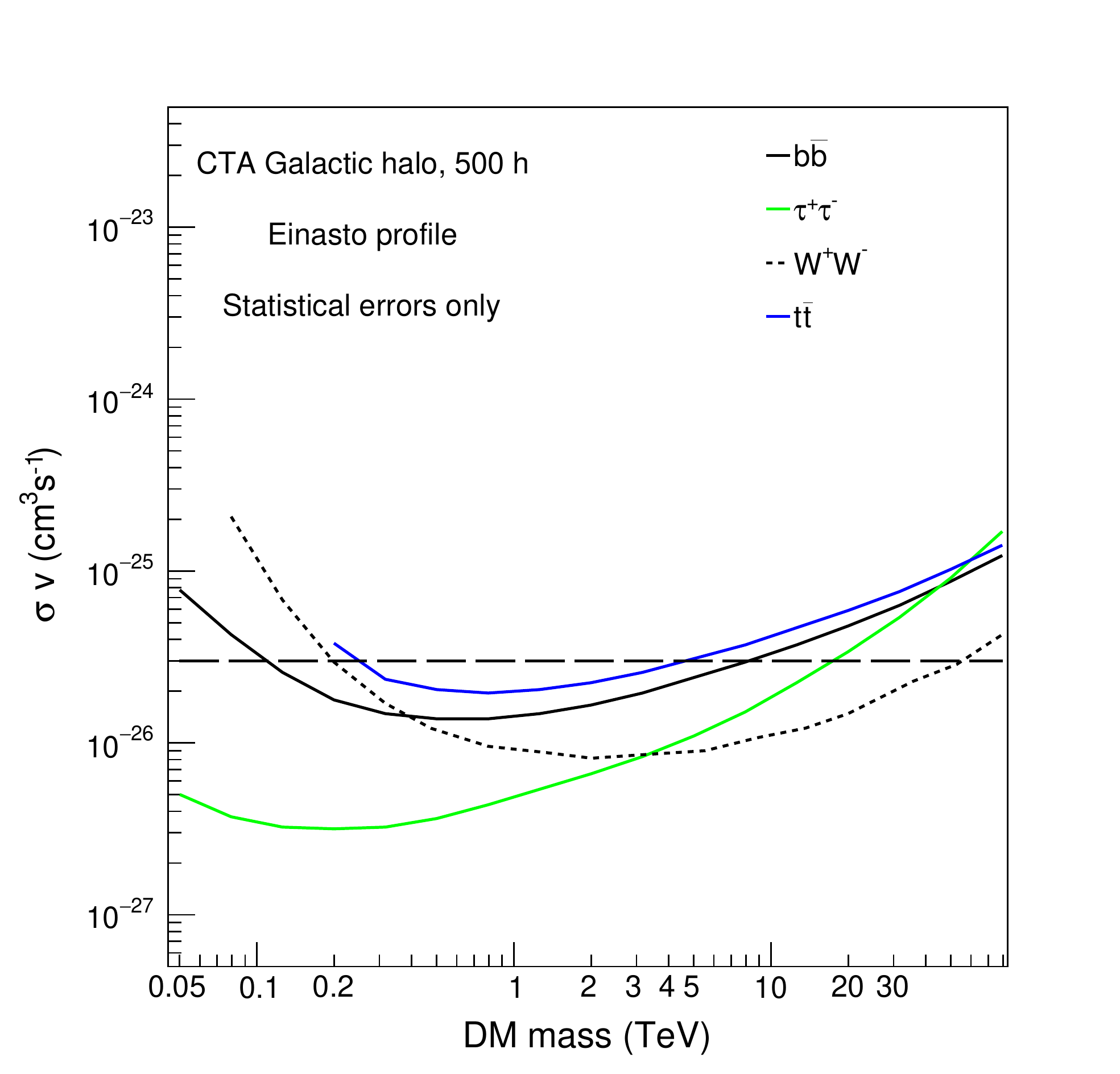}
\includegraphics[width= 7.4cm]{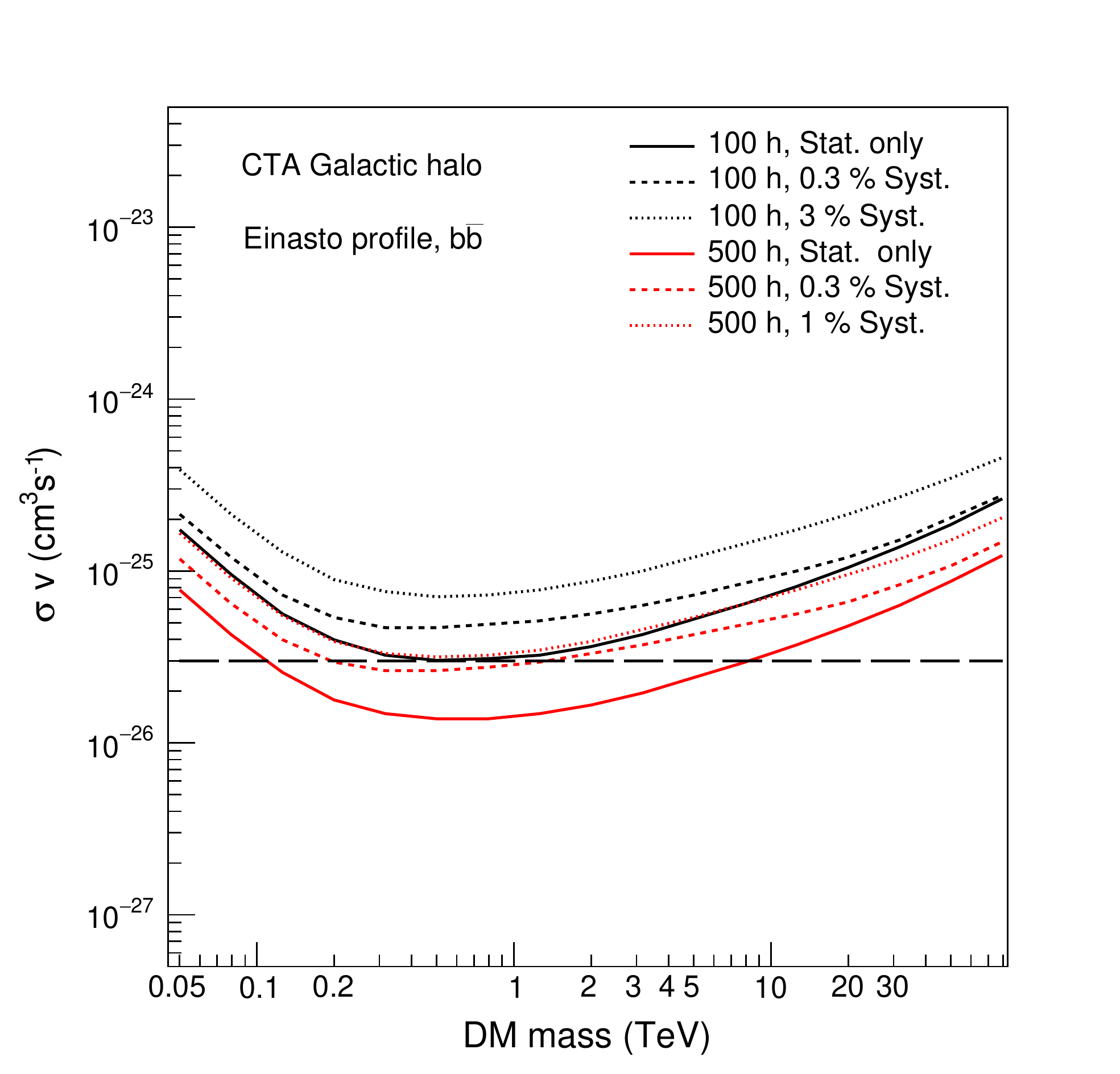}
\caption{Left: CTA sensitivity for $\langle \sigma v\rangle$ from observation of the Galactic halo for different annihilation modes as indicated.
Right: CTA sensitivity for $b\bar{b}$ annihilation modes for different conditions, black is for 100 hours of observation and red is for 500 hours. The solid lines are the sensitivities only taking into account the statistical errors while 
the dashed and dotted curves take into account systematics as indicated. The dashed horizontal lines approximate the level of the thermal cross-section of $\rm 3 \times 10^{-26} cm^3s^{-1}$.}
\label{fig:gcsens}
\end{centering}
\end{figure}

\begin{figure}[!ht]
\begin{centering}
\includegraphics[width= 8.0cm]{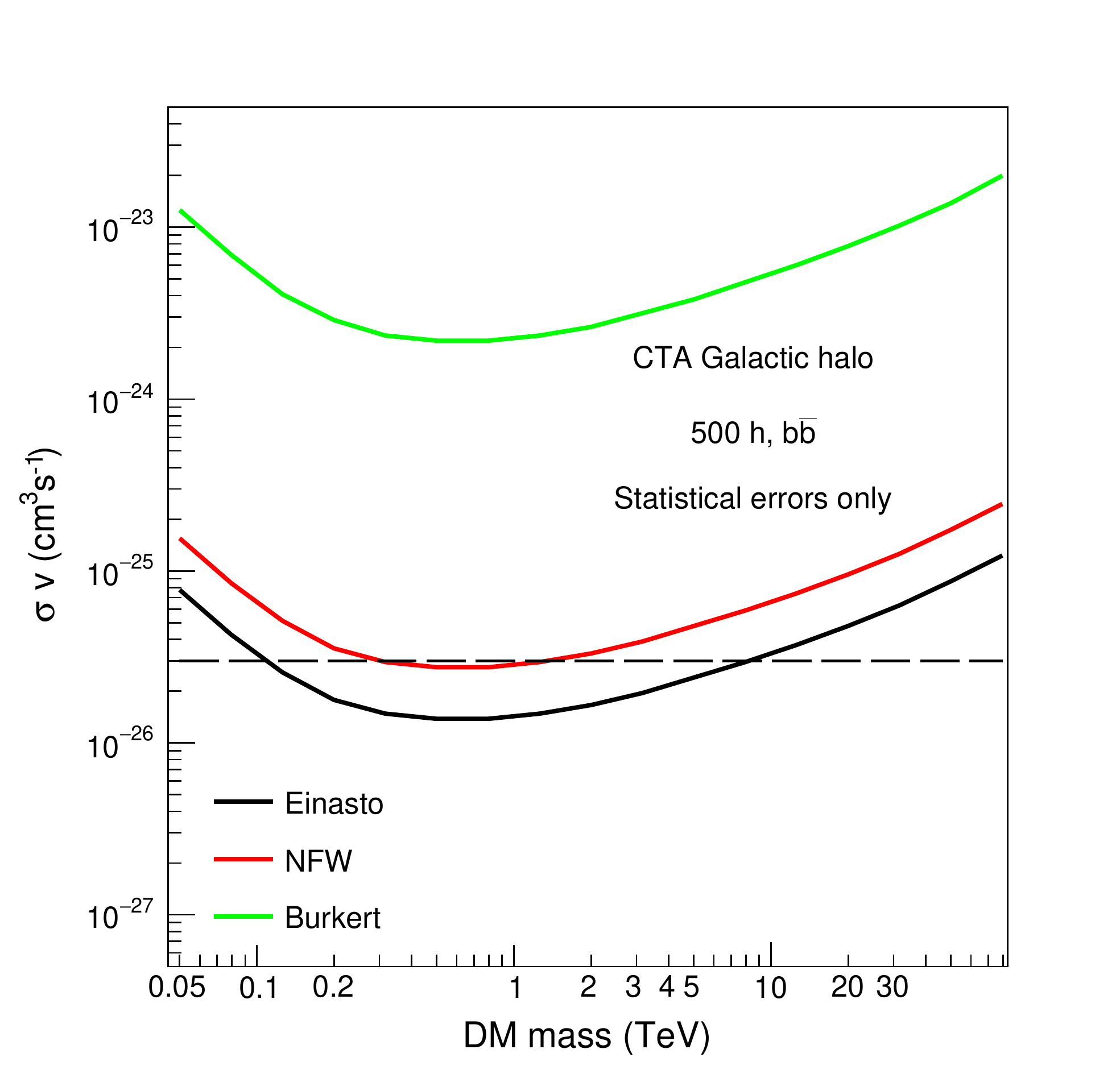}
\caption{CTA sensitivity for $\langle \sigma v\rangle$ on the Galactic halo for cupsy (NFW, Einasto) and cored (Burkert) dark matter halo profiles. The sensitivities are plotted for 500 h observation, the $b\bar{b}$ annihilation channel, and for statistical errors only. The dashed horizontal lines indicate the level of the thermal cross-section of  $\rm 3 \times 10^{-26} cm^3s^{-1}$.}
\label{fig:gcsensprofile}
\end{centering}
\end{figure}

\begin{figure}[!ht]
\begin{centering}
\includegraphics[width= 12cm]{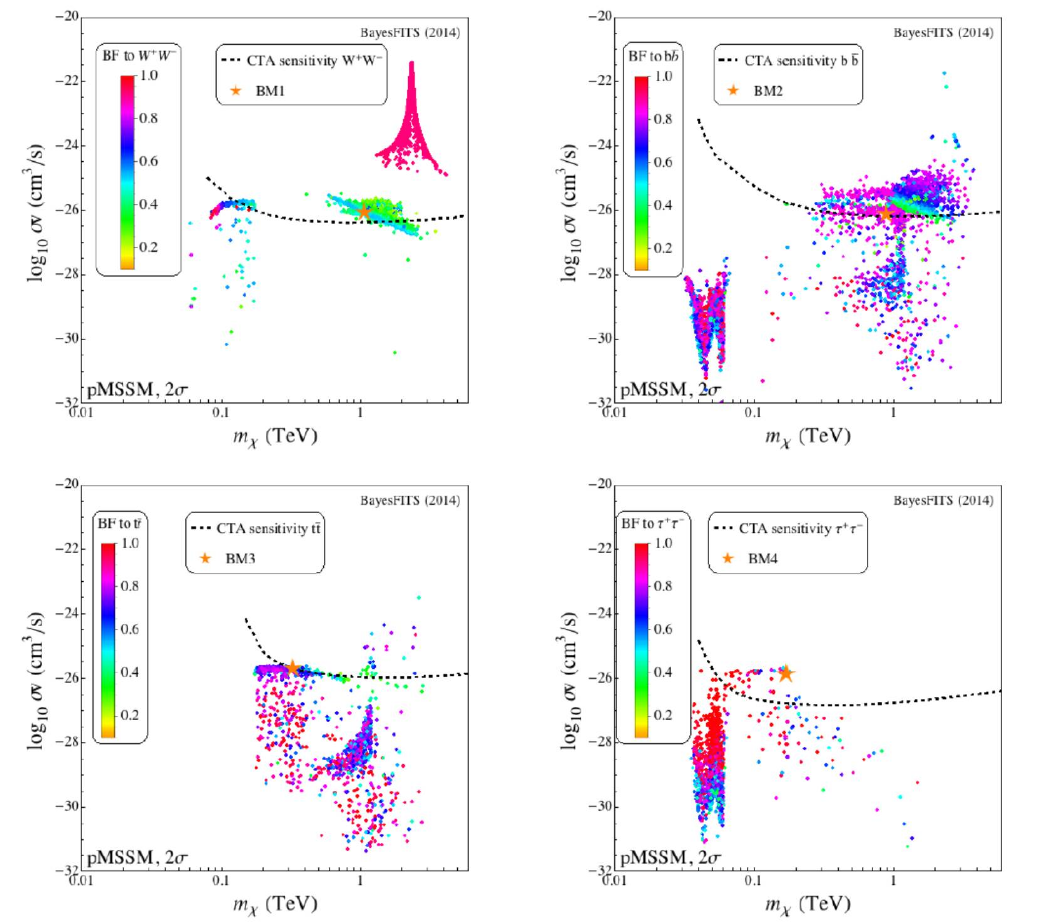}
\caption{Comparison of allowed models from~\cite{Roszkowski15} for each the dominant modes: W$^+$W$^-$, $b\bar{b}$, $t\bar{t}$ and $\tau^+\tau^-$ in the four panels as indicated with the corresponding sensitivities as calculated in their paper.  The colour code shows the value of dominant branching fraction for each point (The stars mark the particular benchmark points discussed in Ref.~\cite{Roszkowski14}).}
\label{fig:gc_allowed}
\end{centering}
\end{figure}

\paragraph{Performance}
The quest for dark matter requires a deep and uniform exposure over several degrees around the central black hole Sgr A* to allow for both spectral and spatial morphological studies, a deep understanding of the instrumental and observational systematics, and precise determinations of the standard astrophysical emissions. The expected CTA energy and angular resolutions are key ingredients to disentangle a dark matter signal from standard astrophysical background. 
The sensitivity predictions for observations in the Galactic Halo are shown in Figure~\ref{fig:gcsens}. The left-hand plot shows the sensitivity for different annihilation modes ($b\bar{b}$, $\tau^+\tau^-$, 
${\rm W}^+{\rm W}^-$, $t\bar{t}$) and the right-hand plot for different observation times with a method~\cite{Lefranc:2015pza} to include systematic uncertainties on the residual cosmic-ray background as indicated in the caption.
Figure~\ref{fig:gcsensprofile} shows the CTA sensitivity for various dark matter halo profiles satisfying stellar dynamics. Even in the case of a pessimistic dark matter distribution at the Galactic Centre , 
{\it e.g.} a Burkert profile, the sensitivity of CTA is comparable to what is expected for a classical dwarf galaxy (see Figure~\ref{fig:Sculptor}).
Figure~\ref{fig:gc_allowed} compares the CTA Galactic halo sensitivity limit predictions with the pMSSM model scan of Ref.~\cite{Roszkowski14}. Each panel shows the branching fraction of the primary annihilation channels for a given model. Similar studies can be found in 
Ref.~\cite{Cahill-Rowley:2013dpa,Catalan:2015cna}. It can be seen that for models with M$_{\chi}>$ 500 GeV CTA will be the only experiment able to probe the vast majority of models.

Similar studies have been carried out in the recent literature on the CTA sensitivity prospects 
towards the Galactic Centre~\cite{Pierre14,Silverwood14,Lefranc:2015pza}.  
A careful examination of these works reveals  differences in the dark matter distribution modeling and/or 
in the sensitivity computation that significantly impact the expected sensitivity. Among the
differences  are the
instrument's response functions together with the residual background used for CTA and the dark matter distribution in the innermost kpc of the Galactic Center. 
A description of the instrument response functions used for this study can be found in the
Appendix.

\begin{figure}[!ht]
\begin{centering}
\includegraphics[width= 10cm]{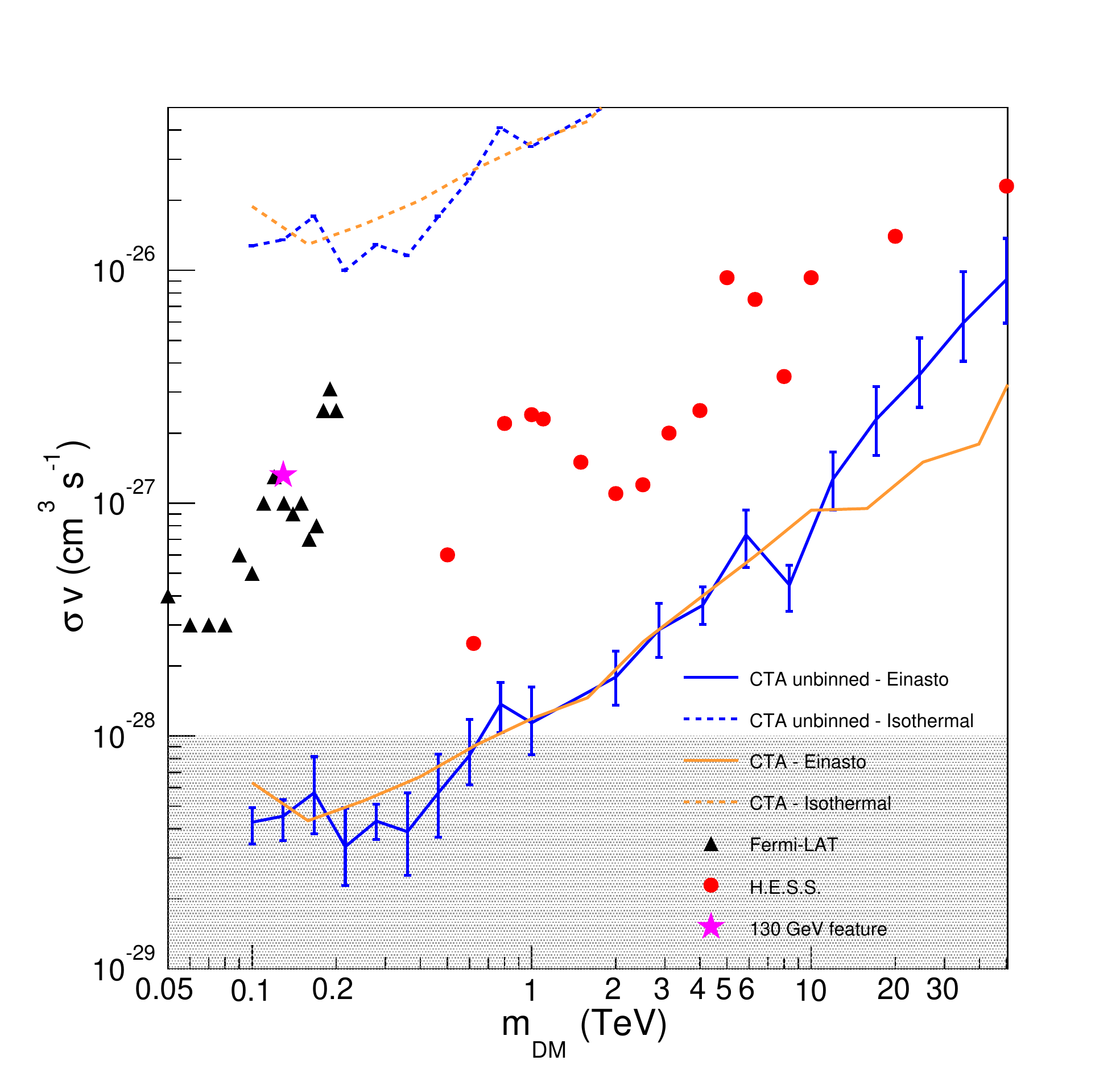}
\caption{Sensitivity of CTA to monochromatic gamma-ray signals from
dark matter annihilation, with E = M$_{\chi}$, after 500 h of observation of a region with 1$^{\circ}$ 
radius around the Galactic Centre using an unbinned likelihood analysis (blue line)
and a differential sensitivity analysis (orange curve) assuming an
Einasto profile. For comparison, the currently best limits from Fermi~\cite{Ackermann12a} (black triangles)
and H.E.S.S.~\cite{Abramowski13c} (red dots) are also shown, along with the much discussed line-like
feature at around 130 GeV~\cite{Bringmann12,Weniger12} (magenta star). The dashed lines also show the mean upper-limits obtained in case of a Burkert profile. The natural scale for monochromatic gamma-ray signal is highlighted as a black shaded area.  
}
\label{fig:gc_line}
\end{centering}
\end{figure}
The estimate of CTA sensitivity in the monochromatic line search is shown in Figure~\ref{fig:gc_line}, where the data are analysed in a circle of 1$^{\circ}$ radius encompassing the Galactic Centre 
position assuming two different dark matter distribution profiles (Einasto and isothermal). The modeling of the standard astrophysical emission is obtained by taking into account the already detected gamma-ray emission by H.E.S.S. in that region convolved with the
instrument response functions as well as the residual rate of charged cosmic rays.
Using the energy-dependent variation of the energy resolution, an unbinned analysis of Monte-Carlo simulations was performed. A sliding search window of four times the size of the energy resolution centred at the position of the expected line is used. The results in Figure~\ref{fig:gc_line} show this method as well as an analysis cross-check, assuming a 500 hours exposure and an Einasto dark matter distribution profile. While the current analyses do not include systematic uncertainties, these are expected to be small. As no known standard astrophysical sources would produce such sharp energetic features, the discrimination against background is easier in line searches compared to the continuum emission search.

\subsubsection{Dwarf Spheroidal Galaxies and Dark Clumps}
\paragraph{Description}
The dwarf spheroidal galaxies (dSphs) of the Local Group could give a clear and unambiguous detection of dark matter. They are gravitationally bound objects and are believed to contain up to $\mathcal{O}$(10$^3$) times more mass in dark matter than in visible matter, making them widely discussed as potential targets. 
Being small and distant, many of the dwarf galaxies will appear as point sources to CTA and hence the nuisance of the instrumental background is much reduced. Although less massive than the Milky Way or the LMC, they are also environments with a favourably low astrophysical gamma-ray background making the unambiguous identification of a dark matter signal easier compared to the Galactic Centre or LMC. Neither astrophysical gamma-ray sources (supernova remnants, pulsar wind nebulae,...) nor gas acting as target material for cosmic rays have been observed in these systems. 

The search program selects dwarf galaxy targets on the basis of both their J-factors and relative J-factor uncertainties. Due to the larger available sample of spectroscopically measured stars, the classical dwarf galaxies such as Draco, Ursa Minor, Sculptor, and Fornax have significantly smaller uncertainties on the
J-factor than the ultra-faint dwarf galaxies \cite{Ackermann13b}.  However, several of the ultra-faint galaxies ({\it e.g.} Ursa Major 2) have J-factors which are larger than the J-factors of the best classical dwarfs which, 
to some degree, 
outweighs the larger J-factor uncertainties for these objects. A recent evaluation of dwarf galaxy J-factors was presented in Ref.~\cite{Martinez13} which used a Bayesian hierarchical modeling analysis to constrain the dark matter mass and scale radius in 18 dSph galaxies with good spectroscopic data. Using these mass models we can compare the CTA detection prospects of the dwarf galaxies in this sample on the basis of the J-factor integrated within an angular region of radius 0.1$^{\circ}$ (J$_{\rm 0.1}$), where the angular scale of the integration region has been chosen to match
the CTA point spread function at 100~GeV.  This study shows that two of the best classical dwarf candidates are Draco in the northern hemisphere with $\rm log10(J_{\rm 0.1}/GeV^2cm^{-5}) = 18.69 \pm 0.16$ and Sculptor in the southern hemisphere with $\rm log10(J_{\rm 0.1}/GeV^2cm^{-5}) = 18.47 \pm 0.18$.  The most promising known ultra-faint galaxies are best observed from the northern hemisphere although discovery of new ultra-faints in the southern hemisphere is anticipated within the next 5-10 years as new optical surveys come online. ultra-faint dwarf candidates observable from the northern hemisphere include Segue 1, Coma Berenices and Ursa Major 2.
Very recently, the Dark Energy Survey (DES) data revealed eight Milky Way satellites~\cite{2015arXiv150302584T,2015arXiv150302079K} in the southern hemisphere likely to be ultra-faint dwarf galaxies. In particular, Reticulum II is the most attractive of these, given both its proximity and high dark matter content~\cite{Bonnivard15}.  
With forthcoming in-depth studies of their dark matter content and distribution, these  ultra-faint dSph candidates may be well-motivated targets for CTA.

Structure formation predicts gravitationally bound dark matter clumps down to much lower masses than observed for dSph galaxies. The low-mass cutoff of the clump distribution is related to the free-streaming scale of dark matter particles in the early universe and is expected to be between 10$^{-12}$ M$_{\odot}$ and 10$^{-3}$ M$_{\odot}$ for typical WIMP scenarios~\cite{Bringmann09}. The number-counts distribution of clumps in Milky-Way-like galaxies has been investigated with numerical N-body simulations, finding a power law: dN/dM$\propto$ M$^{\alpha}$  in clump mass M, with a high-mass cutoff at M $\sim$ 10$^{10}$ M$_{\odot}$ and an index $\alpha$ between -1.9 and -2.0~\cite{Diemand08,Springel08}. 
For the Milky Way, this would imply a clumpiness factor of $\sim$18\% (referring to the relative contribution of the total mass in clumps to the total mass of the main halo). The spatial distribution of dark matter clumps is biased away from the smooth central dark matter distribution of the host halo, {\it i.e.}, the majority of clumps populate outer halo regions (r $>$ 100 kpc). 

In the near future, experiments will be able to probe the subhalo population above 10$^5$ M$_{\odot}$ using gravitational milli-lensing of background objects~\cite{Zackrisson09}, while the statistical properties of lighter clumps (down to mass scales below a Solar mass) in the Solar neighbourhood can be measured from gravitational nano-lensing~\cite{Chen10,Garsden11}. Additional statistical techniques to estimate the presence and properties of moving sources contributing to a diffuse background emission are being
investigated as well~\cite{GeringerSameth10}. Dark clumps orbiting in the solar vicinity also affect the kinematics of stellar streams. A peculiar distribution of gaps would be imposed on local stellar streams by impacting clumps of all mass scales. First studies of local streams have already hinted at clump impacts~\cite{Carlberg13}, while further studies are ongoing or proposed~\cite{Grillmair13,Grillmair14,Hargis14a,Sesar14,Erkal14}. 

Lacking identification in optical surveys, the prime channels of detecting dark clumps are VHE gamma rays emerging from annihilations or decays of dark matter particles in the clump. Clumps would show the unique spectral gamma-ray signature of dark matter annihilation or decay~\cite{Pieri07,Buckley10,Zechlin11}. 
The Smith Cloud is an unique nearby Galactic substructure which was detected as a cold cloud of neutral hydrogen (HI) and is characterised by its large peculiar velocity~\cite{Smith1963}. The dark matter content of such objects is still under debate~\cite{Saul12,Hill09}, but it is plausible that the Smith Cloud hosts a large dark matter halo.

\paragraph{Evolution of knowledge}
Our knowledge of the dark matter distribution in the so-called ``classical'' dSph satellites of the Milky Way is based on dynamical modeling of their internal stellar kinematics (for recent works, see, for example, Refs.~\cite{Bonnivard14,GeringerSameth14}). The availability of multi-object spectrographs on a number of large telescopes combined with the dSphs' proximity to Earth has made it possible to obtain data sets of individual stellar velocities ranging in size from $\sim$170 (Leo II~\cite{Koch07}) up to $\sim$2500 (Fornax~\cite{Walker08}) stars. Ref.~\cite{Walker13} provides an excellent review of the currently available data.

The next step change in the size of dSph kinematic data sets is expected once instruments such as the Prime Focus Spectrograph~\cite{pfs11} mounted on the 8.2m Subaru telescope are available, providing thousands of fibres over a large field of view. Based on the numbers of red giant stars in the classical dSphs down to magnitude 22, it is reasonable to expect that on a 5-10 year timescale there will be samples of at least 1000 stars for all of the classical dSphs, with Sculptor and Fornax yielding particularly rich samples of $\sim$5,000 and $\sim$10,000 stars, respectively.
These new data will need to be complemented by more advanced modeling in order to constrain their dark matter profiles. A number of groups are working on a range of approaches and there is reason to be optimistic that community efforts such as the Gaia Challenge will drive further developments in this area by comparing existing modeling techniques, identifying systematic errors and biases in current algorithms, and facilitating the development of new ones. Studies are underway to estimate the impact that increased data sets and improved modeling can be expected to have on the ability of CTA to constrain the nature of dark matter, building on the earlier work of \cite{Charbonnier11}.

Since the recent discovery of a new population of satellite galaxies around the Milky Way, the ultra-faint dSphs, significant attention has been focused on their potential as targets in which to search for dark matter annihilation signals~\cite{Strigari07b}. The paucity of stars in these objects means that current sample sizes range from $\sim$200 stars in CVnI to $\sim$70 in Segue 1 and as low as 5 in the case of Leo\,V (see~\cite{Walker13} for references), with limited prospects for significantly larger samples in the near future.
These objects with a limited number of stars are prone to higher systematic uncertainties when modeling their dark matter content (see, for instance, Ref.~\cite{Bonnivard14}).
An additional complication is that the nature of these objects is still uncertain. If they represent a population of objects which have been strongly affected by the tidal field of the Milky Way, their dark matter content may have been over-estimated by simple equilibrium models. Further work is required to establish whether the potential constraints which these satellites may place on dark matter, see {\it e.g.},~\cite{Strigari13}, justify the risk that they may not contain sufficient dark matter to be detectable. 
New surveys such as Pan-STARRS and SkyMapper are expected to yield numerous additional ultra-faint dSphs, e.g.~\cite{Koposov07,Tollerud08}. Indeed, a new population of low-luminosity dSphs with a peak surface brightness lower than 30 mag arcsec$^{-2}$ ({\it i.e.} below current detection thresholds) is expected~\cite{Bullock10}. Toy models applied to numerical N-body simulations predict the
discovery of 3-13 L $\gtrsim$ 10$^3$ L$_{\odot}$ and 9-99 L $\lesssim$ 10$^3$ L$_{\odot}$ ({\it i.e.}, ultra-faint) dwarfs with DES, and 18-53 L $\gtrsim$ 10$^3$ L$_{\odot}$ and 53-307 L $\lesssim$ 10$^3$ L$_{\odot}$ dwarfs with LSST~\cite{Hargis14b}.  ultra-faint dSphs are likely to be hosted by light dark subhalos 
($\lesssim$ 10$^6$ M$_{\odot}$) which could be orbiting in the vicinity of the Solar System. These future surveys will extend significantly the knowledge of dwarf galaxies in the southern sky where CTA will have the highest sensitivity. 

\paragraph{Observational Strategy}
The observations of a dwarf spheroidal galaxy will be started in the first year of the Dark Matter Programme.
Any hints of dark matter signals or unknown sources would guide the plans for future observations. 
In the absence of signals, a programme of observation on the most promising dSph 
would be taken. As discussed earlier, according to existing data, the ultra-faint dwarf galaxies have the highest J-factors and so the best of these at the time would be adopted with the criterion of maximizing the chance of discovery. In the time before full CTA operations, it is expected, as discussed previously,
that much more information will be available to make the final choice. If new astrophysical data on stellar dynamics for a given dSph provide further motivation for pursuing the search, the observation strategy will be modified accordingly.  
Following a full evaluation of all results, the observing strategy is to acquire 100~h of observations per year 
on the best candidate dSph at that time. 

\paragraph{Performance}
Dwarf Galaxies have been principle targets for dark matter searches in existing gamma-ray experiments. Many strategies are considered for CTA with the spirit of being flexible to observe the most promising targets known when CTA is operational following information from the new surveys discussed previously or from any other resource.

\begin{figure}[!ht]
\begin{centering}
\includegraphics[width= 7.4cm]{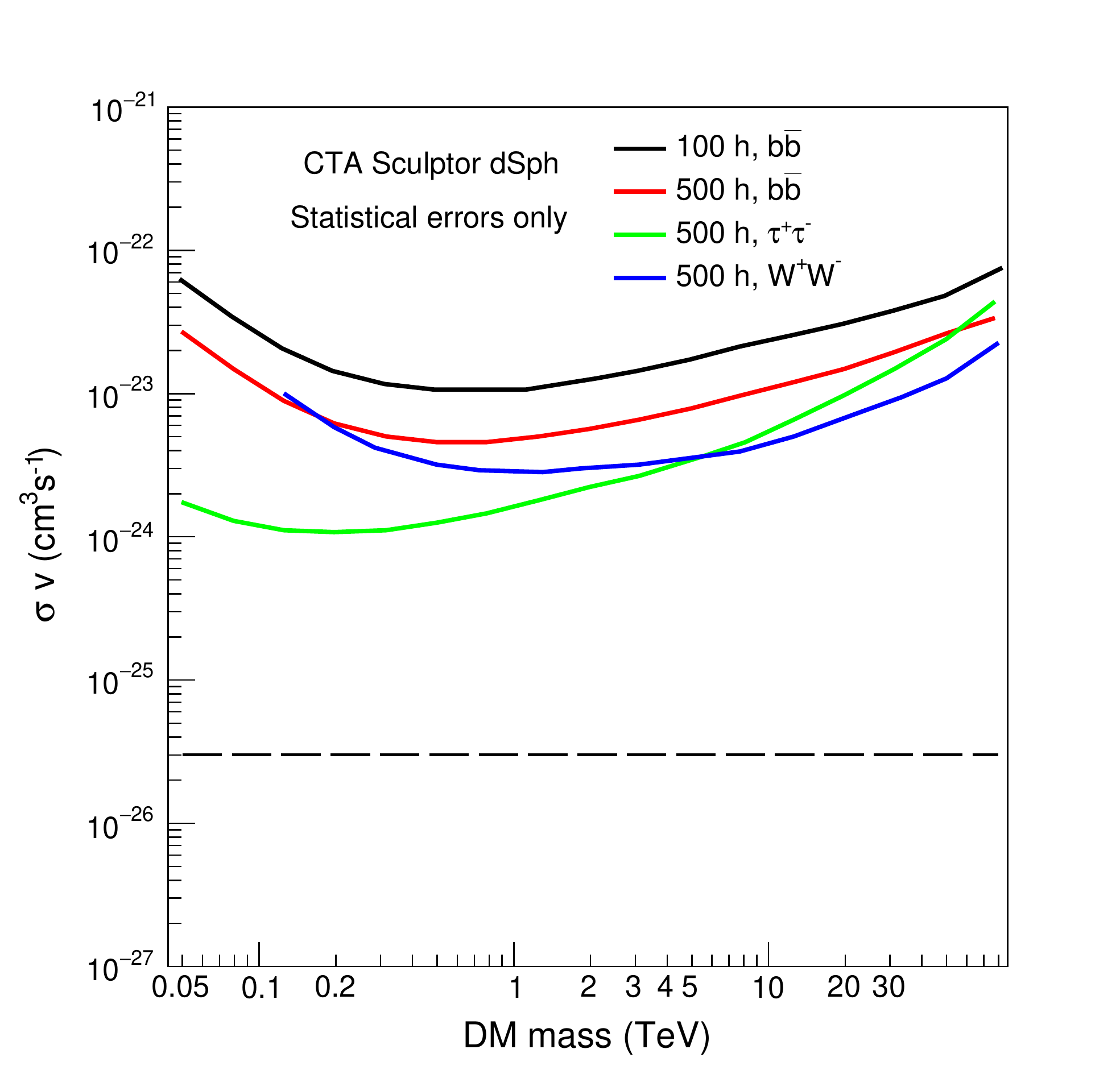}
\includegraphics[width= 7.4cm]{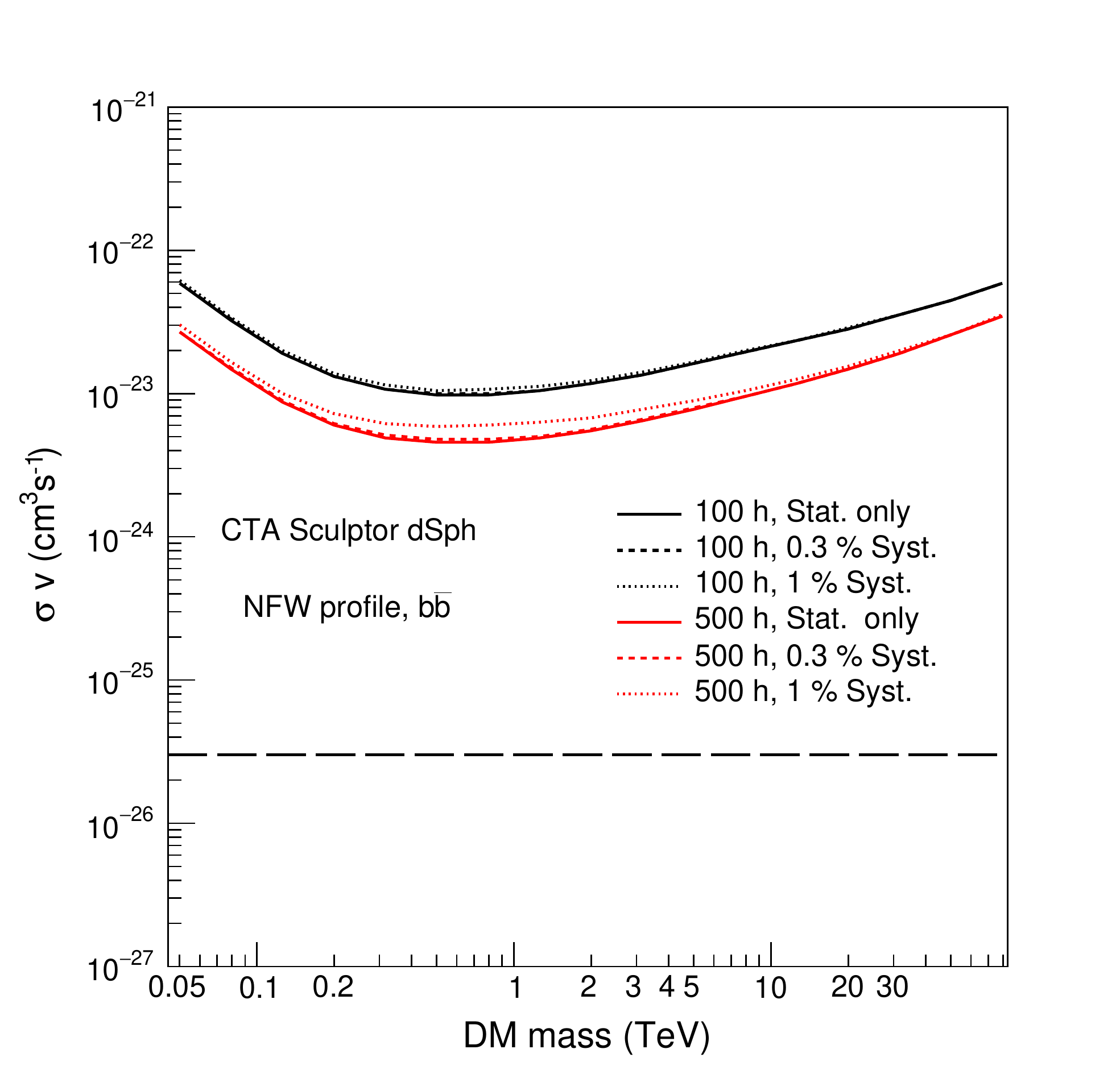}
\caption{Left: CTA sensitivity for $\langle \sigma v\rangle$ from observation of the classical dwarf galaxy Sculptor for different annihilation modes as indicated. Right: CTA sensitivity for $b\bar{b}$ annihilation modes for different conditions; black line is for 100 h of observation and red line for 500~h. The solid lines are the sensitivities only taking into statistical errors while dashed and dotted curves take into account systematics as indicated in the figure. The dashed horizontal line shows the thermal cross-section of $\rm 3 \times 10^{-26} cm^3s^{-1}$.}
\label{fig:Sculptor}
\end{centering}
\end{figure}

An example of the sensitivity which could be obtained by observations of a classical dwarf galaxy is 
shown in Figure~\ref{fig:Sculptor}, using the same analysis methodology~\cite{Lefranc:2015pza} as that performed for the Galactic halo so the results can be directly compared. In making this comparison with Figure~\ref{fig:gcsens}, it can be seen that the sensitivity is a factor 100 worse for this classical dwarf galaxy; however, the effect of systematics is drastically reduced for this small source compared to the extended Galactic Halo, explaining the significant interest in observations of dwarfs. Interestingly, observations of a classical dSph provide comparable sensitivity to the Galactic Centre for the cored dark matter profile. 

\begin{figure}[!ht]
\begin{centering}
\includegraphics[width= 7.6cm]{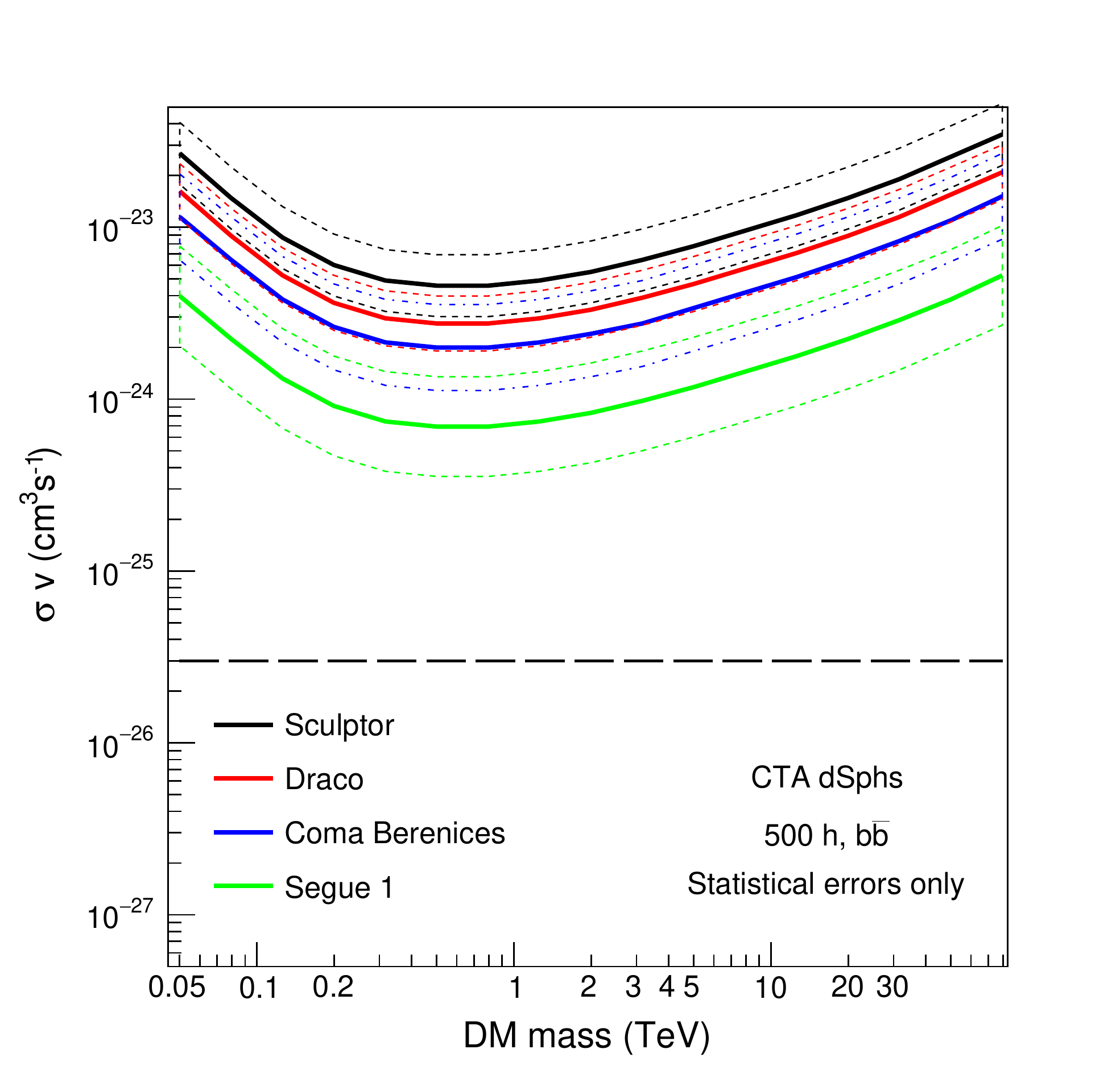}
\caption{CTA Sensitivity for $\langle \sigma v\rangle$ from 500 h observation of the classical dSphs Draco and Sculptor, and the ultra-faint dwarf galaxies Segue 1 and Coma Berenices as indicated. 
Dashed lines correspond to one standard deviation uncertainties on the J-factors. 
Sensitivity is computed assuming the $b\bar{b}$ annihilation mode and statistical errors only are taken into account. The dashed horizontal line shows the thermal cross-section of $\rm 3 \times 10^{-26} cm^3s^{-1}$.}
\label{fig:threedSph}
\end{centering}
\end{figure}

Figure~\ref{fig:threedSph} compares the sensitivity in the $b\bar{b}$ annihilation channel 
for 500 h observation time  on the ultra-faint dSph Segue 1 and Coma Berenices to the sensitivity on classical dSphs Draco and Sculptor. Ultra-faint dSphs provide stronger constraints compared to classical dSphs though 
higher uncertainties are expected the dark matter distribution. 
One standard deviation uncertainties are shown as dashed lines~\cite{Ackermann13b}.  

\subsubsection{Large Magellanic Cloud}
\paragraph{Description}
The Large Magellanic Cloud (LMC) is a nearby satellite galaxy at high Galactic latitude 
and it has the shape of a disk seen nearly face-on. At a distance of only $\sim50$~kpc, and with a large dark matter mass of $\sim 10^{10}$~M$_\odot$, the LMC is a candidate for indirect dark matter searches which has
long been recognized as a potentially favorable target~\cite{Tasitsiomi03}. 
The mass of the LMC of 1\% of the Milky Way makes it interesting even though its distance, at six times further away than the Galactic Centre, gives a large signal reduction. 
The LMC is an extended source for CTA.  Most of the emission lies within the CTA field of view, enabling the full galaxy to be observed in detail.  Like the Milky Way Galactic halo, astrophysical backgrounds are a significant challenge for the LMC.  However, the J-factor of the LMC 
has been claimed to be 
as high as $\rm log_{10}(J/GeV^2cm^{-5}) \sim 20.5$~\cite{Tasitsiomi03,Buckley15}, making it an attractive target for indirect searches.  
However, the spatial extent of the LMC and its significant astrophysical gamma-ray emission
lead to both analysis uncertainties and systematics.

The LMC hosts many interesting astrophysical sources: the largest star-forming region in the Local Group of galaxies, one of the densest stellar clusters known, the most massive stars ever observed, several tens of HII regions, more than a dozen super-bubbles, numerous giant shells, more than 50 catalogued supernova remnants, 
and the remnant of the recent naked-eye supernova SN 1987A. The LMC was detected in high-energy gamma rays and characterised by Fermi-LAT after just one year of data~\cite{Abdo10d}. Recently, powerful VHE emitters have been detected by H.E.S.S.~\cite{Abramowski15a}.
The astrophysical motivation for observations of the LMC are discussed in the LMC 
Key Science Project.

The dark matter distribution uses benchmark halo models of the LMC for which updated modeling was performed in Ref.~\cite{Buckley15} using HI rotation curve data~\cite{Kim98} and stellar velocity data~\cite{vanderMarel13}. The rotation curve was decomposed into a dark matter component, the HI mass, and the stellar mass. At all stages of the analysis, choices were made to maximize the baryonic contribution to the rotation curve and to minimize the dark matter contribution, leading to conservative estimates of the dark matter content. The dark matter distribution was fit to an isothermal profile, representing the conservative case of a cored inner density profile, and an NFW profile, representing a more typical choice for the assumed dark matter distribution. The analysis was performed assuming the best-fit position for the LMC centre based on HI and proper motion studies. The largest source of uncertainty in the rotation curve determination is the inclination of the LMC, so the fit for each profile was performed for extreme values of the inclination angle to bracket this source of uncertainty.  

The rotation curve data place a robust lower limit on the $J$-factor of the LMC of $\sim5 \times 10^{19}$ GeV$^2\,$cm$^{-5}$ (integrated over the LMC), although we emphasize that the rotation curves are also consistent with much larger $J$-factors. The J-factor values integrated within 15$^\circ$ of the LMC centre range from $\sim 5 \times 10^{19}$ to $\sim 5 \times 10^{20}$ GeV$^2$\,cm$^{-5}$ and, for both profiles, the emission profile is peaked towards the centre with more than half of the emission originating from within a few degrees of the centre. Eventually, detectability depends not only on the J-factor but also the dark matter distribution and on how well it can be discriminated from instrumental background and astrophysical emission. The level of expertise for this analysis motivates this program to be conducted by the CTA Consortium.

\paragraph{Evolution of knowledge}
The complexity of the dynamical state of the LMC, and the associated difficulty of modeling it, is reflected in the relatively few attempts at building a comprehensive picture of the mass distribution in its inner regions. 
Ref.~\cite{vanderMarel02} presented a detailed study of the stellar kinematics in the LMC, and recently updated modeling was performed in Ref.~\cite{Buckley15}.
As in the case of spiral galaxies, the baryonic mass in the disk and bar of the LMC contributes a significant fraction of the gravitational potential in the inner regions, making the LMC more difficult to model than the dSphs in which the stellar component can, to first order at least, be treated as a massless tracer population. Furthermore, the LMC is interacting with the Small Magellanic Cloud and has a central stellar bar which is not properly understood dynamically. Both these complications, coupled with the $>$20 degree extension of the LMC on the sky, make it very challenging to construct a precise picture of its inner mass distribution.  
Interestingly, despite all the uncertainties and complications noted above, the range of J values for the LMC is just a factor of 20 when the halo profile is varied from NFW to cored isothermal. By comparison, in the centre of the Milky Way, the J-factor varies by a factor of 100 when the halo is varied over the narrower range of an NFW halo and a halo with inner log slope of 0.5~\cite{Pierre14}. Under such consideration, the LMC appears to be a robust target, despite the lower nominal value of J relative to that of the Milky Way.

Spectroscopic studies continue to expand the size of radial velocity samples in the LMC (see e.g.~\cite{vanderMarel13} for an overview with new instruments ({\it e.g.} MUSE on the VLT)) providing opportunities for detailed studies of the stellar kinematics of the inner degree. Proper motions obtained using Hubble Space Telescope observations~\cite{vanderMarel13} have complemented the line-of-sight velocities, and it is the Gaia satellite that promises the next major step in understanding by mapping out the proper motion velocity field in much greater detail. With 3D velocities, it may become possible to unravel the inner dark halo slope, although considerable further modeling effort for the stellar and gas components will be needed. Finally, gravitational microlensing studies may assist in this by constraining the line of sight depth of the LMC. 

\paragraph{Observational Strategy}
For the LMC, observations will be taken with several pointings to cover the full galaxy. 
A total of 340 h of observations are proposed for both dark matter and
astrophysical motivations (see LMC Key Science Project). 

\begin{figure}[!ht]
\begin{centering}
\includegraphics[width=9cm]{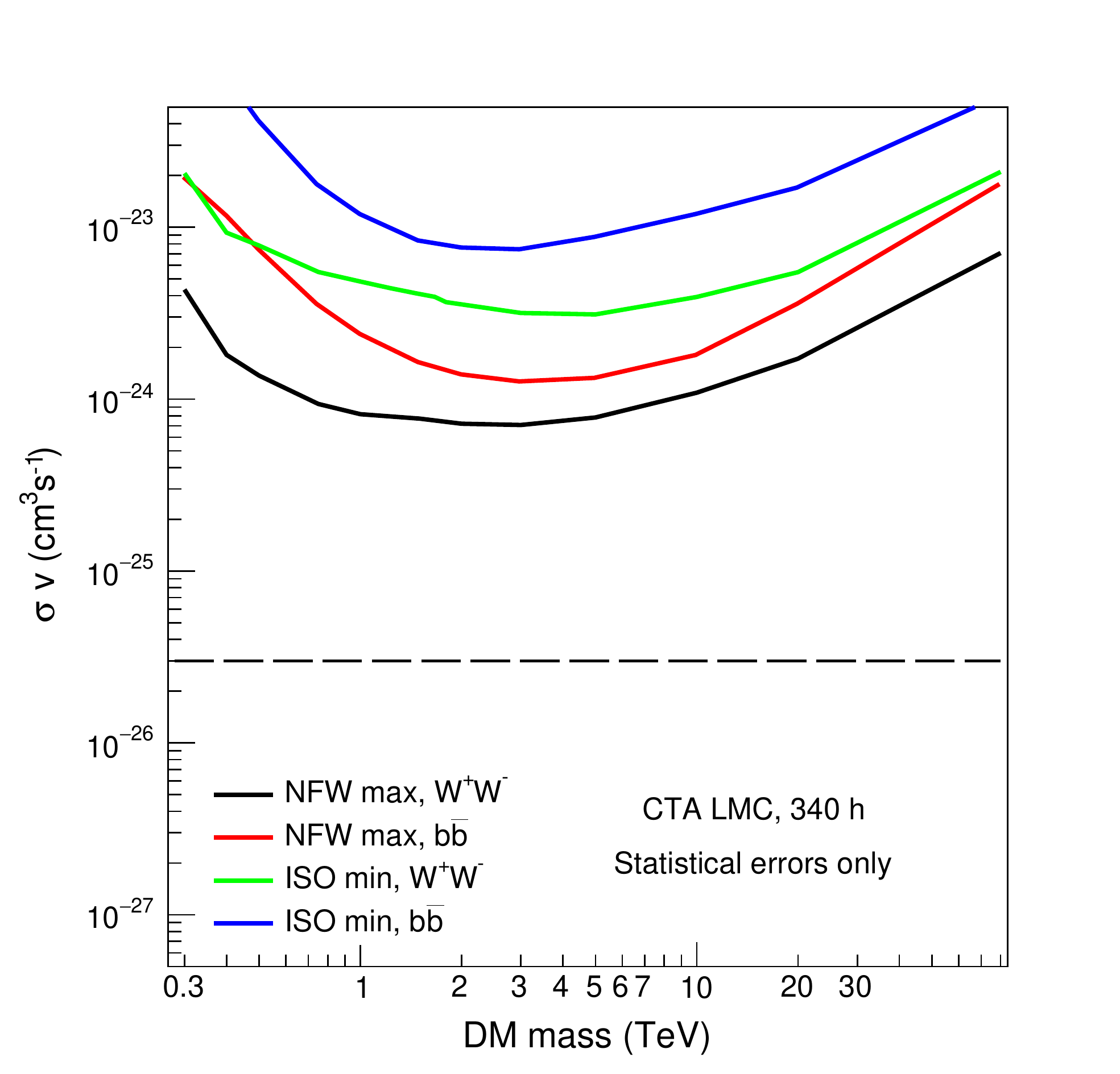}
\caption{CTA sensitivity on $\langle \sigma v\rangle$ from observation of the LMC for 340 hours of observation in the $b\bar{b}$ and W$^+$W$^-$ annihilation channels for both NFW and
isothermal (ISO) dark matter profiles, as shown in the legend. 
The sensitivities are computed with a 200~GeV energy threshold assuming statistical errors only. 
}
\label{fig:lmc}
\end{centering}
\end{figure}

\paragraph{Performance}
The sensitivity is computed for two benchmark annihilation channels, $b\bar{b}$ and W$^{+}$W$^{-}$, with
the results shown in Figure~\ref{fig:lmc}. 
The curves represent the 95\% confidence level upper limits that would be obtained on the dark matter annihilation cross-section as a function of dark matter particle mass in the case that no emission associated with a dark matter spatial template is detected. 
The minimum energy considered in the analyses is 200\,GeV due to the minimum zenith angles allowed for LMC observations.
The strongest sensitivity is achieved for the NFW profile with the maximum rotation curve and maximum allowed density within uncertainties in the inclination angle of the LMC, while the minimum rotation curve with the isothermal profile yields the weakest sensitivity. In the most optimistic case, the expected sensitivity is about a factor of twenty above the natural cross-section. 
The difference in the testable annihilation cross-section between the extreme cases is a factor of $\sim 10$. The astrophysical backgrounds in the LMC and their uncertainties differ from those in the Galactic Centre making it a complementary dark matter search target. 

\subsubsection{Clusters of galaxies}
Clusters of galaxies are the largest and most massive gravitationally bound systems in
the universe, with radii of a few Mpc and total masses of 10$^{14}$ to10$^{15}$ M$_{\odot}$, of which galaxies, gas and dark matter
contribute roughly 5\%, 15\% and 80\%, respectively. 
They present very high mass-to-light ratio environments and should be also considered 
promising targets for indirect dark matter searches, both for decay and annihilation. In particular, the
 dark matter decay rate is directly proportional to the enclosed mass making clusters the best targets together with our Galactic Centre~\cite{2012PhRvD85f3517C}. 
The dark matter halo of galaxy clusters harbour an abundance of dark matter substructures which contribute to the overall dark matter luminosity of the clusters. In principle they provide substantial contribution to the dark matter annihilation  signal. However, large uncertainties in the substructure boost factors remain~\cite{Sanchez-Conde:2013yxa} making them less favoured environments for annihilating dark matter than previously thought~\cite{Pinzke:2009cp,Pinzke11}.  

\begin{figure}[!ht]
\begin{centering}
\includegraphics[width= 9.0cm]{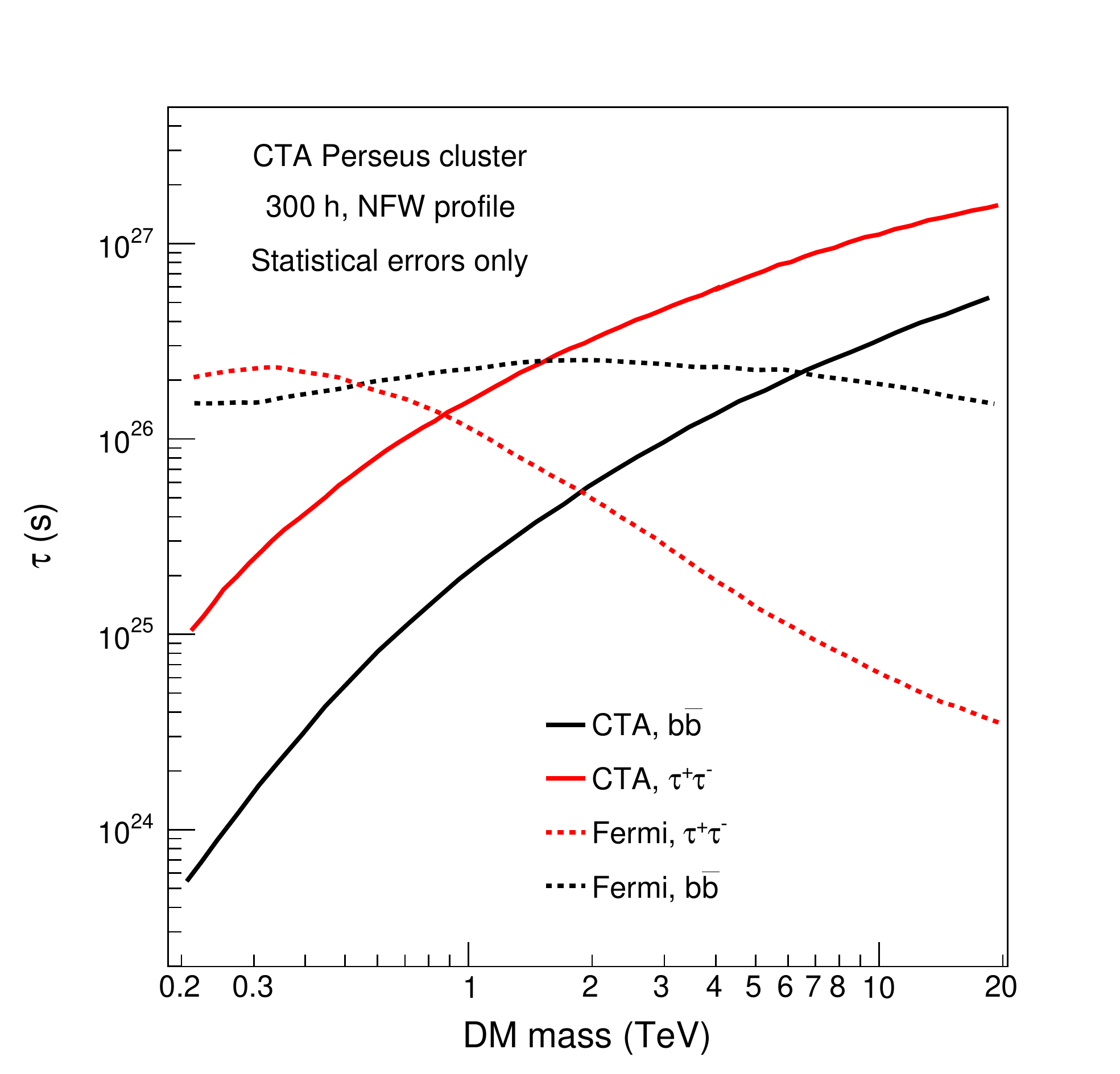}
\caption{
Expected CTA 
sensitivity to the dark matter decay lifetime for 300 h of observation of the Perseus cluster. We 
assume a dark matter profile as in Ref.~\cite{2011JCAP12011S} and adopt the full-likelihood analysis of Ref.~\cite{2012JCAP10032A}, while the adopted dark matter spectra are from Ref.~\cite{2011PhRvD83h3507C}. 
The size of the signal integration region (0.3$^{\circ}$ radius) has been optimised taking into account the expected source extension and the performance for off-axis observations. We assume five off regions. We compare our CTA predictions with the results from the Galactic halo by Fermi~\cite{2012ApJ76191A}.
}
\label{fig:decayingDM}
\end{centering}
\end{figure}

Galaxy clusters are a promising target for decaying dark matter (see, for instance, Ref.~\cite{Cirelli12}).  While the signal originating from annihilating dark matter is proportional to the square of the dark matter density, for decaying dark matter the dependence is on the first power. As a consequence, dense dark matter concentrations outshine the astrophysical backgrounds if annihilation is at play, but remain comparatively dim if dark matter is decaying. Decaying dark matter wins instead, generally speaking, when large volumes are considered.  
Figure~\ref{fig:decayingDM} shows predictions for the case of the Perseus cluster for 300 h of observation. 
We assume a dark matter profile in the cluster as in~\cite{2011JCAP12011S} while adopting the full-likelihood analysis of~\cite{2012JCAP10032A}. 
We consider an integration radius of 0.3$^{\circ}$. As is clear from the figure, CTA can do much better than Fermi~\cite{2012ApJ76191A} at the TeV scale.

The astrophysical motivations for observation of Perseus presented in the Galaxy Cluster KSP complement and extend
the science case for dark matter presented here.

\subsubsection{Summary of Targets}

The Dark Matter Programme comprises ten years of observations dedicated to dark matter targets.  The calendar for the observations is shown in Table~\ref{tab:obsschedule}. The first three years are devoted to deep observations of the Galactic Centre together with observation of the best ultra-faint dwarf galaxy. The observations towards the Galactic Centre will represent an important achievement and the result, eagerly awaited by the scientific community, is an expected deliverable of CTA. 

\begin{table}[!h]
\caption{Strategy for dark matter observations over ten years with CTA. The first three years are devoted to the deep observation of the Galactic Centre (GC) together with the observation of the best ultra-faint dwarf galaxy. In case of non-detection of the GC, observations starting in 
the fourth year focus on the most promising target at that time to provide legacy constraints.
}
\begin{center}
\begin{tabular}{llccccccccccc}
\hline
\hline
Year  & 1 & 2 & 3 & 4 &  5 & 6 & 7 & 8 & 9 & 10\\
\hline
Galactic halo & 175 h & 175 h & 175 h& & & & & & &\\
\hline
Best dSph & 100 h& 100 h& 100 h& \\
\hline
&&&&\multicolumn{6}{c}{{\it in case of detection at GC, large} $\sigma v$}\\
Best dSph & \multicolumn{3}{c}{} & 150 h& 150 h& 150 h& 150 h& 150 h& 150 h& 150 h\\
Galactic halo & \multicolumn{3}{c}{} & 100 h& 100 h& 100 h& 100 h& 100 h& 100 h& 100 h\\
\hline
&&&&\multicolumn{6}{c}{ {\it in case of detection at GC, small} $\sigma v$}\\
Galactic halo & \multicolumn{3}{c}{} & 100 h& 100 h& 100 h& 100 h& 100 h& 100 h& 100 h\\
\hline
&&&&\multicolumn{6}{c}{\it in case of no detection at GC }\\
{\it Best Target}& \multicolumn{3}{c}{} & 100 h& 100 h& 100 h& 100 h& 100 h& 100 h& 100 h\\

\hline
\hline
\end{tabular}
\end{center}
\label{tab:obsschedule}
\end{table}

Two different observation scenarios are considered depending on the outcome of the first three years of 
observations, {\it i.e.} whether there is a detection of a signal or not in the Galactic Centre dataset: 
\begin{itemize}
\item
In case of a detection and a relatively high annihilation cross-section (a few times $10^{-25}$ cm$^3$s$^{-1}$), the schedule will focus the observation time on the ultra-faint dwarf galaxy with a high J-factor to probe the annihilation cross-section in a cleaner environment. Simultaneously, follow-up observations of the Galactic halo would be scheduled to allow for in-depth study of the morphology of the dark matter distribution. 
For a relatively small annihilation cross-section (a few times $10^{-26}$ cm$^3$s$^{-1}$), the following years will be devoted to follow-up observations of the Galactic Centre in order to double at least the photon statistics to accurately characterise the faint detected signal.    
\item In case of no detection at the Galactic Centre, we will focus our observation time towards the {\it best 
target} at that time, on {\it e.g.} an ultra-faint dwarf galaxy or dark clump target, from years 4 to 10. This observation time will be devoted to provide long-standing legacy constraints.
\end{itemize}

\subsection{Data Products}
The following data products will be provided to the community:
\begin{itemize}
\item For each target: data-cubes (binned in the two spatial dimensions around the source and in energy) with the measured number of excess events over the background, together with the statistical significance of the excess.
\item For each detected source: energy spectrum and surface brightness of the emission, with a binning determined by its intensity and spatial extension.
\item A model of diffuse VHE gamma-ray emission in the field of view of each target.
\item For each target: the value of the likelihood function versus gamma-ray flux, in bins of energy, following the prescription used in ~\cite{Ackermann13b}. This information is sufficient to compute the total likelihood (individually for each source or globally for all of them) for any given dark matter model.
\item Using the previous information, constraints will be placed on the parameters of a set of dark matter annihilation and decay channels, including those producing gauge boson, quark, lepton and photon pairs, and any other channels favoured by the theory and other experimental constraints at the time of the data analysis. 
\end{itemize}
These results will be produced and released whenever a new source will be observed and/or when a 
significant amount of new data will justify an update of the results. In the case of no signal, we will publish our results after 100 h and 500 h of observations (for any given source).

\newcommand{\GCCENEXP}{525\,h}
\newcommand{\GCEXTEXP}{300\,h}

\section{KSP: Galactic Centre}
\label{sec:ksp_gc}

\begin{figure}[h]
  \centering
  \includegraphics[width=0.85\linewidth]{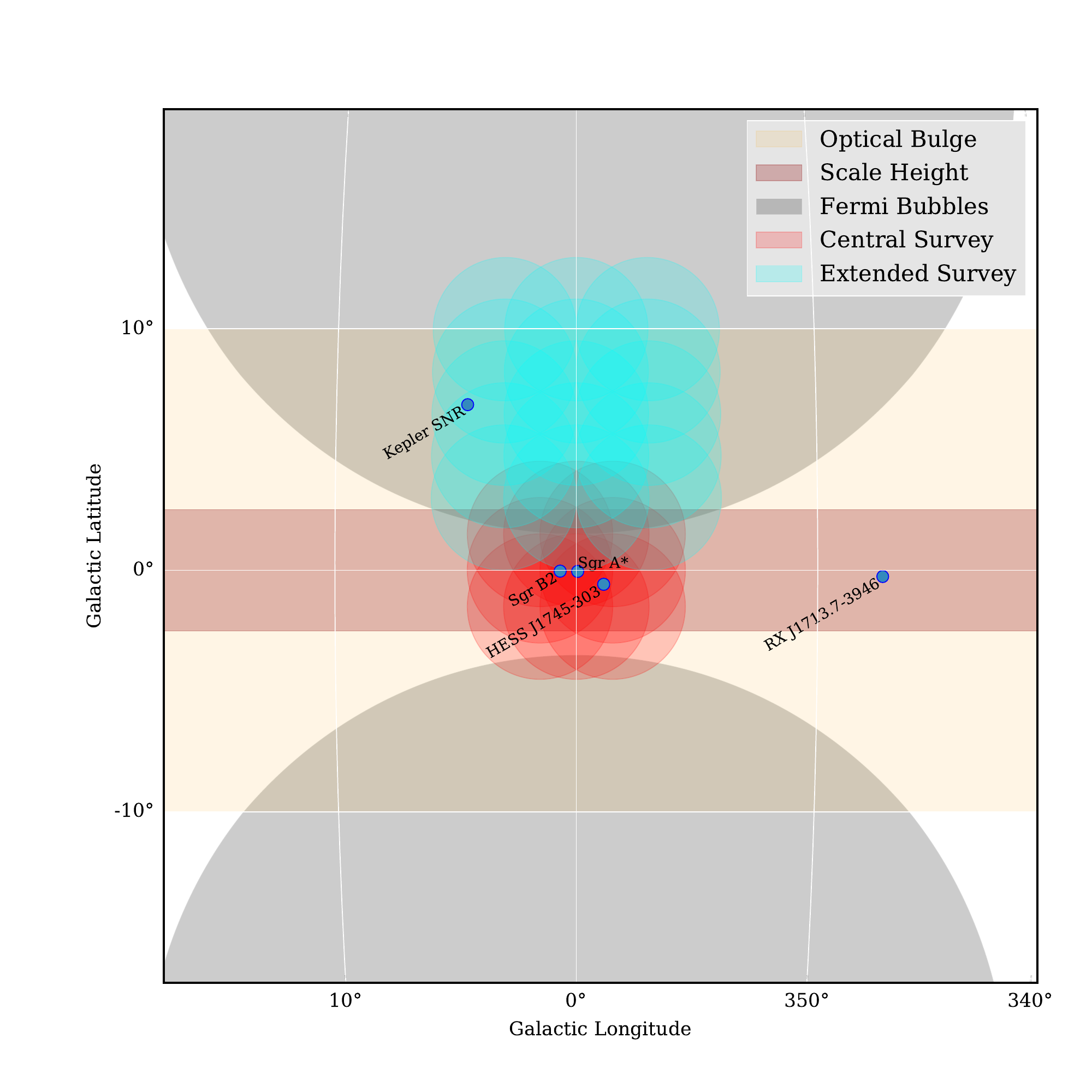}
  \caption{A schematic representation of the Galactic Centre
    KSP. This figure shows one possible observation strategy for
    CTA. The deep survey region is shown in red, with the Galactic
    bulge extension shown in cyan (with each circle representing a
    $6^\circ$ field of view for a typical CTA configuration). Several
    object positions are overlaid with blue dots for reference, in
    particular Sgr A*, the supermassive black hole that lies at the
    geometric center of the galaxy. }
  \label{fig:gc-strategy}
\end{figure}

The Galactic Centre Key Science Project is comprised of a deep
exposure of the inner few degrees of our Galaxy, complemented by an
extended survey to explore the regions not yet covered by existing
very high energy (VHE) instruments at high latitudes to the edge of the
bulge emission. A schematic representation is shown in Figure
\ref{fig:gc-strategy}, with details of the observation strategy and
possible options given in Section \ref{sec:gc-strategy}.

\begin{figure}[h]
  \centering
  \includegraphics[width=0.65\linewidth]{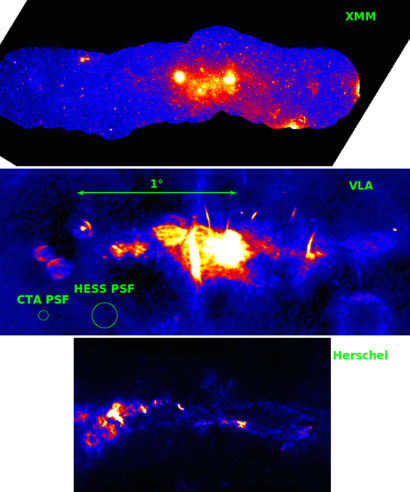}
  \caption{Multi-waveband view of the inner Galactic Centre region,
    showing the wide variety of diffuse emission. The scale is set
    such that the bright point source associated with Sgr A* in radio
    and X-rays is clipped to accentuate the surrounding
    emission. \emph{Top:} the X-ray emission detected by
    XMM-Newton, from \cite{Ponti15}. \emph{Middle:} the radio view, showing diffuse emission, SNR
    shells, and various arc-like radio features, from \cite{LaRosa00}. 
    The CTA point spread
    function is shown in comparison with that of the presently
    operating H.E.S.S. telescope to illustrate the possibility of
    resolving structures with CTA that are point-like with existing
    instruments. \emph{Bottom:} Infrared (IR) view of the gas (target
    material for cosmic rays to produce pion-decay gamma rays),
    showing the toroidal feature in the central molecular zone, from \cite{Molinari11}.}
  \label{fig:mwl}
\end{figure}

\begin{figure}[h]
  \centering
  \includegraphics[width=0.6\linewidth]{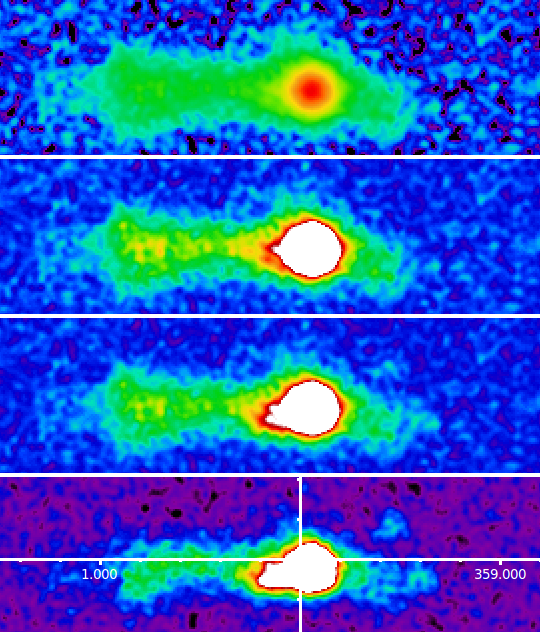}
  \caption{\label{fig:gc-simulation} A simulated view of the Galactic
    Centre region as seen by CTA (excess events above 800 GeV after
    cosmic-ray background subtraction, with Gaussian smoothing).
    \emph{Top and upper-middle:} the reference model that includes a
    point-like source at the position of Sgr\,A* with an
    exponential-cutoff power-law spectrum matching
    H.E.S.S. measurements \cite{Aharonian09gc}, a similarly modeled point-like supernova
    remnant G0.9+0.1 \cite{Aharonian05b}, diffuse emission from a template based on the
    dense matter distribution deduced from HCN molecule observations
    \cite{Jones2012}, and an extended circum-nuclear ring
    template. The top image is in log scale and the upper-middle is
    clipped to show the details of the diffuse emission (thus the
    point-like central source which is an order of magnitude brighter
    is washed out).  \emph{Lower-middle:} the reference model plus a
    catalogue of energetic pulsar-wind-nebulae \cite{Johnson2009}; here the central
    source clearly appears extending towards the left. \emph{Bottom:}
    an alternative model consisting of an extended circum-nuclear ring model
    assuming the source is produced by the interaction of isotropic
    cosmic rays with the circum-nuclear ring gas (which has a $1.5'$
    diameter), diffuse emission from HCN maps as above, and closeby
    pulsar wind nebulae \cite{Johnson2009} (modeled as
    point-sources). The central source and G\,0.9+0.1 are excluded.
    Cosmic-ray background is subtracted using a model generated from
    cosmic-ray events; the same background model is used in all
    cases.}
\end{figure}

The region within a few degrees of the Galactic Centre contains a wide
variety of possible high-energy emitters, including the closest
supermassive black hole, dense molecular clouds, strong star-forming
activity, multiple supernova remnants and pulsar wind nebulae,
arc-like radio structures, as well as the base of what may be
large-scale Galactic outflows (commonly referred to as the \emph{Fermi
  bubbles}); it is also one of the best places to look for dark
matter, as outlined in Chapter \ref{sec:DM_prog}.  The central few
degrees of our Galaxy is arguably one of the most studied regions of
the sky in nearly every wave-band 
(see Figure \ref{fig:mwl}).  In VHE gamma rays alone
(for a review see \cite{vanEldik15}), the region has yielded major
scientific discoveries including an unidentified central point-like
gamma-ray source \cite{Aharonian09gc,Archer14} that may be associated
with Sgr A* and a complex pattern of diffuse emission
\cite{Aharonian:2006au,Archer16} that may be an indication of local
PeV cosmic-ray acceleration in the recent past
\cite{HESS16}. Precision measurements with CTA will allow us to study
this complex region in unprecedented spatial and spectral detail, and
they may allow for the identification of the central source, the
disentangling of models that have been put forth to explain the
extended emission, and perhaps the deeper understanding of the
capability for cosmic-ray acceleration in our Galaxy and the history
thereof.  A simulated CTA view of the Galactic Centre region that
could be accomplished with the observations proposed in this KSP is
shown in Figure~\ref{fig:gc-simulation} for various astrophysical
scenarios; the differences among the scenarios can be clearly
identified. Note that the angular resolution of the diffuse template used
in these simulations is poorer than that of CTA, therefore small-scale
features possible with CTA are not well represented beyond the
included point-like sources.

A deep Galactic Centre survey is considered key science for both
scientific and technical reasons.  The scientific results will
encompass a wide variety of targets and topics and will require a
large time commitment spread over multiple years.  The proposed
observations will provide the deepest exposure of the Galactic Centre
region ever produced in the TeV domain and will provide significant
scientific output on a variety of topics for years to come.  As part
of this project, the CTA Consortium will organize coordinated observations in other
wavebands that will permit detailed variability studies and constitute
an invaluable legacy data set.

The realisation of an extremely deep and high-precision survey of the
Galactic Centre region is non-trivial from a technical standpoint due
to the complexity and confusion of the gamma-ray and optical emission
in the region, the difficulty of modeling the background when there
is an expectation of sources that are larger or similar to the
instrumental field of view, the necessity to handle strong background
optical light that varies by a factor of over ten within the
field of view, and the need to push the systematic errors to extremely
low levels to produce meaningful limits on models. Such a study will
therefore require a deep knowledge of the instrument and atmosphere
and will perhaps even require specialised knowledge of the
calibration, reconstruction, and discrimination of Cherenkov events
that only the CTA Consortium can readily provide.

The main science topics that can be explored with this rich data set are described
below, followed by the observational strategy, data products and expected return.

\subsection{Science Targeted} 

This key science program of the Galactic Centre region covers a wide
range of scientific topics due to the rich environment within the
central few degrees of our Galaxy.

\subsubsection{Scientific Objectives}

\paragraph{Revealing the nature of the central gamma-ray source} 
CTA can provide the answer to what is producing the point-like source of the 
VHE gamma-ray emission at the dynamic centre
of the Milky-Way.

The central VHE source was first detected in 2004
\cite{Kosack:2004ri,Tsuchiya:2004wv}, and although well studied by
current-generation atmospheric Cherenkov telescopes (H.E.S.S. and
VERITAS), it still remains unidentified due to source confusion and
limited sensitivity to variability and small-scale morphology. The
emission may originate from close to the supermassive black hole
Sgr A*, from winds driven out from it, from a background pulsar
wind nebula, or from a variety of other possibilities.  CTA, with its
arc-minute resolution at high energies and dramatic improvement
in sensitivity for rapidly variable phenomena, will have the
opportunity to resolve this question.

\paragraph{Diffuse VHE emission: particle acceleration in the vicinity
  of the Galactic Centre}

The diffuse gamma-ray emission along the Galactic ridge provides both
the best case for studying gamma rays produced by hadronic
interactions and to understand the acceleration history of the central
engine.  This diffuse region is the sole VHE source presently
known that is generally acknowledged to be predominantly hadronic in
nature. H.E.S.S.  observations revealed the presence of a ridge of
hard (photon spectral index, $\Gamma \sim 2.3$) VHE emission, extending over
1.5 degrees along the Galactic plane \cite{Aharonian:2006au}. The
spectrum is similar to that of the central source implying a possible
connection, but it so far shows no evidence of a high-energy 
cutoff -- a
fact that may be explained by the diffusion of cosmic rays from a
central accelerator \cite{HESS16} that reaches PeV energies.  The diffuse 
emission spatially correlates with the inner dense clouds, and
although it establishes that active or recently active cosmic-ray
accelerators are present in the central degree, the origin remains
uncertain.  All current studies are limited by statistics and
field of view, and CTA will thus provide in an incredible wealth of
new information.

Detailed spectro-morphological analyses of the diffuse VHE
emission are required. CTA deep coverage will enable us to properly
determine the level of hadronic cosmic-ray induced emission by
distinguishing clouds from individual sources seen at other
wavelengths, thanks to a largely improved angular resolution with
respect to current experiments.  It will provide a unique measure of
cosmic-ray propagation in the central molecular zone (CMZ), a region
rich in molecules that contains the Galactic Central Radio Arc.  The
large number of photons that can be detected across this region --
particularly at high energies where current instruments suffer from
low statistics -- will allow detailed spectral characterisation.  The
improved statistics with CTA will allow, in particular, the study of
spectral changes associated with diffusion from a central accelerator
that are hinted at with current observations \cite{HESS16}.

\paragraph{Exploring large-scale outflows} 
An extended exposure with CTA will provide, for the first time, the
ability to search for VHE diffuse emission away from the Galactic
plane ($|b| > 0.3^\circ$) in the region of the Galactic Centre.

Beyond the supermassive black hole (SMBH) at the gravitational
centre, the central 30 pc region harbours up to 10\% of the massive
stars ($>20 M_\odot$) in the Galaxy \cite{Figer:2003tu}, implying a
star-formation rate per unit volume two orders of magnitude larger
than that in the rest of the disk. The kinetic power released by
supernova explosions is therefore expected to be as large as
$10^{40}\;\mathrm{erg/s}$ \cite{Crocker:2010qn}. This should produce
very sustained particle acceleration as well as a strong outflow
advecting cosmic rays into the Galactic halo, thus possibly feeding
the large Fermi bubbles
(e.g. \cite{Crocker:2010qn,Yoast-Hull:2014cra}). A presumed period of
high activity of the SMBH has also been invoked to explain the bubbles
(e.g. \cite{Yang:2012fy} and others). Evidence of on-going cosmic-ray
escape from the Galactic Centre region would support this picture.

An interesting feature in this respect is the Galactic Centre lobe
which extends over 150~pc north of the Galactic Centre. It is an
energetic outflow from the inner regions. The energy required to
inflate it is similar to that released by 50 supernovae, and the outflow is 
presumably powered by the sustained star formation activity in the Galactic
Centre \cite{Law:2009wv}. Similarly, a bright extended region of
recombining plasma has been recently discovered at $b=-1.4^\circ$ which 
might be the relic of an energetic event at the Galactic Centre some
$10^5$ years ago \cite{Nakashima:2013yra}. Detection of gamma-ray emission
associated to these features might help understand the connection
between the activity in the central 100~pc and the large-scale Fermi
bubbles.

\paragraph{Supernova Remnants, Pulsar Wind Nebulae, and Molecular Clouds}

A large number of supernova remnants (SNRs) and pulsar-wind nebulae
(PWNe) will be covered by a deep exposure of the Galactic Centre
region (see the radio image in Figure~\ref{fig:mwl}), providing a
significant increase in the number detected of these objects, which
are known VHE accelerators.  Though some nearby SNRs like
G\,0.9+0.1 have already been detected and studied at very high
energies \cite{Aharonian05b}; others remain too weak for current
instruments or suffer from significant source confusion that could be
mitigated by the improved sensitivity and smaller point spread
function of CTA. For example, the inner Galactic Centre field of view
contains G\,1.9+0.3 which may be the youngest Galactic SNR, the
expansion of which can be seen in radio and X-rays, see
e.g. \cite{Borkowski:2014voa}. Since very young SNRs are thought to be
the most likely objects to be producing TeV cosmic rays, G\,1.9+0.3 is
definitely a major target for CTA observation. The region harbours
also a number of older SNRs, some of which may be interacting with
molecular clouds and producing the gamma-ray emission, e.g. the source
known as H.E.S.S.~J1745$-$303
\cite{Aharonian:2005kn,Aharonian:2008gw}. The improved
point spread function (PSF) of CTA will allow morphological
studies of some of these objects that are currently seen as point-like
(see Figure \ref{fig:mwl}).  In particular, about 20 filaments of
non-thermal X-ray emission are visible in the central degree. A
significant number of them appear to be PWNe and could therefore have
non-negligible TeV emission. CTA will be able to resolve several of
them from the truly interstellar emission.

\subsubsection{Context / Advance beyond State of the Art}

While the Galactic Centre key science project will allow us to
explore new domains in both space and energy, it will also drastically
improve upon the existing observations from current or past gamma-ray
telescopes. In particular it will build upon the knowledge gained from
Fermi-LAT at GeV energies and from H.E.S.S. and VERITAS observations
at higher energies. 

Fermi-LAT has successfully explored the entire gamma-ray sky from
100\,MeV up to $\sim$100\,GeV.  In particular, it has measured the
diffuse emission in the Galactic Centre region.  At higher energies
above 100 GeV, however, the limited size of Fermi-LAT severely
constrains the number of detected gamma rays. Moreover, the instrument
suffers from high source confusion due to its limited angular
resolution and from poor sensitivity to short-term variability.

Above 100 GeV, the discovery by imaging atmospheric Cherenkov
telescopes of VHE gamma-ray emission from the region was a
major success \cite{vanEldik15,Aharonian09gc,Archer14},
highlighted by the strong
detection of an unknown point-like source coincident with the
supermassive black hole Sgr\,A* that was
inconsistent with a dark matter interpretation
\cite{Aharonian:2006wh}. 
H.E.S.S. \cite{Aharonian:2006au}, and later VERITAS \cite{Archer16},  
further identified significant
extended VHE diffuse emission in the inner $\pm 1^\circ$. The apparent
correlation of this emission with dense molecular clouds suggests an
illumination of the gas from recently accelerated hadrons.  However,
these results are limited by the sensitivity, angular
resolution, and systematics of
current instruments; further exploring the science of this
region requires an improvement in all three.

\paragraph{Central engine}

\begin{figure}[h]
  \centering
  \includegraphics[width=0.7\linewidth]{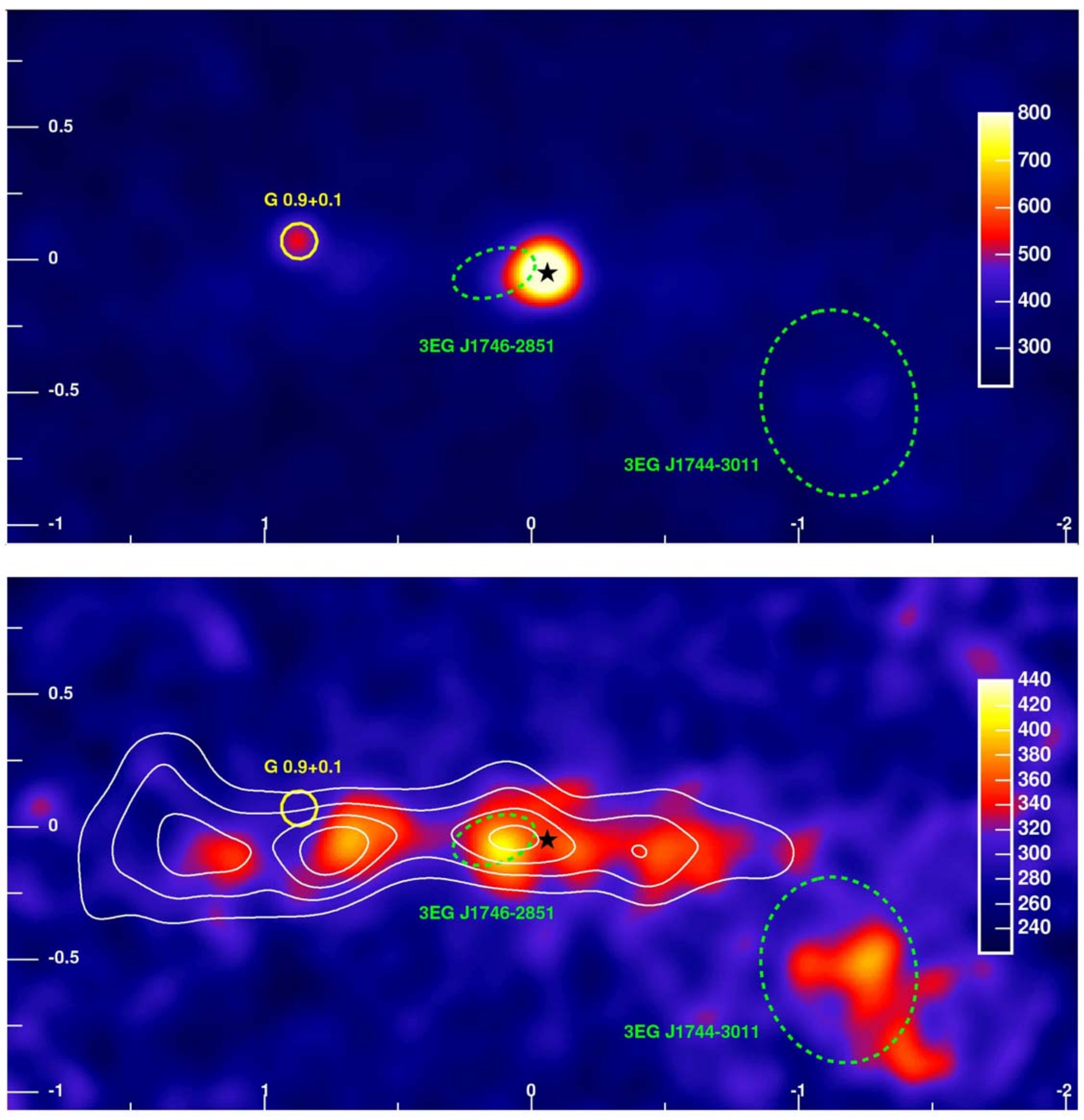}
  \caption{The VHE gamma-ray view of the Galactic Centre with H.E.S.S.,
    reproduced from~\cite{Aharonian:2006au}.
    On the top is the total emission, dominated by the central
    unidentified point source, while the bottom shows the emission
    after the point-source and SNR G\,0.9+0.1 have been subtracted. The
    residual signal shows the diffuse gamma-ray emission along the
    ridge, which roughly matches the morphology of local molecular
    clouds (white contours).}
  \label{fig:H.E.S.S._gc}
\end{figure}

\begin{figure}[h]
   \centering
   \includegraphics[width=0.5\linewidth]{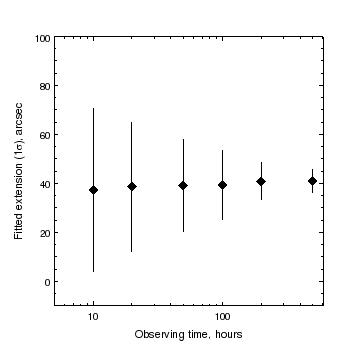}
   \caption{Expectation for the fitted size of the central source
     (assuming a Gaussian shape) made by CTA as a function of
     observing time.  Each point in the curve is generated by 100
     Monte-Carlo simulations. The error bars show the typical spread
     of the simulated measurements.  With 200~h, CTA should reach a
     detection of the source extent with a statistical significance of
     approximately four standard deviations. The error bars are
     accurate to within $\pm 10$\%.}
   \label{fig:gc-extension}
\end{figure}

The central VHE source has been well studied with 
existing atmospheric Cherenkov telescopes, 
but still remains unidentified due to source confusion and
limited sensitivity to variability and small-scale morphology. It is
coincident with Sgr A* within the pointing accuracy of
H.E.S.S. ($<7''$), but it cannot be definitively associated with the
SMBH; for example, there is an X-ray emitting pulsar wind nebula
detected by Chandra that is within the same error circle.

If the central VHE source can be associated with accretion-related
processes around Sgr A*, it will provide the best laboratory for
studying the details of particle acceleration from accretion or jet
emission in a SMBH that is not just nearby, but covered better with
multi-waveband instruments than any other similar source. Sgr A* is seen as a
bright point-source in radio, near-IR, and X-ray wavelengths, and
it exhibits flaring activity on half-hour timescales in both IR and
X-rays, pointing to rapid changes in accretion rate, caused by
anything from clumps in the surrounding gas to an Oort cloud of
asteroids \cite{Zubovas2012:gcasteroid}.  The dynamics of the region
are extremely well measured, with the orbits of stars around the black
hole tracked with high precision.

Starting in 2013, the approach and tidal disruption of a dense gas
cloud (G2) on a trajectory toward the black hole has been
tracked with IR observatories, possibly leading to a detectable,
long-term increase in accretion. So far, no variability of the VHE
source has been seen on short or long timescales. However, current
instruments are only sensitive to changes of flux of roughly 
a factor of two, and systematics may limit long-timescale
studies.

Deep observations of this object with CTA will provide:

\begin{itemize}
\item an angular resolution sufficient to image the arc-minute scale
  VHE source, helping to resolve its nature (Figure
  \ref{fig:gc-extension} shows the ability of CTA to measure the size
  of the central source),
\item the possibility to search for variability of the central source,
  coinciding with X-ray or IR flaring, which would enhance our
  understanding of the emission processes in SMBH, and
\item sufficient spectral sensitivity and energy coverage to determine
  the maximum energy reached by accelerated cosmic rays in this region
  and to measure any possible spectral variation 
  during accretion events.
\end{itemize}

\paragraph{Advances in the study of Diffuse Emission}

The ridge of hard (photon spectral index, $\Gamma \sim 2.3$) VHE emission at
the Galactic Centre extending over 1.5 degrees along the Galactic
plane (seen modeled in Figure \ref{fig:gc-simulation} and in reality
in Figure \ref{fig:H.E.S.S._gc}) is well correlated with the dense
clouds present in the central 200 pc that are traced with line
emission from CS molecules~\cite{Aharonian:2006au}. The photon 
spectral index of this emission is somewhat harder than the cosmic-ray
spectrum measured at Earth. Under the hadronic origin assumption, the
density of multi-TeV cosmic rays is significantly larger in the
central region than in the local medium. Therefore, active or recently
active cosmic-ray accelerators are likely present in the central
degree, but their nature is still very uncertain.

SNR Sgr A East was seen as possible source of these cosmic rays
\cite{Aharonian:2006au}, but the influences of the starburst
environment of the central 200 pc, or of the supermassive black hole
itself, have to be understood. A key question here is to determine how
cosmic rays propagate in the complex environment of the CMZ.
For instance, the large poloidal magnetic fields traced by non-thermal
filaments might limit the diffusion of particles along the Galactic
plane. Similarly, the presence of an advecting wind is required by
several models of the star formation activity of the CMZ
\cite{Crocker:2010qn,Yoast-Hull:2014cra}. The idea of a single
accelerator injecting cosmic rays in the Galactic Centre has therefore been
questioned (see, for example, \cite{Wommer:2008we}), and distributed
turbulent acceleration in the inter-cloud medium has been proposed by
several authors \cite{Melia11,Amano:2011kw}. The starburst activity
and the associated supernovae should also provide enough energy to
create the observed excess and even power the large scale Fermi
bubbles \cite{Crocker:2010qn}. It is therefore fundamental to properly
measure how cosmic rays are distributed and how they diffuse in the central 200
pc. This requires a detailed spatially resolved spectral analysis of
the region that only CTA can provide.

Deep observations of the CMZ with CTA will provide:
\begin{itemize}
\item an angular resolution sufficient to resolve diffuse emission down to the
  arc-minute scale. Comparing with molecular matter surveys, this will
  help resolve the truly interstellar emission from distinct features
  in the region and thus precisely measure the level of hadronic
  cosmic rays in the CMZ. The ability of specific objects (e.g. the
  young and massive Arches cluster, the radio arc, etc.) to accelerate
  particles to very high energies and emit gamma rays will be
  precisely tested;

\item very high photon statistics, making spectral extraction possible
  on scales of a few arc-minutes and for energies above 10 TeV. This
  will allow a detailed study of the cosmic-ray distribution in the
  region that will provide invaluable information on cosmic-ray
  propagation and their penetration into very dense clouds. The
  Galactic Centre region is probably one of the very few regions in
  the Galaxy where such a study will be possible;

\item excellent spectral sensitivity at very high energy to determine
  the maximum energy reached by cosmic rays in the central 200 pc.
\end{itemize}

\subsection{Strategy} \label{sec:gc-strategy}

There are several possible observation strategies to achieve the goals
outlined earlier. In general, the exposure can be broken into two
small survey regions: a deep exposure close in to the central source,
centered on Sgr A*, and an extended region starting at the edge of the
deep exposure and extending to the edge of the Galactic bulge as
delineated by optical emission (approximately $\pm 10^\circ$ in Galactic
latitude). This strategy is shown schematically in \ref{fig:gc-strategy}.  The
exact pointing strategy will be optimised once the final
characteristics of CTA are known, and it may also be adjusted to
provide, for example, better coverage of interesting features of the
Fermi bubbles based on new analyses of GeV data.

Nominally, the exposures can be described as follows (see also
Table~\ref{tab:gc_targets}):

\begin{itemize}
\item \textbf{central survey region}: a deep exposure of \GCCENEXP,
  centered on Sgr A*, with pointings centered on grid points covering
  all combinations of
  $l={\pm1.0^\circ,\ 0^\circ}$ and $b={\pm1.0^\circ,\ 0^\circ}$. The
  grid spacing of this survey
  may be optimised, depending on the final field of view 
  and acceptance characteristics of CTA.  These observations
  should be completed in the first three years of operation because
  of the high importance of the dark matter search (see Chapter 4). 
  In the
  first year, the data should be sufficient to publish an updated
  analysis of the central source. After three years, a detailed
  study of the extended/diffuse emission will be possible.  This
  exposure covers the:
  \begin{itemize}
  \item Galactic Centre central source,
  \item centre of the dark matter halo,
  \item all of the known diffuse emission, including H.E.S.S. J1745$-$303,
  \item multiple SNRs (G\,0.9+0.1, G\,1.9+0.3, ...),
  \item multiple PWNe (Mouse G359.2-0.8), and
  \item central radio lobes (central $\pm 1^\circ$) and arc features.
  \end{itemize}

\item \textbf{extended survey region}: \GCEXTEXP\ of exposure covering
  a large region to the south or north of the Galactic Plane Survey
  (Chapter 6)
  region out to $10^\circ$ in latitude. Since large-scale emission is
  expected to be symmetric, it is likely not necessary to extend observations
  to both north and south, with north being preferred due to lower
  optical brightness. These observations can be taken after the deep
  exposure, i.e. after the third year of operation, though a first
  pass may be interleaved into the first few years to help develop
  the analysis and to look for new bright targets.  Note that the
  observations are shown currently as a box, but could be skewed to
  better cover interesting regions like the edge of a bubble. 
 This exposure covers the:
  \begin{itemize}
  \item edge of the Galactic bulge,
  \item base of Fermi bubbles,
  \item radio ``spurs'', and
  \item the Kepler SNR.
  \end{itemize}
\end{itemize}

\begin{table}[h]
  \centering
  \begin{tabular}{r | l l l}
    & Deep Exposure & Extended Survey & Monitoring+multi-waveband \\ \hline Time
    requested & \GCCENEXP\ & \GCEXTEXP\ & (coordinated with other instruments) \\ Priority & 1 & 3 & 2
    \\ Strategy & survey & survey & periodic + coordinated\\ Site &
    S & S & S \\ Sub-array & Full & Full & Full
    \\ Zenith Range & $<40^\circ$ & $<50^\circ$ &
    $<40^\circ$\\ Atmosphere Quality & high & high & medium \\ Targets
    Covered & multiple & multiple & multiple \\
  \end{tabular}
  \caption{Exposure summary for the Galactic Centre KSP.}
  \label{tab:gc_targets}
\end{table}

\subsubsection{Timeline and Sub-array Choice} 

\begin{figure}[ht!]
  \centering
  \includegraphics[width=0.8\textwidth]{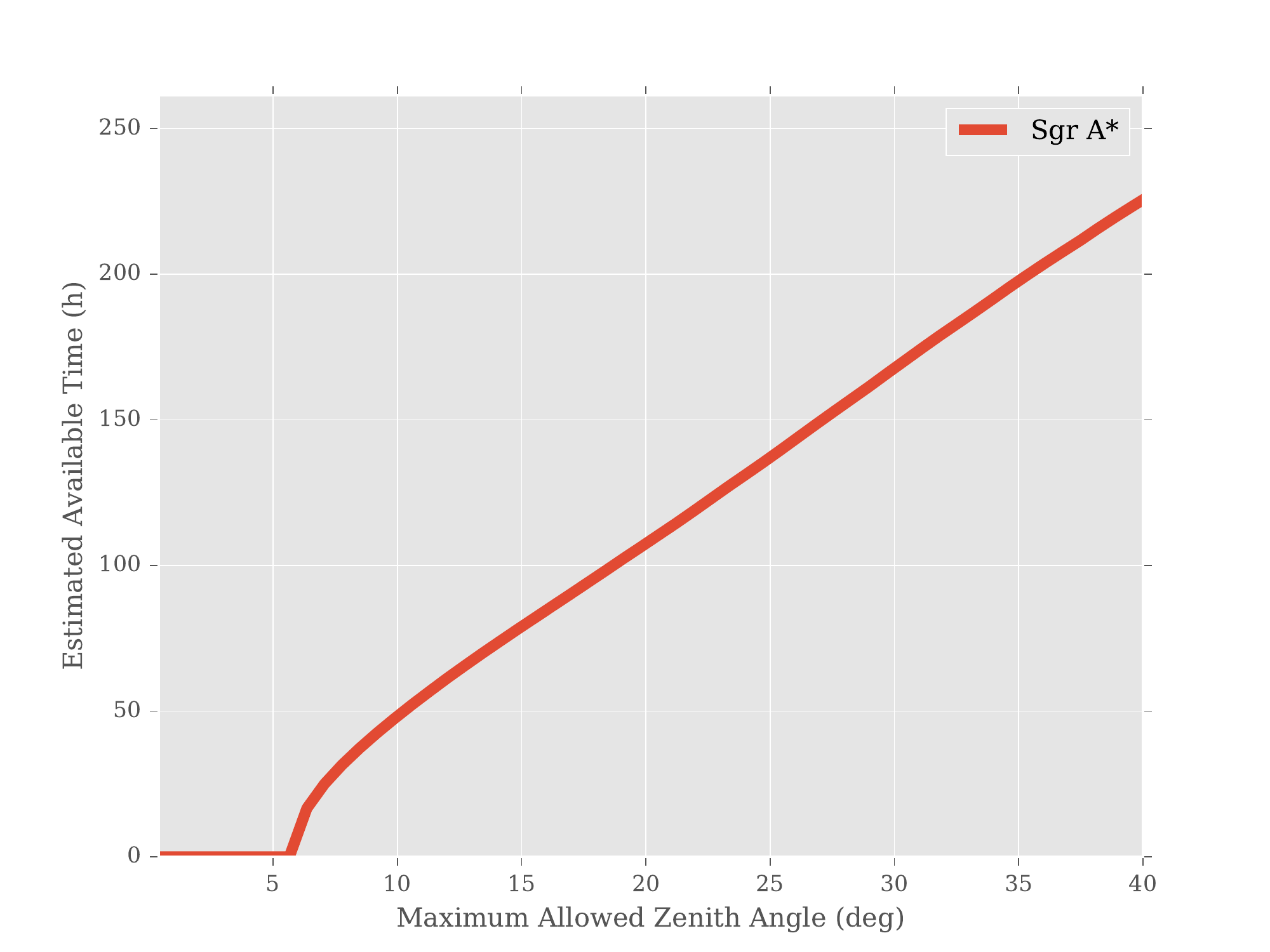}
  \caption{\label{fig:sgra-available} Available time at the position
    of Sgr A* as a function of zenith angle for CTA-South. 
    The available time includes only
    moonless dark time and is adjusted for a typical data taking
    efficiency of 70\%. In order to achieve the time needed per year,
    the maximum allowed zenith angle must be at least $35^\circ$. }
\end{figure}

The Galactic Centre observations should be taken primarily with the southern array
due to the declination of the region of interest $\simeq
-30^\circ$. Some fraction of the observation time should be
coordinated with other instruments (e.g. X-ray or radio) to facilitate
variability studies.  Small zenith angle observations should be
prioritised to give the
low-energy threshold needed for variability studies.  
However, since high-energy observations are also useful,
a fraction of the time could be taken with the northern array,
using large zenith angle observations.
Taking data with the full array (i.e. LST+MST+SST)
is also recommended, since
the interesting energies span the full CTA energy range.  LST
data are required to search for variability at low energies, where CTA
will have a large advantage in sensitivity over Fermi. In
addition, the study of outflows related to the Fermi bubbles requires
low-energy response, since the spectrum exhibits a cutoff around
100 GeV in the Fermi data. Conversely, the high-energy response is
important for mapping the diffuse emission and for determining the
maximum energy of the various accelerators in the region.  One caveat
is that since there is more signal at low energies and thus the
low-energy sensitivity may be achieved faster, it may be possible to
optimise the energy coverage by taking some data with the MST/SST
sub-array only. This must be studied once the final configuration of
CTA is known.

The Galactic Plane Survey (Chapter 6) 
will contribute to observations needed in
this KSP in the first two years, providing a
depth of approximately 15 hours in survey mode.  The deep observations
required for this KSP imply the need to take an additional 160 hours every year during
the first three years, followed by 45 hours per year during the
following seven years. Figure \ref{fig:sgra-available} shows the
estimated observation time available for a single year as a function
of zenith angle, which typically should be as low as
possible. Redistricting the zenith angle to under 35 degrees provides
enough time per year for this KSP, under the assumption of no other
competing sources in the same band of right ascension. 

\subsubsection{Relation to other KSPs}

The Galactic Centre KSP is closely connected to the Dark Matter
  Programme (Chapter~\ref{sec:DM_prog}), the Galactic Plane Survey 
(Chapter~\ref{sec:ksp_gps}), and the Extragalactic Survey (Chapter 8), in the case
that the region of interest is connected to the Galactic Centre.  For
this reason, some coordination between these KSPs will be useful when
determining the detailed observation strategy.

\subsubsection{Analysis Strategy} 

Due to the complexity of this region, careful consideration will be
required for the analysis. Several different analysis approaches will
be applied to the data, particularly because each science case has
different goals. This includes standard survey-like analyses, along
with searches for time variability at multiple scales. It implies both
imaging, likelihood modeling, and spatial/temporal region-based signal
extraction. All science cases require extremely low systematic errors,
ideally better than those required by CTA as a whole, which may be difficult to
achieve without detailed knowledge of the system.

The analysis challenges and possible solutions are as follows:

\begin{itemize}
\item \textbf{Exposures near systematic limits} are likely in the
  central survey region, where the goals include a search for weak
  variability, and detailed spectral and morphological studies. This
  is particularly challenging in this region where the optical
  brightness can vary over an order of magnitude, requiring careful
  consideration in the background determination to avoid biasing.  CTA
  should provide sufficient information to correct for
  observation-to-observation atmosphere variations (i.e. seeing and
  transparency), but this KSP will be one of the first to push them to
  their limits. 
\item \textbf{Strong source confusion} is already a known challenge
  for this region with current telescopes. A detailed likelihood
  modeling analysis will be required in order to disentangle emission
  components, implying the need for iterative modeling.
\item \textbf{Emission that is of similar scale or larger than the
    field of view} is a key part of several of the science goals,
  namely large-scale outflows and the diffuse emission. In all current
  analysis techniques where background is normalised using off-source
  regions in the field of view, such emission is subtracted due to the
  methodology. Therefore new techniques will need to be developed to
  deal with large scale targets, requiring better methods for using
  external information to normalise background rates and the detailed
  spatial shape of the detection efficiency.
\item \textbf{Development of a robust high-energy diffuse model} will
  be required at the lowest energies to separate emission from local
  acceleration from emission from cosmic-ray interactions that is seen
  at GeV energies. To date there is no good model for this emission
  that extends to the core CTA energies in the Galactic Centre region.
\end{itemize}

These challenges imply considerable work to be done by the CTA
Consortium, which may require the development of new science tools
and techniques that should eventually be propagated to Guest Observers.

\subsection{Data Products}

The data products produced by this KSP will include:

\begin{itemize}
\item maps/data-cubes of the full region of interest (excess, flux,
  spectral hardness),
\item spectra of the central source, at various regions in the diffuse
  emission, and at the position of known SNRs, PWNs, and interesting
  sources within the field of view,
\item light curves of the Galactic Centre central source and any other
  point-like sources,
\item a catalogue of the Galactic Central region, including all
  detected sources with morphological, spectral, and variability
  measures; this may require special consideration above what is in a
  general CTA catalogue or for the Galactic Plane Survey due to the
  complexity of the emission region, and
\item a model of diffuse VHE gamma-ray emission in the region (needed for
  dark matter, central source, or
  background object studies); ideally this should provide components
  expected from cosmic-ray interactions and
  from local acceleration.
\end{itemize}

These data products will be produced by the CTA Consortium and will be
made available to the public when the KSP is published.

\subsection{Expected Performance/Return}

\paragraph{Determination of the nature of the central source}

The angular resolution and source localization capabilities of CTA
will provide strong constraints on the extension of the central source
(see Figure \ref{fig:gc-extension}). This will help discard many
scenarios regarding its nature. For example, if the emission arises
from cosmic rays interacting with the circum-nuclear ring (CMR) about
Sgr A*, then we expect a larger extension (up
to $1.5'$, ignoring ballistic propagation) than that produced by the
possible counterpart PWN G359, which would give an extension of
$<10''$ given the magnetic and radiation fields near the black
hole. The proposed observations will very clearly discriminate between
these two options.

\paragraph{A detailed view of the diffuse VHE emission}

The increase in sensitivity over existing instruments will also
provide the ability to study the energy-dependent morphology of the
diffuse region and will enable the extraction of detailed spectra at
various points, possibly showing spectral variations expected if the
diffuse emission was caused by a past, point-like outburst. The
comparison of spectra in various regions of the central 100~pc will
also help in evaluating whether cosmic rays can penetrate the densest cloud
cores present in the CMZ, such as Sgr\,B2, and will provide more
information about possible PeV cosmic-ray production in the region.

\paragraph{Resolving new, previously undetectable sources}

Simulations carried out verify that with the exposure proposed here CTA will be
able to resolve point sources above the diffuse VHE emission at flux
levels from 2 to 5 mCrab, depending on their location in the
region. This will likely lead to the discovery of new sources that are
presently not resolvable, including young massive stellar clusters
such as the Arches and Quintuplet clusters, known SNRs such as Sgr\,C
and SNR G\,1.9+0.3, and over 10 PWN candidates.

\paragraph{Search for variability in the VHE source near Sgr A*} 

The X-ray variability of Sgr A* is on timescales down to an hour,
which are too short for current VHE instruments to resolve. CTA
will provide sensitivity to detect the central source on minute
timescales, allowing for the detection of strong variability on
similar scales and for weaker variability over hour
timescales. Therefore, it should be possible to detect a flare and
measure spectral variability on $<$30 minute scales.  The monitoring
involved in this KSP will also allow CTA to study any long-term flux
variations that may be due to changes in accretion around the black
hole.

\paragraph{Studying the interaction of the central source with
  neighbouring clouds} 

Simulations show that the improvement in the point spread function and in
sensitivity over current instruments will allow CTA to resolve the
Galactic Centre source from neighbouring clouds and to provide a
detailed view of clumps in the central molecular zone. The CTA
point-source localization accuracy of $<3''$ will be sufficient to rule out
some counterparts, and CTA's excellent angular resolution will enable
the search for an extension of the central source that might be
expected in a PWN scenario or in the case of hadronic interactions
with nearby gas.

\paragraph{Science impact}

This KSP will produce the deepest exposure of the Galactic Centre
region, the richest known environment of VHE gamma-ray emission. The
scientific outcome on the different scientific cases discussed here is
enormous. The legacy data set provided by this KSP will benefit many
multi-wavelength and multi-messenger studies.

\section{KSP: Galactic Plane Survey}
\label{sec:ksp_gps}

Astronomical surveys of the Galaxy provide essential, large-scale data sets
that form the foundation for Galactic science at all photon energies.  Here, we
introduce a Galactic Plane Survey (GPS) for CTA that will fulfill a number of
important science goals, including: 1) providing a census of Galactic
very-high-energy (VHE) gamma-ray source populations, namely supernova remnants
(SNRs), pulsar wind nebulae (PWNe) and binary systems, 
through the detection of hundreds of new
sources, substantially increasing the Galactic source count and permitting more
advanced population studies, 2) identifying a list of promising targets for
follow-up observations, such as new gamma-ray binaries and PeVatron candidates,
to be carried out by the Consortium within the Key Science Projects (KSPs) or
to be proposed through the Guest Observer (GO) programme, 3) determining the
properties of the diffuse emission from the Galactic plane, 4) producing a
multi-purpose, legacy data set, comprising the complete Galactic plane at very high 
energies,
that will have long-lasting value to the entire astronomical and astroparticle
physics communities, and 5) discovering new and unexpected phenomena in the
Galaxy, such as new source classes and new types of transient and variable
behaviour.

The GPS KSP will carry out a survey of the full Galactic plane using both the
southern and northern CTA observatories; the survey will be graded so that more
promising regions (especially the inner Galactic region of
$-60^{\circ} < l < 60^{\circ}$) will receive significantly more observation time
than other regions.  Our knowledge from the current Galactic VHE source
population (e.g. log $N$ -- log $S$ extrapolations) and from other high-energy
(HE) surveys (e.g. Fermi-LAT) motivate a target sensitivity for the CTA
GPS at the level of a few mCrab
($1 \mathrm{mCrab} = 5.07 \times 10^{-13}$~ph~cm$^{-2}$~s$^{-1}$ for a minimum
energy threshold of 125 GeV) in order to achieve the aforementioned science
goals.  As demonstrated here, the GPS will achieve a sensitivity\footnote{All
target sensitivities presented in this KSP are a result of the combined
simulations of individual telescope sub-arrays: Small Sized Telescopes (SSTs),
Medium Sized Telescopes (MSTs), and Large Sized Telescopes (LSTs).} (for
detecting a point-like source at a statistical significance of $5\sigma$)
better than 4.2~mCrab over the entire Galactic plane and 1.8~mCrab in the inner
Galactic region.  Thus, the CTA GPS will be a factor of 5~--~20 more sensitive
than surveys carried out by earlier or existing atmospheric Cherenkov
telescopes (HEGRA, H.E.S.S. and VERITAS).  In the northern hemisphere, CTA will
complement and extend observations made by the water Cherenkov telescope HAWC;
CTA will go deeper by a factor of 5--10 in the core energy range of 100 GeV to 10 TeV 
compared to the HAWC
5-yr sensitivity \cite{Abeysekara13,Abeysekara2017}, at much lower energy and with
substantially better angular resolution (e.g. factor of 5 better at 1~TeV).  A
comparison of the instrumental differential sensitivities is shown in
Fig.~\ref{fig:cta_sens}.

There are several strong motivations to carry out the GPS as a KSP of CTA.  The
GPS will provide a valuable legacy product to the entire astronomical 
community -- the first sensitive VHE scan of the entire Galactic plane.
It is expected that many sources will be discovered, resulting in follow-up
observations and a large scientific output.  The periodic data releases,
comprising sky images (maps) and source
catalogues, will strongly shape the observing programme of CTA as well as other
observatories, both in space and on the ground.  In addition, the GPS will have
significant synergies with other multi-messenger facilities to enhance the
scientific profile and output of CTA (see Sect.~6.1.3).  The Consortium, with
its detailed knowledge of the atmospheric Cherenkov technique and CTA-specific
technical information, will be in a good position to develop the non-standard
data analysis tools to account for a variety of observational and instrumental
effects present in exceptionally large data sets.  Of particular importance
will be the issues of analysis of sources extended beyond the CTA point spread
function (PSF), proper treatment of diffuse emission and source confusion and
the production of a coherent source catalogue.

The GPS will provide important pathfinder information for other CTA KSPs,
including the \emph{Galactic Centre KSP} (Sect.~\ref{sec:ksp_gc}), the
\emph{Transients KSP} (Sect.~\ref{sec:ksp_trans}), the
\emph{Cosmic Ray PeVatrons KSP} (Sect.~\ref{sec:ksp_acc}) and the
\emph{Star Forming Systems KSP} (Sect.~\ref{sec:ksp_sfs}).  It will also
support the preparation of the GO programme since it will guide the community
towards interesting targets via the production of source catalogues and sky
maps.

The GPS is divided into an early-phase programme (Years~1~--~2) and a
long-term programme (Years~3~--~10) with a total requested observation time of
1020~h (CTA-South) and 600~h (CTA-North).  For comparison, the total amount of
dark time available for all CTA observations will be 1100--1300~h/year at each site.
Observations will be done during dark time, under good weather conditions, and
at zenith angles less than $45^{\circ}$.  A double-row pointing strategy will be
utilised, with a nominal pointing separation distance of $3^{\circ}$ and with a
cadence set to optimise sensitivity to periodic phenomena on a variety of time
scales.  Follow-up observations of promising regions and sources revealed by
the GPS will mostly be carried out in the GO programme.  Because of their
scientific importance and timeliness, follow-up observations of transients and
PeVatron candidates will be done in the context of the relevant KSPs.

Since the GPS will be carried out over the entire ten-year nominal lifetime of
CTA, it is of high importance that regular data releases to the community are
made.  In the early phase, we expect an intermediate data release on a time
scale of 12--18~months after the start of data taking, with the main data
release for the first two years of observations to be made at the end of
Year~3.  Successive data releases during the long-term programme will take
place every $\sim$2~years.  The data products available to the community via
a searchable online archive will include: source catalogue(s), sky maps,
scientific analysis tools and GO tools (e.g. exposure maps, observation
schedules, etc.).  

With the uniform, comprehensive and mCrab-deep survey proposed here, the GPS
will greatly expand the discovery potential of CTA and will also produce a
great deal of guaranteed science.  Hundreds of sources\footnote{In this KSP, a
``source'' is defined as an astrophysical source, i.e. a discrete source of VHE
gamma rays which have a common astrophysical origin.} will be detected. 
For each of these sources the GPS will provide spectral and morphological
information which can be used in conjunction with multi-wavelength (MWL) data
and comparison to theoretical models to identify the origin of the VHE gamma
rays as well as the favoured scenarios for particle acceleration ultimately
responsible for the gamma-ray production.  With data taken over ten years, the
GPS will also be an important source of serendipitous discoveries.  The
sensitivity achieved by the GPS compared to the current state of the art will
enable a major step forward in understanding the VHE Galactic source population
and will thus permit a better understanding of the origin of cosmic ray.

\subsection{Science Targeted}\label{sec:science}

\subsubsection{Scientific Objectives}\label{subsec:science_obj}

Galactic plane surveys have been carried out in practically every wavelength band,
and new observatories regularly incorporate a GPS into their core observation
programme to take advantage of improvements in telescope capabilities
compared to previous generations and as an efficient way of achieving their
diverse scientific objectives.

Generally speaking, a GPS provides: 

\begin{itemize}
\item important source discovery potential,
\item the critical, first-look data which guide deeper, more specific
  observations,
\item a complete and systematic view of the Galaxy to facilitate our
  understanding of Galactic source populations and diffuse emission, 
\item a comprehensive data set and catalogue which comprise a key part of the
  scientific legacy of an observatory, and
\item upper limits where no significant emission is detected, to constrain
  theoretical emission models
\end{itemize}

\begin{table}
\begin{tabular}{cccccc}
\hline
Telescope & Hemisphere & Galactic Plane & Energy & Sensitivity &
Reference \\
                   &            & Coverage       & (GeV)  & (mCrab)     &  \\
\hline
Fermi-LAT 2FHL & (space) & full plane & $> 50$ & $\sim$30 -- 40 &
\cite{Ajello15} \\
H.E.S.S.-I & S & $-95^{\circ} < l < 60^{\circ}$, $|b| \lesssim 2^{\circ}$ &
$\gtrsim 300$ & 4 -- 20 & \cite{Carrigan13b} \\
VERITAS & N & $67^{\circ} < l < 83^{\circ}$, $-1^{\circ} < b < 4^{\circ}$  &
$\gtrsim 300$ & 20 -- 30 & \cite{Ong13} \\
ARGO-YBJ & N & northern sky & $> 300$ & 240 -- 1000 & \cite{Bartoli13} \\
HEGRA & N & $-2^{\circ} < l < 85^{\circ}$, $|b| < ~1^{\circ}$ & $> 600$ &
150 -- 250 & \cite{Aharonian02s} \\
Milagro & N & northern sky & $> 10,000$ & 300 -- 500 & \cite{Atkins04} \\

\hline 
\end{tabular}
\caption{Compilation of previous VHE surveys of the Galactic plane.  For each
  telescope, the spatial coverage, energy threshold, approximate point-source
  sensitivity and reference are given.  The telescopes are listed in order of
  increasing energy threshold.  For a discussion of the ``mCrab'' unit of flux
  sensitivity, see Sect.~\ref{sec:performance} and \cite{HEGRACrab}.}
\label{tab:vhesurveys}
\end{table}

The major scientific objectives for the CTA GPS include the following:

\begin{itemize}
\item discovery of new and unexpected phenomena in the Galaxy.  These would
  include completely new source classes, new types of transient and variability
  behavior, or new forms of diffuse emission.
\item discovery of PeVatron candidates that are of key importance in our search
  for the origin of cosmic rays.  These candidates will likely require deeper,
  follow-up observations to confirm and characterise their PeVatron nature.
\item detection of many new VHE Galactic sources (of order 300 -- 500, see
  Sect.~\ref{subsec:science_context}), particularly PWNe and SNRs, to
  increase the Galactic source count by a factor of five or more.  The
  substantially increased statistics and more uniform sensitivity will allow
  more advanced population studies to be performed.  The ultimate goal is
  to significantly advance our understanding of the origin of cosmic rays.
\item measurement of the large-scale diffuse VHE gamma-ray emission
  \cite{Abramowski14diffuse}, to better understand its origin in terms of
  inverse-Compton, $\pi^0$ decay, and unresolved source components.
\item discovery of new VHE gamma-ray binary systems, a unique class of objects
  with periodic emission on varying timescales, where physical processes are
  observed from different vantage points depending on each system's orbital
  inclination.  Only five such systems are currently known in the Galaxy.
\item production of a multi-purpose legacy data set comprising sky images and
  source catalogues of the complete Galactic plane at very high energies.  This dataset will
  have long-lasting value to the entire astronomical and astroparticle physics
  communities, far beyond the lifetime of CTA.
\item the GPS will produce and periodically release sky maps and catalogues.
  These will be important road maps to guide further Galactic observations made
  by:
\begin{itemize} 
  \item the CTA GO community,
  \item the broader astronomical communities, and
  \item the CTA Consortium via other KSPs.
\end{itemize}
\end{itemize}

In the context of the defined science themes of CTA (Fig.~\ref{fig:KSPmatrix}),
the goal of the GPS is to substantially improve our understanding of the origin
of cosmic rays by answering the following distinct questions:

\begin{itemize}
\item[1.1] How and where are protons and nuclei accelerated to PeV energies?
\item[1.2] How are particles accelerated in relativistic shocks?
\item[1.4] What is the impact of cosmic rays on the interstellar medium (ISM),
  and how do they propagate?
\item[2.1] What is the role of external photon fields, jet content, and
  geometry in distinguishing jet sources, such as pulsars and microquasars?
\item[2.2] Where and how do pulsar complexes accelerate high-energy
  particles?
\end{itemize}

The current motivation for the CTA GPS builds upon the published work of
others, in particular Dubus et~al. 2013~\cite{Dubus13}.  This KSP document is
consistent with that work, but has gone further by using up-to-date sensitivity
calculations, simulations and observation planning.  Additional comparisons
with observatories at complementary energies, namely Fermi-LAT and HAWC,
have also been carried out.

\begin{figure}[t]
\centering
\includegraphics[width=0.70\textwidth]{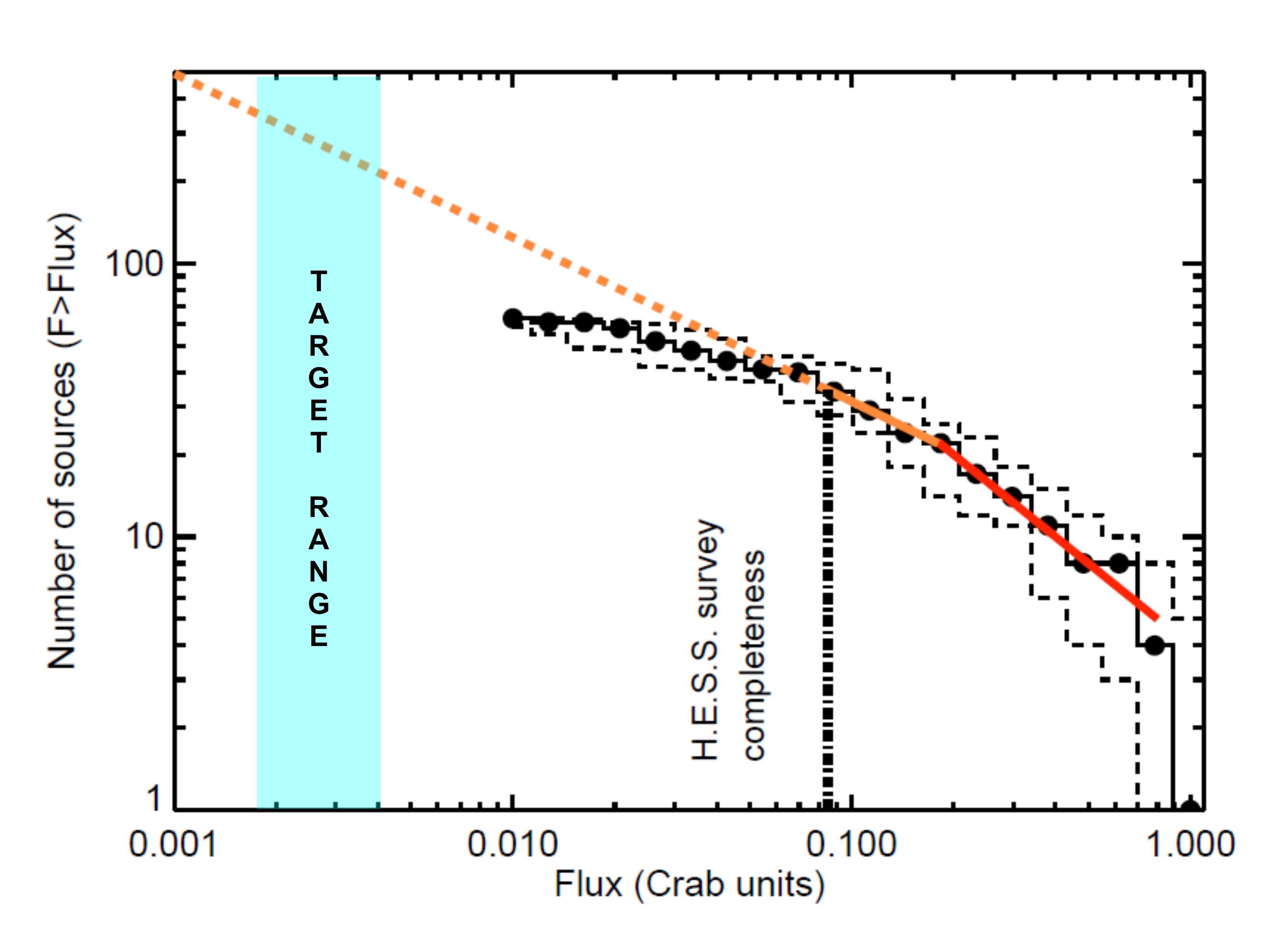}
\caption{Cumulative Galactic source count as a function of VHE gamma-ray flux.
  Adapted from Renaud 2009~\cite{Renaud09} to show the target CTA GPS
  sensitivity range (cyan-shaded region).}
\label{fig:sourcecount}
\end{figure}

\subsubsection{Context / Advance beyond State of the Art}\label{subsec:science_context}

Our current knowledge of the Galaxy at very high energies is based on both surveys and pointed
observations made by ground-based gamma-ray telescopes using the atmospheric
Cherenkov, water Cherenkov and air-shower techniques as well as space-based
gamma-ray telescopes, such as Fermi-LAT and AGILE.  Specifically, surveys have had a major
impact on VHE source detections in the Galactic plane, including the H.E.S.S.
GPS, the Milagro survey of the northern sky, the HEGRA survey of the northern
Galactic plane, the VERITAS survey of the Cygnus region and the
Fermi-LAT catalog of hard sources.  Table~\ref{tab:vhesurveys} provides
a compilation of the previous VHE surveys of the Galactic plane.

A total of $\sim$100 Galactic sources of VHE gamma rays have now been detected.
Their distribution in Galactic latitude peaks sharply along the Galactic plane
($b \simeq 0^{\circ}$), with more than 90\% located at latitudes
$b < 2.0^{\circ}$, although there may be some bias due to the worsening
sensitivity of the H.E.S.S. survey off-plane.  The largest source class is that
of PWNe, followed by SNRs and gamma-ray binaries.  About two-thirds of known
sources are not yet firmly identified; most have multiple plausible
associations that are challenging to disentangle, although some appear to be
dark accelerators that are not detected at lower energies.  Only a handful of the VHE
Galactic sources are point-like in nature (largely binary systems). The large
majority of sources have extended VHE emission, with a typical angular size of
$\sim 0.1^{\circ} - 0.2^{\circ}$ (in radius), and a few are considerably larger
than this.  The reconstructed spectra are generally well fit by power-law
spectral models, with typical differential spectral indices in the range of
$\Gamma \sim 2.3 - 2.7$.

We can estimate the expected number of VHE sources (predominantly PWNe and
SNRs) to be detected in the CTA GPS from the known population of $\sim$100 VHE
Galactic sources.  For example, in Renaud 2009~\cite{Renaud09}, the log $N$ -
log $S$ distribution of the VHE Galactic population (including all detections
made by imaging atmospheric Cherenkov telescopes but dominated by H.E.S.S.
detections) was used to predict 300~--~500 sources for an instrument achieving
a sensitivity of 1~--~3~mCrab. Figure~\ref{fig:sourcecount} shows the
cumulative source count as a function of VHE flux.  Similarly, a few different
source population models were used in Dubus et~al. 2013 \cite{Dubus13} to
estimate the source count for the CTA GPS at 20~--~70 SNRs and 300~--~600
PWNe; however, in this case the models assume point-source sensitivities.  It
is important to note that sensitivity will be worse for extended sources (see
Sect.~\ref{subset:targets}) and the actual detected numbers of sources may be
correspondingly reduced.  Alternatively, these extrapolations could also be
viewed as conservative, since they are based on the currently known VHE source
populations.  The discovery of new source classes would increase the total
number of Galactic sources detectable by CTA.

\begin{figure}[t]
\centering
\includegraphics[width=1.00\textwidth]{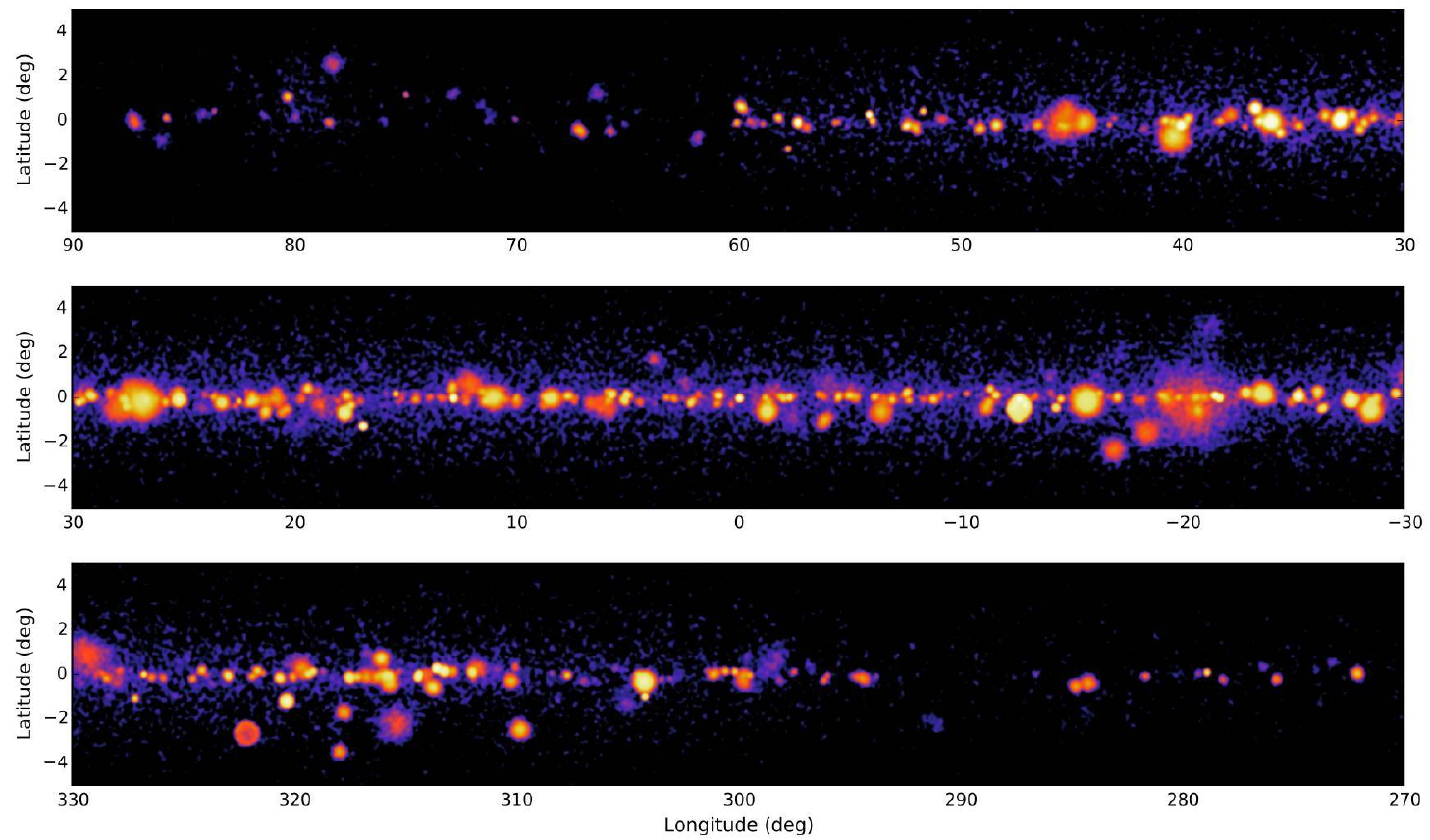}
\caption{Simulated CTA image of the Galactic plane for the inner region
  $-90^{\circ} < l < 90^{\circ}$, adopting the actual proposed GPS observation
  strategy, a source model incorporating both supernova remnant and pulsar wind nebula 
  populations and diffuse emission.}
\label{fig:ctagalplane}
\end{figure}

Figure~\ref{fig:ctagalplane} shows a simulated image of what could result from
a CTA survey of a portion of the Galactic plane using a model that incorporates
SNR and PWN source populations as well as diffuse emission.
The current work is an extension of
earlier simulations described in Ref. \cite{Dubus13}.  As can be seen in this figure, 
the source density in the innermost regions of the Galaxy
($| l | <  30^{\circ}$) could approach 3~--~4 sources per square degree and thus
source confusion is likely to be a concern (see
Sect.~\ref{subsec:source_conf}).  Preliminary work has 
estimated the number of sources expected by calculating the intrinsic VHE
luminosities of the existing sources (with known distances) and then assuming a
disc-like distribution of sources to estimate the increased number of more
distant sources with comparable intrinsic luminosities detectable by CTA.  This
relatively simplistic but straightforward estimate yields $\sim 300$ sources
and is likely conservative because it does not consider possible sources
that are intrinsically dim, i.e. those sources with intrinsic luminosities
comparable or dimmer than Geminga, the source that currently has the lowest
intrinsic VHE luminosity \cite{Abdo09ee}.  \emph{The conclusion that can be
drawn from these population estimates is that CTA can expect the detection of
many hundreds of Galactic sources that would follow from a survey of the plane
with a sensitivity at the level of a few mCrab.}

\begin{figure}
\centering
\includegraphics[width=0.49\textwidth]{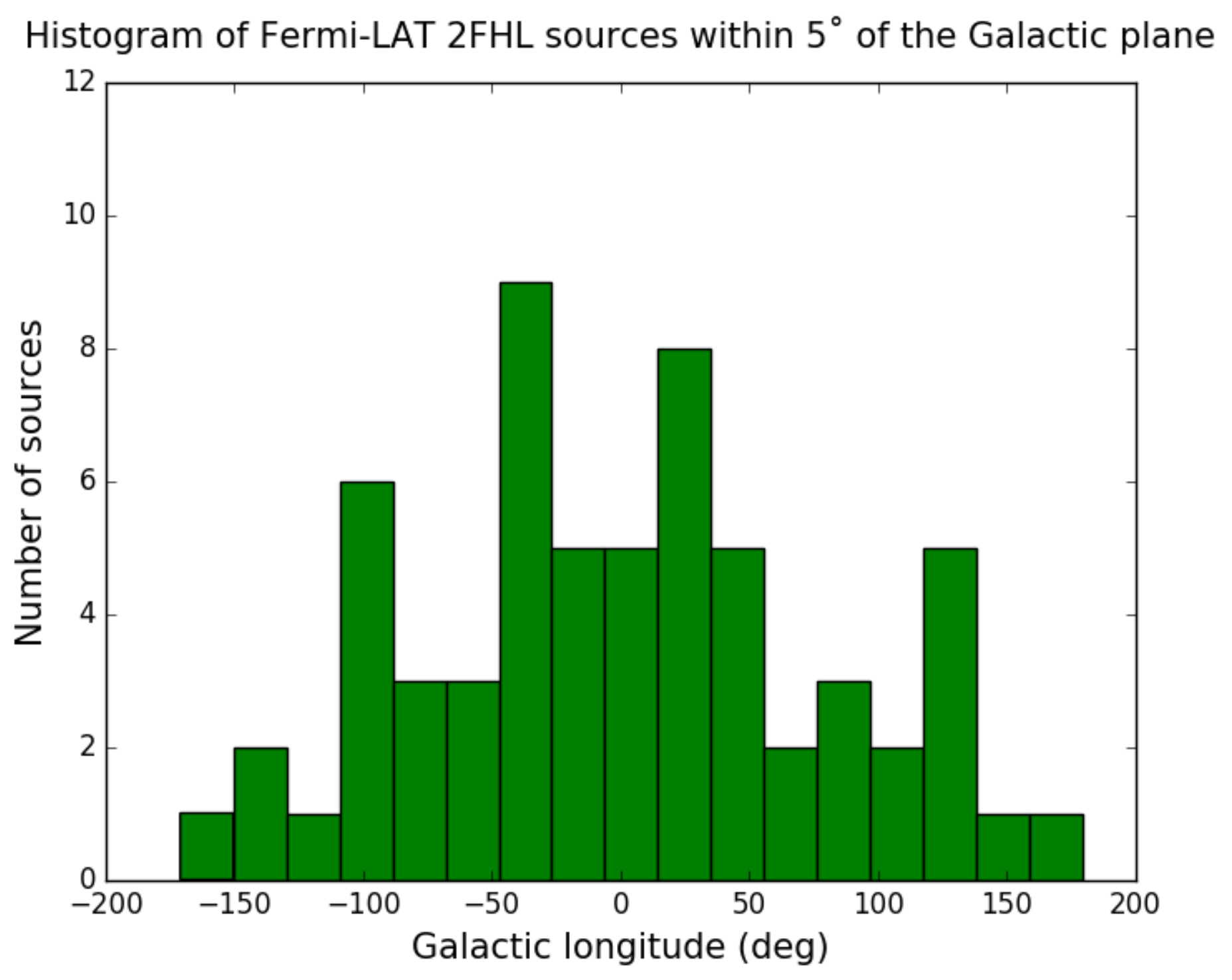}
\includegraphics[width=.95\textwidth]{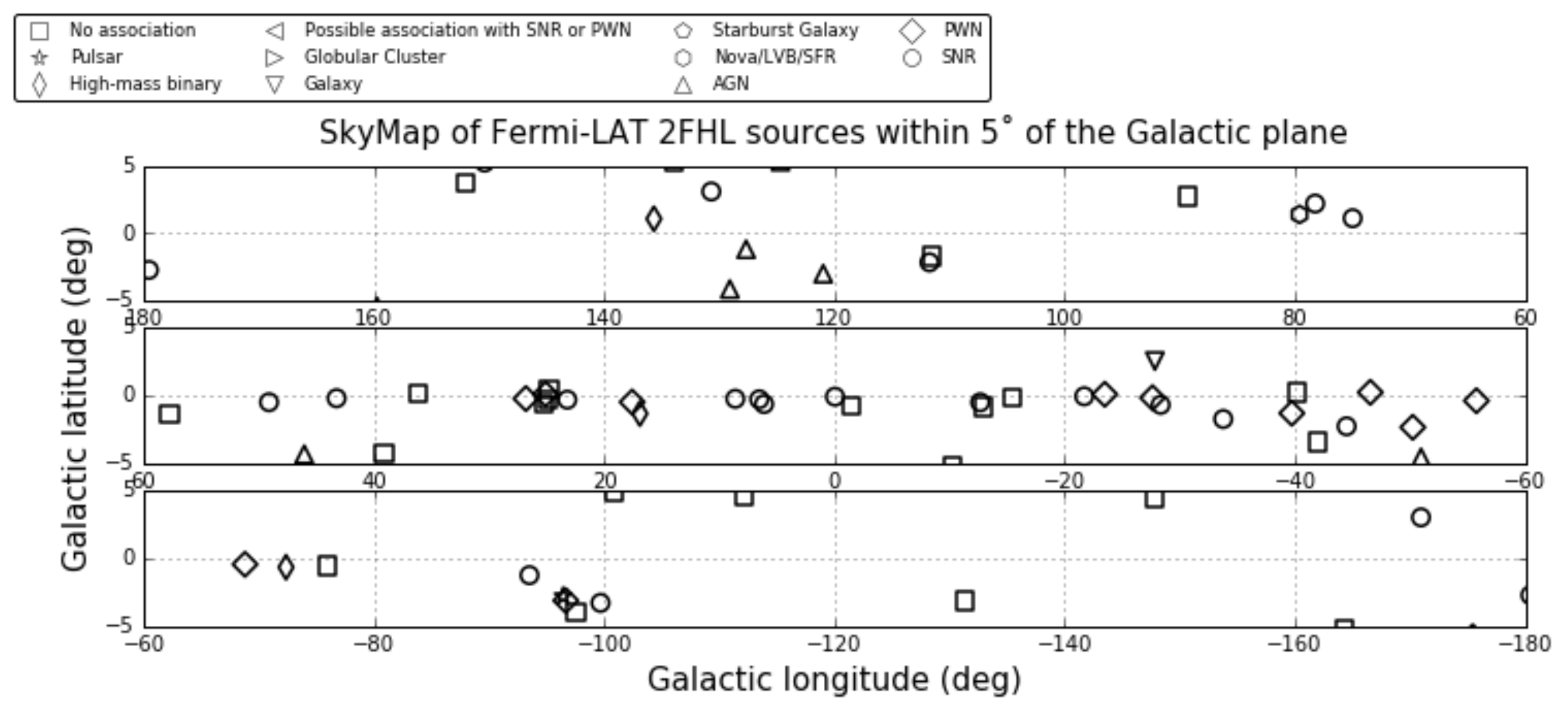}
\caption{(Top Right) histogram of Galactic longitude of Fermi-LAT 2FHL
sources (from ~\cite{Ajello15})
within 5$^{\circ}$ of the Galactic plane; (Bottom) sky map of
Fermi-LAT 2FHL sources within 5$^{\circ}$ of the Galactic plane.}
\label{fig:fermi2fhl}
\end{figure}

For a somewhat different approach, one can use knowledge from Fermi-LAT
to estimate sources that could be expected in the VHE band.  In one study, the
spectra of probable Galactic sources from the Fermi-LAT 2FGL catalogue
were extrapolated to VHE to predict that CTA would detect more than 70 of the
Galactic 2FGL sources \cite{Dubus13}.  The spatial (Galactic coordinate)
distribution of the Fermi-LAT sources has been studied further for this
KSP, using primarily the 2FHL~\cite{Ajello15} catalogue.  Increased exposure by
Fermi-LAT has enabled the production of the latter catalogue, containing
sources of gamma rays with energies $E > 50$~GeV, of which 62 are within
$|b| < 10^{\circ}$.  Of these, 15 are spatially coincident with SNRs, 13 with
PWNe, 6 with SNR/PWN complexes, 3 with X-ray binaries, 1 with a pulsar and 1
with the Cygnus region \cite{Ajello15}.  Figure~\ref{fig:fermi2fhl} shows a
projection in Galactic coordinates of the 2FHL sources, along with a histogram
of the distribution in Galactic longitude.

It can be seen that the 2FHL source distribution is relatively broad in
longitude, which provides support for a VHE survey that is likewise broad.  It
is important to mention several caveats when considering extrapolations from
the Fermi-LAT catalogues to higher energies: a) the extrapolations
assume continuous HE-VHE source spectra with no cutoff or break, b) the largest
VHE source class, PWNe, are bright at VHE but generally faint at HE (therefore
suggesting that extrapolation from the HE regime may not be a fair indicator of
VHE source numbers), c) the second largest source class of SNRs often exhibit
spectral breaks between the HE and VHE regimes, and d) the Fermi-LAT
catalogues have reduced sensitivity in the inner regions of the Galaxy where
source confusion is significant.  Thus, we conclude that information from
Fermi-LAT is a useful indicator from an adjacent energy band but does
not provide unambiguous predictions.

It is challenging to predict with better precision the number of sources that
CTA can expect to detect in the GPS.  While there may be larger-than-expected
source populations in the outer Galaxy and at higher latitudes, it is clear
that the highest concentration of VHE sources lies in the inner Galaxy.
Therefore, a full-plane GPS, with graded sensitivity goals that peak in the
inner Galaxy, is the most efficient way to probe the VHE source population.
Aiming for a factor of approximately ten improvement in sensitivity compared to
the current state-of-the-art will:

\begin{enumerate}
\item enable a major step forward in understanding the VHE Galactic source
population (and hence the origin of cosmic rays),
\item open a large amount of discovery space for new and unexpected phenomena,
and
\item provide an excellent baseline sensitivity in regions of the plane of
great interest to CTA and the overall community (e.g. the Galactic Centre,
Cygnus, Vela, etc.) that will set the stage for even deeper follow-up
observations.
\end{enumerate}

All of these various considerations motivate a target sensitivity for the GPS
at the level of a few mCrab, approximately a factor of ten better than the
H.E.S.S. and VERITAS surveys.

Thus, to summarise the advances that the CTA GPS will bring to the state of the
art:
\begin{itemize}
\item in the south, CTA will go deeper in the inner region
  ($| l | < 60^{\circ}$) by a factor of 5 -- 20 compared to H.E.S.S., and CTA
  will cover the entire accessible Galactic plane, and
\item in the north, CTA will go deeper by a factor 5 -- 10 compared to HAWC
  in the energy range 100 GeV -- 10 TeV and with a factor of five better angular resolution
  (at 1~TeV).  In Cygnus, CTA will reach a factor of ten better sensitivity than
  that achieved by VERITAS.
\end{itemize}

\subsubsection{Multi-wavelength / Multi-messenger Context}

\paragraph{Synergy with HAWC}

The HAWC water Cherenkov array \cite{Abeysekara13,Abeysekara2017} has recently started
operation at Sierra Negra, Mexico.  Larger than its predecessor, Milagro,
and at a higher altitude of 4,100~m, it is substantially more sensitive.  It is
also farther south (latitude = $19^{\circ}$~N) than Milagro (latitude =
$36^{\circ}$~N) which will permit a better view of the inner regions of the
Galaxy.  Table~\ref{tab:ctahawc} shows a comparison of the relevant
capabilities of HAWC and CTA for surveying the Galactic plane.  The table shows
that CTA and HAWC are complementary telescopes.  HAWC will survey the entire
(northern) sky above an energy threshold of $\sim$2~TeV.  CTA will survey the
Galactic plane in both the north and south with an order of magnitude lower
energy threshold, a factor of 5 -- 10 better sensitivity and a superior angular
resolution compared to HAWC.  CTA's angular resolution in particular will allow
it to probe morphological details not apparent in the larger HAWC sources.  

\begin{table}
\begin{tabular}{cccccc}
\hline
Observatory & Hemisphere & Energy Thresh. & Ang. Resolution & Pt. Source
Sensitivity & Ref. \\
\hline
CTA  & N, S & 125 GeV & $\sim0.07^{\circ}$ at 1 TeV & 2 -- 4 mCrab &
\cite{Acharya13} \\
HAWC & N    & 2 TeV   & $0.30^{\circ}$              & 20 mCrab     &
\cite{Abeysekara13} \\
\hline 
\end{tabular}
\caption{Comparison of CTA and HAWC for surveying the Galactic plane at VHE.
The angular resolution is defined as the 68\% confinement radius.  The HAWC
sensitivity assumes a livetime of five years.  The sensitivity estimates for
both instruments assume a power-law spectrum ($E^{-2.3}$ or $E^{-2.4}$) with no
cutoff.  Note that if a source cuts off at 5~TeV, the HAWC sensitivity
degrades to $\sim 50$~mCrab while the CTA sensitivity is not greatly changed.
The sensitivities for the CTA GPS were calculated for an energy threshold of
125~GeV; for a discussion of the unit ``mCrab'' unit of flux sensitivity, see
Sect.~\ref{sec:performance} and \cite{HEGRACrab}.}
\label{tab:ctahawc}
\end{table}

\paragraph{Synergies with Other Instruments}

The CTA GPS will have significant synergies with other MWL and multi-messenger
facilities expected to operate contemporaneously with CTA, greatly enhancing
the profile and scientific output of CTA.  
To help in the interpretation of the CTA results, it is necessary to measure
the distribution of the interstellar gas (both atomic and molecular)
in order estimate the diffuse gamma-ray component arising from
cosmic ray interactions on the gas.
The gas is measured directly using radio and millimetre-wave telescopes
(see Section 2.1) 
and thus collaboration between CTA and such instruments
 in both the southern and northern hemispheres is essential.
On the particle detection side, the
increasingly better localisation of sources of neutrinos and cosmic rays
detected by, e.g., IceCube, Km3Net and the Pierre Auger Observatory, makes a new
searchable catalogue of TeV sources a valuable tool for multi-messenger
analysis of potential cosmic-ray sources.  The GPS will also be very relevant
for the new and upcoming (on the same timescale as CTA early and full
operations) optical and radio ``Transient Factories'' (e.g., \emph{Gaia}, iPTF,
ZTF, LSST, plus SKA and its pathfinders: LOFAR, MeerKAT, ASKAP, and MWA).  For
instance, a key step in the necessary automated decision-tree pipelines for
further observation is to check coordinates against existing catalogues in
order to identify the source.  Given the propensity of VHE sources to vary also
in radio and optical, it is important to provide access to some results of the
GPS as soon as possible, ideally without waiting for the official public
releases described above.  This will require agreements between CTA and other
facilities in order to share key source information from the GPS.  Finally, GPS
observations, spanning a large area of the sky and taken over a period of 10
years, could also prove useful in finding counterparts to sources of
gravitational waves detected by LIGO and VIRGO.

\subsection{Strategy}\label{sec:strategy}

\subsubsection{Observation Requirements}\label{subsec:obsrec}

One of the most critical aspects of the GPS is the sensitivity goal for the
observations -- the sensitivity achieved will directly impact the number of
detected sources and the quality of the determination of the source parameters
(position, morphology, energy spectrum and ability to detect variability).  It
will thus also impact the quality of the source identification.  Because of the
survey technique (i.e. many, separate pointings on a regular grid, where
possible sources could be present anywhere in the field of view), a second
key aspect is the requirement for good stability of CTA performance over
different epochs.  A third, critical aspect is to have the lowest-possible
energy threshold (to measure spectra over the widest range and to reconstruct
possibly soft-spectra sources; e.g. transients with durations of minutes, and
pulsars).  These various key aspects motivate the following general
requirements for observing conditions:

\begin{itemize}
\item dark time (moonless) observations only,
\item good (cloudless and low aerosol) weather conditions,
\item zenith angles below 50$^{\circ}$, 
\item use of both CTA-South and CTA-North, and
\item full array observations for both CTA-South and CTA-North.  Such operations are
  currently defined by having 80\% of each telescope type deployed on a site.
\end{itemize}

\subsubsection{Targets, Observation Strategy, and
  Follow-ups}\label{subset:targets}

For the GPS, the ``target'' is in principle the entire Galactic plane.  Thus,
when planning the GPS, the following parameters are of overall
strategic importance:

\begin{itemize}
\item[](a) Galactic longitude and latitude,
\item[](b) required sensitivity and observation time,
\item[](c) yearly schedule for observations and reviews,
\item[](d) pointing strategy,
\item[](e) pointings, cadence, and schedule of observations, and
\item[](f) follow-up observations: transients and PeVatrons.
\end{itemize}

We consider each of these aspects in turn.

\textbf{(a) Galactic longitude and latitude}

We consider longitude coverage first.  Based on the scientific objectives and
context presented in Chapter~\ref{sec:science}, there is good motivation for a
survey of the entire Galactic plane, at least to a moderate depth.  This
approach has the benefits of maximizing the discovery potential of CTA and also
providing the long-lasting legacy of a full-plane survey.  However, it is also
clear that some regions of the Galactic plane (e.g. the inner region of
$| l | < 60^{\circ}$, the Sagittarius-Carina arm in the south and the
Cygnus-Perseus arms in the north) will likely be more fruitful in VHE source
discovery.  This assumption is based on the density of currently known (e.g.
PWNe, SNRs, binary systems, etc.) and prospective (e.g. globular clusters,
star-forming regions, OB associations, novae,  etc.) source types, as well as
on our current knowledge from HE and VHE instruments.  Thus,
the survey should be carried out in both the southern and northern hemispheres
and a graded approach should be implemented, i.e. the full plane should be
surveyed to some modest sensitivity while certain regions of the plane should
receive significantly more observation time to achieve a deeper sensitivity.

In terms of Galactic latitude, we note that the large majority of current VHE
sources lie within $|b| < 2^{\circ}$.  With the double-grid pointing strategy,
discussed in d) below, we can achieve relatively uniform sensitivity within
$4^{\circ}$ of the Galactic plane and additional, reduced sensitivity out to
$\sim$6$^{\circ}$ due to CTA's relatively large field of view.  Thus, the default
double-grid pointing strategy should be sufficient for most of the Galactic
plane.

\textbf{(b) Required sensitivity and observation time}

The scientific considerations in Sects.~\ref{subsec:science_obj}
and~\ref{subsec:science_context} motivate that the CTA GPS achieve a
point-source sensitivity in the 2 -- 3~mCrab range.  A minimal value of
4~mCrab is set by being able to achieve a substantial improvement over HAWC
(see Table~\ref{tab:ctahawc}).  As shown in Sect.~\ref{sec:performance}, our
current estimates predict a sensitivity of better than $\sim 3$~mCrab over the
entire Galactic plane in the south, with better than 2~mCrab in the inner
region of $| l| < 60^{\circ}$.  In the north, we can reach a sensitivity better
than 4~mCrab over the entire plane, and better than 3~mCrab in the Cygnus and
Perseus regions.

To reach these sensitivities and achieve the scientific goals, a minimum of
1020 hours is needed in the south and 600 hours in the north, as determined by
simulations (see Sect.~\ref{sec:performance}).  These times do not include
follow-up time (e.g. pointed observations) for specific interesting regions of
the plane.

The sensitivities quoted here are those for point sources, and it is important
to note that the sensitivity will degrade for extended sources.  For example,
the sensitivity will be $\sim$3 times worse for an extended source with a
radius of $0.25^{\circ}$.  There will also be degradation in
sensitivity from source confusion.

\textbf{(c) Yearly schedule for observations and reviews}

The GPS will be an important pathfinder for many other KSPs and the probability of discovering
many new VHE sources is very high.  These considerations motivate a relatively
rapid start to the GPS programme in the first two years of CTA, which will lead
to important early scientific results for the project.  On the other hand, we
have also demonstrated that there is great scientific potential in going deeper
to reach a few mCrab in sensitivity, and it is thus essential to continue the
GPS throughout the nominal 10-year lifetime of the project.  We define the
short-term programme (STP) as Years 1 -- 2 and the long-term programme (LTP) as
Years 3 -- 10.  The STP starts in Year 1 and the LTP starts in Year 3.  In
Sect.~\ref{sec:performance}, we give the projected sensitivities achieved in
the STP and in the LTP as a function of Galactic longitude.

The high profile and importance of the GPS require both periodic assessments of
the programme and regular data releases.  The reviews are needed to assess the
progress of the GPS and to consider other information gained from other CTA
KSPs and from other instruments (e.g. HAWC, IceCube, X-ray and space gamma-ray
missions, etc.).  Scheduled reviews, possibly including external scientists,
will be needed on a regular basis, e.g. approximately every two years.  There
may also be internal and more informal periodic assessments.  We would also greatly
benefit from input received from the broader GO community.
In Sect.~\ref{sec:dataprods}, we discuss the requirements and cadence of public
data releases.

\vspace{0.7cm}

\textbf{(d) Pointing Strategy}

\begin{figure}
\centering
\includegraphics[width=0.3\textwidth]{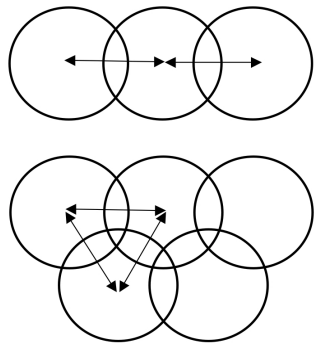}
\caption{Pointing schemes considered for the Galactic Plane Survey.  Single-row
  scheme (top) and double-row scheme (bottom).  The grid spacing is illustrated
  by the arrows in both schemes.}
\label{fig:gpspointing}
\end{figure}

We considered two possible pointing strategies: a single-row raster scan and a
double-row, offset raster scan.  These schemes are described in Dubus et al.
2013~\cite{Dubus13}.  As shown in Fig.~\ref{fig:gpspointing}, the single-row
scheme uses a uniformly-spaced single row of pointings that lie along the
Galactic plane ($b=0^{\circ}$, or slightly adjusted in latitude, if need be).
The double-row scheme uses pointings that lie above and below the $b=0^{\circ}$
line and that have a uniform spacing between adjacent points in the same row
and between nearest points in different rows.  The grid spacing in this scheme
corresponds to the distance between nearest pointings that form an equilateral
triangle. 

Our simulations demonstrate that both pointing strategies will be satisfactory
to achieve the desired sensitivities.  However, the double-row scheme is
preferred because it offers better performance at larger Galactic latitudes
($|b| > 2^{\circ}$) and more uniform sensitivity along the plane, and it may be more
robust during background subtraction.

\textbf{(e) Pointings, cadence, and schedule of observations}

As outlined in the previous section, the pointing strategy will use the
double-row scheme.  We have simulated various separation distances (2$^{\circ}$,
3$^{\circ}$, and 4$^{\circ}$) and find reasonable results with each of these
possibilities; we select 3$^{\circ}$ as a good default choice.  In determining a
realistic schedule for the GPS observations, we keep the following guidelines
in mind:

\begin{itemize}
\item Observations are carried out as close to culmination as possible, to
  minimise zenith angles.  This will permit the best overall sensitivity and
  lowest energy threshold.
\item We revisit fields on a recurring basis (e.g. on day, week, month, and
  year time scales) to improve our sensitivity to detecting periodic phenomena. 
\end{itemize}

A tentative schedule has been developed for the first two years (STP) of the
GPS.  For the southern observatory, a total of 300~h spread over 120$^{\circ}$
of Galactic longitude has been simulated; for the north, it is a total of
180~h over 90$^{\circ}$ of Galactic longitude (see Table~\ref{tab:pssens_gps}).
The scheduling study indicates that the STP observations can be achieved during
the first two years of CTA operations.  Work is continuing on this topic, in
particular to plan for re-scheduling of GPS observations affected by bad
weather or data quality.

\textbf{(f) Follow-up observations: transients and PeVatrons}

The GPS will discover many sources and regions of the plane for which follow-up
observations will be highly motivated.  Some of the follow-up observations will
be carried within the KSP, but we expect that the majority will be done
through the GO programme.

We define transients as sources whose VHE fluxes change by a significant amount
over a relatively short period of time.  Transients can occur anywhere in the
field of view during any CTA observation in any portion of the sky.  The
definition for what a transient is (i.e. for the required flux change and the
appropriate time interval) is defined globally for CTA and is discussed in
detail in the \emph{Transients KSP} (Chapter 9).  Thus, in this regard, although the GPS
and the Extragalactic Survey cover many more pointings than other KSPs, they
are no different than other programs in how transients should be considered.
Any transients discovered during a KSP pointing will be followed-up (or not),
as discussed in Chapter 9.  

It is expected that general follow-up observations of non-transient sources
(including both steady sources and variable sources such as binaries) will be
proposed and carried out in the GO programme.  The one exception to this
general strategy relates to PeVatron sources because of their high importance;
these sources are discussed below.  

One of the main science goals of the CTA Observatory is to discover the origin
of Galactic cosmic rays.  This goal is closely connected to the search for
elusive PeVatrons, putative cosmic accelerators that accelerate particles up to
PeV energies.  The current theoretical consensus is that there should exist a
small number ($\sim$ 2 -- 3) of currently active PeVatrons somewhere in the
Galaxy (see e.g.~\cite{Schure13}).  The CTA GPS will provide an unparalleled
data set which can be used to carry out this search.  With its full-plane
coverage and deep sensitivity -- especially at multi-TeV energies, thanks to
the inclusion of SSTs in the CTA-South array -- it should be possible to
identify at least several PeVatron candidates from the GPS data alone, by
looking for source candidates with hard power-law spectra that extend up to
$\sim$50~TeV and beyond, without evidence for a spectral cutoff.  This work
will be carried out in the context of of the Cosmic Ray PeVatrons KSP, which
will follow-up with deeper, pointed observations to confirm the candidates'
PeVatron nature.

\subsubsection{Relation / Importance to other KSPs}\label{subset:kspsynergy}

As a pathfinder for a number of other KSPs, the GPS 
will cover some of the same sky using observations that
will, in some cases, be taking place in the early phase of CTA operations.  
The synergies with other KSPs are briefly listed below:

\begin{itemize}
\item \textbf{Galactic Centre KSP}: the GPS will survey the region close to
  (i.e. within a few degrees of) the Galactic Centre, achieving a sensitivity
  of 2.7~mCrab in the first two years and 1.8~mCrab over ten years.  The region
  will be surveyed to a much deeper level in the context of the
  \emph{Galactic Centre KSP}.
\item \textbf{Cosmic Ray PeVatrons KSP}: the GPS will find the first evidence
  for PeVatrons which will be followed up by the
  \emph{Cosmic Ray PeVatrons KSP} to confirm and subsequently characterise
  their nature in detail.
\item \textbf{Star Forming Systems KSP}: the GPS will survey the Carina and
  Cygnus star-forming regions and the Westerlund~1 star cluster at a
  sensitivity level of a few mCrab.  In particular, the GPS coverage will
  result in a complete survey of the Cygnus region
  ($65^{\circ} <  l < 85^{\circ}$, $|b| < 3^{\circ}$), with a sensitivity of
  4.2~mCrab in the first two years and 2.7~mCrab over ten years.  Specific
  regions-of-interest will be surveyed to deeper levels in the context of the
  \emph{Star Forming Systems KSP}.
\item \textbf{Transients KSP}: as discussed above, transients detected during
  GPS observations will be followed-up in the context of the
  \emph{Transients KSP}.  Additionally, any AGN detected during the GPS will be
  of interest to the \emph{Active Galactic Nuclei KSP}.
\item \textbf{Extragalactic Survey KSP}: the spatial coverage of the
  Extragalactic Survey is such that it will be contiguous with the GPS, i.e. it
  will extend down to latitudes $b \sim 5^{\circ}$, leaving no un-surveyed zone
  between them.  This will provide information about possible Galactic sources,
  the Fermi Bubbles, and diffuse emission at high latitudes, which are
  less affected by Galactic multi-wavelength backgrounds (or foregrounds).
\end{itemize}

\subsection{Data products}\label{sec:dataprods}

It is essential for the CTA GPS to make regularly scheduled data releases.
First, we discuss the results from the STP (Years 1 and 2). The CTA data rights
policy has the philosophy that data should be released to the community by, or
before, one year after the completion of the programme associated with those
data.  It is natural to define the first two years of data taking as the
initial program.  We thus expect the main GPS data release for the STP to be
done at the latest by the end of Year 3; doing it earlier would be even better.
However, three years is too long a time to hold off on releasing results from
the GPS to the community.  Thus, we need to also prepare for an intermediate
data release of first GPS results on a time scale of 12 -- 18 months after the
start of data taking (i.e. in the middle of Year 2).  This intermediate release
would be similar in intent to the Fermi-LAT bright gamma-ray source
list~\cite{Abdo09d}.

Moving on to the LTP, if we define each successive two-year period as a
successive program, subsequent data releases can be naturally made every two
years, i.e. at the end of Years 5, 7, 9, and 11 (assuming a 10-year nominal
lifetime for CTA).  If one of the CTA sites comes online significantly earlier
than the other, the initial releases may only contain data from the
first-completed array.

Data products available to the outside community via an online searchable
archive will include:
\begin{itemize}
\item source catalogue, including source position, morphology, spectrum and
  light curve, all with appropriate errors,
\item sky maps (FITS format), and
\item scientific analysis tools (including \emph{exclusion} maps, diffuse
  emission templates, etc.).
\end{itemize}

The catalogue/archive will provide information about sources extended beyond
the CTA PSF, in terms of their preferred morphology (obtained with a likelihood
analysis).  For unidentified sources, the catalogue/archive will provide
information about possible counterparts.  The catalogue/archive will be
published and available online, perhaps with web-browsing tools (e.g. along the
lines of TeVCat \cite{tevcat}).  The number and type of
sky maps have not yet been determined, but they will likely include test
significance maps and flux maps (units of
photons~cm$^{-2}$~s$^{-1}$~pixel$^{-1}$).  Legacy data products (FITS format) to
the astronomical community will include high-resolution images, spectra and
light curves.

As discussed in Sect.~\ref{sec:strategy}, results from the CTA GPS will be of
high importance to the general astronomical and astroparticle physics
communities.  Accordingly, prior to carrying out the GPS, we expect to develop
coordinated efforts between CTA and other observatories and data archive
centres to best leverage the worldwide expertise in multi-wavelength science and data
preservation.  For more information on CTA's multi-wavelength programme, see
Sect.~\ref{sec:sci_synergies}.

\subsection{Expected Performance/Return}
\label{sec:performance}

\subsubsection{Performance of the CTA GPS}
\label{subsec:gps_performance}

The first sensitive VHE survey of the Galactic plane was carried out by
H.E.S.S.~\cite{Aharonian05s}, resulting in the discovery of an unexpectedly
large number of VHE gamma-ray sources.  It is almost certain that the CTA GPS
will also lead to the discovery of a larger (and likely even more diverse)
population of VHE sources.  Similarly, the full CTA GPS data set will comprise
data taken during a period of ten years.  In total, 1620~h of observation time
are requested for this programme.

\begin{table}
\centering
\begin{tabular}{|c|cc|c|cc|}
\hline
\multirow{2}{*}{} & \multicolumn{2}{c|}{STP} & LTP &
\multicolumn{2}{c|}{Total} \\
& \multicolumn{2}{c|}{(Years 1 -- 2)} & (Years 3 -- 10) &
\multicolumn{2}{c|}{(Years 1 -- 10)} \\
\hline
\textbf{Galactic Longitude} & \textbf{Hours} & \textbf{Sensitivity} &
\textbf{Hours} & \textbf{Hours} & \textbf{Sensitivity} \\
\hline
\textbf{SOUTH} & & & & &\\
\hline
300\degr -- 60\degr, Inner region & 300 & \textbf{2.7 mCrab} & 480 & 780 &
\textbf{1.8 mCrab} \\
240\degr -- 300\degr, Vela, Carina & & & 180 & 180 & \textbf{2.6 mCrab} \\
210\degr -- 240\degr & & & 60 & 60 & \textbf{3.1 mCrab} \\
\hline
 & & &  & 1020 & \\
\textbf{NORTH} & & & & & \\
\hline
60\degr -- 150\degr, Cygnus, Perseus & 180 & \textbf{4.2 mCrab} & 270 & 450 &
\textbf{2.7 mCrab} \\
150\degr -- 210\degr, anti-Centre, etc. &  &  & 150 & 150 &
\textbf{3.8 mCrab} \\
\hline
 & & &  & 600 & \\
\hline 
\end{tabular}
\caption{Estimated point-source sensitivity reach of the CTA Galactic Plane
  Survey for various regions of the Galactic plane.  These sensitivities
  correspond to an energy threshold of 125~GeV.  For the approximate effective
  exposure ``on-axis'' for each region, see Table \ref{tab:pssens_onaxis}.}
\label{tab:pssens_gps}
\end{table}

\begin{table}
\centering
\begin{tabular}{|c|cc|cc|}
\hline
& \multicolumn{2}{c|}{STP (Years 1 -- 2)} &
\multicolumn{2}{c|}{Total (Years 1 -- 10)} \\
\hline
\textbf{Galactic Longitude} & \textbf{Sensitivity} & \textbf{Eq. Exposure} &
\textbf{Sensitivity} & \textbf{Eq. Exposure} \\
\hline
\textbf{SOUTH} & & & &\\
\hline
300\degr -- 60\degr, Inner region & \textbf{2.7 mCrab} & \textbf{11.0 h} &
\textbf{1.8 mCrab} &  \textbf{28.6 h} \\
240\degr -- 300\degr, Vela, Carina & & &  \textbf{2.6 mCrab} &
\textbf{13.2 h} \\
210\degr -- 240\degr & & & \textbf{3.2 mCrab} & \textbf{8.8 h} \\
 & &  &  & \\
\textbf{NORTH} & & & & \\
\hline
60\degr -- 150\degr, Cygnus, Perseus & \textbf{4.2 mCrab} & \textbf{6.3 h} &
\textbf{2.7 mCrab} & \textbf{15.8 h} \\
150\degr -- 210\degr, anti-Centre, etc. &  &  & \textbf{3.8 mCrab} &
\textbf{7.9 h} \\
\hline 
\end{tabular}
\caption{Estimated point-source sensitivity reach of the CTA Galactic Plane
  Survey and equivalent exposure times for various regions of the Galactic
  plane (within 2$^{\circ}$ of the Galactic plane for each portion of the
  survey).}
\label{tab:pssens_onaxis}
\end{table}

The target point-source sensitivity is less than (better than) 4~mCrab
throughout the plane generally, with a goal of achieving better sensitivity in
key regions in the plane, i.e. less than 3~mCrab in the north and less than
2~mCrab in the south.  Table~\ref{tab:pssens_gps} details the expected
performance for the current plan, for both the south and north and for both the
STP and LTP.  For the approximate effective exposure \emph{on-axis} for each
region, see Table \ref{tab:pssens_onaxis}.

The sensitivities given in Table~\ref{tab:pssens_gps} were determined by
updated simulation calculations.  Two different methods to estimate sensitivity
were used.  The first (parametric) method uses the off-axis sensitivity
distributions to estimate the effective exposure time for various
representative positions in, and near, the Galactic plane.  The sensitivities
for these positions are then estimated from the curve of point-source
sensitivity versus observation time.  The second (full simulation) method uses
the effective areas and background rates to simulate detected source and
background events smoothly distributed in Galactic longitude and latitude.  The
detected events are fit to a combined source and background model, and the
source flux is varied until a five standard deviation detection level is
reached.  The sensitivities are estimated for representative points on a grid
of spacing 0.25$^{\circ}$).  The sensitivities estimated by the two methods are
consistent with each other.

The ``milliCrab'' (mCrab) flux unit is used commonly in the VHE gamma-ray
community but is not yet well-known outside that community.  For these
sensitivity calculations, the Crab flux determined by the HEGRA instrument was
used.  The trigger and reconstruction requirements resulted in an energy
threshold of 125~GeV.  The HEGRA Crab flux and the normalization of the
``mCrab'' unit are given in \cite{HEGRACrab}.

\begin{figure}
\centering
\includegraphics[width=1.00\textwidth]{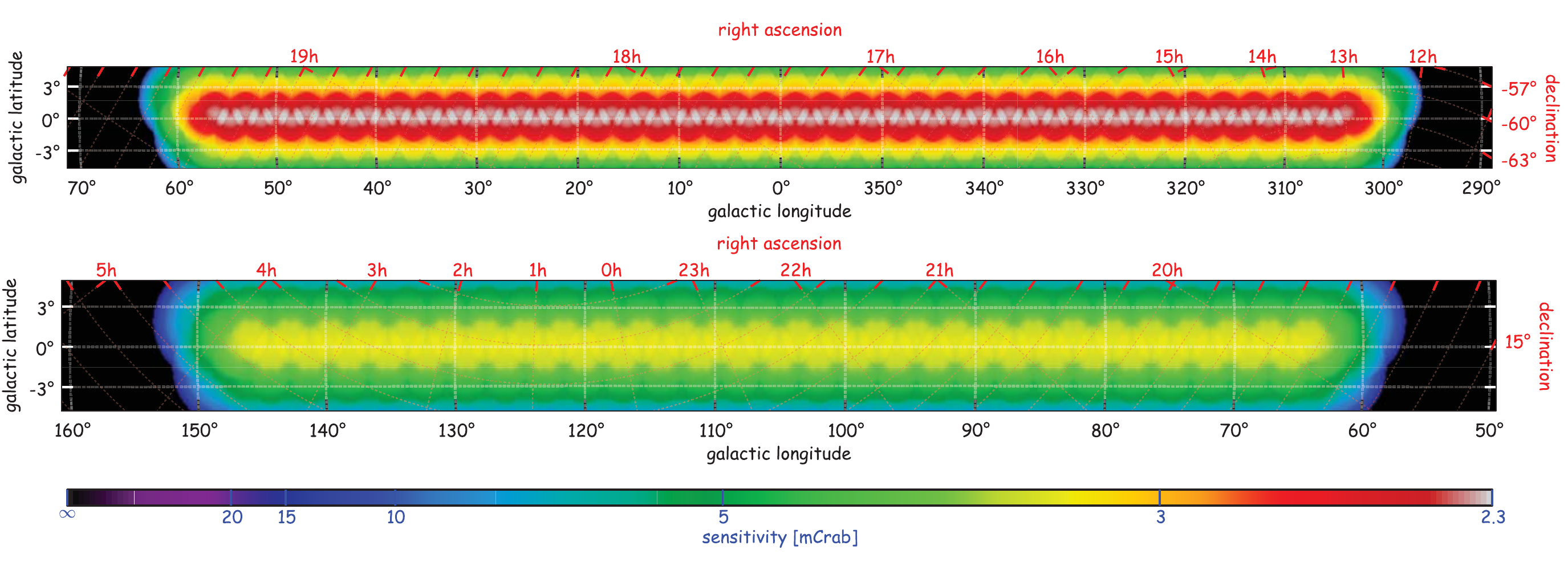}
\caption{Point-source sensitivities (colour scale, in mCrab) achieved in the short-term
  programme (STP; Years 1 -- 2) of the CTA Galactic Plane Survey by the
  north (top) and south (bottom) observatories.}
\label{fig:pssens_stp}
\end{figure}

In this GPS  implementation, during the first STP, we will achieve good
sensitivity in the inner region $|l| < 60^{\circ}$, as well as in Cygnus and
Perseus.  The STP sensitivities are shown graphically in
Fig.~\ref{fig:pssens_stp}.

\begin{figure}
\centering
\includegraphics[width=0.85\textwidth]{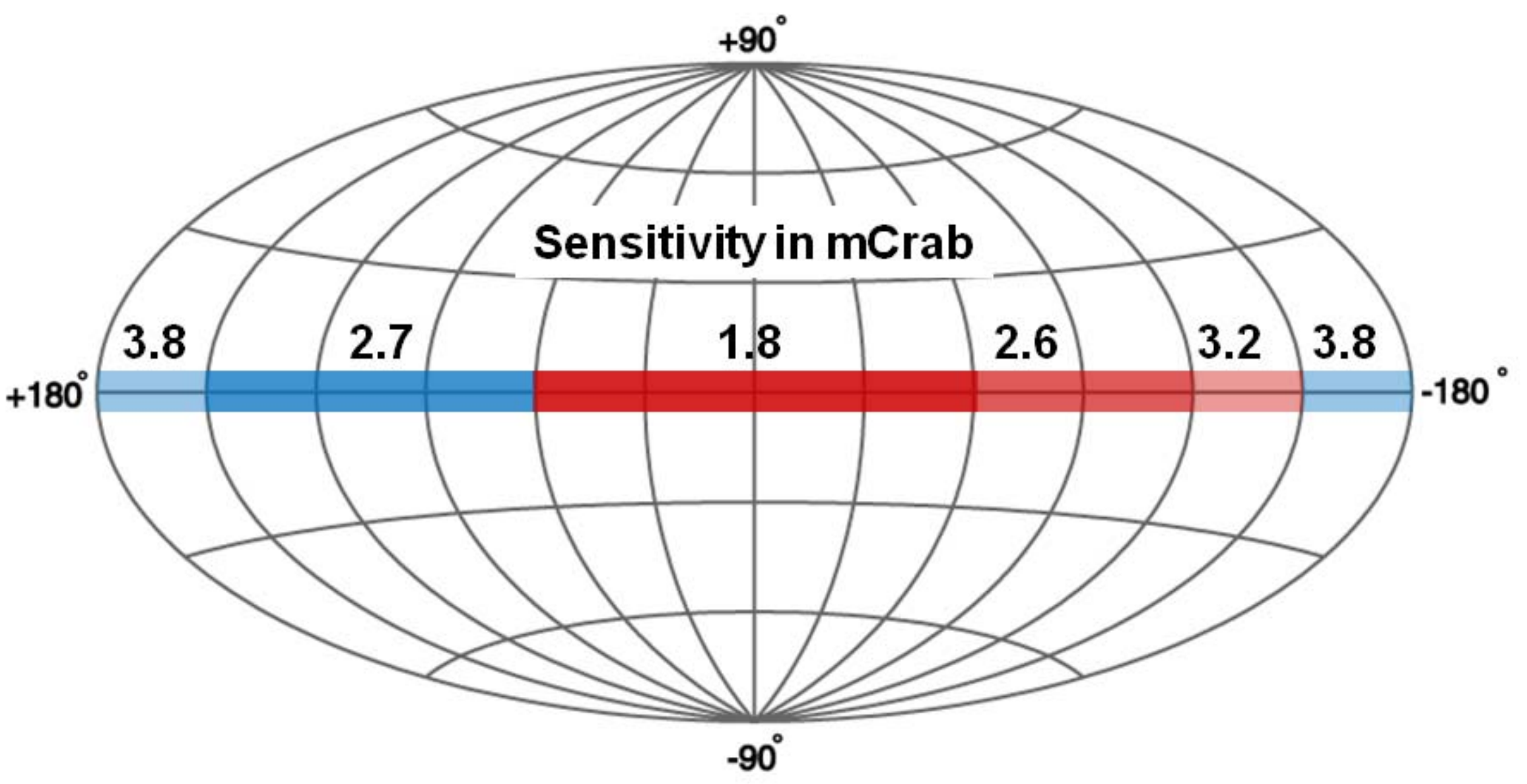}
\caption{Point-source sensitivities (in mCrab) achieved in the full ten-year
  programme of the CTA Galactic Plane Survey for various regions along the
  Galactic plane.  The survey carried out by the southern (northern) array is
  indicated by the red (blue) segments.}
\label{fig:pssens_ltp}
\end{figure}

By the end of the LTP, very deep sensitivity will be achieved in the inner
region, deep sensitivity is achieved in the Cygnus, Perseus, and the
Sagittarius-Carina regions, and finally moderate sensitivity is achieved in the
$150^{\circ} < l < 240^{\circ}$ (extended anti-Centre) region.
Figure~\ref{fig:pssens_ltp} shows the achieved sensitivity for the various
regions of Galactic longitude.  In these figures, the sensitivity appears
more uniform than it will be in reality, since the effects of diffuse emission
and source confusion were not taken into account in the simulations.

\subsubsection{Source Confusion}
\label{subsec:source_conf}

As mentioned in Sect.~\ref{subsec:science_context}, we expect source confusion
to be a significant consideration, especially in the inner regions of the
Galaxy.  The main issues to consider are the unknown shapes of the sources, the
unknown level of diffuse emission, the high source density in the inner Galaxy
(so that many CTA sources will overlap), and the dependency of source
identification on the analysis methods (e.g. maximum likelihood source
detection and de-blending criteria).  The added value of MWL information and
the improved angular resolution of CTA will help resolve many instances of
source confusion, but a large fraction of the sources detected in the GPS will
naturally be relatively weak and thus inherently very difficult to resolve and
identify.

Initial studies have been carried out to try to estimate the expected level of
source confusion.  In one of these, a Galactic population of sources was
simulated based on our current understanding of the VHE source population,
using an extrapolation of the source count (i.e. log~$N$ -- log~$S$),
distributions of source spectral indices and sizes consistent with existing
data, and an assumed spatial distribution of sources around the Galactic
Centre.  No diffuse emission was included (except for the Galactic Centre
ridge) and two different extrapolations of the log~$N$ -- log~$S$ distribution
were used to bracket the range of the expected source density.  A position in
the sky was considered confused if there was more than one simulated source
within a radius of 1.3 times the CTA angular resolution.  Using these
assumptions leads to an approximate lower limit to the amount of source
confusion. The estimated confusion lower limits range from 13 -- 24\% at
100~GeV to 9 -- 18\% at 1~TeV, for the region $|l| <  30^{\circ}$ and
$|b| <  2^{\circ}$.  Work on simulations will continue 
to better quantify this aspect of the CTA GPS.

\subsubsection{Summary}
\label{subsec:summary_performance}

The CTA GPS will produce guaranteed high-impact science returns and legacy
products, as well as greatly expanded discovery potential in VHE astrophysics,
due to its uniform, mCrab-deep coverage of the entire Galactic plane.  PWNe and
SNRs comprise the bulk of currently known VHE Galactic sources, and they will
likely dominate the Galactic source populations for CTA as well.  The log~$N$
-- log~$S$ curves and representative models for source spectra and morphology
can be used to estimate ranges in the expected number of source counts, as
discussed in Sect.~\ref{subsec:science_context}.  Hundreds of sources will be
detected and, for each of these sources, the GPS will provide spectra that can
be used, in conjunction with MWL data and comparisons to theoretical models, to
identify the particle acceleration scenarios most likely relevant for the VHE
gamma-ray emission.  Pulsars and binary systems will also feature prominently
in the GPS data set, with their phaseograms and light curves, respectively,
bringing additional physics to bear.  With data taken over ten years, the GPS
will also be an important source of serendipitous discovery.  Even
non-detections will be of great importance to the astronomical and
multi-messenger communities, with many constraining VHE upper limits
expected.

All of this rich data will be provided to the broader community on a regular
basis and will be an indispensable resource for follow-up proposals to the CTA
GO program.  The GPS will provide a source catalogue (listing e.g. flux and
spectral index) and high-level sky maps, all accessible to the worldwide
community.  These data products will allow us to (a) perform VHE source
population studies, (b) investigate VHE diffuse emission, and (c) carry out
in-depth studies of individual sources through follow-up observations made by
CTA and other observatories.

\section{KSP: Large Magellanic Cloud Survey}
\label{sec:ksp_lmc}

The Large Magellanic Cloud (LMC) is a unique galaxy hosting extraordinary objects, including the star-forming region 30 Doradus (the most active star-forming region in the local group of galaxies), the super star cluster R136 (an exceptional cluster with a large concentration of very massive O and Wolf-Rayet stars), supernova SN1987A (the closest supernova in modern times), and the puzzling 30 Dor C superbubble (a rare superbubble with non-thermal emission). 
As a satellite of the Milky Way, it is one of the nearest star-forming galaxies, and a very active one; it has one tenth of the star formation rate of the Milky Way, distributed in only about two percent of its volume. 
This activity is attested by more than 60 supernova remnants (SNRs), dozens to hundreds of HII regions, and bubbles and shells observed at various wavelengths, all of which promise fruitful gamma-ray observations. 
The LMC is seen nearly face-on at high Galactic latitude, and hence source confusion, line of sight crowding, and interstellar absorption do not hamper these studies, in contrast to the case for the plane of our Galaxy. 
It is therefore a unique place to obtain a significantly-resolved global view of a star-forming galaxy at very high energies. In addition, the distance to the LMC is known to the few percent level, thus allowing precise luminosity measurements to be made, something which is often very difficult for Galactic sources.

The current Fermi-LAT and H.E.S.S. instruments have opened the way for a study of the LMC by CTA.
Observations with these telescopes have revealed a small number of sources, some of uncertain nature. 
With CTA we would have a unique opportunity to further and deeper explore the entire LMC. 
With its unprecedented sensitivity and angular resolution, CTA will complement and extend these early results, and it will
allow us to probe the origin of the VHE emission of a galaxy and its connection to global galactic properties.

This KSP has as many scientific objectives as an entire star-forming galaxy can offer: population studies on SNRs and pulsar wind nebulae (PWNe), transport of cosmic rays from their release into the interstellar medium to their escape from the system, and the search for signatures of the elusive dark matter component of the universe. 
These studies will be complementary to those done and planned for the Milky Way because they feature an unusual global perspective, different astrophysical settings due to the specific conditions in the LMC galaxy and data analysis with different uncertainties and systematics.

To make a significant contribution to 
the above-outlined science topics, the LMC KSP consists of an initial deep scan over a circular region of radius 3.5$^\circ$. 
This will be achieved over the first four years of CTA from a small number of pointings with the southern array, for a total of 340~h of observations. 
Then, if SN1987A is detected in this deep scan, a second part of the project will consist of the monitoring of SN1987A over the following six years at a level of 50 h every 2 years. 

The total of about 500~h spread over one decade is probably beyond what can be granted to individual Guest Observers. 
In addition, the observation of SN\,1987A, if detected after the first deep scan, should be carried out over a decade on a regular basis, requiring a long-term granted observing program.
The size of the LMC is about the size of the field of view of the CTA MSTs and SSTs, and the emission will likely be composed of several contributions, from point-like sources to extended and very extended objects. 
The determination of the best observing strategy and the analysis of the resulting observations will likely be very challenging.
Therefore, this project is well suited to being led by the CTA Consortium. 
In return, the Consortium commits itself to the release of a source catalog, spectra, light curves, and a complete emission model to the community at large.

The interpretation of TeV gamma-ray observations of the LMC will benefit from a large and comprehensive multi-wavelength context, and, in return, some results of the project may reach beyond the usual VHE topics, e.g. analyses of cosmic-ray populations in the LMC may be of interest for studies of the properties and evolution of the interstellar medium done from radio and infrared observations. Considering this, a deep scan of the LMC by CTA would be an appropriate and highly appreciated legacy to the astronomical community.

\subsection{Science Targeted}
\label{sec:ksp_lmc_sci}

The LMC is a nearby satellite galaxy that is visible from the southern hemisphere at a Galactic latitude of $b=-32.9^\circ$ and hence clearly off the Galactic plane. It is located at a distance $d=$50\,kpc (with an uncertainty of $\sim$ 2\%, see \cite{Pietrzynski:2013}) and it has the shape of a disk seen nearly face-on, with a small inclination angle of $i=30-40^\circ$ \cite{vanderMarel:2006}. Considering the typical angular resolution of gamma-ray telescopes, the LMC is perhaps the only object which can provide us with a global and significantly resolved view of an external galaxy at gamma-ray
energies.
Surveys of the LMC over the past few decades and across the electromagnetic spectrum have revealed a very active system in terms of star formation; it has one tenth of the star formation rate of the Milky Way, distributed in only about two percent of its volume \cite{Hughes:2007}. The LMC hosts:
\begin{itemize}
\item 30 Doradus, the largest star-forming region of the local group of galaxies \cite{Walborn:2014},
\item SN\,1987A, the remnant of the nearest naked-eye supernova since Kepler in 1604 \cite{McCray:1993},
\item about 60 well-established and 20 good-candidate SNRs (\cite{Bozzetto17}),
\item one of the densest stellar clusters known: R136 \cite{Crowther:2010},
\item the most massive stars known \cite{Crowther:2010},
\item hundreds of HII regions \cite{Lawton10},
\item more than a dozen superbubbles \cite{Dunne:2001},
\item about 20 supershells and a hundred giant shells \cite{Kim:1999},
\item two of the most powerful pulsars known and their nebulae \cite{Marshall:1998,Seward:1984}, and
\item a well-studied population of star clusters, with ages from a few Myr up to 10Gyr \cite{de-Grijs:2006}.
\end{itemize}

Since most of the gamma-ray emission of our Galaxy is due to massive star evolution, the activity and proximity of the LMC ensures fruitful gamma-ray observations. Indeed, after just one year of all-sky survey observations, Fermi-LAT detected significant emission from the LMC with several source components \cite{Abdo10d}. 
Using more than six years of data, Fermi-LAT revealed a much richer picture comprising four point-like sources, all being extreme objects, large-scale emission from the full extent of the LMC, and a handful of regions featuring extended emission of unclear origin \cite{Ackermann16}. Similarly, H.E.S.S. detected the first extragalactic PWN based on about 50 h of observations of the LMC \cite{Abramowski12c}. 
With a data set of about 200~h available now, H.E.S.S. also detected SNR N132D and superbubble 30~Doradus~C \cite{Abramowski15a} and has started to probe the richness of a star-forming galaxy at TeV photon energies from an external viewpoint. In the future, CTA is expected to make a deeper exploration of the origin of the very-high energy emission of a galaxy and on how it connects to its global properties.

\subsubsection{ Scientific Objectives}
\label{sec:ksp_lmc_sci_obj}

The science objectives of this KSP can be grouped into three key questions:
\begin{itemize}
\item What are the processes and sites in which the bulk of the cosmic rays are accelerated?
\item How do cosmic rays propagate away from sources and interact with the interstellar medium?
\item What is the nature of dark matter?
\end{itemize}
The science case is introduced below in more detail. Quantitative prospects are addressed in Section \ref{sec:ksp_lmc_exp} for cosmic-ray origin and transport, and in Chapter \ref{sec:DM_prog} for dark matter.

\textbf{Cosmic-ray origin}: SNRs are thought to be the main source of cosmic rays up to PeV energies, however the whole picture is still incomplete. The maximum particle energy attainable, the influence of the SN progenitor and its environment on the non-thermal population produced, and the transport of particles away from the accelerator are all still active research topics. The very young SNR SN\,1987A in the LMC is a unique target for studying the production and temporal evolution of the gamma-ray emission from cosmic-ray acceleration during the earliest SNR stages. The temporal evolution of this object was monitored from the very beginning, including the properties of the progenitor star. Given its nearby location, this is the only SNR spatially resolved by modern instruments at different wavelengths so early after the explosion (see \cite{McCray:1993} for a review). The non-thermal radio and hard X-ray emission from the remnant is evidence of particle acceleration in the supernova blast wave, which is impacting on the dense shell produced by the progenitor star in the current epoch \cite{Chevalier:1995,Zanardo:2010}. 
Long-term VHE monitoring of the target will thus provide a unique opportunity for studying the interaction between the SN shock wave and the circumstellar medium. 

For studies of SNRs at a later stage in their evolution, the LMC harbours a rich population of 60 well-established SNRs and an additional 20 good SNR candidates (M. Filipovic, private communication; \cite{Maggi16}). Such a rich sample enables the study of the impact of various properties on remnant evolution and its ability to accelerate particles. These properties include explosion types (thermonuclear versus core-collapse), environmental conditions (interaction with interstellar clouds or circumstellar material versus expansion in cavities), 
and the age of the remnant. 
From radio, optical, and X-ray analyses, several LMC objects seem to be very similar to famous Milky Way remnants. Studying the former could be a useful test of our models for the latter.

\textbf{Cosmic-ray propagation}: Milky Way studies of cosmic-ray origin and propagation have provided a wealth of information, from direct particle measurements in the Solar System to radio and gamma-ray diffuse emission observations from the entire Galaxy. Yet, we lack a global and resolved perspective on the phenomenon. What happens to cosmic rays freshly escaping from 
their sources? Are they confined for some time, and potentially reaccelerated in the highly turbulent medium of bubbles and superbubbles (such as hinted at by the detection of the Cygnus cocoon, see \cite{Ackermann11})? Are they advected into the Galactic halo after bubble breakout, thus contributing to the Galactic wind? In that context, some additional questions are specific to the LMC. Does the LMC have a galactic wind and how is it connected or affected by cosmic-ray transport \cite{Barger16}? Can cosmic-ray transport be influenced by the wind of the Milky Way sweeping away particles? 

Without line-of-sight confusion and with accurate estimates for the distance of the galaxy, observations of the LMC will allow us to test our understanding of the processes that rule the injection of cosmic rays and their subsequent propagation in the system. The usefulness of the LMC in that context was already demonstrated from Fermi-LAT observations; whereas the interstellar gamma-ray emission of the Milky Way is observed to be strongly correlated with the gas, the gamma-ray emission from the LMC is very poorly correlated with it. 
Apart from a large-scale emission component spread across the LMC disk, extended emission from smaller-size regions seems to be correlated with cavities in the interstellar medium and there is currently no explanation for this phenomenon. 
Continuing observations with 
Fermi-LAT will help to refine the picture, but CTA could well provide essential data needed to make real progress. 
Fermi-LAT observations probe $\sim$GeV cosmic rays that accumulate in a galaxy over a 1-10\,Myr duration, and are well suited to study the large-scale transport of cosmic rays. In contrast, CTA will probe $\sim$TeV cosmic rays having a shorter residence time and will thus provide a view of cosmic-ray injection into the interstellar medium (ISM) and the small-scale transport around sources. The two 
instruments are therefore complementary to one another and should allow us to better understand the life cycle of cosmic rays, from injection and possible confinement around sources and in superbubbles to escape into the intergalactic medium.

\textbf{The nature of dark matter}: The LMC is an interesting and complementary target in the 
search for gamma rays from dark matter annihilation. The expected dark matter flux depends on the particle properties of dark matter and the dark matter distribution. The most favorable targets are nearby and have large concentrations of dark matter; this information about the dark matter distribution is expressed in terms of the ``J-factor'', which is the integral of the dark matter density squared along the line of sight. Targets with larger J-factors produce larger dark matter signals. The J-factor of the Galactic Center can be as large as $\log_{10}(J) \sim 23$ (integrated in a cone with opening angle 0.5$^{\circ}$ in units of GeV$^2$ cm$^{-5}$), but the central point source and other sources of gamma rays represent a large astrophysical background to a dark matter search. The J-factor of the LMC has been claimed to be as high as $\log_{10}(J) \sim 20.5$ \cite{Buckley15}. Dwarf galaxies of the Milky Way, which have been considered extensively in dark matter searches because of their known locations, high dark matter densities, and low backgrounds, typically have smaller J-factors of $\log_{10}(J) \sim 18-19$. This simple comparison suggests that the LMC could yield a strong dark matter signal. 
However, the LMC is spatially extended and it has significant astrophysical
gamma-ray emission. 
Nevertheless,
a LMC survey will 
complement the dark matter searches that will be performed in the Galactic Centre and in  
dwarf galaxies by providing a different target, albeit with larger analysis 
uncertainties and systematics.

\textbf{Other science}: Beyond the science topics addressed above, there may be side benefits from deep observations of a $\sim10^\circ$ patch 
of the sky that includes an entire galaxy. 
For example, the scan of the LMC may reveal one or two additional gamma-ray binaries, which could be a useful addition to the handful we know already. The recent discovery in the LMC of the most luminous gamma-ray binary \cite{Corbet:2016} is promising in that respect.
The LMC survey should also lead to the detection of several active galactic nuclei (AGN). There are half a dozen Fermi-LAT sources in the LMC field, with some having hard spectra and at least one being an AGN with a high level of variability (PKS\,0601-70, see \cite{Ackermann16}).

\subsubsection{ Context / Advance beyond State of the Art}
\label{sec:ksp_lmc_contxt}

The LMC was already covered by many surveys at different wavelengths: radio (ATCA, Parkes, Mopra, Planck, Herschel), infrared (AKARI, Spitzer, VISTA), optical, X-rays (ROSAT, XMM-Newton), MeV (INTEGRAL), GeV (Fermi-LAT), and TeV (H.E.S.S.). CTA observations and studies of the LMC will therefore benefit from a highly homogeneous multi-wavelength context.

The continuation of Fermi-LAT observations until 2018 (at least), combined with the continuous improvement in event reconstruction, will no doubt provide a better grasp of the emission morphology and the separation between diffuse and discrete objects. 
However, as mentioned previously, CTA will be complementary to Fermi for two reasons: first, it probes higher-energy cosmic rays and thus a different stage of the life cycle of cosmic rays (because higher-energy cosmic rays may be able to escape the system more easily, especially protons since they suffer much less severe radiative losses than electrons); second, with its better angular resolution, CTA will likely lead to a better and more complete picture of the gamma-ray emission of the LMC. 
A full understanding over the entire gamma-ray range will be achieved with a joint Fermi-LAT and CTA analysis (see Section \ref{sec:ksp_lmc_exp}).

In the meantime, the Planck collaboration will release the analysis of the dust and synchrotron polarized emission from the LMC. This will be complemented 
later by results from the second PILOT balloon flight (2016-2017), with a better sensitivity and angular resolution than Planck. The Square Kilometer Array (SKA) and its precursors will also contribute to these science topics. 
All instruments will provide information about the large-scale magnetic field structure and the leptonic component of the cosmic-ray population. They will also allow a deeper census of the populations of pulsars and PWNe in the LMC \cite{Gelfand15,Keane15}, which could be responsible for extended emission as unresolved gamma-ray sources. 
The interpretation of the Fermi-LAT and 
CTA observations will necessarily have to include the information and constraints from these radio measurements.

ALMA is already operational and can indirectly probe the lower-energy cosmic rays ($<$1\,GeV) because the latter ionize the gas and this process affects the chemistry and molecular line emission. Although no global project dedicated to the LMC yet exists for ALMA, SN\,1987A was among  the first objects to be imaged \cite{Indebetouw14} and other LMC proposals were made. It is likely that by the beginning of CTA operations, we will have estimates of low-energy cosmic-ray density in at least a few prominent places such as 30~Doradus. This again will have to be considered in the analysis and interpretation of the CTA data.

\subsection{Strategy}
\label{sec:ksp_lmc_strat}

\begin{figure}[t!]
\begin{centering}
\includegraphics[width= 11cm]{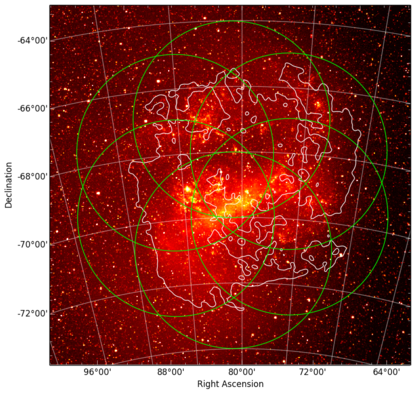}
\caption{Illustration of a possible pointing pattern for the LMC deep scan. The pattern consists of six 
pointings evenly distributed around the LMC centre at a separation
distance of 2.0$^\circ$ (green circles, for a typical field of view radius of 3$^{\circ}$). The spacing of the pointings will depend on the actual off-axis performance of CTA and should provide enough empty field regions for sufficient control of the instrumental background.
Credits: optical skymap from A. Mellinger \cite{Mellinger09}; contour of H I distribution from \cite{Kim98,Kim03}.
}
\label{fig:ksp_lmc_scan}
\end{centering}
\end{figure}

This KSP has one target: the LMC, which we define here as a disk centred on the position 
($\alpha$, $\delta$) = ($80.0^\circ$, $-68.5^\circ$) and having a radius of $3.5^\circ$. In order to maximize the scientific return and the possibility for discoveries, and to take full benefit of the global view that the LMC provides, it is planned to build a uniform sensitivity exposure
over that region, and not just over the inner regions or 30~Doradus. This will be achieved with a small number of pointings
(i.e. less than ten). A possible layout of the scan is shown in Figure~\ref{fig:ksp_lmc_scan}, but 
the layout used will depend on the off-axis performance of CTA and it may need to be revised once the array is fully commissioned. As explained below, there could be a second phase of the project in order to monitor SN\,1987A, in which case the observations would consist of a single pointing centred on SN\,1987A (or wobble observations, or any other pattern motivated by discoveries made during the first phase of the program).

The project comprises two phases over a ten-year period, with 
the second one being optional (depending on whether SN\,1987A is detected):
\begin{enumerate}
\item a deep scan of 340~h, 
ideally performed over the first four years in order to reach an effective exposure of 250~h over the entire LMC disk, as defined above, and
\item long-term monitoring of SN\,1987A, totaling 150~h, at the level of 50~h every two years if the object is detected in the first phase.
\end{enumerate}
Since the deep scan covers a region that is slightly larger than the effective field of view of the CTA MSTs and SSTs, reaching an exposure of 250~h over the LMC disk requires a total of about 340~h of observing time (assuming a field of view radius of 3$^{\circ}$ and a differential sensitivity above 400\,GeV that is at least 60-70\% of the on-axis value).  
The observations are planned to be performed by the full southern array after commissioning in order to get the best sensitivity for this very extended object. Current Monte-Carlo simulation results indicate that the energy threshold would be around 200\,GeV.

For CTA-South, the LMC is observable above an elevation of 40$^\circ$ for a total of about 280 h/yr, with the peak of the time distribution being in the November to January period. Performing the deep scan with CTA in four years imposes observing the LMC for 85 h/yr. This is far below the total available time, but there may be constraints from the schedule or from bad weather. 

\subsection{Data Products}
\label{sec:ksp_lmc_data}

The LMC in TeV gamma rays currently consists of a few objects: PWN N157B, SNR N132D, and the 30 Doradus C superbubble. Extrapolating the known Galactic TeV source population of the Milky Way to the LMC indicates that the 250 h exposure may provide the detection of $\sim$10 objects (see section \ref{sec:ksp_lmc_exp}). If this estimate turns out to be correct, a coherent catalog of LMC sources listing their spectra, fluxes, and light curves should be produced; such a catalog should also include upper limits on potential VHE emitters. If diffuse emission of interstellar origin can be detected and mapped, a complete emission model including the morphological and spectral properties of the various components should be released to the community; this may trigger transverse collaborations with groups interested in interstellar medium studies. Lastly, the long-term monitoring of SN\,1987A
may deserve two specific releases, at 250 h and 400 h of exposure, in the form of light curves and time-dependent spectra.

The schedule for data products and releases to the community is:
\begin{enumerate}
\item An initial release of results is planned based on the first 100 h of exposure, in order to inform the community and to keep a high interest in the LMC.
\item Once the 250 h effective exposure is achieved, products on individual sources, diffuse emission, and SN\,1987A should be released within a 
year, or as soon as possible, to trigger and feed Guest Observer proposals.
\item Optionally, 
a final release of results on SN\,1987A after ten years of monitoring, along with an update on the other topics based on the additional 150~h exposure
is envisioned.
\end{enumerate}

\subsection{Expected Performance/Return}
\label{sec:ksp_lmc_exp}

We present below some of the simulations that were performed to assess the potential of the LMC as a target for CTA and 
to quantify the required amount of observing time. Cosmic-ray origin and propagation aspects are addressed here, while 
the prospects for dark matter searches are discussed in Chapter \ref{sec:DM_prog}.

\begin{figure}[t!]
\begin{centering}
\includegraphics[width= 11cm]{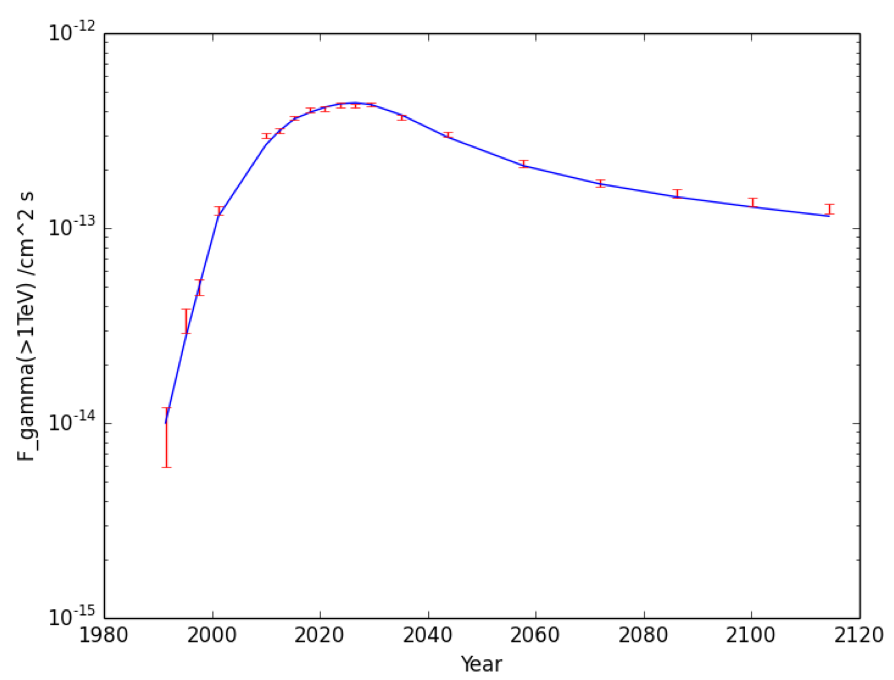}
\caption{Predicted gamma-ray light curve for SN\,1987A (blue, from E.G. Berezhko, private communication) and anticipated detections with CTA for several 50~h
 observations distributed over decades (red points). The recent upper limit from H.E.S.S. observations, at a level of $5 \times 10^{-14}$\,cm$^{-2}$\,s$^{-1}$
 \cite{Abramowski15a}, contradicts this picture; it suggests that the rise to the maximum may be delayed and steeper or that the overall level of the emission is substantially lower.}
\label{fig:ksp_lmc_sn1987a}
\end{centering}
\end{figure}

\textbf{Cosmic-ray origin}: Regarding the gamma-ray emission of SN\,1987A, a time-dependent prediction of the VHE gamma-ray flux was proposed, based on non-linear shock acceleration theory \cite{Berezhko:2011}. In this model, the VHE gamma-ray flux is expected to be rising (since the shock has entered the equatorial ring) and to have reached a level of F($>$ 1TeV) $\simeq$ $2.5 \times 10^{-13}$\,cm$^{-2}$\,s$^{-1}$ (as of 2010), dominated by emission of hadronic origin. Given the uncertainties in the theoretical model and those in the knowledge of the ambient target density structure, the predicted flux has an uncertainty of at least a factor of two. The time evolution of the flux strongly depends on the spatial distribution of the circumstellar medium. The flux is predicted to be increasing in the current epoch and to become a factor of two higher in the next twenty years, hopefully permitting the detection of this source. The extension of the SN shell of a few arc-seconds is much smaller than the achievable angular resolution of CTA . The target is thus expected to be point-like for CTA. Figure \ref{fig:ksp_lmc_sn1987a} shows the $>$1\,TeV detection level provided by multiple 50 h observations distributed over several decades. The recent upper limit from H.E.S.S. observations, at a level of $5 \times 10^{-14}$\,cm$^{-2}$\,s$^{-1}$ \cite{Abramowski15a} challenges this picture; it suggests that the rise to the maximum may be delayed and steeper or that the overall level of the emission is lower \cite{Berezhko:2015}. The prospects for the detection of TeV emission from SN\,1987A will need to be reassessed, especially in view of the lower energy threshold of CTA. In either case, it shows the potential of deep observations of SN\,1987A to constrain models.

For more evolved accelerators, the expected number of TeV-emitting sources can be estimated from the known Galactic VHE gamma-ray source population and the sensitivity of CTA. For an integral energy flux sensitivity of 3 $\times$ 10$^{-14}$ erg cm$^{-2}$ s$^{-1}$ above 1 TeV, in 250 h gamma-ray sources as luminous as 9 $\times$ 10$^{33}$ erg/s would be detectable with CTA at the LMC distance of 50\,kpc. Of all Galactic TeV emitters known in the Milky Way for which distance estimates exist, about 30 sources have a luminosity greater than this value. Correcting for the lower star-formation rate in the LMC ($\sim$1/10 of the Milky Way), and assuming that $\sim$30\% of the Milky Way has been surveyed to this luminosity limit with H.E.S.S., $\sim$10 LMC sources are expected to emit in VHE gamma rays at a level detectable by CTA. Note however that the environment in the LMC is different from the Milky Way. 
For instance, the radiation field energy densities and gas densities are higher, which would result in a higher gamma-ray flux on average, implying more sources to be detectable. N157B, the most luminous PWN known so far, is a prime example in that respect \cite{Abramowski12c}. 

The current population of known SNRs in the LMC amounts to 60 objects, plus 20 plausible candidates. Not all of them are expected to be detected by CTA, but a selection based on simple criteria can provide a first-order quantitative estimate of the number of promising targets. Based on our knowledge of the Galactic population of TeV-emitting SNRs -- namely the GeV-hard, TeV-bright young and isolated SNRs and the GeV-bright, TeV-soft middle-aged, interacting SNRs -- we restrict ourselves to young to intermediate SNRs with an age $<$4000 yrs (i.e. similar to the largest age estimates for Vela~Jr and HESS~J1731-347) and to interacting SNRs whatever their age (given that those known to emit TeV gamma rays in the Galaxy span nearly two orders of magnitude in age). 
A non-exhaustive list of candidates for TeV detection is given below:
\begin{enumerate}
\item Young/intermediate SNRs: SNR 0540-6919 (an O-rich SNR hosting the ``Crab-twin'' pulsar and its nebula),  SNR 0509-67.5 and SNR 0519-69.0 (400 and 600\,yr type Ia remnants with inferred efficient cosmic-ray acceleration, see \cite{Williams11}), DEM L71 (a $\sim$4000\,yr type Ia remnant), N63A (a $\sim$4000\,yr core-collapse remnant candidate).
\item Interacting SNRs: N132D (an O-rich SNR with similarities to Cas A and Puppis A), N49 (the third X-ray brightest in LMC, very energetic explosion, see \cite{Park12}), possibly N49B, N103B (a 870\,yr type Ia SNR with circumstellar interaction, a possible Kepler's older cousin, see \cite{Williams14}), DEM L249 and DEM L238 (10000-15000\,yr type Ia remnants with denser than expected environment, see \cite{Borkowski:2006}).
\end{enumerate}
These remnants have typical sizes below 3 arc-minutes and would likely appear as point-like objects even for CTA.
We emphasize the very interesting opportunity to study type Ia remnants, from very young ones to those that seem to be interacting with circumstellar medium and might be the result of prompt explosions from young $<$100\,Myr progenitors \cite{Borkowski:2006}.

Regarding PWNe, the prospects are not as good, at least based on our present knowledge of the population of pulsars and their nebulae. We currently know 5 PWNe in the LMC: N157B, N158A, B0453-685, B0532-710, DEM L241 (the first one being already detected by H.E.S.S., the second one being associated with the Crab twin pulsar, also named B0540-693 \cite{Brantseg:2014}, and the last now being confirmed as a gamma-ray binary). Prospects are good for the detection of N158A by CTA, especially considering the very high spin-down luminosity of its pulsar ($1.5 \times 10^{38}$\,erg/s; see also \cite{Martin:2014}). Beyond the exceptional pulsars in N157B and N158A, there are 17 other detected pulsars in the LMC \cite{Bozzetto:2012}. For those listed in the ATNF catalog, all have a spin-down luminosity $\leq 5 \times 10^{34}$\,erg/s. This leaves little hope for the detection of the corresponding nebulae, but there is still the possibility to discover TeV PWNe associated with currently unknown pulsars.

\textbf{Cosmic ray propagation}: The current knowledge we have about the hadronic cosmic-ray population of the LMC comes from Fermi-LAT observations in the GeV range. The latest analysis revealed a background of cosmic rays spread over most of the disk of the LMC and having a density which is about three times lower than the local Galactic value.
On top of that average, a few regions may have increased cosmic-ray densities by a factor of 2-3 or more and likely harder spectra, possibly resulting from a recent injection of cosmic rays \cite{Ackermann16}. As a consequence, we considered several objectives for the cosmic-ray population and associated interstellar emission in the CTA energy range. On large spatial scales, we would like to know: i) whether we can detect the background interstellar emission from the large-scale cosmic-ray sea at TeV energies, ii) if this emission is hadronic in origin or if the inverse-Compton process takes over at some point, and iii) if we can gain knowledge about the efficiency with which 10\,TeV cosmic-ray particles escape or diffuse away from the galaxy and how that efficiency compares with what we think takes place in the Milky Way. On smaller scales, we would like to constrain: i) the level at which we can we probe the inhomogeneities of the LMC cosmic-ray population, ii) the knowledge we can get on the transport 
of cosmic rays in and away from active star-forming regions, iii) the escape of the higher-energy cosmic rays from their sources or their confinement around them, and iv) the relation of the gamma-ray emission to star-formation activity and/or ISM conditions.

\begin{figure}[!t]
\begin{centering}
\includegraphics[width= 13.5cm]{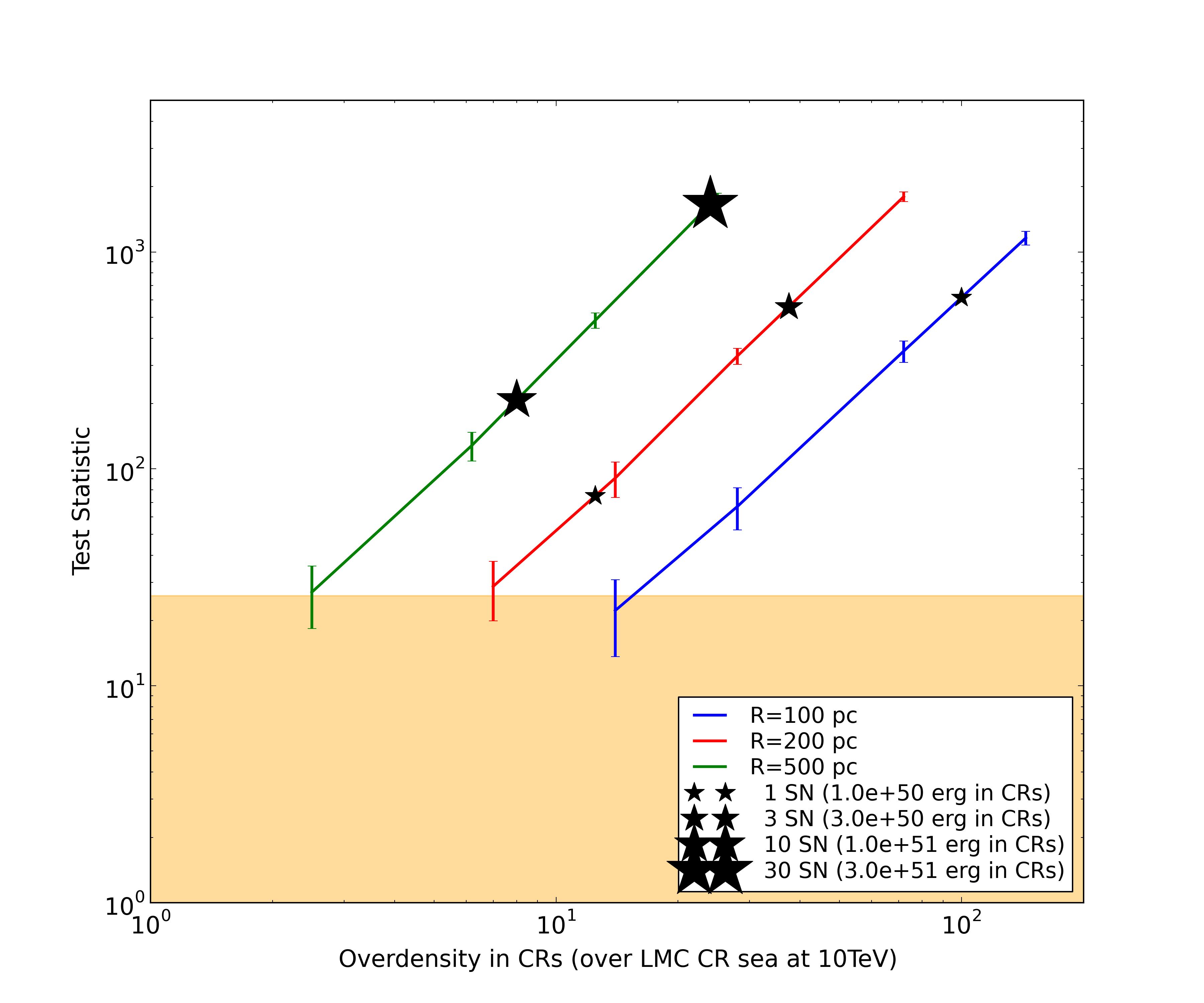}
\caption{Detectability of a region in the LMC as a function of its size (100, 200, 500\,pc) and cosmic-ray density (given by reference to the large-scale background level of cosmic rays at 10 TeV). For a given region size and density, the black stars give the equivalent cosmic-ray density in numbers of supernovae, assuming each event releases 10\% of 10$^{51}$\,erg into the kinetic energy of cosmic rays. The shaded area marks the area of detection significance below five standard deviations.
The high significance reached for 100 and 200\,pc-sized regions enriched in cosmic rays by one or two supernova explosions suggests that spectral and morphological studies will be possible beyond the simple detection.}
\label{fig:ksp_lmc_regiondet}
\end{centering}
\end{figure}

We simulated different scenarios to evaluate the potential of CTA for the study of cosmic-ray propagation and determine the observing time needed for a rich scientific potential. 
The properties of the background interstellar emission from the large-scale cosmic-ray sea at TeV energies are currently unknown and two options were therefore considered. As a first possibility, we considered hadronic emission, where the source spatial model was a gas column density map of the LMC and the source spectral model was a power law with a spectral index of 2.7, normalised to the level of the emission of the LMC disk at 10 GeV (as measured by the Fermi-LAT) and under the assumption that the emission results from old, accumulated cosmic rays interacting with interstellar gas. Alternatively, we considered leptonic emission, where the source spatial model consisted of a smeared out version of the IRAS far-infrared emission distribution, to account for the diffusion of short-lived $>1$\,TeV cosmic-ray electrons around star-forming regions, and the source spectral model was taken from a model of the non-thermal emission from star-forming galaxies applied to a small-sized system similar to the LMC \cite{Martin:2014b}. With an effective 250 h exposure over the LMC, the average detection significance obtained is above three standard deviations for the hadronic model, and above seven standard deviations for the leptonic model. The background interstellar emission from the large-scale cosmic-ray sea could therefore be detected if the inverse-Compton process dominates the emission in the CTA energy range. 
However, the actual spatial and spectral distributions of this possible emission component are unknown and they would have to be determined from the data, which may eventually complicate the detection of such an extended signal.

\begin{figure}[t!]
\begin{centering}
\includegraphics[width=10cm]{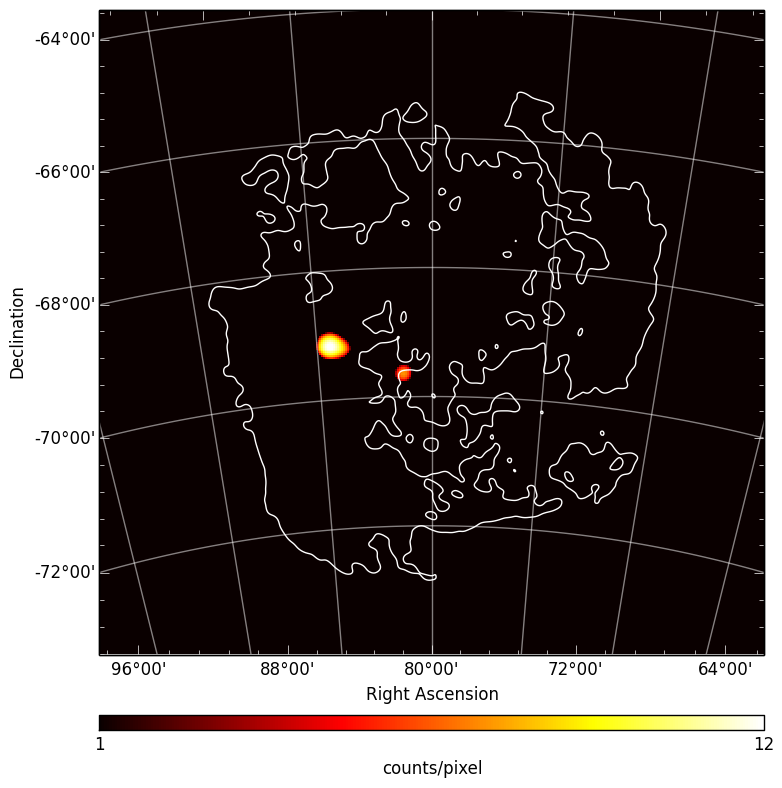}
\includegraphics[width=10cm]{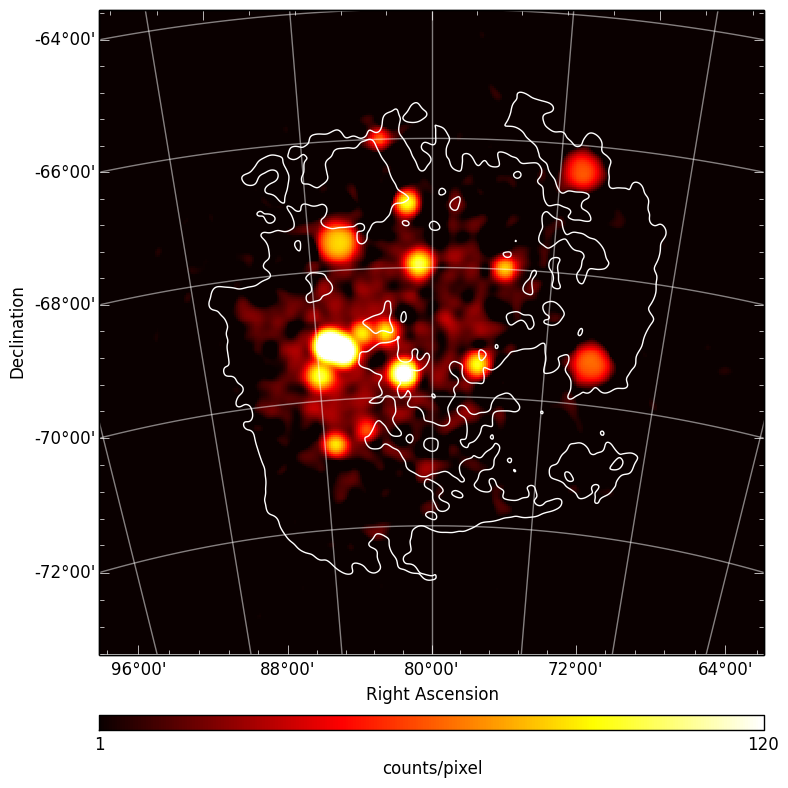}
\caption{Comparison of the LMC as viewed in VHE gamma rays with current instruments (top) and 
with CTA (below). The panels show smoothed residual count maps after subtraction of the instrumental background counts to the simulated events. The emission model includes detected sources (N157B, 30DorC, and N132D), ten point-like sources with $>1$\,TeV luminosities of $\sim10^{34}$\,erg\,s$^{-1}$, and a handful of regions enriched in cosmic rays. The top panel mimics the current H.E.S.S. view of the LMC and was obtained from a simulated 16 h of CTA observations using a single pointing and selecting events with energies $>$800\,GeV. The bottom panel is a simulation of the full CTA survey of the LMC involving six pointings and 340 h of observations for energies $>$200\,GeV, plus an additional source for the brightening SN\,1987A. The diffuse glow in the bottom panel results from the large-scale emission.
Credit: contour of H I distribution from \cite{Kim98,Kim03}.}
\label{fig:ksp_lmc_simumaps}
\end{centering}
\end{figure}

A parametric study was performed to determine the extent to which CTA could probe active star-forming regions where young cosmic rays are released. Simulations were carried out for various region sizes (for radii of 100, 200, 500~pc) 
and for different densities for the cosmic-ray population in these regions. We used as reference the cosmic-ray sea density at 10~TeV, which was determined under the assumption that it has one-third of the local Galactic value at 10 GeV and follows a power-law spectrum with a spectral index of 2.7 up to at least 10 TeV. The emission was assumed to be of hadronic origin with an average hydrogen number density, $N_H$, over the region of $10^{22}$\,cm$^{-2}$ (the densest, molecular regions in the LMC have $N_H$ of a few $\times 10^{22}$\,cm$^{-2}$ while the average over the LMC is $\sim 3 \times 10^{21}$\,cm$^{-2}$). The spectrum of the emission was assumed to be that of a recently-injected cosmic-ray population that homogeneously fills the region and does not diffuse away significantly; we therefore assumed a power-law spectrum with a spectral index of 2.2. The results are shown in Figure~\ref{fig:ksp_lmc_regiondet}. It turns out that in 250~h of effective exposure, CTA would significantly detect regions of 100-200\,pc in radius if they were recently enriched in cosmic rays by one supernova or more (a supernova is associated with 10$^{51}$\,erg of kinetic energy, of which 10$^{50}$\,erg are assumed to go into cosmic rays). Larger regions of 500\,pc in radius would be significantly detected if enriched by about ten supernovae at least, which is probably a rare occurrence. CTA could therefore probe inhomogeneities in the cosmic-ray population down to small scales and few events. The high significance reached for 100 and 200\,pc-sized regions enriched in cosmic rays by one or two supernova explosions suggests that spectral and morphological studies will be possible beyond the simple detection. Such numbers of supernovae in such volumes are consistent with the characteristics of the 103 giant shell candidates listed in \cite{Kim:1999}. The estimated mechanical luminosities that inflated the shell structures are consistent with the numbers of SNe given above for the detectability of their non-thermal gamma-ray emission: shells with observed sizes of 100 and 170\,pc require a mechanical luminosity corresponding to about 1-5 and 3-15 SNe, respectively (see Figure~8 in \cite{Kim:1999}). Yet, the dynamical age of these structures is typically several Myr, a duration over which cosmic rays may have diffused away. On the other hand, there is still ample speculation about the role of such superbubbles in confining, accelerating, and reaccelerating cosmic rays \cite{Bykov14}. With its ability to detect enhancements in the cosmic-ray population down to small scales and few events, CTA may therefore provide a valuable view of what happens in the early stages of cosmic-ray propagation. The recent detection of superbubble 30~Doradus~C with H.E.S.S. indicates the promise of this approach.

Figure \ref{fig:ksp_lmc_simumaps} illustrates the breakthrough that a deep CTA survey of the LMC would allow, compared to the current view provided by existing instruments (H.E.S.S.). Observations were simulated based on an emission model comprising: the three sources detected so far, hadronic and leptonic large-scale diffuse emission, hadronic small-scale diffuse emission from a handful of regions enriched in freshly released
cosmic rays, ten point-like sources with $>1$\,TeV luminosities of $\sim10^{34}$\,erg\,s$^{-1}$, and
SN\,1987A (in the case of CTA observations only, due to the rising flux of the source).
For a simulation of a CTA survey using six pointings and a total of 340~h of observations, almost all sources appear in a residual counts map, even the relatively faint large-scale diffuse emission. The contrast with the simulated current H.E.S.S. view, which probed only the most extreme objects of the LMC, is striking.

\section{KSP: Extragalactic Survey}
\label{sec:ksp_eg}

This Key Science Project consists of a blind survey of 25\% of the total sky. 
The survey is aimed primarily at extragalactic science
with the main objective to construct an unbiased 
very high energy (VHE) extragalactic source catalogue with an 
integral sensitivity limit of $\sim$6\,mCrab above 125\,GeV.
At the moment there are about 60 extragalactic sources seen with the imaging atmospheric Cherenkov
telescopes (IACTs) H.E.S.S., MAGIC and VERITAS,
most of these being BL Lacertae (BL Lac) objects. 
In addition there are five radio galaxies,
six flat spectrum radio quasars (FSRQs) 
and two starburst galaxies (NGC\,253 and M\,82).
The sample is, however, strongly biased since most of observations were motivated by high averaged flux at lower
frequencies in optical, X-ray or gamma-ray wavebands. Moreover, about half of the detections
have been made when alerted that the sources were in a flaring state. 
A second objective of this KSP is to provide a high resolution 
map of the extragalactic sky at gamma-ray energies between 50\,GeV and 10\,TeV.
A third objective is to search for unexpected and serendipitous VHE phenomena over
a large portion of the sky.
The area covered by the Extragalactic Survey
will connect to the Galactic Plane Survey
(GPS) so that no Galactic latitude is left un-surveyed.

As shown in Section~\ref{subsec:performance},
with 1000~h of observation CTA will reach a flux sensitivity for point-like sources at the level of 
6\,mCrab flux for any sky point in the survey.
For the definition of the mCrab unit see Ref. \cite{HEGRACrab}.
This flux limit is on the level of the weakest sources detected with IACTs so far.
When compared to Fermi-LAT (10-year exposure) and HAWC (5-year exposure),
the CTA extragalactic survey will be unique in the energy range between 100\,GeV and 10\,TeV.

The
extragalactic survey of a large portion of the sky is expected to 
be considered as one of the main legacies
of CTA.
Given its scale and scientific importance, the survey fits well within
the KSP concept and its results will have a large beneficial impact on the
broader astronomical community. 
The survey also has strong synergies with fundamental
physics (e.g. the dark matter search and electron spectrum anisotropy).

With the predicted sensitivity, it will be possible to construct an unbiased BL Lac sample
in the nearby universe, up to a redshift of $z \approx 0.2$. Sources in quiescent as well as in flaring states
will be detected and, depending on the number density (\textit{N}) of sources as a function of their flux (\textit{S})
(the log $N$ - log $S$ distribution, also called the luminosity function, LF) for the BL Lacs (which is largely unknown), 30-150 sources are expected to result from the survey; see below in Section~\ref{sec:objectives} for justification of these numbers. 
Moreover, more distant ($z \gtrsim 0.2$) BL Lacs and FSRQs will be detected, 
but only when they will be in elevated or flaring flux states because in the quiescent flux state they would typically remain below the CTA detection threshold
for short exposures.
Also several new radio galaxies will be discovered in VHE gamma rays.
The list of potential discoveries is long and the highlights include:
\begin{itemize}
 \item unbiased determination of the yet unknown log $N$ - log $S$ of gamma-ray AGN (BL Lacs and possibly FSRQs),
 \item discovery of extreme blazars peaking in the $\sim$100 GeV - 1 TeV region,
 \item serendipitous detection of fast flaring sources, not detectable in short observation time (hours) by lower sensitivity observatories like Fermi-LAT and HAWC,
 \item discovery of gamma-ray emission from yet undetected source classes such as Seyfert galaxies, ultraluminous infrared galaxies (ULIRGs), etc.,
 \item discovery of dark sources with no astrophysical counterpart, which would be a possible 
signal from dark matter annihilation,
 \item the possible detection of a gamma-ray burst (GRB) in the prompt phase (the probability for this
       is 2--4 times higher if the survey is performed in divergent pointing mode, see below), and
 \item the study of large scale anisotropies in the electron spectrum at energies between 100\,GeV and few TeV.
\end{itemize}

To ensure the uniformity of the survey we propose to perform it with the full CTA Observatory in a 
period of time of not more than three years. 
In the present Monte Carlo simulations, the sensitivity of the southern array is about a factor of 
1.5 better than the northern array 
in the energy range above 100\,GeV because of the larger number of MSTs and the presence of
the SSTs in the south. A factor of approximately 2.2 
longer exposure is needed to compensate for this difference in sensitivity. 
Nonetheless, it is proposed to utilise both CTA sites for the extragalactic survey
since it will allow us to complete the survey to the required sensitivity
within three years. To achieve a uniform sensitivity, it is proposed to
cover 15\% of the sky 
in the south using 400~h and the remaining
10\% of the sky  
in the north using 600~h.  Such coverage
would ensure a 6\,mCrab sensitivity over the whole area of the survey after 1000 h
of observations in total.

It is anticipated that the divergent pointing mode (see Section~\ref{divpointing}) will be more effective for this survey than the standard mode. 
Because of its greatly improved capabilities for serendipitous flaring sources,
the divergent pointing mode will be used for this KSP if it can be shown
that the required flux sensitivity can be achieved within the same time (even if 
this results in a moderate worsening of the spectral and angular resolutions).
However, detailed studies on the divergent mode are still ongoing and the estimates here are based on the
performance of the normal pointing mode, which is a conservative choice.

\subsection{Science Targeted} 

Through the unbiased survey of a significant part of the extragalactic sky we mainly target a population study
of the local (z$<$0.2) universe in gamma rays in the energy range of 100\,GeV to 10\,TeV.  
It will be the first time that such a large portion ($\sim25\%$) of the sky is observed 
uniformly and with high sensitivity at these energies. 
The uniqueness of the survey also enables 
the search for new source classes, 
as well as the search for large scale structures in the electron spectrum as outlined below. 
At the same time, such a survey will lead to an unbiased picture of the flaring sky in the 100\,GeV -- 10\,TeV regime,
which is vital for understanding the largely unknown duty cycles of blazar flaring activities.
The other strength of this KSP 
is the ability to look for the unexpected discoveries (e.g. dark gamma-ray sources, new source classes, etc.).

\subsubsection{Scientific Objectives}
\label{sec:objectives}

{\textbf {Blazars:}} 

\noindent Luminosity functions (or luminosity distributions, and their possible dependence on the redshift) 
are fundamental for understanding the main physics drivers within the sources and their evolution.
An unbiased survey with a uniform exposure allows us to measure the luminosity
function and the number density of sources depending on their flux (the
so-called log $N$ - log $S$ distribution). To construct the log $N$ - log $S$ distribution one
needs a sufficient number of sources ($>50$), which are representative for the
underlying population.
The main source class of gamma rays in the
extragalactic sky are blazars (BL Lacs and FSRQs), which are believed to
produce gamma rays inside relativistic jets with large bulk Lorentz factors
($\sim10$ or larger) beaming the emission towards the observer. The mechanism
of the gamma-ray production in blazars is not completely understood, with evidence 
that most of the emission has a leptonic origin and that flaring activity is
associated with freshly accelerated electrons or rapid changes in the magnetic
field.  Determining the log $N$ - log $S$ of blazars is of fundamental importance as it is
also crucial for determining the total gamma-ray background.
The measurement of the extragalactic gamma-ray background
(which does not yet exist for energies $>$700\,GeV) 
is the ultimate way to verify the theory of gamma-ray production in AGN.
Assuming the unified
scheme of the AGN is correct --- 
that the phenomenology of AGN mainly follows our viewing angle with
 respect to the jet orientation ---
the jets of the blazars have a small angle to the
observer's line of sight, 
which means that blazars account for a few $\%$ of the total AGN gamma-ray emission
\cite{UrryPadovani1995a, HenriSauge2006a}.

The number of detected blazars will determine the quality of the blazar LF.
However, the problem for measuring the LF is not just the number of objects but the bias
introduced in targeting. 
This bias can be solved with a wide survey. The number of
sources will affect the uncertainties, i.e. the ability to discriminate
between different LF distributions and eventually
the contribution of the different sub-classes (FSRQ, BL Lac).

\begin{figure}[h!]
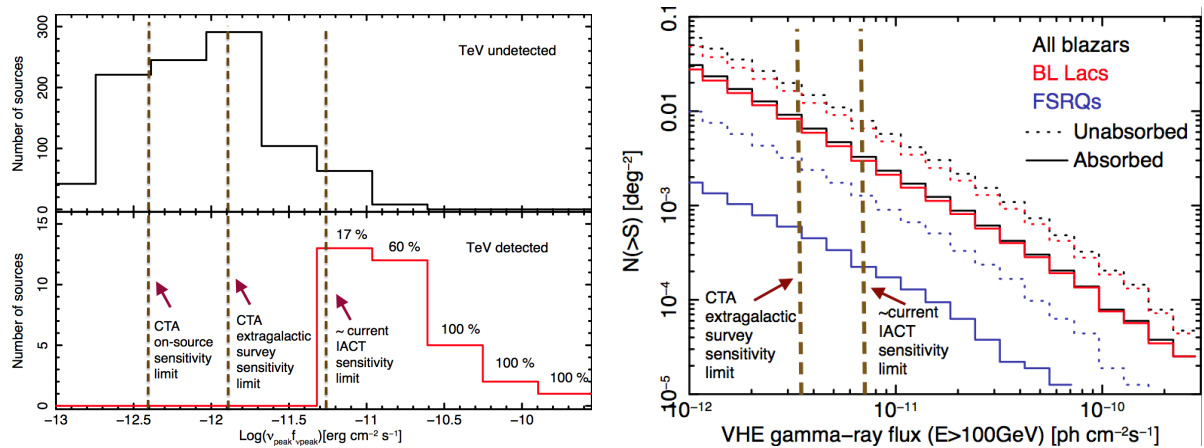

\begin{centering}
  \includegraphics[trim=0 3 2 2, clip=true, width=3.0in]{KSP_EG/Giommi_v4}
\hspace{0.2cm}
\includegraphics[trim=0 2 1 0, clip=true, width=3.0in]{KSP_EG/GiommiLogNLogS_v2}
\caption{Predictions for the number of blazars on the sky in the GeV--TeV gamma-ray domain. 
Left: Expected source counts as a function of the integral gamma-ray flux above 100\,GeV in 27,000\,deg$^{2}$. 
The upper panel shows predictions by \cite{Arsioli14} together with the current and envisioned sensitivity limits of 
imaging atmospheric Cherenkov telescopes (IACTs).
The lower panel shows detected AGN with IACTs. 
The numbers above the red histogram give the fraction of expected sources detected by current
instruments.
Right:
Simulated log $N$ - log $S$ distribution from \cite{Padovani14}.
The dashed (solid) lines represent the expected distributions without (with) taking into account the absorption by the EBL.
According to this study, with the 6\,mCrab sensitivity during the proposed survey CTA should 
detect around 100 sources in 10,000\,deg$^{2}$.
}
 \label{fig:logNlogS}
\end{centering}
\end{figure}

We foresee detecting 30-150 blazars within the survey.
If we extrapolate Fermi-LAT sources with known redshifts into the CTA energy range assuming a reasonable intrinsic cutoff of 1 TeV on average,
we obtain some 30--40 sources to be detected with the proposed sensitivity.
When instead of extrapolating from Fermi-LAT, we use predictions by \cite{Arsioli14},
who used IR and X-ray data to construct the log $N$ - log $S$ of blazars, we obtain 
approximately 75 sources.
In Figure~\ref{fig:logNlogS}, 
the sensitivity of the survey is shown and compared to the 1~mCrab CTA 50\,h sensitivity as well as the flux limit of the current 
VHE gamma-ray telescopes.
The histogram in the upper left plot is for 27,000\,deg$^2$, 
leading to a factor of 2.7 smaller predicted numbers for a 10,000\,deg$^2$ (25\% of the sky) survey. 
However, the catalogue is incomplete and can easily result in double as many sources, i.e.\ close to 150. Similarly high numbers
are obtained by \cite{Inoue10} 
based on the EGRET luminosity function \cite{Inoue09}.
There is also the possibility of detecting distant objects ($z \sim 1$) at energies 
$\gtrsim 300$\,GeV by including secondary gamma-ray emission from blazars in the
presence of intergalactic magnetic fields of ($10^{-17} - 10^{-15}$)\,G
\cite{Inoue14}.
In
\cite{Padovani14},
the expected number of blazars emitting above 100 GeV is estimated 
 based on Monte Carlo simulated surveys that reproduce the 
Fermi-LAT results well. 
They predict that the CTA extragalactic survey will detect 
between 110 and 180 blazars depending on different assumptions on the VHE spectrum (see Figure~\ref{fig:logNlogS}, right plot).

Even if many of the sources detected will be the known ones, the slope of the log $N$ - log $S$ distribution can only be determined through an unbiased survey. 
Some sources will be detected during flaring gamma-ray states introducing a bias when studying the quiescent state of sources only.
However, we estimate the bias to be in the order of $\sim10\%$\footnote{
The uncertainty in the number of sources found in a flaring state is large due to unknown VHE gamma-ray duty cycles. A rough
estimate can be made using the monthly duty cycles of Fermi-LAT. High-flux events above 1.5 
standard deviation significance have a duty
cycle of about 5 to 10\% for FSRQs and BL Lacs in the GeV band \cite{Ackermann2011}. 
Therefore, we assume that during the proposed survey 10\% of the sources will be in a flaring state.},
and the result can be corrected for it. 
In addition, the detected unbiased flaring episodes (within the survey) 
are as important to characterize the gamma-ray sky as the sources detected in the quiescent state.
One of the major unanswered questions of modern AGN physics is the existence or not of the blazar sequence. If the sequence exists,
there should be no, or only a few, detections of high luminosity blazars that have their synchrotron peak in optical and UV wavebands.
The survey will be able to probe this prediction.

{\textbf {Extreme blazars:}}
 
\noindent The survey will reveal a population of extreme blazars, i.e.\ sources with hard spectra having their 
gamma-ray peak 
in the range of 100\,GeV to more than 10\,TeV. 
Extreme blazars are very interesting because of their use in studies
of the extragalactic background light (EBL) and the intergalactic magnetic field (IGMF),
as well as due to the fact that the gamma-ray emission in leptonic scenarios 
must be produced in a deep Klein-Nishina regime 
(which is unusual) or have a hadronic origin. 
The sources are usually too weak to be detected with Fermi-LAT even in ten years exposure,
making it hard to make predictions for CTA. 
In the local universe, two bright extreme blazars are known: Mrk\,421 and Mrk\,501
(at z$\sim$0.03). 
A third extreme blazar is 1ES\,0229+200 (z=0.14).
Typical VHE gamma-ray fluxes for Mrk\,421 and Mrk\,501 are on the order of 200\,mCrab when in 
a quiescent state.
Taking the envisioned sensitivity of 6\,mCrab as the detection
limit, sources 40 times weaker can be detected
during the survey. This leads to an increase in distance by a factor of $\sim6$ (out to z$\sim$0.2), 
or to an increase in the accessible volume to
Mrk-like sources by a factor of 200. The extrapolation from two sources has naturally
a large uncertainty, but making a conservative estimate 
and taking EBL absorption into account (which is not relevant
for energies below 300\,GeV for these redshifts), we expect to detect 30--150 extreme blazars.

{\textbf {Radio galaxies:}}
 
\noindent Near-by radio galaxies have also been found to be sources of gamma rays.
However, the mechanisms by which radio galaxies produce their gamma-ray
emission are not well understood, with several explanations currently 
possible. Gamma rays are possibly produced near the nucleus or in the knots
along the jet, but radio lobes are also discussed as candidate sites. Only five
radio galaxies have been detected in VHE
gamma rays so far, and the flaring duty cycles are poorly known. The survey we
suggest will help detecting many more radio galaxies, which would boost our
understanding of the emission mechanisms and the contribution of radio galaxies
to the total production of gamma rays. 
Using the fact that the five radio galaxies have their TeV fluxes in the order of 20\,mCrab when in
a quiescent state, we estimate that within the survey sensitivity we will detect a few new radio galaxies.

{\textbf {Starburst galaxies:}} 

\noindent Two close-by starburst galaxies (M\,82 and NGC\,253) have been detected in VHE gamma rays. 
Both show weak emission at the flux level of 6\,mCrab.
With the suggested extragalactic survey, we will reach the 
sensitivity to detect further starburst galaxies at any location in the survey 
if their gamma-ray flux is at a similar level.
However, since M\,82 and NGC\,253 are probably the 
brightest gamma-ray starburst galaxies, we foresee at most only a few 
new detections within the survey.

{\textbf {New source classes:}}

\noindent An unbiased survey of $\sim$25\% of the extragalactic sky enables the 
detection of new source classes.  
Clusters of galaxies, Seyfert 2 galaxies and ULIRGs have been proposed to emit VHE gamma rays
but none have been detected so far. 
At GeV energies, Fermi-LAT did detect the Seyfert 2 galaxy Circinus \cite{Hayashida2013}.
Beyond this, 
the survey has 
chances to detect gamma-ray sources without clear association with known objects (dark sources)
or sources where no strong non-thermal emission is detected. This can be a clear signature of unknown 
physics, for example, the decay of a new particle, such as dark matter, into gamma rays.

{\textbf {Gamma-Ray Bursts and other transients:}}

\noindent Gamma-ray bursts (GRBs)
are usually distant events with redshift z$>$1,
meaning that most of their VHE gamma-ray emission will be
absorbed by the extragalactic background light. However,
the universe is basically transparent to gamma rays with
energies E$<$30 GeV and absorption is not severe up to 50\,GeV even for large cosmological 
distances. This means that with the low-energy response 
of CTA there will be a good chance
to catch GRBs even if their redshift is $z>>$1, provided their
central engine emits to sufficiently high energies. For more details
on the prospects on GRB detection see Chapter~\ref{sec:ksp_trans}.

There is a chance of catching a GRB in the onset of the 
prompt phase if the burst occurs within the observed field.
If the survey is performed in the normal pointing mode, the chances to catch a GRB
during the survey are the same as during any other pointed observations and are 
quite low, $\sim 0.08$\,yr$^{-1}$ in the 8\,deg diameter field of view (FoV) of the MSTs.
However, if the extragalactic survey can be conducted in divergent 
pointing mode with an instantaneous FoV
considerably larger than in normal pointing, it would significantly enhance the prospects
for studying cosmic transients at very high energies 
without relying on triggering the observations with an external alert. In particular, carrying
out the survey
in divergent pointing mode could enable:
\begin{itemize}
 \item the detection of GRBs from their onset, including the prompt phase of short GRBs and realizing the associated improvements for probing Lorentz invariance violation,
 \item unbiased searches for VHE transients in general, and
 \item multi-wavelength and/or multi-messenger
spatio-temporal correlation studies with other observatories with wide FoV and/or limited localization capabilities.
\end{itemize}
In the most optimistic case, CTA can cover up to 1,000\,deg$^2$ simultaneously when observing in
divergent pointing mode,
which is about a factor of 20 higher than the FoV when using normal pointing. 
This would boost the rate of GRBs observable by CTA in the prompt phase to about 2 per year.
For the purpose of studying GRBs or 
transients, a FoV as large as possible is favored, but this needs
to be balanced with the sensitivity and energy threshold required for the other aims of the survey.
More details on the unique prospects for studying transients with a divergent pointing survey
are discussed in the Transients KSP (Chapter~\ref{sec:ksp_trans}).

{\textbf {Large scale electron anisotropy:}} 

\noindent The sensitivity of CTA is sufficient to detect the
diffuse electron component of the cosmic-ray
spectrum above 100\,GeV in every survey 
pointing\footnote{We assume here a single pointing of 30\,min and normal pointing mode.}. 
Thus, systematic errors permitting,
the survey will allow the study of large-scale anisotropies in the electron background, 
which has not yet been possible at these energies.

{\textbf {Diffuse gamma-ray background:}}

\noindent Though, as stated before, the beamed gamma rays from blazars are estimated to 
account for about $\sim1\%$ of the total AGN gamma-ray emission,
the extragalactic gamma rays with E$>$100\,GeV are mainly from blazars. 
Therefore, through the construction of the log $N$ - log $S$ distribution of blazars, 
the survey will allow us to resolve the average diffuse 
gamma-ray background~\cite{DiMauro13,Inoue16}
and compare it to the measurement of the gamma-ray background at lower energies. 
Such a comparison is vital to understand the completeness of the sample detected and thus the total amount of extragalactic gamma rays produced.
A direct measurement of the gamma-ray background at very high energies will
also be attempted by CTA using the survey data,
but due to the electron background this will be 
a very challenging task, as discussed in \cite{Sol13}.

{\textbf {Dark Matter clumps:}}

\noindent Structure formation predicts gravitationally bound dark
matter clumps down to much lower masses than for dwarf spheroidal
galaxies. The total number of clumps within Galactic halos
can be as high
as $10^{15}$. It is likely
that a large population of lighter dark clumps exists with highly suppressed
(or even negligible) baryonic content, hiding from detection in
currently operating sky surveys.
Because of the difficulty in the identification of clumps in optical surveys, the prime
channel of detecting dark clumps are high-energy and VHE gamma rays resulting
from annihilations or decays of dark matter particles in the clump. Such
objects, if bright enough, could appear as unidentified sources in the
proposed survey. Because of the limited sensitivity of Fermi-LAT, if the dark matter mass is
above a few hundred GeV, these objects could have been
missed by Fermi-LAT and could be detectable by CTA. 

\subsubsection{Context / Advance beyond State of the Art}

\begin{figure}[h!]
\begin{centering}
\includegraphics[width=6.0in]{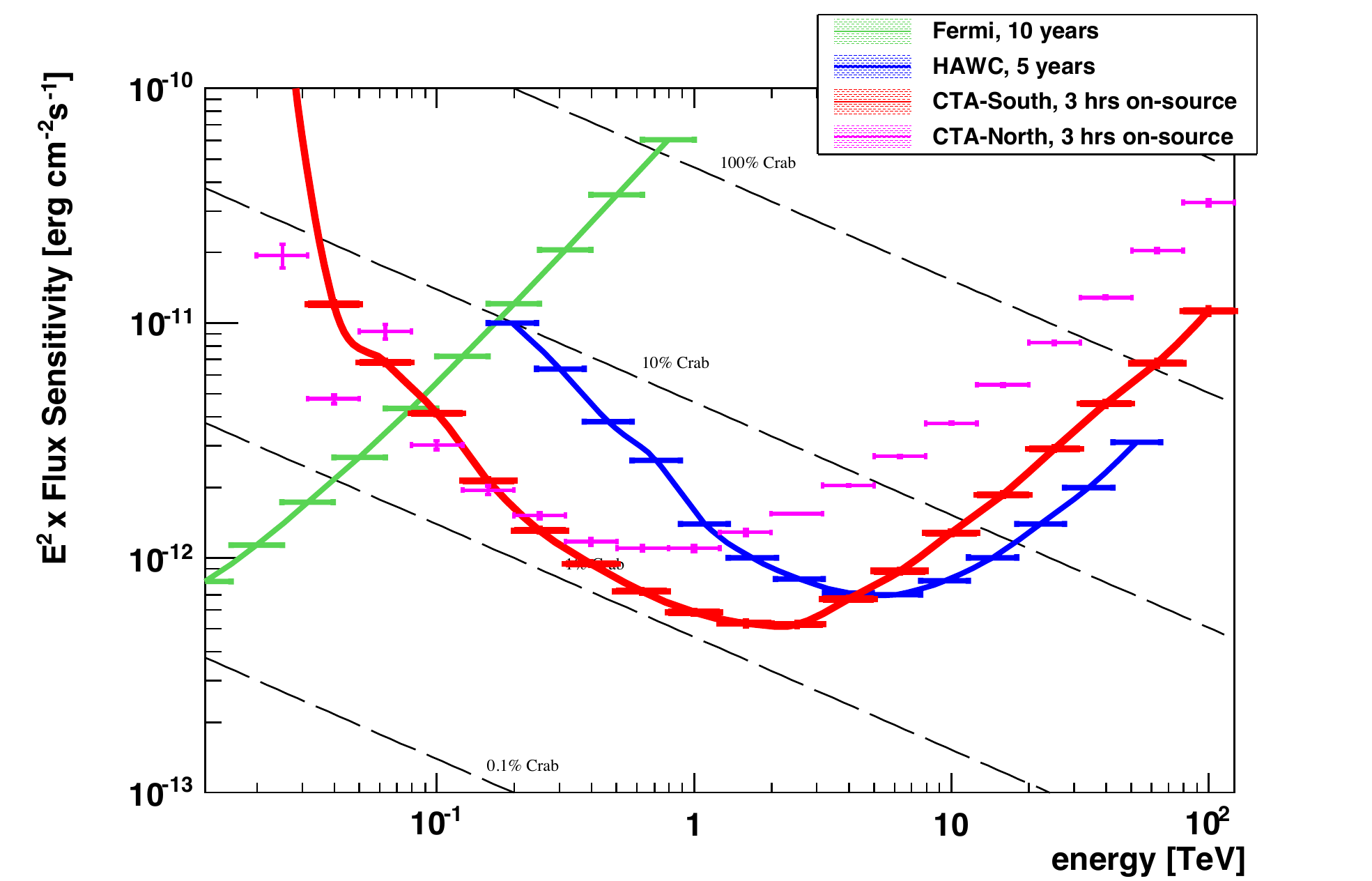}
\caption{Differential sensitivities of instruments in the
GeV--TeV range together with the CTA extragalactic survey sensitivity.
The Fermi-LAT sensitivity is shown for ten years exposure (dark green, \cite{Atwood2013a}).
HAWC is shown for a five-year exposure (blue, \cite{Abeysekara13}). 
The CTA target sensitivity for the extragalactic 
survey is 6~mCrab and is similar to
the 3~h sensitivity of the southern array (red). Sensitivity for the northern CTA array is also shown (magenta).
See text for details.
}
 \label{fig:sensitivity}
\end{centering}
\end{figure}

There are about 70 extragalactic sources detected with the current generation of IACTs: H.E.S.S., MAGIC and VERITAS.
The majority of the sources belongs to the AGN subclass called BL Lacs. 
The sample is, however, strongly biased since most of the
observations were motivated by high averaged flux at lower
frequencies in the optical, X-ray or gamma-ray wavebands. Moreover, about half of the detections
have been made following alerts that indicated that the sources were in a flaring state. 
The performance of the current generation of IACTs does not allow for an unbiased 
survey of a large portion of the extragalactic sky with a reasonable flux sensitivity.
Assuming a survey time of 1000 h and a FoV for the current telescopes of around 10\,deg$^2$,
one obtains a 1~h exposure for every sky point in a 10000\,deg$^2$ survey ($\sim 25$\% of the
sky), which would allow the detection of only 
$\sim$10 of the brightest sources.
Therefore, with the current IACTs no systematic survey can be performed over a large portion of 
the sky and spending 1000~h on such a survey cannot be justified.

For CTA the situation will be quite different because CTA will be much more capable than
the present day instruments.
In Figure~\ref{fig:sensitivity}, the differential sensitivities of 
CTA and the current wide-field instruments (Fermi-LAT and HAWC) are shown. 
With a 6~mCrab integral sensitivity above 125\,GeV 
(which would roughly correspond to the red curve in the plot for sources with Crab-like spectra),
the CTA extragalactic survey will make a unique measurement between 100\,GeV and 10\,TeV.
It is worth mentioning that the integral sensitivity of Fermi-LAT after an exposure of ten
 years is expected to reach a level 
of 20~mCrab 
above 200\,GeV, which makes Fermi-LAT competitive with CTA for steady sources up to that energy.
In the southern sky, which HAWC does not access, the uniqueness of the survey extends to higher energies.
In the north, HAWC is complementary to CTA at energies above 10\,TeV. 
It is clear that, the CTA-North array, with a smaller number of telescopes and
no SSTs, will require a longer exposure to obtain the same target sensitivity.

\subsection{Strategy}

The CTA extragalactic survey is proposed to take 1000~h and cover
25\% of the overhead sky.
The survey is unique and in that it opens a new window for the search of 
extragalactic sources as well as it being the first attempt for a complete
log $N$ - log $S$ study of close-by blazars in VHE gamma rays.  

The normal, conservative, pointing scheme would be the default one.
In case the simulations show that the divergent pointing mode is
more sensitive than normal
pointing for the survey, the survey would be performed using divergent pointing.
In this case, the complexity of the event reconstruction and data analysis would
require the set-up of a special analysis pipeline and the unique expertise
of the CTA Consortium.

\subsubsection{Possibility of several pointings for a given field of view}

It will be shown later (Section~\ref{subsec:performance}) 
that the optimal strategy for the survey is a sequence of pointings
at grid points of the survey area with a 3$^{\circ}$ separation between the points.
This pointing strategy would result
in approximately 3\,h effective observation time for every point on the sky within
the boundary of the survey. The default plan would be to observe at each grid point
for the required amount of time (i.e. 0.51 h in the south and 1.11 h in the north,
see Table~8.2).
However, it may be better to have shorter pointings 
(e.g.\ 10\,minutes in the south and 22\,minutes in the north) 
for each grid point and to return to each
grid point three times during the survey. This would
increase the chances of catching flaring activity,
while keeping the total observation time the same.
Also, shorter exposures would enhance CTA's ability to detect
short, bright transient phenomena.
We will keep this possibility in mind for the optimization of the survey
strategy once the time needed for re-pointing CTA is known.

\subsubsection{Shallow survey versus deep survey}

We investigated different survey strategies. In particular, we focused on the comparison between 
a shallow and wide survey versus a deep and narrow one. 
According to the study made by \cite{Padovani14},
which is based on Monte-Carlo simulations of blazar populations,
observing a four times narrower field for, consequently, four times longer time would result
in a detection of about 50\% less sources. We, therefore, argue for a wide area shallow survey.
However, the shallow survey should be deep enough to obtain a significantly (more than a factor of ten)
better sensitivity than Fermi-LAT and HAWC for steady sources. A quarter-of-the-sky survey is 
a good compromise 
(see Section~\ref{sec:results} for a discussion of the achievable sensitivities).
We stress, however, that the survey area can be further optimized 
 using results on the number of serendipitous sources found in pointed 
observations during the CTA construction phase (see Section~\ref{sec:serendipitous}).

\subsubsection{Targets}

\begin{figure}[h!]
\begin{centering}
\includegraphics[width=6.6in]{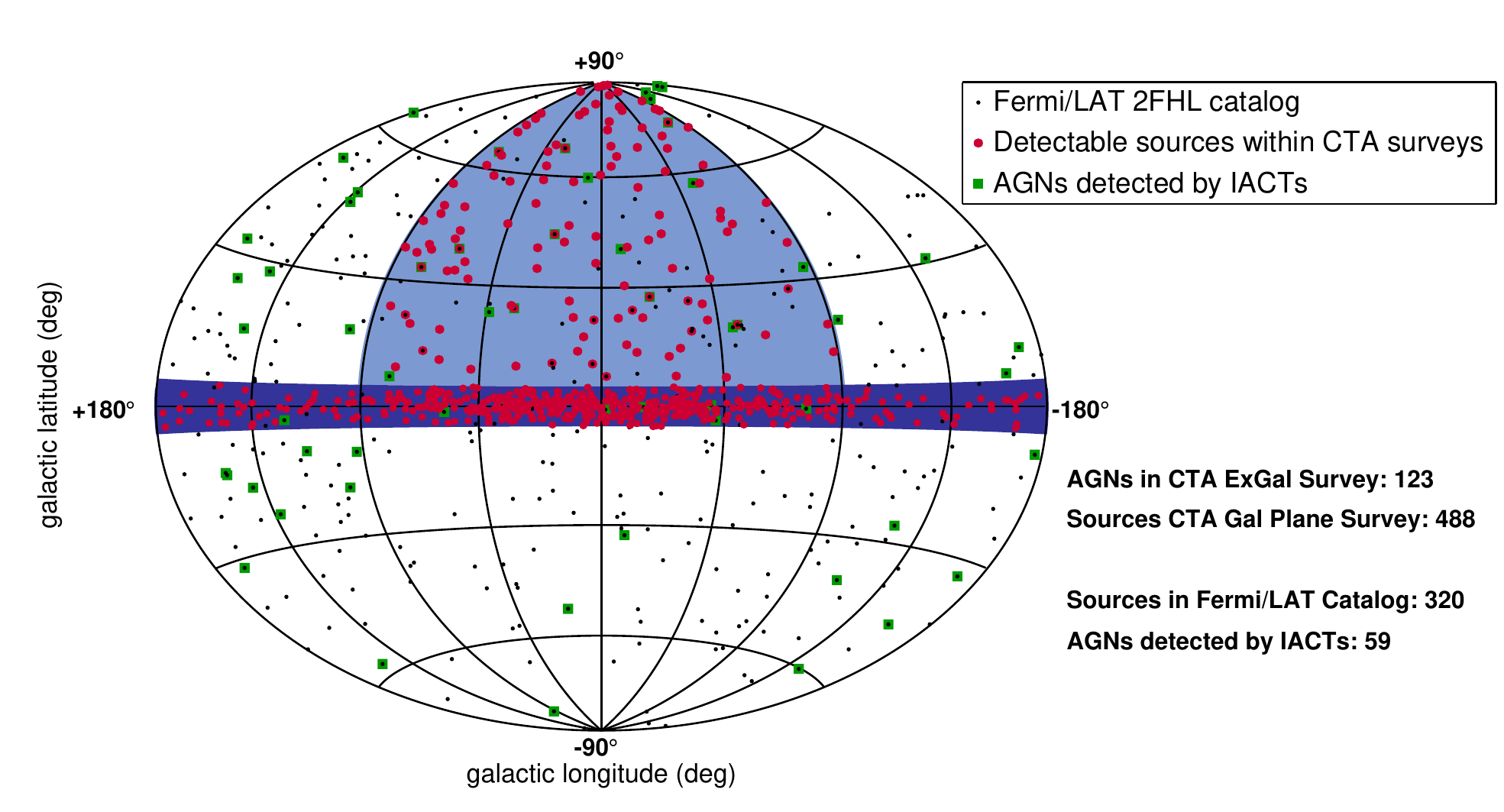}
\caption{Proposed region of the extragalactic survey in Galactic coordinates: $b > 5^{\circ}; -90^{\circ} < l < 90^{\circ}$, 25\% of the sky, marked in a light blue.
The Galactic Plane Survey is indicated by a darker blue. 
Red points show a hypothetical example of the sources to be detected in
the extragalactic and Galactic CTA surveys.
Extragalactic and unidentified Fermi-LAT hard-spectrum sources 
(2FHL catalogue \cite{Ajello15}) are displayed as black dots whereas green points show the AGN that have
been detected so far by IACTs \cite{tevcat}.}
 \label{fig:skymap}
\end{centering}
\end{figure}

We propose an extragalactic survey as shown by the light blue area 
of Figure~\ref{fig:skymap}. 
The survey would connect with the Galactic Plane Survey ($|b|<5^{\circ}$, dark blue area)
and cover $\sim$25\% of the sky, 
over Galactic longitude $-90^{\circ} < l < 90^{\circ}$. The proposed survey would be performed using both CTA arrays 
for zenith angles of observations smaller than 45$^{\circ}$ to ensure uniformity in the 
energy threshold and resulting sensitivity.
Several highly interesting regions, such as Cen A (south) and
the Virgo cluster, Coma cluster, and Fermi bubbles (north),
will be covered by the proposed survey. 

A hypothetical result of the survey is illustrated Figure~\ref{fig:skymap} by the red dots.
The black dots represent 2FHL sources (Galactic sources are not shown) from \cite{Ajello15}.
The red dots result from
CTA simulations extrapolating 2FHL sources for a CTA exposure of 6 hours,
assuming an averaged flux state and that 5\% of the sources will be found in a flaring state.
The Galactic sources are simulated to follow the spatial distribution of pulsars
in the ATNF pulsar catalogue.
The green points show extragalactic sources already detected in the TeV gamma-ray regime
from \cite{tevcat}.

\subsection{Data Products }

This KSP will produce an {\bf extragalactic gamma-ray catalogue} of detected sources.
For each source, the catalogue will contain the:
\begin{itemize}
\item differential energy spectrum,
\item significance of the detection,
\item source location,
\item integral flux,
\item variability index,
\item extension of the source, if applicable, and
\item association with known objects, if made.
\end{itemize}
\noindent The catalogue will also contain the time intervals (modified Julian dates)
for the observation periods and the time intervals for any detected flares.

The catalogue will be released one year after completion of the survey.
In case of detection of a significant transient event,
a public alert will be issued at the earliest possible moment (via, e.g.,
the Virtual Observatory network using VOEvent protocol, see
the Transients KSP in Chapter~\ref{sec:ksp_trans}).

Interrupting the survey for self-triggers and/or external triggers will be possible
but the conditions for such events should be carefully studied and
covered by a separate proposal.
Follow-up observations are not part of this KSP and are expected to be largely
carried out through the GO Programme.

In order to maximize the scientific output of the extragalactic survey,
we are planning to organize an extensive multi-wavelength effort
to accompany the survey. We foresee obtaining simultaneous optical data and snapshots
in the radio band. Since the optical and radio observations have a much smaller FoV 
than the CTA telescopes, we plan to observe potential sources within the CTA FoV only
or make shallow scan observations. For the new detected TeV sources without 
well determined redshifts, we will launch 
follow-up optical spectroscopy observations to obtain the redshifts.
In the X-ray band, we will launch ToO programmes for follow-ups on detected gamma-ray flares.
We will complement our multi-wavelength efforts by the public Fermi-LAT data in the GeV
gamma-ray band.
It is planned that all available simultaneous multi-wavelength data will be published in the 
extragalactic gamma-ray catalogue as additional information.

\subsection{Expected Performance/Return }
\label{subsec:performance}

In this section we estimate the expected survey sensitivity. All numbers are
given for a survey size of $25\%$ of the sky. 
We focus on the estimation of what can be
achieved through parallel pointing of the telescopes (i.e. where during
each observation all
telescopes are pointing in the same direction). The prospect of using the divergent
pointing mode is also briefly discussed.

\subsubsection{Method}

The survey pattern considered is similar 
to that presented in an earlier paper on surveys with CTA \cite{Dubus13}.
We consider a pattern where the observation pointing directions
are uniformly distributed on a grid of equilateral triangles.
The observations are simulated using standard software used
in the gamma-ray band (ctools and GammaLib)
and the instrument response functions come from CTA simulations. A
representative portion of the survey is considered and the sensitivity at each
point is calculated by simulating a point source with 0.25 degree steps in the
source position. The simulated source has a Crab-like spectrum, and its flux is
adjusted so that the source is detected at a significance of five
standard deviations above the background.
That flux then corresponds to the detection sensitivity at that location in the survey. 
The sensitivities are given in comparison with the Crab nebula flux, as discussed
in Ref. \cite{HEGRACrab}.

\subsubsection{Serendipitous discoveries during the construction phase}
\label{sec:serendipitous}

\begin{figure}[h!]
\begin{centering}
\includegraphics[width=5.0in]{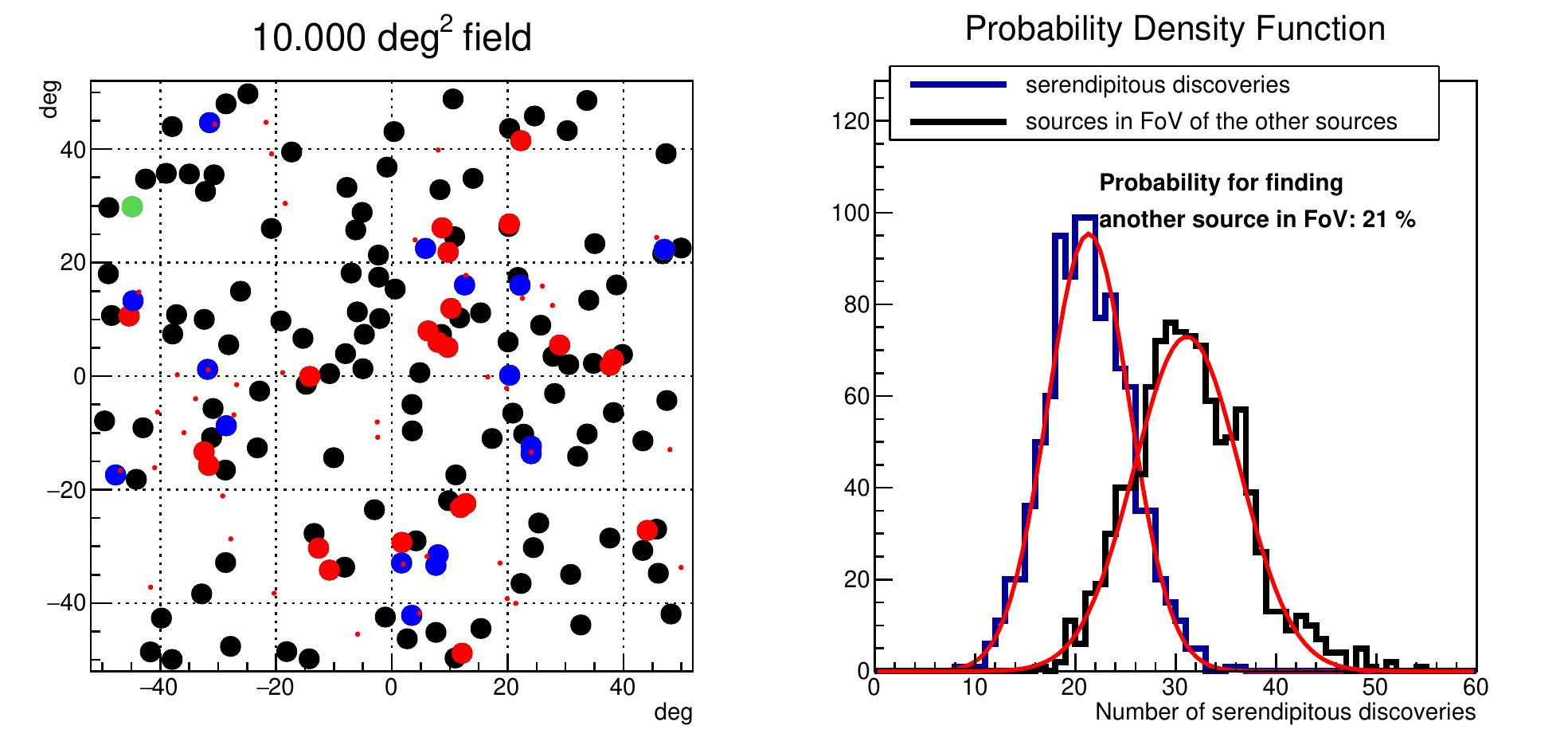}
\includegraphics[width=5.0in]{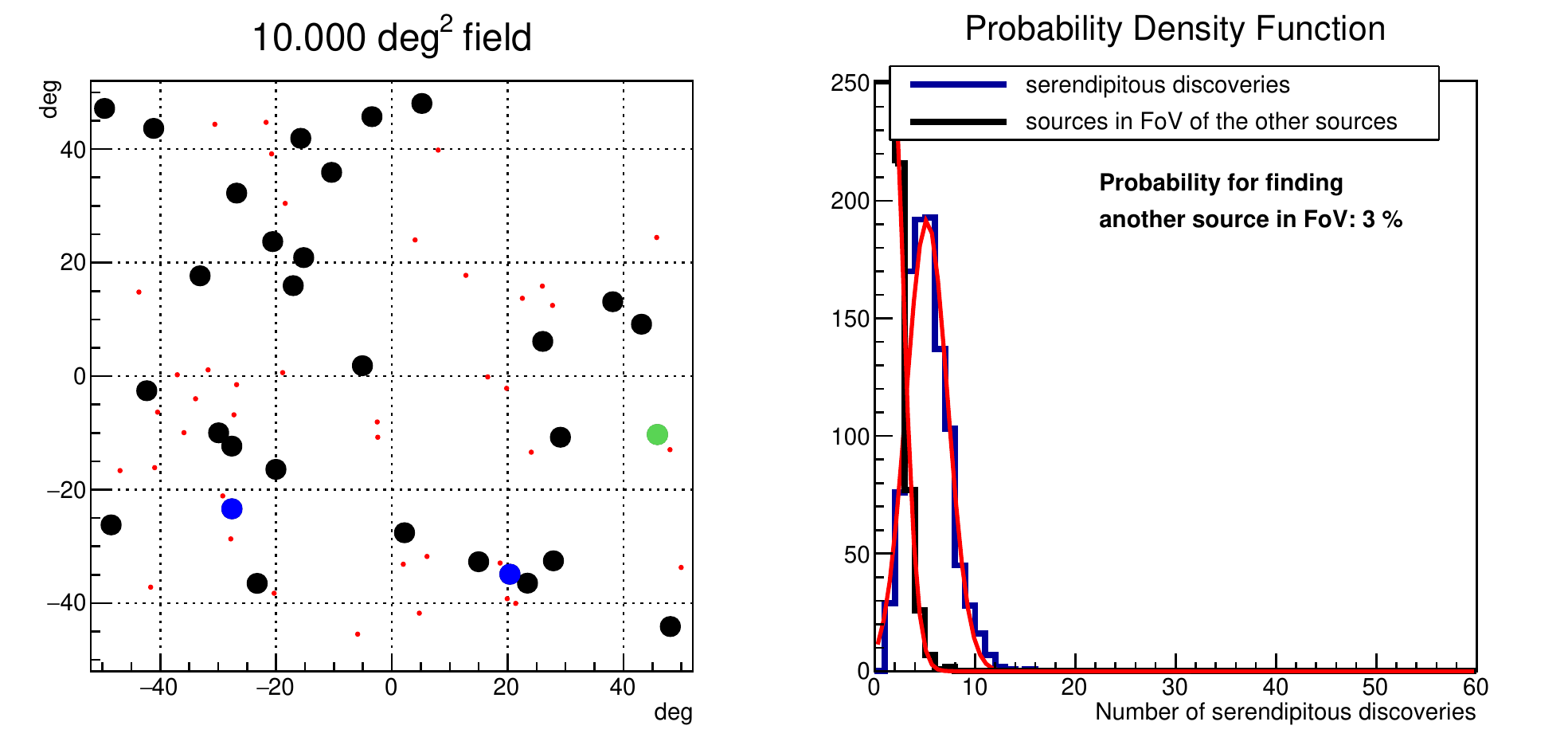}
\caption{Results of the Monte-Carlo simulation for serendipitous discoveries during the CTA construction phase.
In the optimistic case (upper plots), 20-30 sources can be discovered serendipitously,
whereas in the pessimistic case (lower plots), only 2-5 serendipitous discoveries can be made.
See text for details.
}
 \label{fig:toyMC}
\end{centering}
\end{figure}

As stated above, there is a large spread in the theoretical predictions 
on the source numbers and their class types for the envisioned extragalactic survey with CTA.
We investigated the chances for serendipitous discovery of new gamma-ray sources during the construction
phase of CTA. We considered on-source observations that will be performed 
after completion of half of the CTA array and before the start 
of observations with the full array. 
It is reasonable to assume that 
there will be some 50 extragalactic fields observed with this half-CTA array, for
20\,h each. Taking a reduced sensitivity of the half-array into account we assume
that within a 20~h observation, a 6\,mCrab integral sensitivity level will be achieved
within a $3^{\circ}$ radius of the centre of the field of view.
Using these assumptions we performed two Monte-Carlo simulations:
\begin{itemize}
 \item optimistic scenario: assuming that there are 150 extragalactic sources detectable with CTA 
(flux higher than 6\,m Crab) in 10,000\,deg$^{2}$, and
 \item pessimistic scenario: assuming that there are 30 extragalactic sources detectable with CTA 
(flux higher than 6\,m Crab) in 10,000\,deg$^{2}$.
\end{itemize}
In each case a population of sources was uniformly simulated in 10,000\,deg$^{2}$.
Two cases of CTA pointings were considered:
in the first one, all 50 pointings are random and not correlated with the simulated source positions.
In a second one,
we assumed that the 50 pointings are performed towards a subset of the simulated population
(if there are more pointings than sources, the rest of the pointings are random). 
Each simulation was 
repeated 1,000 times and the mean value in the number of detections, as well as the 
spread, were obtained.

The results of the simulations are shown in Fig.~\ref{fig:toyMC}.
In the plots on the left we show one out of 1,000 simulations.
In these plots, the small red dots indicate the random pointing positions of CTA.
We count serendipitous discoveries in following cases:
\begin{itemize}
\item serendipitous detection of a source in a random pointing (Fig.~\ref{fig:toyMC}, big blue dots);
\item serendipitous detection of a second source in a random pointing (Fig.~\ref{fig:toyMC}, big green dots);
\item serendipitous detection of a second source in a pointing towards another source (Fig.~\ref{fig:toyMC}, big red dots);
\end{itemize}
In the histograms on the right the distributions of the serendipitous detections are plotted.
Depending on the assumptions made, we find that, in the optimistic case, 20-30 sources can be discovered serendipitously to pointed observations.
In the pessimistic case, 2-5 sources will be discovered serendipitously.

These Monte Carlo simulations show us that with a partial CTA array during the construction phase we will be able to obtain a reasonable estimate
on the number of expected sources in the extragalactic survey, which can be used to refine the survey strategy. 

\subsubsection{ Results after the completion of the survey}
\label{sec:results}

\begin{figure}[t!]
\begin{centering}
\includegraphics[width=4.0in]{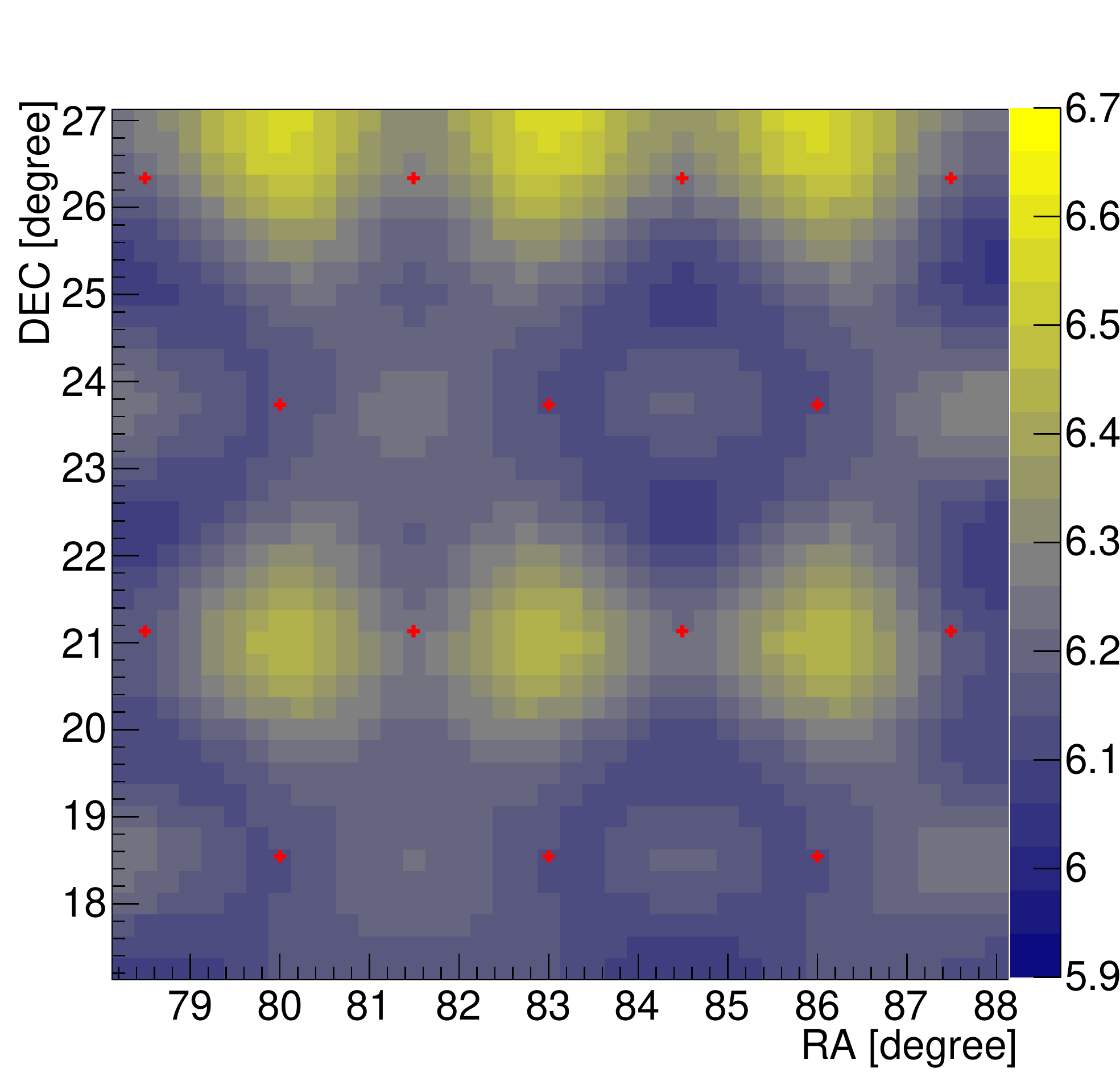}
\caption{Sensitivity map (colour scale, integral sensitivities in 
units of mCrab for energies above 125 GeV) from simulations for a portion of the survey, assuming 
the southern array for 600~h and covering $10000$\,deg$^2$. 
The grid is in approximate RA/Dec coordinates.
Note that the colour scale is zero-suppressed to better 
illustrate sensitivity variations across the survey. 
Red crosses denote the actual pointing positions.
}
 \label{fig:pattern}
\end{centering}
\end{figure}

Sensitivities have been calculated for the south and north arrays using the CTA
simulations. 
A preliminary study has shown that the instrument response
functions obtained using a set of cuts optimized in 50 h observations and a 
Crab-like spectrum give the best results in terms of survey integral sensitivity compared to
other available sets of cuts. The results presented here make use of the
instrument response functions (IRFs) derived from these cuts. Nevertheless, since
the cuts have not been optimized for the survey, the
sensitivity quoted here may not be optimal. 

For the CTA-South and CTA-North arrays, we use the configurations with
4 LSTs, 24 MSTs and 35 SSTs, and
4 LSTs and 15 MSTs, respectively.
The same arrays without LSTs have also been
considered.  An optimization is performed considering the spacing between the
pointings (the smaller the spacing the more uniform the sensitivity) and 
the exposure per pointing, keeping the total exposure of 
the 10,000\,deg$^2$ constant.
Three different spacings have been considered: 2, 3
and 4 degrees.  An example of the resulting sensitivities and their variation is
shown in Figure~\ref{fig:pattern}.
This particular example shows the performance
of the south array with a 3$^{\circ}$ spacing between the pointings.  The
integral sensitivities are colour-coded and the grid is in RA/Dec coordinates.

\begin{table}
\centering
\begin{tabular}{|l|l ||c|c||c|c||c|c| }
\hline
\multicolumn{2}{|c||}{} & \multicolumn{6}{c|}{Spacing between the observations}\\
\cline{3-8} 
 \multicolumn{2}{|c||}{ARRAY}   & \multicolumn{2}{c||}{4 degree}      & \multicolumn{2}{c||}{3 degree}      & \multicolumn{2}{c|}{2 degree}      \\ 
 \multicolumn{2}{|c||}{}  & \multicolumn{2}{c||}{0.83h / obs.} & \multicolumn{2}{c||}{0.46h / obs.} & \multicolumn{2}{c|}{0.21h / obs.} \\ 
\cline{3-8} 
 \multicolumn{2}{|c||}{}  & $S$ & $\Delta\,S$ & $S$ & $\Delta\,S$ & $S$ & $\Delta\,S$ \\
\hline 
 \multicolumn{2}{|c||}{South-noLST}  & 5.4  &  0.9               & 4.8  &   0.4               & 5.0                & 0.5                \\ 
\hline                                                                                                                                   
 \multicolumn{2}{|c||}{North}   & 8.61 &  1.2               & 8.0 &   0.8               & 8.1               & 0.8                \\ 
\hline
\end{tabular}
\caption{Estimation of the survey sensitivity for a total of 600~h of
observations and a coverage of 25\% of the sky, for the south and north arrays
and for various grid spacings (in degrees).
The sensitivity, $S$, 
is given in units of mCrab, see \cite{HEGRACrab}.
$\Delta\,S$ represents the survey sensitivity fluctuation; this is the standard deviation of
the sensitivity distribution over the sampled survey field of view.
}
\label{EXSURV:tab1}

\end{table}

\begin{table}
\centering
\begin{tabular}{|l|l ||c|c| }
\hline
\multicolumn{2}{|c||}{ARRAY} & Time per pointing (h) & {Integral sensitivity}\\
 \multicolumn{2}{|c||}{}  &  & $S \pm \Delta\,S$ \\
\hline
 \multicolumn{2}{|c||}{South-noLST}  & 0.51 &  5.1  $\pm$   0.5                \\ 
\hline                                                                                                                                   
 \multicolumn{2}{|c||}{North}  & 1.11 &  5.4  $\pm$   0.6                \\ 
\hline
\end{tabular}
\caption{Same as Table~\ref{EXSURV:tab1}
but for the three degree spacing with the sensitivity numbers scaled for a 
total of 1000~h shared between CTA-South (400~h) 
and CTA-North (600~h), covering 60\% and 40\% of the survey area, respectively.
}
\label{EXSURV:tab2}

\end{table}

The results of the study for 600\,h and 25\% of the sky are summarized in Table~\ref{EXSURV:tab1}.
Note that here we assume that entire 600\,h are spent in South or in North.
A spacing of three degrees between the observations gives a better mean sensitivity compared to four or two degrees. 
Using this spacing and pointing scheme, every sky point in the survey has an effective observation time
about 6 times larger than that of a single pointing. 
In the south, the survey integrated sensitivity is on the order of 5\,mCrab above 125 GeV. 
In the north, a sensitivity of 8\,mCrab can be achieved.
These numbers will only improve with an optimized array layout and advanced analysis techniques.

In order to obtain a uniform sensitivity between North and South, we propose to
cover 15\% of the sky (60\% of the survey) from the south in 400\,h and the remaining
10\% of the sky (40\% of the survey) from the north in 600\,h.  Such coverage
would ensure a 6~mCrab sensitivity over the area of the whole survey after 1000\,h
observations from south and north in total. 
Having both CTA-S and CTA-N participating in the survey
allows one to cover several regions of particular interest (e.g. the Coma cluster, Fermi bubbles, Cen A, and
Virgo cluster)
and complete the extragalactic sky survey within the first two years of operation, which would not be possible
with one site alone. The sensitivities for the proposed shared survey for a total of 1000\,h are shown in Table~\ref{EXSURV:tab2}.
We note that with improved data analysis techniques, a better 
sensitivity than the targeted one can be achieved in the same observation time.

\subsubsection{Participation of LSTs} 
\label{lstinexgalsurvey}

Including the LSTs into the extragalactic survey has some crucial advantages, 
including:
\begin{itemize}
\item  a higher sensitivity to transients as most of them are distant sources with soft spectra (but the improvement factor is unclear and our best guess is that the number of detected transients would increase by a factor of two),
\item  the potential discovery of short-term variability for sources 
at energies below 100\,GeV, as the LSTs provide several orders of magnitude
better sensitivity than Fermi-LAT for exposures of less than one hour (see Figure~\ref{fig:sci_fermicta}), 
\item  the measurement of the low-energy lever arm in the spectra for most of the detected sources, and
\item  the detection of a factor two more Fermi-LAT known sources (this is a difference between the 25\,GeV threshold energy of the LSTs and 80\,GeV for the MSTs).
\end{itemize}
There are, however, some caveats. One of them is that including LSTs into the survey does not change the
overall sensitivity much 
because we evaluated the sensitivity above 125\,GeV, which is dominated by the MSTs.
A second point is that the LSTs have a narrower field of view than the MSTs which means
that the survey could be done more rapidly without the LSTs.
Including the LSTs also adds some non-uniformity of the exposure.
Finally, if the LSTs were excluded from the survey, they could be devoted to other programmes (done in parallel to the survey) that require a low-energy threshold 
and do not require a high sensitivity above 1~TeV. 
Nevertheless, we recommend to include the LSTs into the extragalactic survey arguing that the advantages outweigh the disadvantages.

\subsubsection{Prospects for divergent pointing}
\label{divpointing}

In the divergent pointing mode, each telescope in the array is pointed
to a location on the sky that is 
slightly offset from its neighbour in order to cover a larger portion of the sky at once.

Here we briefly summarise the results of a preliminary study based on
CTA simulations with no LSTs.
The divergent pointing pattern considered resulted in a field of view of 14
degrees in diameter with sensitivity fluctuations below a factor two across the 
field of view. 
Comparing divergent pointing and normal pointing indicates that a  sensitivity gain of roughly 
1.5 can be achieved if divergent pointing is used to do the survey.

An independent study on the prospects of the divergent pointing mode was 
made by \cite{Szanecki15} using early simulations of an array of 23 MSTs. 
The authors showed that 
the divergent mode will be significantly superior to the normal pointing mode for source
detections, i.e. it will have a superior flux sensitivity. They calculated that
by separating the telescope pointings by up to 6$^\circ$ 
(from the pointing of the inner telescopes to the outermost ones),
the needed time for a given flux sensitivity can be reduced by a factor of 2.3,
which confirmed the gain in sensitivity seen in the study discussed earlier.
As expected, however, the angular and energy reconstruction accuracy for the divergent  pointing mode is up to a factor of about two 
worse than for the normal pointing. Still, such an increase in 
flux sensitivity is very attractive for the survey, especially because of the increased
chances of observing GRBs that occur within the field of view.

The studies of the divergent pointing mode are very promising 
and will be continued with the latest simulations using 
the final array layout for the northern and southern arrays 
in order to reach a robust conclusion on the use of this mode.

\section{KSP: Transients}
\label{sec:ksp_trans}

The universe hosts a diverse population of astrophysical objects, within our Galaxy and beyond, that explode or flare up in dramatic and unpredictable fashion across the electromagnetic spectrum and over a broad range of timescales spanning milliseconds to years. Collectively designated ``transients'', many are known to be emitters of high-energy gamma rays and are also potential sources of non-photonic signals that include cosmic rays, neutrinos and/or gravitational waves (GWs). They are of great scientific interest, being associated with catastrophic events involving relativistic compact objects such as neutron stars (NSs) and black holes (BHs) that manifest the most extreme physical conditions in the universe. However, their dynamic nature has often hindered detailed observational characterization and robust physical understanding. One of the key strengths of CTA is the unprecedented sensitivity in very high-energy (VHE) gamma rays for transient phenomena and short-timescale variability \cite{Funk13}, which can revolutionise our knowledge of cosmic transients. Its relatively large field of view (FoV) is also a crucial asset in discovering transient events on its own, as well as in following up alerts of transients issued by monitoring instruments. 

In this context, we propose follow-up observations of six classes of targets (denoted A-F) triggered by external or internal alerts, together with an unbiased survey for transients utilizing divergent pointing observations (G).
\begin{spacing}{0.5}\end{spacing}
A) {\bf Gamma-ray bursts} (GRBs), based on external alerts from monitoring facilities.
Thought to be triggered by special types of stellar collapse and merger events involving NSs and/or BHs, these highly luminous and distant explosions in the universe are also one of its most mysterious phenomena, with many basic aspects still poorly understood \cite{Meszaros13, Inoue13, Kumar15}. In addition to addressing key issues regarding 
the physics of GRBs, CTA will use GRBs as probes of cosmic-ray physics, observational cosmology and fundamental physics \cite{Inoue13, Mazin13, Ellis13}.
\begin{spacing}{0.5}\end{spacing}
B) {\bf Galactic transients}, based on external alerts from monitoring facilities.
A wide range of compact objects in our Galaxy exhibit different types of jets and winds that accelerate high-energy particles in sporadic outbursts, whose production mechanisms can be greatly clarified through CTA observations \cite{Bednarek13, deOna13, Paredes13}. These include flares from pulsar wind nebulae (PWNe; relativistic outflows driven by rotating NSs) \cite{deOna13, Buehler14}, 
flares from magnetars (NSs with anomalously high magnetic fields),
jet ejection events from microquasars and other X-ray binaries (NSs or BHs accreting matter from a stellar companion) \cite{Paredes13, Dubus15}, novae (explosions on the surfaces of white dwarfs) \cite{Dubus15}, etc.
\begin{spacing}{0.5}\end{spacing}
C) {\bf X-ray, optical and radio transients}, based on alerts from ``transient factory''  facilities.
Large numbers of X-ray, optical and radio transient phenomena will be newly identified by current and upcoming transient factories capable of regularly monitoring large areas of the sky in these wavebands \cite{Obrien13}, including tidal disruption events (TDEs) \cite{Komossa15}, supernova shock breakout (SSB) events \cite{Brown15} and fast radio bursts (FRBs) \cite{Katz16}. Observing a selected sample of such alerts with CTA offers new strategies for elucidating various known types of transients, as well as the potential for discovering completely new source classes.
\begin{spacing}{0.5}\end{spacing}
D) {\bf High-energy neutrino transients}, based on alerts from neutrino observatories.
Cosmic high-energy neutrinos are clear indicators of hadronic cosmic-ray production \cite{Halzen13} and have begun to be detected by current facilities \cite{Aartsen13Ta}, although their origin is yet unclear \cite{Ahlers15}. CTA follow-up of appropriately selected alerts can determine their origin \cite{Aartsen13Td, Aartsen15a}
and can possibly give insight on extragalactic and/or Galactic cosmic rays as well.
\begin{spacing}{0.5}\end{spacing}
E) {\bf GW transients}, based on alerts from GW observatories.
GWs are most prominently expected from cosmic transients and were directly detected for the first time from binary BH merger events\,\cite{Abbott16a, Abbott16e} without any clear evidence of associated electromagnetic signals \cite{Abbott16c} (see however \cite{Connaughton16}). More GW detections are expected in the coming years, including those of NS mergers accompanied by electromagnetic 
emission \cite{Fernandez15}, albeit often with large localisation uncertainties. Follow-up by CTA with suitable strategies can play a unique and essential role for identifying and understanding their sources \cite{Doro13, Bartos14}.
\begin{spacing}{0.5}\end{spacing}
F) {\bf Serendipitous VHE transients}, identified via the CTA
real-time analysis (RTA) during scheduled CTA observations.
The RTA can recognize new transients or flaring states of known sources at 
very high energies anywhere in the FoV and automatically issue alerts within 30 sec \cite{Bulgarelli15, Fioretti15}. As with transient factory events, follow-up of a selected sample will greatly advance studies of known and unknown transients.
\begin{spacing}{0.5}\end{spacing}
G) {\bf VHE transient survey}, utilizing divergent pointing and in conjunction with the CTA Extragalactic Survey KSP (Chapter~\ref{sec:ksp_eg}).
As a novel capability of CTA, observations in divergent pointing mode covering a large instantaneous FoV could offer not only more efficient surveying of the extragalactic sky \cite{Dubus13, Szanecki15, Gerard15}, but also unique prospects for a VHE transient survey not biased by alerts. The potential discovery space includes detection of GRBs from their onset and consequently improved tests of Lorentz invariance violation (LIV), searches for new classes of VHE transients, and simultaneous multi-wavelength (MWL) and/or multi-messenger (MM) studies with other wide FoV facilities of short-duration transients such as SSBs and FRBs.

Key science questions addressed by this KSP include:

$\bullet$ What are the physical mechanisms that drive jets and winds around neutron stars and black holes?

$\bullet$ What are the physical mechanisms that drive GRBs, the most luminous explosions in the universe?

$\bullet$ What is the origin of the ultra-high energy cosmic rays (UHECRs), the highest energy particles in the universe?

$\bullet$ What is the origin of the recently discovered cosmic high-energy neutrinos?

$\bullet$ What is the origin of the recently discovered fast radio bursts?

$\bullet$ What are the sources of GWs and the physical mechanisms that drive them?

$\bullet$ Is Einstein's theory of special relativity correct?

$\bullet$ Are there unknown types of explosive phenomena in the universe?

{\bf Validity as a key science project (KSP):}

\noindent Suitability as a KSP is clear for all targets, as the observing programme proposed here demands:

- expertise of the CTA Consortium to optimize:
i) for A-E, the response to external alerts and MWL/MM coordination with other large-scale collaborations, often bound by memoranda of understanding (MoUs), ii) for F, internal data communication, and iii) for G, the array pointing mode and coordination with the extragalactic survey strategy,

- starting observations using partial arrays, for A-F, and

- persistent, dedicated efforts for achieving positive detections for A-E and G, and possibly F as well.

{\bf Observing strategy and required time:}

\noindent The strategies and required times 
are given below for each target class, to be considered provisional guidelines subject to revision in the light of actual findings. At present, we anticipate aggregate observing times of 390 h/yr/site for the two-year early phase, 125 h/yr/site for the first two years of full-array operation, and 95 h/yr/site from the 3rd year onwards.

{\bf A) Gamma-ray bursts:} We propose that all alerts during dark time with zenith angle less than 70 deg be followed up by the full array for 2 hours each; if a positive detection is achieved, the
observations are 
extended for as long as the target remains detectable. Together with some high-energy alerts not promptly accessible by CTA, the total estimated observing time is 50 h/yr/site, to be distributed equally for each site and each year, starting from the operation of the first Large-Sized
Telescopes (LSTs).

{\bf B) Galactic transients:} 
Different observing strategies (e.g. trigger criteria, observing times, site requirements, etc.)
are warranted depending on the type of object, as summarised and prioritised. In all, 150 h/yr/site is proposed for the early phase, plus 30 h/yr/site for the first two years of full operation. Further continuation is contingent on the discovery of new sources with fast variability, except for automatic follow-up of magnetar giant flares as part of A).

{\bf C) X-ray, optical and radio transients:} Their follow-up hold great scientific promise but currently involves various unknowns that preclude the determination of explicit strategies. Nevertheless, to conduct exploratory science along with requisite tests of the alert system, 50 and 10 h/yr/site are proposed for the early and full phases, respectively.

{\bf D) High-energy neutrino transients:} Follow-up is proposed for alerts from IceCube and other high-energy neutrino observatories of candidate muon neutrino-induced track events, either multiplets of events that arrive sufficiently close together in time and sky position or single, well-reconstructed events that can be identified as neutrinos with high confidence. The observing time should not exceed 20 h/yr/site for the early phase and 5-10 h/yr/site for the full phase.

{\bf E) GW transients:} Follow-up of low-latency GW alerts can entail covering large areas of the sky via tiling or divergent pointing for a modest number of expected events, albeit subject to sizable uncertainties. As with neutrinos, we propose 20 h/yr/site for the early phase and 5-10 h/yr/site afterwards.

{\bf F) Serendipitous VHE transients:} Unpredictable by definition, they constitute important exploratory targets whose follow-up prospects depend on the performance of the RTA. Accounting for different possibilities and tests of the system, we propose 100 and 25 h/yr/site for the early and full phases, respectively.

{\bf G) VHE transient survey:} To be performed via divergent pointing and concurrently with parts of the extragalactic survey, and the associated observing time will be accounted for in the Extragalactic Survey KSP. Detailed plans for its implementation are to be decided after more comprehensive Monte Carlo predictions for the expected performance become available.

\begin{table*}
\begin{center}
\caption[]{Summary table of proposed maximum observation times for follow-up targets in the Transients KSP. Observations of Galactic transients could be extended beyond Year 3 of regular operations if new source classes with fast variability are discovered. The early phase, prior to array completion, is assumed to last for two years.}
\label{table:obstimes}
\smallskip
\begin{tabular}{llllll}
\hline \hline
			& \multicolumn{5}{c}{Observation times (h~yr$^{-1}$~site$^{-1}$)} \\
Priority & Target class		& Early phase	& Years 1--2	& Years 3--10 & Years 1--10 \\
\hline
1 & GW transients					& 20			& 5			& 5		& \\
2 & HE neutrino transients				& 20			& 5			& 5		& \\
3 & Serendipitous VHE transients		& 100		& 25			& 25		& \\
4 & GRBs					& 50			& 50			& 50		& \\
5 & X-ray/optical/radio transients		& 50			& 10			& 10		& \\
6 & Galactic transients				& 150		& 30			& 0(?)	& \\
\hline
& Total per site (h~yr$^{-1}$~site$^{-1}$)	& 390	& 125	& 95		& \\
& Total both sites (h~yr$^{-1}$)		& 780	& 250	& 190	& \\
& Total in different CTA phases (h)		& 1560	& 500 	& 1520	& 2020 \\
\hline
\end{tabular}
\end{center}
\end{table*}

{\bf Key data products:}

\noindent Spectra and light curves for each positively detected object on timescales depending on the target class, plus upper limits for a fraction of alerts for A-E. For the most part, data rights can follow the proposed protocol of being proprietary for one year, but selected information should be communicated rapidly in the form of Gamma-ray burst Coordinate Network (GCN) notices, Astronomer's Telegrams, IAU circulars, etc. to ensure MWL/MM follow-up.

\vspace{0.7cm}

{\bf Summary of simulations:}

\begin{figure}[!htb]
\begin{centering}
\includegraphics[width=0.84\textwidth]{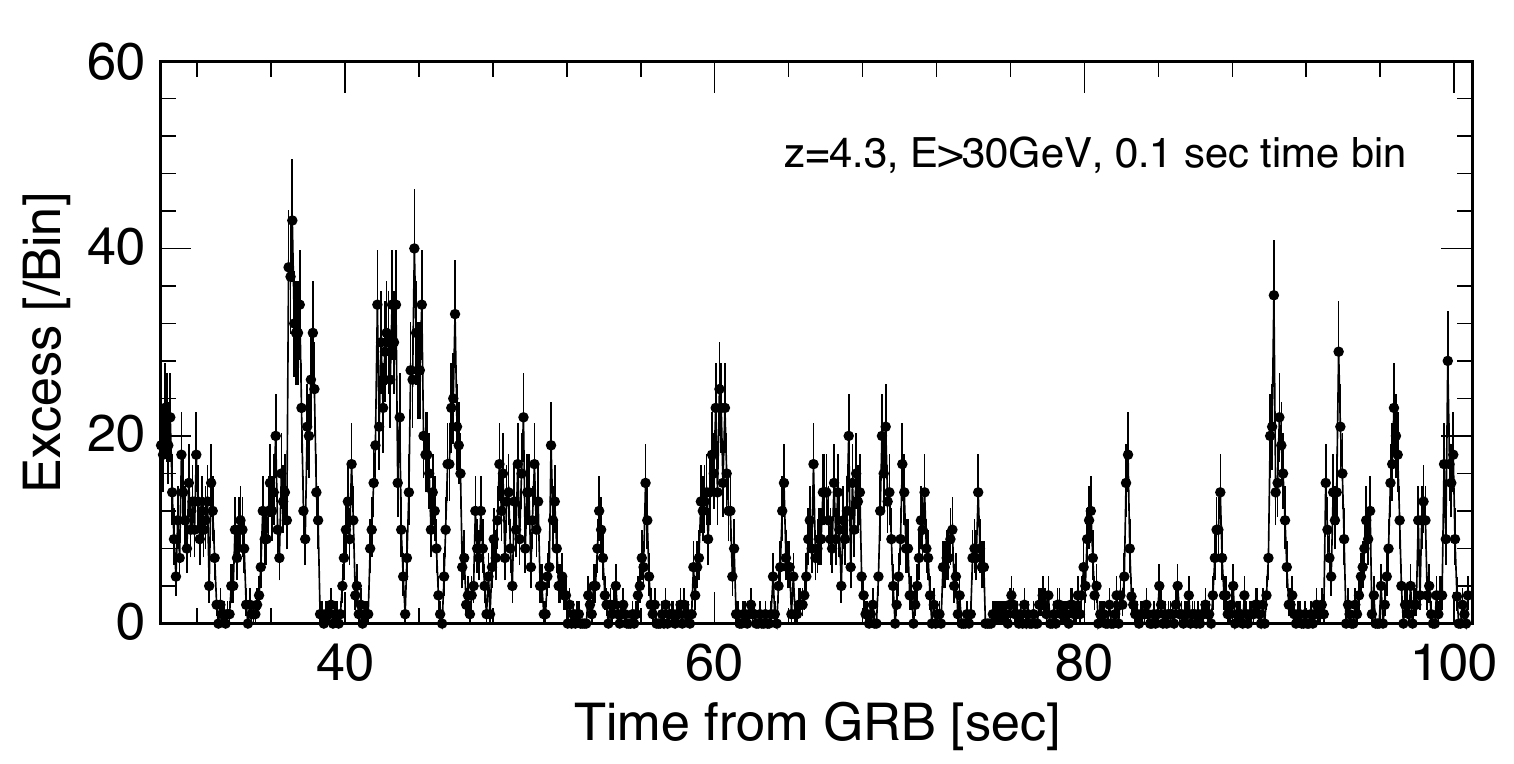}
\caption{
Simulated CTA light curve of GRB 080916C at $z$=4.3, for observed photon energies above 30 GeV with 0.1 sec time binning and plotted from $t_0=\,$30 seconds after burst onset. The assumed template is the measured Fermi-LAT light curve above 0.1 GeV for this burst, extrapolating the intrinsic spectra to very high energies with power-law indices as determined by Fermi-LAT in selected time intervals~\cite{Abdo09a} and taking the EBL model of Ref. \cite{Finke10}. For more details, see Ref.~\cite{Inoue13}.
\label{fig:GRBlightcurve}
}
\end{centering}
\end{figure}

\noindent Detailed simulations demonstrate that CTA detections of bright GRBs allow measurements of their VHE light curves (Figure~\ref{fig:GRBlightcurve}) and spectra (Figure~\ref{fig:GRBspectra}) in unprecedented detail, from which invaluable information is expected concerning radiation mechanisms, hadronic cosmic-ray signatures, constraints relevant to cosmology and fundamental physics, etc. Simulated observations of PWN flares (Figure~\ref{fig:crabflare}) and X-ray binary jet outbursts (Figure~\ref{fig:cygflare}) exemplify the power of CTA for probing the pertinent mechanisms of emission and particle acceleration.

\subsection{Science Targeted}
\label{sec:ksp_trans_sci}

\subsubsection{Scientific Objectives}
\label{sec:ksp_trans_sci_obj}

{\bf A) Gamma-ray bursts:}
The most luminous cosmic explosions after the Big Bang are also one of the most enigmatic classes of transients~\cite{Meszaros13, Inoue13, Kumar15}. Phenomenologically, they are defined by a ``prompt'' emission phase that is prominent in the MeV band with durations in the range 
$T_{90} \sim\,$0.01--1000~s and rapid, irregular variability, followed by an ``afterglow'' phase with emission that decays gradually over hours to weeks or longer and spans all wavebands from radio to high-energy gamma rays. Two populations can be distinguished in the distributions of duration and prompt spectra: ``long, soft GRBs'' with $T_{90} \gtrsim$ 2 s and ``short, hard GRBs'' with $T_{90} \lesssim$ 2 s \cite{Kouveliotou93}. Both are likely to originate from ultra-relativistic jets, the former triggered by certain types of stellar core collapse events that form NSs or BHs, and the latter widely discussed to be triggered by merger events of NS-NS or NS-BH binaries. However, many of the basic physical properties of GRBs remain poorly understood, such as the nature of the central engine and the mechanisms of jet formation, particle acceleration and radiation, particularly for the prompt and early afterglow phases. Known to occur at cosmological distances (median redshifts $\bar{z} \sim 2$ and $\bar{z} \sim 0.5$ for long and short GRBs, respectively), they can also serve as important probes of the extragalactic background light (EBL), intergalactic magnetic fields (IGMF)~\cite{Inoue13, Mazin13}, and the fundamental nature of space-time~\cite{Ellis13, Doro13}. They are also one of the leading candidates for the currently mysterious sources of UHECRs, the highest energy particles known to exist in the universe~\cite{Halzen13}.

GeV-TeV gamma rays provide crucial insight into all of these issues. Specific science goals that can be addressed by CTA observations of GRBs include:

$\bullet$ determining the velocity of the jet and location of the emission site via intrinsic 
gamma-gamma absorption features and short timescale variability,

$\bullet$ determining the mechanisms of particle acceleration and radiation for the prompt emission via time-resolved, broadband spectra,

$\bullet$ determining the mechanisms of particle acceleration and radiation for the early afterglow emission,

$\bullet$ testing the GRB origin of UHECRs by revealing hadronic gamma-ray signatures in time-resolved spectra,

$\bullet$ clarifying the global evolution of stars and supermassive black holes (SMBHs) in the universe via gamma-gamma attenuation features due to the EBL over a large range of redshifts, potentially beyond the reach of active galactic nuclei (AGN) at $z>2$~\cite{Sol13}, and

$\bullet$ testing Lorentz invariance violation (LIV) with high precision via energy dependence of photon arrival times.

Thus, outstanding contributions can be expected not only to the astrophysics of GRBs, but also for cosmic-ray physics, observational cosmology and fundamental physics. For more details, see 
Section~\ref{sec:ksp_trans_ret}A and \cite{Inoue13}.

Note that automated follow-up of GRBs by CTA can occasionally hit upon completely different types of transients such as giant flares of magnetars~\cite{Mereghetti15} (see B), magnetar-like behaviour of gamma-ray binaries~\cite{Torres11}, relativistic SSB events~\cite{Brown15} or relativistic TDEs~\cite{Komossa15} (see E), benefiting the studies of such source classes as well.

{\bf B) Galactic transients:}
Various types of compact objects in our Galaxy are seen to produce jets, winds or other kinds of outflows and then convert their energy into high-energy particles and radiation, whose physical mechanisms are often not well understood. Although certain classes of such objects are persistent and/or periodically variable emitters, others, listed below, are transients that exhibit irregular and unpredictable variability in different wavebands. Some of the latter are known high-energy
(HE; GeV band) gamma-ray sources, but none have been clearly detected so far in VHE gamma rays. CTA observations of these Galactic transients are expected to provide new, critical insight into the underlying physical processes \cite{Bednarek13, deOna13, Paredes13}.

$\bullet$ Pulsar wind nebula (PWN) flares:
PWNe are bubbles of relativistic plasma that are energised by magnetically-driven winds emanating from rotating NSs. While many are observed as steady sources spanning the radio to VHE bands, the unexpected discovery of bright, HE gamma-ray flares from the Crab nebula have revealed that they can also be transients \cite{Tavani11, Abdo11b, Buehler14}. CTA observations can discriminate different models that have been proposed and probe the physics of energy dissipation and particle acceleration in pulsar winds \cite{Bednarek13, deOna13, Buehler14} 
(Section \ref{sec:ksp_trans_ret}B).

$\bullet$ Magnetar giant flares:
Magnetars are NSs thought to be powered by the energy of their anomalously high magnetic fields rather than their rotation \cite{Mereghetti15}. On rare occasions, they trigger giant flares, primarily in MeV photons, which are the brightest bursts of extra-solar radiation ever observed. To date, only three such flares (one each in 1979, 1998 and 2004) are known, and (to our knowledge) none have been properly observed above GeV energies, despite some predictions for HE afterglow emission \cite{Taylor06}. CTA follow-up may provide the first VHE coverage of such flares and offer fresh clues to their mysterious origin.

$\bullet$ Microquasars:
NSs or BHs activated by accreting matter from their companion stars can be classified according to their companion mass into high mass or low mass X-ray binaries (HMXBs or LMXBs). A subset of them are observed to generate collimated jets of plasma in sporadic outbursts and are called microquasars \cite{Paredes13, Dubus13b, Dubus15}. Some HMXB microquasars such as Cyg~X-3 and Cyg~X-1 are known sources of HE 
gamma rays \cite{Abdo09d, Tavani09, Zanin16} whose emission mechanisms (e.g. leptonic or hadronic) are still unclear, while some LMXB microquasars such as GRS 1915+105 have long been predicted to be gamma-ray emitters \cite{Dubus13b}. MWL observations featuring CTA will provide not only crucial tests of emission models (Section \ref{sec:ksp_trans_ret}B), but also new insight into the long-standing problem of astrophysical jet formation \cite{Paredes13, Dubus15}.

$\bullet$ Transient binary pulsars:
Some X-ray binary pulsars that display transient behaviour may contain pulsar winds interacting with material from their stellar companions and emit GeV-TeV gamma rays in a way analogous to some known, periodic gamma-ray binaries \cite{Dubus15}. These include Be/X-ray binary pulsars that show aperiodic X-ray outbursts \cite{Acciari11Ta} and transitional pulsars that switch between rotation-powered and accretion-powered phases \cite{Stappers14}. Their study with CTA will shed new light on the physics of pulsar winds, accretion onto NSs, and binary pulsar evolution.

$\bullet$ Novae:
Under certain conditions, matter accreting onto white dwarfs undergoes thermonuclear explosions and give rise to novae, with thermal optical emission lasting days to weeks. Unexpectedly, HE gamma rays were detected from several novae and interpreted as emission from particles accelerated by shocks in their expanding ejecta \cite{Ackermann14T, Dubus15}. Many uncertainties remain concerning the mechanisms of radiation and particle acceleration that can be clarified by CTA observations \cite{Dubus15, Metzger16}.

$\bullet$ Unidentified HE transients:
Wide-field instruments operating at GeV energies, such as Fermi-LAT, AGILE, DAMPE or upcoming missions,
are capable of discovering new HE transients in the Galactic plane that may not be immediately identifiable with known objects \cite{Ackermann13Ta}. Follow-up with CTA can play a crucial role in clarifying their nature or possibly unravelling new types of Galactic transients.

\vspace{0.5cm}

{\bf C) X-ray, optical and radio transients:}
Wide-field monitoring facilities operating in the X-ray to soft gamma-ray bands (hereafter simply ``X-rays'' unless indicated otherwise), as exemplified currently by Swift and INTEGRAL and to be succeeded by SVOM, have proven to be effective transient factories, capable of discovering a large variety of transient phenomena over a broad range of timescales \cite{Gehrels13}. New and upcoming optical transient factories such as GAIA, Pan-STARRS, iPTF, ZTF and LSST \cite{Kulkarni12} as well as radio transient factories such as SKA and its pathfinders LOFAR, MeerKAT, MWA, and ASKAP \cite{Fender11} (see details in Chapter~\ref{sec:sci_synergies}) guarantee a revolution in our physical understanding of the transient universe~\cite{Obrien13}. Besides GRBs and the Galactic transients mentioned above, thousands of sources of a very diverse nature are expected to be discovered, from nearby flaring stars with extreme non-thermal variability \cite{Fender15} to orphan GRB afterglows at cosmological distances \cite{Ghirlanda15}. 
Thus, following up properly selected alerts from X-ray/optical/radio transient factories with CTA will pave a new and unexplored path to study the universe in VHE gamma rays, with great promise for exciting results in widely disparate areas of astrophysics. Some particularly noteworthy classes of expected transients are as follows.

$\bullet$ Tidal disruption events (TDEs):
All sufficiently massive galaxies are expected to harbour SMBHs in their central regions. They can occasionally brighten in the optical to soft X-ray bands via a transient accretion event lasting weeks to months caused by tidal disruption of an approaching star or gas cloud \cite{Komossa15}. A fraction of them are relativistic TDEs that are transient versions of blazars, co-producing jets with prominent non-thermal radio to hard X-ray emission \cite{Bloom11, vanVelzen16}. HE/VHE emission has been predicted for such events as well as for non-relativistic TDEs \cite{Chen16}, but the predictions are yet to be critically tested. CTA follow-up observations can bring forth fresh insight into the physics of tidal disruption, as well as new perspectives on jet formation by SMBHs that are complementary to observations of AGN \cite{Sol13}.

$\bullet$ Supernova shock breakout (SSB):
Besides their emission powered by radioactive decay on timescales of weeks, early UV to X-ray flashes are expected from supernovae due to shock-heated ejecta on timescales of hours. A subset of such events found by Swift on minute-timescales involve relativistic ejecta and may (or may not) be associated with low-luminosity GRBs \cite{Campana06, Soderberg08, Brown15}. Detection by CTA of correlated VHE emission \cite{Kashiyama13}, either via rapid follow-up or via a simultaneous VHE survey, can probe particle acceleration in radiation-rich environments, supernova explosion dynamics and its connection with GRBs.

$\bullet$ Fast radio bursts (FRBs):
Radio surveys with wide FoV and high time resolution have unexpectedly uncovered GHz-frequency, ms-duration FRBs \cite{Lorimer07, Thornton13}. Apart from redshifts $z \sim 0.2-1.3$ inferred from their large dispersion measures, their origin is unclear and a plethora of models have been proposed \cite{Katz16}. MWL follow-up of one FRB alert has led to the possible identification of a radio afterglow and an elliptical host galaxy, implying a progenitor akin to NS-NS mergers \cite{Keane16b} and an intriguing potential link between FRBs, short GRBs and GW events (see A, E above). However, doubts have been raised for this particular case \cite{Williams16}. Another FRB that is so far unique in exhibiting burst repetition \cite{Spitler16} was recently localised to sufficient accuracy to be reliably associated with a dwarf host galaxy at $z \sim 0.2$ and a persistent radio counterpart \cite{Chatterjee17}, potentially pointing a young NS origin. There may possibly be more than one class of FRBs. Notwithstanding the current ambiguities, CTA follow-up of selected FRBs can test their supposed connection with young NSs or NS-NS mergers and help to solve their mysterious origin. Note that correlated, millisecond VHE bursts expected in some models \cite{Lyubarsky14} can only be realistically tested by a simultaneous radio and VHE survey (see G).

{\bf D) High-energy neutrino transients:}
Several types of transients covered in this KSP are also promising candidate sources of high-energy neutrinos, whose detection with currently operating or upcoming neutrino facilities would reveal their inner workings in novel ways that are complementary to photons and GWs. 
Follow-up of neutrino alerts can address the long-standing question of the origin of hadronic cosmic rays if they are transient sources \cite{Halzen13}. A new mystery, possibly independent of the directly observed hadronic cosmic rays, is the origin of cosmic neutrinos with TeV-PeV energies discovered by IceCube~\cite{Aartsen13Ta, Ahlers15}. While they do not appear to be correlated with bright GRBs or AGN \cite{Aartsen15a}, it is possible that they are generated by fainter GRBs/transients or flaring states of persistent objects, some of them perhaps Galactic, that may give rise to VHE gamma rays via processes concurrent with neutrino production. Spatial and temporal coincidence studies between neutrinos and gamma rays can identify their sources, and unequivocally prove hadronic cosmic-ray acceleration therein \cite{Aartsen13Td, Aartsen15a}.

{\bf E) GW transients:}
Cataclysmic transient events such as mergers of binary NSs and/or BHs are predicted to be the strongest sources of GWs. Their direct detection opens up a completely new window to probe the dynamical behaviour of relativistic compact objects, in a way complementary to photons or neutrinos. This long-awaited goal was recently achieved for BH-BH mergers by the LIGO-VIRGO collaboration \cite{Abbott16a, Abbott16e}. With the upcoming network of new-generation observatories comprising LIGO, VIRGO, KAGRA and INDIGO, many more detections are expected including NS-NS or NS-BH mergers \cite{Andersson13}. These are also leading candidates for the progenitors of short GRBs~\cite{Berger14, Fernandez15} and probable VHE gamma-ray emitters under certain conditions, the nearest of which could occur within the expected detection horizon of GW observatories~\cite{Doro13, Bartos14}. However, for their first years of operation, the localisation errors of GW events are expected to be very large, typically spanning 100-1000 deg$^2$, as was indeed the case for the first detections \cite{Abbott16c, Abbott16e}. Even after the detectors reach full performance, large localisation errors can still be the case for events near the sensitivity limit. Thanks to the relatively large FoV, especially if divergent pointing can be employed, CTA follow-up of GW candidates may offer better localisation prospects compared to other wavebands through more efficient searches over larger areas of the sky for their electromagnetic counterparts. This will provide the impetus for further, extensive MWL follow-up that may reveal the source, distance, and energetics of these events.

{\bf F) Serendipitous VHE transients:}
One of the key advantages of CTA is its high instantaneous sensitivity combined with its large FoV of several degrees. This offers the tantalizing possibility of serendipitously discovering transient VHE sources within the FoV during any observation. In this context, the RTA system can identify flaring sources in the FoV in less than 30~s from data taking, with a sensitivity at most a factor of three worse than the final analysis \cite{Fioretti15}. Therefore, the RTA system will provide serendipitous discoveries of new VHE gamma-ray sources in the FoV, which should be observed more deeply by extending the observations being conducted at any time. The RTA will also automatically issue alerts to the outside community to promote MWL follow-up. The nature of such serendipitous VHE transients can be any of those described above, or, more interestingly, of a new and unexpected class.

{\bf G) VHE transient survey:}
By virtue of the large number of its constituent telescopes, CTA offers the novel capability of observations in divergent pointing mode, whereby the telescopes (primarily MSTs) are pointed in progressively offset directions to allow simultaneous coverage of a considerably larger FoV compared to normal pointing mode \cite{Dubus13, Szanecki15, Gerard15}. While this entails some loss in energy resolution and angular resolution and some increase in energy threshold, the enlarged FoV can compensate for these losses and enable more efficient surveys for persistent point sources, a potentially crucial asset for the Extragalactic Survey (see Chapter~\ref{sec:ksp_eg}). Furthermore, it can provide enhanced prospects and unique discovery space for elucidating cosmic transients at very high energies without relying on follow-up of alerts, in particular:

$\bullet$ detection of GRBs from their onset, including the prompt phase of short GRBs and associated improvements for probing LIV,

$\bullet$ unbiased searches for VHE transients in general, including new classes of transients, and

$\bullet$ MWL and/or MM spatio-temporal correlation studies with other observatories with wide FoV and/or limited localisation capabilities, including simultaneous MWL observations of short-duration transients such as SSBs \cite{Brown15, Kashiyama13} and FRBs \cite{Keane16a, Lyubarsky14}.

\subsubsection{Context / Advance beyond State of the Art}
\label{sec:ksp_trans_sci_con}

{\bf A) Gamma-ray bursts:}
Fermi-LAT has been detecting GeV-band emission from GRBs at a rate of $\gtrsim$10 yr$^{-1}$, totalling $\sim$120 as of December 2016, revealing a rich phenomenology~\cite{Ackermann13Tb}:
i) most MeV-bright GRBs are accompanied by GeV emission whose flux is generally consistent with extrapolations of the measured MeV spectra,
ii) the GeV emission often extends up to $\sim$10-30 GeV, and occasionally to $\sim$100 GeV, with no conspicuous spectral cutoffs (with one exception \cite{Ackermann11T}),
iii) the few brightest GRBs clearly exhibit a hard spectral component that significantly exceeds simple extrapolations of the MeV spectra,
iv) GeV emission is generally observed during the prompt as well as the afterglow phase, lasting up to a few thousand seconds, and occasionally to $\sim$1 day,
v) GeV emission is seen in both long and short GRBs, and
vi) current observations are not inconsistent with the majority of GRBs possessing such high-energy emission, including fainter GRBs (although $\sim$20 \% of MeV-bright GRBs show indirect evidence for some sort of spectral break between MeV and GeV \cite{Ackermann12T}).

These observations, particularly of photons at a few GeV and above, have led to some important physical insight, including significantly higher velocities of the emission zone than had been inferred previously~\cite{Abdo09a}, evidence for non-trivial mechanisms of emission and/or particle acceleration during the early afterglow~\cite{Ackermann14G, Kouveliotou13}, constraints on the EBL at the highest redshifts so far~\cite{Atwood13}, and the strongest limits to date on the violation of Lorentz invariance~\cite{Abdo09b, Vasileiou13}. Nevertheless, in many cases, the limited photon statistics achievable with Fermi-LAT at tens of GeV have prevented firm conclusions, and many competing theoretical models remain viable. Imaging atmospheric Cherenkov telescope (IACT) observations of GRBs can potentially provide significant progress through much higher photon statistics (see Figure \ref{fig:GRBlightcurve}, Figure \ref{fig:GRBspectra} and
Section~\ref{sec:ksp_trans_ret}A for more details). However, the sensitivity and energy threshold of current IACTs imply a detection rate that is considerably less than $\sim$1 GRB yr$^{-1}$, and no GRB has been clearly detected so far despite efforts over the last decade~\cite{Acciari11Tb, Aleksic14t, Abramowski14Ta}. CTA can lead to a major breakthrough by detecting GRBs at an appreciable rate of $\gtrsim$1 yr$^{-1}$ with far superior photon statistics compared to Fermi-LAT in the pivotal energy range above 10 GeV~\cite{Inoue13, Gilmore13, Kakuwa12}.

HAWC may be able to detect GRBs at a rate of $\sim$1 yr$^{-1}$ without external alerts from their onset, albeit with much higher energy threshold and lower sensitivity \cite{Abeysekara12, Taboada14}.  CTA follow-up offers significant advantages with more sensitive observations down to considerably lower energies for studying long GRBs and the afterglows of short GRBs, the latter of which will also be key for identifying GW sources (see D below). Even the prompt emission of short GRBs may be accessible to CTA if they can be caught during a survey with divergent pointing, with particularly interesting implications for testing LIV (see G below). On the other hand, an intriguing prospect is follow-up of GRBs detected and alerted by HAWC, which would be almost guaranteed to be detectable by CTA as long as the alert is fast enough.

{\bf B) Galactic transients:}
None of our proposed target classes have been unambiguously detected as transients at very high energies. Those observed by Fermi-LAT and/or AGILE as transient sources of HE gamma rays on timescales of hours to days include PWN flares from the Crab nebula \cite{Tavani11, Abdo11b}, the HMXB microquasars Cyg~X-3 \cite{Abdo09d, Tavani09} and Cyg~X-1 \cite{Malyshev13, Bodaghee13, Zanin16}, several novae \cite{Ackermann14T}, and the transitional pulsar PSR~J1023+0038 \cite{Stappers14}. A few unidentified HE transients have also been seen at low Galactic latitudes, although some, or all, of them may be extragalactic \cite{Ackermann13Ta}. No signals have been seen yet in IACT observations of Crab nebula flares \cite{Mariotti10, Ong10, Abramowski14Tb}, Cyg~X-3 \cite{Aleksic10T} or novae \cite{Ahnen15} (see however tentative evidence for Cyg~X-1 \cite{Albert07}). Except for some novae, current IACT upper limits are insufficient to constrain spectral turnovers between the HE and VHE bands, so more sensitive observations with CTA are warranted to characterise the maximum energy and nature (leptonic or hadronic) of the radiating particles.

Despite high expectations on both theoretical and empirical grounds \cite{Dubus13b}, LMXB microquasars such as GRS~1915+105 have so far not been detected above GeV energies \cite{Bodaghee13}. Nor have transient Be/X-ray binary pulsars such as 1A~0535+262 \cite{Acciari11Ta}. To our knowledge, no proper HE or VHE observations have been conducted yet for the rare giant flares from magnetars \cite{Taylor06}. CTA observations of these objects may allow the first detailed exploration of their behaviour in the gamma-ray domain. 

HAWC may be able to detect Crab nebula flares at multi-TeV energies ~\cite{Abeysekara13} and possibly provide alerts to CTA for more sensitive follow-up observations. However, other types of transients may be more difficult for HAWC, especially if they have VHE spectral breaks that are expected for some objects. Through detailed coverage of the highest energy electromagnetic window, MWL studies of Galactic transients featuring CTA could clarify the relevant mechanisms of particle acceleration and radiation, as well as the associated physical conditions concerning magnetic fields, matter and radiation density.

{\bf C) X-ray, optical, and radio transients:}
So far, dozens of candidate TDEs have been observed in the optical and/or X-ray bands, but only a handful of them have been identified as relativistic TDEs with jets 
\cite{Komossa15,Bloom11,vanVelzen16}. 
Although searches for a few such objects in the GeV \cite{Bloom11, Peng16b} and TeV 
\cite{Aliu11T, Aleksic13T} bands have been negative, systematic and comprehensive studies of TDEs at these energies have yet to be conducted. Only a small number of SSB events have been discovered to date in the optical and/or X-ray bands \cite{Brown15}, with proper coverage lacking above GeV. The situation is expected to improve with new optical and X-ray transient factories that can detect and provide alerts for many more TDEs and SSBs by the time of full CTA operation.

Real-time alerts of FRBs have begun to be issued only recently, with typical latency of a few hours \cite{Keane16b}. A follow-up observation was conducted $\sim$15 hours afterwards by H.E.S.S., yielding the first upper limits on VHE afterglow emission from an FRB \cite{Abdalla17}. Upcoming radio facilities with improved capabilities for FRB identification should allow more efficient follow-up for a larger number of events, critically constraining FRB models and helping to solve their mysterious origin.

More generally, new types of X-ray, radio and optical transients are likely to be discovered by transient factories that will come online during the following years; these are not yet providing alerts for current IACTs (although some tests have already been conducted, e.g. LOFAR-MAGIC). The follow-up of sources triggered by transient factories will basically be a MWL effort, where CTA will play a key role in elucidating the high-energy end of the electromagnetic spectrum, significantly improving upon the capabilities of Fermi-LAT, AGILE and/or HAWC for the detection of fast transient sources.

{\bf D) High-energy neutrino transients:}
The IceCube observatory reported the first evidence for extraterrestrial high-energy neutrinos in the energy range of 30 TeV -- 2 PeV, with a diffuse flux significantly in excess of the expected atmospheric neutrino background ~\cite{Aartsen13Ta, Ahlers15}. Their sky distribution is consistent with isotropy, and no significant correlations with known astrophysical objects have been found so far.  The lack of a correlation 
is not surprising as the majority of the detections are cascade-type events with large positional uncertainties of $\sim$10 deg. The neutrinos are likely produced in inelastic interactions between hadrons accelerated to $>$PeV energies and ambient low-energy photons and/or hadrons. VHE photons are inevitably co-produced in these interactions and may be observable as long as they escape the source and propagate unattenuated. Detection of such components would not only provide critical means to identify and elucidate the neutrino sources, but also greatly contribute to solving the long-standing puzzle of the origin of UHECRs and/or Galactic cosmic rays. If the sources are transient, follow-up by CTA of appropriate alerts from neutrino facilities can achieve much better sensitivity compared to Fermi-LAT or HAWC~\cite{Funk13}.

MWL follow-up programmes of neutrino alerts from IceCube and ANTARES have been carried out by telescopes in the optical, X-ray and VHE gamma-ray bands, including MAGIC \cite{Aartsen15a}, VERITAS \cite{Santander15}, and H.E.S.S. \cite{Schuessler15}, albeit with no positive detections to date. ANTARES alerts rely on real-time event reconstruction of single, upgoing candidates, whose detection threshold is tuned to give acceptable rates for optical or X-ray follow up \cite{AdrianMartinez16a}. The first IceCube alert program, in operation since 2011, has been based on a pre-defined list of sources, mostly AGN, when an increase in the rate of upgoing neutrino candidate events (multiplets or clusters) is seen at a source position above a certain threshold, which is adjusted to yield a tolerable false alert rate of a few per year \cite{Aartsen13Td, Aartsen16}. Since 2016, new alert channels were activated based on real-time, online event reconstruction that identifies single-muon neutrino-induced track events with high confidence and localisation accuracies of order of one degree or better, caused by either extremely high-energy through-going events \cite{Aartsen13Td} or high-energy starting events with contained interaction vertices \cite{Aartsen15a}. Under consideration is an unbiased, full-sky scan for upgoing multiplets without relying on a pre-defined source list. Availability of an equally sensitive installation in the northern hemisphere (e.g. KM3NeT~\cite{KM3NeT} and GVD~\cite{GVD}) is still uncertain but would increase the sensitivity towards the southern sky and permit lower energy thresholds for Galactic sources. A substantial expansion of IceCube is also being proposed for the future \cite{Aartsen15b}.

{\bf E) GW transients:}
The momentous discovery of GW150914 by LIGO-VIRGO heralded the birth of GW astronomy \cite{Abbott16a}. Quite surprisingly, this was a BH-BH merger event \cite{Abbott16b}, in contrast to most predictions that the first GW detections would be coalescence events of NS-NS or NS-BH binaries. The event was followed up by numerous facilities covering the entire electromagnetic spectrum \cite{Abbott16c}, as well as HE neutrinos \cite{AdrianMartinez16b}, with almost all follow-ups yielding null results, as had been expected for BH-BH mergers. However, the report of a low-significance, time-coincident signal by Fermi-GBM \cite{Connaughton16} has provoked a number of new theoretical suggestions for potential electromagnetic signals from binary BH mergers (see, e.g., \cite{Perna16}) which must be tested through further observations.

Eventual GW detections of NS-NS or NS-BH mergers are also highly anticipated \cite{Andersson13}, as are associated electromagnetic signals \cite{Fernandez15}. These events are also the most promising progenitors of short GRBs, from which VHE afterglow emission is predicted for the fraction of events with their GRB-emitting jets pointing close to our line of sight~\cite{Abdo09b}, in contrast to the GW emission that should be nearly isotropic. Although GW alerts are expected to have large localisation errors with areas having non-trivial topology and spanning 100-1000 deg$^2$, especially for the first few years of LIGO-VIRGO, follow-up should still be manageable with CTA through either tiling or divergent pointing strategies \cite{Bartos14}. 

In view of the difficulty of wide-field instruments at other wavebands in effectively covering sky regions as large as 1000 deg$^2$ and the relative brightness of the high-energy afterglow from short GRBs compared to its optical counterpart, CTA may provide the best localisation prospects and enable subsequent MWL follow-up of a manageable region of the sky. Compact binary mergers can also eject a significant amount of fast material in directions away from the GRB jet axis \cite{Lehner14}. The interaction of that material with ambient media may possibly induce additional high-energy emission, even for non-GRB, off-axis GW events \cite{Kyutoku14}. The detection of such components will provide unique information concerning the dynamics of these events, complementing potential electromagnetic counterparts at other wavelengths (e.g. optical/infrared ``kilonovae''~\cite{Tanvir13}). Follow-up is therefore warranted for all GW alerts that are accessible to CTA.

{\bf F) Serendipitous VHE transients:}
To our knowledge, there has not been any new VHE transient that occurred serendipitously in the FoV during observations by existing IACTs. A systematic search in the Fermi-LAT data for variable sources on weekly timescales has identified numerous flaring sources, most of which can be associated with blazars and with no strong evidence for new Galactic transients beyond novae and the Crab nebula \cite{Ackermann13Ta}. HAWC may discover VHE transients without external alerts by virtue of its wide FoV and high duty cycle, particularly GRBs and PWN flares (see A, B above). If alerts for such HAWC detections are available sufficiently rapidly, they can be followed up by CTA with its higher sensitivity and lower energy threshold to obtain much more information.

{\bf G) VHE transient survey:}
Divergent pointing observations by CTA will be an unprecedented endeavour for IACTs. It will offer the prospects for an unbiased survey of GRBs and other transients at very high energies, beyond the energy range of Fermi-LAT and at much better sensitivity and lower energy threshold compared to HAWC \cite{Dubus13, Szanecki15, Gerard15}. This capability will be of special importance in the currently burgeoning era of time domain astronomy, with numerous other facilities already operating, or soon coming online, that are geared to surveying large areas of the sky for transient phenomena, particularly in the optical and radio bands (Chapter~\ref{sec:sci_synergies}). An intriguing possibility is coordinated, simultaneous MWL observations of the same regions of the sky by CTA and such facilities, which could enable the first comprehensive studies of transients with short durations.

\subsection{Strategy}
\label{sec:ksp_trans_strat}

It is highly desirable to begin transient follow-up observations with CTA as soon as possible, including with partial arrays during construction. 
An early start is warranted due to the infrequent nature and uncertain flux of the transients, for example one should not miss a once per decade transient event due to incomplete telescope array commissioning and verification. Such observations are very suitable for early high-risk execution (unlike for example, precision measurements of known systems) and can be effectively managed by the CTA Consortium due to its extensive scientific and technical knowledge. It may be necessary to invest a significant amount of time before a successful detection, and trigger conditions will need to be modified based on experience. All of this makes the transient programme well suited to execution as a Key Science Project.

Described below are the observing strategies for each target class. In case follow-up demands arise simultaneously for different classes, we prioritize those that are expected to be rarer, 
less time consuming, and of potentially higher scientific impact, in the following order: 1. GW transients, 2. HE neutrino transients, 3. serendipitous VHE transients, 4. GRBs, 5. X-ray/optical/radio transients, and 6. Galactic transients. The proposed strategies and prioritization are provisional and subject to change depending on what is actually observed and how the fields evolve scientifically.

{\bf A) Gamma-ray bursts:}
Alerts during operation of CTA are expected primarily from soft gamma-ray instruments such as Swift, Fermi-GBM and SVOM, the latter planned to be launched no later than 2021~\cite{Godet14}. Additional, albeit rarer, alerts can come from wide-field instruments in high-energy gamma rays such as Fermi-LAT, DAMPE, HAWC, and LHAASO, and possibly also from instruments in other wavebands such as GAIA, LSST, etc. As cosmologically distant objects, they should be uniformly distributed across the entire sky with equal rates for the CTA southern and northern sites. In view of EBL attenuation typically expected at higher energies, the LSTs will be vital for follow-up, having the lowest energy threshold and the fastest slewing capabilities of the three CTA telescope types. Nonetheless, the full array including MSTs and SSTs should be slewed to maximize the sensitivity at all energies. The full array will be particularly important if the redshift turns out to be $z \lesssim 1$ (see, e.g., Ref,~\cite{Ackermann14G}), in which case the detection of even multi-TeV photons may be feasible for a bright event \cite{Inoue13}.

We propose the following strategy for GRB observations, as summarized in Table~\ref{table:GRB}:

1. Prompt follow-up by the full array of all `accessible' GRB alerts, i.e. those occurring during
dark time and having zenith angles less than 70 degrees,
with exposure of 2 hours for each alert. The expected alert rates are $\sim$5/yr/site for Swift or SVOM and $\sim$10/yr/site for Fermi-GBM~\cite{Gilmore13, Kakuwa12, Inoue13}, totaling $\sim$12/yr/site when accounting for some overlap. 
For Fermi-GBM alerts with localisation errors larger than the LST FoV, some form of scanning or multiple-pointing (tiling) observation may be advantageous~\cite{Finnegan11}, whereas the alternative possibility of employing divergent pointing of LSTs has been deemed less effective from preliminary studies. 
All available telescopes, particularly all LSTs, should always be employed to guarantee maximum sensitivity at the lowest energies, as the detailed properties of GRBs can vary greatly from burst to burst.

2. Extended observations for detected GRBs with the full array. The RTA system will clarify whether VHE photons are detected with a latency of 30 sec, in which case the observation should continue for as long as the target is visible and detectable. The 
predicted GRB detection rates are of order $\sim$1/yr/site~\cite{Gilmore13, Kakuwa12}.

3. Late-time follow-up of high-energy GRBs not accessible promptly with the full array. Cases may be expected where a bright GRB is detected by Fermi-LAT, DAMPE, HAWC or LHAASO but is not accessible immediately to CTA by occurring on the other side of the Earth at trigger time. These should be followed up as soon as the target becomes visible, for which we estimate a rate of $\sim$1/yr/site.

\begin{table*}
\begin{center}
\caption[]{Summary of GRB follow-up strategy and observing time for one array site. The 
numbers are equal for the CTA-South and CTA-North sites.}
\label{table:GRB}
\smallskip
\begin{tabular}{@{~~}l@{~~}l@{~~}l@{~~}l@{}}
\hline \hline
Strategy							& Expected event	& Exposure per		& Exposure per\\
								& ~rate (yr$^{-1}$)	& ~follow-up (h)	& ~year (h yr$^{-1}$)\\ 
\hline
Prompt follow-up of accessible alerts	& $\sim$12		& 2				& 25\\
Extended follow-up for detections		& 0.5--1.5			& 10--15			& 10--15\\
Late-time follow-up of HE GRBs		& $\sim$1			& 10				& 10\\
~not accessible promptly				& & &\\
\hline
\end{tabular}
\end{center}
\end{table*}

Some important caveats concerning the detection rate predictions quoted above are discussed in Section~\ref{sec:ksp_trans_ret}A. Note that additional follow-up observations during partial moon time is feasible and can increase the detection rate by $\sim 50$\%. Analysis of GRB data may be amenable to looser selection
cuts for background rejection and correspondingly lower energy thresholds compared to standard analysis criteria, allowing access to spectral regions less affected by EBL absorption. On the other hand, some data may be taken while the telescope is slewing; 
such data will need to be carefully calibrated on the basis of a good understanding of the behaviour of the telescopes under such conditions.

Contemporaneous MWL follow-up with available facilities covering the entire electromagnetic spectrum should be utilized whenever possible to obtain complementary data. A prime motivation is good characterization of the time-dependent, broadband spectra of the afterglow, in order to pin down key physical quantities such the total burst energy, jet collimation angle, 
and ambient density. 
Of utmost importance is the determination of the GRB redshift by optical/infrared telescopes, the failure of which would seriously compromise the science return, especially in view of the modest expected detection rates. For this purpose, in addition to maximal cooperation with external observers, it is desirable to have an on-site telescope dedicated for CTA follow-up, which can localise a majority of the afterglows to sufficient accuracy so that larger telescopes can be alerted for spectroscopic follow-up.  
Infrared coverage is preferable to cope with optically dark GRBs that constitute a large fraction of all afterglows~\cite{Greiner11}, but may only be viable with external telescopes in view of cost limitations. 

The demand for ground-based localisation may become somewhat less urgent once SVOM is launched, whose on-board optical telescope with red coverage can localise up to $\sim$70\% of afterglows by itself. Note that the probability of GRB redshift determination is currently higher in the north than the south by $\sim$50\%, thanks to the prevalence of telescopes and astronomical communities available for follow-up~\cite{Bagoly09}. 
GRBs may also be detected by GW or neutrino observatories (Sections~\ref{sec:ksp_trans_sci_obj}C and \ref{sec:ksp_trans_sci_obj}D). 
Low significance (sub-threshold) detections may be useful for coincident searches among multiple observatories, and thus 
CTA should participate in relevant networks such as the Astrophysical Multi-messenger Observatory Network (AMON)~\footnote{http://amon.gravity.psu.edu/index.shtml}.

{\bf B) Galactic transients:}
The targets proposed for this KSP are listed in Table~\ref{table:galactic}. We have tentatively provided prioritization among them, in case multiple source classes happen to be active at the same time.
Magnetar giant flares are placed first due to their rarity, short duration, extraordinary properties and lack of precedent gamma-ray coverage. HE-detected source classes follow, roughly in order of their typical duration of activity: PWN (Crab) flares, HMXB microquasars (Cyg~X-3 and Cyg~X-1), and unidentified HE transients. LMXB microquasars are placed next in view of their high expectations despite not yet being clearly detected at high energies (see however \cite{Loh16}). Novae and transitional pulsars are given lower priority on account of their expected spectral turnovers above the HE band, followed by Be/X-ray binary pulsars. 
We emphasize that this ordering is provisional and is open to change in response to what is actually observed and how the field develops during the course of this KSP.

Since we are exploring the unknown for all of these sources, trigger criteria are not provided in detail here and will be determined based 
on the actual evolution of the transient phenomena. Approximate trigger rates for CTA follow-up, their urgency and duration of activity are given, together with the total number of hours of observation per transient episode and sites to be used. For sources with priority rank 5 to 8 we propose to trigger only in 2--3 cases with good enough sensitivity. A total observation time of 150 h/yr/site is estimated for Galactic transients during the early phases of CTA, and of 30 h/yr/site during the first two years of regular operations. Extension of this observation time into years 3 to 10 of full operations is contingent on the discovery of new Galactic source classes that show fast variability. The general requirements for observations will be zenith angles below 50$^\circ$ (70$^\circ$ for PWN flares) and any weather conditions. 

\begin{sidewaystable}
\begin{small}
\begin{center}
\caption[]{Summary table of Galactic transients proposed within the Transient KSP during the early phase of CTA. The codes for the last column are:
S (south), N (north, A (any), and B (both, if possible).}
\label{table:galactic}
\smallskip
\begin{tabular}{@{}l@{~~}l@{~~}l@{~~}l@{~~}l@{~~}l@{~~}l@{~~}l@{~~}l@{~~}l@{}}
\hline \hline
Follow-up	& Target class	& Detected	& Trigger		& Rate	& Urgency	& Activity			& Obs. time (h)	& Total	& Site \\
\ priority	&			& \ @ HE		&			& \ (yr$^{-1}$)	&		& \ duration		& \ /night   	& \ time (h)	& \\
\hline
1 & Magnetar giant flares			& --	& MeV		& 0.1	& 1 min	& 1--2 d			& Max. 1	& 10   & A/B \\
2 & PWN flares: Crab nebula  			& Y	& HE			& 1		& 1 d	& 5--20 d (HE)		& 4 		& 50   & S\&N \\
3 & HMXB microquasars: Cyg~X-3	& Y	& HE/X-ray	& 0.5	& 1 d	& 50--70 d (HE)	& Max. 1	& 50   & N \\
   &			                Cyg~X-1	& Y	& HE/X-ray	& 0.2	& 1 d	& 1--10 d ?		& Max. 1	& 30   & N \\
4 & Unidentified HE transients		& Y	& HE			& 1		& 1 d	& ?				& 2 		& 20   & A/B \\
5 & LMXB microquasars			& ?	& X-ray/radio	& 1		& 1 d	& Weeks			& 2 		& 20   & A/B \\
6 & Novae					& Y	& HE/opt.		& 2		& 1 d	& Weeks			& 2 		& 20   & A/B \\
7 & Transitional pulsars			& Y	& Radio/opt.	& 0.5	& 1 d	& Weeks			& 2 		& 20   & A/B \\
8 & Be/X-ray binary pulsars		& N	& X-ray		& 1		& 1 d	& Weeks			& 2 		& 20   & A/B \\
\hline
\end{tabular}
\end{center}
\end{small}
\end{sidewaystable}

Simultaneous MWL follow-up observations should be arranged for all targets to maximize the scientific output. The best approach would be to collaborate with existing teams of experts (internal or external to CTA) that already have granted observing time in multiple facilities. Efforts are warranted to establish collaborations with such teams prior to the early science phase. Furthermore, a dedicated, on-site, 0.5-1 m class optical telescope can provide important benefits, including coordinated optical photometry to clarify various thermal and non-thermal processes and possibly polarimetry to obtain valuable complementary information on magnetic fields, density of ambient gas, etc.

{\bf C) X-ray, optical, and radio transients:}
To exploit the potential of X-ray, optical, and radio transient factories, we suggest 50 h/yr/site during the early phase to test and tune the filter and trigger response system based on responding to promising bright transients. 
Thereafter, we consider 10 h/yr/site during regular operations to use the tuned filters to follow up on the many new transient alerts we expect to receive. 

The systems for responding to X-ray alerts from Swift, INTEGRAL or SVOM are already well established for current IACTs and can be generally followed for CTA as well. Based on existing data, we do not expect the rate of alerts for transients of interest such as TDEs or SSBs to exceed those of GRBs. On the other hand, planning concrete strategies at this moment for follow-up to optical or radio transient factory alerts is challenging, due to numerous uncertainties concerning the actual performance of each facility, the latency for receiving different types of alerts, 
and the extent of the information that will be available. This is especially the case for fast transients such as FRBs. Nevertheless, a sketch of some possible strategies can be outlined for the relatively slower, Galactic transients. For example, optical transient factories are expected to discover large numbers of novae. Several novae have been detected by Fermi-LAT in HE gamma rays \cite{Ackermann14T}, and current IACTs are searching for VHE gamma rays from these objects, as predicted in some hadronic models. We expect that observations conducted by these and other high-energy facilities during the following years will provide a clearer picture as to what the best trigger conditions may be for follow-up with CTA. Possible filtering criteria could be based on optical magnitude (to select nearby objects), nova type (nature of the companion and expected physical behaviour), properties of the HE gamma rays detected by Fermi, etc. Similar exercises will also be conducted for other types of objects during the following years and prior to the early science of CTA, such that the optimal methodology can be developed for selecting the particular types of transient factory alerts that are most interesting for our aims.

{\bf D) High-energy neutrino transients:}
Follow-up with the full CTA is proposed for three different types of neutrino alerts, which are already being issued by IceCube for current IACTs:

{\it 1. Upgoing multiplets:} The strategy that is already being pursued by IceCube and present IACTs is based on event multiplets for a pre-defined list of known objects, consisting mostly of AGN in the northern hemisphere \cite{Aartsen16, Schuessler15}. Alerts to IACTs are issued whenever IceCube observes an increase in the rate of upgoing neutrino candidate events from the positions of the listed sources above a certain threshold, whose significance is adjusted to around
three standard deviations so that the false alarm rate remains at a tolerable level of a few/yr. The search algorithms analyze all time intervals ranging from seconds up to 3 weeks, covering the known range of AGN flare durations. 

{\it 2. Extremely high energy events:} A real-time, online system in operation since July 2016 that can identify through-going, track-type events induced by muon neutrinos of $>$160 TeV energy
with very high confidence \cite{Aartsen13Td}.
It is capable of issuing alerts with a latency of $\sim$30 seconds and localisation accuracy of a fraction of a degree (0.22 deg median angular resolution). The expected event rate is $\sim$4 per year from the whole sky, 
about half of which will be of atmospheric origin.

{\it 3. High-energy starting events:} An alternative system operating since April 2016 that is real-time, online and with relatively low background distinguishes track-type events due to $\gtrsim$100 TeV muon neutrinos whose primary interaction vertices are well-reconstructed within the IceCube detector volume, utilizing the veto technique that enabled the first identification of astrophysical high-energy neutrinos \cite{Aartsen13Ta}. The alert latency is similar to extremely high energy events while the localisation accuracy is 0.4-0.6 deg (median angular resolution).
The event rate is expected to be around 4 per year in all regions of the sky but with a higher expected atmospheric event rate of $\sim$75 \%.

Analogous to the strategy for GRBs, we propose follow-up of all alerts during dark time with zenith angle (at the CTA site) less than 70 degrees for at least 2 hours each. The exposure would be 
extended if RTA reveals a transient signal. The rate of all accessible neutrino alerts will be a few/yr/site, consisting mostly of Types 2 and 3
with a small fraction of Type 1. Note that the frequency of alerts can be increased by relaxing the relevant threshold. 
In view of various uncertainties, during the early phase of CTA, we propose a total of 20 h/yr for each site in order to develop and fine tune the follow-up strategies with the IceCube collaboration. After full operations, a total of 5-10 h/yr for each site is deemed sufficient.

Depending on the extent of the information that will be publicly disclosed for the alerts, MoUs between IceCube and the CTA Consortium may be warranted. 
If the RTA uncovers a gamma-ray signal, further alerts should be sent out rapidly to the community in order to identify the source through MWL follow-up in all possible wavebands (although some X-ray and optical facilities that have MoUs with IceCube may directly follow-up the alerts from IceCube).

Other strategies for follow-up of neutrino alerts from IceCube may be implemented, such as an unbiased, full-sky scan of upgoing multiplets without recourse to a pre-defined source list or online identification of UHE tau neutrino events. Several scenarios are under discussion to enhance the IceCube detector itself, many of which invoke an augmented fiducial volume to increase the sensitivity to TeV-PeV neutrinos \cite{Aartsen15b}, which would also increase the number of issuable triggers. Construction of northern observatories such as KM3NeT or GVD would also
improve the prospects substantially. The CTA neutrino follow-up program should evolve in accordance with such future developments.

{\bf E) GW transients:}
The CTA Consortium will receive GW alerts from LIGO-VIRGO as stipulated in the MoU that has already been signed. Follow-up strategies for GW alerts may involve tiling or divergent pointing strategies for localisations with large uncertainties, to be assessed in detail once Monte Carlo simulations of the corresponding sensitivity become available. For very large localisation errors, an alternative strategy may be to target nearby galaxies within the regions of highest probability for GW emission. Although the most plausible prediction for VHE emission is from short GRB afterglows in the fraction of NS-NS or NS-BH mergers with their GRB jet axes close to the line of sight~\cite{Bartos14}, it is generally not possible to know before the alert whether this will be the case or not. Some VHE emission may also be possible from non-GRB, off-axis GW events due to interaction of fast ejecta and ambient media. Following the discovery of GW150914,
various theoretical proposals have been put forth for electromagnetic signals accompanying BH-BH mergers and these should be observationally tested. This motivates the follow-up of all accessible GW alerts. The all-sky detection rate in GWs of binary NS mergers is quite uncertain, but is expected to be 40 /yr (with a range of 0.4 - 400 /yr) for full LIGO-VIRGO \cite{Abadie10}. Initial LIGO results may suggest even higher rates for binary BH mergers \cite{Abbott16d};
these rates should become clearer within the next few years.

As with GRBs and neutrinos, we consider follow-up of all GW alerts during dark time with zenith angles less than 70 degrees for 2 hours each, with additional exposure in the case of positive detections. Taking only the higher range of rate estimates for binary NS mergers, the rate of real alerts would be $\sim$1-10 /yr/site, with a corresponding observing time of $\sim$2-20 h/yr/site. The expected detection rate for on-axis, short GRB afterglows jointly with GWs is only $\lesssim$ 0.03/yr~\cite{Bartos14}, while that for off-axis GW events is more difficult to predict. More relevant, and difficult
to estimate at this stage, is the
rate of false GW alerts. In a spirit similar to neutrino follow-up, we propose a total of 20 and 5-10 h/yr/site for GW follow-up in the early and full CTA phases, respectively.

{\bf F) Serendipitous VHE transients:}
VHE transients in the FoV will be identified by the RTA system~\cite{Bulgarelli15, Fioretti15}, which will explore the data on different timescales from seconds up to hours. 
The sensitivity will vary as a function of the timescale considered and it will also depend on the VHE sources present in the FoV and the baseline flux in the case of a known source that is flaring. Once a transient source is detected, the short-term scheduler of CTA will allow extension of the observations. MWL follow-up should also be promoted via alerts to the community with a latency depending on its measured duration, especially for newly discovered objects. In the first stage, the serendipitous VHE transients should be observed for as long as possible during the same night. The decision on the continuation of the observations on
the following night should be motivated with the obtained CTA and MWL data of the particular VHE transient. We note that, in general, the CTA on-site analysis performed $\sim$10 hours after data taking could also provide serendipitous VHE transients in the FoV. In such cases, follow-up observations could be conducted during the subsequent night. 

If numerous serendipitous VHE transients are found, different patterns in the flux and spectral variability can be identified and the observation strategy can be correspondingly 
modified (e.g. number of hours observed in the first night, use of sub-arrays, etc.). Assuming five  transients per site and 5 h of observations for each transient on average, we estimate a total observation time of 25 h/yr/site during regular operations. During the early phase, 
100 h/yr/site will be needed to account for tests, possible fake alerts with the preliminary RTA system, etc. In general, the observations should be extended up to zenith angles of 60$^\circ$ in any weather conditions. After gaining some experience, this limit could be changed, and CTA sub-arrays could eventually be used for the follow-up observations.

In Table~\ref{table:obstimes} we summarize the observation times proposed for each class of 
targets (A-F) in the different phases of CTA. The proposed times are equal for the northern and southern arrays. In total, we request 390 h/yr/site during the early phase, 125 h/yr/site during Years 1 and 2 and 95 h/yr/site during Years 3 to 10 (2020 h during Years 1 to 10 including both sites). 

{\bf G) VHE transient survey:}
The VHE transient survey via divergent pointing could be conducted in conjunction with some fraction of the Extragalactic Survey KSP, and the associated observing time is considered to be
part of that KSP. In order to be effective for the aims of both the transient and survey KSPs, a judicious compromise must be struck between the gain in the FoV and the concomitant decrease in sensitivity and increase in energy threshold. The chances of a GRB serendipitously occurring in the 8-deg diameter FoV of the MSTs during normal pointing observations is once per
$\sim 10^4$ hours of dark time observations~\cite{Inoue13}. If the instantaneous FoV can be enlarged by a factor of a few, the probability of finding GRBs during the prompt phase in
$\sim$1000 hours of blind survey observations starts to become plausible. 
For example, a recent Monte Carlo simulation study \cite{Gerard15} (see also \cite{Szanecki15}) showed that assuming an array of 18 MSTs and 56 SSTs with a suitable divergent pointing pattern, the survey efficiency for steady point sources is comparable to
that achieved using normal pointing. In the same configuration, the sensitivity at a 10 deg offset remains within a factor of 3 of that at the center of the FoV, so it may be feasible to nominally consider a $\sim$ 20 deg diameter FoV for searches of relatively bright transients.

The annual all-sky rates for GRBs of the types detected by Swift-BAT and Fermi-GBM are respectively $\sim$800 and $\sim$600 when accounting for their instrumental FoV and the fraction of time capable of triggering. Thus the number of such GRBs occurring in the aforementioned $\sim$ 20 deg diameter FoV during dark time with 10\% duty cycle is $\sim$1 per year \cite{Inoue13}. The rate of GRBs actually detectable by CTA in this mode requires more detailed knowledge of the corresponding sensitivity and threshold on short exposure times. It is possible that pointing modes that are divergent enough to be more interesting for GRBs may not be the ideal choice for other aims of the extragalactic survey. 
Discussions on precisely how to implement the survey will continue and will be helped by
Monte Carlo predictions for the expected performance throughout the relevant parameter space.

Note that unlike follow-up of external alerts, no time delays are involved due to communication with alert facilities or telescope slewing, and inadequate localisations are not a concern. 
Although GRBs are expected to occur with equal probability anywhere in the sky, targeting regions at high Galactic latitudes is strongly favoured, as those near the Galactic plane are affected by interstellar absorption that hampers X-ray and optical follow-up required for good localisation and redshift determination~\cite{Bagoly09}. 
The RTA will play a critical role by automatically identifying new transients, rapidly alerting the community to foster MWL follow-up and allowing the option of switching to follow-up observations with normal pointing if the detected signal meets certain criteria. The scientific prospects can be significantly enhanced if the timing and/or FoV of the observation can be arranged to coincide with comparable surveys at other wavelengths. Close coordination with upcoming transient factory facilities may be particularly valuable, potentially allowing simultaneous MWL studies of short-duration transients for the first time.

\subsection{Data Products}
\label{sec:ksp_trans_data}

The data products expected from the Transients KSP are outlined below.

{\bf A) Gamma-ray bursts:} 
Measurements are foreseen of spectra and light curves for $\sim$1 GRB/yr/site with more than 100 photons above 30 GeV per event~\cite{Inoue13, Gilmore13, Kakuwa12}. In addition, upper limits can be expected for $\sim$10 GRB/yr, also offering valuable information for constraining GRB physics, the EBL, and LIV, as long as the redshift of the GRB is determined. 
Data rights should follow the proposed protocol, but
it is also essential that some information be disseminated immediately through GCN notices, particularly in the case of detections, so as to encourage MWL follow-up observations and maximise the science return. For example, similar to GCN notices from Fermi-LAT, 
we may communicate information such as {\it ``XXX photons are detected above XXX GeV''}.

{\bf B) Galactic transients:} Information on CTA observations of Galactic transients will be published in Astronomer's Telegrams as soon as possible after the observations are made, typically on the timescale of a day to several days. Once the CTA performance is established, light curves and spectra may be published even during the early phase.

{\bf C) X-ray, optical, and radio transients:} The data products resulting from triggers by X-ray/optical/radio transient factories will be light curves and spectra (or upper limits). The schedule for the data release will depend on the agreements reflected in the corresponding MoUs. 

{\bf D) High-energy neutrino transients:} Some types of alerts may be distributed via non-public channels, in which case the details of the specific requirements and data rights will be addressed in dedicated MoUs. 
These MoUs should provide the possibility to issue VOEvent alerts, GCNs, or Astronomer's Telegrams as soon as possible after the detection of a transient gamma-ray source in the vicinity of the neutrino trigger in order to ensure MWL coverage.

{\bf E) GW transients:} The most critical part of the CTA detection of counterparts to GW candidates is the source localisation. The CTA Consortium 
has signed an MoU with LIGO-VIRGO to permit the free exchange of information regarding GW detections and their localisations, counterpart detections by CTA, and counterpart detections by other follow-up partners who are signatories to the MoU. 
Joint GW-CTA observations of a GW candidate will be published according to the publication policy laid out in the MoU. 

{\bf F) Serendipitous VHE transients:} The detection of serendipitous VHE sources in the CTA FoV should be announced to the community as soon as possible through the use of tools such as VOEvent. 

{\bf G) VHE transient survey:} As with F, RTA will be essential to autonomously identify new transients and promptly alert the community for their MWL follow-up. 

\subsection{Expected Performance/Return}
\label{sec:ksp_trans_ret}

Here we discuss the expected performance of the Transients KSP and potential scientific return from
observations of the various target categories.

{\bf A) Gamma-ray bursts:}
Simulations for selected aspects of GRB observations by CTA have been presented in~\cite{Inoue13} and~\cite{Mazin13}; below we provide a brief summary of the results.

\begin{figure}[!htb]
\begin{centering}
\includegraphics[width=0.65\textwidth]{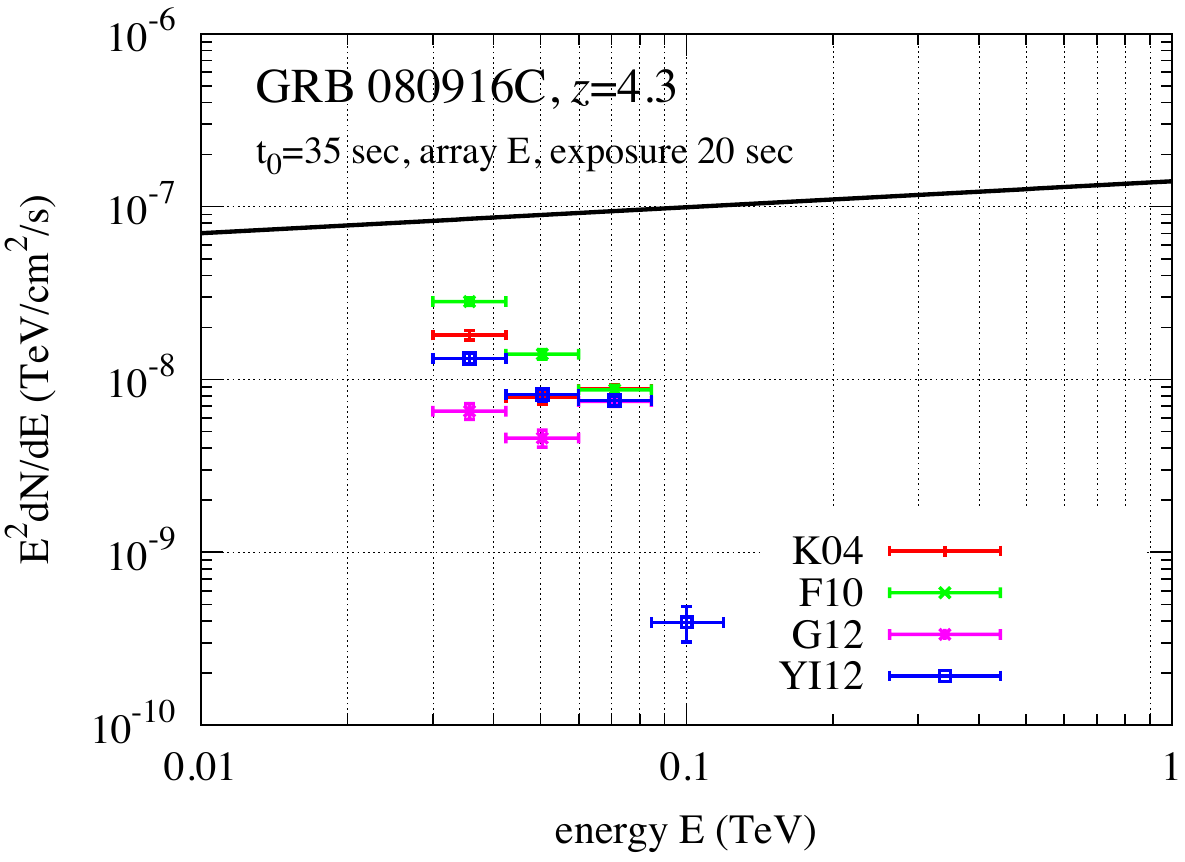}
\caption{
Simulated CTA energy spectrum of GRB 080916C at $z=4.3$, with an assumed intrinsic flux $dN/dE = 1.4 \times 10^{-7} (E/{\rm TeV})^{-1.85} {\rm cm^{-2} s^{-1} TeV^{-1}}$ (black line) corresponding to the time interval 55-100 s after burst onset~\cite{Abdo09a}, for an exposure time of 20 s. Compared are expectations for different EBL models: Kneiske et al. 2004 `best fit' (K04, red) \cite{Kneiske04}, Finke et al. 2010 (F10, green) \cite{Finke10}, Gilmore et al. 2012 (G12, magenta) \cite{Gilmore12}, and Inoue et al. 2013 (YI12, blue) \cite{InoueY13}. See~\cite{Inoue13, Mazin13} for more details.
\label{fig:GRBspectra}
}
\end{centering}
\end{figure}

{\it Spectra.}
Choosing as templates a few prominent Fermi-LAT bursts whose spectra and variability were relatively well characterized up to multi-GeV energies, we assume that their intrinsic spectra extend to very high energies with the power-law indices measured by LAT in specific time intervals.
Selected models are used to account for the effect of the EBL. 
Figure~\ref{fig:GRBspectra} shows an example of the distant but bright GRB 080916C at redshift $z=4.35$~\cite{Abdo09a}, with particular attention on the effect of the EBL. Despite the high redshift, even for exposure times as short as 20 sec, the
detection of several hundred photons above 30 GeV can be expected, with significant differences in the number detected depending on the assumed level of the EBL. The VHE spectra of such GRBs will be extremely valuable for probing the uncertain EBL at high redshifts through their consequent attenuation features, potentially beyond the range feasible through AGN observations, which would provide unique information on the cosmic history of star formation, black hole accretion and intergalactic reionization. Further physics insights can be obtained in combination with variability information, as discussed below.

{\it Light curves.}
As for simulations of spectra, we use 
as templates the Fermi-LAT data of selected bursts extrapolated into the CTA band, for which the example of GRB 080916C at $z=4.35$ is shown in Figure~\ref{fig:GRBlightcurve}. Despite the time delay required for follow-up (hence the plot being displayed only for times later than 30 sec after burst onset), 
CTA is potentially capable of resolving the light curve in exquisite detail for such bright bursts (and to a lesser extent for bursts with more moderate brightness), especially if it can begin observations during the prompt phase. 
Of particular value for extracting crucial physics information is the energy-dependence of the light curves, within the CTA band as well as within other bands from keV to GeV, which could: i) elucidate the origin of the poorly understood mechanism of GRB prompt and/or early afterglow emission, ii) reveal definitive signatures of hadronic emission processes through their characteristic delays at high energies, iii) distinguish intrinsic spectral cutoffs from those due to EBL attenuation (of which the latter should remain time-independent), and iv) probe LIV effects through their unique, expected energy dependence.

{\it Detection rates.}
Already discussed earlier, the predictions for the detection rate for CTA follow-up observations are based on a GRB population model tuned to match Swift observations, combined with assumptions on their VHE spectra from extrapolations of Fermi-LAT observations \cite{Gilmore13, Kakuwa12, Inoue13}. The general expectation is of the order of one CTA detection per year per site, 
the majority of which are for the early afterglow phase. One cautionary remark that must be made is that Fermi-LAT-detected GRBs reflect only the high-luminosity end of the GRB luminosity distribution \cite{Ackermann13Tb}. CTA may have better chances of detecting GRBs with more moderate luminosities, whose high-energy properties are yet to be observationally constrained. Depending on whether their true power at very high energies relative to the MeV band is higher or lower than that deduced from LAT-detected bursts, the aforementioned detection rates could be either underestimates or overestimates, respectively. Given the lack of VHE detections so far, they are unlikely to be serious underestimates, although the reality will only become clear with further observations.

\begin{figure}[!h]
\begin{centering}
\includegraphics[width=0.58\textwidth]{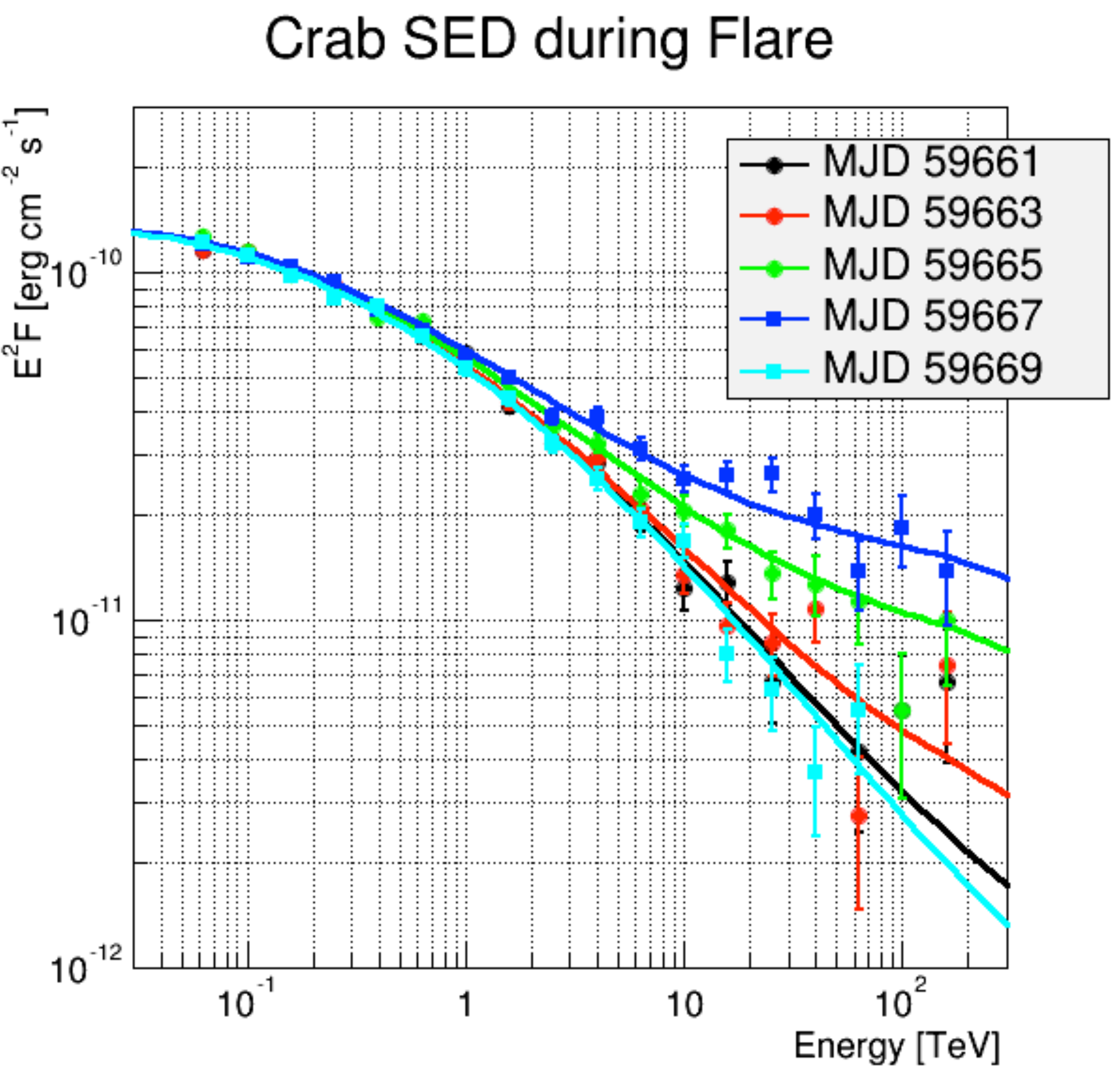}
\caption{
Simulated energy spectra of a Crab nebula flare observed with CTA. The model presented in \cite{Kohri12}, assuming a Lorentz factor $\Gamma$ of 70,  has been used to simulate the inverse-Compton component of the 2011 April flare observed with Fermi-LAT. A total of ten pointings of 4 h each separated by one day have been used. The variable tail from 10 to 100~TeV is clearly detectable.
\label{fig:crabflare}
}
\end{centering}
\end{figure}

{\bf B) Galactic transients:}
Using simulations of the CTA performance, we have estimated the observation times needed to detect selected types of known Galactic transient sources. Assuming that the Crab nebula high-energy flares are due to synchrotron emission, to detect the variable inverse-Compton component at multi-TeV energies we would need to monitor the source for 4~h/night during approximately 10 nights. With this strategy we could unveil the nature of these flares and, according to the model presented in \cite{Kohri12}, we could constrain the bulk Lorentz factor $\Gamma$ of the putative moving plasma blobs. An example of different spectra obtained in the case of $\Gamma=70$ is shown in Figure~\ref{fig:crabflare}. 
The observations with the southern array would cover the multi-TeV emission, while with the northern array we could check if there is any contribution from the high-energy end of the synchrotron spectrum at low zenith angles and look for the multi-TeV variability at high zenith angles.

\begin{figure}[!h]
\begin{centering}
\includegraphics[width=0.48\textwidth]{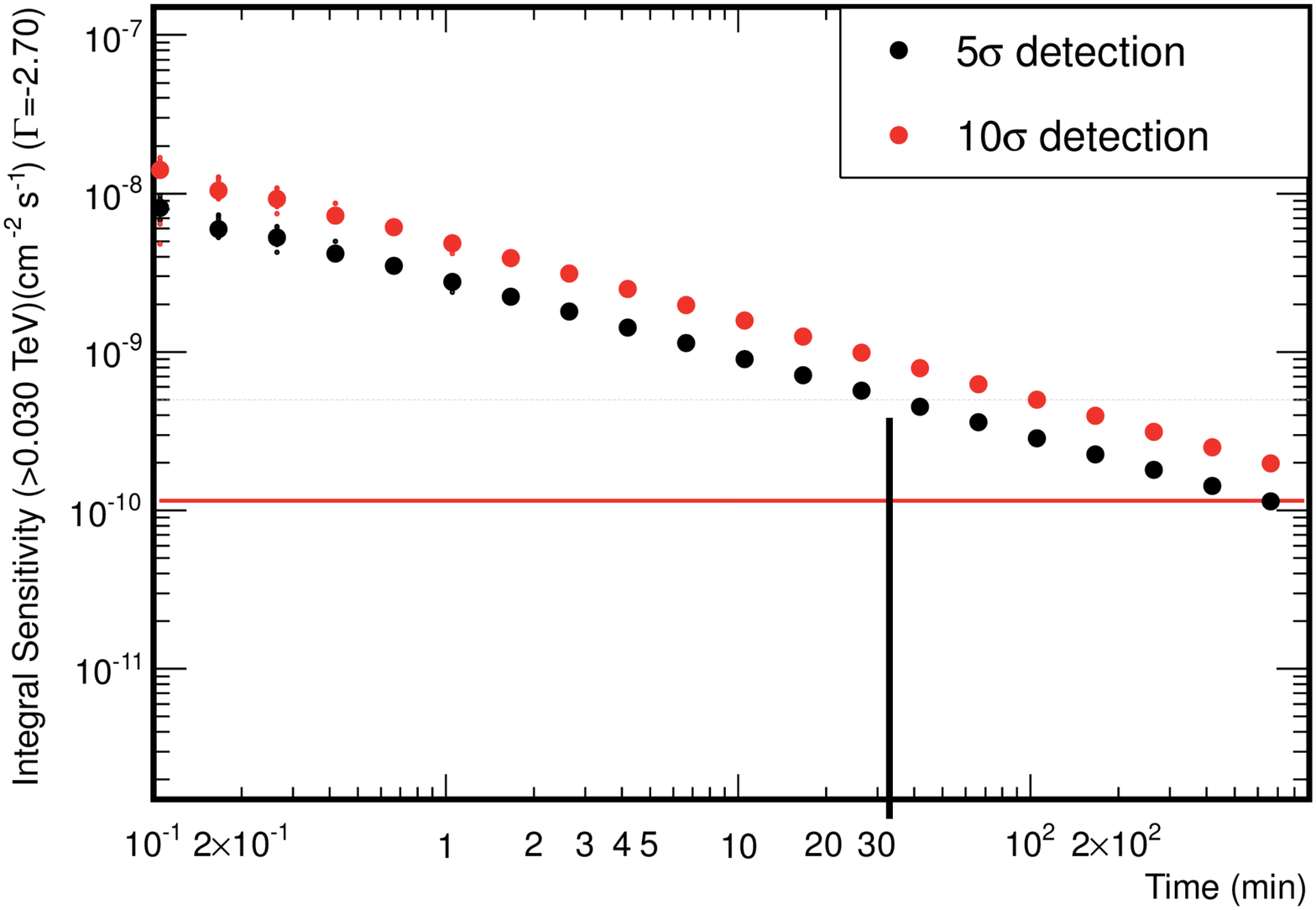}
\hspace{0.3cm}
\includegraphics[width=0.48\textwidth]{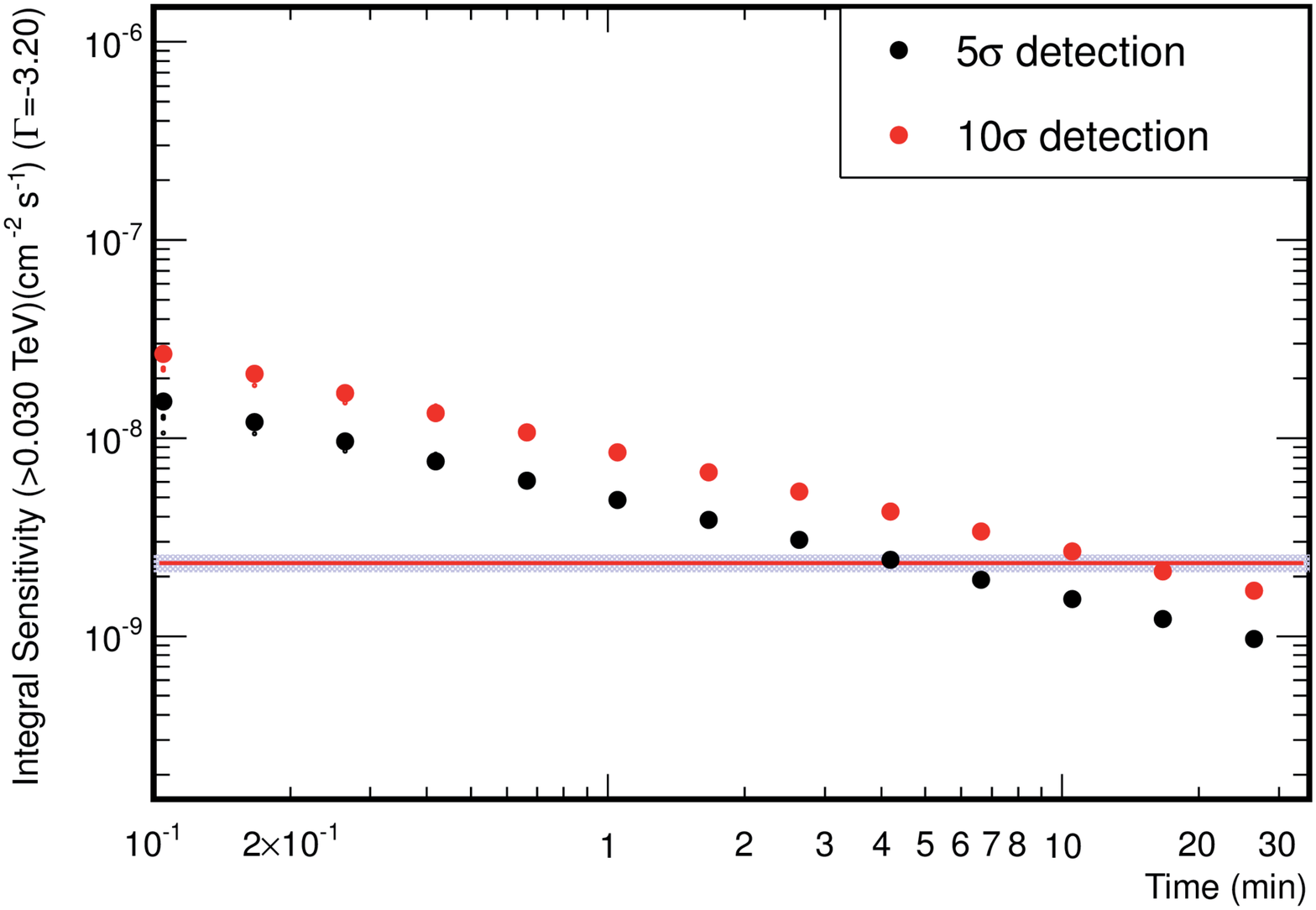}
\caption{
Left: Simulated integral sensitivity above 30~GeV of a flare from Cygnus~X-3 observed with the CTA northern array. The Fermi-LAT high flux reported in \cite{Abdo09c} has been extrapolated assuming a photon index of $-2.7$ and is represented by the horizontal red line. A five standard deviation detection could be obtained in $\sim$10 hours of observations, while the determination of the spectrum could take up to 30 hours of observations, denoted by the black vertical line.
Right: Same as left panel but for Cygnus~X-1.
The possible VHE spectrum of Ref. \cite{Albert07} has been used as a template (pink horizontal line), resulting in a five standard deviation detection in 4 minutes and a good spectrum in $\sim$15 minutes for CTA (see also \cite{Zanin16}).
\label{fig:cygflare}
}
\end{centering}
\end{figure}

For Cygnus~X-3, extrapolating the Fermi-LAT spectrum reported in \cite{Abdo09c} during the high flux state and assuming a photon index of $-2.7$, the northern array of CTA could detect the 
source at a five standard deviation confidence level in $\sim$10 hours of observation (see Figure~\ref{fig:cygflare}). Obtaining a spectrum could require up to 30 hours. If the spectrum is harder during the flares these numbers could decrease significantly: a photon index of $-2.2$ would provide a detection in $\sim$1.5 hours and a spectrum in $\sim$5 hours. 
For Cygnus~X-1, we use the possible VHE detection \cite{Albert07} to estimate that a 
five standard deviation detection could be obtained in 4 minutes and a good spectrum in $\sim$15 minutes (see Figure~\ref{fig:cygflare}; also Fig. 7 of \cite{Zanin16}). 
In summary, for some Galactic transients simulations have been conducted, but we would be mostly exploring the unknown. 

{\bf C) X-ray, optical, and radio transients:}
No simulations have been conducted for follow-up of X-ray, optical or radio sources from transient factories, since in general, we will be exploring predominantly new phase space. However, as in the other transient cases, there is the potential to provide high-impact results with a moderate amount of observation time.

{\bf D) High-energy neutrino transients:}
Detection of VHE gamma rays associated with high-energy neutrinos would be instrumental for revealing and understanding the sources of the neutrinos. If the parent hadrons can also escape from the sources with reasonable efficiency and contribute to the observed cosmic rays, the neutrino plus gamma-ray signature could constitute the long-sought smoking gun of the sources of UHECRs. 
One model-independent way to estimate values for the VHE flux from potential sources of the IceCube neutrinos detected so far is discussed in~\cite{Padovani14b}. Assuming that each detected neutrino originates from a distinct astrophysical object via $pp$ or $p\gamma$ interactions, time-averaged fluxes of concomitant VHE gamma rays can be estimated for each object to be at the level of $0.6 - 1.7 \times 10^{-8}\ \mathrm{GeV\ cm^{-2}\ s^{-1}}$ above several tens of TeV. At face value, this would be readily detectable in a few hours of CTA observation. However, this neglects EBL attenuation that can be severe depending on the photon energy and source redshift \cite{Mazin13}, as well as internal
gamma-gamma absorption that can be significant in cases of $p\gamma$ production \cite{Ahlers15}.
The true VHE gamma-ray flux can be higher if the sources are transient with durations shorter than the relevant IceCube exposure time and they will be lower if they are actually more numerous than the currently detected number of neutrinos. 

{\bf E) GW transients:}
The impact of CTA detecting a counterpart to GW radiation cannot be overstated. First, it would strengthen the significance of the GW candidate, particularly if the temporal coincidence was strong (as expected for short GRB afterglows). 
The association of a short GRB-like afterglow with a GW event can test the compact binary merger model for short GRBs and possibly discriminate between NS-NS or NS-BH progenitors. A good localisation of the VHE counterpart would permit follow-up observations by optical telescopes that could reveal the redshift and hence the energetics of these mergers. The sensitivity of CTA to on-axis GW events may be the same as that to short GRBs (see A above), but the effect of tiling and divergent pointing strategies that can improve the sensitivity to both has yet to be simulated and optimized. If VHE emission is limited to short-GRB-like beams, detection can only be expected for less than a few percent of all follow-up observations. 
However, as mentioned above (Section~\ref{sec:ksp_trans_sci_con}D), NS mergers may also possibly induce off-axis high-energy emission as a consequence of fast mass ejection in non-polar directions and associated particle acceleration \cite{Kyutoku14}. Finally, observations of binary BH mergers will test newly proposed models for associated electromagnetic emission. Detection of such events will constitute a unique discovery and provide novel information on the dynamics of binary coalescence events.

{\bf F) Serendipitous VHE transients:}
As stated earlier, there have been no new serendipitous VHE transients discovered so far in the FoV of currently existing VHE telescope. On the other hand, transient sources are regularly discovered in the large fields of view of Fermi-LAT or AGILE, albeit being very bright objects 
(flaring AGN or GRBs in most cases).
With our current limited knowledge, we can provide very approximate estimates for
 the probability for CTA to detect such objects in any given observation. The chances for serendipitous occurrence of GRBs during normal pointing observations is quite low, once per $\sim$35(13) yr in the 4.5(8) deg diameter FoV of the LST(MST) array~\cite{Inoue13}.
This rate can be increased with divergent pointing (see Section~\ref{sec:ksp_trans_strat}G). 
Of this, the fraction detectable by CTA may be still less, depending on the uncertain high-energy properties of GRBs with moderate to low luminosities that have not been well constrained by existing observations (see remarks in Section~\ref{sec:ksp_trans_strat}A). 
For flaring AGN, the prospects depend on the sensitivity reached in the particular observation. For example, at the sensitivity of 5 mCrab as envisaged for the Extragalactic Survey, we can scale the number of expected AGN detections for the survey and assume a 10\% flare duty cycle to obtain a 
2-10 (6-30)\% probability of finding a flaring AGN in the FoV of the LST (MST) array, subject to uncertainties on the AGN luminosity function and duty cycle at very high energies. 
Expectations for other types of serendipitous transients are even more uncertain. 
The true situation should become clearer as actual observations proceed for the Extragalactic Survey as well as the flare programme of the AGN KSP (Chapter~\ref{sec:ksp_agn}). Note that strategies for the Galactic Plane Survey KSP are currently being considered whereby the same regions of the Galactic plane are revisited over a range of timescales, enabling improved probes of variability in known objects as well as searches for new Galactic transients (Chapter~\ref{sec:ksp_gps}). In any case, in view of the unprecedented sensitivity and FoV of CTA, discoveries of new types of VHE transients may be feasible and impactful.

{\bf G) VHE transient survey:}
Divergent pointing observations would greatly increase the FoV of CTA, at the cost of
reduced sensitivity and increased energy threshold compared to follow-up observations with
normal pointing. However, these two techniques are complementary and would provide
complete time coverage of the prompt emission phase of 
both long and short GRBs from their onset, which would address important questions concerning the physical mechanisms of jet formation, particle acceleration and radiation in these objects. 
Divergent pointing would be the only way to detect or set limits on the VHE prompt emission of short GRBs, of great importance in view of their potential connection with NS-NS mergers, the most promising sources of GWs. 
While Fermi-LAT observation of the short GRB 090510 has set the first limit beyond the Planck scale for LIV with a linear dependence on the mass scale~\cite{Abdo09b}, the much better photon statistics potentially offered by detecting a bright, short GRB with CTA in divergent mode can lead to a further, dramatic improvement by up to three orders of magnitude.

A VHE transient survey by CTA also opens up significant discovery space. As mentioned earlier, 
Fermi-LAT GRBs correspond only to the most luminous ones 
(i.e. the ``tip of the iceberg'') in the GRB luminosity distribution \cite{Ackermann13Tb}, and the high-energy properties of the great majority of GRBs with lower luminosities still remain unexplored. (For reference, blazars are well known to have spectral energy distributions that differ strongly and systematically with luminosity, with those of lower-luminosity BL Lac objects extending to much higher photon energies compared to the higher-luminosity flat-spectrum radio quasars.) Other potential, fast extragalactic VHE transients include relativistic SSB events~\cite{Kashiyama13}. The most exciting prospect would be the discovery of new and completely unexpected types of VHE transient sources.

MWL coordinated surveys between CTA and upcoming transient factories can further advance the horizons of transient research. A prime candidate may be the recently discovered fast radio bursts (FRBs), i.e. bursts at GHz frequencies with millisecond duration of unknown origin but with clear signs of being extragalactic through radio dispersion effects, and with inferred all-sky rates approximately 
three orders of magnitude higher than GRBs \cite{Keane16a}. Some models for FRBs predict correlated VHE bursts on similar timescales~\cite{Lyubarsky14}, which would be testable only by simultaneous observations involving CTA and suitable radio facilities. Note that $\sim$2 FRBs per day could be occurring in the FoV of CTA using the divergent pointing mode.

\section{KSP: Cosmic Ray PeVatrons}
\label{sec:ksp_acc}

Cosmic rays are primarily energetic nuclei
which fill the Galaxy and carry on average as much energy per unit
volume as that in starlight, in interstellar magnetic fields, or
in motions of interstellar gas \cite{Gaisser90}. The 
confinement time of cosmic rays within the Galaxy is of the order of $\sim$10 million 
years \cite{Gaisser90}, and this implies that cosmic-ray sources must provide 
$\sim 10^{41}$~erg/s in the form of accelerated particles in order to maintain 
the cosmic-ray intensity at the observed level. Moreover, the spectrum of cosmic rays observed 
at Earth is largely dominated by protons up to the so called {\it knee} in the spectrum at
an energy of a few PeV. Above that energy, the differential power-law spectrum 
steepens from $\sim E^{-2.7}$ to $\sim E^{-3.0}$, where $E$ is the particle energy,
and its chemical composition becomes heavier \cite{Antoni2005}. Thus, the search for cosmic-ray sources 
has been focused on objects able to satisfy these requirements, i.e. 
{\it powerful cosmic-ray proton PeVatrons}.

It is well-known that supernova remnants (SNRs) are able to satisfy the 
cosmic-ray energy requirement if they can somehow convert $\sim$10\% of the supernova 
kinetic energy into accelerated particles \cite{Hillas05}. In this context, 
particles are accelerated via diffusive shock acceleration at the expanding SNR
shocks \cite{Drury83}. The acceleration of cosmic rays at SNR shocks is accompanied by 
an amplification of the magnetic field that can boost the acceleration of 
protons up to the energy of the knee and even beyond \cite{Bell04}. However, the 
details of such an amplification mechanism are not completely understood, and 
thus it is still unclear whether or not SNRs can act as cosmic-ray PeVatrons. To 
conclude, one hundred years after their discovery, the question of the origin of cosmic rays
is still open. 

A tight connection exists between cosmic-ray physics and very high-energy (VHE) gamma-ray 
astronomy. This comes about because if SNRs indeed are the sources of cosmic rays, they 
should also be bright VHE gamma-ray sources, due to the decay of neutral pions 
produced in the interactions between the accelerated cosmic rays and the gas swept up 
by the shock \cite{Drury94}. 
In fact, at GeV energies, clear evidence for the pion bump in the gamma-ray spectra
of several SNRs has now been demonstrated \cite{Ackermann13}.
At very high energies, a number of SNRs are now
well-established
gamma-ray emitters \cite{Aharonian13}, but current observations do 
not allow us to unambiguously ascribe the gamma-ray emission to hadronic 
processes. This is because electrons can also be accelerated at SNR shocks and
produce gamma rays through inverse-Compton scattering off soft ambient photons.
An essential way to make progress would be to observe SNRs in the almost unexplored 
$>10\,$TeV energy domain. The detection of an SNR whose spectrum extends without
any appreciable attenuation up to energies of $\sim$100 TeV would imply that: 
i) the emission is hadronic, because the leptonic emission is strongly 
suppressed at such high energies due to reduced Compton losses in the Klein-Nishina 
regime and ii) the SNR is a PeVatron, because $\sim$100 TeV photons are
produced by $\sim$PeV protons. It is thus of paramount importance to 
identify gamma-ray sources whose spectra extend, without any appreciable 
suppression, well beyond $\sim$10 TeV. This task is challenging for current 
atmospheric
Cherenkov telescopes, due to their limited sensitivity in the $\gg$10 TeV energy
domain, but it is well within the reach of CTA\footnote{While this document was finalised, the H.E.S.S. Collaboration reported on the discovery of a cosmic PeVatron in the Galactic Centre region, possibly identified with the supermassive black hole Sgr~A* \cite{HESS16}. Such a detection reveals possible new scenarios for the acceleration of PeV cosmic rays in the Galaxy and strengthens the case for deep searches with CTA.}.

This KSP proposes an observational strategy to search for PeVatrons with CTA. 
First, we suggest to perform deep observations of known sources with 
particularly hard spectra (i.e. not much steeper than $\approx E^{-2}$) and with hints for a possible spectral extension 
into the multi-TeV energy domain. Second, we also suggest to search 
for diffuse gamma-ray emission from the vicinity of prominent gamma-ray bright
SNRs. This is motivated by the belief that PeV particles are expected to 
quickly escape the SNR shock \cite{Bell13b} and then propagate diffusively in 
the ambient medium surrounding the remnant. The interactions of such runaway 
PeV particles with the ambient gas produce gamma rays with a characteristic 
hard spectrum extending up to $\sim$100 TeV \cite{Gabici07}. An obvious target
for this second kind of investigation is the SNR RX~J1713.7$-$3946, which is one of the 
best-studied supernova remnants at very high energies. 
Though the leptonic or hadronic origin of 
its gamma-ray emission is still debated \cite{Ellison10,Gabici14}, the 
steepening clearly observed in the gamma-ray spectrum at energies of 
$\sim$10 TeV suggests that the highest energy particles might have escaped the 
remnant. If this is the case, the diffuse gamma-ray emission one would expect 
to surround the SNR has been shown to be well within the reach of 
CTA \cite{Casanova10}.

The deep investigation of SNRs using
CTA at full sensitivity, 
with special emphasis on the high-energy range, 
will allow us to address the basic questions listed above 
and will provide crucial information to the scientific community, in 
understanding both particle acceleration mechanisms and supernova remnants. The 
proposed observations also require an excellent understanding of the 
instrument, including an optimal angular resolution, high control of systematic
effects at the highest energies in deep observations and a good pointing strategy
based on preliminary results of the CTA Galactic Plane Survey 
(GPS, see Chapter~\ref{sec:ksp_gps}).
Given the high-risk/high-gain and exploratory nature of these observations 
and its strong interest in developing the Small-Sized Telescope (SST) array,
the CTA Consortium is in a good position to devote sufficient resources
in the context of a Key Science Project 
to tackle this important unresolved question of cosmic-ray origins.

\subsection{Science Targeted} 

\subsubsection{Scientific Objectives}

The scientific goals of this KSP appeal directly to the 
fundamental question of the origin of the cosmic-ray sea that fills our Galaxy. 

\begin{enumerate}
\item Where and how in the Galaxy are cosmic rays accelerated up to PeV energies? 
\item Are we sitting in a particular location of the Galaxy, or do the cosmic rays form a 
uniform sea within the whole Galaxy?
\item What is the distribution of PeVatrons in the Galaxy? 
\item Do young shell-type SNRs accelerate hadronic cosmic rays up to PeV energies?
\item If so, up to which energies and how effective is this acceleration?

\end{enumerate}

\subsubsection{Context / Advance beyond State of the Art}

\paragraph{Hadronic mechanisms and the connection with cosmic-ray origin} 

Very tight connections exist between cosmic-ray studies and gamma-ray astronomy, due to
the fact that cosmic-ray protons can undergo hadronic interactions with the 
interstellar medium, producing neutral pions that in turn decay into gamma rays
(see, e.g., \cite{Stecker71,Dermer86}). This is of particular relevance for the 
identification of cosmic-ray sources, since the production of gamma rays is expected, 
at some level, during cosmic-ray acceleration. 

At GeV energies, the 
acceleration of hadrons at SNR shocks has now been clearly established 
\cite{Ackermann13}. At TeV energies, atmospheric Cherenkov telescope 
observations of SNRs 
interacting with molecular clouds show very strong hints of hadrons around them 
\cite{Aharonian08}. 
Finally, the $>10\,$TeV energy domain is currently an 
important unexplored regime in the gamma-ray section of the electromagnetic spectrum, and opening this observational window would allow us to establish a link between 
the observed PeV Galactic cosmic rays and their astrophysical accelerators. 

Below we propose a number of observations to be performed by CTA
that will shed light on the question of the origin of PeV cosmic rays:

\begin{itemize}

\item {\bf PeVatrons:} Only a few bright VHE
 sources have been detected above $\sim$20 TeV so far. 
Although the newly-commissioned 
HAWC air shower detector \cite{Abeysekara13,Abeysekara2017} will have significantly
improved sensitivity relative to its predecessor Milagro,
CTA will be 
able to reach the same sensitivity as five years of HAWC in about 50 hours and
with much better spectroscopic capability ($\Delta E/E < 0.1$, 68\% containment),
thanks to the capabilities of the Small Sized Telescopes (SSTs) in the southern array. 
This unique capability of CTA will enable the detection of the 
most energetic photons ever observed and the sampling of the highest energy particles
in the Galaxy up to the PeV scale. Finding the extreme accelerators 
powering those particles is a goal that is within the reach of CTA.

Moreover, at energies above 50 TeV, the problematic ambiguity between leptonic 
and hadronic origin is nearly completely resolved, since 100 TeV photons 
are produced preferentially by hadronic processes.  This is a result of the Klein-Nishina
effect, where the cross-section for inverse-Compton electron-photon interactions 
decreases very quickly above a few tens of TeV. 
The hadronic cosmic rays capable of producing photons 
of $\sim$100 TeV should have energies approximately an order of magnitude higher, 
i.e. in the PeV regime. 
Discovering the sites of cosmic-ray acceleration up to 1 PeV
within the Galaxy would confirm the hypothesis that cosmic rays below the 'knee' can be
accelerated in the Galaxy and would finally shed light on the origin of Galactic cosmic rays. 
Within the SNR scenario for the origin of cosmic rays, only a handful of such PeVatrons are expected
to be currently active in the Galaxy; here we propose to follow up promising
high-energy candidates first seen in the CTA GPS. 

We propose to trigger dedicated observations for sources detected in the CTA GPS which exhibit hard power-law spectra
that extend up to extremely high energies, i.e. by requiring 
a detection at the three standard deviation level above 50 TeV. 
With the dedicated observations we will: i) measure the
extension of the spectrum to the highest energies CTA can reach and ii) provide 
a precise localisation and measurement of the source size (extension beyond the
point spread function (PSF), if any) to identify the multi-wavelength counterpart responsible for producing such 
radiation and thereby pinpointing the origin of the PeVatron. Even the 
discovery of only one Galactic PeVatron would represent a major breakthrough 
and allow us to understand the physics of the most extreme accelerators in the
Galaxy.

\item {\bf RX~J1713.7$-$3946:} is one of the brightest VHE gamma-ray sources detected.
At a distance of 1 kpc \cite{Fukui03}, it is an ideal target to study the 
acceleration of cosmic rays in SNRs. Besides being the brightest SNR, numerous 
theoretical and observational studies at many different wavelengths have been 
performed, making RX~J1713.7$-$3946 an ideal laboratory to understand hadronic
acceleration in SNRs. 

At present, young SNRs are the most probable candidates for being the major 
accelerators of cosmic rays. The detections of non-thermal X-rays have already provided
evidence for the acceleration of electrons to ultra-relativistic energies at 
SNR shocks \cite{Koyama95}, and previous VHE gamma-ray observations confirmed the 
existence of high-energy particles in the shocks of young SNRs (see e.g. 
\cite{Enomoto02,Aharonian04,Aharonian05,Acciari2011,Albert2007}). However, the leptonic or hadronic 
nature of those particles is still uncertain \cite{Gabici14,Acero13}. 

For RX~J1713.7$-$3946, some evidence has been put forward 
to support a leptonic origin of the TeV gamma-ray emission \cite{Abdo11a}. 
However, a leptonic origin stands in contradiction with the high magnetic field
required to explain the observed width of X-ray filaments. Radio observations 
of CO and HI gas have revealed, on the other hand, a highly inhomogeneous 
medium surrounding the SNR, such as clumpy molecular clouds 
\cite{Fukui13,Sano14}, whereas X-ray observations have revealed details of the particle acceleration (e.g. magnetic field amplification) via spectro-morphological characterisation of the remnant 
\cite{Uchiyama07}. Spectro-morphological characterisation at very high energies, along with a complete comparison with the synchrotron X-ray emission and the CO material, is not yet feasible due to the lack of statistics 
and limited angular resolution of the present generation of atmospheric Cherenkov
telescopes. Moreover, in the diffusive shock acceleration picture, the most 
energetic cosmic rays leave the shell at the beginning of the Sedov phase, 
while the less energetic ones are still confined today.
These high-energy runaway protons can produce extended enhanced gamma-ray 
emission when colliding with atomic and molecular gas in the vicinity of the 
SNR. The well-studied molecular gas surrounding of RX~J1713.7$-$3946 provides 
the perfect tracer to investigate this scenario, together with the improved CTA
sensitivity and angular resolution \cite{Casanova10}. Note that the presence of gamma-ray emission extending beyond the SNR shell of RX~J1713.7$-$3946 has been recently reported by the H.E.S.S. Collaboration \cite{Abdalla16a}.

Simulations show that CTA will provide clear measurements of: i) the extension of 
the gamma-ray bright shell, constraining the emission of the high-energy 
runaway particles interacting with the dense medium, ii) the radial profile, 
which is expected to differ if the gamma rays are produced mainly by electrons 
or hadrons, and iii) the spectral distribution in different regions of the shell, 
sampling differences in magnetic fields and/or in dense, clumpy regions.
The spectral measurements will determine the 
maximum energy reached in the source, providing an unprecedented spectral 
characterisation to be compared with the low-energy radiation part. 

\end{itemize}

\subsection{Strategy}

Deep observations of several selected sources are requested to investigate one
of the basic questions that has driven the VHE gamma-ray field over the last few years, 
in particular, to understand the acceleration of Galactic cosmic rays to PeV
energies. 
Despite considerable
observation time devoted by the current generation of Cherenkov telescopes, it
is clear that a significant breakthrough can be achieved 
only with the sensitivity of CTA over a large energy range and 
with superior angular resolution. 
In a conservative approach, and considering a supernova rate in our Galaxy of at most a few per century and
the hypothesis that only young sources in the very early stages of
their evolution (i.e. in the first $\sim$100 years) 
are able to accelerate particles to PeV energies, we expect a limited number
of PeVatrons in the Galaxy in the present epoch.
Accordingly, we propose to re-observe the 
five most promising hard-spectrum 
candidates, in which no evidence of a cutoff is detected from the shallower
Galactic Plane Survey observations. 
These PeVatron candidates will be observed for 50\,h each, 
increasing by a factor of 3-4 the exposure expected to be reached in the 
Galactic Plane Survey (see Chapter~\ref{sec:ksp_gps}).
The trigger criteria rely strongly on early results from the GPS observations and 
follow-up observations will be carried out
on those sources that display an extension of their 
spectrum up to several tens of TeV. 

Deep observations of the SNR RX\,J1713.7$-$3946 are also proposed based on its 
particular location and characteristics. 
RX\,J1713.7$-$3946 is one of the best 
studied SNRs at gamma-ray energies and a considerable body of theoretical
work has been devoted to it.
Moreover, it is embedded in a dense, 
well-studied region that provides a large amount of target material. 
Thus, the source is an optimal one in which to 
investigate the interactions of cosmic rays, which are believed 
to escape from the acceleration region and collide with these clouds. 
This type of study requires a good understanding 
of the imaging atmospheric Cherenkov technique and
of the instrument point spread function,
when imaging the SNR surroundings at a few parsecs resolution.

\subsubsection{Targets}

Table~\ref{tab:acc_targets} summarises the targets and required
exposure for this Key Science Project.

\begin{table*}[h!]
  \centering
  \begin{tabular}{l|ccccc}
    \hline
    Target & Type & Exposure (h) & Array  & Year &
    Configuration \\\hline\hline
    RX\,J1713.7--3946 & SNR & 50 & S  & $1-3$ & Full array\\ 
    PeVatrons & Unknown & 5$\times$50 & S  & $>$3 & MSTs + SSTs \\
    \hline
\hline
  \end{tabular}
  \caption{Summary of objects proposed for observations in this Key Science Project.}
  \label{tab:acc_targets}

\end{table*}

\subsection{Data Products}

The data products produced in this KSP will consists of maps/datacubes and 
spectral information. We will provide large maps of 10-degree size with 
information on each selected source, along with morphological fits, 
including detailed analysis of any shell-type SNRs. With respect to the
spectro-morphological studies, we will also provide spectral information in 
different regions of the (extended) sources, including spectral 
information in a few-arcminute-size wedges to investigate cosmic-ray
escape in extreme accelerators. The pointing strategy depends on the 
preliminary results of the GPS observations. 
We will follow-up those sources for which an extension of a 
hard spectrum is detected, increasing the exposure to achieve improved 
statistics above 50 TeV. The results of these follow-up
observations will be released 
simultaneously with the GPS ones, if possible.  

\subsection{Expected Performance/Return}

To evaluate the capability of CTA to detect PeVatrons, we simulated
the capability of CTA to detect 50-TeV photons in 
$\sim$15~h 
(an approximation for the effective observing time achieved
in the inner Galaxy in the first several years of the GPS).
For the differential energy spectrum of the PeVatron, we used a simple power-law with no 
cutoff, $\phi = \phi_{\rm 0} (E /{\rm TeV})^{-\Gamma}$, with 
$\phi_{\rm 0} = 2.1 \cdot 10^{-11}$ TeV$^{-1}$ cm$^{-2}$ s$^{-1}$ (similar to the 
flux level of RX\,J1713.7$-$3946, a prototypical, bright VHE source) and 
took the spectral index
$\Gamma = 2$ as proxy and calculated the minimum percentage of the flux 
$\phi$ detectable above 50 TeV at a significance of three
standard deviations in a 15~h observation time. This significance level 
was chosen as a reasonable indication for a real excess which can 
be investigated with further observations. 
Simulations were performed assuming that the PeVatrons under examination are point sources and  taking the background rate from the Monte Carlo simulations performed by the CTA Consortium to evaluate the instrument response.
The results show
that after 15~h of observation CTA will be able to detect enough photons to reconstruct a point at 
50 TeV having a flux level of $\sim$7\% of the 
flux $\phi_{\rm 0}$, that is, of the order of $\sim$10$^{-12}$ TeV$^{-1}$ cm$^{-2}$ 
s$^{-1}$, if there is no cutoff in the spectrum. If a cutoff is included in the 
spectrum, we estimate that a minimum of 50~h observation time is required to 
obtain enough statistics to determine an energy cutoff of $\sim$60 TeV for a 
source with a flux 10\% of $\phi_{\rm 0}$. 
\begin{figure}[h!]
\centering
\includegraphics[width=5.0in, trim= 0.0in 0.0in 0.0in 0.0in, clip=true]{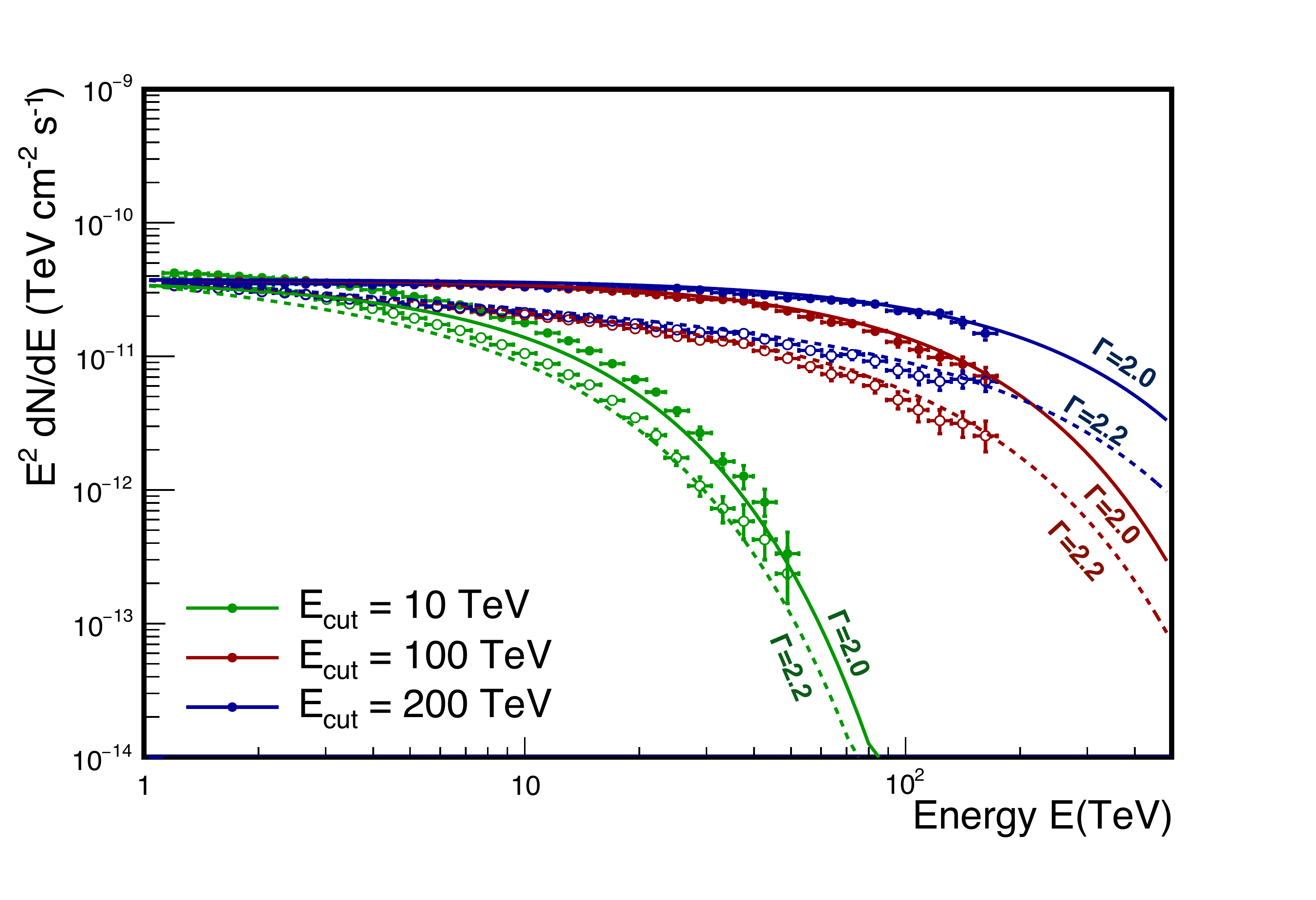}
\caption{Simulated reconstructed spectra for CTA for a PeVatron source with a flux
equal to the Crab nebula, using two photon indices (solid: 2.0, dashed: 2.2).
Three different exponential energy cutoff values are used, as indicated by the colors.}
\label{fig:ksp_pevatronsspec}
\end{figure}

To test the capability of CTA to estimate spectral features, we show in Figure~\ref{fig:ksp_pevatronsspec} the results of simulations of a powerful Crab-like source characterized by two hard photon indices (2.0 in solid lines and 2.2 in dashed lines). The different spectral features are clearly reconstructed even for the most extreme case, assuming an energy cutoff of 
200 TeV (best-fit values are 203$\pm$23 TeV and 188$\pm$25 TeV for photon indices of
2.0 and 2.2, respectively). 
To compare with the case described above, the same simulations for a 7\% of $\phi_0$ spectrum in 50 h show that spectral points could be reconstructed up to 200 TeV, reproducing an energy cutoff as high as $135 \pm45$ and $110 \pm  35$ TeV for the same 2.0 and 2.2 photon indices.
More details on the impact of the different CTA 
configurations can be found in two dedicated papers \cite{deOna13,Acero13}. 

\begin{figure}[h!]
\centering
\includegraphics[width=0.95\linewidth]{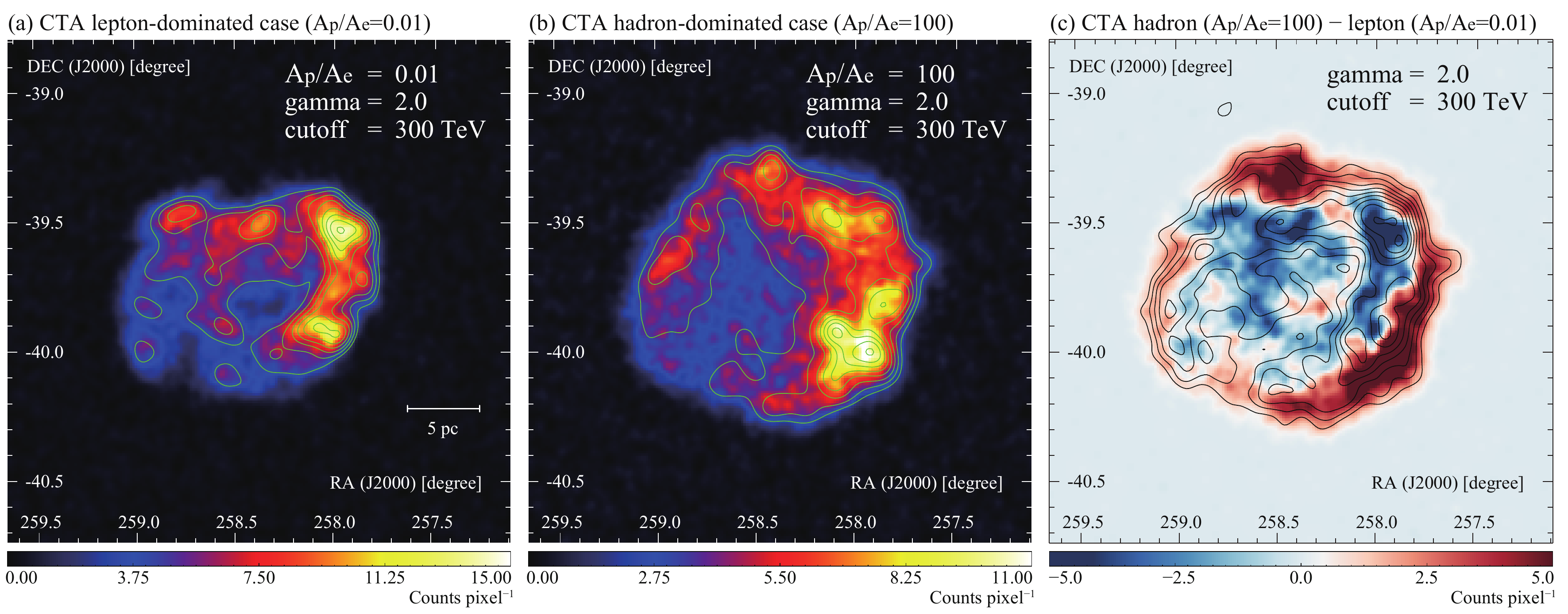}
\caption{Simulated gamma-ray images of RX\,J1713.7$-$3946 in different 
scenarios: (a) $A_{\mathrm{p}} / A_{\mathrm{e}} = 0.01$ (lepton-dominated case) and 
(b) $A_{\mathrm{p}} / A_{\mathrm{e}} = 100$ (hadron-dominated case), with
$\Gamma_{\mathrm{p}} = 2.0$ and $E_\mathrm{c}^{\mathrm{p}} = 300$ TeV. 
Reproduced from~\cite{Nakamori14}.
The green 
contours show: (a) \emph{XMM-Newton} X-ray intensity \cite{Berezhko09} and (b) 
total interstellar proton column density \cite{Sano14}, smoothed to match the 
PSF of CTA. The subtracted image of (a)$-$(b) is shown in (c). The black  
contours in (c) correspond to the 
VHE gamma-ray emission as reported by H.E.S.S. \cite{Aharonian04}.
}
\label{fig:ksp_j1713morph}
\end{figure}

We also performed simulations of RX\,J1713.7$-$3946, the bright SNR proposed to 
study cosmic-ray acceleration. The gamma-ray emission from this object was simulated by 
assuming different emission mechanisms, supported by the extensive, existing 
multi-wavelength observations. To evaluate leptonic emission, we used an X-ray image of 
RX\,J1713.7$-$3946 from \emph{XMM-Newton} observations as a template that 
traces the gamma-ray morphology. We considered a simplified case where 
the gamma-ray spectrum is spatially independent. For the hadronic case, 
we obtained the total target gas distribution based on CO and HI observations and
used it as a template that traces the gamma-ray morphology. 
To evaluate
different levels of hadronic and leptonic distributions, we considered several 
cases with different values of $A_{\mathrm{p}} / A_{\mathrm{e}}$,
where
$A_{\mathrm{e}}$ and $A_{\mathrm{p}}$ are the
leptonic and hadronic normalisation
parameters, respectively.
The absolute values of these normalisation parameters
were determined by requiring that the integration of 
the sum $N_{\mathrm{e}}(E) + N_{\mathrm{p}}(E)$ over the remnant 
equaled the total photon flux measured by H.E.S.S. 

Figure~\ref{fig:ksp_j1713morph} (left and middle) show 
the images for $A_{\mathrm{p}} / A_{\mathrm{e}} =$ 0.01 and 100, respectively.  Each
image assumes 50~h observations with CTA. 
The lepton-dominated case (Figure~\ref{fig:ksp_j1713morph}, left) shows a 
gamma-ray image similar to the X-ray image, and the hadron-dominated image is 
(Figure~\ref{fig:ksp_j1713morph}, middle) similar to the interstellar proton distribution
including both CO and HI. The difference between the two cases
is significant as shown in the 
subtracted image (Figure~\ref{fig:ksp_j1713morph}, right).
We note here that another obvious target for morphological investigations would be the SNR Vela Junior \cite{Abdalla16b}.

\begin{figure}[t!]
\centering
\includegraphics[width=0.75\linewidth]{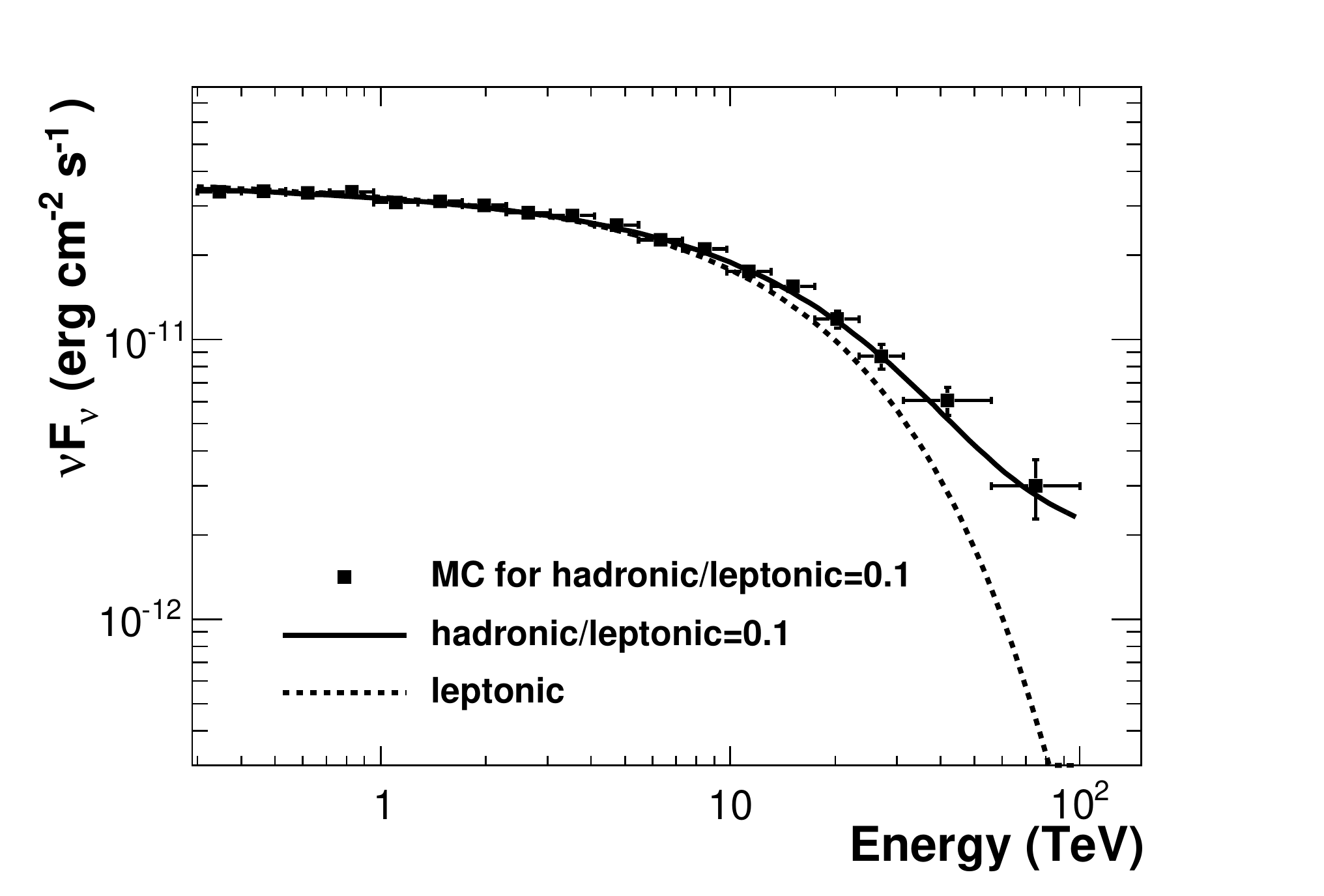}
\resizebox{0.70\columnwidth}{!}{\includegraphics[clip=true]{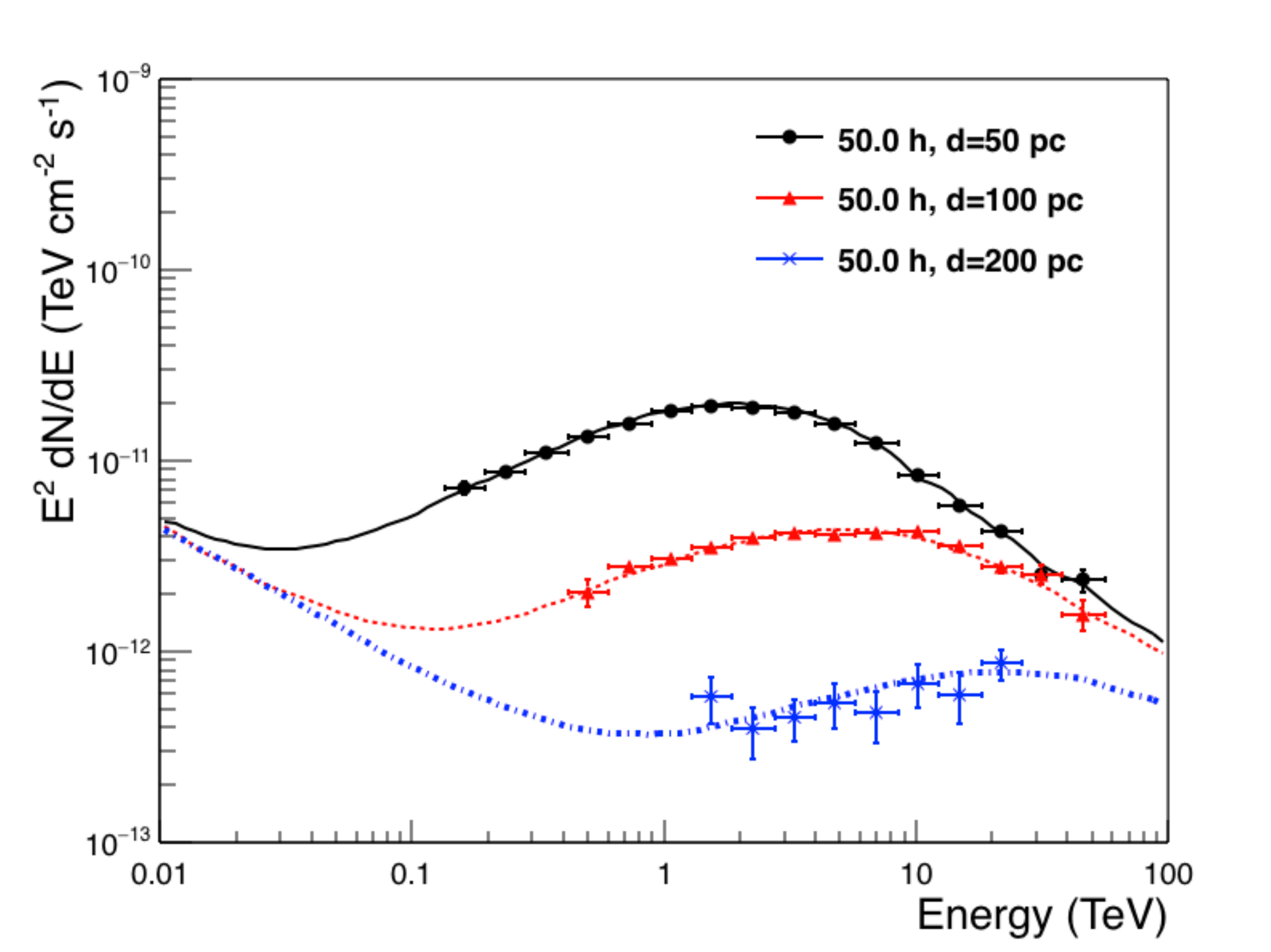}} 
\caption{\emph{Top:} Simulation of the different spectral shapes resulting from
different scenarios considered for SNR RX\,J1713.7$-$3946. 
Reproduced from~\cite{Nakamori14}.
The black 
squares show the total of the fluxes for the leptonic and hadronic spatial 
templates. The solid line shows the input spectra of the gamma-ray simulation. The 
dotted line is for the model where the emission is only due to leptonic processes.
\emph{Bottom:} Gamma-ray spectra for a molecular cloud illuminated by cosmic rays coming from a
nearby SNR (lines) and simulated observations (data points) for a 50~h CTA
observation. See Ref. \cite{Acero13}.
Black solid, red dashed and blue dotted lines 
correspond to distances between the SNR and the cloud of 50, 100, and 200 pc, 
respectively. The distance of the cloud to Earth is 1 kpc, the cloud mass is 
$10^5 \mathrm{M}_{\odot}$, and the SNR age is 2000 years.}
\label{fig:ksp_j1713spec}
\end{figure}

In addition to the morphological studies, the expected spectra for different 
scenarios were simulated for RX\,J1713.7$-$3946 \cite{Acero13}  
(Figure~\ref{fig:ksp_j1713spec}, top). The results show a clear discrepancy 
detected at high energies according to different origins of the gamma-ray 
radiation in the supernova.  

As a further study, we simulated the response of CTA to the gamma-ray 
emission from a massive cloud ($\sim 10^5 \mathrm{M}_{\odot}$) illuminated by a 
nearby, young SNR which injected in the interstellar medium $3 \cdot 10^{50}$ erg of 
kinetic energy in cosmic rays.
The cloud was assumed to be located at different distances
(50, 100 and 200 pc) from the SNR itself.  
Cosmic rays escaping from the accelerator can 
diffuse in the turbulent interstellar magnetic field and 
illuminate the cloud to produce gamma rays due to proton-proton collisions.
Apart from being an indirect way to identify cosmic-ray sources, studies of illuminated
molecular clouds are notable in that they can be used to estimate the cosmic-ray diffusion 
coefficient in the vicinity of the accelerators, since the properties of the 
expected gamma-ray emission depend on this value as well as its energy 
dependence \cite{Gabici07}. Figure\,\ref{fig:ksp_j1713spec} (bottom) shows results from these simulations, performed assuming a typical cosmic-ray diffusion coefficient equal to $10^{28} (E / \mathrm{GeV})^{0.5}$ and a distance to the SNR from Earth of 1 kpc. The
expected spectra exhibit a concave shape that varies in a broad energy range 
between 100 GeV and tens of TeV and depends on the distance between the SNR and the
cloud(s). The spectrum is further dependent on the 
SNR age and the assumed value of the 
diffusion coefficient. These dependencies can be parameterised and compared 
with future observations in order to extract or constrain physical 
parameters such as the cosmic-ray 
diffusion coefficient and the source's cosmic-ray 
acceleration efficiency \cite{Pedaletti:2013}.
For a nearby source ($d \sim 1$ kpc), the cloud is detectable if located within a few hundred parsecs from the cosmic-ray 
accelerator. This corresponds to an angular distance of $\approx 6^{\circ} (l/100~{\rm pc}) (d/{\rm kpc})^{-1}$, where $l$ is the distance between the SNR and the cloud and
$d$ is the distance to Earth.
This angular scale
is of the same order as the field of view of CTA.
Thus, for nearby SNRs, illuminated clouds should be searched for in the GPS data and 
then repointed to with follow-up observations to 
obtain a deeper exposure and better determination of the spectrum and morphology.

\section{KSP: Star Forming Systems }
\label{sec:ksp_sfs}

Cosmic rays are believed to be an important regulator of the
star-formation process. It is hence important to understand where
cosmic rays are being accelerated, how they propagate, and where they
interact in the interstellar medium (ISM). While travelling through
the ISM, cosmic rays interact with the ambient gas and radiation
fields to produce gamma rays, which trace directly the parental
cosmic-ray population. Gamma rays are thus among the best tools to
study cosmic-ray properties in star-forming environments. CTA
observations of star-forming systems over six orders of magnitude in
the star-formation rate (SFR) will help to unveil the relationship
between high-energy particles and the star-formation process. The
study of individual star-forming regions and star-forming galaxies
will furthermore help to disentangle source-specific properties from
global ones. By studying the gamma-ray emission in these systems, we
can measure the fraction of interacting high-energy particles as a
function of the SFR and hence investigate to which extent cosmic rays,
magnetic fields and radiation are in equipartition.

Within the Galaxy, observations of the Carina and Cygnus regions and
the most massive stellar cluster Westerlund~1 will allow us to: a)
constrain the fraction of mechanical stellar wind energy transferred
into gamma rays down to a level of $10^{-8}$, b) study particle
acceleration in Galactic stellar clusters and superbubbles, and c)
search for signs of cosmic-ray propagation and interaction with the
ISM via spectra-morphological studies.

Outside the Galaxy, the Large Magellanic Cloud (LMC) is the only other
galaxy for which CTA will be able to resolve the very high-energy
(VHE) gamma-ray source population and study in detail the similarities
and differences to our own Galaxy. Although important for the purpose
of this KSP, the LMC is discussed separately in
Chapter~\ref{sec:ksp_lmc} and is only addressed here for modeling
purposes. Observations of the Andromeda galaxy will provide important
measurements and estimates of cosmic-ray properties and diffusion in
the nearest spiral galaxy.

Long observations of the two starburst galaxies NGC\,253 and M\,82
will enable us to test how cosmic rays traverse the ISM,
to distinguish between the truly diffuse emission and individual
source populations such as pulsar wind nebulae (PWNe), and to possibly
resolve the starburst nucleus in VHE gamma rays. Observations of the
only ultraluminous infrared galaxy (ULIRG) likely within the reach of
CTA, Arp\,220, will for the first time allow us to test cosmic-ray
properties in a system where all accelerated particles are expected to
interact -- i.e. a calorimetric system -- and might establish ULIRGs
as a new source class in VHE gamma rays \footnote{see e.g. the recent
  tentative detection of GeV gamma rays from the direction of Arp\,220,
  reported in \cite{Peng16, Griffin16}.}.

Deep observations of at least one object per decade in the SFR are
required to investigate the relationship between star formation and
gamma-ray emission and to study the impact of cosmic rays on their
environment from the smallest to the largest scales. Table
\ref{tab:sfstargets} summarises the list of objects proposed for
observations within this KSP, of which four are also part of other
KSPs. The observational program is well suited to be a KSP to be
carried out by the CTA Consortium for a number of reasons.  The total
proposed observation time of 720 hours could be difficult to obtain in
a single Guest Observer (GO) program.  Splitting observations into
separate GO projects, with possible different foci and too low an
exposure would require re-observation of targets and would significantly
delay these timely studies. Furthermore, the proposed targets require
excellent control of instrumental systematics. For example, the
predicted emission from M\,31 and Arp\,220 is at the sensitivity limit
of CTA and the central molecular zones of NGC\,253 and M\,82 have
sizes close to the CTA point spread function (PSF).  The expected
extension of sources in Cygnus covers a large fraction of the
field of view (FoV) of CTA and mandates an excellent understanding of
the residual cosmic-ray background.  Moreover, the level of optical
background light in the Carina region and in M\,31 varies by more than
an order of magnitude and will require the development of specialized
analysis tools.

This KSP will provide legacy data products for the entire astronomical
community such as catalogues of sources, flux maps and data cubes
(gamma-ray excess maps binned in energy) for the Cygnus and Carina
regions. These will be useful to community members to prepare deeper
observations for later, open time proposals.

\begin{table*}
  \centering
  \begin{tabular}{l|ccc}
    \hline
    Target & Type & Extension & SN rate (yr$^{-1}$) \\\hline\hline
    Carina$^\dagger$ & star-forming region & $\sim$$1.5^\circ$ & $1.7\times 10^{-4}$\\
    Cygnus$^\dagger$ & star-forming region & $\sim$$2.5^\circ$ & $1.0\times 10^{-4}$\\
    Wd~1$^\dagger$ & stellar cluster & $\sim$$2.2^\circ$ & $1.0\times 10^{-4}$\\ \hline
    M\,31 & galaxy & $\sim$$2.0^\circ$ & $2.5\times 10^{-3}$\\\hline
    NGC\,253 & starburst & $\sim$$0.1^\circ \times 0.5^\circ (20'' \times 45'')$ & 0.03 \\
    M\,82 & starburst & $\sim$$4.8' \times 4.8' (15'' \times 50'')$ & 0.2\\\hline
    Arp\,220 & ULIRG & $\sim$$2.5'' \times 1''$ & 4\\\hline
  \end{tabular}
  \caption{Summary of targets proposed for observations in this
    Key Science Project, along with the expected extension of the gamma-ray emission 
    and estimated supernova (SN)
    rates. SN rates in Galactic objects have been derived from the star-formation rate 
    (Carina) and/or stellar remnants (Cygnus, Wd~1). The sizes for
    NGC\,253 and M\,82 are for the entire galaxy and the sizes of the starburst
    regions are in parentheses.}
  \label{tab:sfstargets}
\end{table*}

\subsection{Science Targeted}

Understanding the fundamentals of the star-formation process is one of
the great challenges in modern astrophysics (see
e.g. \cite{Kennicutt12}). Through the ionisation of ISM material,
cosmic rays affect astrochemistry and mediate the interaction of ISM
material with magnetic fields, both of which are important influences
on molecular cloud structure and star formation. Very massive stars
undergo supernova (SN) explosions at the end of their lives and enrich
the ISM with heavy elements, necessary for the evolution of life on
Earth. Supernova remnants (SNRs), the leftovers of SN explosions, are
believed to be a major source of cosmic rays, which ionise the
surrounding medium of SNRs, form a significant pressure component in
the ISM and amplify magnetic fields. Thereby, they can suppress or
enhance star-formation in individual molecular clouds and entire
galaxies \cite{Socrates08, Jubelgas08}. The acceleration, propagation
and interaction of cosmic rays are hence crucial to 
understanding the evolution of the building blocks of the universe on
all scales: from stars to stellar clusters, from giant molecular
clouds to galaxies and clusters of galaxies. Cosmic rays are believed
to amplify magnetic-field fluctuations upstream of the blast wave
shock \cite{Bell04}, and strong magnetic fields may hamper the
star-formation process. Cosmic rays escape their acceleration sites
and low-energy cosmic rays are especially a dominant source of
ionisation in the Galaxy. They penetrate deep into the dense cores of
molecular clouds - much deeper than ultraviolet radiation. There they
initiate a complex chemistry and influence the star-formation process
\cite{Ceccarelli11, Papadopoulos13}. There is also increasing evidence
that cosmic rays are dynamically important in galaxy formation and
need to be considered in cosmological simulations \cite{Booth13,
  Salem14b}.

SNR shells are not the only sources discovered at TeV gamma-ray
energies. The population of Galactic VHE gamma-ray sources to date
comprises a variety of other object types: SNRs interacting with
molecular clouds, pulsars, PWNe, and gamma-ray binary systems. All
these objects are associated with late stages of stellar evolution,
and they cluster tightly along the Galactic mid-plane with a scale
height similar to that of the molecular gas and of the regions of star
formation. Moreover, gamma-ray emission from massive star-forming
regions and stellar clusters has been detected \cite{Aharonian02,
  Abramowski12a}, which can be seen as an indication for cosmic-ray
acceleration in strong stellar winds as proposed in, e.g., Ref. \cite{Casse80}.

\begin{figure}[thbp!]
\begin{centering}
  \resizebox{0.95\columnwidth}{!}{\includegraphics{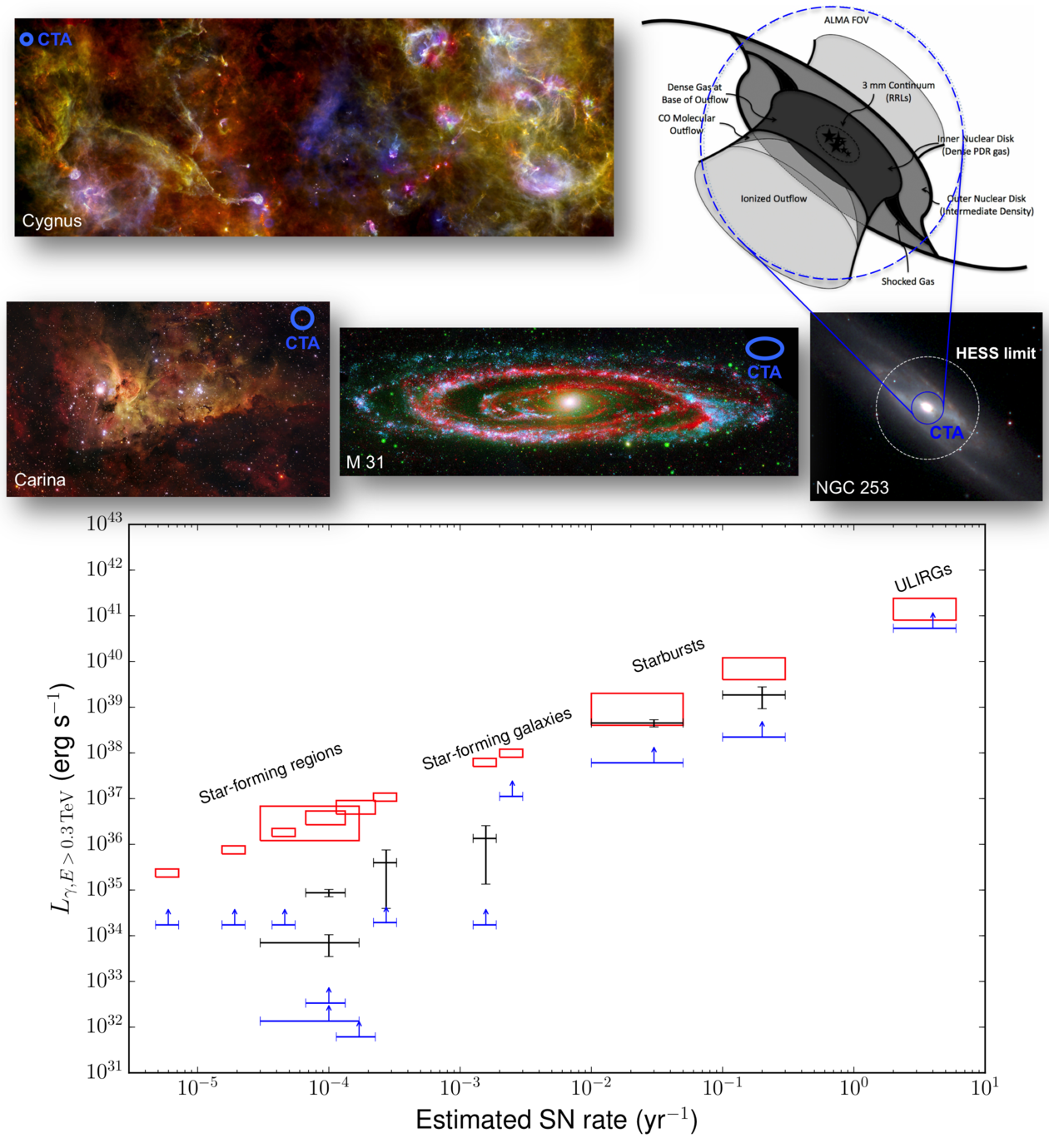}}
  \caption{ {\bf Top:} Representative multi-wavelength (MWL) images of
    four objects addressed in this KSP and the expected CTA
    performance. For Carina and Cygnus, blue circles indicate the CTA
    resolution; for M\,31 they indicate the maximum extension CTA will
    be able to detect and for NGC\,253 the minimum extension CTA will
    be able to resolve. For NGC\,253, the dashed white circle is the
    H.E.S.S. extension limit. Image credits: Cygnus - ESA/PACS/SPIRE,
    Martin Hennemann \& Fr\'ed\'erique Motte; Carina - ESO/IDA/Danish
    1.5 m/R.Gendler, J-E. Ovaldsen, C. Th\"one, and C. Feron; M\,31 -
    NASA/JPL-Caltech/K. Gordon (Univ. of Arizona) \& GALEX Science
    Team; NGC\,253 - 2MASS, WISE, \cite{Meier15}. {\bf Bottom:} The
    expected calorimetric gamma-ray luminosity of star-forming
    regions, stellar clusters, star-forming galaxies, starbursts, and
    ULIRGs in red. The sizes of the boxes represent uncertainties in
    the SFR and the estimated calorimetric gamma-ray flux. Blue arrows
    indicate the expected CTA sensitivity for the anticipated
    observation time. Black points indicate measurements in the TeV
    domain, or, where objects are only detected in GeV gamma rays,
    extrapolations to VHE gamma rays, based on the Fermi-LAT
    spectra. In case SN rate estimates do not exist, the 70\,$\mu$m
    flux is used to infer the SFR \cite{Lawton10} and subsequently
    translated into a SN rate based on the scaling relation
    $\nu_{\rm SN} = (0.010 \pm 0.002)\,\cdot\,{\rm SFR} ({M}_{\odot} /\rm{yr})$. \\\\
    {\it 1) The 30\,Doradus and LMC estimates have been derived by
      extrapolating the Fermi-LAT measurement with a broken power-law
      ($\Gamma_1 = 2.0, \Gamma_2=2.4$, $E_{\rm c} = 1$\,{\rm TeV}) and
      assuming a
      90\% error on the integral flux.\\
      2) The M\,82 flux has been estimated by combining the VERITAS
      and Fermi-LAT measurements and assuming a 50\% error on
      the integral flux.\\
      3) The Cygnus flux has been derived by extrapolating the
      Fermi-LAT measurement, assuming a power-law spectrum with index
      $\Gamma=2.2$ and a 50\% error on the integral flux.}  }
  \label{fig:SFR}
\end{centering}
\end{figure}

Most of the light of young massive stars is radiated at ultraviolet
wavelengths, absorbed by dust and re-emitted at infrared
wavelengths. Kennicutt \cite{Kennicutt98} has shown that the
infrared luminosity and the star-formation rate of galaxies follow a
linear relation. A similar scaling relation between infrared
luminosity (and radio luminosity) and high-energy gamma-ray luminosity
in a sample of nearby galaxies has recently been found 
\cite{Ackermann12}, which suggests a direct connection between the
process of star-formation and the high-energy particles. Although this
relation is striking, the spread in gamma-ray luminosity of one order
of magnitude indicates that the efficiency with which gas is
channelled via massive stars into high-energy particles 
is different in individual objects. This points towards a complex
interplay between star formation, particle acceleration and escape and
feedback with the ISM in these systems.

\subsubsection{ Scientific Objectives}

Gamma rays provide a powerful tool to study cosmic-ray properties in star-forming
environments and to thereby unveil the relationship between
high-energy particles and the star-formation process. Although
theoretical work has made major progress in recent years, in-depth
studies of star forming systems on all spatial scales with the next generation CTA are
required to answer the most pressing questions, including:
\begin{enumerate}
\item What is the relationship between star-formation and particle
  acceleration in systems on all scales? Does a universal
  far-infrared/TeV luminosity relationship exist?
\item How does the calorimetric fraction change as a function of the SFR
  and does equipartition hold in star forming systems?
\item What is the contribution of different source classes to the
  cosmic-ray population in star forming systems? Where and when are particles
  accelerated, how do they leave and what is their impact on the
  surrounding ISM?
\end{enumerate}

CTA observations of Galactic and extragalactic massive stellar
clusters and star-forming regions, of spiral and starburst galaxies,
and ULIRGs will allow us to address these questions. The study of
individual star-forming regions and entire systems will furthermore
help to disentangle source-specific properties from global ones. For
example, the starburst episode in starburst galaxies and ULIRGs only
lasts for tens of millions of years, but it plays a key role in galaxy
evolution and the star-formation history of the universe. By studying
the gamma-ray emission in these systems we can measure the fraction of
high-energy particles that interact as a function of the SFR and hence
investigate to what extent cosmic rays, magnetic fields and radiation
are in equipartition. Figure~\ref{fig:SFR} illustrates the potential
of CTA to study star forming systems on all scales and compares the predicted
calorimetric flux from existing measurements to the sensitivity of
CTA. To estimate the calorimetric limit, we follow Ref. \cite{Abramowski12b}
and assume that $10^{50}$\,erg per SN explosion are
transferred into protons, which interact and channel 1/6 of their
energy into gamma rays that follow a power-law spectrum with differential
index $\Gamma=2.2$. The fraction that is then radiated at energies
$E>0.3$\,TeV is 5\%. For the LMC and 30\,Doradus the 70$\mu$m flux is
used to infer the SFR \cite{Lawton10} and subsequently the SN
rate. For Cygnus and Westerlund~1 the SN rate is derived from the number and
ages of stellar remnants and for Carina from the SFR alone.

\subsubsection{ Context / Advance beyond State of the Art}

So far, only very few star forming systems have been discovered at TeV
gamma-ray energies, and those that could be detected are at the limit
of current-generation instruments. These systems are either at the sensitivity
limit of H.E.S.S., MAGIC, or VERITAS, too extended for detailed
spectro-morphological studies, or are not yet resolvable. Furthermore,
several of the star forming systems seen at GeV energies with
Fermi-LAT are as yet undetected in the TeV regime, with predicted flux
levels at or just below the sensitivity of current-generation
Cherenkov telescopes. For example, the starburst galaxy NGC\,253 is
the weakest VHE gamma-ray source detected so far \cite{Abramowski12b},
the massive stellar cluster Wd~1 is one of the largest TeV sources in
the sky \cite{Abramowski12a} and the ULIRG Arp~220 has a predicted
flux level which requires a long exposure, even with CTA.

Compared to observations at GeV energies, available for all the
systems discussed here, observations with CTA are highly
complementary. CTA will probe higher gamma-ray energies and hence
higher particle energies that are crucial to understand the overall cosmic-ray
population in star-forming systems. Thanks to the much higher photon statistics and
superior point spread function above $\simeq$50\,GeV, CTA will provide much more
detailed gamma-ray spectra and high-resolution images than Fermi-LAT. Fermi-LAT is
sensitive to large-scale interstellar emission in the Milky Way and
nearby galaxies, whereas CTA will probe younger particles close to
their acceleration sites.

\paragraph{Star-forming regions}
The Carina nebula and the Cygnus region are exceptionally luminous and
massive Galactic star-forming regions, and Wd~1 is the most massive
stellar cluster in the Galaxy. As such they are very unique systems,
different in many respects, and their study at GeV and TeV energies
will help to understand where particles in star-forming regions are
accelerated and how they escape, and, along with MWL observations,
the observations will reveal what their impact on the ISM is.

The {\bf Carina nebula} is one of the largest, most active and best
studied H{\sc ii} regions in our Galaxy and is a place of ongoing star
formation. It harbours eight open stellar clusters with more than 60
O-type stars, three Wolf-Rayet stars and Eta Carinae, the only
colliding-wind binary (CWB) firmly established to emit GeV gamma rays
(see e.g. \cite{Reitberger15}). The age estimates of the Carina
clusters, Tr14, Tr15 and Tr16, indicate several past episodes of star
formation in the northern region and more recent star formation
ongoing in the southern part of the nebula \cite{Preibisch11}. The
study of extended X-ray emission as seen with Suzaku, XMM-Newton and
Chandra \cite{Hamaguchi07, Ezoe09, Townsley11} suggests that the
diffuse plasma originates in one or several unrecognised SNRs and/or
may be attributed to stellar winds from massive stars. Based on the
gas content in the Carina nebula and a limit on the gamma-ray flux,
H.E.S.S. constrained the cosmic-ray enhancement factor in this region
to be less than one order of magnitude above the cosmic-ray sea
\cite{Abramowski12g}.

The Carina arm region also hosts the very young, but as yet undetected
at gamma-ray energies, starburst cluster NGC\,3603, where no SN
explosion is expected to have occurred. The second most massive
Galactic stellar cluster, Westerlund~2, is associated with a TeV source,
and it harbours many massive stars and remnants of a few stellar
explosions \cite{Abramowski11}. Westerlund~2, NGC\,3603 and Eta
Carinae provide complementary information on how much kinetic stellar
wind energy is channeled into cosmic rays. The proposed CTA
observations will provide the required spatial coverage and flux
sensitivity (in a comparably ``clean'' environment) to constrain the
fraction of input mechanical stellar wind energy that is converted
into gamma-ray emission in stellar clusters to a level of
$2 \times 10^{-8}$ in NGC\,3603, $5 \times 10^{-8}$ in Tr16, and
$2 \times 10^{-7}$ in Tr14.  Furthermore, the observations will allow
us to probe the high-energy end of the Eta Carinae spectrum and to
answer the question to what extend CWBs contribute to the Galactic
cosmic-ray population.

{\bf Cygnus} can be seen as a small-scale version of a starburst,
harbouring all types of known Galactic TeV sources. Cygnus is one of the
most promising parts of the Galactic plane to address questions
related to particle acceleration and high-energy phenomena in massive
star-forming regions. Cygnus has a total mass in molecular gas of a
few million solar masses and a total mechanical stellar wind energy
of $\gtrsim$10$^{39}$\,erg\,s$^{-1}$. This corresponds to
several per cent of the kinetic energy input by SNe into the entire
Galaxy. A wealth of gamma-ray sources discovered by the Fermi-LAT,
HEGRA, Milagro, VERITAS, MAGIC, HAWC, and ARGO-YBJ collaborations
prove that ongoing particle acceleration proceeds in the Cygnus region
up to hundreds of TeV energies. The first ever unidentified source
discovered at TeV energies, TeV~J2032+4130, is located in the
direction of the Cygnus OB2 association \cite{Aharonian02}. Fermi and
VERITAS discovered GeV and TeV emission towards the Gamma Cygni
SNR, respectively. Most importantly, Fermi-LAT data revealed the presence of a
cocoon of freshly-accelerated cosmic rays stretching between Cygnus
OB2, the NGC 6910 open cluster and Gamma Cygni \cite{Ackermann11}. The cocoon provides evidence of particles escaping
a close-by accelerator and being confined by the magnetic fields in
the extreme environment of massive stellar clusters. The ARGO-YBJ
collaboration claimed the identification of the source ARGO~J2031+4157
(associated with the Milagro source MGRO J2031+41) with the cocoon
\cite{Bartoli14}.

Next to Orion, Cygnus is possibly the best studied star-forming region
in the Milky Way across the entire electromagnetic spectrum, and it
has been targeted in large Herschel, Spitzer and Chandra
programmes. At VHE gamma-ray energies, a survey of the region to a flux level
of a few percent of the Crab nebula was made
by VERITAS \cite{Popkow15}. 
Once CTA is operational, HAWC will have collected
$4-5$ years of data on the Cygnus region and should reveal even more
TeV emitters. CTA will provide the necessary angular resolution, flux
sensitivity and energy coverage above $\sim$50\,GeV to disentangle the
gamma-ray sources that pile up and overlap in this region and to identify
their low-energy counterparts. CTA will furthermore study the
propagation of cosmic rays in the ISM, map the spatial cosmic-ray
energy density distribution and thereby study the correlation between
cosmic rays and active star-formation sites. Given the proximity of
Cygnus to Earth, CTA observations will provide the best linear
resolution and energy flux sensitivity compared to any other
star-forming region in the Galaxy or the LMC.

{\bf Westerlund~1 (Wd 1)} is the most massive stellar cluster in the
Milky Way and contains a very rich population of massive stars,
including a magnetar. H.E.S.S. observations have revealed an emission
region twenty times larger than the size of the cluster. The detailed
spectral and morphological studies demonstrated that a significant
part of the emission may well originate from Wd~1 and that protons are
most likely responsible for the emission \cite{Abramowski12a}. The
apparent size of the emission region also supports the idea that
particles are accelerated in the cluster, and then escape and interact with the
surrounding material and a nearby H{\sc ii} region. This scenario is
further supported by the morphology of the H{\sc i} gas in the region,
which shows a bubble-like structure where dense gas partly overlaps
with regions of strong TeV emission \cite{Kothes07,
  Abramowski12a}. The analysis of Fermi-LAT data revealed an extended
emission region, offset from Wd~1, and only partly overlapping with
the TeV emission \cite{Ohm13a}. This could imply that energy-dependent
and anisotropic diffusion, and advection, are at work in this system,
or that this region harbours multiple particle accelerators. The very
large amount of energy in cosmic rays of $10^{51}$\,erg required to
explain the H.E.S.S. data can easily be provided by the stellar winds
and multiple supernovae that went off in Wd~1.

Deep CTA observations of this region will provide the required spatial
and energy coverage to perform a high-resolution study of the
energy-dependent morphology of the emission and to compare Wd 1
to the Cygnus~X superbubble (which is detected at GeV energies) and to
30\,Dor~C (the TeV detected superbubble in the LMC). Furthermore, the
VHE gamma-ray spectrum shows no indication of a cutoff up to 10\,TeV,
although the spectrum could only be reconstructed with limited
precision given the large size of the emission region. CTA
observations are crucial to reconstruct the spectrum with much
higher precision, determine the maximum energy to which particles can
be accelerated in stellar clusters, and probe if these objects are
potential PeV accelerators. In the same field of view, the most
luminous TeV-emitting Galactic SNR, HESS~J1640$-$465,
HESS~J1641$-$463, and the gamma-ray pulsar PSR~B1706$-$44 are
located. Early observations of the Wd~1 region will demonstrate the
great potential of CTA for the study of very extended objects with
complex morphologies and for the separation of close-by sources.

\paragraph{Star-forming galaxies} 

The {\bf Andromeda galaxy (M\,31)} is the nearest spiral galaxy at a
distance of $\sim$780\,kpc to Earth. It has a similar SFR as the Milky
Way and can hence be considered as its twin. Studying the similarities
and differences in gamma rays will provide important insights on the
global cosmic-ray properties of M\,31 and our own Galaxy. 
M\,31 
is detected by Fermi-LAT and is consistent with a
point-like source but with an indication of a marginal spatial
extension (see Fig.~\ref{fig:SFR} for the CTA capabilities to detect
such an extension) \cite{Abdo10f}. The GeV spectrum is noticeably
harder than for the Milky Way, and the ratio of gamma-ray luminosity
over SFR is higher by a factor of three. The clear advantages compared
to our own Galaxy are the outside, unbiased view and the fact that the
SFR is much better known for M\,31. Observations of the Andromeda
galaxy with CTA within this KSP will provide invaluable insights into
cosmic-ray properties in a spiral galaxy, allowing us to extend the
measured gamma-ray spectrum to TeV energies and to search for truly
diffuse emission and for diffuse emission from individual source
populations such as PWNe.

\paragraph{Starburst galaxies}

The enhanced star-formation activity seen in starburst galaxies is
often triggered by a close fly-by of galaxies, a galaxy merger or
galactic bar instabilities. The extremely high thermal and non-thermal
particle pressure in the central regions of starburst galaxies often
leads to the formation of a galactic wind that enriches the
intra-cluster medium with heavy elements and that is proposed to be a
re-acceleration site of particles to energies beyond $10^{15}$\,eV
\cite{Dorfi12}. The high star-formation activity results in a highly
increased SN rate. This increased SN rate is predicted to lead to an
increased production of cosmic rays, which can then interact in the
starburst region to produce gamma rays \cite{Voelk89}. Indeed, the
emission of GeV and TeV gamma rays from the starburst galaxies {\bf
  NGC\,253 and M\,82} was detected, demonstrating that cosmic rays and
the star-formation process are connected on galactic scales
\cite{Acciari09, Abdo10, Abramowski12b}. These observations finally
provided evidence for proton acceleration in the two archetypal
starburst galaxies. The gamma-ray luminosity of starbursts depends on
the efficiency with which energy in cosmic rays is converted into
gamma-ray emission. If all the hadronic cosmic-ray energy is lost in
proton-proton collisions, the system can be considered {\it
  calorimetric}. Non fully-calorimetric systems on the other hand may
indicate a strong influence of winds. Indeed, only $\sim$$40\%$
of the cosmic rays in NGC\,253 and M\,82 seem to interact in the
starburst nucleus, with the remainder being advected away in the
starburst wind. The study of starburst galaxies at TeV energies will
test the details of the cosmic-ray physics.

Measuring the calorimetric fraction as a function of the SFR in
different starbursts and in ULIRGs tells us how much gamma-ray
luminosity is produced per given input cosmic-ray and inner starburst
gas density. Gamma rays are especially well-suited for this study
since starburst galaxies are an order-of-magnitude brighter in gamma
rays than in radio. A second key question is whether or not
equipartition holds in starburst environments and what the mechanism
is if it does. It is expected that the denser the starburst the less
in equipartition the environment will be. Equipartition may perhaps
hold in M\,82 and NGC\,253 (e.g. \cite{Domingo05, Rephaeli10}), but
it might not be realised in the ULIRG Arp\,220 (e.g. \cite{Thompson06,
  Lacki11}). Thirdly, CTA observations of star forming systems will
also shed light on how mixing of cosmic rays with ambient gas
works. Models of non-thermal emission from starbursts assume that
cosmic rays experience the volume-averaged gas density. But, this can
only happen if the cosmic rays leave the hot phase and enter the
molecular gas and H{\sc ii} regions. Additionally, we know that the
ISM, the magnetic field, and the injection of cosmic rays are not
necessarily homogeneous. Some works have expanded from these
assumptions by correlating radio emission, gamma-ray emission, and
star-formation to constrain cosmic-ray diffusion \cite{Murphy12}, by
creating three-dimensional steady-state models of cosmic rays in M\,82 and NGC\,253
\cite{Persic08, Rephaeli10} or by considering the time- and
space-dependent contributions of individual injectors of cosmic rays
\cite{Torres12}.

CTA observations will help to determine how cosmic rays traverse the
different gas phases in starbursts and which one regulates its
transport, or provides target material for hadronic interactions, or
contains the magnetic fields responsible for the radio
emission. Observations of both starbursts are important, as M\,82 is
brighter, located in the northern hemisphere and is unique as it
exhibits a strong wind. NGC\,253 on the other hand is located in the
south, has a weaker wind and offers the possibility to detect the
disk. For both galaxies, we can potentially spatially resolve the
starburst core. Most importantly, the two starbursts have
star-formation rates (and SN rates) 
that differ by a factor of ten and hence they probe
different regimes in the radio-gamma-ray relation. Finally, the
detailed spectral study of the brightest nearby starbursts M\,82 and
NGC\,253 may help distinguishing the diffuse starburst emission from
gamma-ray emission produced by individual populations of sources. This
is of particular importance for the more abundant Galactic TeV sources
such as PWNe, for which spectral features can appear \cite{Mannheim12,
  Ohm13b}.

\paragraph{ULIRGs} ULIRGs are mergers of galaxies, where much of the
gas from the spiral disks falls into a common centre of gravity. The
merger creates an extreme molecular environment $(>1000$\,cm$^{-3})$
within a small region of a few hundred parsecs size, triggering a huge
burst of star formation \cite{Sanders96}. The interstellar medium in
an ULIRG is similar to the inner parts of giant molecular clouds 
and photon energy densities in the infrared can reach up to
$1000$\,eV\,cm$^{-3}$, leading to strong radiative losses. ULIRGs are
relatively rare objects \cite{Sanders96}, with {\bf Arp\,220} being
the only ULIRG within 100\,Mpc ($z = 0.018$) of Earth. Arp\,220 is
also a very extreme object; a SN explosion is expected to occur every
six months, compared to $\sim$1 per century in the Milky Way or in
Andromeda (e.g. \cite{Pavlidou02}), and several individual SNRs are
visible at any time \cite{Smith98}. The SN rate in Arp\,220 is the
highest known in the local universe, making it the perfect object to
study cosmic-ray production, interaction with gas, escape, and
environmental feedback on galactic scales. Arp\,220 would be the first
object of this source class to be detected at gamma-ray energies (and,
in fact, at any energy beyond 10\,keV). The radio spectrum of Arp\,220
is explained as synchrotron and free-free emission and absorption from
primary and secondary electrons \cite{Torres04, Torres05}. The
inferred magnetic fields can reach the mG range in the most
extreme star-formation regions. The predicted gamma-ray emission of
Arp\,220 is on the verge of current instruments. Fermi-LAT
\cite{Ackermann12}, MAGIC \cite{Albert07a} and VERITAS
\cite{Fleischhack15} have reported upper limits. For the former, the
galaxy is expected to be detected within the 10-year mission
duration, and indeed \cite{Peng16, Griffin16} report a tentative
detection using more than 7 years of Fermi data. Moreover, all
correlations between star-formation indicators and gamma-ray
luminosity place Arp\,220 in a detectable range for Fermi-LAT and in
line with measured values for the Small Magellanic Cloud, the LMC,
M\,31, M\,82, NGC\,253, and others given sufficient integration
time. At TeV gamma-ray energies, CTA should detect Arp\,220 after a
deep exposure, although its distance will limit the detection to a
point-like source.

The timeline of CTA matches the launches of recent and upcoming major
instruments in radio, optical, millimetre and X-ray wavelengths. The
next generation of radio surveys in H{\sc i} and CO transition lines
with the upgraded ATCA system and the SKA pathfinder ASKAP are
underway; they will map the three-dimensional neutral gas distribution
of our Galaxy with unmatched spatial and velocity resolution and will
cover the Galactic stellar clusters and star-forming regions discussed
in this KSP. Millimetre and sub-millimetre instruments such as APEX,
IRAM, and ALMA will provide measures of the ionisation level of dense
molecular clouds induced by cosmic rays \cite{Ceccarelli11}.  
All these instruments will add information about the thermal energy
content of the hot gas and about the energy input from massive stars;
in this way, they will complement CTA in the study of
ultra-relativistic electrons via synchrotron radiation.

\subsection{Strategy}

A summary of the proposed targets, their exposure and observation
conditions are given in Table~\ref{tab:obscond}. Given the angular 
size of the
expected gamma-ray emission, the proposed targets will typically fit
into the CTA field of view, but 
more extended, Galactic star-forming regions such as Cygnus
may require a mini-scan.

\begin{table*}
  \centering
  \begin{tabular}{l|cccccc}
    \hline
    Target & Exposure (hrs) & Array & Year & Zenith & Moonlight
                                                      fraction & \\\hline\hline
    Carina$^\dagger$ & 100 & S & $1-3$ & $<45^\circ$ & 0\% \\
    Cygnus (OB1/OB2)$^\dagger$ & 130 & N & $1-2$ & $<50^\circ$ & 0\% \\
    Wd~1$^\dagger$ & 40 & S & $0-1$ & $<50^\circ$ & 25\% \\ \hline
    M\,31 & 150 & N & $2-5$ & $<45^\circ$ & 0\% \\\hline
    NGC\,253 & 100 & S & $1-3$ & $<40^\circ$ & 0\% \\
    M\,82 &  100 & N & $1-3$ & $<55^\circ$ & 0\% \\\hline
    Arp\,220 & 100 & S/N & $2-5$ & $<50^\circ$ & 0\% \\\hline
  \end{tabular}
  \caption{Summary of observation times and conditions for the
    proposed targets. Observation times for objects that are also part of the 
    Galactic Plane Survey (GPS)
    KSP (marked with a $^\dagger$; Chapter~\ref{sec:ksp_gps}), have
    been corrected for the effective exposure achieved by the GPS in
    the respective years.}
  \label{tab:obscond}
\end{table*}

\subsection{Data Products}

The proposed deep observations of star forming systems within this KSP
will provide numerous data products that will, according to the
proposed CTA data rights policy, be released to the public one year
after the data have been taken. Although the entire program will
be only finished by the end of year 5, data products on individual
targets will be released sooner. Since Wd~1 is a strong source,
observations of the Wd~1 region will be performed early on and could
be used to publish first science results in year 2, demonstrating the
potential of CTA. Data will be released to the public in the form of
maps (flux, hardness ratio) and data cubes (gamma-ray excess). The
Carina and Cygnus region have been observed in basically all
wavelength bands. CTA observations will provide rich legacy products
that are of great interest to a broad scientific community and will be
released in year 4. CTA data will also in this case be released in the
form of maps (flux) and data cubes (gamma-ray excess). Additionally
source catalogues, including spectral and morphological information,
and spectra and upper limits for source populations (PWNe, SNe, H{\sc
  ii} regions) will be released. Data obtained in observations of
M\,82 and NGC\,253 will provide spectral and morphological information
that will be released to the community in year 4 in the form of maps
and data cubes. The same data products will be released for M\,31 and
Arp\,220 towards the end of the program in year 6.

\subsection{Expected Performance/Return}

The expected performance and science return has been evaluated with
simulations, utilising internal CTA tools for M\,31 and using
the gammalib/ctools framework~\cite{Knodlseder16} for all other
sources.

\paragraph{Star-forming regions} 

\begin{figure}[thbp!]
\centering
\resizebox{0.5\columnwidth}{!}{\includegraphics{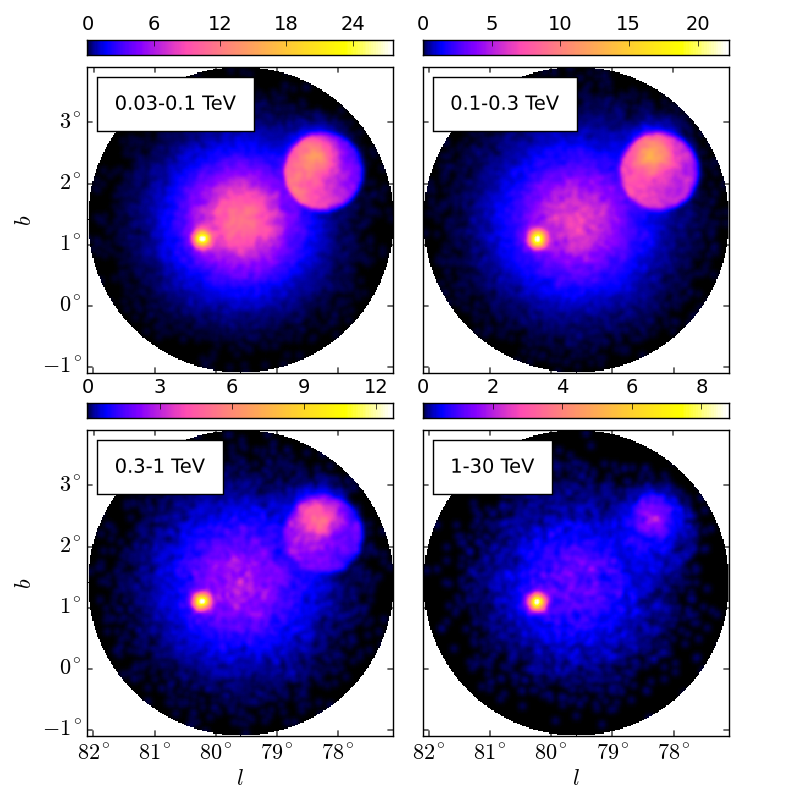}}
\resizebox{0.8\columnwidth}{!}{\includegraphics{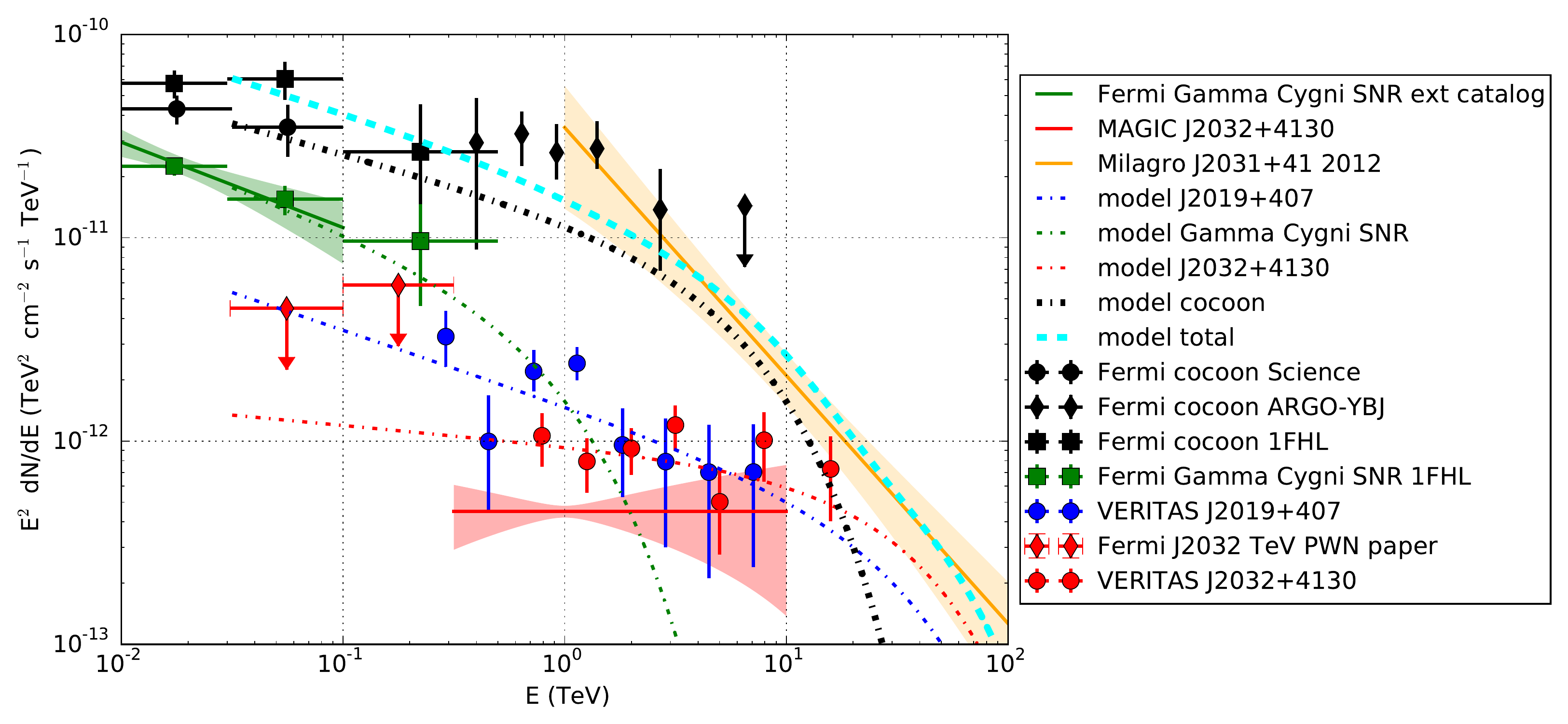}}
\caption{Simulated gamma-ray count map for CTA of the Cygnus OB2 region
  in four different energy bands (top) and spectra for gamma-ray
  sources in Cygnus OB2 (bottom), based on the observations proposed in
  this KSP.
   \label{fig:sims_cyg}}
\end{figure}

Simulations of two deep Cygnus pointings towards: a) Cygnus~OB2 and
the cocoon of freshly accelerated cosmic rays
and b) Cygnus~OB1 and VER~J2019+407 have been performed. With the
proposed observation time and based on the known sources as shown in
Figure~\ref{fig:sims_cyg}, the diffuse cocoon component can be
detected at the 18 standard deviation significance
level. This exposure is necessary, but also sufficient, to study the
energy-dependent morphology of this complex region in detail. It
furthermore provides the gamma-ray statistics required to detect
sub-structures in the cocoon, to separate faint sources from the truly
diffuse emission and to determine maximum particle energies in all
objects. The Cygnus~OB1 region shows strong diffuse emission and is
proposed for a medium-deep exposure to find the counterparts to the
Milagro and VERITAS sources and to investigate their relationship to
the numerous stellar clusters and H{\sc ii} regions. Simulations show
that all sources can be detected at level that will allow for a detailed
spectro-morphological study and a search for faint sources, embedded
in the strong diffuse emission. Cutoff energies of the known sources
can be reconstructed with high precision; in the case of the very hard
source VER~J2019+368 (spectral index $\Gamma=1.75$), the cutoff energy
can be reconstructed even up to energies of 80\,TeV.

The {\bf Carina region} hosts several stellar clusters and a huge
reservoir of gas. With the proposed CTA observations, cosmic-ray
enhancement factors in this region down to the cosmic-ray sea that
fills the Galaxy will be probed. Additionally, these observations will
test how much energy is channelled from stellar winds into gamma-ray
emission. For individual clusters, the fraction of wind-kinetic energy
going into gamma-ray emission will be measured down to a level of
$2 \times 10^{-8}$ for NGC\,3603, $5 \times 10^{-8}$ for Tr16, and
$2 \times 10^{-7}$ for Tr14. Simulations have been performed to
investigate the CTA potential to determine the Eta Carinae spectral
cutoff. Although a detection is possible within less than the proposed
observation time, multi-year coverage is required to constrain the
phase-dependent cosmic-ray escape into the ISM and the maximum energy
in this CWB.

\begin{figure}[thbp!]
\centering
\resizebox{0.95\columnwidth}{!}{\includegraphics{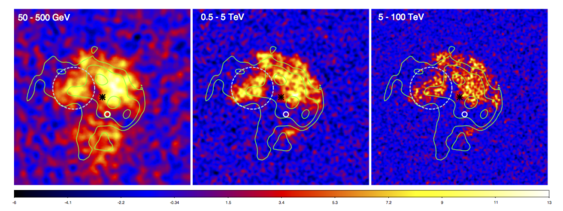}}
\caption{Expected Westerlund~1 gamma-ray excess maps for CTA in three different energy
  bands, smoothed with the CTA point spread function. H.E.S.S. contours are shown in
  green \cite{Abramowski12a}, the black star indicates the optical stellar cluster 
  and the magenta circle is the Fermi-LAT-detected
  PSR~J1648$-$4611. The light blue dashed circle shows the star-forming
  region G\,340.2$-$0.2.  See also \cite{Ohm13a}.
  \label{fig:Wd1sim}}
\end{figure}

Simulations for {\bf Westerlund~1} have been performed using the H{\sc
  i} data in that region as a tracer for the TeV emission, adding a
two-dimensional Gaussian source component as found in the Fermi-LAT energy band
\cite{Ohm13a}. As indicated in Figure~\ref{fig:Wd1sim},
CTA should be able to detect an energy-dependent morphology, resolve
substructures and, if the GeV and TeV emissions are indeed connected,
probe cosmic-ray propagation effects. The gamma-ray spectrum can be
reconstructed with significant spectral points extending up to
100\,TeV, if the Wd~1 spectrum has no intrinsic cutoff. The
observations can hence establish whether or not Wd~1 is a Galactic
PeVatron (see e.g. \cite{Bykov14}).

Additionally, the mechanical stellar wind energy input of the
star-forming regions correspond to $\sim$$0.1$\%,
for Carina, $\sim$1\%
for Wd~1 and a few percent for Cygnus~OB2 of the total mechanical energy
input by SNe in the entire Galaxy. If no emission is detected,
the CTA observations will put strong constraints on the contribution
of stellar winds as accelerators of Galactic cosmic rays.

\paragraph{Star-forming galaxies} 

Simulations of {\bf M\,31} have been performed assuming different
spectra and morphologies. If the Fermi-LAT measurement is extrapolated
to TeV energies, a point-like signal can be detected by CTA for
gamma-ray spectral indices of $\Gamma = (2.2 - 2.4)$ in the proposed
observation time. A maximum extension of $(0.1 \times 0.2)^\circ$ can
be detected only in the case that the diffuse emission has an index of
$\Gamma=2.2$ (or harder). If the diffuse gamma-ray spectrum of M\,31
extends to higher energies, has a steeper index, or exhibits a cutoff
(as, e.g. seen in the Milky Way), CTA will not be able to measure the
true diffuse emission. However, it will be possible to measure an
extended emission region of up to $(1.2 \times 0.3)^\circ$, in the
case of an additional source population that peaks in the TeV domain
(such as PWNe) and contributes to the gamma-ray signal (see
Fig.~\ref{fig:SFR}). In this case, an integrated signal of 4\% of the
Crab nebula flux with average source index $\Gamma_s = 2.2$ could be
detected.

\paragraph{Starburst galaxies} 

\begin{figure}[thbp!]
\centering
\resizebox{0.54\columnwidth}{!}{\includegraphics{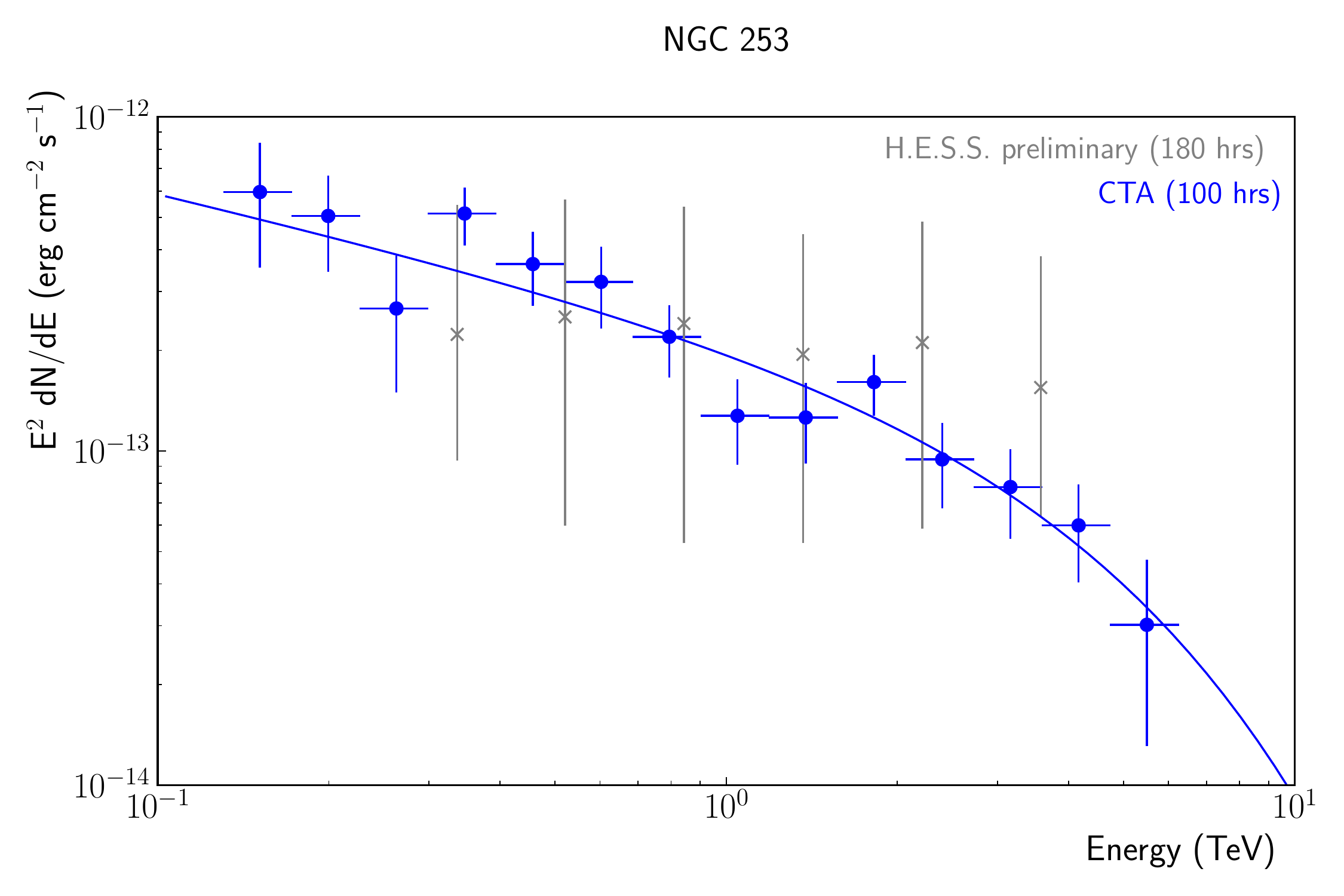}}
\resizebox{0.45\columnwidth}{!}{\includegraphics{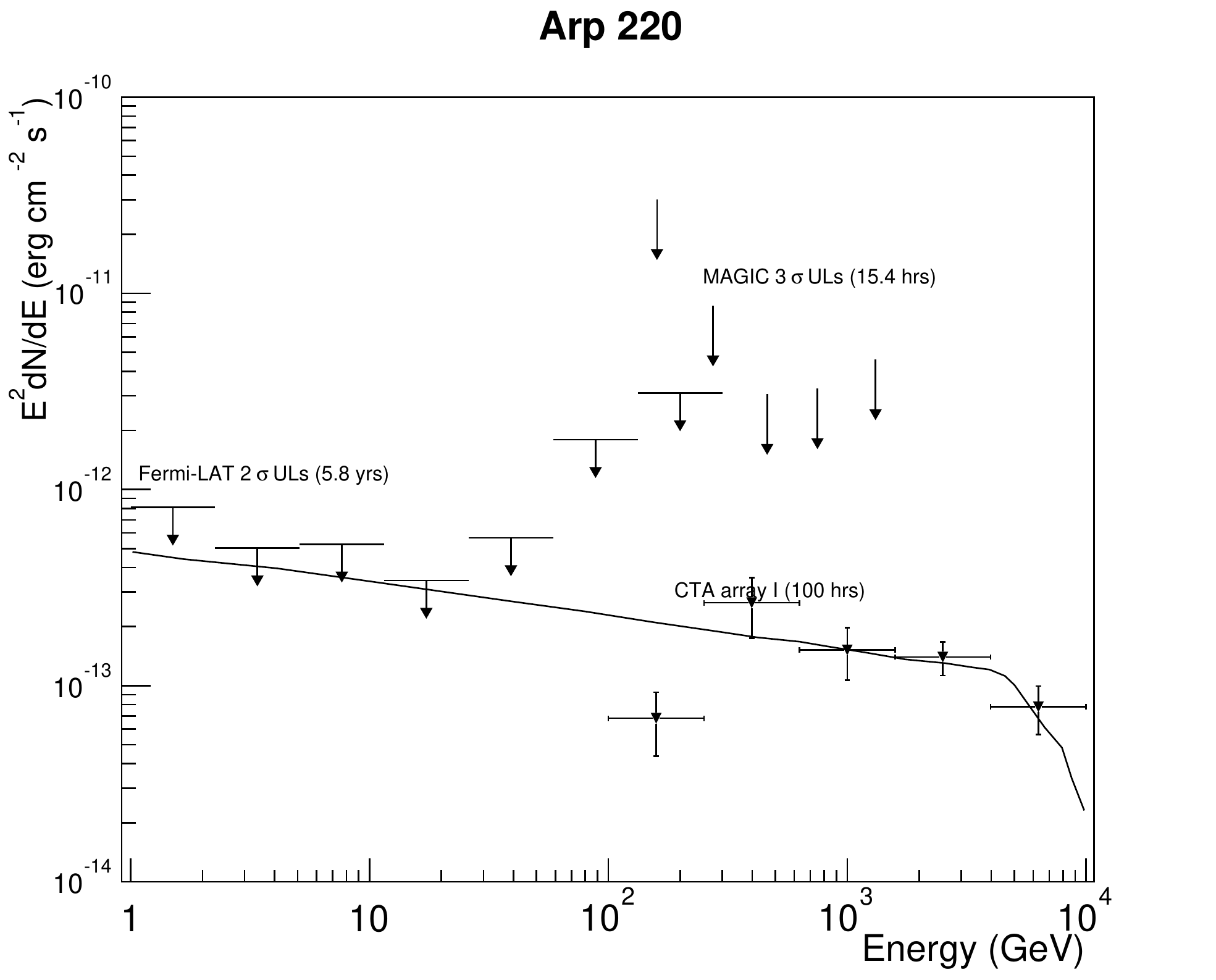}}
\caption{ Expected performance for CTA observations of NGC\,253 (left)
  and Arp\,220 (right). H.E.S.S. data are shown for NGC\,253 as gray points 
  (H.E.S.S. Collaboration, paper in preparation). For Arp\,220, the MAGIC limits
  and theoretical model presented in \cite{Albert07a} are depicted. The
  Fermi-LAT limits are obtained using 5.8 years of data.
  \label{fig:sims}}
\end{figure}

Simulations of NGC\,253 have been performed to study the CTA potential
to measure the extent of the starburst nucleus. ALMA and SMA
observations of the core of NGC\,253 indicate an extension of the
central molecular zone of $\sim$$0.35'\times0.75'$.
Resolving the starburst region in NGC\,253 would prove that the
gamma-ray emission correlates with regions of dense gas and hence
increased star-formation and SN activity. With the proposed
observations, a gamma-ray emission region as small as the ALMA field
of view, as indicated in Figure~\ref{fig:SFR}, could be detected with
three standard deviation 
statistical significance. This estimate does not include systematic
uncertainties, but it is also not optimised for a high-resolution
analysis. The same study will be performed for M\,82, which has a
slightly higher flux than NGC\,253, but it will need to be observed
with the northern array (where the sensitivity is somewhat worse than
in the south). Simulations have also been performed to test if the
disk component of NGC\,253 could be resolved. For an asymmetric
Gaussian-shaped region with corresponding widths $\sigma_1=13.8'$
and $\sigma_2=3.4'$,
the disk can be detected if it exhibits a flux higher than 50\% of the
core emission as seen by H.E.S.S. The cutoff (or gamma-gamma
absorption feature) in the gamma-ray spectra of NGC\,253 and M\,82
will be probed with high precision up to energies of $\sim5$\,TeV
-- the energy where gamma-gamma absorption features should appear
(Figure~\ref{fig:sims}, left; \cite{Acero13, Inoue11}).

\paragraph{ULIRGs} 

Simulations of {\bf Arp\,220} have been performed and reveal that if
the system is fully calorimetric as predicted by theory, the source
will be detected with CTA. A simulated energy spectrum of Arp\,220 is
shown in Figure~\ref{fig:sims} right. 

In summary, the simulations show that the CTA observations proposed in
this KSP will significantly expand our knowledge of cosmic-ray
accelerators in the Milky Way and in nearby galaxies and hence allow us to
study the connection between particle acceleration and star formation
as function of SFR and SN rate.

\FloatBarrier

\clearpage
\section{KSP: Active Galactic Nuclei }
\label{sec:ksp_agn}

Very high energy (VHE) observations of active galaxies harbouring supermassive black holes (SMBHs) and ejecting relativistic outflows represent a unique tool to probe the physics 
of extreme environments, including accretion physics, jet formation, interaction of the black-hole magnetosphere with the accretion disk corona, relativistic interaction processes, and general relativity. The same observations
also allow us to search for signatures of ultra-high energy cosmic rays (UHECRs) and to characterise the evolution and differentiation (through intrinsic diversity or interaction within the host galaxy) of some of the brightest cosmic sources through space and time. 
The use of gamma-loud active galactic nuclei (AGN) as beacons provides insights into the cosmological evolution of star and galaxy formation through constraints on photon fields and magnetic fields along the line of sight.
In addition, the study of VHE signals from extragalactic sources has a strong impact on the search for new fundamental physics. 

AGN are known to emit variable radiation across the entire electromagnetic spectrum up to multi-TeV energies, with fluctuations on timescales from several years down to a few minutes. 
At present, AGN make up about 40\% of the $\sim$180 sources detected at very high energies with ground-based telescopes. Apart from five nearby radio galaxies, all VHE AGN are blazars, i.e.\ their jets are closely aligned with the line of sight to Earth. The non-thermal multi-wavelength emission from blazars is characterised by two broad spectral bumps peaking in the optical to X-ray range and in
 the MeV to VHE gamma-ray range. Almost three quarters of blazars emitting in the VHE band are classified as 
high-frequency peaked BL Lac objects (HBLs), but there are also a few VHE blazars of other classes: flat-spectrum radio quasars (FSRQs), low- and intermediate frequency peaked BL Lac objects 
(LBLs and IBLs, respectively) and a newly defined class of ultra-high-frequency peaked BL Lac objects (UHBLs, EHBLs or ``extreme blazars''), with spectral peaks of the high-energy bump above $\sim$1\, TeV. 
The highest redshift of this sample of VHE detected sources is z$\sim$0.9, and there is some evidence of the detection of photons above 100\,GeV for redshifts as large as 1.5 \cite{Armstrong16}. 
The currently known population of VHE AGN is still very limited with respect to the coverage of different classes and redshifts. 

CTA has the potential to substantially improve this coverage and increase the population of VHE sources at high redshifts.
Extrapolations of averaged spectra from the high-energy (HE) gamma-ray band covered with Fermi-LAT show that more than 200 blazars of different classes and several radio galaxies should be detectable with acceptable exposure times, up to redshifts of z$\sim$2 \cite{Sol13}. A much larger number should be accessible when accounting for sources with very hard spectra (see also the 
Extragalactic Survey Key Science Project, Chapter~\ref{sec:ksp_eg}) 
and sources in flaring states. The targeting of flaring states will increase considerably our access to low-frequency peaked objects and help us
constrain their emission process, which is currently supposed to differ from the processes in HBLs due to interactions with stronger photon fields in LBLs. 
In addition, CTA will allow us for the first time to study the VHE variability of numerous sources during their quiescent or low-flux states, of which very little is known up to now. 

\begin{figure}[htb!]
\begin{centering}
\resizebox{0.9\columnwidth}{!}{\includegraphics{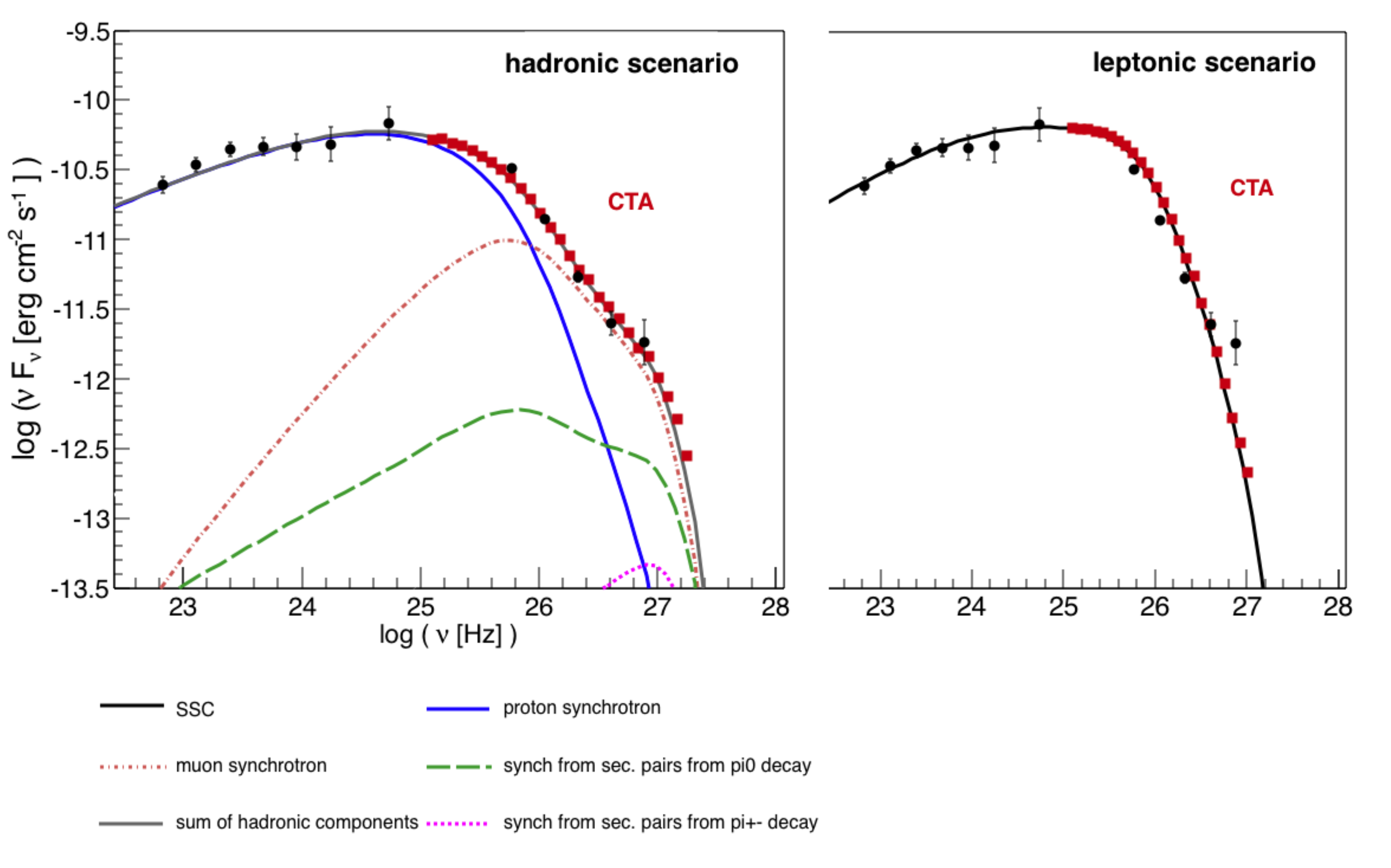}}
\caption{A comparison of the expected CTA spectra for two specific (simple) emission models for the blazar PKS\,2155-304. A hadronic 
scenario, where high-energy emission is caused by proton- and muon-synchrotron photons and secondary emission from proton-photon interactions, is shown on the left, and a standard leptonic synchrotron self-Compton (SSC) model on the right. 
The exposure time assumed for the simulations (33\,h) is the same as the live time for the H.E.S.S. observations (black data points above 3$\times$10$^{25}$\,Hz). 
The statistical uncertainties in the CTA data points are smaller than the red squares. For more details see~\cite{Zech2013, Cerruti2015, Zech2016}.
   \label{fig:kspAGNmodels}
}
\end{centering}
\end{figure}

The AGN Key Science Project plays a crucial role in addressing the three CTA key science themes: 

- {\bf Probing Extreme Environments} :
  Data from the AGN KSP, together with multi-wavelength (MWL) and possibly multi-messenger (MM) data, will bring us closer to a comprehensive understanding of the different types of blazars and their 
  supposed parent population of radio galaxies, through the exploitation of a reference sample of high-quality spectra and light curves from different AGN classes.
   The signals detected from AGN over a large range of redshifts will also be used as beacons for a precise measurement of the extragalactic background light (EBL) \cite{Mazin13} and to constrain the strength of the intergalactic magnetic field (IGMF) \cite{Sol2013igmf}.

- {\bf Understanding the Origin and Role of Relativistic Cosmic Particles} :  
  This KSP aims at probing the nature of the gamma-ray emitting particles in AGN, by comparing leptonic and hadronic emission models against steady and time-resolved spectra over an unprecedented energy 
  range. High-quality VHE spectra of a population of sources and complementary MWL data will enable searches for signatures of UHECRs that will be distinguishable from other propagation or internal effects due to an extensive coverage of 
  redshifts, classes and activity states.
   
- {\bf Exploring Frontiers in Physics } :  
   VHE observations from AGN will provide important data for searches for Lorentz invariance violation (LIV) and axion-like particles (ALPs) which should leave discernible signatures in the gamma-ray spectra
   and light curves, with repercussions on our knowledge of general relativity, quantum gravity and dark matter.

\begin{figure}[htb!]
\begin{centering}
\resizebox{0.7\columnwidth}{!}{\includegraphics{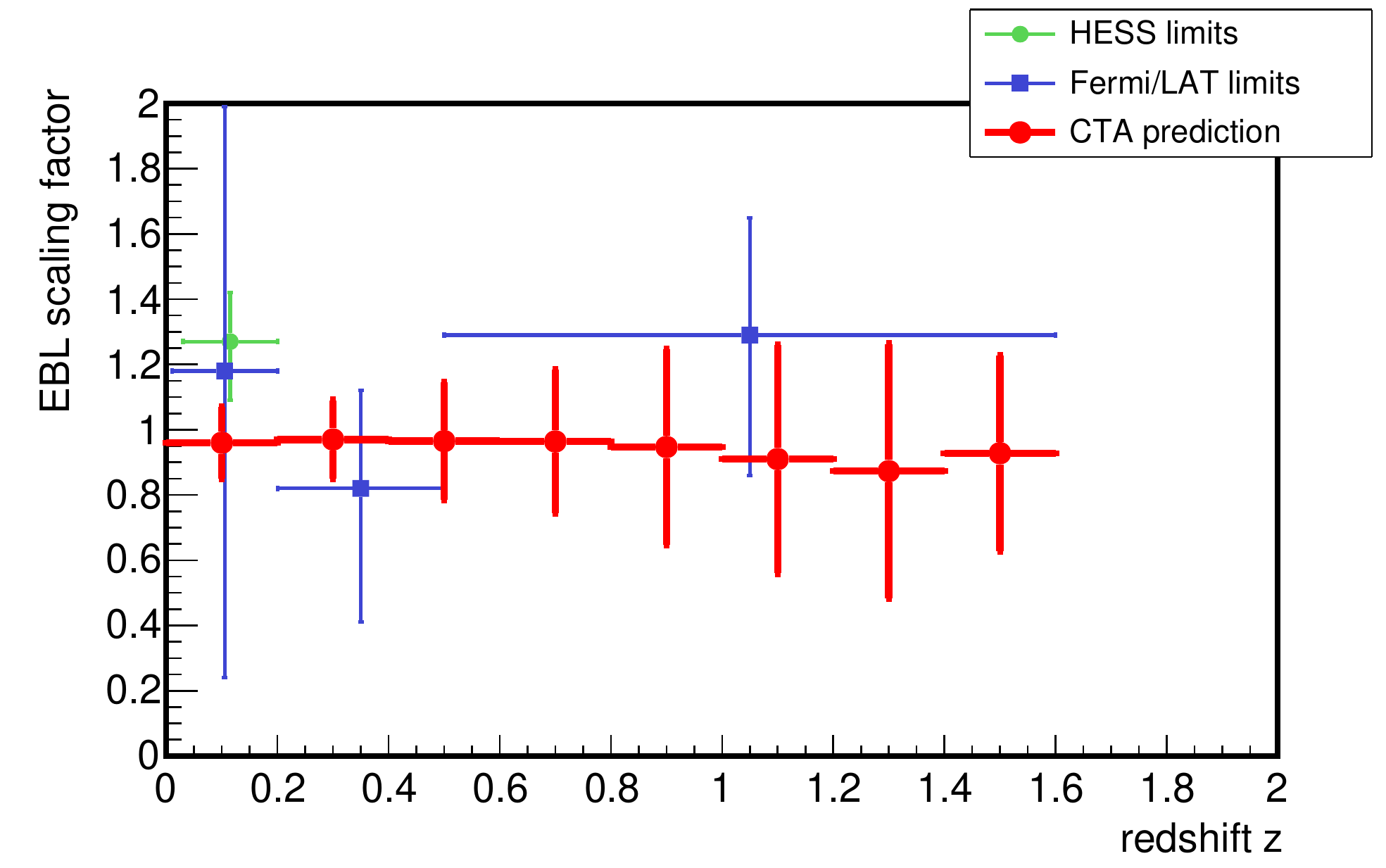}}
\caption{
Potential for CTA to resolve the extragalactic background light (EBL)
density. The normalization of the EBL density with respect to a
state-of-the-art EBL model \cite{Franceschini2008} is reconstructed
as a function the distance of the sources used as gamma-ray beacons.
Assumed are quiescent and flare states of ten sources per redshift bin
and an average flux level of 25\% of the Crab nebula at 100 GeV prior to
absorption. The assumed exposure time takes into account that at higher
redshifts the CTA data will be dominated by short flare states. Results
obtained by H.E.S.S. \cite{Abramowski2013a} and Fermi-LAT
\cite{Ackermann2012} are shown for comparison. It should be noted
that so far only a handful of sources above a redshift of 0.4 have been
detected with ground-based gamma-ray telescopes.
     \label{fig:kspAGNebl}
}
\end{centering}
\end{figure}

In terms of guaranteed science, data from this KSP will provide a wealth of information on the physics of gamma-loud AGN, with direct implications on
our understanding of e.g.\ acceleration and emission processes, characteristics of relativistic jets, and accretion regimes of the SMBH. As an  example, Figure~\ref{fig:kspAGNmodels} shows 
the expected CTA spectra for two simple emission scenarios --- a leptonic and a hadronic model --- in comparison with the currently available data. A set of high-quality spectra from different blazar types and different redshifts is
needed to confidently distinguish intrinsic spectral features of 
multi-component models, such as the hadronic model shown here, from external absorption 
on standard leptonic models. 
The AGN KSP will also lead to a precision measurement of the EBL spectrum at z$\sim$0, down to the far-infrared and to a determination of its evolution up to z$\sim$1 (cf. Figure~\ref{fig:kspAGNebl}). These data will also place strong constraints on the strength of the IGMF, informing us on conditions in the early universe and on predictions of fundamental physics.

In addition, the AGN KSP carries a great discovery potential for new VHE AGN classes, e.g.\ Narrow Line Seyfert 1 (NLSy1) galaxies or radio-quiet AGN classes, such as Seyfert galaxies or low-luminosity AGN (LLAGN).
Estimates of the detectability of extended emission from the Centaurus\,A kpc jet, from its radio lobes or from the radio lobes of M\,87 are model dependent~\cite{Hardcastle11,Sol13}. If detected, such an extended signal would be a major breakthrough in the study of emission mechanisms in these sources and in the unification of different radio-loud AGN in general. 
Naturally, the possible detection of signatures from UHECR, the IGMF (in the form of ``pair halos'' or ``pair echoes''), ALPs or LIV are of great interest for the wider scientific community. 

To this aim, we want to observe a reference sample of VHE AGN, which should cover all different gamma-loud classes and a large range of redshifts in a homogeneous way (in terms of required signal strength for
each source and observation mode). The currently available VHE data on AGN have clearly shown the necessity of accessing not only high-quality spectra, but also flux and
spectral variability on all timescales, together with simultaneous or contemporaneous (depending on the timescales under study) MWL coverage.
Carrying out these observations in the form of a Key Science Project 
will provide the community with a 
homogeneous data set from a reference sample of VHE AGN, selected using well-defined criteria and including various archetypal AGN. The AGN KSP includes multi-annual observations (up to at least ten years for long-term monitoring of a few selected sources), which are difficult to propose and follow up for individual observers. 
With the aim of self-triggering the CTA arrays on AGN flares, special ``snapshot'' observation modes, i.e.\ very short exposures on a large number of candidate flaring sources with small CTA sub-arrays, will be tested and optimised.

There are several arguments for carrying out this scientific programme in the context
of a Key Science Project done by the CTA Consortium.
As mentioned above, the AGN KSP will provide legacy data products (spectra, light curves, etc.) 
from a homogeneous sample of AGN that will be of great value to the community.
The KSP is designed to guarantee coverage of a minimum number of sources at different redshifts
and of different classes, as well as to ensure the 
long-term monitoring of a few prominent sources for at least
ten consecutive years.  It is important that data from the long-term monitoring are released
promptly to constitute a useful data set for the community.
In addition, the snapshot programme requires very frequent short observations with multiple
sub-arrays, where it can profit from flexible scheduling.  
To optimize the scheduling of these observations and their rapid evaluation, and to
guarantee the timely emission of alerts, this programme is ideally suited to be carried out by
the CTA Consortium. Finally, it is scientifically important that a maximum number
of observations of flaring events be done with simultaneous Fermi-LAT coverage;
therefore, an essential part of the AGN KSP is concerned with observations made in the pre-operational
and early phases of CTA.  The organization of the proposed observations in the context of a KSP
will greatly facilitate the necessary coordination with many different MWL and MM instruments.

Key data products will include legacy data sets of high-quality spectra, long-term light curves, 
time-varying spectra and high-resolution flare light curves. Furthermore, 
CTA will be able to send outgoing alerts for AGN flares to MWL facilities based on the snapshot programme, which monitors the VHE flux state of a large number of AGN.

Both the northern and southern arrays are needed to successfully implement this observational programme, otherwise the sample of accessible sources would be too limited, especially in the case of sources with soft spectra or at high redshifts which are difficult to detect.
Observations will be carried out in dark time and under partial moonlight.
Part of the programme can begin with partially complete arrays. It is strongly recommended to begin the targeted observations and the flare programme as soon as possible (i.e. in the period before the start of operations
of the full CTA facility) to profit from simultaneous coverage with the Fermi-LAT detector before the end of its mission.

\subsection{Science Targeted} 
\label{sec:AGNscience}

\subsubsection{ Relativistic jets from supermassive black holes } 

{\bf What are the relevant particle acceleration and emission processes in VHE blazars ? \\How are different blazar types related ?}

AGN studies in the nearby universe have indicated that their diversity may be understood by varying a few physical parameters (e.g.\ jet orientation and power; properties of accretion disk,
corona and torus; spin and mass of the SMBH). 
Theoretical models predict that the jet power in AGN arises from the spin and mass of the central SMBH, as well as from the magnetic field at its horizon \cite{Blandford1977}.
Although some VHE observations seem to favour a link between accretion physics, jet formation and VHE gamma-ray production, no universal connection has been established so far. Measurements of the gamma-ray spectra of blazars can be used to estimate their jet power. This provides insight into the energetics of the source and can be compared to the estimated accretion rate to confirm or dismiss potential connections (see, e.g., \cite{Ghisellini2014}). 
VHE data in particular can complete our present understanding with
information from low-luminosity, high-frequency peaked blazars, which are supposed to probe a different accretion regime and possibly a different black-hole spin range than the softer sources with spectral peaks in the
Fermi-LAT range.

Results from modelling of blazar spectral energy distributions (SEDs) indicate that the presence or absence of thermal photons, emitted from the accretion disk and reflected from the dust torus or broad line region (BLR) and thus linked to the accretion process, could be responsible for the different shapes of the SED observed from different classes (see, e.g., \cite{Ghisellini2010, Boettcher2013, Cerruti2013, Dermer2014}). High-quality spectra at very high energies 
are crucial to distinguish between the components of the SED due to inverse-Compton up-scattering of these different photon fields and to search for discernible spectral features from photon-photon
absorption (see e.g.\ \cite{Poutanen2010, Senturk2013}). The interaction of VHE gamma rays with thermal photons also constrains the possible locations of their emission, e.g. to be within or outside the BLR (see, e.g., \cite{Brown2013, Abeysekara2015}). 
Thanks to the improved capabilities of CTA (i.e. wider energy covered, improved energy resolution
and improved sensitivity) and with
complementary multi-wavelength information, we will be able to better constrain the 
theoretical models and discriminate between the different seed photon sources.

At present, the available emission models are generally not sufficiently well constrained, although in certain cases the basic one-zone synchrotron self-Compton (SSC) model has already shown 
its limitations when confronted with available data (e.g.\ discussion in \cite{Cerruti2015}). 
To distinguish between different acceleration mechanisms, source parameters (such as the size of the emitting region and the magnetic field strength) and underlying particle populations, both spectral and temporal information are needed. Detection of flares, as well as long-term monitoring of sources, together with simultaneous MWL data, enables the reconstruction of time-dependent SEDs, which would put our current models to a serious test. 

High-quality VHE spectra of different sources, as part of carefully organised MWL coverage, will help us characterise the nature of different blazar classes (addressing blazar unification and redshift evolution) and allow us to disentangle the origin (leptonic- or hadronic-dominated emission) and sites of the non-thermal radiation seen at highest energies, with direct implications for {\it in-situ}  
particle acceleration mechanisms (shock, shear, turbulence, magnetic reconnection), as well as jet power and jet dynamics.

Given the large dominance of low-redshift HBLs in the currently known VHE AGN sample, we want to widen the VHE coverage to include more low-frequency peaked objects
(FSRQs, LBLs, and IBLs), as well as more sources at
higher redshifts. Except for a few cases, FSRQs are difficult to detect during quiescent states, but their characteristically large VHE flux variations, up to a factor of 100 on short time scales (days/hours), make them good targets for detections during flaring states (see, e.g., \cite{Lindfors2013}). There is increasing evidence that many LBLs and maybe IBLs share common properties with FSRQs (see, e.g., \cite{Abdo10b}).  
Observing flaring events from such objects can help to understand the links between FSRQs and the bulk of the VHE blazars known to date and to pinpoint the conditions for efficient particle acceleration and radiation within blazar jets. In addition, it will also be necessary to observe some objects of each class during their quiescent states to gain unbiased access to their emission mechanisms, which might well be different between quiescent and high states. 

On the other end of the supposed ``blazar sequence'', we also want to gain access to a larger sample of ``extreme blazars'' or UHBLs \cite{Costamante2001, Senturk2013, Bonnoli2015}. This new class of BL Lac objects, 
of which only a handful of sources have been detected so far, is characterised by a very hard intrinsic VHE spectrum, with a peak of the high-energy spectral bump around 1\,TeV or higher, and little or no flux variability. The interpretation of their multi-band emission with the standard SSC scenarios requires extreme parameter values (see, e.g., \cite{Katarzynski2006}). More complex leptonic scenarios (see, e.g., \cite{Lefa2011, Asano2014}) or
hadronic scenarios (see, e.g., \cite{Murase2012, Cerruti2015}) are required for a convincing description of their emission.

{\bf What causes the observed variability in AGN from time scales of a few years down to a few minutes ?}

Variability at all wavelengths is one of the defining properties of AGN. The most rapid variations
in gamma rays are on the scale of only a few minutes.
The very rapid 
variability of flares puts strong constraints on the size of the emitting region and its bulk velocity due to light crossing-time arguments.
The power spectra of AGN light curves show flicker-noise or red-noise behaviour on time scales ranging from minutes to years, which points to a stochastic origin of their emission. The cause, additive or multiplicative processes, and the source of these variations, intra-jet or disk-modulated, remain largely debated (see, e.g., \cite{Biteau2012}). In both source scenarios, flaring events are rare, but contribute significantly to the total time-averaged emission of the source. Monthly observations with Fermi-LAT indicate that the most probable value of the duty cycle of high-flux events (i.e.\ above 1.5 standard deviations) is about 5 to 10\% for BL Lac objects and FSRQs~\cite{Ackermann2011}. The long-term variability and duty cycle of AGN in VHE gamma rays are largely unknown. Most of the known VHE AGN are detectable on daily time scales only during flares for current instruments. Light curves with shorter time intervals are only available for a handful of sources in extreme flaring states.

One can roughly distinguish three typical time scales relevant for AGN physics:

- {\bf Slow: annual timescale}\\
The annual time scales are related to the duty cycle of the sources, and  
the claimed periodicities and quasi-periodicities have yearly timescales. 
There does not seem to be a significant difference between 
gamma-ray (detected by Fermi-LAT) and non gamma-ray blazars (see\ e.g.\ the Roma BZB catalog\footnote{\url{http://www.asdc.asi.it/bzcat}}), indicating that the duty cycles are long ($>$6 years of Fermi-LAT observations) also in gamma rays. 
The (quasi-)periodicities could be related to binary supermassive black holes, jet precession or processes in the accretion disk.
Finally, breaks in the power spectra of X-ray light curves have been observed on such long time scales for a wide range of black-hole masses and have been shown to scale with the ratio of black hole mass to the accretion rate (see e.g.\ \cite{McHardy2011}). On the other hand, the absence of variability even on relatively long time scales could be an indication in favour of scenarios where gamma-ray emission stems from cosmic-ray induced
cascades that develop outside of the source, as will be discussed below.

- {\bf Intermediate: timescale of days, weeks, months}\\
These time scales are related to the macrophysics of the emission region in the AGN jet such as its size, location, geometry, and dynamics. A better understanding of the emission region is directly linked to information about the possible acceleration and emission mechanisms. Sampling of the light curves at these time scales is crucial for comparison with light curves at other wavelengths to distinguish different emission components.

\begin{figure}[htb!]
\begin{centering}
\resizebox{0.8\columnwidth}{!}{\includegraphics{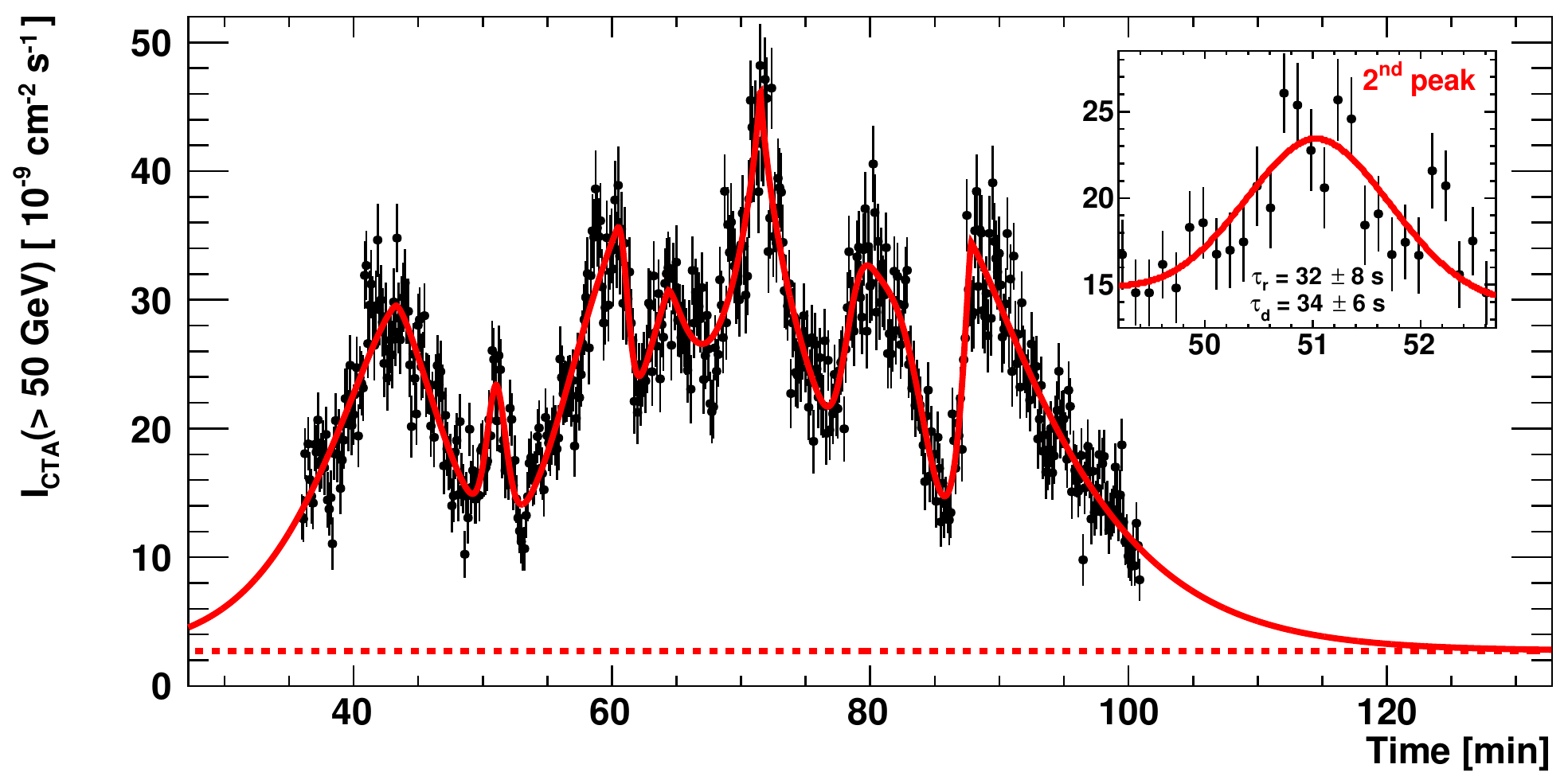}}
\caption{
   Simulated light curve based on an extrapolation of the power spectrum of the 2006 flare from PKS\,2155-304 \cite{Sol13}. Error bars indicate the statistical uncertainty at
one standard deviation. The doubling rise time and decay time are indicated for the second peak in insert.
   \label{fig:kspAGNflare}
}
\end{centering}
\end{figure}

-  {\bf Rapid: timescale of hours, minutes}\\
Sampling blazar fluxes below the light-crossing time scale of the SMBH, T$_G \sim$ 3\,h $\times$ (M/10$^9$M$_{\odot}$) and during the whole visibility window of a night is a key strategy to understand the flickering behaviour of blazars on short time scales. Such measurements put strong constraints on the bulk Doppler factor, as well as on particle acceleration and cooling
processes. Very rapid variability 
has so far only been studied in a few extreme flares, which have shown the limitations of standard emission models and given rise to more sophisticated scenarios, e.g.\ 
turbulent, multi-zone emission~\cite{Marscher2014} or magnetic reconnection (see, e.g., \cite{deGouveia2010, Giannios2013}) inside the jet or in the jet launching region (see, e.g., \cite{Kadowaki15, Singh15, Khiali15}), or pulsar-like acceleration in the black-hole magnetosphere 
(see, e.g., \cite{Osmanov2010, Levinson2011}). To illustrate the expected performance of CTA, Figure~\ref{fig:kspAGNflare} shows the simulated CTA light curve of a hypothetical AGN
flare modeled on a flare detected from the blazar PKS\,2155-304 in 2006. As can be seen, CTA will for the first time probe sub-minute timescales.\\

The observations of the AGN KSP cover flux variations on these three different time scales. 
The slow and intermediate time scales will be probed with the 
``long-term monitoring'' programme, which will provide light curves with regular (weekly) sampling over several years. The intermediate variability will also be tested with observations of relatively bright sources as part of the ``high-quality spectra'' programme.    
Rapid variability will be probed with the dedicated ``AGN flares'' programme. In conjunction with other MWL data, these programmes will provide unprecedented data sets for detailed time series analyses.

{\bf From where does the VHE emission of radio galaxies originate?}

In standard AGN unification models, blazars are thought to be a sub-class of a larger parent population of radio galaxies, distinguished by a close alignment of their jets with the line of sight to Earth.
The observation of radio galaxies at very high energies is of special interest with respect to  understanding the unification of different blazar and radio galaxy classes. Since their jets are not pointing directly towards us, 
radio galaxies can be mapped in the radio band and at other frequencies, and they provide data which are less biased by strong relativistic beaming effects.

The number of radio galaxies detected at very high energies is still very small (this is also true for the GeV band). Only the most nearby sources have been detected at very high energies 
so far, namely M\,87, Centaurus\,A (``Cen A''), NGC\,1275 and PKS\,0625-35; 
these are all FR-I radio galaxies, which are often interpreted as ``misaligned BL Lacs''.~\footnote{The VHE emission from IC\,310, another radio galaxy candidate, is most likely linked to the blazar core of this source \cite{Aleksic14a}.}
The small sample of known TeV radio galaxies do not exhibit any common features at wavelengths below the gamma-ray band \cite{Sol13}. This might indicate that the VHE emission is not directly related to a specific property of those sources
and is rather a general characteristic of radio galaxies.  
It raises the question whether the VHE emission in radio galaxies is generated in the jet, as is generally assumed for blazars, or in the core region, possibly due to magnetic processes. This 
latter possibility
would naturally correlate their radio and VHE properties with those of micro-quasars 
(see, e.g., \cite{Khiali15b}). 

Results from Fermi-LAT observations of Cen~A have shown that the extended emission from the giant lobes in radio galaxies is visible in gamma rays~\cite{Abdo10c}\footnote{Extended emission has also been
recently detected with Fermi-LAT from the radio galaxy Fornax\,A \cite{Ackermann2016}.}. Even if a simple extrapolation of the Fermi-LAT spectrum from this emission does not seem easily accessible for CTA, an additional VHE emission component is not ruled out and prospects for another nearby radio galaxy, M\,87, are more promising~\cite{Sol13}. In Cen~A, emission from the kpc jet provides another challenging target that could lead to an important VHE discovery~\cite{Hardcastle11}. 

To indisputably determine the exact location of the VHE emission, simultaneous or contemporaneous VLBI coverage of VHE flares, with an angular resolution able to image the flaring region, will be essential. The power of such observations is very well illustrated by the observations of M\,87 \cite{Acciari09b}. VLBI observations of this radio galaxy have shown that different episodes of VHE flares may be related to different emission regions, in this
case the radio core and the HST-1 radio knot \cite{Cheung07}. Such unexpected results strengthen the case for further studies.

An open problem is also presented by the mismatch seen between the Fermi-LAT and H.E.S.S. spectra from the core of Cen~A ~\cite{Abdo10e}. A high-quality measurement of the spectrum in the overlap region will be able to verify the indication of a spectral hardening towards the high energy end of the Fermi-LAT spectrum \cite{Sahakyan2013,Brown2016} and to distinguish between different leptonic or lepto-hadronic scenarios trying to explain the origin of the GeV and TeV emission 
 \cite{Abdo10e, Sahu2012, Petropoulou2014, Cerruti2016}.

{\bf Do other classes of AGN emit VHE gamma rays ?}

After the recent detections with Fermi-LAT~\cite{Abdo09dd, DAmmando2012, Foschini2011, DAmmando2013}, it is clear that NLSy1s are a class of AGN that should be within the reach of CTA during 
flaring states. While TeV blazars and radio galaxies tend to be hosted by giant ellipticals with a SMBH of 10$^{8-9}$M$_\odot$, NLSy1s are thought to be hosted by spiral galaxies and to lie in the low range of the SMBH mass spectrum (10$^{6-8}$M$_\odot$), while exhibiting very efficient accretion (but see~\cite{Marconi2008, Calderone2013}). 
VHE gamma rays detected from NLSy1s could provide crucial insights into the unification of various classes of AGN and on the conditions for jet formation as a function of the SMBH mass and 
accretion rate, as well as the role of the host galaxy. Several NLSy1s are included in our list of targets to survey for flares.  

Even more challenging targets are Seyfert galaxies or low-luminosity AGN (LLAGN). Several models for the direct production of gamma-ray emission in the vicinity of SMBH have been proposed 
(see, e.g., \cite{Neronov07, Rieger2008, Istomin2009, Osmanov2010, Levinson2011} and references therein). No 
VHE gamma-ray emission has been detected from such objects so far, although at GeV energies Fermi-LAT has
detected the Seyfert II galaxy Circinus \cite{Hayashida2013}. 
Given the weakness of the expected signal from such sources, a stacking analysis of several candidates seems the most promising approach, with nearby galaxy clusters being the best targets. The deep observation of M\,87, included in the AGN KSP, will at the same time provide coverage of the centre of the Virgo cluster, while observations of other clusters are foreseen within the dedicated Cluster of Galaxies Key Science Project (Chapter~\ref{sec:ksp_clust}). In addition,
a stacking analysis can also be carried out on the fields of view of all other observations of AGN.

\subsubsection{ Blazars as probes of the universe } 

{\bf What is the spectrum of the EBL at redshift z$\sim$0 and how does it evolve at higher redshifts ?}

The EBL encodes important information on the star formation history of the universe and 
it constitutes a major research area in observational gamma-ray cosmology. 
Direct measurements of the EBL are difficult, due to strong foregrounds from our Solar System and the Galaxy, with current constraints from galaxy counts and direct observations leaving the absolute level of the EBL density uncertain by a factor of two to ten. The excellent ``spectral lever arm'' of CTA, covering both intrinsic (with optical depth $\tau=0$) and absorbed ($\tau\approx3$) ranges of the energy spectra from blazars, will allow an unprecedented (indirect) precision measurement of the EBL density, particularly in the poorly constrained low frequency range (mid to far infrared).

The potential of EBL studies with CTA has been discussed in detail \cite{Mazin13}. 
With CTA, we aim for a substantial improvement over current measurements of 
the EBL performed with Fermi-LAT~\cite{Ackermann2012} and current atmospheric Cherenkov telescopes (see e.g.~\cite{Abramowski2013a,Biteau2015}). We will take advantage of the unique capability of 
CTA to measure simultaneously and with high precision both the unabsorbed intrinsic (GeV) and attenuated (TeV) parts of the blazar spectra and thereby disentangle intrinsic physical processes from external absorption features, which will be essential for a precise EBL density determination.

For this purpose, we will observe a large sample of blazars located at different redshifts. 
Many of the targets are already discovered TeV emitters, others are seen with Fermi-LAT so far only, and others will only be seen by CTA when in flaring states.
In addition, the exceptional statistics from extreme flaring states of, e.g., Mrk\,421, Mrk\,501 or PKS\,2155-304, is of utter importance for the study of pair-creation on the far-infrared EBL, for which gamma rays are needed above 10\,TeV, requiring observations of nearby hard-spectrum sources. The detection of high-redshift sources, more likely during flares, would allow us to measure the evolution of the cosmic optical background. Flares also help us distinguish the effects of EBL absorption from internal spectral variations.

The first goal of this part of the AGN KSP is a measurement of the EBL spectrum from the mid ultraviolet to the far infrared at z$\sim$0 with an uncertainty of 10\% on the overall flux, which can be reached with a sufficient number of sources, evenly spread over redshifts up to z$\sim$1. The expected ``scaling factor'', a measure of the sensitivity to differences between the assumed and actual 
EBL density, for measurements with CTA is shown in Figure~\ref{fig:kspAGNebl}. The second goal is the characterisation of the EBL evolution up to a redshift of z$\sim$1, which can be directly compared to 
models of galaxy evolution. 

In addition, we note that the measurement of gamma-ray absorption on the EBL may also be used for the determination of the expansion rate of the universe, i.e.\ the Hubble constant, as has been shown in~\cite{Biteau2015,Dominguez2013}. The value adopted for the Hubble constant has an influence on the evolution of the EBL density and on the gamma-ray propagation that should lead to a measurable effect on 
the detected spectrum. Sources at redshifts between z$=$0.04 and z$=$0.1 seem particularly constraining for such a measurement.
This method would present an approach that is independent from more established methods such
as Cepheids, supernovae, the cosmic microwave background and baryon acoustic oscillations.
Its accuracy will increase with an improved understanding of the EBL and new VHE data.

{\bf What is the strength of the IGMF ?}

The origin of the IGMF, be it structure formation, inflation, or primordial phase transitions, is still widely discussed. For recent overviews see, for example, \cite{Widrow2002, Kulsrud2008, Kandus2011, Widrow2012,Ryu2012}.
The actual detection of a non-zero IGMF, if primordial, could shed new light on conditions in the early universe and could 
complete the dynamo description for the origin of cosmic magnetic fields, by providing magnetic seed fields for dynamo amplification processes in turbulent flows during the formation of large scale structures. On the other hand, an origin of the IGMF through astrophysical mechanisms, e.g.\ bulk outflows of magnetized material from radio galaxies, could explain young magnetized large-scale structures, with little time for dynamo growth, such as the magnetic bridge identified in the Coma supercluster \cite{Kim1989}. 
The importance of a characterisation of the IGMF for our understanding of the evolution of the universe and the development of galactic magnetic fields has been outlined (see, e.g., \cite{Sol13} and the references therein). 

\begin{figure}[htb!]
\begin{centering}
\resizebox{0.45\columnwidth}{!}{\includegraphics{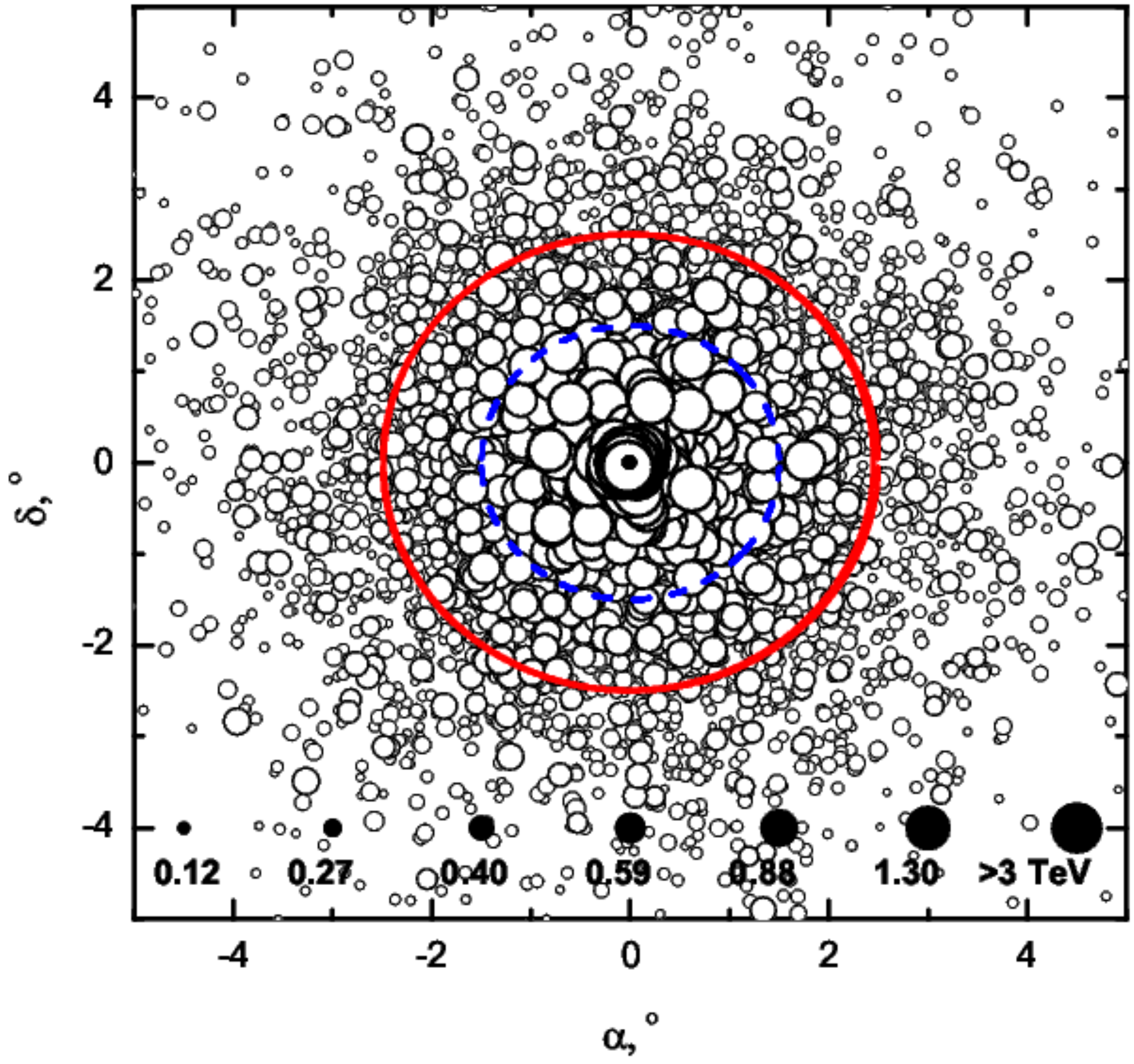}}
\resizebox{0.53\columnwidth}{!}{\includegraphics{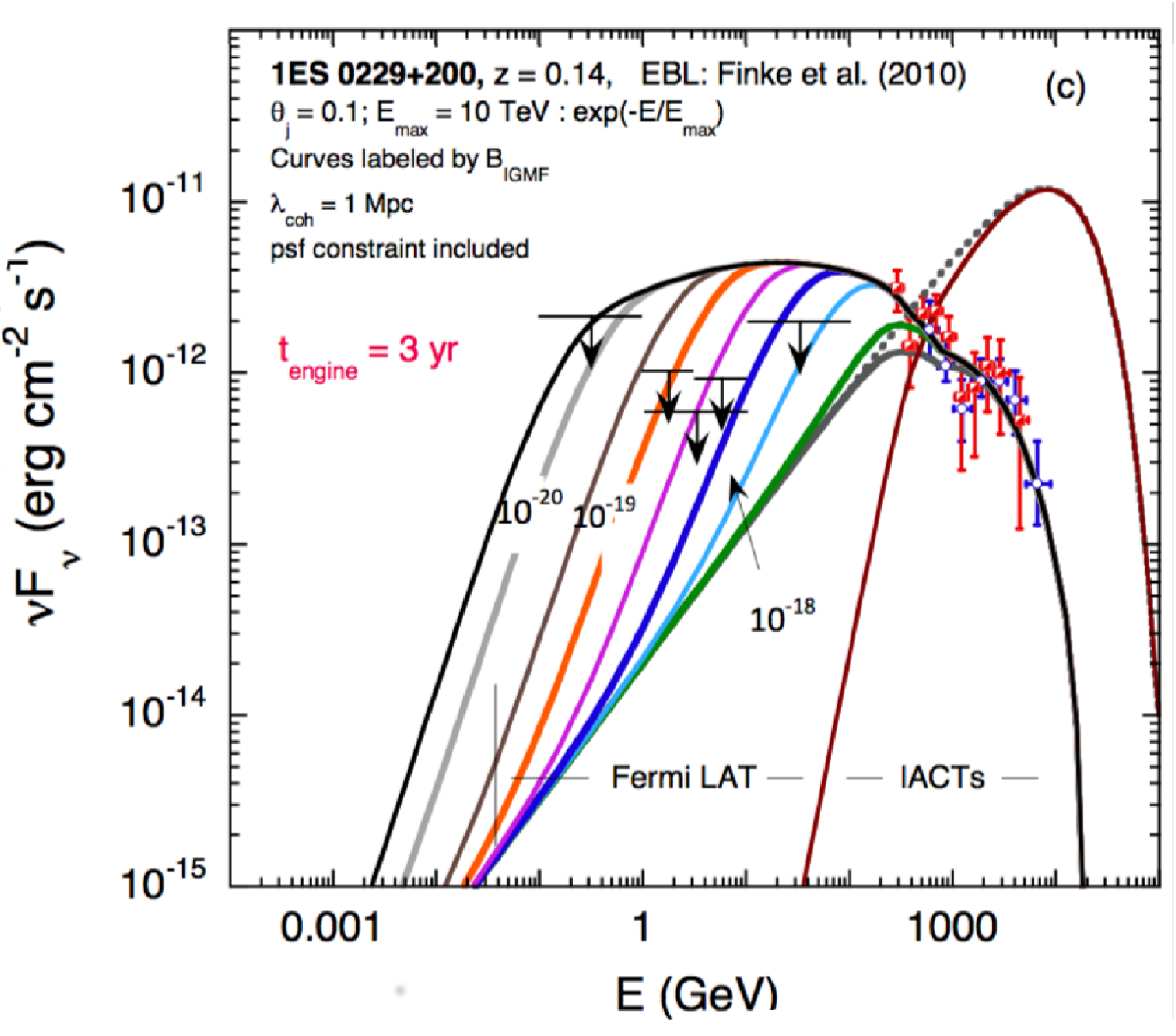}}
\caption{
Left: The arrival directions of primary and secondary gamma rays (black circles) from a source at a distance D$=$120\,Mpc with an IGMF strength of 10$^{-14}$\,G. The sizes of the black circles are proportional to the photon energies. The AGN intrinsic gamma-ray spectrum is described as a power law with an exponential cutoff. The blue dashed and red solid circles are radii of 1.5$^{\circ}$ and 2.5$^{\circ}$, respectively, to indicate the extension of the pair halo image. Figure taken from \cite{Elyiv2009}.  
Right:  A model of the cascade radiation spectrum in a pair echo, applied to observations of the blazar 1ES\,0229+200.  
The cascade spectra assume persistent TeV emission for different values of the magnetic field strength and coherence length. Figure taken from \cite{Dermer2011}.
   \label{fig:kspAGNigmf}
}
\end{centering}
\end{figure}

Indirect detection methods, using a subset of the target selection for the EBL measurement, are our best approach towards a first measurement of the strength of the IGMF.
There are two possible strategies to adopt when studying the IGMF using gamma rays above a few TeV: 

- {\bf Imaging analysis} searches for extended ``pair halos'' around blazars, which are expected for IGMF strengths $\gtrsim$10$^{-16}$\,G. Such searches will profit from the improved angular resolution and wide field of view of CTA (see Figure~\ref{fig:kspAGNigmf}, left panel). A first hint for the existence of pair halos in stacked Fermi-LAT data of blazars has recently been reported \cite{Chen15}.

- {\bf Time-resolved spectral analysis} explores a different parameter space, for IGMF strengths $\lesssim$10$^{-16}$\,G, which lead to so-called ``pair echoes'' (see Figure~\ref{fig:kspAGNigmf}, right panel) in the intergalactic medium, resulting in delayed signals at reduced energies. Rapid flux variations should be washed out by the cascade-like reprocessing of gamma rays towards lower energies. 
The ability of CTA to disentangle spectral components that are time-dependent from those that are constant, especially in the low-energy range, will be of special importance.

\subsubsection{UHECRs and fundamental physics}
\label{sec:agn_fund_phys}
{\bf AGN as potential sources of UHECRs } 

In a similar way to the important advances that were made in the search for Galactic cosmic-ray emitters with GeV and TeV observations of supernova remnants and molecular clouds, one may expect that CTA will lead to an insight into the origin of UHECRs through high-quality data from extragalactic sources. Indirect evidence for cosmic-ray acceleration in such sources, from observations of gamma rays and possibly neutrinos, might indeed be the only way to solve this question, given the difficulties of direct searches for UHECR sources due to the low event statistics 
and the deviation of charged particles in extragalactic and Galactic magnetic fields.

\begin{figure}[h!]
\begin{centering}
\resizebox{0.75\columnwidth}{!}{\includegraphics[angle=-90]{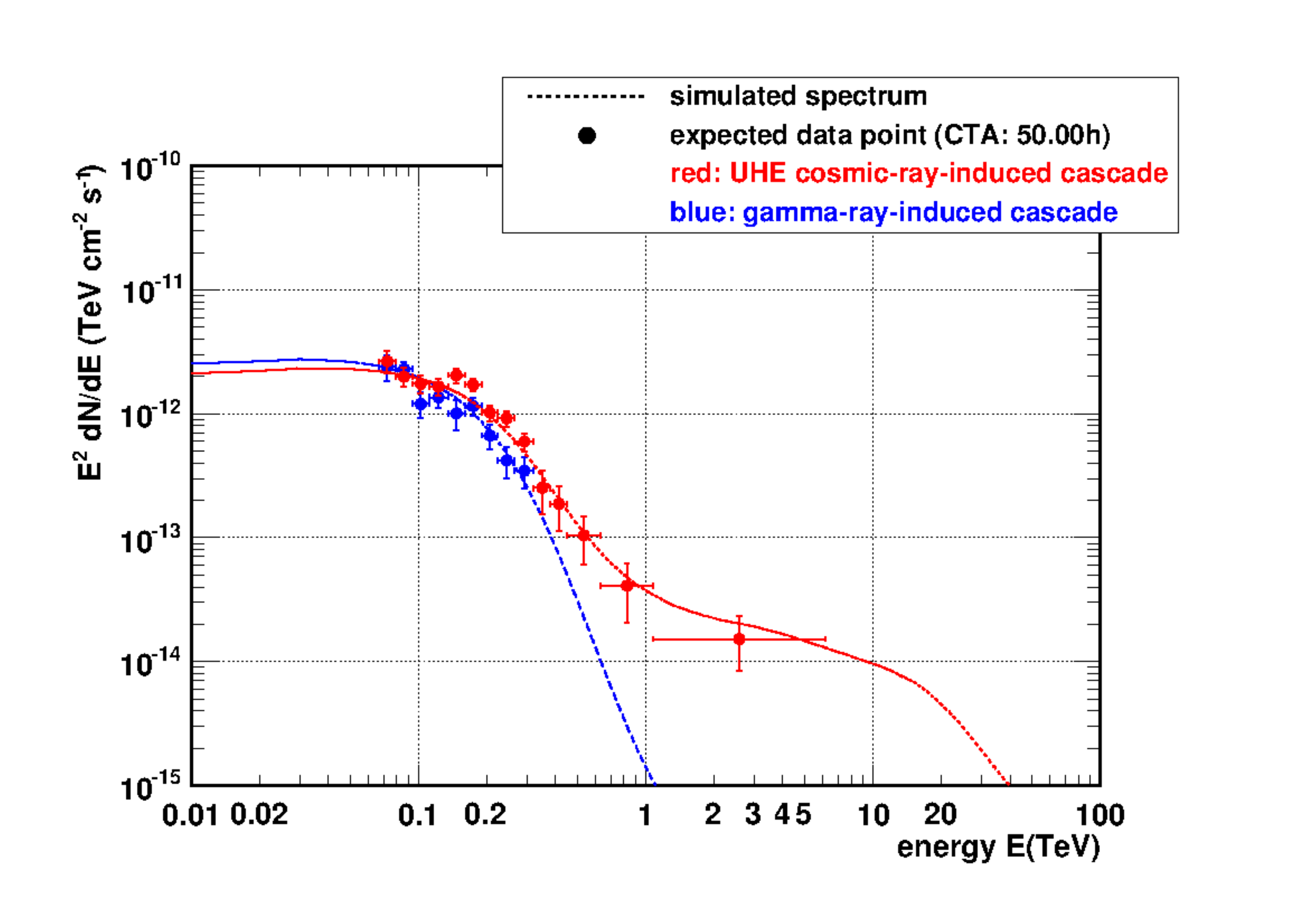}}
\caption{
  The expected CTA spectrum for the distant blazar KUV0031-1938, based on an extrapolation from current measurements and the addition of gamma-ray induced cascades (blue curve), is compared to the spectrum from the same source when including UHECR induced cascade emission (red curve). Error bars indicate the statistical uncertainty of one
standard deviation. See also \cite{Murase2012,Takami2013}.
   \label{fig:kspAGNuhecr}
}
\end{centering}
\end{figure}

Although the detection of very rapid variability of the high-energy flux during flares from a few bright sources such as PKS\,2155-304, Mrk\,421 or Mrk\,501 seems to favour leptonic emission scenarios 
in blazars (but see, e.g., \cite{Barkov2010}), a significant contribution from hadrons during low states is not excluded. We will collect information on the occurrence of rapid flares in different AGN classes, as well as on the flux variability during low states, which will constrain current hadronic emission models. Simultaneous multi-wavelength observations will be crucial for such studies.
A possible signature of UHECRs in gamma rays is a hard spectrum at energies beyond the characteristic gamma-ray absorption energy by the EBL. UHECRs interact with photons of the cosmic microwave background and the EBL during propagation in intergalactic space to produce UHE photons and leptons, which trigger electromagnetic cascades. If the IGMF is sufficiently weak, their cascaded secondary photons contribute to the gamma-ray flux from the source. Models based on this scenario reproduce successfully the SED of hard-spectrum blazars 
(see, e.g., \cite{Essey2010, Essey2011, Murase2012}). Figure~\ref{fig:kspAGNuhecr} shows the expected spectral hardening from UHECR induced intergalactic cascades for a high-redshift blazar observable by CTA.

Another signature for the presence of relativistic hadrons in AGN is the occurrence of synchrotron-pair cascades in the emission region, which can lead, under certain conditions, to a visible spectral ``bump'' in the TeV range (\cite{Zech2013, Abdo11c, Boettcher2013,Zech2016}, see Figure~\ref{fig:kspAGNmodels}). 
In addition, models of gamma-ray emission close to the central SMBH, which can best be probed in LLAGN, provide yet another scenario for the acceleration of hadrons to ultra-high energies.

The search for such signatures requires high-quality spectra covering a wide spectral range for a sample of different VHE AGN classes and redshifts, together with information on the source variability. This will be achieved
with the data gathered from the ``high-quality spectra'' and ``AGN flare'' programmes that are detailed below.
In addition, several emission scenarios based on different acceleration sites have been proposed for radio galaxies, especially Cen~A. These scenarios can be tested with the deep VHE observations we propose for this source (see, e.g., \cite{Joshi2013, Sahu2012, Rieger2009, Yang2012} and many others).

{\bf Can we find signatures for the existence of axion-like particles ?}

An issue with wide ranging implications for fundamental physics that can be probed in several ways with measurements of VHE gamma rays from distant sources is the existence of 
ALPs (for more information, see Chapter~\ref{sec:sci_intro}).
EBL absorption of VHE photons would be considerably reduced by $\gamma \leftrightarrow$ ALP
oscillations taking place in ambient magnetic fields \cite{deAngelis2007,Simet2008,Sanchez2009,deAngelis2009,deAngelis2011,Dominguez2011,deAngelis2013,Galanti2015}.
First hints for a reduced gamma-ray opacity might have been observed in the form of a pair-production anomaly at the four standard deviation level~\cite{Horns2012}
and of unusually hard gamma-ray spectra compared to what is expected from standard EBL absorption \cite{Horns2013,Rubtsov2014}. (These hints do not seem to be confirmed
however, see \cite{Sanchez:2013lla,Biteau2015,dominguez2015}).
Photon-ALP oscillations would provide a natural solution to these problems if they are confirmed~\cite{Simet2008,Dominguez2011,Horns2012a}.

Alternatively, the $\gamma \to$ ALP conversion could occur inside the blazar or in its direct environment, and the ALP$\to \gamma$ reconversion could take place in the Milky Way magnetic field \cite{Simet2008}.
One possibility is that the blazar resides inside a group of galaxies or a galaxy cluster,
where the $\gamma \to$ALP conversion is triggered by the cluster magnetic field~\cite{Horns2012a,Meyer2014}.
It should be noted that these environments will be extensively studied by SKA using Faraday rotation measurements to achieve a better parametrization of the magnetic fields.

As far as completely isolated BL Lac objects are concerned, the $\gamma \to$ALP conversion probability strongly depends
on the  uncertain position of the VHE gamma-ray emission region along the jet and on the strength of the magnetic field therein \cite{Tavecchio2014,Meyer2014}.
On the other hand, the turbulent magnetic field in radio lobes at the termination of AGN jets could lead to a significant proportion of ALPs in the beam \cite{Tavecchio2014}.
Photon-ALP mixing could also explain the observation of VHE gamma ray in FSRQs
if the emission zone is located inside the BLR (e.g.\ \cite{Tavecchio2012}).

Due to the expected spectral signatures of photon-ALP mixing, a search for ALPs suggests an observational strategy closely related to that of EBL studies. In order to detect a reduced opacity,
observations at high optical depths are necessary.  Dedicated studies \cite{Meyer2014,Meyer2014a}
show that flaring episodes of distant blazars at redshifts $0.2 \lesssim z \lesssim0.5$ are best suited for this work.
At the same time, the intrinsic spectrum emitted by the blazar has to be determined as precisely as possible. 
To this end, MWL coverage, especially with X-ray telescopes, is of importance since it can provide insights on 
the particle distribution inside the source and thus help constrain the intrinsic spectrum in the
CTA energy range.

Observations of bright FSRQs at any redshift offer an additional opportunity to search for ALPs and to test the hypothesis
that ALPs facilitate the escape of gamma rays from the BLR. Promising targets coincide with the
sources selected for EBL studies.
The search for oscillations in the spectra are best carried out with bright sources located in regions
where the magnetic field is known to be strong. Thus, bright AGN in galaxy clusters are ideal targets \cite{Abramowski2013b,Ajello16}.
Examples are PKS\,2155-304, M\,87 and NGC\,1275 (see also the Clusters of Galaxies KSP, 
Chapter~\ref{sec:ksp_clust}). Blazars in isolated environments could serve as a control 
sample in case indications for ALP-induced oscillation features in spectra are found.

{\bf Can we rule out Lorentz Invariance Violation ?}  

The significance of LIV searches as a motivation in the development of quantum gravity theories has been discussed in Chapter~\ref{sec:sci_intro}. CTA observations of AGN would enable multiple tests of LIV to break the degeneracy between intrinsic source physics and propagation-induced effects, given the balance of variability, distance and very high energy emission \cite{Doro13}. The duration of AGN flares is sufficient for atmospheric Cherenkov telescopes to slew to them, but still retains the potential to observe rapidly-varying features to provide constraining limits on LIV effects. The AGN we target are regularly spaced in redshift so that propagation-induced versus source intrinsic dispersion effects can be tested for.

Recent observations of GRBs with Fermi-LAT have provided stringent constraints on the 
linear term in generic models that break Lorentz invariance \cite{Fermi2009}, and measurements of VHE photons from AGN flares with H.E.S.S. \cite{Abramowski11a} have shown that atmospheric Cherenkov telescope data from AGN flares can put strong constraints on the quadratic term, which is not yet well constrained.

Even if the AGN light curve does not have sufficiently rapid features to determine dispersion for any single flaring episode, a LIV induced dispersion will mean that the higher energy photons will always be shifted with respect to the lower energy ones in the light curve. The accumulation of long-term monitoring data means that we can still potentially determine time delays with high confidence through the use of cross-power spectral analysis methods \cite{Doro13}. This will be the first time that routine AGN observations, i.e.\ not at exceptional flux levels, will 
allow us to constrain LIV parameters.

The LIV that modifies the dispersion relation for gamma rays could also affect the kinematics in the pair-production with the EBL, changing the threshold for interaction and allowing VHE photons to be detected that would not normally arrive at the observer \cite{Kifune:1999ex,Fairbairn2014,Biteau2015}. A study with deep observations of suitably hard-spectrum, distant AGN testing for any changes in EBL absorption, akin to the ALPs study, would enable a probe of new physics with relevance to LIV studies also.

\subsubsection{Advance beyond State of the Art}

One of the main difficulties in all the studies discussed above with current-generation instruments lies in the separation of intrinsic spectral features from propagation-induced effects. 
The strength of the observing strategy of CTA's AGN KSP lies in the observation of a large set of sources of different classes and at different redshifts with high gamma-ray statistics. In addition, variability leads to different patterns for intrinsic and propagation-induced effects. With the AGN KSP, we will for the first time gather information from a selection of sources at all time scales.
This will allow us to distinguish signatures of the EBL, IGMF and modified fundamental interactions from source physics. %

AGN observations will profit greatly from the improved performance of CTA compared to current instruments. CTA will provide better access to low-frequency peaked sources (FSRQs, LBLs, IBLs, 
and probably NLSy1s) due to its wider energy coverage. The improved sensitivity will allow us to measure variability, even in quiescent source states and to record time-resolved spectra for many  bright sources. Measurements of the EBL will, for the first time, make use of the access to absorbed and un-absorbed components in a single
gamma-ray spectrum, which will reduce 
greatly the systematic uncertainties. 
Similarly, IGMF measurements will use the original and reprocessed components in a single spectrum. Improved energy resolution and sensitivity will lead to high-quality spectral measurements. The improved angular resolution will be useful for the search for extended emission from 
radio galaxies and halos. There is also, for the first time, the possibility of carrying out snapshots with sub-arrays to increase the probability of detecting flares. 

A complementary approach to CTA in the VHE band is represented by ground-based 
air-shower detectors such as HAWC, Argo YBJ or the future LHAASO. 
These detectors have the advantage of a very large duty cycle 
and wide sky coverage, which will be useful for continuous monitoring of bright AGN, but 
they suffer from a higher energy threshold and poorer sensitivity at a comparable exposure time for a
given field of view.  

When comparing its expected performance with that of Fermi-LAT, the great advantage of CTA in the overlap region of energies (a few tens of GeV) 
lies in its much better sensitivity to rapid variability.
Furthermore, the higher energy reach of CTA is important to detect features at the end of the observed spectrum, such as breaks or cutoffs, which can originate from acceleration and radiation 
mechanisms or from absorption effects. CTA has limited sensitivity to the majority of FSRQs and LBLs in low states, but, on the other
hand, the high-energy emission from UHBLs, which are very difficult to detect with Fermi-LAT, falls almost exclusively into CTA's energy range. IBLs and HBLs will be well covered due to the wide
energy range of CTA. The overall accessible source statistics will be less than for Fermi-LAT, given the difference in the energy bands and the steeply declining spectra of AGN, but observations with
CTA will provide much higher resolution in light curves and variable spectra for the detected sources.

\subsection{Strategy}

Current advances in AGN physics rely to a large extent on the simultaneous measurement of the AGN spectral energy distribution (SED) and its time variation across all wavebands, from radio to VHE. 
In the context of the AGN KSP, the CTA Consortium will:
\begin{itemize}
\item set up coordinated - where appropriate long-term - programmes of AGN observations, involving observation proposals, target of opportunity (ToO) and MoU-based co-operations with other facilities and instruments, enabling coherent and consistent multi-wavelength coverage and joint data analysis (see Section~\ref{sec:AGNMWL} for details),
\item ensure effective and consistent follow-up of AGN flares reported by other facilities and instruments, under MoUs where appropriate, and set up programmes for flare monitoring and alert generation, and
\item help to implement observation modes and scheduling for the most efficient monitoring of variable sources, by flexibly dividing the arrays into multiple subsystems as appropriate.
\end{itemize}

The following observation programmes will be carried out under the AGN KSP:
\begin{enumerate}
\item {\bf Long-term monitoring} of a few prominent VHE AGN up to at least ten years:\\
-{\it guaranteed science}: long-term light curves and time resolved spectra for all studies involving AGN variability on medium and long time scales, addressing quiescent states, duty cycles, and spectral variability. \\
-{\it discovery potential}: disk-jet connection, (quasi-)periodic oscillations, and LIV.\\

\item Search for, and follow-up of, short-term {\bf AGN flares} based on external ToOs and internal snapshot monitoring of a large number of AGN:\\
-{\it guaranteed science}:  effectively sampling the high states of AGN and high-statistics, high-temporal resolution spectra for studies of short-term variability; 
  EBL / IGMF measurements at z $>$0.5.\\
-{\it discovery potential}: TeV emission from NLSy1s and others; LIV and ALPs.\\

\item {\bf High-quality spectra} for a systematic coverage of redshifts and AGN typology:\\
-{\it guaranteed science}:  a comprehensive data set, obtained under uniform conditions, for AGN classification and evolution studies; high-precision spectra for a precise measurement of the EBL, studies of emission scenarios for VHE blazars and of the IGMF. \\
- {\it discovery potential}: extended VHE emission from two nearby radio galaxies; UHECR signatures; LIV and ALPs.\\ 
\end{enumerate}

Data from long-term monitoring will be released immediately in the form of light curves. The ``AGN flare'' programme will produce regular alerts to the community, e.g.\ in the form of VOEvents. 
High-quality spectra will be compiled in a catalogue that will be available to the community after the usual proprietary period. 

The long-term source monitoring and snapshot programmes can make use of observations under moonlight. 
In general, extraction of high-quality spectra with energy coverage down to the lowest energies will require observations taken at small zenith angles.
Specific developments are under way for an efficient online real-time analysis during the snapshot programme.

\subsubsection{Target Selection}
\label{sec:agntargets}

{\bf 1) Long-term monitoring.} 

In the long-term monitoring programme, which will provide long-term VHE light curves, we want to cover all known types of VHE AGN (UHBL, HBL, IBL, LBL, FSRQs, and radio galaxies). 
To arrive at a representative sample, we want to observe two to three sources per class. As an example, a list of 15 potential targets, including some of the most prominent TeV sources, is given in Table~\ref{tab:AGNlongterm}. 

\begin{table}[!h]
\begin{tabular}{| l | l |}
\hline
 UHBLs & 1ES\,0229+200 (N) , 1ES\,1426+428 (N) , 1ES\,1101-232 (S) \\
\hline
 HBLs & Mrk\,421 (N) , Mrk\,501 (N) , PKS\,2155-304 (S) \\ 
\hline
 IBLs & 1ES\,1011+496 (N) , 3C\,66A (N) , W\,Comae (N)\\ 
\hline
 LBLs & AP\,Librae (S) , BL\,Lacertae (N)  \\
 \hline
 FSRQs & PKS\,1510-089 (S) , PKS\,1222+216 (N)   \\
\hline
 Radio Galaxies & M\,87 (N) , NGC\,1275 (N)  \\
\hline
\end{tabular}
\caption{Example list of targets for long-term monitoring. The labels ``(N)'' and ``(S)'' indicate observations with the northern and southern array, respectively.}
\label{tab:AGNlongterm}
\end{table}

Each target should be observed on average $\sim$30\,min once a week during its period of detectability with the full array (possibly less for the bright UHBLs and HBLs and more
for soft-spectrum sources).   
This would result in $<$12\,h per year per target, leading to greater than
seven standard deviation
($>$7$\sigma$) detections of all sources in their quiescent state, sufficient for spectral measurements. For the brighter sources,
spectra can be extracted on a weekly or monthly basis. The regular observations will permit tracking of the flux variability of all sources in the form of long-term light curves. If a source is
found in a flaring state, more intensive coverage can be triggered as part of the ``AGN flare programme'' (see below).

The total exposure time for 15 fields of view would be $<$180\,h per year.  With the currently proposed list of targets, this time would 
be split into 132\,h of yearly observation time for the north and 48\,h for the south. Observations would fall naturally into dark time ($\sim$50\%) and moon time ($\sim$50\%).
Note that even though the main objective of this programme, i.e.\  studying the long-term flux variation, can be achieved with LSTs and MSTs, an additional (currently not foreseen) SST component 
in the north would be very useful for extending the spectral coverage of HBLs and UHBLs during flaring states above 10\,TeV with good sensitivity, 

The list of targets will be reviewed after five years and reduced to the ten most interesting objects (in terms of variability patterns) for further monitoring over at least ten years, and ideally over the full CTA lifetime.

{\bf 2) AGN flare programme.}

Given the central importance of variable phenomena for VHE research with CTA, an extended programme needs to be set up to target AGN flares. The expected number of external alerts, based on our experience 
from current experiments (with, e.g., Fermi-LAT or Swift), would be roughly 25 per year, about half of which will 
be followed up with a full CTA array. This would lead to $\sim$10$-$15 external flare alerts to be followed up per year. An on-site optical facility with photometric and polarimetric facilities would
be extremely useful in this respect, as discussed below (cf. Section~\ref{sec:AGNMWL}). Future ``transient factories'' as well as ground-based VHE detectors, discussed in Section~\ref{sec:AGNMWL} and Chapter~\ref{sec:sci_synergies},
might increase this rate slightly. One should note that ground-based VHE detection will provide rapid alerts only for very bright hard-spectrum sources. The alerts from 
optical and radio transient factories will be more numerous, but as the time scales of the 
flares from the transient factories
are very different than VHE flares, only a few of these alerts will be followed with the full
CTA arrays. However, these alerts can be used as a selection criterion for potential targets for the snapshot programme. 

Thus, although they are very important, it is not sufficient to rely on external alerts from MWL facilities, since different variability behaviour is seen even when comparing VHE data from current atmospheric Cherenkov telescopes to HE data 
from Fermi-LAT.
In addition to external alerts, plus some flares expected from the long-term monitoring targets and from the selection of targets for high-quality spectra (see below), we will carry out snapshots, 
i.e.\ very short exposures of a large list of targets, using CTA sub-arrays, to self-trigger full-array observations in the case of flares. This method is already being used successfully by current 
atmospheric Cherenkov telescopes.
\footnote{Alternatively, one might consider to extend the lifetime of certain of the current 
atmospheric Cherenkov telescopes 
to serve as dedicated VHE monitoring facilities for CTA.}

The selection of potential targets for snapshots should be large, given the limited rate of detectable flares. An example of sources that should be included are:
\begin{itemize}
\item FSRQs and soft-spectrum BL Lac type objects (LBL, IBL). A selection can be based on Fermi-LAT average spectra, by computing the ``flux enhancement'' needed to reach 30\% of the 
Crab nebula flux in the CTA energy range. When the enhancement factor is within the typical variation for such sources (factor of 10 to 40), the source is selected,
\item a selection of $\sim$20 known VHE gamma-ray emitting HBLs/UHBLs (see\ TeVCat~\cite{tevcat}), giving preference to bright sources with well established redshifts,
\item a selection of $\sim$10 radio galaxies detected with Fermi-LAT, and 
\item NLSy1s detected with Fermi-LAT (eight sources with high significance so far \cite{Dammando2015,Dammando2016}).
\end{itemize}

The above selection of $\sim$80 AGN will be followed with CTA sub-arrays with a sensitivity of $\sim$20\% of the Crab nebula flux.

- {\bf Soft-spectrum sources} (FSRQs, NLSy1s, LBLs, certain IBLs, certain and radio galaxies) will be observed with the Large-Sized Telescopes (LSTs).
 Setting the threshold for triggering a flaring alert to 20\% of the flux from 3C\,279 during its 2006 flare~\cite{Magic2008} (corresponding to roughly 20$\%$ of the Crab nebula flux), 
 four LSTs will need $\sim$10\,min of exposure time plus an overhead 
 (pointing, stopping and starting the observation) of $\sim$2\,min per pointing.\footnote{The use of sub-arrays with two LSTs is less efficient in terms of the total time needed.}
 Five targets could be covered in one hour, i.e.\ covering $\sim$20 targets twice a week would require $\sim$4\,h per week per site.  A total of $\sim$200\,h of LST observation time would be needed per 
 year per site to cover 40 sources~\footnote{Each source is assumed to be visible for about half a year.}. 
 
- {\bf Hard-spectrum sources} (HBLs, UHBLs, certain IBLs, and certain radio galaxies) will be observed preferably with sub-arrays of Medium-Sized Telescopes (MSTs), or Small-Sized Telescopes (SSTs) in the case of very bright, very hard sources observed from the southern hemisphere. sub-arrays of 
four MSTs are estimated to provide a sensitivity of 20\% of the Crab nebula flux in roughly 3\,min. If CTA is divided into ten such sub-arrays (in total, 4 for the northern plus 6 for the southern 
array\footnote{Given the currently simulated configurations, one of the northern sub-arrays would contain only 3 MSTs and would be pointed to the brightest targets.}), only $\sim$10\,min would be 
needed to cover 20 sources. This would lead to an estimated average observation time of $\sim$20\,min per week per site and $\sim$17\,h per year per site for the full MST sub-array, to 
cover 40 sources.

Observations with the northern site will be given preference, where possible, due to the smaller overlap with Galactic observation programmes. With this in mind, the total observation time 
could be split roughly into 300\,h per year with the northern LSTs, 22\,h per year with the northern MSTs, 100\,h per year with the southern LSTs, and 11\,h per year with the southern 
MSTs. The requirement for observation time with the LSTs is rather high and one should try to target most sources, except for the ones with the softest spectra, with MST sub-arrays. 
Work on such an optimization is under way.

To estimate the overall rate of triggers from snapshots, we assume the flaring probability for the brightest events to be $\sim$1\%, based on past observations. 
If we monitor $\sim$20 sources twice a week at each site, this corresponds to roughly 7 sources observed each night per site and to less than one hour per night per site. 
For any night, the probability to catch a flare would be about 7$\%$, leading to about 20 flares per year per site triggered with snapshots. As will be discussed below, the number of snapshots 
will be reduced after the first two years of full operation, which should lead to a reduction to roughly half of this number in the third year and to a further reduction in the following years.

Once a CTA array has been successfully triggered, the average observation time to follow up on a trigger is estimated to be roughly 4\,h. 
Based on these estimates, we foresee the total follow-up observing time per year with the full arrays (sum for two sites) for snapshot triggers, triggers from the other observation programmes of 
the AGN KSP, and accepted external triggers to be roughly 200\,h, with a larger fraction for the northern site.

In addition, some more observing time needs to be allotted to sub-array observations that verify external triggers and to perform periodically spaced observations to verify the state of the source
once an external trigger has been issued, to search for delayed flares. The verification should require about 10\,min for each of the estimated 25 external triggers, i.e.\ $\sim$4.5\,h per year. 
The follow-up verifications, on average 10 per year, would add $\sim$40\,h of observing time per sub-array per year. 

Given the nature of these observations, a large fraction will be carried out during moon time. The impact on data quality, especially on the low-frequency peaked objects due to the higher energy
threshold during observations with moonlight, will be evaluated during the first year of observations.

{\bf 3) High-quality spectra.}

This last observation programme consists of two distinct groups of targets: 
 
{\bf 3a) Coverage of redshifts and VHE AGN classes.}

A set of targets has been selected to maximise the population in redshift space with VHE AGN of different classes. 
We have extracted spectra of known TeV emitters and promising source candidates from the 
Fermi-LAT 1FHL catalog \cite{Ackermann2013} and extrapolated their spectra to the TeV range, 
while applying a redshift- and energy-dependent absorption on the EBL following the model of \cite{Franceschini2008}. 
For those sources for which no redshift was provided in the catalogue, we have used the lower limits determined by~\cite{Shaw2013} and~\cite{Pita2014a}.  
The resulting spectra were then compared against the CTA performance curves. 

We selected our targets among those sources with an expected significance above 5$\sigma$ in 20\,h of observations in the energy region where EBL absorption is significant, i.e.\ at energies where the optical depth is larger than 1. To sample both the intrinsic and absorbed parts of the spectrum, we additionally required at least five spectral points, with at least two points in the absorbed ($\tau>$1) regime and at least one point with $\tau>$2, and a significance of 20$\sigma$ above 60\,GeV for most targets (10\,$\sigma$ above 60 GeV for a few sources with relatively low fluxes). For very hard sources, we introduced a spectral break at 500\,GeV. 

\begin{table}
\begin{tabular} {| l | l | l | l | l | l | l | l | l |}
\hline
object & RA & DEC & class & $\sigma$(20\,h)& redshift & TeVCat& t [h] (20$\sigma$)& N/S \\	
\hline
\hline
IC 310 &  49.169 &   41.322 & HBL & 35.63 & 0.019 & yes &   6 & N\\
Mrk 421 (lt) & 166.121 & 38.207 & HBL & 212.50 & 0.031 & yes &  (1) & N \\
Mrk 501 (lt) & 253.489 & 39.754 & HBL & 101.46 & 0.034 & yes &  (1)  & N\\
1ES 2344+514 & 356.759 & 51.705 & HBL & 66.95 & 0.044 & yes &  2  & N\\
Mrk 180 & 174.100  &   70.159 & HBL & 27.55 & 0.046& yes & 11 & N \\
1ES 1959+650 & 300.007 & 65.158 & HBL & 28.60 & 0.047 & yes &  10 & N \\
PKS 0521-36 &	80.85 &	-36.34 &		RG	 & 9.25 &	0.06 &	no &	23 (10$\sigma$) & S \\
PKS 0548-322 & 87.669  &  -32.260 & HBL & 14.34  & 0.069 & yes & 39 & S\\
PKS 0625-35 & 96.704 & -35.479 & RG & 26.14 & 0.055 & yes &  12 & S \\
PKS 1440-389 &  221.011 &    -39.145 & HBL & 27.28 & 0.065 (?) & no & 11 & S \\
PKS 2005-489 & 302.360 & -48.830 & HBL & 60.83 & 0.071 & yes &  2 &S \\
\hline
MS 13121-4221 & 198.749 & -42.696 & HBL & 85.52 & 0.108 & yes &  1 & S \\
VER J0521+211 & 80.449 & 21.220 & IBL & 53.45 & 0.108 & yes &  3 & N \\
VER  J0648+152 & 102.225  &  15.281 & HBL & 18.10 & 0.179 & yes & 24 & N \\
PKS 2155-304 (lt) & 329.721 & -30.219 & HBL & 131.71 & 0.116 & yes &  (1) & S\\
1ES 1215+303 & 184.460  &  30.104 & HBL & 22.34 & 0.130 (?) & yes & 16 & N\\
1H 1914-194 & 289.441 & -19.365 & HBL & 25.37 & 0.137 & no &  12  & S\\
1ES 0229+200 (lt) & 38.20250 & 20.29 & UHBL & - & 0.14 & yes & (1) & N \\
TXS 1055+567 & 164.666 & 56.459 & IBL & 22.78 & 0.143 & no &  15 & N \\
PG 1218+304 & 185.337 & 30.194 & HBL & 32.22  & 0.184 & yes &  8 & N \\
1ES 0347-121 & 57.3458 & -11.97 & UHBL & - & 0.188 & yes & 1 & S \\
\hline
1H 1013+498 & 153.773 & 49.427 & HBL & 21.59 & 0.212 & yes &  17 & N \\
PKS 0301-243 & 45.868 & -24.128 & HBL & 39.46 & 0.260 & yes &  5 & S \\
PMN J1936-4719 & 294.214 & -47.356 & BLL & 20.10 & 0.265 & no &  20 & S \\
PMN J0816-1311 &  124.091 &  -13.177  & HBL &  18.17 &  0.290 (ll) & no &   24 & S \\
3C 66A (lt) &  35.669 &  43.035 & IBL  & 49.64 & 0.33 (ll) & yes &  (3)  & N\\
1ES 0502+675 & 77.038 & 67.624 & HBL & 41.65 & 0.340 & yes &  5 & N \\
TXS 0506+056 & 77.394  & 5.714 & IBL & 13.17 & 0.210 (ll) & no & 46 & N\\
MS 1221.8+2452 & 186.146 & 24.628 & HBL & 13.09 & 0.218 & yes & 47 & N \\
\hline
PG 1553+113 & 238.942 & 11.190 & HBL & 67.27 & 0.43 (ll) & yes &  2 & N\\
4C +21.35 & 186.220 & 21.377 & FSRQ & 13.74 & 0.434 & yes & 42 & N \\
1ES 0647+250 & 102.712 & 25.081 & HBL & 24.75 & 0.490 (ll) & yes &  13 & N \\
KUV 00311-1938 & 8.407 & -19.361 & HBL & 22.07 & 0.506 (ll) & yes &  16 & S \\
PMN J1610-6649 & 242.726  & -66.849 & HBL & 14.70 & 0.447 (ll) & no & 37 & S \\
\hline
PKS 1424+240 & 216.766  &   23.790 & HBL & 29.33 & 0.600 (ll) & yes & 9 & N \\ 
PKS 1958-179 &	300.28 &	-17.87 &		FSRQ &	7.0 &	0.65	 &no	 &41 (10$\sigma$) & S \\
B3 0133+388 & 24.147 & 39.100 & HBL & 14.22 & 0.750 (ll) & yes &  40 & N \\ 
\hline
PKS  0537-441 & 84.714  & -44.088 & LBL/FSRQ & 19.33 & 0.892 & no & 21  & S\\
4C +55.17 & 149.421  & 55.377 & FSRQ & 8.53 & 0.899 & no & 27 (10$\sigma$) & N \\  
PKS 0426-380 & 67.178  &  -37.937 & LBL/FSRQ & 10.63 & 1.111 & no & 18 (10$\sigma$) & S \\ 
\hline
\end{tabular}
\caption{Example selection list of sources for high-quality spectra. Sources are ordered by redshift. 
Those sources that are already covered in the selection for long-term monitoring are labelled ``(lt)''.
Lower redshift limits are marked with ``(ll)'' and uncertain redshifts with ``(?)''. The column labelled ``TeVCat''
indicates if a source is an already known TeV emitter or not \cite{tevcat}.}
\label{tab:AGNspectra}
\end{table}

\begin{figure}[htb!]
\begin{centering}
\resizebox{0.85\columnwidth}{!}{\includegraphics{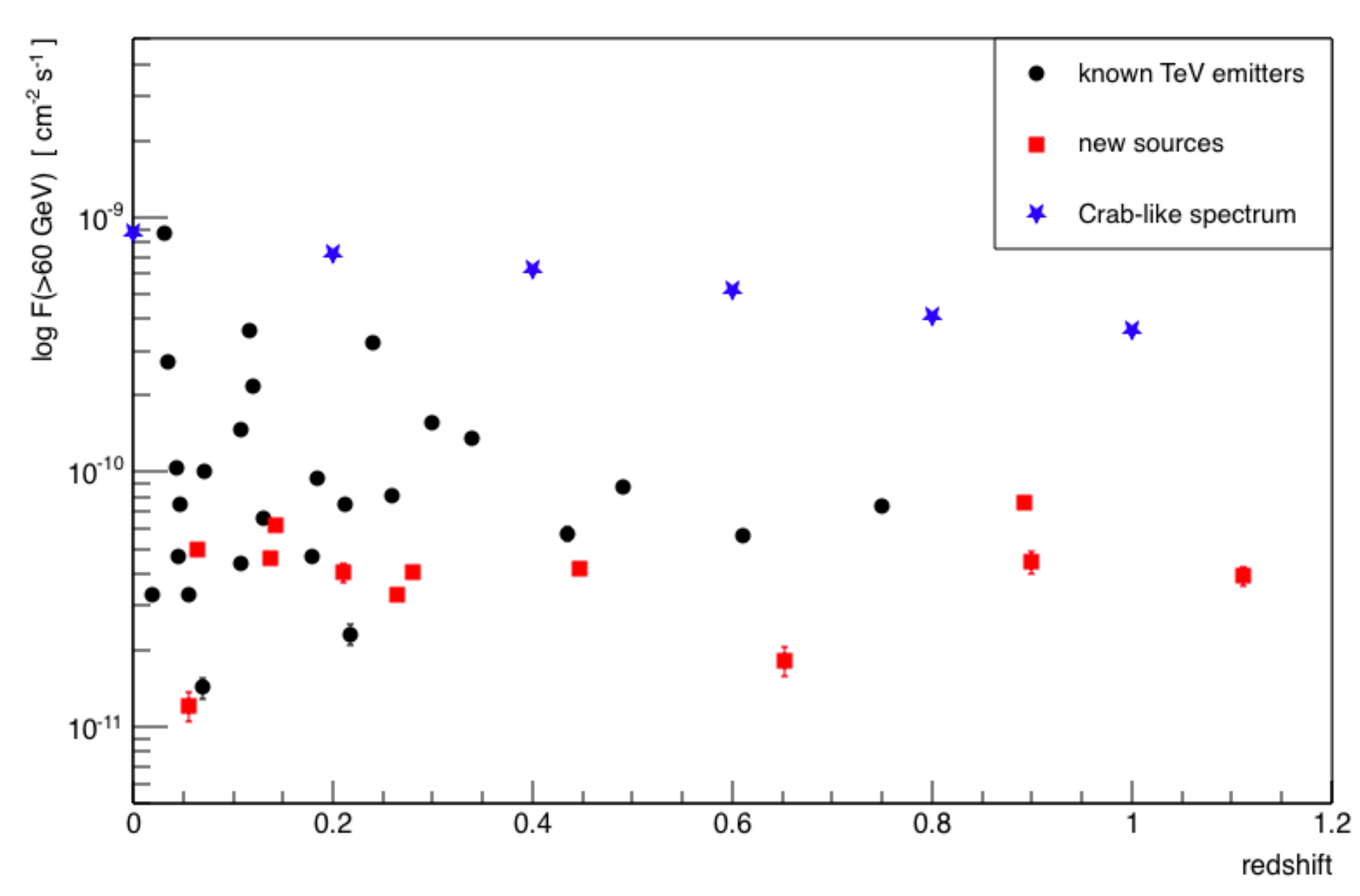}}
\caption{Expected integral fluxes above 60\,GeV 
(from extrapolations of Fermi-LAT spectra) versus redshift for the selection of sources for the high-quality spectra programme.
The flux from a source with a Crab nebula-like spectrum, shifted to different redshifts, is shown for comparison. For studies of the EBL, IGMF and blazar evolution, this sample will be
completed with observations from the AGN flare programme at high redshifts.
 \label{fig:AGNfluxes}
}
\end{centering}
\end{figure}

An example selection list
of such sources is given in Table~\ref{tab:AGNspectra}. Due to unpredictable long-term variations in the blazar flux states, this list is preliminary and will be 
updated with the most promising targets once CTA operations begin. If a source shows no signal after the exposure time expected for a detection at a predefined significance, it will be replaced 
with a new source in the same redshift band and of the same class, if possible. The distribution of expected integral fluxes (above 60\,GeV) against redshift can be seen in Figure~\ref{fig:AGNfluxes}.

About ten sources have been selected in each of the two intervals of $\Delta$z$=$0.1 from z$=$0 to z$=$0.2; these sources are particularly 
constraining for far-infrared measurements of the EBL and a determination of the Hubble constant. Ideally, at least five sources would then be needed in each interval of $\Delta$z$=$0.2 from z$=$0.2 to 
z$=$1.0 for a precision measurement of the EBL and its evolution and for a study of the evolution of blazars with redshift. Care was taken to include sources of different gamma-ray loud AGN classes,
wherever available.
For the IGMF study, we will select a subset of sources for which the spectral region where EBL absorption is significant ($\tau >$ 1) is located at TeV energies, to ensure that the bulk of the reprocessed emission 
is detectable with CTA. 

Several hard-spectrum, high-redshift sources in this selection have been identified to be particularly well suited for ALP searches.
We also foresee deep observations of at least two hard-spectrum sources from the above selection for the study of LIV effects in the interactions between gamma rays and diffuse extragalactic background radiation 
and for the search for UHECR-induced cascade features. A source with z$<$0.15 and with an intrinsic spectral cutoff above 10\.TeV would be required to obtain limits of above 10$^{11}$\,GeV on a quadratic LIV term. It should be noted that the energy range above 10\,TeV will be targeted mostly with southern 
sources, due to the SST component in the southern array. 

The total estimated observation time would be $\sim$343\,h  for the northern array and $\sim$283\,h for the southern array, given the current source selection. It should be noted that the difference between CTA-North and CTA-South sensitivity (roughly a factor 4 in sensitivity above 
10\,TeV) was taken into account in the target selection and required observation time estimation.
Given the importance of observing high-quality spectra over a maximum energy range, we ask for low zenith angle observations using the full array, except for the soft-spectrum targets (FSRQs and LBLs),
where VHE emission above a few TeV is not expected and the SST component can be used to observe other targets.
Observations should be performed where possible within one month to avoid mixing different states. Given the amount of exposure time required, this campaign can begin during the (late) construction 
phase of CTA, where we would focus on those targets that require the least exposure, and 
it could extend over the first three years of full operations. We estimate that roughly half of the total observation time could
be covered during the phase before the start of full CTA operations.

\begin{figure}[htb!]
\begin{centering}
\resizebox{0.6\columnwidth}{!}{\includegraphics{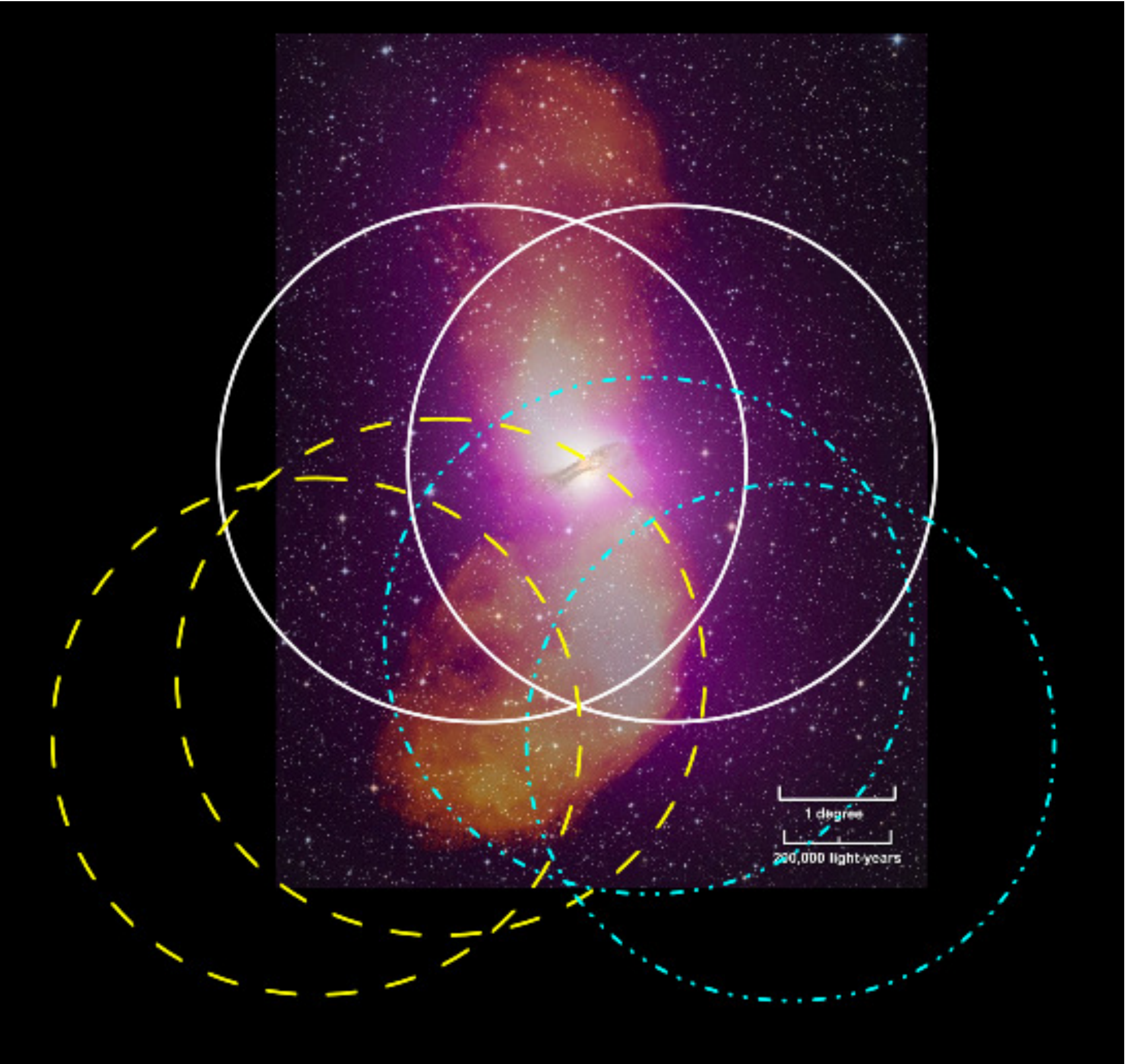}}
\caption{Scheme of a possible pointing pattern for a scan of the Cen\,A central core and southern lobe. The circles correspond to the size of the LST field of view (4.5 deg in diameter), which is 
the smallest field of view of the three CTA telescope types.
Different colours and line styles indicate the three different exposures with pointings in wobble mode
(see text for discussion). Credits for the skymap: NASA/DOE/Fermi-LAT Collaboration, Capella Observatory, and Ilana Feain, Tim Cornwell, and Ron Ekers (CSIRO/ATNF), R. Morganti (ASTRON), and N. Junkes (MPIfR).
 \label{fig:AGNcenascan}
}
\end{centering}
\end{figure}

{\bf 3b) Deep exposures of two radio galaxies.}

The two nearest radio galaxies, both known TeV emitters, are proposed for deep observations to extract high-quality spectra and to search for extended emission:

- {\bf Cen A:} To probe extended emission from Cen~A, either from its radio lobes or its kpc jet, a deep scan is foreseen. Three deep exposures with the southern CTA array will be 
necessary to completely cover the central part of Cen A, including the kpc jet and the southern extended radio lobe, with sufficient overlap between the fields of view and allowing for a significant exposure away from the lobe for background subtraction.
The three fields of view will be centred on the central core and on the lateral edges of the southern radio lobe. Data will be taken in wobble mode, where the telescopes are pointed at small offsets from the 
targeted position, usually 0.5 to 1.5 deg, in alternating directions to facilitate background subtraction (cf. Figure~\ref{fig:AGNcenascan} for a schematic view of the pointings).

The exposure of the core region will provide a high-quality spectrum of the emission from the central source and will allow us to distinguish between different proposed interpretations of the 
mismatch seen between the Fermi-LAT and H.E.S.S. spectra. 
If this first exposure shows an indication for a signal from the lobes, two dedicated exposures of the southern lobe will be carried out. 
We choose the southern lobe since it is seen to be brighter than the northern one at other wavelengths. An observing time of 50\,h (dark time) will be needed for each of the
three exposures, resulting in a total of 150\,h of observing time with the southern array. All telescopes should be included in these observations. In the case of discovery of extended emission from the 
southern lobe, this observational programme could be extended to both radio lobes, either in the form of an updated KSP or in the form of a GO proposal, which would profit from the technical expertise 
that we will have acquired, especially concerning estimates of systematic errors.\\

- {\bf M\,87:} Given its smaller extension in the sky, M\,87 can be covered with a single deep observation. A total observing time of 100\,h 
of dark time is foreseen for the full northern array. This observation will produce a high-quality spectrum and might lead to the detection of extended emission from the radio lobes \cite{Sol13}. 
In addition, a possible extension of the hard power-law of M\,87 to a few tens of TeV 
would allow for a very good probe of the EBL
in the regime of a few tens of $\mu$m, although significant
statistics in this energy range would probably require data from
flaring states due to the foreseen absence of an SST component in the northern array. 
It should be noted that M\,87 is also a prime target for monitoring, due to its high variability. Exposure time spent on the source with the long-term monitoring programme during the first three years will contribute to the deep exposure.

\vspace{0.2cm}

{\bf Time distribution model for the AGN KSP}

A summary of the required observation times for the different parts of the AGN KSP is given in Table~\ref{tab:AGNtimes}. 
The current time distribution model foresees that the high-quality spectrum, long-term monitoring and the AGN flare programmes 
begin during the construction phase of CTA. Spectrum measurements will be completed by the end of the third year of operations with the full arrays. 
However, with the estimated rate of flares it will take about five years to collect the 30 to 50 flares needed to cover the high-redshift bins for the EBL measurement.
It should be noted that $<$50\% of flares are estimated to provide sufficiently high-quality spectra to be used for this purpose. 
For FSRQs, NLSy1s and high-redshift sources, the programme should continue beyond the first five years. After the first two years, the trigger threshold will be increased to 
progressively reduce the observing time dedicated to ToOs and snapshots.
Long-term monitoring should continue for at least ten years, with a reduction in the number of targets to the ten most interesting ones (given their past record of observations) after the first five years.

\begin{table}[h!]
\begin{centering}
\begin{tabular} { | l | l | l | l | l | }
\hline
Programme & total N [h] & total S [h] & duration [yr] & observation mode \\	
\hline
\hline
{\bf Long-term monitoring} & 1110  & 390 & 10 \,\, $\dagger$ & full array \\
\hline
{\bf AGN flares} & & & & \\
snapshots  & 1200  & 475 & 10 \,\, $\ast$ & LSTs \\
snapshots  & 138 & 68 & 10 \,\, $\ast$ &  MSTs (assuming 10 sub-arrays)\\
verification ext. trig. & 300 & 150 & 10 \,\, $\ast$ & LSTs or MST sub-arrays \\
follow-up of triggers &  725 & 475 & 10 \,\, $\ast$ &  full array \\
\hline
{\bf High-quality spectra} & & & & \\
redshift sample &  195 & 135 & 3 & full array \\
M\,87 and Cen\,A & 100 & 150 & 3 & full array \\
\hline
\hline
\end{tabular}
\end{centering}
\caption{Summary of required observing times for the northern site (``N'') and the southern site (``S'') for the different parts of the observation programme. 
The total duration of each programme is given in the fourth column, where
a ``$\ast$'' (``$\dagger$'')  indicates a reduction of the yearly exposure time after 2 (5) years.
}
\label{tab:AGNtimes}
\end{table}

\begin{figure}[h!]
\begin{centering}
\resizebox{0.7\columnwidth}{!}{\includegraphics{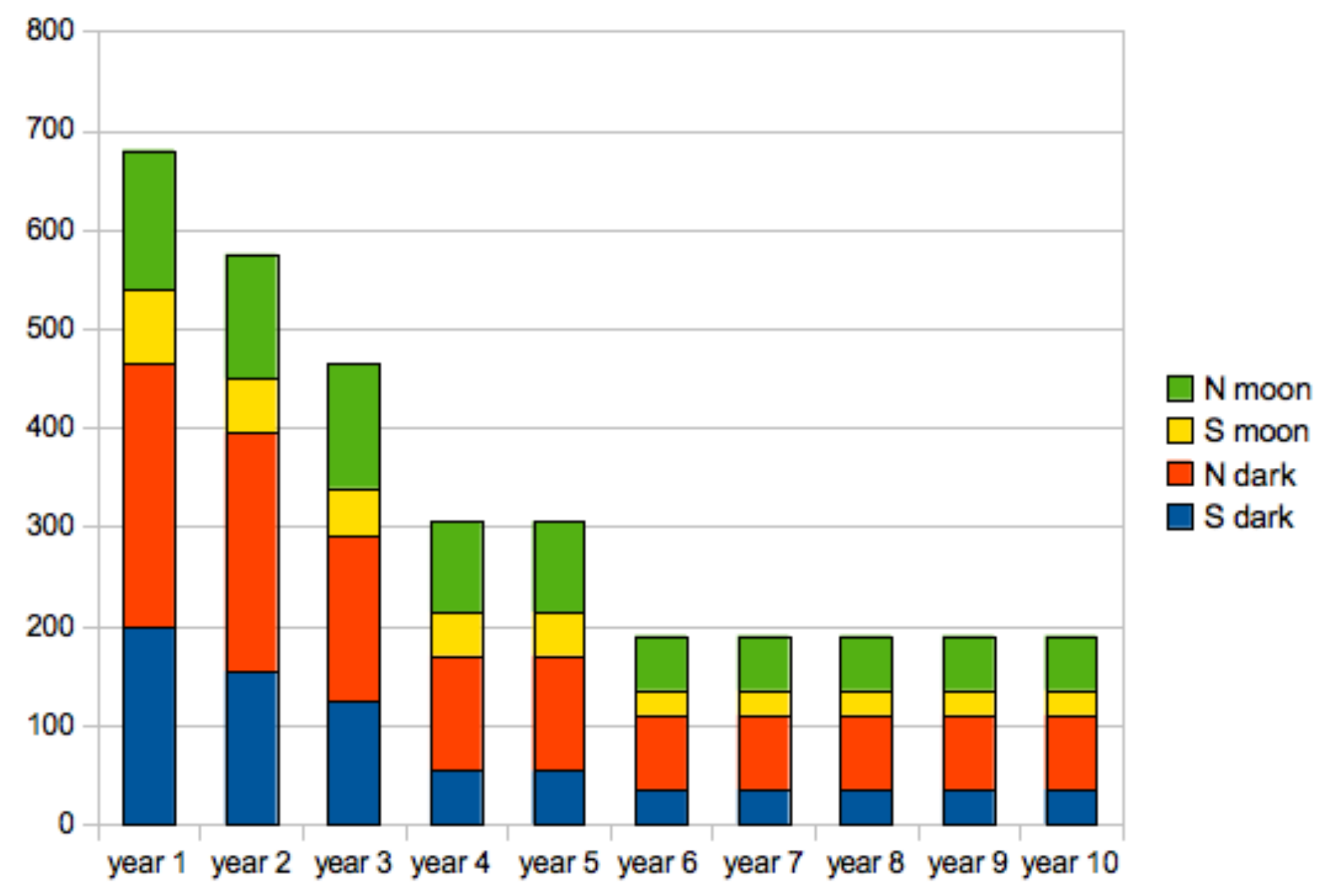}}
\caption{Distribution of the required full-array observation times (in hours) for the AGN KSP over the first ten years of full operation, including all observation programmes, 
except for snapshots and verification of external triggers, which will not necessitate the full array (cf. Table~\ref{tab:AGNtimes}).
The different colours in each column indicate, from top to bottom, the exposure during moon time for the northern and the southern arrays
and the exposure during dark time for the northern and the southern arrays.
 \label{fig:kspAGNtimedistribution}
}
\end{centering}
\end{figure}

Based on ten years of operation with the full arrays, the AGN KSP would require a total of about 3300\,h full-array exposure, not accounting for observations with sub-arrays (i.e.\ snapshots and verification
of external triggers). 
However, the use of moonlight time for the observation of hard-spectrum sources would reduce the amount of required dark time for full-array observations to about 2000\,h. 
For example, targets from the 
EBL/IGMF selection with redshifts z$<$0.2 can in general be observed with moderate moonlight, given the existence of data points at several TeV.
About 1000 additional hours of observations could be carried out during the construction phase, split in equal parts between the high-quality spectra programme, the 
long-term monitoring programme and the AGN flare programme. Data for several targets (e.g.\ PKS\,2155-304, PG\,1553+113, etc.) will also be available from the 
science verification observations. 
A time distribution model for exposures with the full array during the first ten years of full operations is shown in Figure~\ref{fig:kspAGNtimedistribution}.

\begin{figure}[h!]
\begin{centering}
\resizebox{0.7\columnwidth}{!}{\includegraphics{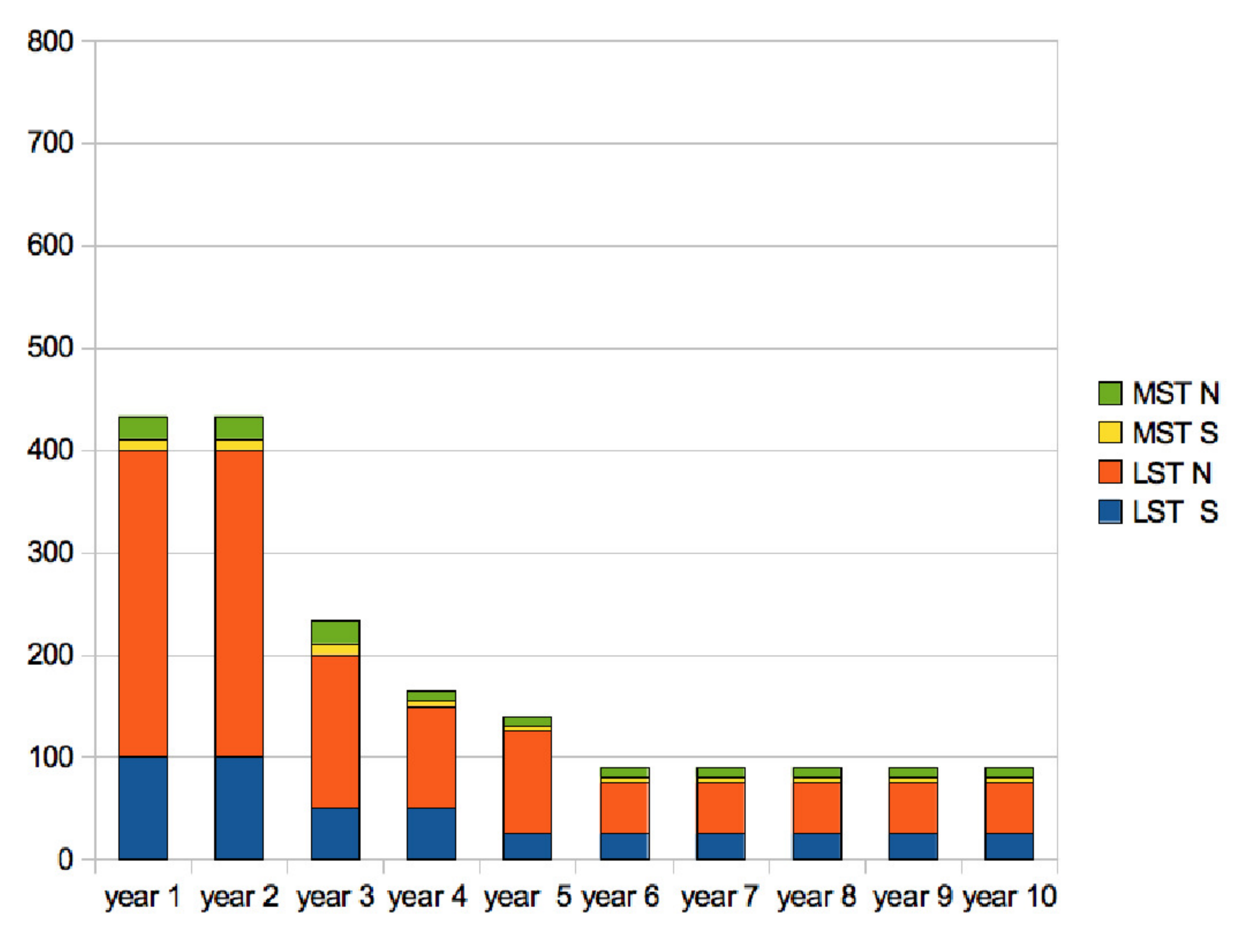}}
\caption{Distribution of the proposed 
sub-array observation times (in hours) for the snapshot programme over the first ten years of full operation (see Table~\ref{tab:AGNtimes}).
The different colours in each column indicate, from top to bottom, the exposure with MST sub-arrays with the northern site and 
southern site and the exposure with LST sub-arrays with the northern site and southern site. It should be noted that about half of these observations will make use of moonlight.
 \label{fig:kspAGNtimedist_snapshots}
}
\end{centering}
\end{figure}

The observation time with partial arrays must also be considered. 
An estimated 1350 hours (550 hours) of LST exposure time and 290 hours (140 hours) of MST exposure time are required over ten years for the 
northern (southern) array for snapshots and verification of external triggers. About half of these observations will be carried out under moonlight, leading to a total of about 950\,h of required dark time for 
the LST sub-array 
and about 220\,h of required dark time for the MST sub-arrays over ten years of full operation.
Snapshots with LSTs could be carried out while the MST and SST component of the array are used for observations that do not require very low energy reach. 
MST snapshots could also be run in parallel with other observations. A study on optimizing the usage of LST sub-arrays is under way. A time distribution model for snapshot exposures with LST and MST sub-arrays during the first ten years of full operations is shown in Figure~\ref{fig:kspAGNtimedist_snapshots}.

\subsubsection{Multi-wavelength \& multi-messenger coverage}
\label{sec:AGNMWL}

The unique advantage of observations of AGN in the VHE band is that they provide information on the most energetic particles inside those sources and probe the shortest variability time scales. 
However, to maximise the scientific return from such observations, a combination with data from other wavebands and messengers is of great importance. 
Simultaneous MWL observations have proven of great value in the past to constrain emission scenarios through correlated flux variations and detailed SEDs. 
The main reason is that different components in the blazar spectra vary on different time scales and are thought to be produced in different places along the jet. 

The availability of dedicated optical instruments in close proximity to the two CTA sites, with photometric and polarimetric capabilities, would be an important advantage for this KSP. As discussed 
in Chapter~\ref{sec:sci_synergies}, this would permit a continuous monitoring of all targeted sources, providing complementary data simultaneous to the VHE observations. Candidate 
sources in the search for flaring states could be monitored for high optical states, which seem generally linked to increased VHE activity (see, e.g., \cite{Aleksic2012, Abramowski12f}). 

In addition, the optical polarization monitoring of blazars has proven to be a  
powerful tool for locating the emission region within the jet in luminous quasars 
(see, e.g., \cite{Marscher2008}, \cite{Marscher2010}). Polarization traces the ordered magnetic field of the jet and can
be modified by shocks. Rotations of the optical polarization vector or shifts in the polarization fraction have been found to coincide with gamma-ray outbursts and can 
be used to trigger VHE observations. 
A good example is the RoboPol project\footnote{\url{http://robopol.org}} that monitored a sample of 100 gamma-loud blazars in the optical band, requiring 60\% of the time of the 
dedicated facility \cite{Pavlidou14}. A large statistical sample of sources in
different activity states enabled the assessment of the statistical connection
between the optical polarization and high-energy events. During its
three first seasons, RoboPol detected 40 optical polarization angle
rotations indicating a physical connection to gamma-ray flares, possibly
based on several distinct underlying mechanisms \cite{Blinov2016}.
Based on these results it is already clear that a dedicated instrument
is needed to provide triggers for CTA based on optical rotations and to
systematically monitor AGN observed with CTA.

All of the observation programmes of this KSP require at least some MWL support to guarantee an optimal impact of the resulting data:
\begin{itemize}

\item Long-term monitoring of a few prominent AGN

Well sampled, simultaneous light curves at different wavelengths (X-ray, optical, radio) are necessary to allow us to search for correlations and time-lags between different bands.
This is most efficiently organized via dedicated long-term monitoring programmes. Current examples are the Tuorla blazar monitoring (in optical) and Mets\"{a}hovi Radio Observatory programmes, the data of which are being used 
by the MAGIC collaboration, or the systematic follow-up of H.E.S.S. observations with the on-site optical ATOM telescope and joint observation programmes with the Nan\c{c}ay Radio Telescope. 
For CTA, the optical flux and polarization monitoring 
may make use of dedicated telescopes.
In the radio band, several facilities exist and coordination of long-term programmes seems feasible. 
While no direct correlations between the low-frequency radio band and the higher frequency bands are expected, follow-up at low frequencies (e.g.\ with MWA, LOFAR,...) is still interesting to study long-term variations.
Organizing the long-term coverage in the X-ray band is more challenging and needs to be explored in the near future.

\item AGN flare programme
  
MWL coverage is crucial, at least for the most prominent flares, for variability studies and dynamical spectral modelling of flaring states.
The most relevant instruments for CTA are discussed in Chapter~\ref{sec:sci_synergies}.
The organisation of MWL campaigns should include: spectral information from X-ray telescopes (currently operational: Chandra, XMM, Swift, NuStar; planned: Athena+ ...);
photometry and polarimetry from optical telescopes; flux densities and polarimetry from radio and sub-millimeter telescopes (currently operational: OVRO, Mets\"{a}hovi and Nan\c{c}ay radio telescopes, GMRT, JVLA, SMA, ASKAP, MeerKAT, eMerlin, ATCA, ALMA, VLBI facilities; soon operational: SKA, Event Horizon Telescope...).  
Simultaneous observations with Fermi are of great interest for flaring AGN and should be given high priority during the first years of operation of CTA, as long as Fermi is operational. 

A network of exchanges with MWL and MM facilities will also be very important for incoming and outgoing alerts associated with AGN flares. In addition to wide-field VHE detectors (especially HAWC and the future LHAASO) and
Fermi, X-ray instruments with 
very wide field of view, e.g. the ASTROSAT telescope or the planned SVOM satellite, are very interesting for incoming triggers. 
Optical telescopes will provide another source of alerts triggered by high flux states and changes in polarization. Although such external triggers will be very useful, the fact that variability patterns are not
the same in different wavebands, and the limited sensitivity and modest low-energy reach of the wide-field VHE detectors, still mean that the internal snapshot programme remains essential to fully profit from the timing capabilities of CTA. 

On the other hand,
CTA will provide outgoing alerts to the MWL and MM communities.  
Exchanges with MM experiments (currently operational: IceCube, ANTARES, VIRGO, LIGO, Pierre Auger Observatory, TA; planned: KM3Net, eLISA, JEM/EUSO...) to trigger mutual alerts already exist for the current generation of atmospheric Cherenkov telescopes (e.g.\ MoUs with IceCube and ANTARES) and should be continued and intensified with CTA.

\item High-quality spectra 

 MWL coverage is needed for the interpretation of SEDs, but since we are dealing mostly with low states, simultaneous observations will 
 not always be crucial and contemporaneous/archival data will suffice for some sources. High coverage optical data will be very useful to compare the state of sources at low energies against archival data.
 In the case of significant flux variations, however, MWL campaigns for simultaneous coverage will need to be arranged quickly, as discussed for
 the AGN flare programme.
Simultaneous Fermi-LAT data would be very complementary to CTA data on the intrinsic high-energy spectra for EBL and IGMF studies. 
 
Contrary to the low-state blazar observations, the deep exposures of the radio galaxies M\,87 and Cen\,A should be accompanied by MWL campaigns. Spectra from instruments with high angular resolution (X-rays: Chandra, later Athena+,...; optical: the future JWST; sub-mm and radio: ALMA and VLBI facilities) are especially interesting to probe different scenarios for the location of the VHE emission region(s), 
as has been seen in previous campaigns on M\,87.

\end{itemize}

Another aspect of complementary MWL observations concerns the early organisation of redshift campaigns. A significant fraction of the AGN detected with CTA, especially BL Lac objects, will have no spectroscopic redshift 
and sometimes only poor photometric redshift estimates. For example, about 50\% of AGN from the 1FHL catalogue~\cite{Ackermann2013} that will be detectable with CTA, have currently no reliable redshift. This  puts a serious limit on our selection of sources for the ``high-quality spectra'' programme and will also be a challenge for the exploitation of AGN flares. The problem is more acute at redshifts z$>$0.2. To increase the 
fraction of AGN with well-known redshifts, efforts are currently underway 
to set up dedicated campaigns.

To summarize, the organization of MWL observations can be divided into two types of campaign: long-term monitoring and targeted campaigns (both ToOs and pre-planned observations). For the latter, 
we envisage the use of proposal-driven observatories and time granted via ToOs and MoUs. Guaranteeing long-term monitoring observations via proposal-driven observatories is much more challenging. In the optical band, a dedicated automatic telescope at each CTA site would be ideal, while in the radio band, MoUs with existing telescopes (such as Mets\"ahovi, Nan\c{c}ay, MWA, ...) could be sufficient. The possibility of long-term campaigns including X-ray facilities will need to be explored.

\subsection{Data Products}

The data products that will result from the different parts of the AGN KSP are the following:

\begin{enumerate}

\item {\bf Long-term monitoring} will provide high-resolution light curves and time-resolved spectra in different source states for 15
archetypal sources of different classes. The light curves will be released immediately in an online catalogue and will be continuously updated.
All data will be released in such a format as to allow external users to extract spectra for time periods of their choice.

\item Data from the {\bf AGN flare programme} will result in time-resolved spectra and light curves from flaring states of all different types of VHE AGN.
Flares detected with CTA snapshots will trigger alerts to inform the community and to enable MWL follow-ups. Such alerts can take the form of Astronomer's 
Telegrams or VOEvents. Data from flares that are detected under this programme will generally be released after the proprietary period for each flare observation. Mutual
agreements with other collaborations will permit joint publications and the exchange of data from specific flares before the end of the proprietary period.

\item {\bf High-quality spectra} of about 40 blazars and radio galaxies will be published, together with the complete data sets, after the proprietary period for this programme.
The spectra will form 
a homogeneous catalogue to be used for source statistics and modeling by the wider community. 

\item Deep exposures of {\bf M\,87} and {\bf Cen\,A} will potentially result in maps tracing the VHE emission in those sources, if an extended 
component is found. These maps will be published together with the complete data sets from these observations after the proprietary period for this programme. 

\end{enumerate}

\subsection{Expected Performance/Return}

The expected scientific results from the AGN KSP can be separated into guaranteed results and potential discoveries.\\

{\bf Guaranteed science return}

\begin{figure}[htb!]
\begin{centering}
\resizebox{0.9\columnwidth}{!}{\includegraphics{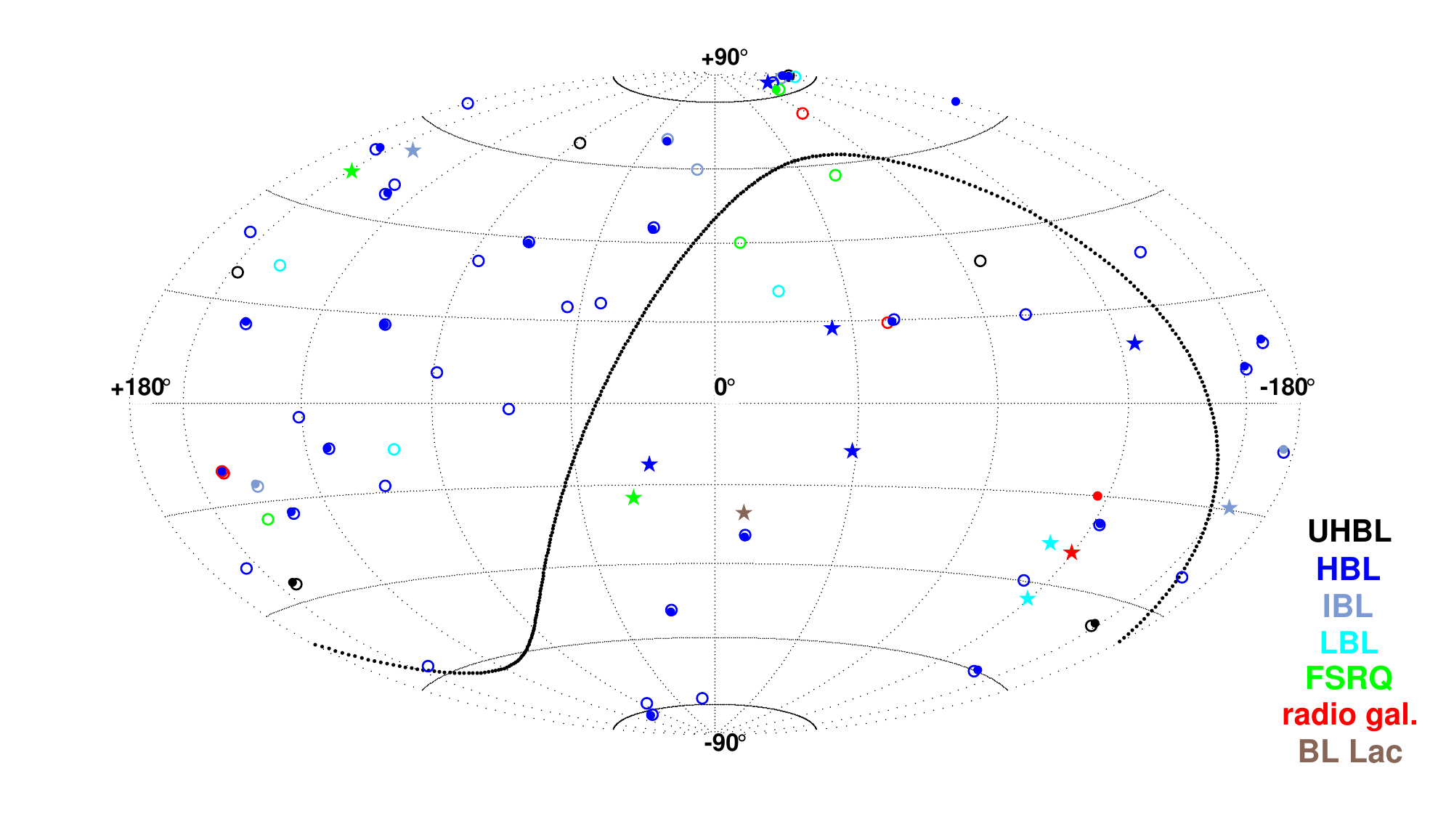}}
\caption{
  Skymap of potential sources from the ``high-quality-spectra'' programme in Galactic coordinates.
  In the map, the already known VHE AGN
 are marked with open circles. Targets for the AGN KSP are marked with
  full circles (known VHE emitters) or stars (new candidate sources). Different colours in the map correspond to different AGN classes.  
  The solid black line indicates the celestial equator.
 \label{fig:kspAGNskymap}
}
\end{centering}
\end{figure}

\begin{figure}[htb!]
\begin{centering}
\resizebox{0.8\columnwidth}{!}{\includegraphics{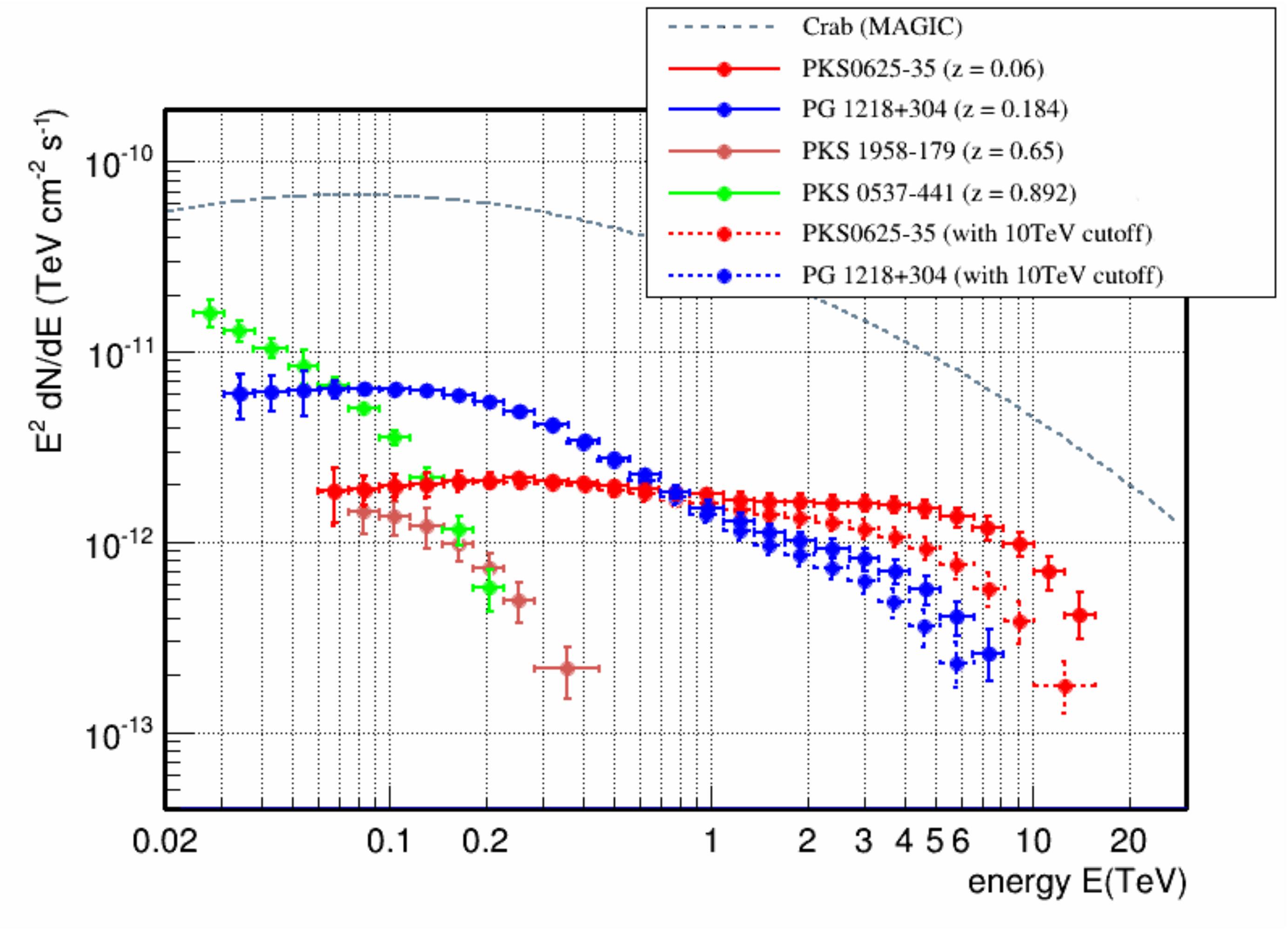}}
\caption{
  Four examples of simulated 
  CTA spectra at different redshifts based on extrapolations from Fermi-LAT and 
  absorption by the EBL \cite{Franceschini2008}. For the two hardest source spectra, alternative scenarios with an exponential cutoff at 10\,TeV 
  are shown for comparison. Error bars indicate the statistical uncertainty 
  of one standard deviation. 
 \label{fig:kspAGNspectra}
}
\end{centering}
\end{figure}

The main deliverable of the AGN KSP will be a homogeneous collection of high-quality spectra and light curves, which will be exploited for spectral fitting, variability studies, and comparison against emission models.
As was seen with the results from Fermi-LAT and the current
generation of atmospheric Cherenkov telescopes,
the knowledge gained from such an increased quality and quantity of available data will
be profound and not restricted to a single science
question. Figure~\ref{fig:kspAGNskymap} shows a sky map of the current target selection for the extraction of high-quality spectra.
Simulated spectra for a few sources are shown in Figure~\ref{fig:kspAGNspectra} (as well as in Figures~\ref{fig:kspAGNmodels} and \ref{fig:kspAGNuhecr}) to illustrate how different physical assumptions on the mechanism
of the observed emission can be distinguished using data from the AGN KSP.

The systematic search for flaring events with snapshots will lead to the detection of the order of 200 flares with fluxes above 20$\%$ of the Crab nebula flux 
from all different known types of gamma-loud AGN during the first ten years of full operation. External triggers will lead to an expected number
of about 10 to 15 additional flare detections per year.  
This advance in time-domain VHE observations will be unequalled by other instruments. VHE air shower arrays, such as HAWC and LHAASO, have a much larger 
duty cycle and sky coverage, but will only be sensitive to the brightest flares from hard-spectrum sources. The source populations probed with CTA and with
air shower arrays are thus complementary.

A certain number of serendipitous source detections (on the order of ten) is also guaranteed, based on our experience with the current generation of atmospheric Cherenkov telescopes. 
It should be noted that the ``high-quality spectrum'' observations alone 
will cover a non-negligible fraction of the sky. If CTA points to $\sim$40 separate fields of view 
with the MSTs, a total geometrical field of view of more than 4$\%$ of the sky will be covered.

A precision measurement of the EBL and its evolution with an unprecedented 
accuracy of $\sim$10\%  
will be carried out, thanks to the observations of steady sources and flares, (see Figure~\ref{fig:kspAGNebl}). Observations with 
atmospheric Cherenkov telescopes are among 
the only means for making such a measurement, which is
closely linked to our understanding of the evolution of the universe and its star-forming history.

\vspace{0.5cm}

{\bf Discovery Potential}

Based on Fermi-LAT observations of 
flaring events, the high-energy end of the spectrum from NLSy1s will be accessible to CTA
with high probability and a few of these
sources should be detectable by CTA. The measurement of their spectra will profit from the very good time resolution of CTA and will constrain the conditions 
in the emission region of these still poorly understood objects.

Other AGN types, such as 
radio-quiet Seyfert galaxies or Low-Luminosity AGN, will be probed for VHE emission through a  stacking analysis. 
In the absence of a signal, such studies will put stringent upper limits on theories that require particle acceleration in the magnetosphere linked to the SMBH.

Extended VHE emission from radio galaxies would be a major discovery as it would prove that particles can be accelerated to very high energies in the extended
jets or radio-lobes of AGN, thus constraining the energetics and the magnetic fields in these structures and providing us with a first direct view of the location of the 
VHE emission regions in AGN.

\begin{figure}[h!]
\begin{centering}
\resizebox{0.6\columnwidth}{!}{\includegraphics{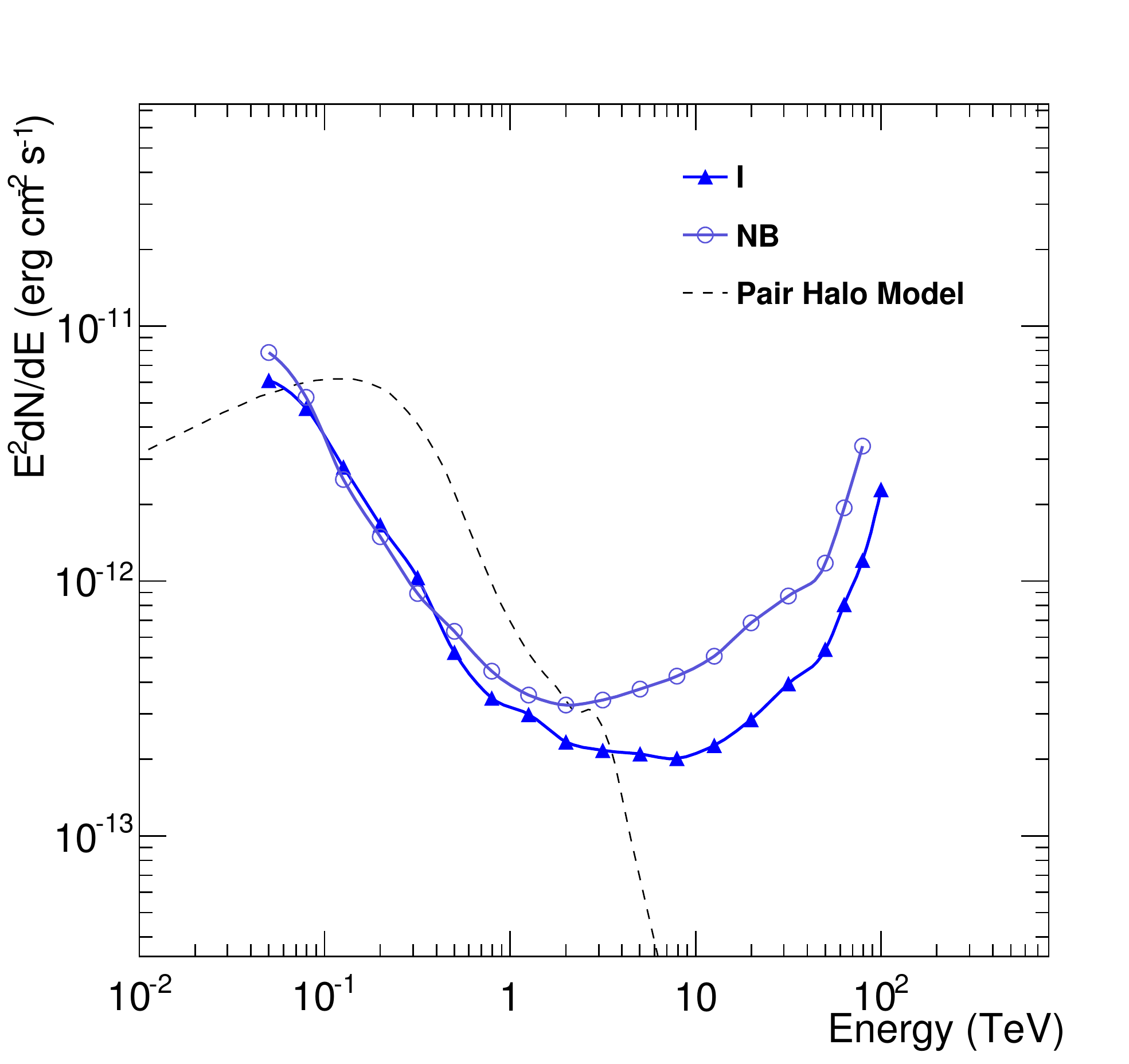}}
\caption{
  Estimate of the flux for the expected pair halo emission compared to sensitivity curves 
for the southern (``I'') and northern (``NB'') sites. 
  A differential angular distribution of a pair halo at z$=$0.129 and E$_{\gamma} >$ 100\,GeV was used for the theoretical model and an observation time 
  of 50\,h was assumed. For more details see \cite{Sol13}. 
     \label{fig:kspAGNhalo}
}
\end{centering}
\end{figure}

Another major discovery would be the detection of gamma-ray signatures from the IGMF in the form of pair halos or pair echoes. This would be the first (indirect)
measurement of the strength of the IGMF, leading to important implications on its origin. Clearly, the derivation of reliable constraints will already be an important step.
Figure~\ref{fig:kspAGNhalo} shows the expected signal from a pair halo model compared to CTA sensitivity curves.

\begin{figure}[h!]
\begin{centering}
\resizebox{\columnwidth}{!}{\includegraphics{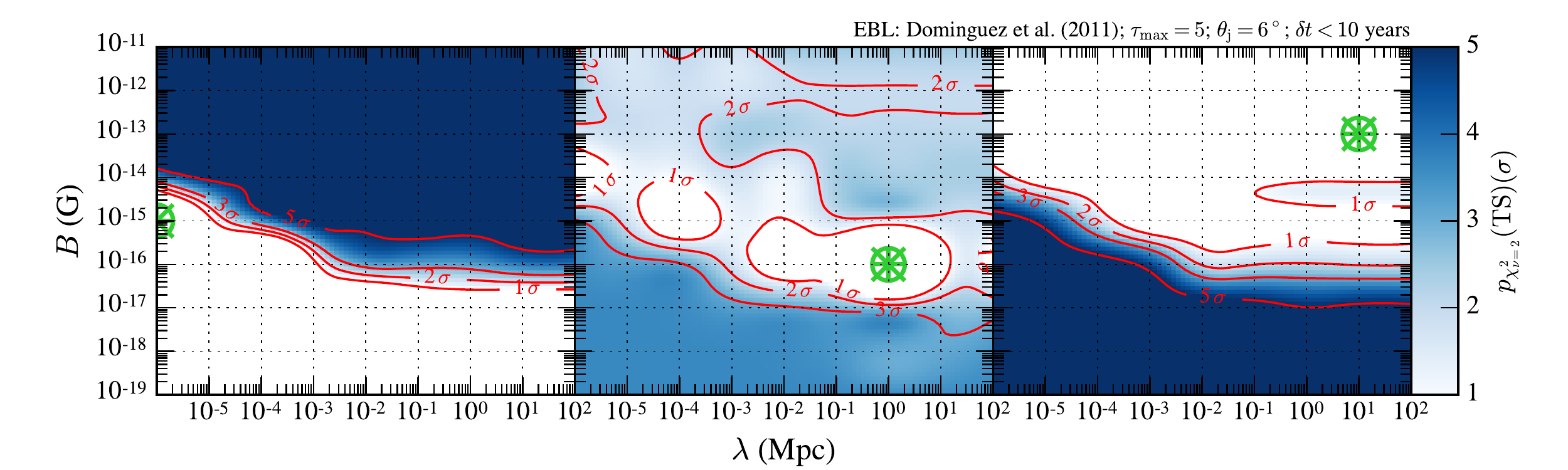}}
\caption{
Possible constraints on the IGMF field strength (B) and coherence length ($\lambda$) for three different assumptions on the true IGMF (green markers).
Colours and contour lines give the significance with which IGMF parameters can be rejected in units of the standard deviation ($\sigma$). 
The constraints are derived from a combined likelihood analysis of simulated CTA spectra of four UHBLs using the non-observation of a pair echo. 
Only cascade photons that arrive within the 80\% containment radius and a time delay of less than ten years are included in the analysis. For more
information, see \cite{Meyer16}. 
\label{fig:kspAGNecho}
}
\end{centering}
\end{figure}

Expected results of searches for pair echos with CTA are shown in Figure~\ref{fig:kspAGNecho} for three different assumptions on the magnetic field strength
and coherence length of the IGMF. These results are derived for the observation of several known blazars with hard spectra in the TeV band (\cite{Meyer16}).

CTA might also identify the sources of UHECRs, through indirect signatures of protons or nuclei in the gamma-ray spectra. The long-standing
question of UHECR origin 
has been difficult to tackle using data from UHECR particle detectors alone. 
Apart from neutrino detectors, gamma-ray telescopes 
seem to provide the only other means of attacking this open problem. 
A possibility to investigate physics beyond the Standard Model is given by the search for ALPs, the discovery of which would have profound implications for particle physics. 
The existence of such particles would also imply that the ``gamma-ray horizon'' for TeV detection is far less constraining than what is currently assumed. On the other hand, 
if no ALP signatures are found, strong constraints on their parameters are possible that could rule out certain ALP dark-matter scenarios. 

The search for LIV will lead either to the verification of the theory of general relativity at a new level of sensitivity or possibly to the discovery of new physics, 
again with profound implications for fundamental physics. 
With data from this KSP, we will match the Fermi-LAT sensitivity 
at the Planck scale on the linear term 
with a different type of astrophysical object. We will also improve the sensitivity on 
the quadratic term by a factor of two compared to current results, i.e. at a scale of 
a few times 10$^{11}$\,GeV. These results will be largely independent of the source physics since they will be based on data from sources at different 
redshifts and in different states. Whilst transient events will still provide the best constraints, the observations in the AGN KSP will be capable of providing competitive 
sensitivity independent of flaring events, through a mixture of routine 
observations of many sources and deeper exposures of a few suitable hard-spectrum sources.

\FloatBarrier

\section{KSP: Clusters of Galaxies }
\label{sec:ksp_clust}

Galaxy clusters are expected to be reservoirs of cosmic rays accelerated by structure formation processes, galaxies and active galactic nuclei (AGN). The detection of diffuse synchrotron radio emission in several clusters confirms the presence of cosmic-ray electrons and magnetic fields permeating the intra-cluster medium (ICM). While there is no direct proof for proton acceleration yet, gamma rays can prove it as cosmic-ray protons can yield high-energy gamma-ray emission through neutral pion decay. Additionally, some of the galaxies hosted in clusters could be detected in gamma rays individually, if harbouring AGN, or via their summed output, e.g.~in the case of star-forming galaxies. Finally, about 80\% of the mass of clusters is in the form of dark matter and therefore they are considered prime targets for indirect dark matter searches.

\begin{figure}[b!]
\centering 
\includegraphics[width=.53\textwidth]{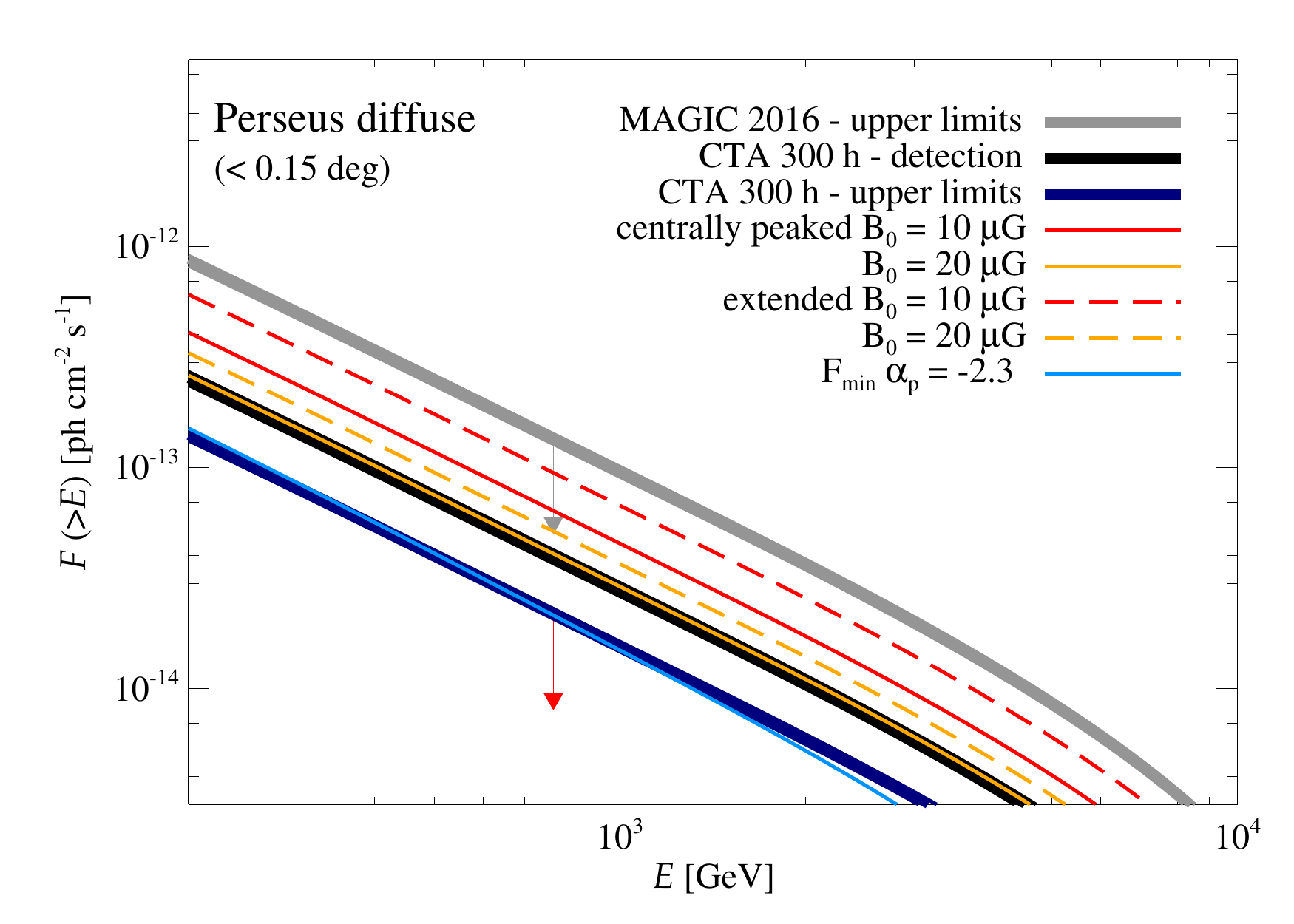}
\includegraphics[width=.455\textwidth]{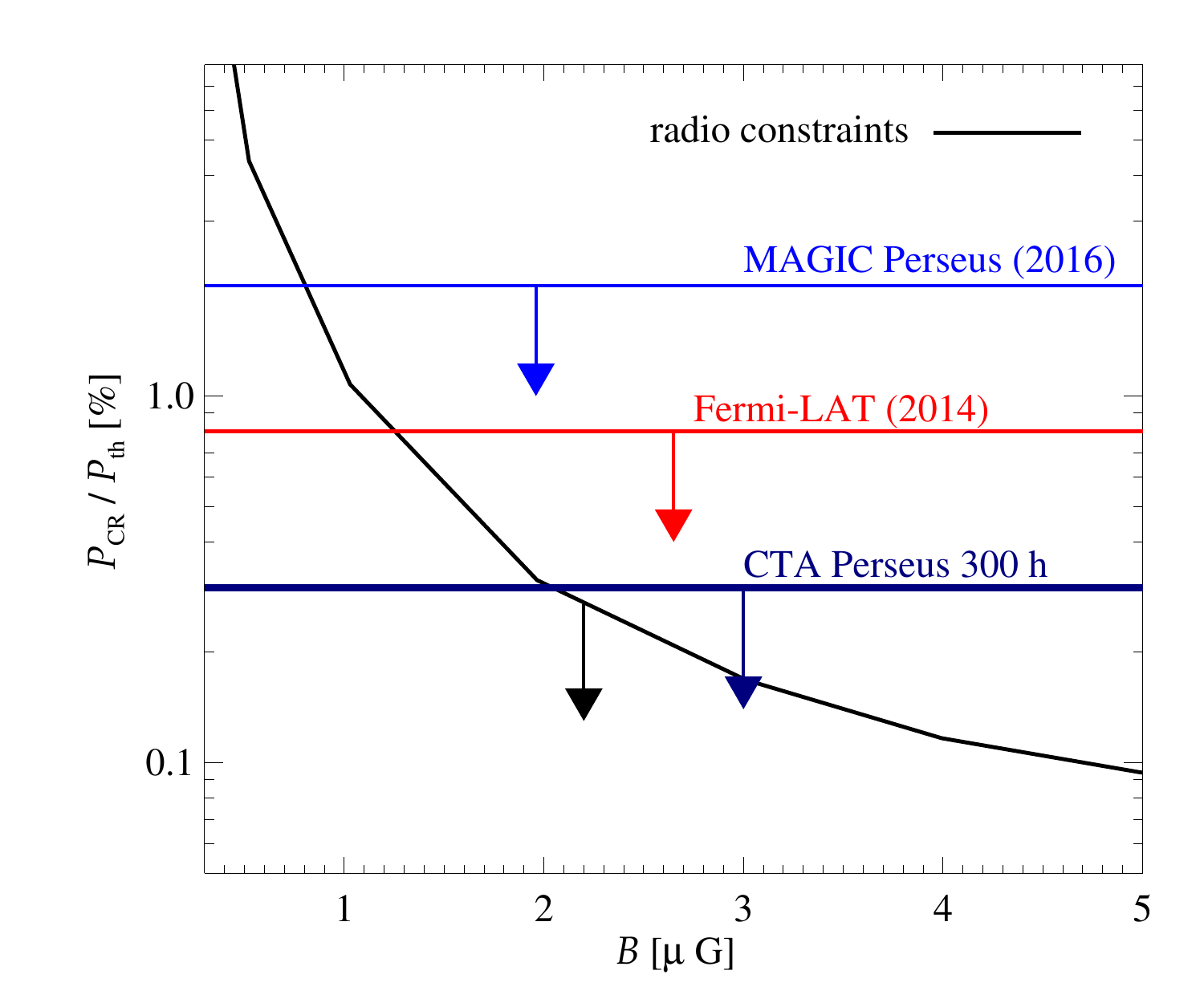}
\caption{\label{fig:pexpectations} 
Left: Theoretical parameter space for the predicted intensity of cosmic-ray-induced gamma-ray emission in the Perseus cluster integrated within a region of radius $0.15^{\circ}$ from the cluster centre. The models refer to centrally peaked (solid line) and extended (dashed line) cosmic-ray profiles \cite{Pinzke10,Zandanel14,Aleksic16a}. We show them for two different values of the central magnetic field for which the models fit the radio mini-halo of Perseus assuming it has an hadronic origin. The grey band shows current upper limits on the diffuse emission from the MAGIC observations \cite{Aleksic16a}. The dark blue band is an estimate of the 95\% upper limit level reachable with 300~h of observation by CTA-North, while the black band represent an estimate of the corresponding detectability level. All these three bands refer to the adopted models, with the upper and lower boundaries corresponding to the extended and centrally peaked models, respectively. We additionally show, with a light blue line, the so-called minimum gamma-ray flux, for a cosmic-ray proton spectral index of $2.3$, assuming that the Perseus radio mini-halo is of hadronic origin \cite{Aleksic16a}.
Right: Upper limits on the cosmic-ray-to-thermal pressure in clusters. We show the upper limits obtained from GMRT radio
observations (black line), which depend on the assumed magnetic field value \cite{Brunetti07b}, together with the latest upper
limits from the MAGIC observations (blue line, \cite{Aleksic16a}) and the Fermi-LAT observations (red line, \cite{Ackermann14}). The thick dark blue line is an estimate of the level reachable with the proposed CTA observation of Perseus. All the gamma-ray limits refer to the case of the centrally peaked cosmic-ray profile.
}
\end{figure}

Focusing on the diffuse gamma-ray emission in clusters coming from proton-proton interactions, and based on both theoretical studies and hydrodynamical simulations, Perseus should be the brightest cluster of galaxies in gamma rays. 
Perseus also hosts two gamma-ray-bright AGN: NGC~1275, one of the few radio galaxies known to emit gamma rays, and IC~310, potentially one of the closest known blazars. These two objects prevent Fermi-LAT from studying the ICM itself in Perseus because of its modest angular resolution (especially at low energies). 
At very high energies, MAGIC performed a deep observation campaign and put constraints on the cosmic-ray-to-thermal pressure in the cluster at the few percent level \cite{Aleksic16a}. Currently the best limits on the proton content in clusters comes from Fermi-LAT observations \cite{Ackermann14}. Our simulations for CTA show that a 300~h observation will potentially allow a detection of the Perseus cluster in gamma rays or, alternatively, set unprecedented limits on the cosmic-ray proton content -- potentially triggering a substantial revision of the current paradigm of proton acceleration and confinement in galaxy clusters (see Figure~\ref{fig:pexpectations}). Additionally, due to the large mass of this cluster, CTA can significantly improve the constraints on decaying dark matter with respect to those from Fermi-LAT observations of the Galactic halo (see Section~4.2.4). 

The high-risk/high-gain character of this work makes it a worthy Key Science Project.
Additionally, the long observation time needed to achieve our scientific goals calls for a Consortium effort. 
Perseus is observable in optimal conditions, i.e.~low zenith angles ($<60^{\circ}$), only by the northern array. Here we refer to the fully complete northern array, as we need the best achievable sensitivity to aim for a detection of the diffuse gamma-ray emission.
The data products will depend on whether we get a detection or not.  
They will range from maps/datacubes (excess, flux, spectral hardness) to morphological analysis of the possible diffuse emission and detailed analysis on point-like sources. Special care will be given to examine the parameter space of cosmic-ray physics in clusters of galaxies.

The detection of diffuse gamma-ray emission from clusters of galaxies would establish 
a new class of gamma-ray sources. 
Thus, it would constitute a significant achievement and would be
potentially groundbreaking for the study of 
cosmic-ray acceleration and its relation to large-scale 
structure formation processes, the ICM and magnetic fields.

\subsection{Science Targeted} 
Clusters of galaxies represent the latest stage of structure formation. 
They are the most massive gravitationally bound systems in the universe, characterised 
by radii of a few Mpc and masses of the order of $10^{14} - 10^{15} M_{\odot}$. About 80\% of their mass 
is in the form of dark matter, while galaxies and gas contribute roughly 5\% and 15\%, respectively \cite{Voit05}. 
While there has been no detection of diffuse gamma-ray emission in clusters so far, these objects are expected to be 
gamma-ray emitters for several reasons that are explained below.

Clusters are being assembled today and are actively developing in this latest, and most energetic, phase of structure formation \cite{Forman03}.
It follows that they should dissipate energies of the order of the final gas binding energy via merger and accretion shocks as 
well as via turbulence (e.g.~\cite{Miniati15b}). These processes are also likely to accelerate electrons and protons to high energies (see \cite{Brunetti14} for a review).
While the presence of non-thermal electrons is confirmed by the detection of diffuse synchrotron radio emission from 
several clusters (see \cite{Feretti12} for a review), there is no direct proof of proton acceleration yet. 
However, cosmic-ray protons in the ICM have a long cooling time and should accumulate in clusters over 
cosmological times \cite{Volk96,Berezinsky97}. Therefore, they are expected to contribute to the relativistic particle content of clusters. 
If protons are present in significant amounts, their hadronic interactions with the ICM will generate gamma-ray emission via neutral pion decays (see e.g.~\cite{Blasi99,Pfrommer08}), which should dominate over inverse-Compton scattering or bremsstrahlung emission \cite{Blasi01,Pinzke10}. Additionally, protons can be accelerated to ultra-high energies by high Mach number shocks, such as accretion shocks around galaxy clusters \cite{Inoue05,Vannoni11}, or they can be directly injected into the ICM by the cluster's AGN \cite{Armengaud06,Kotera09}. The interaction of these ultra-high energy protons with both the ICM and the background (cosmic microwave background and infrared) photons \cite{Kelner08} could produce secondary electrons that would also yield significant gamma-ray emission via inverse-Compton scattering. The production and confinement of cosmic rays at very high energies is uncertain, as is the contribution of galaxy clusters to the high-energy part of the observed cosmic-ray spectrum.

Large-scale diffuse synchrotron radio emission is observed in several galaxy clusters. The existence of these so-called radio
halos and relics proves the presence of relativistic cosmic-ray electrons, as mentioned above, and of 
$\mu$G magnetic fields permeating the ICM \cite{Feretti12}. 
The short cooling length of synchrotron-emitting electrons at radio frequencies ($\sim10$ kpc) represents a challenge for theoretical
models that aim at explaining diffuse radio emission extending over several Mpc.
In the hadronic model, radio-emitting electrons are produced by cosmic-ray protons interacting with the protons 
of the ICM. 
Alternatively, in the re-acceleration model, cosmic-ray electrons can be re-accelerated by merger-induced turbulence and 
produce the observed radio emission during cluster mergers. 
See \cite{Brunetti14} for a review of these processes.

Peripheral radio relics show irregular morphology and highly polarised emission, and they appear to trace merger shocks. 
Radio halos, which are generally characterised by unpolarised radio emission (or with a low level of polarisation), are centred on clusters and show a more regular morphology. 
The latter can be divided in two classes. 
Giant radio halos are typically associated with merging clusters and have large extensions; e.g. the halo in Coma has an extension of about 2~Mpc. 
Radio mini-halos are associated with relaxed clusters that harbour a cool core and typically extend over a few hundred kpcs, e.g.~the Perseus 
radio mini-halo has an extension of about 0.3~Mpc. The observed morphological similarities with the thermal X-ray emission suggests radio halos 
may be of hadronic origin. In fact, cool-core clusters are characterised by high thermal X-ray emissivities and ICM densities that are more peaked 
in comparison to non cool-core clusters that often show signatures of cluster 
mergers (see, e.g.,~\cite{Croston08}).  

We know now that giant radio halos cannot be purely of hadronic origin \cite{Brunetti12,Zandanel14}. The challenge is to determine what the exact contribution of secondaries is to the observed radio emission, if any, and, therefore, the origin of the observed radio-emitting electrons (e.g.~\cite{Pinzke16}). On the other hand, current observations do not allow us to favour any model for radio mini-halos as both turbulent and hadronic scenarios fit current observations well (e.g.~\cite{ZuHone13,Jacob16}). There is also the problem of the connection between giant and mini-halos (e.g.~\cite{Zandanel14}). Understanding the nature of these halos and their possible connection will allow fundamental constraints on cosmic-ray acceleration and transport in galaxy clusters to be derived. Only adequately deep gamma-ray observations of massive nearby systems hosting diffuse radio emission can ultimately break the degeneracy of radio observations in solving the problem of the cosmic-ray electron/proton fraction in the ICM.

In addition to cosmic-ray protons accelerated by structure formation shocks in the ICM, galaxies and AGN activity can 
also inject and accelerate protons into the cluster environment. Based on the current theoretical paradigm, these contributions are 
thought to be about an order of magnitude lower compared to those from 
structure formation shocks \cite{Brunetti14}. However, 
as clusters host hundreds of galaxies, gamma-ray observations could detect their combined emission as a sum of discrete sources, as, e.g.~in the case of star-forming galaxies in clusters \cite{Storm12,Persic12}. 
From estimates of the star-forming galaxy population, 
it is possible to determine the minimum pion-decay-induced gamma-ray emission expected from galaxy clusters. Rough 
estimates \cite{Storm12} give integral flux lower limits above 1~TeV of the order of $10^{-14}$~cm$^{-2}$~s$^{-1}$ for Virgo 
and $10^{-16}$~cm$^{-2}$~s$^{-1}$ for Perseus for typical radii of several
 degrees (about $7.6^{\circ}$ for Virgo and $1.3^{\circ}$ for
Perseus). However, such an extended signal strongly challenges detection possibilities for CTA. 
Additionally, AGN (both aligned and misaligned) in clusters can be individually studied as in the case of NGC~1275 and IC~310 in Perseus 
\cite{Aleksic10,Aleksic10a,Aleksic12,Aleksic14,Aleksic14a}. For all these reasons, CTA observations of clusters of galaxies are imperative;
they will pave the road to very high-energy studies of member galaxies
 and open up a completely new window on clusters. 

Galaxy clusters are also interesting environments in which to study axion-like particles (ALPs) \cite{Wouters13}. For example, the strong magnetic field in the centre of the Perseus cluster (see \cite{Aleksic10} for a discussion) would offer an ideal environment for photon-ALP conversion. Therefore, NGC~1275 is an optimal target to search for irregular features in the gamma-ray spectrum possibly connected to this phenomenon \cite{Ajello16} (see also Chapter 12). 
Finally, galaxy clusters present very high mass-to-light ratio environments and should also be considered as targets for indirect dark matter searches, in particular for dark matter decay (see Chapter~\ref{sec:DM_prog}).

\subsubsection{Scientific Objectives}
The scientific goals of this Key Science Project are to:
\begin{enumerate}
  \item detect, for the first time, diffuse gamma-ray emission from clusters of galaxies,
  \item determine the cosmic-ray proton content of clusters and its dynamical impact on the cluster environment,
  \item study the clusters' cosmic-ray proton acceleration, propagation and confinement properties,
  \item study the origin of the radio-emitting relativistic electrons and the connected particle acceleration mechanisms in galaxy clusters, and
  \item study the magnetic field distribution in clusters, in synergy with radio and X-ray observations. 
\end{enumerate} 

\subsubsection{Context / Advance beyond State of the Art}
There have been many gamma-ray observations of clusters over the last twenty years from EGRET, WHIPPLE, CANGAROO, VERITAS, H.E.S.S., MAGIC and Fermi-LAT \cite{Reimer03,Ackermann10a,Ackermann10b,Jeltema11,Han12,Ando12,Huber13,Zandanel14b,Ackermann14,Prokhorov14,
Vazza14,Griffin14,Selig15,Vazza15,Ackermann15a,Ackermann15b,Perkins06,Perkins08,Aharonian09z1,Domainko09,Galante09,Kiuchi09,Acciari09z,Aleksic10,Aleksic12b,Arlen12,Abramowski12z,Aleksic16a}.
The MAGIC observations of Perseus \cite{Aleksic10,Aleksic12b,Aleksic16a} and the Fermi-LAT results on Coma 
as well as the combined-likelihood analysis of 50 clusters \cite{Ackermann14,Zandanel14b,Ackermann15a} represent 
some of the key accomplishments in this area. 
These observations limit the cosmic-ray-to-thermal pressure in galaxy clusters to 
be less than a few percent, assuming that the spatial distribution of cosmic-ray protons follows that of the ICM. 
They also constrain several models for the production of gamma rays from structure formation processes in clusters. However, a large
amount of parameter space is still available and many fundamental quantities, such as the cosmic-ray spectral and spatial distributions, 
acceleration efficiency and transport properties, remain to be clearly determined.

Detection of gamma rays from galaxy clusters will be of high importance, if realised. Even a null detection will result in exclusion limits significantly more constraining than those currently available. In turn, they will produce a significant change in our view of cosmic-ray proton acceleration and transport in the ICM, and will provide fundamental constraints on the contribution to the proton population in clusters by galaxies and AGN. A precise assessment of the total proton content is of paramount importance also for cosmological studies done using the cluster mass function, as the presence of cosmic rays induces a bias on the hydrostatic mass estimates \cite{Ando08,Aleksic16a}. 
To determine whether CTA observations will be able to bring a major breakthrough in this field, below we compare CTA prospects with the results of Fermi-LAT which is now in its 9$^{\rm th}$ year of observations and which will likely continue to operate for a few more years.

\begin{figure}[t!]
\centering 
\includegraphics[width=.7\textwidth]{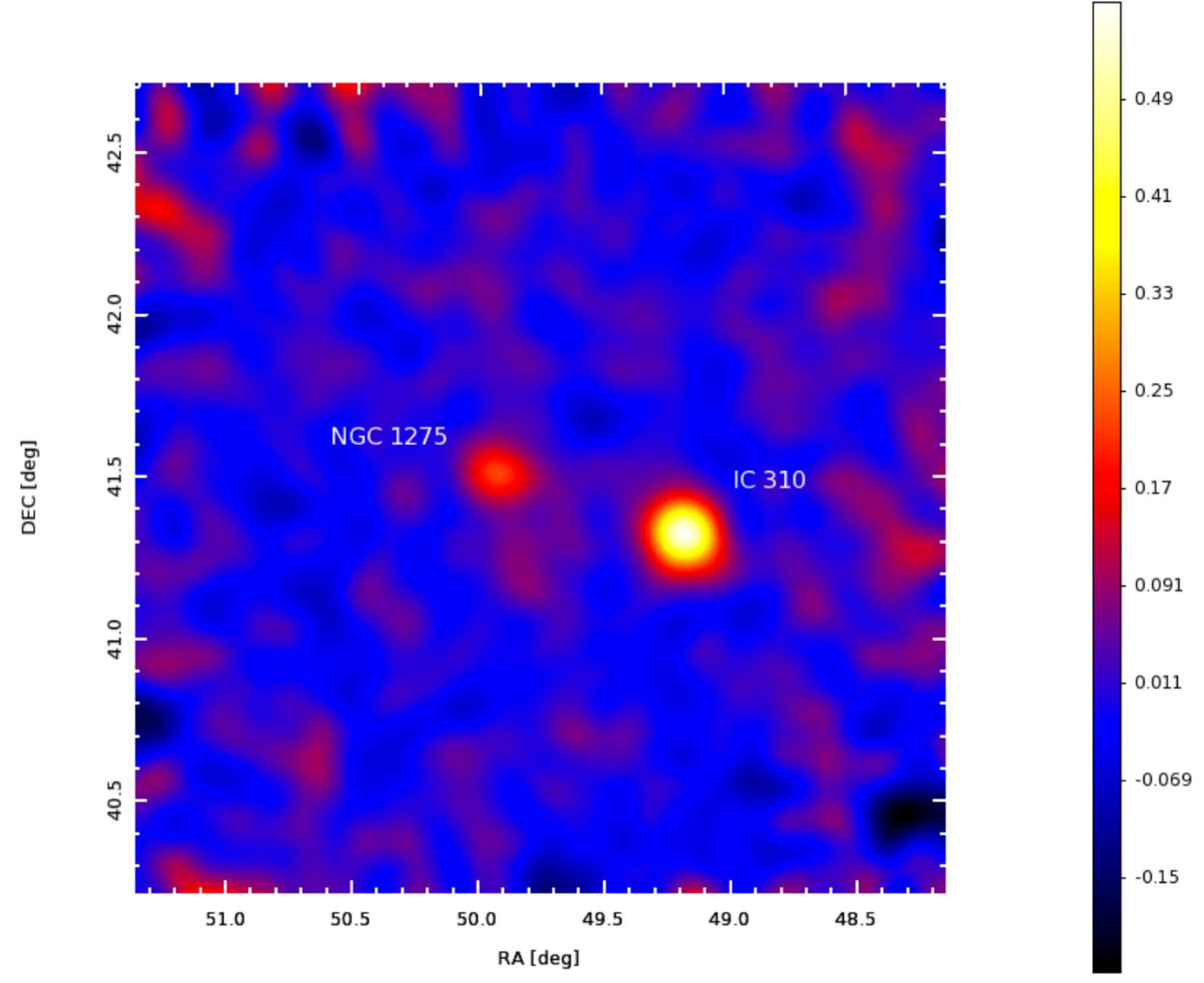}
\caption{\label{fig:skymaps} 
Skymap of the Perseus cluster as observed by MAGIC above 250~GeV (253~h of observation; figure adapted from \cite{Aleksic16a}). The colour scale shows the relative flux in signal-to-background ratio. Both NGC~1275, at the centre of the cluster, and IC~310 can be clearly seen.}
\end{figure}

As explained in the next section, the best target for this program is the Perseus cluster of galaxies. The presence in the cluster field of two gamma-ray bright AGN, NGC~1275 and IC~310, prevents Fermi-LAT from studying the cluster environment itself. This is due to the poor angular resolution of Fermi-LAT at low energies and to its limited sensitivity at very high energies ($>100$~GeV). Therefore, in the case of Perseus, CTA has a clear advantage over Fermi-LAT thanks to its excellent angular resolution. Perseus is the best studied cool-core cluster \cite{Churazov03} and it hosts the archetypal radio mini-halo that is the brightest known to date \cite{Pedlar90,Gitti02}. The high ICM density in the centre of the cluster implies a very high density of target protons for hadronic interactions. CTA is in a unique position to unravel the high-energy properties of the Perseus cluster and its member galaxies. In Figure~\ref{fig:skymaps}, we show the gamma-ray skymap of the Perseus cluster as observed by MAGIC \cite{Aleksic16a}.

The CTA observations of the Perseus galaxy cluster will dramatically improve upon the current MAGIC results. We discuss this in detail in Section~\ref{sec:clusterperf}, where we explain how the proposed observations will improve the current limits by about a factor of six.

Perseus is the most promising target for this program and the proposed CTA observations will represent a major leap forward in our understanding of this cluster and its non-thermal activity. The observation of a second target and its selection is a more complex matter and will be reevaluated over the next few years, particularly taking into account the final results of Fermi-LAT \cite{Charles2016} and advances in radio astronomy.

\newpage

\subsection{Strategy} 
\label{sec:clustersstrategy}

The first step of the target selection is based upon the cluster's distance and mass as measured in X-rays \cite{Reiprich02}. 
Due to the cosmic-ray proton confinement and accumulation over cosmological times, gamma-ray emission should scale with the cluster's mass \cite{Pinzke10,Brunetti14}, although this scaling may have a significant scatter depending on the merging history of a given target \cite{Vazza16}. Here, we consider different predictions for gamma rays generated by structure formation processes in clusters \cite{Kushnir09,Pinzke10,Brunetti12,Pinzke11}, and we obtain a first selection of candidates, in order of preference, as follows: Perseus, Ophiuchus, Coma, Norma and Centaurus, where the first three targets host well-known radio halos. The Virgo cluster is also an excellent candidate, but its large angular size, more than $10^{\circ}$ across, represents a serious observational challenge \cite{Ackermann15b}. Perseus is expected to be the brightest cluster in gamma rays due to a combination of distance, mass and high ICM density in the cluster's centre, implying a high density of target protons for hadronic  interactions.

It is hard to provide precise predictions of the 
gamma-ray emission in clusters due to the uncertainties on cosmic-ray acceleration efficiencies and 
transport properties in these unique environments. Here, we follow two complementary strategies by considering: 
i) expectations based on radio constraints assuming an hadronic origin for the observed diffuse radio emission, where present, and ii) expectations based on hydrodynamical simulations that provide predictions for the cosmic-ray spatial and spectral distributions. In particular, we base our simulations on the hadronic model predictions given in \cite{Pinzke10,Zandanel14,Zandanel14b}. 

The gamma-ray flux induced by secondaries from hadronic interactions must respect the measured radio synchrotron emission, as electrons are also produced from charged pion decays. Therefore, the magnetic field strength and distribution in clusters is also an important ingredient and is 
usually parametrised as $B=B_0\,(n/n_0)^{\alpha_B}$, where $n$ is the ICM gas density. Generally we have limited knowledge of 
cluster magnetic fields, apart from very detailed work on Faraday rotation measurements of the Coma cluster \cite{Bonafede10,Bonafede13}, 
which provide good estimates for the Coma magnetic field ($B_0 \approx 5$~$\mu$G, $\alpha_B \approx 0.5$). 
In this sense, the synergy with radio observations of clusters by the 
Low-Frequency Array (LOFAR; \cite{Rottgering12}) and 
other Square Kilometre Array (SKA) precursors will be crucial. By the same token, SKA itself 
is expected to shed new light on the magnetic field in clusters of galaxies and will
therefore significantly narrow down the available 
parameter space \cite{Govoni13,Bonafede15}.
 
While Ophiuchus seems {\it a priori} to be a good target, the extremely low-surface-brightness radio emission of its central region 
lowers significantly the prediction for gamma rays, independent of the exact value of the magnetic field 
strength \cite{Zandanel14}. 
The same is true for Norma, Centaurus and Virgo, where the absence of bright diffuse radio emission also suggests a 
low level of gamma-ray emission. On the contrary, both Perseus and Coma host very bright diffuse radio emission 
allowing for higher hadronically induced gamma-ray emission. Therefore, in the following, we consider only these last 
two objects for CTA observations.
   
\begin{figure}[t]
\centering 
\includegraphics[width=.46\textwidth]{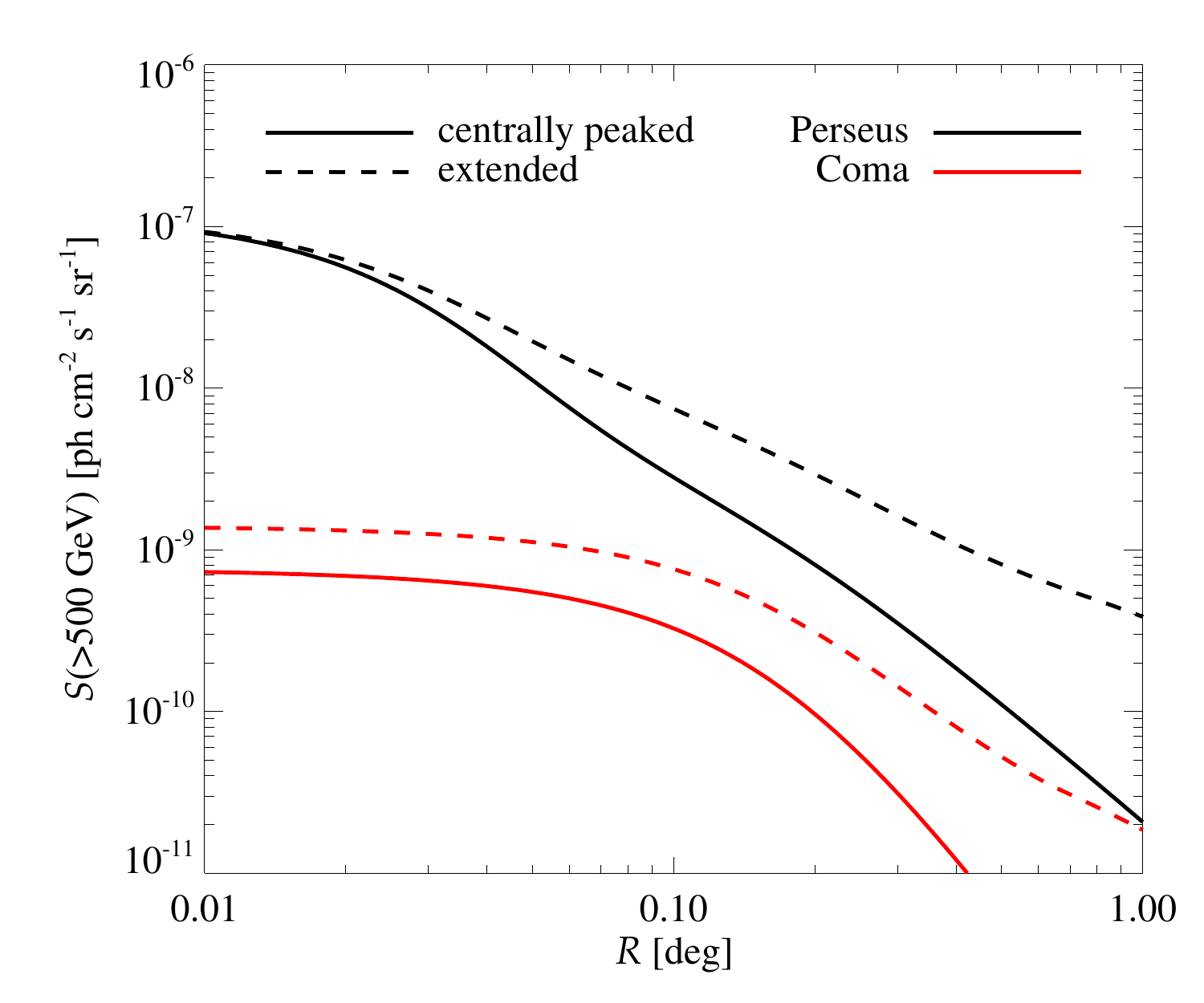}
\includegraphics[width=.53\textwidth]{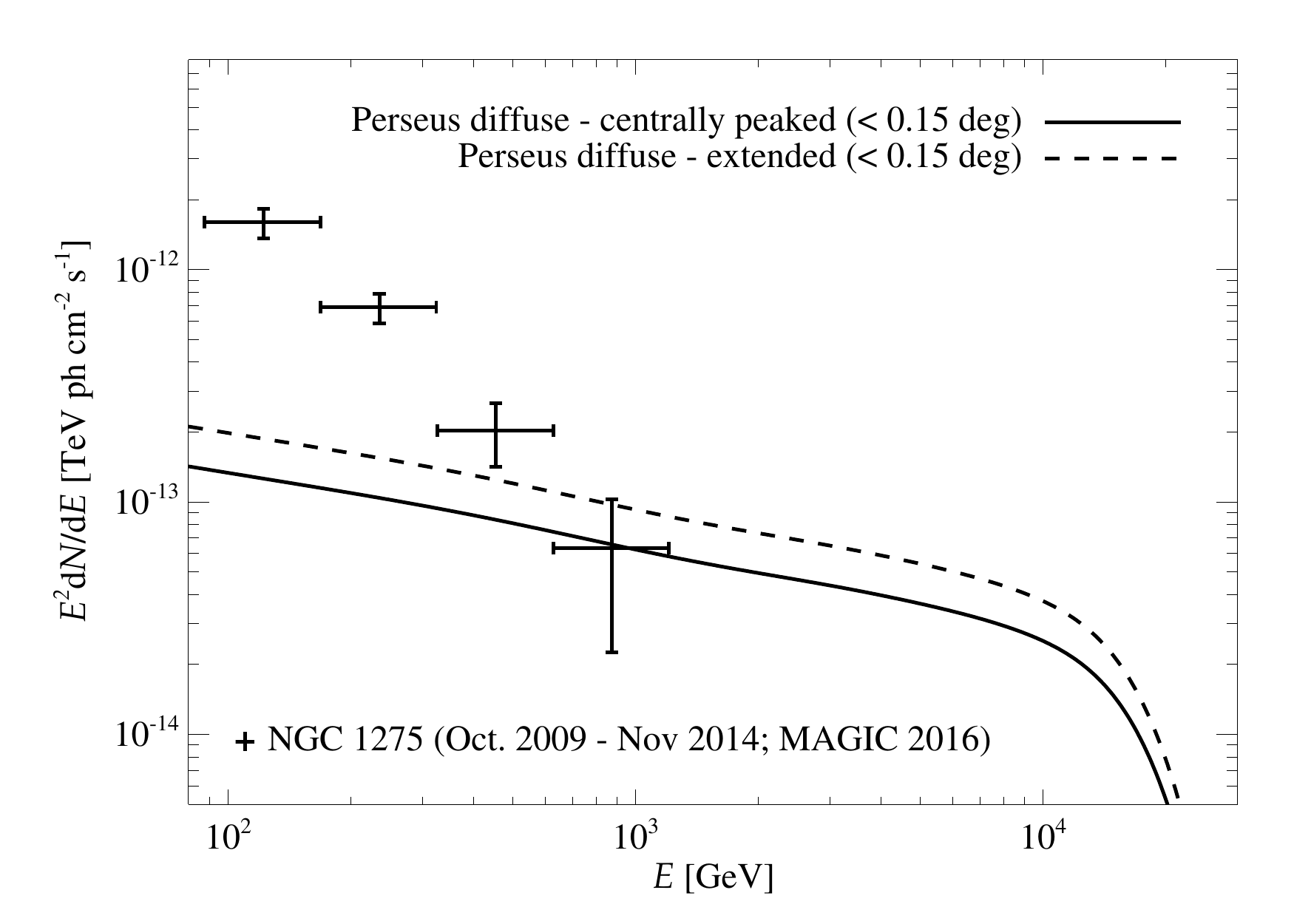}
\caption{\label{fig:profiles} Left: Predicted surface brightness of the hadronic-induced gamma-ray emission (above 500~GeV) from the Perseus and Coma galaxy clusters for two models with centrally peaked and extended cosmic-ray profiles, as described in the text \cite{Pinzke10,Zandanel14,Zandanel14b,Aleksic16a}. The luminosity distance and radius of Perseus are 78 and 1.9~Mpc, respectively, while for Coma they are 101 and 2.3~Mpc, respectively \cite{Reiprich02}. In both cases, the angular  radius is approximately $1.3^{\circ}$. Right: Differential spectra for the two models adopted for the hadronic-induced diffuse gamma-ray emission of Perseus integrated within a radius of $0.15^{\circ}$ from the centre. The effect of the EBL absorption is clear above 10~TeV. Additionally shown is the spectrum of NGC~1275, the central radio galaxy of Perseus, as measured by MAGIC \cite{Aleksic16a}. Note that the higher energy data point of the NGC~1275 spectrum is only marginally significant and is in agreement with the current upper limits on the diffuse emission \cite{Aleksic16a}.}
\end{figure}

In the left panel of Figure~\ref{fig:profiles}, we show the surface brightness of the predicted hadronically induced gamma-ray emission from the Perseus and Coma clusters. For each cluster, two sets of models are shown. The difference in the models comes from
the assumption on how the cosmic-ray protons are distributed in the cluster,
varying from a centrally peaked \cite{Pinzke10} to a more radially extended distribution in order to account for possible variations in the cosmic-ray transport properties \cite{Ensslin11,Wiener13,Zandanel14}. In the case of Coma, the flat cosmic-ray profile is the maximum allowed hadronic contribution that still respects the Fermi-LAT constraints \cite{Zandanel14b}. The peaked one is the maximum possible emission for which the radio counterpart does not overshoot the measured radio flux at the cluster centre (under the above mentioned assumption for the Coma magnetic field) \cite{Zandanel14b}. In the case of the Perseus cluster, both models fit well the radio data of the mini-halo at 1.4~GHz \cite{Pedlar90} assuming a central magnetic field of $B_0 = 10$~$\mu$G and $\alpha_B = 0.3$ and $0.5$ for the centrally peaked and extended models, respectively. Note that, in the case of mini-halos that are in cool-core clusters, current observations favour relatively high values of magnetic fields \cite{Govoni04,Clarke04,Ensslin06,Kuchar11}. 
These two models for Perseus are the same that have been explored in the latest MAGIC publication \cite{Aleksic16a} (the centrally peaked model is referred to there as the semi-analytical model \cite{Pinzke10}). 

The total gamma-ray flux above 500 GeV within the virial radius\footnote{The cluster virial radius is defined here with respect to an average density that is 200 times the critical density of the universe. Values are taken from \cite{Reiprich02}.} predicted for Perseus is $0.2-1.0\times10^{-12}$~cm$^{-2}$~s$^{-1}$, while, for comparison, for Coma it is $1.3-6.3\times10^{-14}$~cm$^{-2}$~s$^{-1}$. We emphasise that these models adopt a cosmic-ray spectral index (equal to the photon spectral index) of about $2.2$ in the energy range of interest for Cherenkov telescopes, as suggested by simulations \cite{Pinzke10}. A harder cosmic-ray component would dramatically increase our detection chances, while a softer one would obviously worsen them. In the case of Coma, where we know that the hadronic contribution represents at most only a part of the observed radio emission, both cases could be realistic. However, in the case of Perseus, we assume that the observed radio emission is of hadronic origin and the adopted spectral distribution matches well the observed radio spectral shape \cite{Pedlar90} (see also the discussion in \cite{Aleksic12b}). Note that, following Ref.~\cite{Dominguez11b}, we apply the correction due to the attenuation by the extragalactic background light (EBL) on our spectra. As a consequence, for example, our Perseus spectrum appears as a power law with an index of $2.33$ above 300~GeV and a cutoff above 10~TeV \cite{Aleksic16a}, as shown in the right panel of Figure~\ref{fig:profiles} for the two adopted models.

\begin{figure}[t]
\centering 
\includegraphics[width=.45\textwidth]{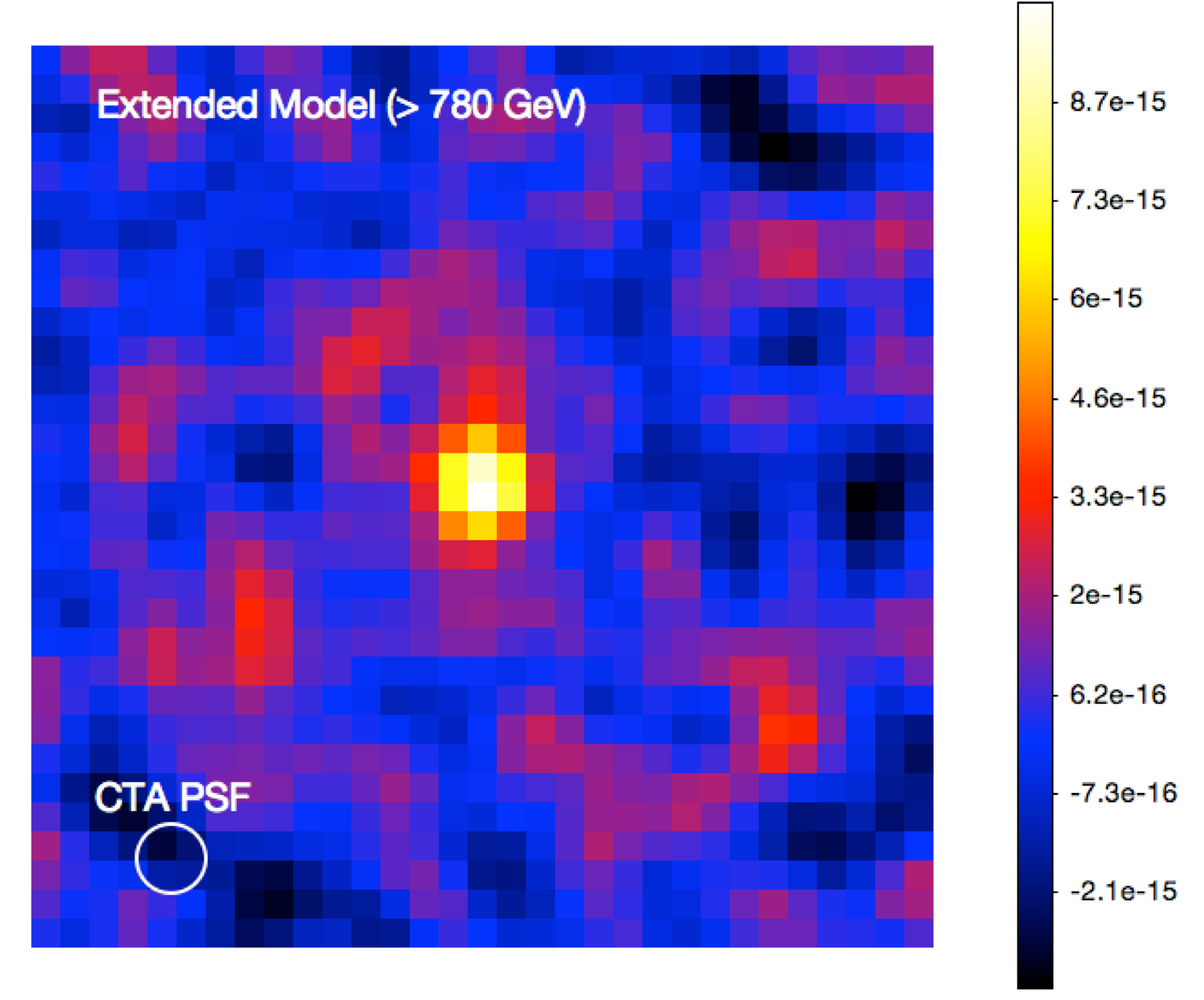}
\includegraphics[width=.45\textwidth]{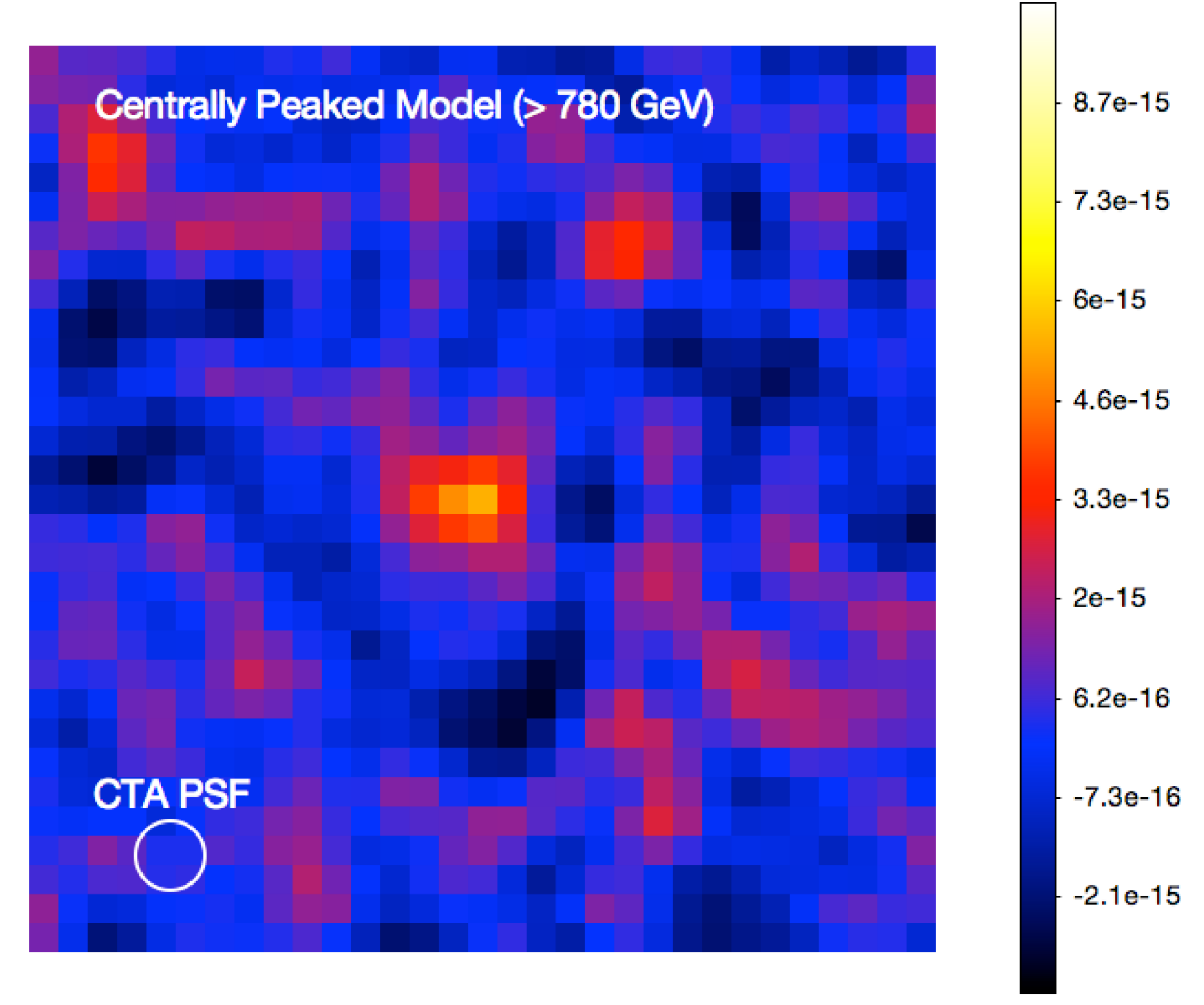}
\caption{\label{fig:simulations} Simulated gamma-ray flux (cm$^{-2}$~s$^{-1}$) skymaps above 780~GeV for 100~h of observation of a region of $1.55^{\circ}\times1.55^{\circ}$ around the Perseus cluster centre. The left panel shows the result for the extended cosmic-ray profile, while the right panel shows the result for the centrally peaked cosmic-ray profile. The white circles indicate the expected CTA 68\% containment radius at 780~GeV (of about $0.06^{\circ}$). We simulated the signal and background event counts in the region with a pixel size of $0.05^{\circ}$. The maps were obtained by calculating the corresponding flux and subtracting the background from the signal (resulting in some areas having negative flux values) and then 
smoothing the final map with a Gaussian function with $\sigma = 0.06^{\circ}$.}
\end{figure}

Our simulations suggest that the models adopted for Coma would need more than a thousand hours of observation to be detectable by CTA, while the models adopted for Perseus are within reach in a few tens of hours.\footnote{The detection criteria is a statistical significance greater than five standard deviations, the number of excess events $>10$ and a signal/background ratio $> 0.03$.} In particular, we search for the integral energy range, starting with energies above 100~GeV, and the angular region size, from 0.05 to 1~deg, that minimize the number of hours needed to reach a detection.
Not surprisingly, the minimum is typically reached using the full energy range, which maximises the total flux, and the smallest angular region size, which minimises the background event counts. In this case, the models predict that Perseus would be detectable with 40~h and 50~h of observation by the northern CTA array for the extended and centrally peaked cases, respectively. However, these numbers are based on the assumption that we can perfectly model the two bright point-sources in the Perseus field of view, NGC~1275 and IC~310. The presence of IC~310 will not be a major obstacle in the search for the diffuse emission as it is located about $0.6^{\circ}$ away from the centre of the cluster from where we expect the bulk of the diffuse emission (see Figure~\ref{fig:skymaps}). However, NGC~1275 is located right in the centre and it dominates the gamma-ray emission up to a few hundred GeV as shown in the right panel of Figure~\ref{fig:profiles}. In order to account for this, we use the conservative strategy of excluding the lowest energies from the detection criteria and find that, considering only energies above 780 GeV where the contribution of NGC~1275 should be negligible, the extended and centrally peaked models should be detectable in 60~h and 100~h, respectively.

We can foresee three ways to disentangle the possible diffuse emission from that of NGC~1275:

\begin{itemize}
\item The first method relies on the spectral characteristics of the two emissions. 
The NGC~1275 spectrum at the energies of interest has a very soft 
index of about $4$ \cite{Aleksic14}, and the corresponding gamma-ray emission 
becomes undetectable above a few hundred GeV. On the contrary, the spectrum of 
the possible hadronic-induced diffuse emission should be much harder and should
show no cutoff except at the highest energies (due 
to EBL absorption). The predicted spectrum is shown in the right panel of Figure~\ref{fig:profiles}.
\item The second method relies on the extension of the emission. NCG~1275 is a 
point-like source 
while the predicted hadronic-induced emission is expected to be 
quite extended, as can be seen in the left panel of Figure~\ref{fig:profiles}. 
The excellent angular resolution of CTA
should permit a clean disentangling of the two.
Figure~\ref{fig:simulations} shows the simulated gamma-ray flux skymaps above 780~GeV for 100~h of observations 
for the two considered models of cosmic-ray-induced emission in the Perseus cluster. The 68\% containment radius 
for the CTA-North array at 780~GeV is about $0.06^{\circ}$.
For both models considered, the detected diffuse emission extends beyond the 68\% containment radius. 
\item The third potential discriminator is variability as the diffuse emission is not expected to exhibit any.
\end{itemize}

In conclusion, if diffuse gamma-ray emission is present at a level suggested by the adopted models, we will have the means 
to disentangle it from the point-like emission. A more complex matter will be to disentangle it from a possible dark matter 
induced component. While the discrimination strategy would be based on the difference in the emission morphology and the 
shape of the energy spectrum, how well the discrimination could be done would critically depend on the significance of the detection. 
We leave a more detailed study of all the gamma-ray components possibly observable by CTA in the Perseus cluster for future work.

We eventually plan to observe the Perseus cluster region for 300~h with CTA-North. This amount of observation time is necessary for two reasons: i) to ensure enough event counts, in the case of detection of the adopted models, for a detailed morphology study that will be mandatory, e.g.~to enable disentangling the point-like and diffuse emissions, and ii) to allow a deeper reach in the parameter space of theoretical models,
in the context of relaxing the strongest assumption made in this section, i.e.~the hadronic origin of the Perseus radio mini-halo. (See further discussion in Section~\ref{sec:clusterperf}).

Note that we used the 1.4~GHz observations of the radio mini-halo of the Perseus cluster \cite{Pedlar90} to calibrate our models. However, results at 327~MHz \cite{Sijbring93,Gitti02} suggest that the high-frequency observations may be missing some of the flux. Depending on the exact value of the magnetic field at the radii where the bulk of the emission is coming from, this could boost the corresponding allowed gamma-ray emission. This boost factor can range from about 1.1, for the magnetic field model adopted here, to more than 1.2, for magnetic field models considering the effect of ICM compression in the cooling-flow region \cite{Gitti02}. Such boosts would imply that CTA would need about 20\%-30\% less observation time to report a detection of the adopted models.

\subsubsection{Targets}
Table~\ref{tab:clust_targets} shows the selected cluster target. We expect the Perseus cluster to host the brightest gamma-ray 
emission coming from proton-proton interactions. Additionally, Perseus hosts two extremely interesting gamma-ray point-sources: 
NGC~1275, one of the few radio galaxies known to emit gamma rays and the galaxy IC~310. As mentioned earlier, the Perseus field of view is observable in optimal conditions, i.e.~at zenith angles $<60^{\circ}$, only by the northern array. Note that while in Table~\ref{tab:clust_targets} we ask for low zenith angle observations to maximise sensitivity, medium and large zenith angle observations could provide a larger effective area at high energies, increasing the corresponding event counts and, therefore, possibly improving morphological studies. We leave the detailed study of this latter possibility for future work.

\begin{table}[h!]
\begin{center}
\begin{tabular}{ccccc}
\hline\hline
\phantom{\Big|}
Target  &  Time [h]  & Hemisphere & Array Configuration &  Special Requirements\\

\hline \\[-0.5em]
Perseus    &      300       &   North &   Full array & Low zenith angles\\
\hline\hline
\end{tabular}
\end{center}
\caption{Galaxy cluster selected for this Key Science Project along with the basic requirements.}
\label{tab:clust_targets}
\end{table}

\newpage

\subsection{Data Products}
The data products produced will clearly depend on whether a significant detection of the diffuse gamma-ray emission in Perseus is made or not.
In the case of a detection, the data products will include maps/data-cubes (excess, flux, and spectral hardness) for the whole cluster, 
morphological analysis of the diffuse emission, spectra and light curves of NGC~1275 and IC~310 and a model for their subtraction from the whole cluster emission. We will additionally provide a detailed analysis of the cosmic-ray physics parameter space matching the observed emission and other possible diffuse contributions from, e.g., dark matter. In the case where no diffuse gamma-ray emission is detected, the analysis will be focused on the constraints that can be made on the cosmic-ray physics in clusters.  

We plan to release the data products related to the diffuse emission in two steps: the first after 100~h of observation have been performed and the second at the end of the programme after 300~h of observation. We note that the results on the point sources, including possible serendipitous discoveries, will be released before the completion of the cluster campaign following the strategy outlined for the AGN KSP (see Section~12.4).

\subsection{Expected Performance/Return }
\label{sec:clusterperf}

We estimate here the parameter space accessible to the proposed CTA observations for the Perseus cluster.~In the left panel of Figure~\ref{fig:pexpectations}, we show the predicted gamma-ray integral spectrum from 
hadronic-induced emission for the adopted models, including the effect of EBL absorption. In particular, 
we show the centrally peaked and extended models corresponding to $B_{0}=10$~$\mu$G which, according to our estimates, should be detectable in 100~h and 60~h of observations, respectively. We also show the 
cases corresponding to $B_{0}=20$~$\mu$G. We estimate the detection level reachable in 300~h of observation 
for these two models simply by scaling by the square root of the time that is obtained with our detection criteria.
The detection level is shown with a black band in the left panel of Figure~\ref{fig:pexpectations} and while it suggests that we 
could be able to detect also the cases corresponding to $B_{0}=20$~$\mu$G, we recall that this optimistically 
assumes no contribution from the central radio galaxy NGC~1275, which will become larger for longer observation times. Nevertheless, if the Perseus mini-halo is (mostly) of hadronic origin and the adopted models are correct, CTA observations will achieve a detection in about $100$~h of observation. In this case, with the proposed 300~h of observation, we will be able to determine critical cosmic-ray parameters such as the spectral and spatial distributions (connected to cosmic-ray transport properties) and to study the acceleration efficiency. Taking into account currently available radio data, and future data from LOFAR and SKA precursors, we will also be able to provide a complementary measure of the magnetic field strength and distribution in the cluster.

The left panel of Figure~\ref{fig:pexpectations} also shows, with a grey band, the current constraints on the parameter space for the centrally peaked and extended models obtained from MAGIC observations \cite{Aleksic16a}. We stress that the figure does
not mean to show the real gamma-ray flux excluded by observations but rather the theoretical parameter space for the diffuse
emission models. In fact, the low energies will always be dominated by the emission from NGC~1275, as shown in the right panel of Figure~\ref{fig:profiles}. For comparison with current upper limits by MAGIC, we show, with a dark blue band, an estimate of the 95\% upper limit level reachable with 300~h of observations by CTA, assuming zero excess events, for the two adopted models. This shows that CTA observations could improve by about a factor of six on the current MAGIC constraints.

We additionally estimate the so-called minimum gamma-ray emission expected from Perseus in the hadronic scenario
(taking the radio flux at 1.4~GHz as a reference value \cite{Pedlar90}).~This lower-limit is obtained by assuming that the 
observed radio emission is of hadronic origin and that the magnetic field is much higher than the equivalent magnetic field strength of the cosmic microwave background throughout the radio-emitting region (see \cite{Aleksic12b,Aleksic16a} for details). A non-detection at this level would rule out the secondary origin of the Perseus diffuse radio emission, independent of the exact value of the cluster's magnetic field. The left panel of Figure~\ref{fig:pexpectations} shows, with a light blue line, the case for a proton spectral index of $\alpha_{\mathrm{p}} = 2.3$. The comparison with the upper limit level reachable with 300~h of observations suggests that a hadronic origin of the Perseus radio mini-halo for $\alpha_{\mathrm{p}} \leq 2.3$ could be excluded independently of magnetic field values.

The proposed 300~h of observation are motivated by the goal of making scientific statements which go beyond the assumption of the hadronic origin of the observed non-thermal emission in Perseus and, instead, aim to measure the cosmic-ray proton spectrum and energy density in the ICM. 
Although the cosmic-ray proton spectrum in clusters is uncertain, indications from hydrodynamical 
simulations and observations of discrete sources in clusters (i.e.~AGN, whose ejecta also contribute to the cosmic-ray proton population in the ICM) suggest reasonable values between $2.0$ to $2.5$  \cite{Brunetti14}. 
Assuming that the cosmic-ray pressure scales as the thermal ICM pressure in the cluster --
as a result of cosmic-ray proton advection on cosmic times and as suggested by several simulations -- and recalling the MAGIC
results on Perseus \cite{Aleksic16a}, the proposed CTA observation will push the cosmic-ray-to-thermal pressure limits down to $0.1$\% -- $2.5$\% for proton spectral indices of $2.1$ to $2.5$, respectively. We show the case for the model with a centrally peaked cosmic-ray profile in the right panel of Figure~\ref{fig:pexpectations}, together with current limits from both gamma-ray and radio observations, noting that the latter constraints depend on the adopted magnetic field values. CTA observations will not only provide unprecedented constraints on the cosmic-ray acceleration efficiency at structure formation shocks, potentially below 
10\% according to \cite{Pinzke10,Ackermann14}, but they will also strongly constrain the contribution to the cluster proton population from galaxies and AGN. In fact, such low values of the cosmic-ray-to-thermal pressure can potentially put constraints on the cosmic-ray electron/proton fraction in the cluster's AGN (NGC~1275 in this case) or, alternatively, on how these protons are transported from the central AGN to the cluster periphery (an issue also connected to the confinement of such protons in AGN bubbles \cite{Brunetti14}).~Last but not least, CTA observations will eventually constrain the cosmic-ray bias on the hydrostatic mass estimates \cite{Ando08,Aleksic16a} down to the $<5\%$ level, independent of assumptions on cosmic-ray spectral and spatial distributions. Models with extended cosmic-ray profiles imply a growing cosmic-ray-to-thermal pressure in cluster outskirts. While the current constraints from MAGIC are only below  $\sim$15\%, the proposed CTA observations could potentially constrain these models to $\leq 3\%$.

To summarise, the observation campaign on Perseus proposed in this Key Science Project will lead to a major step forward in our understanding 
of cosmic rays and non-thermal phenomena in clusters. Additionally, as mentioned above, we will be studying the emission from the 
galaxies NGC~1275 and IC~310 and exploring the nature of dark matter (see Chapter~\ref{sec:DM_prog}) and the existence of ALPs (see Chapter~\ref{sec:ksp_agn}). 
The proposed programme represents a major opportunity for CTA to make a 
significant contribution to galaxy cluster science and it will be 
a very significant step forward compared to the results from Fermi-LAT and current atmospheric Cherenkov telescopes. Therefore, this KSP will lead to a 
long-lived legacy in the study of non-thermal
phenomena in clusters of galaxies.

\section{Capabilities beyond Gamma Rays}
\label{sec:nongamma}

Although designed as a gamma-ray observatory, CTA is a powerful tool for a 
range of other astrophysics and astroparticle physics. For example, CTA can make precision studies of charged 
cosmic rays in the energy range from $\sim$100 GeV up to PeV energies and it can be used as an instrument for optical intensity interferometry, to provide unprecedented angular resolution in the optical for bright sources. 
Below we briefly summarise these possibilities.
Most of the topics we discuss can be explored in parallel with gamma-ray data-taking, without interfering with the major science operations of CTA. Those studies (such as intensity interferometry) which require specific observations, can likely make use of bright moonlight time, thus enhancing the CTA science return without negative impact on the key science goals.

\subsection{Cosmic-ray Nuclei}

The origin of cosmic rays remains one of the most important questions in astrophysics. Over the past few decades, 
a consensus has emerged that supernova remnants (SNRs) are likely to be
the main sources of cosmic rays up to energies of 
the cosmic-ray knee ($\sim$3~PeV). Finding out if SNRs are indeed the dominant accelerators of 
cosmic rays is a major science goal of CTA. 
While gamma-ray observations provide a view of the accelerators themselves, 
CTA will also address this question by directly detecting cosmic rays which reach the Earth.

The elemental composition and energy spectrum of cosmic rays provide 
important clues about their acceleration and diffusion through the Galaxy. 
For example, composition measurements at lower energies ($<$100 GeV) indicate that the cosmic-ray spectrum observed at the Earth is significantly steeper than 
in the sources~\cite{Juliusson72} and that cosmic rays spend most of their lifetime in the 
Galactic halo~\cite{Garcia-Munoz75}. 
Particularly interesting is the elemental composition around the knee of the energy spectrum. 
The steepening of the cosmic-ray spectrum at this point is still not understood; 
it is expected to be related to the maximum energy to which particles can be 
accelerated inside Galactic cosmic-ray accelerators. 
This maximum energy should increase with increasing rigidity of the particle 
(rigidity = momentum/charge) and therefore lead to an enhancement of heavier nuclei with 
increasing energy at the knee. 
Experimental  evidence for this has been reported by the KASCADE collaboration~\cite{Apel13}, although the analysis is challenging due to its dependence upon hadronic interaction models, leading to  significant systematic uncertainties.

Cosmic-ray composition measurements above 100~TeV will give us further insights into this question.  
Balloon and space born measurements have measured the composition of cosmic rays up to $\approx$100~TeV. 
It is difficult to extend these measurements to higher energies
because of the limited collection area and observation time of these
instruments, although the CALET and ISS-CREAM instruments should significantly extend the reach of the satellite technique.
At energies of a few PeV, detection of the secondary shower particles from the ground becomes possible, allowing for larger collection areas and composition measurements at these energies. 
These measurements are, however, affected by severe systematic uncertainties related to the simulation of hadronic showers.

CTA will be able to perform composition measurements in the TeV to PeV
domain by measuring the Cherenkov light emitted by cosmic-ray nuclei
prior to their first interaction~\cite{Kieda01}.
As was shown by the H.E.S.S.~\cite{Aharonian07dc} and VERITAS~\cite{Wissel10} collaborations, this emission can be detected by atmospheric 
Cherenkov telescopes, enabling the reconstruction of the charge of the primary. 
Using this technique, CTA is expected to measure the iron spectrum of cosmic rays up to, and beyond, 1~PeV. For lighter nuclei down to oxygen, the measurement of the spectrum is expected to extend up to a few hundred TeV. Techniques based on shower properties rather than direct Cherenkov emission of the primary will provide further capability for composition measurements, in particular for light nuclei.
Such studies will also be carried out by the LHAASO instrument (see Section 2.5).

\subsection{Cosmic-ray Electrons}
\label{sec:KSP:electrons}

Electrons form only a small fraction of the cosmic rays -- at GeV energies the fraction is around 1\% --
yet they are important for a wide range of scientific questions.
In the very-high-energy (VHE) regime, the lifetime of cosmic-ray electrons is 
severely limited by radiative losses due to synchrotron emission in 
interstellar magnetic fields and inverse-Compton scattering on star light, IR light 
and the cosmic microwave background.
The lifetime of a cosmic-ray electron with energy $E$ can be expressed as: 

$$t \approx 5 \times 10^{5} (E / 1 \mathrm{TeV})^{-1} ((B/5 \mu\mathrm{G})^2 +1.6(w/1 \mathrm{eV cm}^{-3}))^{-1}\ \ \mathrm{years},$$ 

\noindent where $w$ is the energy density in low frequency photons in the interstellar medium and $B$ is the mean interstellar magnetic field \cite{Aharonian08}. This restricts the distances that VHE cosmic-ray electrons can propagate and implies a local origin of TeV electrons ($<1$~kpc distance) as discussed in 
e.g. \cite{Aharonian95} and \cite{Kobayashi04}. 
In the TeV range, therefore, only a few local sources are able to contribute and the spectrum 
can differ dramatically from the power-law form seen at lower energies, revealing the imprint of nearby sources. 
Thus, VHE cosmic-ray electrons offer a unique tool to study our local neighbourhood in terms of single electron accelerators, 
rather than testing the averaged contribution of the whole population of sources in the Galaxy, as visible in cosmic-ray protons.
 
In addition to the limited propagation distance, 
the cooling of cosmic-ray electrons also leads
to a steepening of the electron spectrum 
far from acceleration sites, i.e. as observed at the Earth.
This steepening, from a $\sim$$E^{-2.7}$ to
$\sim$$E^{-3}$ power-law form, is not 
observed in the cosmic-ray protons, and
has the effect of further reducing the 
flux and poses a severe challenge 
for direct balloon or satellite measurements. 
While balloon measurements generally suffer from short exposures, a new generation of satellite experiments 
recently provided high-statistics electron measurements up to several hundred GeV
\cite{Abdo09e,Adriani11,Aguilar14,Abdollahi17}. 
Still, the energy regime above $\sim$2~TeV is not yet accessible to these instruments due to either 
limited detector area or the lack of sufficient detector depth for energy measurement at higher energies.

The multi-TeV energy regime is of particular interest because it has the unique 
feature of reflecting the situation in our local neighbourhood. Exploring this energy range would provide a new perspective on local particle accelerators.
Another frequently discussed possibility is the potential signature of dark matter 
annihilation in the electron spectrum (see e.g. \cite{Malyshev09} and references therein), 
which was fueled by the observation of a rise in the positron fraction by 
PAMELA \cite{Adriani09} and AMS-02 \cite{Aguilar13}.

Imaging atmospheric Cherenkov telescopes (IACTs), such as CTA, detect
VHE cosmic-ray electrons through the same air shower technique that they
detect VHE gamma rays.
Electrons initiate air showers very similar to those originating from gamma rays
and so special data-taking conditions are not required.
Furthermore, due to the isotropic nature of cosmic-ray electrons, 
their measurement does not need targeted observations but 
can be performed with any data set
where significant gamma-ray emission regions can be readily excluded (i.e. away from the Galactic Plane).
This makes a cosmic-ray electron measurement at TeV energies an 
excellent by-product of regular observations made by CTA. 

A proof-of-principle measurement of cosmic-ray electrons with IACTs 
was made by H.E.S.S. \cite{Aharonian08} and later confirmed by 
MAGIC \cite{Borlatridon11} and VERITAS \cite{Staszak15}. 
The H.E.S.S. data extended the measured energy range of cosmic-ray electrons up to 6~TeV and revealed a break in the spectrum at around 1~TeV, with a  
$\sim\,E^{-4}$ spectrum at higher energies.
As there are almost certainly no regions of the sky from which cosmic ray electrons do not arrive,
the contamination of IACT measurements by the residual background of
hadronic cosmic rays must be estimated 
with the aid of simulations rather than control
regions within the field of view.
Limitations of the measurement therefore arise due to 
our limited understanding of hadronic physics in air showers, introducing systematic uncertainties
into the estimation of the level of hadronic contamination.
Additionally, to obtain a sufficiently large data set, long observations are required by the current generation 
of instruments, which introduces significant systematic uncertainties to account for the wide variety of observing conditions.
At the highest energies electron statistics become the limiting factor, due to the steeply declining
spectrum at TeV energies.

\begin{figure}[h!]
\begin{centering}
    \resizebox{0.45\columnwidth}{!}{\includegraphics{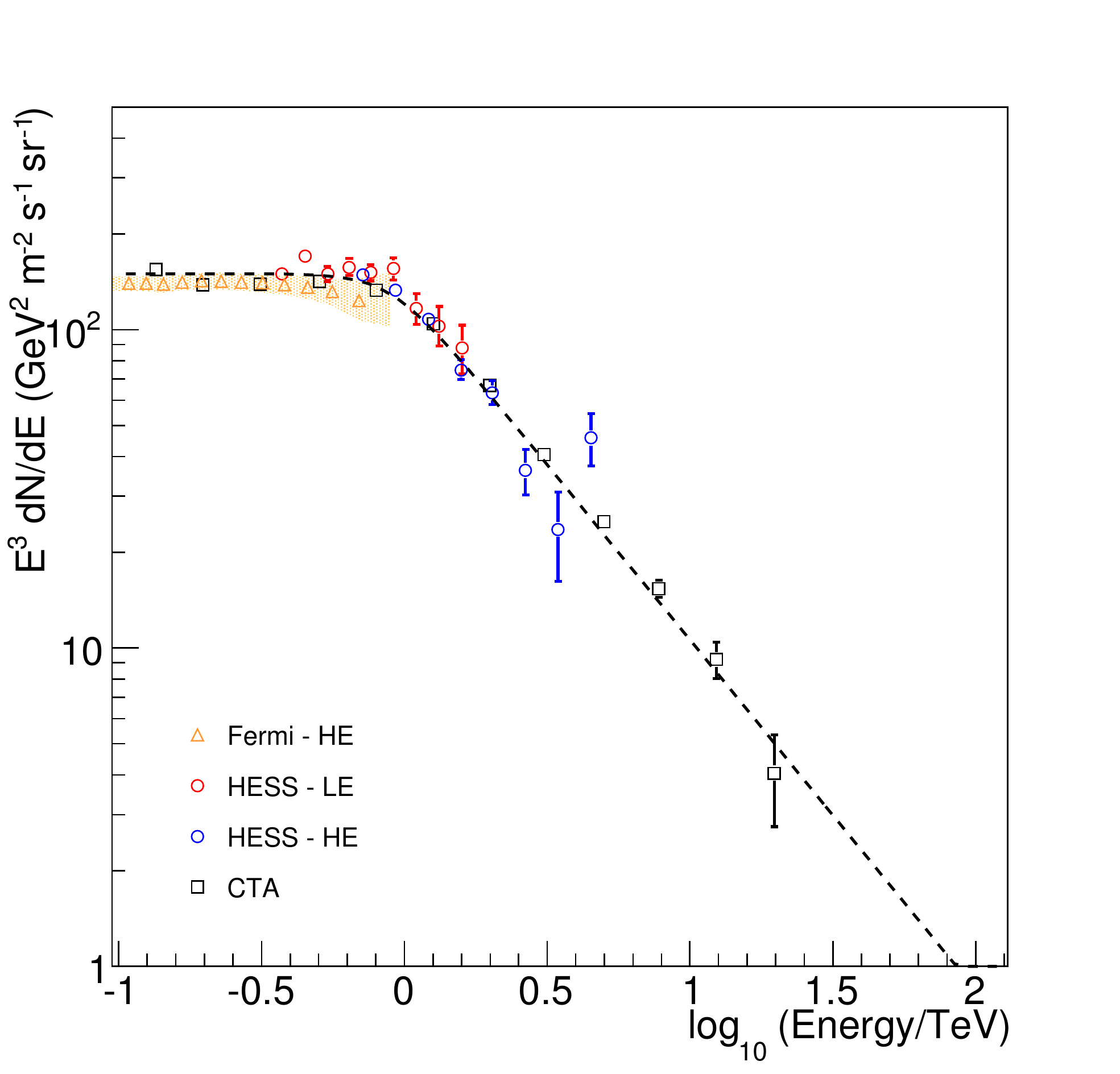}}
  \resizebox{0.45\columnwidth}{!}{\includegraphics{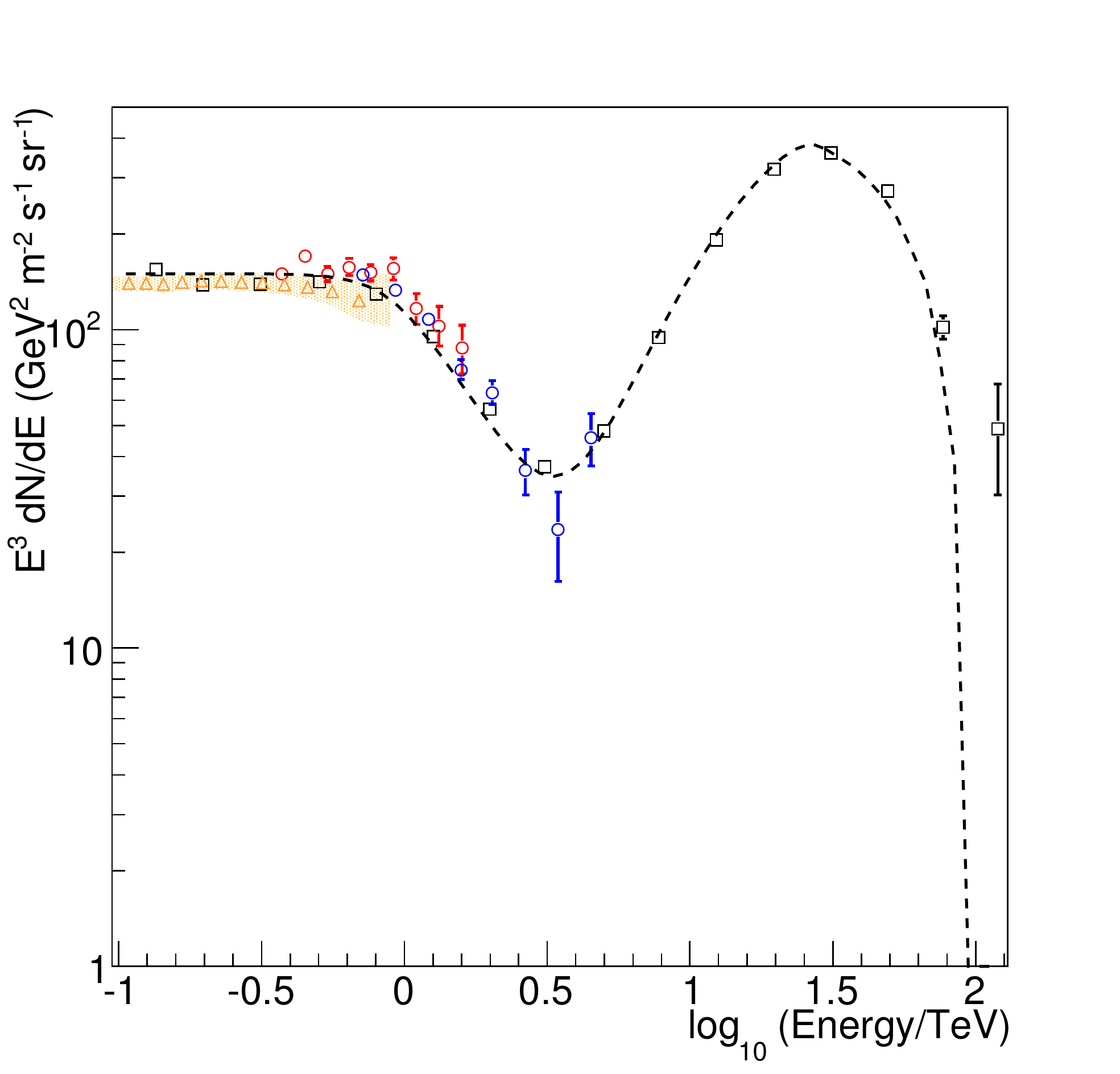}}
      \caption{
        Predictions for the measurement of the high energy electron spectrum using
   1000 h of CTA-South data, without any local source (left) and with an injection of
        $5\times10^{48}$ ergs in cosmic-ray electrons at 500 parsec 
     distance 5000 years ago (right), plausibly comparable to the case of the Vela PWN.
      The differential energy spectrum multiplied by a factor of $E^{3}$ is plotted as a function
      of energy.
        Note that these studies were done with no Galactic diffuse component present.
Reproduced from~\cite{Parsons11}.
      }
  \label{fig:nongamma-electrons}
\end{centering}
\end{figure}

Significant improvements can be expected from a CTA measurement of the
electron spectrum, compared to the results from present instruments.
With the CTA design optimised for a broad energy coverage and high
sensitivity at TeV energies, an electron measurement will cover a
significantly broader energy range, starting at $\sim$100~GeV.  At
high energies a measurement of the spectrum is possible up to 20~TeV
in the case that there is no contribution from a local source and that the
spectrum continues with an index of -4.1. In the case of a
local source contribution (see Figure~\ref{fig:nongamma-electrons}),
the electron spectrum may be measurable with CTA up to $\sim$100~TeV.

In addition to the greatly increased sensitivity and effective area of CTA, the energy resolution will also be significantly enhanced, aiding the identification of local source signatures in the spectrum. Furthermore, 
with the expected level of systematic uncertainties, the measurement of anisotropy in the electron arrival directions 
may be possible with CTA and would provide, together with spectral features at TeV energies, a probe of 
cosmic-ray electron source characteristics and propagation.
A proper understanding of atmospheric variations is the key to such an anisotropy measurement, and again CTA will improve dramatically on the situation for current IACTs. A complete monitoring and correction scheme is under development 
for CTA~\cite{Gaug14}, aiming to greatly reduce systematic variations in measured flux. Once cosmic-ray electron anisotropy is measured or constrained, cosmic-ray electrons are expected to play an important role in the calibration and monitoring of CTA~\cite{Parsons16}.

Current ground-based measurements are dominated by the systematic
errors arising from uncertainties in the simulations of hadronic air
showers (necessary for the discrimination between electrons and the
hadronic background) and from atmospheric conditions. For the case of
the simulations, differences between hadronic interaction models are
currently a leading source of uncertainty. CTA will be able to reduce
these uncertainties by exploiting
a new and (especially at TeV energies) 
improved generation of hadronic models. 
The increase in the centre-of-mass energy of hadronic collisions 
to $\sim$14~TeV for proton-proton collisions at the LHC, is clearly a significant step towards a 
better understanding of this energy regime. 
Furthermore, the experiments LHCf and TOTEM are dedicated specifically to the measurement of cross-sections in the forward direction, thus providing key input to models and hopefully leading to a significant reduction in the uncertainties of the hadronic models used in air showers \cite{dEnterria11}.  

The measurement of the electron spectrum with CTA will
be complementary to observations by the CALET instrument, 
which was recently mounted on the international space station and is able to measure cosmic-ray electrons up to 10 TeV \cite{Marrocchesi15}. In combination, these instruments will provide detailed observations of the electron spectrum at never before seen energies, potentially discovering and characterising powerful cosmic-ray accelerators in the local Galactic neighbourhood.

\subsection{Optical Measurements with CTA}

CTA will provide unprecedented telescopic light-collecting area -- of the
order of 10,000 m$^{2}$ distributed over a few square kilometers -- far exceeding the mirror area of all of the world's large ($>$3 m) optical telescopes combined.
Although mainly devoted to detecting Cherenkov light, other 
applications can be envisioned.  These include searching for 
very rapid events from compact astrophysical sources, observing occultations of 
stars by distant Kuiper-belt objects, searching for optical signals from nearby
civilisations (OSETI), or serving as a terrestrial ground station for 
optical communication with distant spacecraft.  Since most such operations 
could be carried out during nights with bright moonlight which would not normally
be used for regular gamma-ray observations, such additional applications 
would enable a more efficient use of CTA, enhance scientific contacts with other fields, 
and yield unique science not achievable at any other observatory. One area which has been explored in detail for CTA is
operation as an optical intensity interferometer.

Diffraction-limited optical aperture synthesis over kilometre baselines, operating optical telescopes in a similar way as radio interferometers,  is a long-held astronomical vision.
This would enable imaging of stellar surfaces and their environments, explore their evolution 
over time, and reveal interactions of stellar winds and gas flows in binary star systems.  
The current best resolution in optical astronomy is offered by 
amplitude (phase-) interferometers which combine light from telescopes 
separated by up to a few hundred meters.  Several concepts have been proposed to 
extend such facilities to scales of a kilometer or more, but their 
realisation remains challenging either on the ground or in space. 
Limiting parameters include the requirement of optical and 
atmospheric stability to a small fraction of an optical 
wavelength and the need to cover many interferometric baselines, 
given that optical light (as opposed to radio) cannot be copied with retained phase 
but has to be split up by beam-splitters to achieve interference among multiple telescope pairs.

The atmospheric and optical stability requirements of amplitude
interferometers can be circumvented by {\it intensity interferometry}, 
a technique once developed for measuring stellar sizes (originally using
an instrument at Narrabri in Australia \cite{HanburyBrown74, Tuthill14}). 
What is observed is the second-order coherence of light (i.e., that of 
intensity, not of amplitude or phase), by measuring temporal correlations of 
arrival times between photons recorded in different telescopes.  The variation of that 
correlation with increasing telescopic distance provides the second-order spatial 
coherence of starlight, corresponding to the square of the visibility observed in 
any classical amplitude interferometer. Thus one obtains [the square of] the 
Fourier transform of the brightness distribution of the source, 
from which its spatial properties and its two-dimensional image can be retrieved. 

The great observational advantage is that the method is practically 
insensitive to either atmospheric turbulence or to telescope optical 
imperfections, enabling very long baselines, as well as observing at short optical wavelengths. 
Since telescopes are connected only electronically, error budgets and required 
precisions relate to electronic timescales of nanoseconds and to light-travel distances of 
tens of centimetres or meters rather than small fractions of an optical wavelength.  
However, the method requires very good photon statistics and therefore large telescopes, 
even in observing bright stars.  These requirements for large and widely distributed 
optical flux collectors, equipped with fast detectors and high-speed electronics, 
have caused intensity interferometry not to be further pursued in a significant way since the 
early attempts.

The technical specifications for atmospheric Cherenkov telescopes are remarkably similar 
to the requirements for intensity interferometry (and conversely, the original telescopes 
at Narrabri were once used for pioneering experiments in observing Cherenkov light flashes). 
In the original Narrabri instrument, the telescopes were moved during observations to maintain 
their projected baseline; however electronic time delays can now be used instead to 
compensate for different arrival times of the wavefront to various telescopes, 
removing the need for mechanical mobility.  The potential of using modern Cherenkov 
telescope arrays for optical intensity interferometry has been explored recently \cite{LeBohec06, Dravins13}.  
Pairs of 12~m Cherenkov telescopes of the VERITAS array in Arizona were connected to
test electronics for intensity interferometry \cite{Dravins08}; 
theoretically simulated observations, including subsequent reconstruction of source images, 
have been carried out \cite {Nunez12}; 
laboratory measurements with arrays of small telescopes mimicking the CTA array have been 
made \cite{Dravins14}.  

The `easiest' targets for interferometry are bright and hot, single or binary stars of 
spectral types O and B or Wolf-Rayet stars with their various circumstellar 
emission-line structures. The stellar disk diameters of the brightest such objects in the sky are 
typically 0.2 -- 0.5 mas and thus lie (somewhat) beyond what can be resolved with 
existing amplitude interferometers. In addition, 
rapidly rotating oblate stars, circumstellar disks, 
winds from hot stars, blue super-giants and extreme objects such 
as $\eta$~Carinae, interacting binaries, the hotter parts of (super)nova explosions, and pulsating Cepheids 
are clear candidates \cite{Dravins13}.

While initial experiments will likely use only a subset of the telescopes in the CTA array, 
best performance will be reached if the full CTA is equipped for interferometry 
at the shortest optical wavelengths.  In the violet ($\lambda\sim$350~nm), the spatial resolution 
of CTA would then approach $\approx$30\,$\mu$as. 
Such resolutions have hitherto been reached only in the radio and open up significant
discovery space in the most well-established waveband of astronomy.

In addition to this interferometry possibility, CTA's huge collection
area and extremely fast photo-detection and processing results in
significant potential for the detection of transient optical
phenomena, especially on relatively short timescales (of seconds or less).
Examples include detecting counterparts to gamma-ray transients such
as GRBs, radio transients, and gravitational wave events, studying outer Solar 
System bodies \cite{Lacki14}, and the
search for optical SETI signals \cite{Hanna09,Abeysekara2016}.
These capabilities will be explored in detail in the near future.

\section{Appendix: Simulating CTA}
\label{sec:MC}

The expectations for the scientific capabilities of CTA presented in this document are 
based on sensitivity and response functions derived from detailed Monte Carlo 
simulations of the CTA observatory.
The simulations model extensive air showers initiated by high-energy 
primary particles (gamma rays or background cosmic-ray nuclei and electrons), 
the Cherenkov light production by 
the shower particles, and the measurement processes of imaging atmospheric Cherenkov telescopes.

The air shower simulations are based on the CORSIKA code \cite{corsika} and include
realistic assumptions of the atmospheric conditions at the two CTA sites in the southern and northern 
hemispheres. 
The  absorption and scattering of the Cherenkov photons due to 
atmospheric aerosols and molecules and the response of the telescopes 
are simulated with the tool sim\_telarray \cite{Bernlohr08}.
The telescope simulations include the ray-tracing of the photons through 
the optical structure and a detailed simulation of the focal plane detector, trigger system, and detector readout. 
The simulations also include the noise photons coming from the expected night-sky background light at each site, with a level equivalent to dark-sky observations towards an extragalactic field.
The telescope array assumed for the southern site consists of four 
Large-Sized Telescopes, 25 Medium-Sized Telescopes and 70 Small-Sized Telescopes, with 
an area covered by the array of $\sim 4\,$km$^2$. 
The northern array consists of
four Large-Sized Telescopes and 15 Medium-Sized Telescopes, with area covered by the array of  
$\sim 0.4\,$km$^2$.
For a more detailed description of the simulation parameters see \cite{Hassan15}.

For all performance estimations, a gamma-ray source with a power-law shaped energy spectrum with a differential spectral index of 2.57 is assumed. 
The sensitivity calculations require a large number of simulated events from cosmic-ray protons and electrons/positrons, due to the excellent gamma-hadron separation capabilities of CTA.
The simulation set for each CTA site comprises about one billion simulated gamma-ray and electron showers and about 100 billion proton showers. 

Two independent analyses of the Monte Carlo samples have been carried out, yielding consistent results. 
The analysis codes used are derived from tools developed in the MAGIC and the VERITAS collaborations. 
These analyses apply 
imaging cleaning algorithms to remove channels with signals dominated by background light, a geometrical image parameterisation, and stereoscopic methods.
Most events observed by CTA will have shower cores that lie inside the area of the array.
Compared to presently operating observatories,
these so-called contained events will lead to 
much better sampled showers by CTA with a larger number of telescopes and will 
result in significantly improved reconstruction accuracy for the gamma-ray direction and energy.
The gamma-hadron separation will also be much more powerful in CTA, resulting in improved sensitivity in the background-limited energy range of CTA.
The gamma-hadron separation methods applied are based on multivariate methods 
(random forests and boosted decision trees) and separation cuts are optimised for the given observation time that is 
typically 50 hours, 5 hours, or 30 minutes.
All analyses are optimised for the
best point-source sensitivity per differential energy bin (five logarithmic bins per decade of energy); better energy and angular resolution can be expected for dedicated high-resolution selection 
criteria.

\clearpage
\section*{Acknowledgments}
\label{sec:ACK}

\begin{center}

\vspace{-1.0cm}

We gratefully acknowledge financial support from the following agencies and organizations:

\medskip

Ministerio de Ciencia, Tecnolog\'{\i}a e Innovaci\'{o}n Productiva (MinCyT), Comisi\'{o}n Nacional de Energ\'{\i}a At\'{o}mica (CNEA), Consejo Nacional de Investigaciones Cient\'{\i}ficas y T\'{e}cnicas (CONICET), Argentina;
State Committee of Science of Armenia, Armenia;
The Australian Research Council, Astronomy Australia Ltd, The University of Adelaide, Australian National University, Monash University, The University of New South Wales, The University of Sydney, Western Sydney University, Australia;
Federal Ministry of Science, Research and Economy, and Innsbruck University, Austria;
Conselho Nacional de Desenvolvimento Cient\'{\i}fico e Tecnol\'{o}gico (CNPq), Funda\c{c}\~{a}o de Amparo \`{a} Pesquisa do Estado do Rio de Janeiro (FAPERJ), Funda\c{c}\~{a}o de Amparo \`{a} Pesquisa do Estado de S\~{a}o Paulo (FAPESP), Brasil;
The Natural Sciences and Engineering Research Council of Canada and the Canadian Space Agency, Canada;
CONICYT-Chile grants PFB-06, FB0821, ACT 1406, FONDECYT-Chile grants 3160153, 3150314, 1150411, 1161463, 1170171, Pontificia Universidad Cat\'{o}lica de Chile Vice-Rectory of Research internationalization grant under MINEDUC agreement PUC1566, Chile;
Croatian Science Foundation, Rudjer Boskovic Institute, University of Osijek, University of Rijeka, University of Split, Faculty of Electrical Engineering, Mechanical Engineering and Naval Architecture, University of Zagreb, Faculty of Electrical Engineering and Computing, Croatia;
Ministry of Education, Youth and Sports, MEYS LE13012, LG14019, LM2015046, LO1305, LTT17006 and EU/MEYS CZ.02.1.01/0.0/0.0/16\_013/0001403, Czech Republic;
Ministry of Higher Education and Research, CNRS-INSU and CNRS-IN2P3, CEA-Irfu, ANR, Regional Council Ile de France, Labex ENIGMASS, OSUG2020, P2IO and OCEVU, France;
Max Planck Society, BMBF, DESY, Helmholtz Association, Germany;
Department of Atomic Energy, Department of Science and Technology, India;
Istituto Nazionale di Astrofisica (INAF), Istituto Nazionale di Fisica Nucleare (INFN), MIUR, Istituto Nazionale di Astrofisica (INAF-OABRERA) Grant Fondazione Cariplo/Regione Lombardia ID 2014-1980/RST\_ERC, Italy;
ICRR, University of Tokyo, JSPS, MEXT, Japan;
Netherlands Research School for Astronomy (NOVA), Netherlands Organization for Scientific Research (NWO), Netherlands;
The Bergen Research Foundation, Norway;
Ministry of Science and Higher Education, the National Centre for Research and Development and the National Science Centre, UMO-2016/22/M/ST9/00583 and UMO-2014/13/B/ST9/00945, Poland;
Slovenian Research Agency, Slovenia;
South African Department of Science and Technology and National Research Foundation through the South African Gamma-Ray Astronomy Programme, South Africa;
MINECO National R+D+I, Severo Ochoa, Maria de Maeztu, CDTI, MultiDark Consolider-Ingenio 2010, PAIDI, UJA, Spain;
Swedish Research Council, Royal Physiographic Society of Lund, Royal Swedish Academy of Sciences, The Swedish National Infrastructure for Computing (SNIC) at Lunarc (Lund), Sweden;
Swiss National Science Foundation (SNSF), Ernest Boninchi Foundation, Switzerland;
Durham University, Leverhulme Trust, Liverpool University, University of Leicester, University of Oxford, Royal Society, Science and Technology Facilities Council, UK;
U.S. National Science Foundation, U.S. Department of Energy, Argonne National Laboratory, Barnard College, University of California, University of Chicago, Columbia University, Georgia Institute of Technology, Institute for Nuclear and Particle Astrophysics (INPAC-MRPI program), Iowa State University, the Smithsonian Institution, Washington University McDonnell Center for the Space Sciences, The University of Wisconsin and the Wisconsin Alumni Research Foundation, USA.

\medskip

The research leading to these results has received funding from the European Union's Seventh Framework Programme (FP7/2007-2013) under grant agreements No~262053, 317446, and 332350.
This project is receiving funding from the European Union's Horizon 2020 research and innovation programs under agreement No~676134.

\medskip

We gratefully acknowledge the critical efforts of the late Professor Giovanni Bignami in the development of CTA.

\end{center}

\addcontentsline{toc}{chapter}{Acknowledgments}
\markboth{Acknowledgments}{}

\normalsize\small\footnotesize\scriptsize
\bibliographystyle{bibgen}

\bibliography{Bib/CTApubs,Bib/TDRscience,Bib/TDRmisc}	

\begin{thebibliography}{100}

\bibitem{FERMIperf}
\url{http://www.slac.stanford.edu/exp/glast/groups/canda/lat_Performance.htm}

\bibitem{Abeysekara2017}
{Abeysekara} A.U., {Albert} A., {Alfaro} R. {\em et~al.\/} (2017).
\newblock {\em {Observation of the Crab Nebula with the HAWC Gamma-Ray
  Observatory}\/}.
\newblock \apj, {\bf 843}, 39

\bibitem{HESSperf}
{Holler} M., {Berge} D., {van Eldik} C. {\em et~al.\/} (2015).
\newblock {\em {Observations of the Crab Nebula with H.E.S.S. Phase II}\/}.
\newblock arXiv:1509.02902

\bibitem{MAGICperf}
{Aleksi{\'c}} J., {Ansoldi} S., {Antonelli} L.A. {\em et~al.\/} (2016).
\newblock {\em {The major upgrade of the MAGIC telescopes, Part II: A
  performance study using observations of the Crab Nebula}\/}.
\newblock Astroparticle Physics, {\bf 72}, 76

\bibitem{VERITASperf}
\url{https://veritas.sao.arizona.edu/about-veritas-mainmenu-81/veritas-specifications-mainmenu-111}

\bibitem{Bartos14}
{Bartos} I., {Veres} P., {Nieto} D. {\em et~al.\/} (2014).
\newblock {\em {Cherenkov Telescope Array is well suited to follow up
  gravitational-wave transients}\/}.
\newblock \mnras, {\bf 443}, 738

\bibitem{Abbott16a}
{Abbott} B.P., {Abbott} R., {Abbott} T.D. {\em et~al.\/} (2016).
\newblock {\em {Observation of Gravitational Waves from a Binary Black Hole
  Merger}\/}.
\newblock Physical Review Letters, {\bf 116}, 6, 061102

\bibitem{PublicPerformanceCurves}
{\em \url{https://www.cta-observatory.org/science/cta-performance/}\/}

\bibitem{Dubus13}
{Dubus} G., {Contreras} J.L., {Funk} S. {\em et~al.\/} (2013).
\newblock {\em {Surveys with the Cherenkov Telescope Array}\/}.
\newblock Astroparticle Physics, {\bf 43}, 317

\bibitem{Abdo07b}
{Abdo} A.A., {Allen} B., {Berley} D. {\em et~al.\/} (2007).
\newblock {\em {TeV Gamma-Ray Sources from a Survey of the Galactic Plane with
  Milagro}\/}.
\newblock \apjl, {\bf 664}, L91

\bibitem{Amenomori05}
{Amenomori} M., {Ayabe} S., {Chen} D. {\em et~al.\/} (2005).
\newblock {\em {A Northern Sky Survey for Steady Tera-Electron Volt Gamma-Ray
  Point Sources Using the Tibet Air Shower Array}\/}.
\newblock \apj, {\bf 633}, 1005

\bibitem{Arsioli14}
{Arsioli} B., {Fraga} B., {Giommi} P. {\em et~al.\/} (2015).
\newblock {\em {1WHSP: An IR-based sample of \~{}1000 VHE {$\gamma$}-ray blazar
  candidates}\/}.
\newblock \aap, {\bf 579}, A34

\bibitem{Padovani14}
{Padovani} P. \& {Giommi} P. (2015).
\newblock {\em {A simplified view of blazars: the very high energy
  {$\gamma$}-ray vision}\/}.
\newblock \mnras, {\bf 446}, L41

\bibitem{Inoue13}
{Inoue} S., {Granot} J., {O'Brien} P.T. {\em et~al.\/} (2013).
\newblock {\em {Gamma-ray burst science in the era of the Cherenkov Telescope
  Array}\/}.
\newblock Astroparticle Physics, {\bf 43}, 252

\bibitem{deOna13}
{de O{\~n}a-Wilhelmi} E., {Rudak} B., {Barrio} J.A. {\em et~al.\/} (2013).
\newblock {\em {Prospects for observations of pulsars and pulsar wind nebulae
  with CTA}\/}.
\newblock Astroparticle Physics, {\bf 43}, 287

\bibitem{Uchiyama07}
{Uchiyama} Y., {Aharonian} F.A., {Tanaka} T. {\em et~al.\/} (2007).
\newblock {\em {Extremely fast acceleration of cosmic rays in a supernova
  remnant}\/}.
\newblock \nat, {\bf 449}, 576

\bibitem{Bulgarelli13}
{Bulgarelli} A., {Fioretti} V., {Contreras} J.L. {\em et~al.\/} (2013).
\newblock {\em {The Real-Time Analysis of the Cherenkov Telescope Array
  Observatory}\/}.
\newblock arXiv:1307.6489

\bibitem{Funk13}
{Funk} S., {Hinton} J.A. \& {CTA Consortium} (2013).
\newblock {\em {Comparison of Fermi-LAT and CTA in the region between 10-100
  GeV}\/}.
\newblock Astroparticle Physics, {\bf 43}, 348

\bibitem{Begelman08}
{Begelman} M.C., {Fabian} A.C. \& {Rees} M.J. (2008).
\newblock {\em {Implications of very rapid TeV variability in blazars}\/}.
\newblock \mnras, {\bf 384}, L19

\bibitem{Paredes13}
{Paredes} J.M., {Bednarek} W., {Bordas} P. {\em et~al.\/} (2013).
\newblock {\em {Binaries with the eyes of CTA}\/}.
\newblock Astroparticle Physics, {\bf 43}, 301

\bibitem{Picozza13}
{Picozza} P. \& {Boezio} M. (2013).
\newblock {\em {Multi messenger astronomy and CTA: TeV cosmic rays and
  electrons}\/}.
\newblock Astroparticle Physics, {\bf 43}, 163

\bibitem{Dravins13}
{Dravins} D., {LeBohec} S., {Jensen} H. {\em et~al.\/} (2013).
\newblock {\em {Optical intensity interferometry with the Cherenkov Telescope
  Array}\/}.
\newblock Astroparticle Physics, {\bf 43}, 331

\bibitem{Gabici14}
{Gabici} S. \& {Aharonian} F.A. (2014).
\newblock {\em {Hadronic gamma-rays from RX J1713.7-3946?}\/}.
\newblock \mnras, {\bf 445}, L70

\bibitem{Acero13}
{Acero} F., {Bamba} A., {Casanova} S. {\em et~al.\/} (2013).
\newblock {\em {Gamma-ray signatures of cosmic ray acceleration, propagation,
  and confinement in the era of CTA}\/}.
\newblock Astroparticle Physics, {\bf 43}, 276

\bibitem{Nakamori14}
{The CTA Consortium: Acero, F}, {Aloisio} R., {Amans} J. {\em et~al.\/} (2017).
\newblock {\em Prospects for cherenkov telescope array observations of the
  young supernova remnant rx j1713.7−3946\/}.
\newblock The Astrophysical Journal, {\bf 840}, 2, 74

\bibitem{Volk96}
{V{\"o}lk} H.J., {Aharonian} F.A. \& {Breitschwerdt} D. (1996).
\newblock {\em {The Nonthermal Energy Content and Gamma-Ray Emission of
  Starburst Galaxies and Clusters of Galaxies}\/}.
\newblock \ssr, {\bf 75}, 279

\bibitem{Abbott16b}
{Abbott} B.P., {Abbott} R., {Abbott} T.D. {\em et~al.\/} (2016).
\newblock {\em {Astrophysical Implications of the Binary Black Hole Merger
  GW150914}\/}.
\newblock \apjl, {\bf 818}, L22

\bibitem{Abbott16aa}
--- (2016).
\newblock {\em {GW151226: Observation of Gravitational Waves from a
  22-Solar-Mass Binary Black Hole Coalescence}\/}.
\newblock Physical Review Letters, {\bf 116}, 24, 241103

\bibitem{Abbott16e}
--- (2016).
\newblock {\em {Binary Black Hole Mergers in the First Advanced LIGO Observing
  Run}\/}.
\newblock Physical Review X, {\bf 6}, 4, 041015

\bibitem{Sol13}
{Sol} H., {Zech} A., {Boisson} C. {\em et~al.\/} (2013).
\newblock {\em {Active Galactic Nuclei under the scrutiny of CTA}\/}.
\newblock Astroparticle Physics, {\bf 43}, 215

\bibitem{Mazin13}
{Mazin} D., {Raue} M., {Behera} B. {\em et~al.\/} (2013).
\newblock {\em {Potential of EBL and cosmology studies with the Cherenkov
  Telescope Array}\/}.
\newblock Astroparticle Physics, {\bf 43}, 241

\bibitem{Broderick12}
{Broderick} A.E., {Chang} P. \& {Pfrommer} C. (2012).
\newblock {\em {The Cosmological Impact of Luminous TeV Blazars. I.
  Implications of Plasma Instabilities for the Intergalactic Magnetic Field and
  Extragalactic Gamma-Ray Background}\/}.
\newblock \apj, {\bf 752}, 22

\bibitem{Hinton13}
{Hinton} J., {Sarkar} S., {Torres} D. {\em et~al.\/} (2013).
\newblock {\em {A New Era in Gamma-Ray Astronomy with the Cherenkov Telescope
  Array}\/}.
\newblock Astroparticle Physics, {\bf 43}, 1

\bibitem{SKAScience}
Bourke T.L., Braun R., Fender R. {\em et~al.\/} (editors) (2015).
\newblock {\em {Proceedings, Advancing Astrophysics with the Square Kilometre
  Array (AASKA14)}\/}, volume AASKA14. SISSA, SISSA

\bibitem{Rayetal2012}
{Ray} P.S., {Abdo} A.A., {Parent} D. {\em et~al.\/} (2012).
\newblock {\em {Radio Searches of Fermi LAT Sources and Blind Search Pulsars:
  The Fermi Pulsar Search Consortium}\/}.
\newblock proc. Fermi Symposium

\bibitem{Lorimer07}
{Lorimer} D.R., {Bailes} M., {McLaughlin} M.A. {\em et~al.\/} (2007).
\newblock {\em {A Bright Millisecond Radio Burst of Extragalactic Origin}\/}.
\newblock Science, {\bf 318}, 777

\bibitem{Thornton13}
{Thornton} D., {Stappers} B., {Bailes} M. {\em et~al.\/} (2013).
\newblock {\em {A Population of Fast Radio Bursts at Cosmological
  Distances}\/}.
\newblock Science, {\bf 341}, 53

\bibitem{Pavlidou14}
{Pavlidou} V., {Angelakis} E., {Myserlis} I. {\em et~al.\/} (2014).
\newblock {\em {The RoboPol optical polarization survey of gamma-ray-loud
  blazars}\/}.
\newblock \mnras, {\bf 442}, 1693

\bibitem{Blinov15}
{Blinov} D., {Pavlidou} V., {Papadakis} I. {\em et~al.\/} (2015).
\newblock {\em {RoboPol: first season rotations of optical polarization plane
  in blazars}\/}.
\newblock \mnras, {\bf 453}, 1669

\bibitem{Reinthal2012}
{Reinthal} R., {Lindfors} E.J., {Mazin} D. {\em et~al.\/} (2012).
\newblock {\em {Connection Between Optical and VHE Gamma-ray Emission in Blazar
  Jets}\/}.
\newblock Journal of Physics Conference Series, {\bf 355}, 1, 012013

\bibitem{Lindfors2012}
{Lindfors} E. \& {MAGIC Collaboration} (2012).
\newblock {\em {Recent results from MAGIC observations of AGN}\/}.
\newblock Journal of Physics Conference Series, {\bf 355}, 1, 012003

\bibitem{Aleksic2012}
{Aleksi{\'c}} J., {Alvarez} E.A., {Antonelli} L.A. {\em et~al.\/} (2012).
\newblock {\em {Discovery of VHE {$\gamma$}-rays from the blazar 1ES 1215+303
  with the MAGIC telescopes and simultaneous multi-wavelength observations}\/}.
\newblock \aap, {\bf 544}, A142

\bibitem{Tepe12}
{Tepe} A. \& {HAWC Collaboration} (2012).
\newblock {\em {HAWC - The High Altitude Water Cherenkov Detector}\/}.
\newblock Journal of Physics Conference Series, {\bf 375}, 5, 052026

\bibitem{DiSciascio16}
{Di Sciascio} G. \& {on behalf of the LHAASO Collaboration} (2016).
\newblock {\em {The LHAASO experiment: from Gamma-Ray Astronomy to Cosmic
  Rays}\/}.
\newblock arXiv:1602.07600

\bibitem{Aartsen13Ta}
{Aartsen} M. {\em et~al.\/} (2013).
\newblock {\em {Evidence for High-Energy Extraterrestrial Neutrinos at the
  IceCube Detector}\/}.
\newblock Science, {\bf 342}, 1242856

\bibitem{Razzaque13}
{Razzaque} S. (2013).
\newblock {\em {Galactic Center origin of a subset of IceCube neutrino
  events}\/}.
\newblock \prd, {\bf 88}, 8, 081302

\bibitem{Katz06}
{Katz} U.F. (2006).
\newblock {\em {KM3NeT: Towards a km$^{3}$ Mediterranean neutrino
  telescope}\/}.
\newblock Nuclear Instruments and Methods in Physics Research A, {\bf 567}, 457

\bibitem{Aasi13}
{Aasi} J., {Abadie} J., {Abbott} B.P. {\em et~al.\/} (2013).
\newblock {\em {Prospects for Localization of Gravitational Wave Transients by
  the Advanced LIGO and Advanced Virgo Observatories}\/}.
\newblock arXiv:1304.0670

\bibitem{Abadie10}
{Abadie} J., {Abbott} B.P., {Abbott} R. {\em et~al.\/} (2010).
\newblock {\em {Topical Review: Predictions for the rates of compact binary
  coalescences observable by ground-based gravitational-wave detectors}\/}.
\newblock Classical and Quantum Gravity, {\bf 27}, 17, 173001

\bibitem{Punturo10}
{Punturo} M., {Abernathy} M., {Acernese} F. {\em et~al.\/} (2010).
\newblock {\em {The Einstein Telescope: a third-generation gravitational wave
  observatory}\/}.
\newblock Classical and Quantum Gravity, {\bf 27}, 19, 194002

\bibitem{Abdallah16c}
Abdallah H. {\em et~al.\/} (2016).
\newblock {\em {Search for dark matter annihilations towards the inner Galactic
  halo from 10 years of observations with H.E.S.S}\/}.
\newblock Phys. Rev. Lett., {\bf 117}, 11, 111301

\bibitem{Ackermann15c}
{Ackermann} M., {Albert} A., {Anderson} B. {\em et~al.\/} (2015).
\newblock {\em {Searching for Dark Matter Annihilation from Milky Way Dwarf
  Spheroidal Galaxies with Six Years of Fermi Large Area Telescope Data}\/}.
\newblock Physical Review Letters, {\bf 115}, 23, 231301

\bibitem{Roszkowski15}
{Roszkowski} L., {Sessolo} E.M. \& {Williams} A.J. (2015).
\newblock {\em {Prospects for dark matter searches in the pMSSM}\/}.
\newblock Journal of High Energy Physics, {\bf 2}, 14

\bibitem{Roszkowski14}
Roszkowski L., Sessolo E.M. \& Williams A.J. (2014).
\newblock {\em {What next for the CMSSM and the NUHM: Improved prospects for
  superpartner and dark matter detection}\/}.
\newblock JHEP, {\bf 1408}, 067

\bibitem{Abramowski11b}
Abramowski A. {\em et~al.\/} (2011).
\newblock {\em {Search for a Dark Matter annihilation signal from the Galactic
  Center halo with H.E.S.S}\/}.
\newblock Phys.Rev.Lett., {\bf 106}, 161301

\bibitem{Zwicky33}
Zwicky F. (1933).
\newblock {\em {Die Rotverschiebung von extragalaktischen Nebeln}\/}.
\newblock Helv.Phys.Acta, {\bf 6}, 110

\bibitem{Clowe03}
Clowe D., Gonzalez A. \& Markevitch M. (2004).
\newblock {\em {Weak lensing mass reconstruction of the interacting cluster
  1E0657-558: Direct evidence for the existence of dark matter}\/}.
\newblock \apj, {\bf 604}, 596

\bibitem{Bradac08}
Bradac M., Allen S.W., Treu T. {\em et~al.\/} (2008).
\newblock {\em {Revealing the properties of dark matter in the merging cluster
  MACSJ0025.4-1222}\/}.
\newblock \apj, {\bf 687}, 959

\bibitem{Clowe06}
Clowe D., Bradac M., Gonzalez A.H. {\em et~al.\/} (2006).
\newblock {\em {A direct empirical proof of the existence of dark matter}\/}.
\newblock Astrophys. J., {\bf 648}, L109

\bibitem{Alcock00}
Alcock C. {\em et~al.\/} (2000).
\newblock {\em {The MACHO project: Microlensing results from 5.7 years of LMC
  observations}\/}.
\newblock \apj, {\bf 542}, 281

\bibitem{Tisserand06}
Tisserand P. {\em et~al.\/} (2007).
\newblock {\em {Limits on the Macho Content of the Galactic Halo from the
  EROS-2 Survey of the Magellanic Clouds}\/}.
\newblock Astron.Astrophys., {\bf 469}, 387

\bibitem{Planck13}
Ade P. {\em et~al.\/} (2014).
\newblock {\em {Planck intermediate results. XVI. Profile likelihoods for
  cosmological parameters}\/}.
\newblock Astron.Astrophys., {\bf 566}, A54

\bibitem{Planck15}
{Planck Collaboration}, {Ade} P.A.R., {Aghanim} N. {\em et~al.\/} (2016).
\newblock {\em {Planck 2015 results. XIII. Cosmological parameters}\/}.
\newblock \aap, {\bf 594}, A13

\bibitem{Springel08}
Springel V., Wang J., Vogelsberger M. {\em et~al.\/} (2008).
\newblock {\em {The Aquarius Project: the subhalos of galactic halos}\/}.
\newblock \mnras, {\bf 391}, 1685

\bibitem{Diemand08}
Diemand J., Kuhlen M., Madau P. {\em et~al.\/} (2008).
\newblock {\em {Clumps and streams in the local dark matter distribution}\/}.
\newblock Nature, {\bf 454}, 735

\bibitem{Navarro95}
Navarro J.F., Frenk C.S. \& White S.D. (1996).
\newblock {\em {The Structure of cold dark matter halos}\/}.
\newblock \apj, {\bf 462}, 563

\bibitem{Graham06}
Graham A.W., Merritt D., Moore B. {\em et~al.\/} (2006).
\newblock {\em {Empirical Models for Dark Matter Halos. III. The Kormendy
  relation and the $log(rho_e)-log(R_e)$ relation}\/}.
\newblock Astron.J., {\bf 132}, 2711

\bibitem{Navarro08}
Navarro J.F., Ludlow A., Springel V. {\em et~al.\/} (2010).
\newblock {\em {The Diversity and Similarity of Cold Dark Matter Halos}\/}.
\newblock Mon.Not.Roy.Astron.Soc., {\bf 402}, 21

\bibitem{Walker09}
Walker M.G., Mateo M., Olszewski E.W. {\em et~al.\/} (2009).
\newblock {\em {A Universal Mass Profile for Dwarf Spheroidal Galaxies}\/}.
\newblock \apj, {\bf 704}, 1274

\bibitem{Walker11b}
Walker M.G. \& Penarrubia J. (2011).
\newblock {\em {A Method for Measuring (Slopes of) the Mass Profiles of Dwarf
  Spheroidal Galaxies}\/}.
\newblock \apj, {\bf 742}, 20

\bibitem{Zeldovich80}
Zeldovich Y., Klypin A., Khlopov M.Y. {\em et~al.\/} (1980).
\newblock {\em {Astrophysical constraints on the mass of heavy stable neutral
  leptons}\/}.
\newblock Sov.J.Nucl.Phys., {\bf 31}, 664

\bibitem{Blumenthal85}
Blumenthal G.R., Faber S., Flores R. {\em et~al.\/} (1986).
\newblock {\em {Contraction of Dark Matter Galactic Halos Due to Baryonic
  Infall}\/}.
\newblock \apj, {\bf 301}, 27

\bibitem{Merritt03}
Merritt D. (2004).
\newblock {\em {Evolution of the dark matter distribution at the galactic
  center}\/}.
\newblock Phys.Rev.Lett., {\bf 92}, 201304

\bibitem{Merritt06}
Merritt D., Harfst S. \& Bertone G. (2007).
\newblock {\em {Collisionally Regenerated Dark Matter Structures in Galactic
  Nuclei}\/}.
\newblock Phys.Rev., {\bf D75}, 043517

\bibitem{Gondolo99}
Gondolo P. \& Silk J. (1999).
\newblock {\em {Dark matter annihilation at the galactic center}\/}.
\newblock Phys.Rev.Lett., {\bf 83}, 1719

\bibitem{Merritt02}
Merritt D., Milosavljevic M., Verde L. {\em et~al.\/} (2002).
\newblock {\em {Dark matter spikes and annihilation radiation from the galactic
  center}\/}.
\newblock Phys.Rev.Lett., {\bf 88}, 191301

\bibitem{Gnedin08}
Gnedin N.Y., Tassis K. \& Kravtsov A.V. (2009).
\newblock {\em {Modeling Molecular Hydrogen and Star Formation in Cosmological
  Simulations}\/}.
\newblock \apj, {\bf 697}, 55

\bibitem{Wise10}
Wise J.H. \& Abel T. (2011).
\newblock {\em {Enzo+Moray: Radiation Hydrodynamics Adaptive Mesh Refinement
  Simulations with Adaptive Ray Tracing}\/}.
\newblock \mnras, {\bf 414}, 3458

\bibitem{Teyssier01}
Teyssier R. (2002).
\newblock {\em {Cosmological hydrodynamics with adaptive mesh refinement: a new
  high resolution code called ramses}\/}.
\newblock Astron.Astrophys., {\bf 385}, 337

\bibitem{Keres11}
Keres D., Vogelsberger M., Sijacki D. {\em et~al.\/} (2012).
\newblock {\em {Moving mesh cosmology: characteristics of galaxies and
  haloes}\/}.
\newblock \mnras, {\bf 425}, 2027

\bibitem{Ackermann13b}
Ackermann M. {\em et~al.\/} (2014).
\newblock {\em {Dark matter constraints from observations of 25 Milky Way
  satellite galaxies with the Fermi Large Area Telescope}\/}.
\newblock Phys.Rev., {\bf D89}, 4, 042001

\bibitem{Susskind82}
Susskind L. (1984).
\newblock {\em {The Gauge Hierarchy Problem, Technicolor, Supersymmetry, and
  all that. (Talk)}\/}.
\newblock Phys.Rept., {\bf 104}, 181

\bibitem{Bertone04}
Bertone G., Hooper D. \& Silk J. (2005).
\newblock {\em {Particle dark matter: Evidence, candidates and constraints}\/}.
\newblock Phys.Rept., {\bf 405}, 279

\bibitem{Cushman:2013zza}
Cushman P., Galbiati C., McKinsey D. {\em et~al.\/} (2013).
\newblock {\em {Snowmass CF1 Summary: WIMP Dark Matter Direct Detection}\/}.
\newblock arXiv:1310.8327

\bibitem{Tulin13}
{Tulin} S., {Yu} H.B. \& {Zurek} K.M. (2013).
\newblock {\em {Beyond collisionless dark matter: Particle physics dynamics for
  dark matter halo structure}\/}.
\newblock \prd, {\bf 87}, 11, 115007

\bibitem{Cahill-Rowley:2013dpa}
Cahill-Rowley M., Cotta R., Drlica-Wagner A. {\em et~al.\/} (2013).
\newblock {\em {Complementarity and Searches for Dark Matter in the pMSSM}\/}.
\newblock arXiv:1305.6921

\bibitem{Bringmann11}
Bringmann T., Calore F., Vertongen G. {\em et~al.\/} (2011).
\newblock {\em {On the Relevance of Sharp Gamma-Ray Features for Indirect Dark
  Matter Searches}\/}.
\newblock Phys.Rev., {\bf D84}, 103525

\bibitem{Arina09}
Arina C., Hambye T., Ibarra A. {\em et~al.\/} (2010).
\newblock {\em {Intense Gamma-Ray Lines from Hidden Vector Dark Matter
  Decay}\/}.
\newblock JCAP, {\bf 1003}, 024

\bibitem{Bringmann07}
Bringmann T., Bergstrom L. \& Edsjo J. (2008).
\newblock {\em {New Gamma-Ray Contributions to Supersymmetric Dark Matter
  Annihilation}\/}.
\newblock JHEP, {\bf 0801}, 049

\bibitem{Birkedal05}
Birkedal A., Matchev K.T., Perelstein M. {\em et~al.\/} (2005).
\newblock {\em {Robust gamma ray signature of WIMP dark matter}\/}.
\newblock arXiv:0507194

\bibitem{Bergstrom04}
Bergstrom L., Bringmann T., Eriksson M. {\em et~al.\/} (2005).
\newblock {\em {Gamma rays from Kaluza-Klein dark matter}\/}.
\newblock Phys.Rev.Lett., {\bf 94}, 131301

\bibitem{Bergstrom05}
{Bergstrom} L., {Bringmann} T., {Eriksson} M. {\em et~al.\/} (2005).
\newblock {\em {Gamma rays from heavy neutralino dark matter}\/}.
\newblock Phys.Rev.Lett., {\bf 95}, 241301

\bibitem{Bergstrom89}
Bergstrom L. (1989).
\newblock {\em {Radiative Processes in Dark Matter Photino Annihilation}\/}.
\newblock Phys.Lett., {\bf B225}, 372

\bibitem{Toma13}
Toma T. (2013).
\newblock {\em {Internal Bremsstrahlung Signature of Real Scalar Dark Matter
  and Consistency with Thermal Relic Density}\/}.
\newblock Phys.Rev.Lett., {\bf 111}, 091301

\bibitem{Mayer09}
Mayer L. (2010).
\newblock {\em {Environmental mechanisms shaping the nature of dwarf spheroidal
  galaxies: the view of computer simulations}\/}.
\newblock Adv.Astron., {\bf 2010}, 278434

\bibitem{Aharonian06a}
Aharonian F. {\em et~al.\/} (2006).
\newblock {\em {H.E.S.S. observations of the Galactic Center region and their
  possible dark matter interpretation}\/}.
\newblock Phys.Rev.Lett., {\bf 97}, 221102

\bibitem{Iocco11}
Iocco F., Pato M., Bertone G. {\em et~al.\/} (2011).
\newblock {\em {Dark Matter distribution in the Milky Way: microlensing and
  dynamical constraints}\/}.
\newblock JCAP, {\bf 1111}, 029

\bibitem{Schaye14}
Schaye J., Crain R.A., Bower R.G. {\em et~al.\/} (2015).
\newblock {\em {The EAGLE project: Simulating the evolution and assembly of
  galaxies and their environments}\/}.
\newblock \mnras, {\bf 446}, 521

\bibitem{Wyrzykowski14}
{Wyrzykowski} L., {Rynkiewicz} A.E., {Skowron} J. {\em et~al.\/} (2015).
\newblock {\em {The Largest Sample of Microlensing Events and the Structure of
  the Galactic Bulge from the OGLE-III Survey}\/}.
\newblock \apjs, {\bf 216}, 12

\bibitem{Freeman12}
Freeman K., Ness M., de~Boer E.W. {\em et~al.\/} (2013).
\newblock {\em {ARGOS II: The Galactic Bulge Survey}\/}.
\newblock \mnras, {\bf 428}, 3660

\bibitem{Howard08}
Howard C.D., Rich R.M., Reitzel D.B. {\em et~al.\/} (2008).
\newblock {\em {The Bulge Radial Velocity Assay (BRAVA): I. Sample Selection
  and a Rotation Curve}\/}.
\newblock \apj, {\bf 688}, 1060

\bibitem{gaia}
GAIA: \url{http://sci.esa.int/gaia/}

\bibitem{Zoccali14}
{Zoccali} M., {Gonzalez} O.A., {Vasquez} S. {\em et~al.\/} (2014).
\newblock {\em {The GIRAFFE Inner Bulge Survey (GIBS). I. Survey description
  and a kinematical map of the Milky Way bulge}\/}.
\newblock \aap, {\bf 562}, A66

\bibitem{Lefranc:2015pza}
Lefranc V., Moulin E., Panci P. {\em et~al.\/} (2015).
\newblock {\em {Prospects for Annihilating Dark Matter in the inner Galactic
  halo by the Cherenkov Telescope Array}\/}.
\newblock \prd, {\bf 91}, 12, 122003

\bibitem{Catalan:2015cna}
{Cabrera-Catalan} M.E., {Ando} S., {Weniger} C. {\em et~al.\/} (2015).
\newblock {\em {Indirect and direct detection prospect for TeV dark matter in
  the nine parameter MSSM}\/}.
\newblock \prd, {\bf 92}, 3, 035018

\bibitem{Pierre14}
{Pierre} M., {Siegal-Gaskins} J.M. \& {Scott} P. (2014).
\newblock {\em {Sensitivity of CTA to dark matter signals from the Galactic
  Center}\/}.
\newblock \jcap, {\bf 6}, 024

\bibitem{Silverwood14}
Silverwood H., Weniger C., Scott P. {\em et~al.\/} (2015).
\newblock {\em {A realistic assessment of the CTA sensitivity to dark matter
  annihilation}\/}.
\newblock JCAP, {\bf 1503}, 03, 055

\bibitem{Ackermann12a}
Ackermann M. {\em et~al.\/} (2012).
\newblock {\em {Fermi LAT Search for Dark Matter in Gamma-ray Lines and the
  Inclusive Photon Spectrum}\/}.
\newblock Phys.Rev., {\bf D86}, 022002

\bibitem{Abramowski13c}
{Abramowski} A. {\em et~al.\/} (2013).
\newblock {\em {Search for photon line-like signatures from Dark Matter
  annihilations with H.E.S.S}\/}.
\newblock Phys.Rev.Lett., {\bf 110}, 041301

\bibitem{Bringmann12}
Bringmann T., Huang X., Ibarra A. {\em et~al.\/} (2012).
\newblock {\em {Fermi LAT Search for Internal Bremsstrahlung Signatures from
  Dark Matter Annihilation}\/}.
\newblock JCAP, {\bf 1207}, 054

\bibitem{Weniger12}
Weniger C. (2012).
\newblock {\em {A Tentative Gamma-Ray Line from Dark Matter Annihilation at the
  Fermi Large Area Telescope}\/}.
\newblock JCAP, {\bf 1208}, 007

\bibitem{Martinez13}
Martinez G.D. (2015).
\newblock {\em {A Robust Determination of Milky Way Satellite Properties using
  Hierarchical Mass Modeling}\/}.
\newblock {\bf 451}, 2524

\bibitem{2015arXiv150302584T}
{The DES Collaboration}, {Bechtol} K. {\em et~al.\/} (2015).
\newblock {\em {Eight New Milky Way Companions Discovered in First-Year Dark
  Energy Survey Data}\/}.
\newblock \apj, {\bf 807}, 50

\bibitem{2015arXiv150302079K}
{Koposov} S.E., {Belokurov} V., {Torrealba} G. {\em et~al.\/} (2015).
\newblock {\em {Beasts of the Southern Wild: Discovery of nine Ultra Faint
  satellites in the vicinity of the Magellanic Clouds}\/}.
\newblock \apj, {\bf 805}, 130

\bibitem{Bonnivard15}
Bonnivard V., Combet C., Maurin D. {\em et~al.\/} (2015).
\newblock {\em {Dark matter annihilation and decay profiles for the Reticulum
  II dwarf spheroidal galaxy}\/}.
\newblock Astrophys. J., {\bf 808}, 2, L36

\bibitem{Bringmann09}
Bringmann T. (2009).
\newblock {\em {Particle Models and the Small-Scale Structure of Dark
  Matter}\/}.
\newblock New J.Phys., {\bf 11}, 105027

\bibitem{Zackrisson09}
Zackrisson E. \& Riehm T. (2010).
\newblock {\em {Gravitational lensing as a probe of cold dark matter
  subhalos}\/}.
\newblock Adv.Astron., {\bf 2010}, 478910

\bibitem{Chen10}
Chen J. \& Koushiappas S.M. (2010).
\newblock {\em {Gravitational Nanolensing from Subsolar Mass Dark Matter
  Halos}\/}.
\newblock \apj, {\bf 724}, 400

\bibitem{Garsden11}
Garsden H., Bate N. \& Lewis G. (2012).
\newblock {\em {Probing planetary mass dark matter in galaxies: gravitational
  nanolensing of multiply imaged quasars}\/}.
\newblock \mnras, {\bf 420}, 3574

\bibitem{GeringerSameth10}
Geringer-Sameth A. \& Koushiappas S.M. (2012).
\newblock {\em {Detecting unresolved moving sources in a diffuse
  background}\/}.
\newblock \mnras, {\bf 425}, 862

\bibitem{Carlberg13}
Carlberg R.G. \& Grillmair C.J. (2013).
\newblock {\em {Gaps in the GD-1 Star Stream}\/}.
\newblock \apj, {\bf 768}, 171

\bibitem{Grillmair13}
{Grillmair} C.J., {Cutri} R., {Masci} F.J. {\em et~al.\/} (2013).
\newblock {\em {Detection of a Nearby Halo Debris Stream in the WISE and 2MASS
  Surveys}\/}.
\newblock \apjl, {\bf 769}, L23

\bibitem{Grillmair14}
{Grillmair} C.J. (2014).
\newblock {\em {Two New Halo Debris Streams in the Sloan Digital Sky
  Survey}\/}.
\newblock \apjl, {\bf 790}, L10

\bibitem{Hargis14a}
{Hargis} J., {Willman} B., {Sand} D. {\em et~al.\/} (2014).
\newblock {\em {Milky Way Stellar Streams: A Window to Purely Dark
  Subhalos}\/}.
\newblock NOAO Proposal

\bibitem{Sesar14}
{Sesar} B., {Banholzer} S.R., {Cohen} J.G. {\em et~al.\/} (2014).
\newblock {\em {Stacking the Invisibles: A Guided Search for Low-luminosity
  Milky Way Satellites}\/}.
\newblock \apj, {\bf 793}, 135

\bibitem{Erkal14}
{Erkal} D. \& {Belokurov} V. (2015).
\newblock {\em {Forensics of Subhalo-Stream Encounters: The Three Phases of Gap
  Growth}\/}.
\newblock \mnras, {\bf 450}, 1136

\bibitem{Pieri07}
Pieri L., Bertone G. \& Branchini E. (2008).
\newblock {\em {Dark Matter Annihilation in Substructures Revised}\/}.
\newblock \mnras, {\bf 384}, 1627

\bibitem{Buckley10}
Buckley M.R. \& Hooper D. (2010).
\newblock {\em {Dark Matter Subhalos In the Fermi First Source Catalog}\/}.
\newblock Phys.Rev., {\bf D82}, 063501

\bibitem{Zechlin11}
Zechlin H., Fernandes M., Elsaesser D. {\em et~al.\/} (2012).
\newblock {\em {Dark matter subhaloes as gamma-ray sources and candidates in
  the first Fermi-LAT catalogue}\/}.
\newblock Astron.Astrophys., {\bf 538}, A93

\bibitem{Smith1963}
{Smith} G.P. (1963).
\newblock {\em {A peculiar feature at $l^{II}$ = 40$^{\circ}$.5, b$^{II}$ = -
  15$^{\circ}$.0}\/}.
\newblock \bain, {\bf 17}, 203

\bibitem{Saul12}
Saul D.R., Peek J., Grcevich J. {\em et~al.\/} (2012).
\newblock {\em {The GALFA-HI Compact Cloud Catalog}\/}.
\newblock \apj, {\bf 758}, 44

\bibitem{Hill09}
Hill A.S., Haffner L.M. \& Reynolds R.J. (2009).
\newblock {\em {Ionized Gas in the Smith Cloud}\/}.
\newblock \apj, {\bf 703}, 1832

\bibitem{Bonnivard14}
{Bonnivard} V., {Combet} C., {Maurin} D. {\em et~al.\/} (2015).
\newblock {\em {Spherical Jeans analysis for dark matter indirect detection in
  dwarf spheroidal galaxies - impact of physical parameters and
  triaxiality}\/}.
\newblock \mnras, {\bf 446}, 3002

\bibitem{GeringerSameth14}
Geringer-Sameth A., Koushiappas S.M. \& Walker M. (2015).
\newblock {\em {Dwarf galaxy annihilation and decay emission profiles for dark
  matter experiments}\/}.
\newblock \apj, {\bf 801}, 74

\bibitem{Koch07}
Koch A., Kleyna J., Wilkinson M. {\em et~al.\/} (2007).
\newblock {\em {Stellar kinematics in the remote Leo II dwarf spheroidal galaxy
  -- Another brick in the wall}\/}.
\newblock Astron.J., {\bf 134}, 566

\bibitem{Walker08}
Walker M.G., Mateo M. \& Olszewski E. (2009).
\newblock {\em {Stellar Velocities in the Carina, Fornax, Sculptor and Sextans
  dSph Galaxies: Data from the Magellan/MMFS Survey}\/}.
\newblock Astron.J., {\bf 137}, 3100

\bibitem{Walker13}
{Walker} M. (2013).
\newblock {\em {Dark Matter in the Galactic Dwarf Spheroidal Satellites}\/}, p.
  1039

\bibitem{pfs11}
{\em \url{http://sumire.ipmu.jp/pfs/intro.html}\/}.
\newblock (2011)

\bibitem{Charbonnier11}
Charbonnier A., Combet C., Daniel M. {\em et~al.\/} (2011).
\newblock {\em {Dark matter profiles and annihilation in dwarf spheroidal
  galaxies: prospectives for present and future gamma-ray observatories - I.
  The classical dSphs}\/}.
\newblock Mon.Not.Roy.Astron.Soc., {\bf 418}, 1526

\bibitem{Strigari07b}
Strigari L.E., Koushiappas S.M., Bullock J.S. {\em et~al.\/} (2008).
\newblock {\em {The Most Dark Matter Dominated Galaxies: Predicted Gamma-ray
  Signals from the Faintest Milky Way Dwarfs}\/}.
\newblock \apj, {\bf 678}, 614

\bibitem{Strigari13}
Strigari L.E. (2013).
\newblock {\em {Galactic Searches for Dark Matter}\/}.
\newblock Phys.Rept., {\bf 531}, 1

\bibitem{Koposov07}
Koposov S., Belokurov V., Evans N. {\em et~al.\/} (2008).
\newblock {\em {The Luminosity Function of the Milky Way Satellites}\/}.
\newblock \apj, {\bf 686}, 279

\bibitem{Tollerud08}
Tollerud E.J., Bullock J.S., Strigari L.E. {\em et~al.\/} (2008).
\newblock {\em {Hundreds of Milky Way Satellites? Luminosity Bias in the
  Satellite Luminosity Function}\/}.
\newblock \apj, {\bf 688}, 277

\bibitem{Bullock10}
{Bullock} J.S. (2010).
\newblock {\em {Notes on the Missing Satellites Problem}\/}.
\newblock arXiv:1009.4505

\bibitem{Hargis14b}
{Hargis} J.R., {Willman} B. \& {Peter} A.H.G. (2014).
\newblock {\em {Too Many, Too Few, or Just Right? The Predicted Number and
  Distribution of Milky Way Dwarf Galaxies}\/}.
\newblock \apjl, {\bf 795}, L13

\bibitem{Tasitsiomi03}
Tasitsiomi A., Siegal-Gaskins J.M. \& Olinto A.V. (2004).
\newblock {\em {Gamma-ray and synchrotron emission from neutralino annihilation
  in the Large Magellanic Cloud}\/}.
\newblock Astropart.Phys., {\bf 21}, 637

\bibitem{Buckley15}
{Buckley} M.R., {Charles} E., {Gaskins} J.M. {\em et~al.\/} (2015).
\newblock {\em {Search for gamma-ray emission from dark matter annihilation in
  the large magellanic cloud with the fermi large area telescope}\/}.
\newblock \prd, {\bf 91}, 10, 102001

\bibitem{Abdo10d}
{Abdo} A.A., {Ackermann} M., {Ajello} M. {\em et~al.\/} (2010).
\newblock {\em {Observations of the Large Magellanic Cloud with Fermi}\/}.
\newblock \aap, {\bf 512}, A7+

\bibitem{Abramowski15a}
{H.E.S.S.~Collaboration}, {Abramowski} A., {Aharonian} F. {\em et~al.\/}
  (2015).
\newblock {\em {The exceptionally powerful TeV {$\gamma$}-ray emitters in the
  Large Magellanic Cloud}\/}.
\newblock Science, {\bf 347}, 406

\bibitem{Kim98}
{Kim} S., {Staveley-Smith} L., {Dopita} M.A. {\em et~al.\/} (1998).
\newblock {\em {An H I Aperture Synthesis Mosaic of the Large Magellanic
  Cloud}\/}.
\newblock \apj, {\bf 503}, 674

\bibitem{vanderMarel13}
van~der Marel R.P. \& Kallivayalil N. (2014).
\newblock {\em {Third-Epoch Magellanic Cloud Proper Motions II: The Large
  Magellanic Cloud Rotation Field in Three Dimensions}\/}.
\newblock \apj, {\bf 781}, 2, 121

\bibitem{vanderMarel02}
van~der Marel R.P., Alves D.R., Hardy E. {\em et~al.\/} (2002).
\newblock {\em {New understanding of large magellanic cloud structure, dynamics
  and orbit from carbon star kinematics}\/}.
\newblock Astron.J., {\bf 124}, 2639

\bibitem{2012PhRvD85f3517C}
{Combet} C., {Maurin} D., {Nezri} E. {\em et~al.\/} (2012).
\newblock {\em {Decaying dark matter: Stacking analysis of galaxy clusters to
  improve on current limits}\/}.
\newblock \prd, {\bf 85}, 6, 063517

\bibitem{Sanchez-Conde:2013yxa}
Sanchez-Conde M.A. \& Prada F. (2014).
\newblock {\em {The flattening of the concentration-mass relation towards low
  halo masses and its implications for the annihilation signal boost}\/}.
\newblock \mnras, {\bf 442}, 2271

\bibitem{Pinzke:2009cp}
Pinzke A., Pfrommer C. \& Bergstrom L. (2009).
\newblock {\em {Gamma-rays from dark matter annihilations strongly constrain
  the substructure in halos}\/}.
\newblock Phys.Rev.Lett., {\bf 103}, 181302

\bibitem{Pinzke11}
{Pinzke} A., {Pfrommer} C. \& {Bergstr{\"o}m} L. (2011).
\newblock {\em {Prospects of detecting gamma-ray emission from galaxy clusters:
  Cosmic rays and dark matter annihilations}\/}.
\newblock \prd, {\bf 84}, 12, 123509

\bibitem{Cirelli12}
Cirelli M., Moulin E., Panci P. {\em et~al.\/} (2012).
\newblock {\em {Gamma ray constraints on Decaying Dark Matter}\/}.
\newblock Phys. Rev., {\bf D86}, 083506

\bibitem{2011JCAP12011S}
{S{\'a}nchez-Conde} M.A., {Cannoni} M., {Zandanel} F. {\em et~al.\/} (2011).
\newblock {\em {Dark matter searches with Cherenkov telescopes: nearby dwarf
  galaxies or local galaxy clusters?}\/}.
\newblock \jcap, {\bf 12}, 011

\bibitem{2012JCAP10032A}
{Aleksi{\'c}} J., {Rico} J. \& {Martinez} M. (2012).
\newblock {\em {Optimized analysis method for indirect dark matter searches
  with imaging air Cherenkov telescopes}\/}.
\newblock \jcap, {\bf 10}, 032

\bibitem{2012ApJ76191A}
{Ackermann} M. {\em et~al.\/} (2012).
\newblock {\em {Constraints on the Galactic Halo Dark Matter from Fermi-LAT
  Diffuse Measurements}\/}.
\newblock \apj, {\bf 761}, 91

\bibitem{2011PhRvD83h3507C}
{Cembranos} J.A.R., {de La Cruz-Dombriz} A., {Dobado} A. {\em et~al.\/} (2011).
\newblock {\em {Photon spectra from WIMP annihilation}\/}.
\newblock \prd, {\bf 83}, 8, 083507

\bibitem{Ponti15}
{Ponti} G., {Morris} M.R., {Terrier} R. {\em et~al.\/} (2015).
\newblock {\em {The XMM-Newton view of the central degrees of the Milky
  Way}\/}.
\newblock \mnras, {\bf 453}, 172

\bibitem{LaRosa00}
{LaRosa} T.N., {Kassim} N.E., {Lazio} T.J.W. {\em et~al.\/} (2000).
\newblock {\em {A Wide-Field 90 Centimeter VLA Image of the Galactic Center
  Region}\/}.
\newblock \aj, {\bf 119}, 207

\bibitem{Molinari11}
{Molinari} S., {Bally} J., {Noriega-Crespo} A. {\em et~al.\/} (2011).
\newblock {\em {A 100 pc Elliptical and Twisted Ring of Cold and Dense
  Molecular Clouds Revealed by Herschel Around the Galactic Center}\/}.
\newblock \apjl, {\bf 735}, L33

\bibitem{vanEldik15}
{van Eldik} C. (2015).
\newblock {\em {Gamma rays from the Galactic Centre region: A review}\/}.
\newblock Astroparticle Physics, {\bf 71}, 45

\bibitem{Aharonian09gc}
{Aharonian} F., {Akhperjanian} A.G., {Anton} G. {\em et~al.\/} (2009).
\newblock {\em {Spectrum and variability of the Galactic center VHE
  {$\gamma$}-ray source HESS J1745-290}\/}.
\newblock \aap, {\bf 503}, 817

\bibitem{Archer14}
{Archer} A., {Barnacka} A., {Beilicke} M. {\em et~al.\/} (2014).
\newblock {\em {Very-high Energy Observations of the Galactic Center Region by
  VERITAS in 2010-2012}\/}.
\newblock \apj, {\bf 790}, 149

\bibitem{Aharonian:2006au}
{Aharonian} F., {Akhperjanian} A.G., {Bazer-Bachi} A.R. {\em et~al.\/} (2006).
\newblock {\em {Discovery of very-high-energy {$\gamma$}-rays from the Galactic
  Centre ridge}\/}.
\newblock \nat, {\bf 439}, 695

\bibitem{Archer16}
{Archer} A., {Benbow} W., {Bird} R. {\em et~al.\/} (2016).
\newblock {\em {TeV Gamma-Ray Observations of the Galactic Center Ridge by
  VERITAS}\/}.
\newblock \apj, {\bf 821}, 129

\bibitem{HESS16}
{HESS Collaboration}, {Abramowski} A., {Aharonian} F. {\em et~al.\/} (2016).
\newblock {\em {Acceleration of petaelectronvolt protons in the Galactic
  Centre}\/}.
\newblock \nat, {\bf 531}, 476

\bibitem{Aharonian05b}
{Aharonian} F., {Akhperjanian} A.G., {Aye} K.M. {\em et~al.\/} (2005).
\newblock {\em {Very high energy gamma rays from the composite SNR G
  0.9+0.1}\/}.
\newblock \aap, {\bf 432}, L25

\bibitem{Jones2012}
{Jones} P.A., {Burton} M.G., {Cunningham} M.R. {\em et~al.\/} (2012).
\newblock {\em {Spectral imaging of the Central Molecular Zone in multiple 3-mm
  molecular lines}\/}.
\newblock \mnras, {\bf 419}, 2961

\bibitem{Johnson2009}
{Johnson} S.P., {Dong} H. \& {Wang} Q.D. (2009).
\newblock {\em {A large-scale survey of X-ray filaments in the Galactic
  Centre}\/}.
\newblock \mnras, {\bf 399}, 1429

\bibitem{Kosack:2004ri}
Kosack K. {\em et~al.\/} (2004).
\newblock {\em {TeV gamma-ray observations of the galactic center}\/}.
\newblock \apj, {\bf 608}, L97

\bibitem{Tsuchiya:2004wv}
Tsuchiya K. {\em et~al.\/} (2004).
\newblock {\em {Detection of sub-TeV gamma-rays from the Galactic Center
  direction by CANGAROO-II}\/}.
\newblock \apj, {\bf 606}, L115

\bibitem{Figer:2003tu}
Figer D.F., Rich R.M., Kim S.S. {\em et~al.\/} (2004).
\newblock {\em {An extended star formation history for the Galactic Center from
  Hubble Space Telescope / NICMOS observations}\/}.
\newblock \apj, {\bf 601}, 319

\bibitem{Crocker:2010qn}
Crocker R.M., Jones D.I., Aharonian F. {\em et~al.\/} (2011).
\newblock {\em {Wild at Heart:-The Particle Astrophysics of the Galactic
  Centre}\/}.
\newblock Mon.Not.Roy.Astron.Soc., {\bf 413}, 763

\bibitem{Yoast-Hull:2014cra}
Yoast-Hull T.M., Gallagher J. \& Zweibel E.G. (2014).
\newblock {\em {The Cosmic Ray Population of the Galactic Central Molecular
  Zone}\/}.
\newblock \apj, {\bf 790}, 86

\bibitem{Yang:2012fy}
Yang H.Y., Ruszkowski M., Ricker P. {\em et~al.\/} (2012).
\newblock {\em {The Fermi Bubbles: Supersonic AGN Jets with Anisotropic Cosmic
  Ray Diffusion}\/}.
\newblock \apj, {\bf 761}, 185

\bibitem{Law:2009wv}
Law C. (2010).
\newblock {\em {A Multiwavelength View of a Mass Outflow from the Galactic
  Center}\/}.
\newblock \apj, {\bf 708}, 474

\bibitem{Nakashima:2013yra}
Nakashima S., Nobukawa M., Uchida H. {\em et~al.\/} (ApJ).
\newblock {\em {Discovery of the recombining plasma in the south of the
  Galactic center; a relic of the past Galactic center activity?}\/}.
\newblock 2013, {\bf 773}, 20N

\bibitem{Borkowski:2014voa}
Borkowski K.J., Reynolds S.P., Green D.A. {\em et~al.\/} (2014).
\newblock {\em {Nonuniform Expansion of the Youngest Galactic Supernova Remnant
  G1.9+0.3}\/}.
\newblock \apj, {\bf 790}, L18

\bibitem{Aharonian:2005kn}
{Aharonian} F. {\em et~al.\/} (2006).
\newblock {\em {The H.E.S.S. survey of the inner galaxy in very high-energy
  gamma-rays}\/}.
\newblock \apj, {\bf 636}, 777

\bibitem{Aharonian:2008gw}
{Aharonian} F., {Akhperjanian} A.G., {Barres de Almeida} U. {\em et~al.\/}
  (2008).
\newblock {\em {Exploring a SNR/molecular cloud association within HESS
  J1745-303}\/}.
\newblock \aap, {\bf 483}, 509

\bibitem{Aharonian:2006wh}
{Aharonian} F. {\em et~al.\/} (2006).
\newblock {\em {H.E.S.S. observations of the Galactic Center region and their
  possible dark matter interpretation}\/}.
\newblock Phys.Rev.Lett., {\bf 97}, 221102

\bibitem{Zubovas2012:gcasteroid}
{Zubovas} K., {Nayakshin} S. \& {Markoff} S. (2012).
\newblock {\em {Sgr A* flares: tidal disruption of asteroids and planets?}\/}.
\newblock \mnras, {\bf 421}, 1315

\bibitem{Wommer:2008we}
Wommer E., Melia F. \& Fatuzzo M. (2008).
\newblock {\em {Diffuse TeV Emission at the Galactic Centre}\/}.
\newblock \mnras, {\bf 387}, 987

\bibitem{Melia11}
{Melia} F. \& {Fatuzzo} M. (2011).
\newblock {\em {Diffusive cosmic-ray acceleration at the Galactic Centre}\/}.
\newblock \mnras, {\bf 410}, L23

\bibitem{Amano:2011kw}
Amano T., Torii K., Hayakawa T. {\em et~al.\/} (2011).
\newblock {\em {Stochastic Acceleration of Cosmic Rays in the Central Molecular
  Zone of the Galaxy}\/}.
\newblock arXiv:1110.3140

\bibitem{Abeysekara13}
{Abeysekara} A.U., {Alfaro} R., {Alvarez} C. {\em et~al.\/} (2013).
\newblock {\em {Sensitivity of the high altitude water Cherenkov detector to
  sources of multi-TeV gamma rays}\/}.
\newblock Astroparticle Physics, {\bf 50}, 26

\bibitem{Ajello15}
{Ackermann} M., {Ajello} M., {Atwood} W.B. {\em et~al.\/} (2016).
\newblock {\em {2FHL: The Second Catalog of Hard Fermi-LAT Sources}\/}.
\newblock \apjs, {\bf 222}, 5

\bibitem{Carrigan13b}
{Carrigan} S., {Brun} F., {Chaves} R.C.G. {\em et~al.\/} (2013).
\newblock {\em {Charting the TeV Milky Way: H.E.S.S. Galactic plane survey
  maps, catalog and source populations}\/}.
\newblock arXiv:1307.4868

\bibitem{Ong13}
{Ong} R. {\em et~al.\/} (2013).
\newblock {\em {Recent VERITAS Results on VHE Gamma-ray Sources in Cygnus}\/}.
\newblock proc. 33rd ICRC Rio de Janiero, Brazil

\bibitem{Bartoli13}
{Bartoli} B., {Bernardini} P., {Bi} X.J. {\em et~al.\/} (2013).
\newblock {\em {TeV Gamma-Ray Survey of the Northern Sky Using the ARGO-YBJ
  Detector}\/}.
\newblock \apj, {\bf 779}, 27

\bibitem{Aharonian02s}
{Aharonian} F. {\em et~al.\/} (2002).
\newblock {\em {A Search for TeV Gamma-Ray Emission from SNRs, Pulsars and
  Unidentified GeV Sources in the Galactic Plane in the Longitude Range between
  -2 deg and 85 deg.}\/}.
\newblock \aap, {\bf 395}, 803

\bibitem{Atkins04}
{{Atkins}, R and others} (2004).
\newblock {\em {TeV Gamma-Ray Survey of the Northern Hemisphere Sky using the
  Milagro Observatory}\/}.
\newblock ApJ, {\bf 608}, 680

\bibitem{HEGRACrab}
{The flux of very high energy gamma rays from the Crab nebula is set to that
  measured by HEGRA, in Aharonian, F. {\it et al.} (2004), {\em The Crab Nebula
  and Pulsar between 500 GeV and 80 TeV: Observations with the HEGRA
  Stereoscopic Air Cherenkov Telescopes}, ApJ, {\bf 614}, 897. The HEGRA Crab
  nebula spectrum is dN/dE = 2.83 x $10^{-11}$ (E/1
  TeV)$^{-2.62}$\,cm$^{-2}$\,s$^{-1}$\,TeV$^{-1}$. For an energy threshold of
  125 GeV, 1 mCrab = $5.07 \times 10^{-13}$\,cm$^{-2}$\,s$^{-1}\, .$}

\bibitem{Abramowski14diffuse}
{H.~E.~S.~S.~Collaboration}, {:}, {Abramowski} A. {\em et~al.\/} (2014).
\newblock {\em {Diffuse Galactic gamma-ray emission with H.E.S.S}\/}.
\newblock arXiv:1411.7568

\bibitem{Renaud09}
{Renaud} M. (2009).
\newblock {\em {Latest results on Galactic sources as seen in VHE
  gamma-rays}\/}.
\newblock Proceedings of 44th Recontres de Moriond 2009

\bibitem{Abdo09ee}
{Abdo} A.A., {Allen} B.T., {Aune} T. {\em et~al.\/} (2009).
\newblock {\em {Milagro Observations of Multi-TeV Emission from Galactic
  Sources in the Fermi Bright Source List}\/}.
\newblock \apjl, {\bf 700}, L127

\bibitem{Acharya13}
{{Acharya}, B. et al.} (2013).
\newblock {\em {Introducing the CTA concept}\/}.
\newblock Astroparticle Physics, {\bf 43}, 3

\bibitem{Schure13}
{Schure} K.M. \& {Bell} A.R. (2013).
\newblock {\em {Cosmic ray acceleration in young supernova remnants}\/}.
\newblock \mnras, {\bf 435}, 1174

\bibitem{Abdo09d}
{Abdo} A. {\em et~al.\/} (2009).
\newblock {\em {Fermi Large Area Telescope Brght Gamma-Ray Source List}\/}.
\newblock ApJS, {\bf 183}, 46

\bibitem{tevcat}
TeVCat: \url{http://tevcat.uchicago.edu/}

\bibitem{Aharonian05s}
{Aharonian} F. {\em et~al.\/} (2006).
\newblock {\em {The H.E.S.S. Survey of the Inner Galaxy in Very High-Energy
  Gamma-Rays}\/}.
\newblock ApJ, {\bf 636}, 777

\bibitem{Pietrzynski:2013}
{Pietrzy{\'n}ski} G., {Graczyk} D., {Gieren} W. {\em et~al.\/} (2013).
\newblock {\em {An eclipsing-binary distance to the Large Magellanic Cloud
  accurate to two per cent}\/}.
\newblock \nat, {\bf 495}, 76

\bibitem{vanderMarel:2006}
{van der Marel} R.P. (2006).
\newblock {\em {The Large Magellanic Cloud: structure and kinematics}\/}.
\newblock In M.~{Livio} \& T.M. {Brown} (editors), {\em The Local Group as an
  Astrophysical Laboratory\/}, pp. 47--71

\bibitem{Hughes:2007}
{Hughes} A., {Staveley-Smith} L., {Kim} S. {\em et~al.\/} (2007).
\newblock {\em {An Australia Telescope Compact Array 20-cm radio continuum
  study of the Large Magellanic Cloud}\/}.
\newblock \mnras, {\bf 382}, 543

\bibitem{Walborn:2014}
{Walborn} N.R., {Sana} H., {Sim{\'o}n-D{\'{\i}}az} S. {\em et~al.\/} (2014).
\newblock {\em {The VLT-FLAMES Tarantula Survey. XIV. The O-type stellar
  content of 30 Doradus}\/}.
\newblock \aap, {\bf 564}, A40

\bibitem{McCray:1993}
{McCray} R. (1993).
\newblock {\em {Supernova 1987A revisited}\/}.
\newblock \araa, {\bf 31}, 175

\bibitem{Bozzetto17}
{Bozzetto} L.M., {Filipovi{\'c}} M.D., {Vukoti{\'c}} B. {\em et~al.\/} (2017).
\newblock {\em {Statistical Analysis of Supernova Remnants in the Large
  Magellanic Cloud}\/}.
\newblock \apjs, {\bf 230}, 2

\bibitem{Crowther:2010}
{Crowther} P.A., {Schnurr} O., {Hirschi} R. {\em et~al.\/} (2010).
\newblock {\em {The R136 star cluster hosts several stars whose individual
  masses greatly exceed the accepted 150M$_{solar}$ stellar mass limit}\/}.
\newblock \mnras, {\bf 408}, 731

\bibitem{Lawton10}
{Lawton} B., {Gordon} K.D., {Babler} B. {\em et~al.\/} (2010).
\newblock {\em {Spitzer Analysis of H II Region Complexes in the Magellanic
  Clouds: Determining a Suitable Monochromatic Obscured Star Formation
  Indicator}\/}.
\newblock \apj, {\bf 716}, 453

\bibitem{Dunne:2001}
{Dunne} B.C., {Points} S.D. \& {Chu} Y.H. (2001).
\newblock {\em {X-Rays from Superbubbles in the Large Magellanic Cloud. VI. A
  Sample of Thirteen Superbubbles}\/}.
\newblock \apjs, {\bf 136}, 119

\bibitem{Kim:1999}
{Kim} S., {Dopita} M.A., {Staveley-Smith} L. {\em et~al.\/} (1999).
\newblock {\em {H I Shells in the Large Magellanic Cloud}\/}.
\newblock \aj, {\bf 118}, 2797

\bibitem{Marshall:1998}
{Marshall} F.E., {Gotthelf} E.V., {Zhang} W. {\em et~al.\/} (1998).
\newblock {\em {Discovery of an Ultrafast X-Ray Pulsar in the Supernova Remnant
  N157B}\/}.
\newblock \apjl, {\bf 499}, L179

\bibitem{Seward:1984}
{Seward} F.D., {Harnden} Jr. F.R. \& {Helfand} D.J. (1984).
\newblock {\em {Discovery of a 50 millisecond pulsar in the Large Magellanic
  Cloud}\/}.
\newblock \apjl, {\bf 287}, L19

\bibitem{de-Grijs:2006}
{de Grijs} R. \& {Anders} P. (2006).
\newblock {\em {How well do we know the age and mass distributions of the star
  cluster system in the Large Magellanic Cloud?}\/}.
\newblock \mnras, {\bf 366}, 295

\bibitem{Ackermann16}
{Ackermann} M., {Albert} A., {Atwood} W.B. {\em et~al.\/} (2016).
\newblock {\em {Deep view of the Large Magellanic Cloud with six years of
  Fermi-LAT observations}\/}.
\newblock \aap, {\bf 586}, A71

\bibitem{Abramowski12c}
{H.E.S.S.~Collaboration}, {Abramowski} A., {Acero} F. {\em et~al.\/} (2012).
\newblock {\em {Discovery of gamma-ray emission from the extragalactic pulsar
  wind nebula N 157B with H.E.S.S.}\/}.
\newblock \aap, {\bf 545}, L2

\bibitem{Chevalier:1995}
{Chevalier} R.A. \& {Dwarkadas} V.V. (1995).
\newblock {\em {The Presupernova H II Region around SN 1987A}\/}.
\newblock \apjl, {\bf 452}, L45

\bibitem{Zanardo:2010}
{Zanardo} G., {Staveley-Smith} L., {Ball} L. {\em et~al.\/} (2010).
\newblock {\em {Multifrequency Radio Measurements of Supernova 1987A Over 22
  Years}\/}.
\newblock \apj, {\bf 710}, 1515

\bibitem{Maggi16}
{Maggi} P., {Haberl} F., {Kavanagh} P.J. {\em et~al.\/} (2016).
\newblock {\em {The population of X-ray supernova remnants in the Large
  Magellanic Cloud}\/}.
\newblock \aap, {\bf 585}, A162

\bibitem{Ackermann11}
{Ackermann} M., {Ajello} M., {Allafort} A. {\em et~al.\/} (2011).
\newblock {\em {A Cocoon of Freshly Accelerated Cosmic Rays Detected by Fermi
  in the Cygnus Superbubble}\/}.
\newblock Science, {\bf 334}, 1103

\bibitem{Barger16}
{Barger} K.A., {Lehner} N. \& {Howk} J.C. (2016).
\newblock {\em {Down-the-barrel and Transverse Observations of the Large
  Magellanic Cloud: Evidence for a Symmetric Galactic Wind on the Near and Far
  Sides of the Galaxy}\/}.
\newblock \apj, {\bf 817}, 91

\bibitem{Corbet:2016}
{Corbet} R.H.D., {Chomiuk} L., {Coe} M.J. {\em et~al.\/} (2016).
\newblock {\em {A Luminous Gamma-ray Binary in the Large Magellanic Cloud}\/}.
\newblock \apj, {\bf 829}, 105

\bibitem{Gelfand15}
{Gelfand} J., {Breton} R., {Ng} C.Y. {\em et~al.\/} (2015).
\newblock {\em {Pulsar Wind Nebulae in the SKA era}\/}.
\newblock Advancing Astrophysics with the Square Kilometre Array (AASKA14), 46

\bibitem{Keane15}
{Keane} E., {Bhattacharyya} B., {Kramer} M. {\em et~al.\/} (2015).
\newblock {\em {A Cosmic Census of Radio Pulsars with the SKA}\/}.
\newblock Advancing Astrophysics with the Square Kilometre Array (AASKA14), 40

\bibitem{Indebetouw14}
{Indebetouw} R. \& {SN1987A ALMA Cycle 0 Team} (2014).
\newblock {\em {ALMA resolves SN 1987A's dust factory and particle
  accelerator}\/}.
\newblock In {\em American Astronomical Society Meeting Abstracts \#223\/},
  volume 223 of {\em American Astronomical Society Meeting Abstracts\/}, p.
  354.37

\bibitem{Mellinger09}
{Mellinger} A. (2009).
\newblock {\em {A Color All-Sky Panorama Image of the Milky Way}\/}.
\newblock \pasp, {\bf 121}, 1180

\bibitem{Kim03}
{Kim} S., {Staveley-Smith} L., {Dopita} M.A. {\em et~al.\/} (2003).
\newblock {\em {A Neutral Hydrogen Survey of the Large Magellanic Cloud:
  Aperture Synthesis and Multibeam Data Combined}\/}.
\newblock \apjs, {\bf 148}, 473

\bibitem{Berezhko:2011}
{Berezhko} E.G., {Ksenofontov} L.T. \& {V{\"o}lk} H.J. (2011).
\newblock {\em {Expected Gamma-Ray Emission of Supernova Remnant SN 1987A}\/}.
\newblock \apj, {\bf 732}, 58

\bibitem{Berezhko:2015}
--- (2015).
\newblock {\em {Re-examination of the Expected Gamma-Ray Emission of Supernova
  Remnant SN 1987A}\/}.
\newblock \apj, {\bf 810}, 63

\bibitem{Williams11}
{Williams} B.J., {Borkowski} K.J., {Reynolds} S.P. {\em et~al.\/} (2011).
\newblock {\em {Dusty Blast Waves of Two Young Large Magellanic Cloud Supernova
  Remnants: Constraints on Post-shock Compression}\/}.
\newblock \apj, {\bf 729}, 65

\bibitem{Park12}
{Park} S., {Hughes} J.P., {Slane} P.O. {\em et~al.\/} (2012).
\newblock {\em {An X-Ray Study of Supernova Remnant N49 and Soft Gamma-Ray
  Repeater 0526-66 in the Large Magellanic Cloud}\/}.
\newblock \apj, {\bf 748}, 117

\bibitem{Williams14}
{Williams} B.J., {Borkowski} K.J., {Reynolds} S.P. {\em et~al.\/} (2014).
\newblock {\em {Spitzer Observations of the Type Ia Supernova Remnant N103B:
  Kepler's Older Cousin?}\/}.
\newblock \apj, {\bf 790}, 139

\bibitem{Borkowski:2006}
{Borkowski} K.J., {Hendrick} S.P. \& {Reynolds} S.P. (2006).
\newblock {\em {Dense, Fe-rich Ejecta in Supernova Remnants DEM L238 and DEM
  L249: A New Class of Type Ia Supernova?}\/}.
\newblock \apj, {\bf 652}, 1259

\bibitem{Brantseg:2014}
{Brantseg} T., {McEntaffer} R.L., {Bozzetto} L.M. {\em et~al.\/} (2014).
\newblock {\em {A Multi-wavelength Look at the Young Plerionic Supernova
  Remnant 0540-69.3}\/}.
\newblock \apj, {\bf 780}, 50

\bibitem{Martin:2014}
{Martin} J., {Torres} D.F., {Cillis} A. {\em et~al.\/} (2014).
\newblock {\em {Is there room for highly magnetized pulsar wind nebulae among
  those non-detected at TeV?}\/}.
\newblock \mnras, {\bf 443}, 138

\bibitem{Bozzetto:2012}
{Bozzetto} L.M., {Filipovi{\'c}} M.D., {Crawford} E.J. {\em et~al.\/} (2012).
\newblock {\em {Multifrequency study of the Large Magellanic Cloud supernova
  remnant J0529-6653 near pulsar B0529-66}\/}.
\newblock \mnras, {\bf 420}, 2588

\bibitem{Martin:2014b}
{Martin} P. (2014).
\newblock {\em Interstellar gamma-ray emission from cosmic rays in star-forming
  galaxies\/}.
\newblock \aap, {\bf 564}, A61

\bibitem{Bykov14}
{Bykov} A.M. (2014).
\newblock {\em {Nonthermal particles and photons in starburst regions and
  superbubbles}\/}.
\newblock Astron Astrophys Rev, {\bf 22}, 77

\bibitem{UrryPadovani1995a}
{Urry} C.M. \& {Padovani} P. (1995).
\newblock {\em {Unified Schemes for Radio-Loud Active Galactic Nuclei}\/}.
\newblock \pasp, {\bf 107}, 803

\bibitem{HenriSauge2006a}
{Henri} G. \& {Saug{\'e}} L. (2006).
\newblock {\em {The Bulk Lorentz Factor Crisis of TeV Blazars: Evidence for an
  Inhomogeneous Pileup Energy Distribution?}\/}.
\newblock \apj, {\bf 640}, 185

\bibitem{Inoue10}
{Inoue} Y., {Totani} T. \& {Mori} M. (2010).
\newblock {\em {Prospects for a Very High-Energy Blazar Survey by the
  Next-Generation Cherenkov Telescopes}\/}.
\newblock \pasj, {\bf 62}, 1005

\bibitem{Inoue09}
{Inoue} Y. \& {Totani} T. (2009).
\newblock {\em {The Blazar Sequence and the Cosmic Gamma-ray Background
  Radiation in the Fermi Era}\/}.
\newblock \apj, {\bf 702}, 523-536

\bibitem{Inoue14}
{Inoue} Y., {Kalashev} O.E. \& {Kusenko} A. (2014).
\newblock {\em {Prospects for future very high-energy gamma-ray sky survey:
  Impact of secondary gamma rays}\/}.
\newblock Astroparticle Physics, {\bf 54}, 118

\bibitem{Ackermann2011}
Ackermann M. {\em et~al.\/} (2011).
\newblock {\em {The Second Catalog of Active Galactic Nuclei Detected by the
  Fermi Large Area Telescope}\/}.
\newblock ApJ, {\bf 743}, 171

\bibitem{Hayashida2013}
{Hayashida} M., {Stawarz} {\L}., {Cheung} C.C. {\em et~al.\/} (2013).
\newblock {\em {Discovery of GeV Emission from the Circinus Galaxy with the
  Fermi Large Area Telescope}\/}.
\newblock \apj, {\bf 779}, 131

\bibitem{DiMauro13}
{Di Mauro} M., {Calore} F., {Donato} F. {\em et~al.\/} (2014).
\newblock {\em {Diffuse {$\gamma$}-Ray Emission from Misaligned Active Galactic
  Nuclei}\/}.
\newblock \apj, {\bf 780}, 161

\bibitem{Inoue16}
{Inoue} Y. \& {Tanaka} Y.T. (2016).
\newblock {\em {Lower Bound on the Cosmic TeV Gamma-Ray Background
  Radiation}\/}.
\newblock \apj, {\bf 818}, 187

\bibitem{Atwood2013a}
{Atwood} W., {Albert} A., {Baldini} L. {\em et~al.\/} (2013).
\newblock {\em {Pass 8: Toward the Full Realization of the Fermi-LAT Scientific
  Potential}\/}.
\newblock arXiv:1303.3514

\bibitem{Szanecki15}
{Szanecki} M., {Sobczy{\'n}ska} D., {Nied{\'z}wiecki} A. {\em et~al.\/} (2015).
\newblock {\em {Monte Carlo simulations of alternative sky observation modes
  with the Cherenkov Telescope Array}\/}.
\newblock Astroparticle Physics, {\bf 67}, 33

\bibitem{Meszaros13}
{M{\'e}sz{\'a}ros} P. (2013).
\newblock {\em {Gamma ray bursts}\/}.
\newblock Astroparticle Physics, {\bf 43}, 134

\bibitem{Kumar15}
{Kumar} P. \& {Zhang} B. (2015).
\newblock {\em {The physics of gamma-ray bursts and relativistic jets}\/}.
\newblock \physrep, {\bf 561}, 1

\bibitem{Ellis13}
{Ellis} J. \& {Mavromatos} N.E. (2013).
\newblock {\em {Probes of Lorentz violation}\/}.
\newblock Astroparticle Physics, {\bf 43}, 50

\bibitem{Bednarek13}
{Bednarek} W. (2013).
\newblock {\em {High energy {$\gamma$}-ray emission from compact galactic
  sources in the context of observations with the next generation Cherenkov
  Telescope Arrays}\/}.
\newblock Astroparticle Physics, {\bf 43}, 81

\bibitem{Buehler14}
{B{\"u}hler} R. \& {Blandford} R. (2014).
\newblock {\em {The surprising Crab pulsar and its nebula: a review}\/}.
\newblock Reports on Progress in Physics, {\bf 77}, 6, 066901

\bibitem{Dubus15}
{Dubus} G. (2015).
\newblock {\em {Gamma-ray emission from binaries in context}\/}.
\newblock Comptes Rendus Physique, {\bf 16}, 661

\bibitem{Obrien13}
{O'Brien} P.T. \& {Smartt} S.J. (2013).
\newblock {\em {Interpreting signals from astrophysical transient
  experiments}\/}.
\newblock Royal Society of London Philosophical Transactions Series A, {\bf
  371}, 20498

\bibitem{Komossa15}
{Komossa} S. (2015).
\newblock {\em {Tidal disruption of stars by supermassive black holes: Status
  of observations}\/}.
\newblock Journal of High Energy Astrophysics, {\bf 7}, 148

\bibitem{Brown15}
{Brown} P.J., {Roming} P.W.A. \& {Milne} P.A. (2015).
\newblock {\em {The first ten years of Swift supernovae}\/}.
\newblock Journal of High Energy Astrophysics, {\bf 7}, 111

\bibitem{Katz16}
{Katz} J.I. (2016).
\newblock {\em {Fast radio bursts, A brief review: Some questions, fewer
  answers}\/}.
\newblock Modern Physics Letters A, {\bf 31}, 1630013

\bibitem{Halzen13}
{Halzen} F. (2013).
\newblock {\em {Pionic photons and neutrinos from cosmic ray accelerators}\/}.
\newblock Astroparticle Physics, {\bf 43}, 155

\bibitem{Ahlers15}
{Ahlers} M. \& {Halzen} F. (2015).
\newblock {\em {High-energy cosmic neutrino puzzle: a review}\/}.
\newblock Reports on Progress in Physics, {\bf 78}, 12, 126901

\bibitem{Aartsen13Td}
{IceCube Collaboration}, {Aartsen} M.G., {Abbasi} R. {\em et~al.\/} (2013).
\newblock {\em {The IceCube Neutrino Observatory Part I: Point Source
  Searches}\/}.
\newblock arXiv:1309.6979

\bibitem{Aartsen15a}
{The IceCube Collaboration}, {Aartsen} M.G., {Abraham} K. {\em et~al.\/}
  (2015).
\newblock {\em {The IceCube Neutrino Observatory - Contributions to ICRC 2015
  Part I: Point Source Searches}\/}.
\newblock arXiv:1510.05222

\bibitem{Abbott16c}
{Abbott} B.P., {Abbott} R., {Abbott} T.D. {\em et~al.\/} (2016).
\newblock {\em {Localization and broadband follow-up of the gravitational-wave
  transient GW150914}\/}.
\newblock arXiv:1602.08492

\bibitem{Connaughton16}
{Connaughton} V., {Burns} E., {Goldstein} A. {\em et~al.\/} (2016).
\newblock {\em {Fermi GBM Observations of LIGO Gravitational Wave event
  GW150914}\/}.
\newblock \apjl, {\bf 826}, L6

\bibitem{Fernandez15}
{Fern{\'a}ndez} R. \& {Metzger} B.D. (2015).
\newblock {\em {Electromagnetic Signatures of Neutron Star Mergers in the
  Advanced LIGO Era}\/}.
\newblock arXiv:1512.05435

\bibitem{Doro13}
{Doro} M., {Conrad} J., {Emmanoulopoulos} D. {\em et~al.\/} (2013).
\newblock {\em {Dark matter and fundamental physics with the Cherenkov
  Telescope Array}\/}.
\newblock Astroparticle Physics, {\bf 43}, 189

\bibitem{Bulgarelli15}
{Bulgarelli} A., {Fioretti} V., {Zoli} A. {\em et~al.\/} (2015).
\newblock {\em {The On-Site Analysis of the Cherenkov Telescope Array}\/}.
\newblock arXiv:1509.01963

\bibitem{Fioretti15}
{Fioretti} V., {Bulgarelli} A., {Zoli} A. {\em et~al.\/} (2015).
\newblock {\em {Real-Time Analysis sensitivity evaluation of the Cherenkov
  Telescope Array}\/}.
\newblock proc. 34th ICRC, The Hague, Netherlands

\bibitem{Gerard15}
{Gerard} L. (2015).
\newblock {\em {Divergent pointing with the Cherenkov Telescope Array for
  surveys and beyond}\/}.
\newblock arXiv:1508.06197

\bibitem{Abdo09a}
{Abdo} A.A., {Ackermann} M., {Arimoto} M. {\em et~al.\/} (2009).
\newblock {\em {Fermi Observations of High-Energy Gamma-Ray Emission from GRB
  080916C}\/}.
\newblock Science, {\bf 323}, 1688

\bibitem{Finke10}
{Finke} J.D., {Razzaque} S. \& {Dermer} C.D. (2010).
\newblock {\em {Modeling the Extragalactic Background Light from Stars and
  Dust}\/}.
\newblock \apj, {\bf 712}, 238

\bibitem{Kouveliotou93}
{Kouveliotou} C., {Meegan} C.A., {Fishman} G.J. {\em et~al.\/} (1993).
\newblock {\em {Identification of two classes of gamma-ray bursts}\/}.
\newblock \apjl, {\bf 413}, L101

\bibitem{Mereghetti15}
{Mereghetti} S., {Pons} J.A. \& {Melatos} A. (2015).
\newblock {\em {Magnetars: Properties, Origin and Evolution}\/}.
\newblock \ssr, {\bf 191}, 315

\bibitem{Torres11}
{Torres} D.F., {Rea} N., {Esposito} P. {\em et~al.\/} (2012).
\newblock {\em {A Magnetar-like Event from LS I +61 303 and Its Nature as a
  Gamma-Ray Binary}\/}.
\newblock \apj, {\bf 744}, 106

\bibitem{Tavani11}
{Tavani} M., {Bulgarelli} A., {Vittorini} V. {\em et~al.\/} (2011).
\newblock {\em {Discovery of Powerful Gamma-Ray Flares from the Crab
  Nebula}\/}.
\newblock Science, {\bf 331}, 736

\bibitem{Abdo11b}
{Abdo} A.A., {Ackermann} M., {Ajello} M. {\em et~al.\/} (2011).
\newblock {\em {Gamma-Ray Flares from the Crab Nebula}\/}.
\newblock Science, {\bf 331}, 739

\bibitem{Taylor06}
{Taylor} G.B. \& {Granot} J. (2006).
\newblock {\em {The Giant Flare from SGR 1806-20 and its Radio Afterglow}\/}.
\newblock Modern Physics Letters A, {\bf 21}, 2171

\bibitem{Dubus13b}
{Dubus} G. (2013).
\newblock {\em {Gamma-ray binaries and related systems}\/}.
\newblock \aapr, {\bf 21}, 64

\bibitem{Tavani09}
{Tavani} M., {Bulgarelli} A., {Piano} G. {\em et~al.\/} (2009).
\newblock {\em {Extreme particle acceleration in the microquasar CygnusX-3}\/}.
\newblock \nat, {\bf 462}, 620

\bibitem{Zanin16}
{Zanin} R., {Fern{\'a}ndez-Barral} A., {de O{\~n}a Wilhelmi} E. {\em et~al.\/}
  (2016).
\newblock {\em {Gamma rays detected from Cygnus X-1 with likely jet origin}\/}.
\newblock \aap, {\bf 596}, A55

\bibitem{Acciari11Ta}
{Acciari} V.A., {Aliu} E., {Araya} M. {\em et~al.\/} (2011).
\newblock {\em {Gamma-Ray Observations of the Be/Pulsar Binary 1A 0535+262
  During a Giant X-Ray Outburst}\/}.
\newblock \apj, {\bf 733}, 96

\bibitem{Stappers14}
{Stappers} B.W., {Archibald} A.M., {Hessels} J.W.T. {\em et~al.\/} (2014).
\newblock {\em {A State Change in the Missing Link Binary Pulsar System PSR
  J1023+0038}\/}.
\newblock \apj, {\bf 790}, 39

\bibitem{Ackermann14T}
{Ackermann} M., {Ajello} M., {Albert} A. {\em et~al.\/} (2014).
\newblock {\em {Fermi establishes classical novae as a distinct class of
  gamma-ray sources}\/}.
\newblock Science, {\bf 345}, 554

\bibitem{Metzger16}
{Metzger} B.D., {Caprioli} D., {Vurm} I. {\em et~al.\/} (2016).
\newblock {\em {Novae as Tevatrons: prospects for CTA and IceCube}\/}.
\newblock \mnras, {\bf 457}, 1786

\bibitem{Ackermann13Ta}
{Ackermann} M., {Ajello} M., {Asano} K. {\em et~al.\/} (2013).
\newblock {\em {The First Fermi-LAT Gamma-Ray Burst Catalog}\/}.
\newblock \apjs, {\bf 209}, 11

\bibitem{Gehrels13}
{Gehrels} N. \& {Cannizzo} J.K. (2013).
\newblock {\em {High-energy transients}\/}.
\newblock Philosophical Transactions of the Royal Society of London Series A,
  {\bf 371}, 20120270

\bibitem{Kulkarni12}
{Kulkarni} S.R. (2012).
\newblock {\em {Cosmic Explosions (Optical Transients)}\/}.
\newblock arXiv:1202.2381

\bibitem{Fender11}
{Fender} R.P. \& {Bell} M.E. (2011).
\newblock {\em {Radio transients: an antediluvian review}\/}.
\newblock Bulletin of the Astronomical Society of India, {\bf 39}, 315

\bibitem{Fender15}
{Fender} R.P., {Anderson} G.E., {Osten} R. {\em et~al.\/} (2015).
\newblock {\em {A prompt radio transient associated with a gamma-ray superflare
  from the young M dwarf binary DG CVn}\/}.
\newblock \mnras, {\bf 446}, L66

\bibitem{Ghirlanda15}
{Ghirlanda} G., {Salvaterra} R., {Campana} S. {\em et~al.\/} (2015).
\newblock {\em {Unveiling the population of orphan {$\gamma$}-ray bursts}\/}.
\newblock \aap, {\bf 578}, A71

\bibitem{Bloom11}
{Bloom} J.S., {Giannios} D., {Metzger} B.D. {\em et~al.\/} (2011).
\newblock {\em {A Possible Relativistic Jetted Outburst from a Massive Black
  Hole Fed by a Tidally Disrupted Star}\/}.
\newblock Science, {\bf 333}, 203

\bibitem{vanVelzen16}
{van Velzen} S., {Anderson} G.E., {Stone} N.C. {\em et~al.\/} (2016).
\newblock {\em {A radio jet from the optical and x-ray bright stellar tidal
  disruption flare ASASSN-14li}\/}.
\newblock Science, {\bf 351}, 62

\bibitem{Chen16}
{Chen} X., {G{\'o}mez-Vargas} G.A. \& {Guillochon} J. (2016).
\newblock {\em {The Gamma-ray Afterglows of Tidal Disruption Events}\/}.
\newblock \mnras

\bibitem{Campana06}
{Campana} S., {Mangano} V., {Blustin} A.J. {\em et~al.\/} (2006).
\newblock {\em {The association of GRB 060218 with a supernova and the
  evolution of the shock wave}\/}.
\newblock \nat, {\bf 442}, 1008

\bibitem{Soderberg08}
{Soderberg} A.M., {Berger} E., {Page} K.L. {\em et~al.\/} (2008).
\newblock {\em {An extremely luminous X-ray outburst at the birth of a
  supernova}\/}.
\newblock \nat, {\bf 453}, 469

\bibitem{Kashiyama13}
{Kashiyama} K., {Murase} K., {Horiuchi} S. {\em et~al.\/} (2013).
\newblock {\em {High-energy Neutrino and Gamma-Ray Transients from
  Trans-relativistic Supernova Shock Breakouts}\/}.
\newblock \apjl, {\bf 769}, L6

\bibitem{Keane16b}
{Keane} E.F., {Johnston} S., {Bhandari} S. {\em et~al.\/} (2016).
\newblock {\em {The host galaxy of a fast radio burst}\/}.
\newblock \nat, {\bf 530}, 453

\bibitem{Williams16}
{Williams} P.K.G. \& {Berger} E. (2016).
\newblock {\em {No precise localization for FRB 150418: claimed radio transient
  is AGN variability}\/}.
\newblock \apjl, {\bf 821}, L22

\bibitem{Spitler16}
{Spitler} L.G., {Scholz} P., {Hessels} J.W.T. {\em et~al.\/} (2016).
\newblock {\em {A repeating fast radio burst}\/}.
\newblock \nat, {\bf 531}, 202

\bibitem{Chatterjee17}
{Chatterjee} S., {Law} C.J., {Wharton} R.S. {\em et~al.\/} (2017).
\newblock {\em {A direct localization of a fast radio burst and its host}\/}.
\newblock \nat, {\bf 541}, 58

\bibitem{Lyubarsky14}
{Lyubarsky} Y. (2014).
\newblock {\em {A model for fast extragalactic radio bursts}\/}.
\newblock \mnras, {\bf 442}, L9

\bibitem{Andersson13}
{Andersson} N., {Baker} J., {Belczynski} K. {\em et~al.\/} (2013).
\newblock {\em {The transient gravitational-wave sky}\/}.
\newblock Classical and Quantum Gravity, {\bf 30}, 19, 193002

\bibitem{Berger14}
{Berger} E. (2014).
\newblock {\em {Short-Duration Gamma-Ray Bursts}\/}.
\newblock \araa, {\bf 52}, 43

\bibitem{Keane16a}
{Keane} E.F. \& {SUPERB Collaboration} (2016).
\newblock {\em {Fast Radio Bursts: Searches, Sensitivities and
  Implications}\/}.
\newblock arXiv:1602.05165

\bibitem{Ackermann13Tb}
{Ackermann} M., {Ajello} M., {Albert} A. {\em et~al.\/} (2013).
\newblock {\em {The Fermi All-sky Variability Analysis: A List of Flaring
  Gamma-Ray Sources and the Search for Transients in Our Galaxy}\/}.
\newblock \apj, {\bf 771}, 57

\bibitem{Ackermann11T}
{Ackermann} M. {\em et~al.\/} (2011).
\newblock {\em {Detection of a Spectral Break in the Extra Hard Component of
  GRB 090926A}\/}.
\newblock \apj, {\bf 729}, 114

\bibitem{Ackermann12T}
{Fermi Large Area Telescope Team}, {Ackermann} M., {Ajello} M. {\em et~al.\/}
  (2012).
\newblock {\em {Constraining the High-energy Emission from Gamma-Ray Bursts
  with Fermi}\/}.
\newblock \apj, {\bf 754}, 121

\bibitem{Ackermann14G}
{Ackermann} M., {Ajello} M., {Asano} K. {\em et~al.\/} (2014).
\newblock {\em {Fermi-LAT Observations of the Gamma-Ray Burst GRB 130427A}\/}.
\newblock Science, {\bf 343}, 42

\bibitem{Kouveliotou13}
{Kouveliotou} C., {Granot} J., {Racusin} J.L. {\em et~al.\/} (2013).
\newblock {\em {NuSTAR Observations of GRB 130427A Establish a Single Component
  Synchrotron Afterglow Origin for the Late Optical to Multi-GeV Emission}\/}.
\newblock \apjl, {\bf 779}, L1

\bibitem{Atwood13}
{Atwood} W.B., {Baldini} L., {Bregeon} J. {\em et~al.\/} (2013).
\newblock {\em {New Fermi-LAT Event Reconstruction Reveals More High-energy
  Gamma Rays from Gamma-Ray Bursts}\/}.
\newblock \apj, {\bf 774}, 76

\bibitem{Abdo09b}
{Abdo} A.A., {Ackermann} M., {Ajello} M. {\em et~al.\/} (2009).
\newblock {\em {A limit on the variation of the speed of light arising from
  quantum gravity effects}\/}.
\newblock \nat, {\bf 462}, 331

\bibitem{Vasileiou13}
{Vasileiou} V., {Jacholkowska} A., {Piron} F. {\em et~al.\/} (2013).
\newblock {\em {Constraints on Lorentz invariance violation from Fermi-Large
  Area Telescope observations of gamma-ray bursts}\/}.
\newblock \prd, {\bf 87}, 12, 122001

\bibitem{Acciari11Tb}
{Acciari} V.A., {Aliu} E., {Arlen} T. {\em et~al.\/} (2011).
\newblock {\em {VERITAS Observations of Gamma-Ray Bursts Detected by Swift}\/}.
\newblock \apj, {\bf 743}, 62

\bibitem{Aleksic14t}
{Aleksi{\'c}} J., {Ansoldi} S., {Antonelli} L.A. {\em et~al.\/} (2014).
\newblock {\em {MAGIC upper limits on the GRB 090102 afterglow}\/}.
\newblock \mnras, {\bf 437}, 3103

\bibitem{Abramowski14Ta}
{H.E.S.S.~Collaboration}, {Abramowski} A., {Aharonian} F. {\em et~al.\/}
  (2014).
\newblock {\em {Search for TeV Gamma-ray Emission from GRB 100621A, an
  extremely bright GRB in X-rays, with H.E.S.S.}\/}.
\newblock \aap, {\bf 565}, A16

\bibitem{Gilmore13}
{Gilmore} R.C., {Bouvier} A., {Connaughton} V. {\em et~al.\/} (2013).
\newblock {\em {IACT observations of gamma-ray bursts: prospects for the
  Cherenkov Telescope Array}\/}.
\newblock Experimental Astronomy, {\bf 35}, 413

\bibitem{Kakuwa12}
{Kakuwa} J., {Murase} K., {Toma} K. {\em et~al.\/} (2012).
\newblock {\em {Prospects for detecting gamma-ray bursts at very high energies
  with the Cherenkov Telescope Array}\/}.
\newblock \mnras, {\bf 425}, 514

\bibitem{Abeysekara12}
{Abeysekara} A.U., {Aguilar} J.A., {Aguilar} S. {\em et~al.\/} (2012).
\newblock {\em {On the sensitivity of the HAWC observatory to gamma-ray
  bursts}\/}.
\newblock Astroparticle Physics, {\bf 35}, 641

\bibitem{Taboada14}
{Taboada} I. \& {Gilmore} R.C. (2014).
\newblock {\em {Prospects for the detection of GRBs with HAWC}\/}.
\newblock Nuclear Instruments and Methods in Physics Research A, {\bf 742}, 276

\bibitem{Malyshev13}
{Malyshev} D., {Zdziarski} A.A. \& {Chernyakova} M. (2013).
\newblock {\em {High-energy gamma-ray emission from Cyg X-1 measured by Fermi
  and its theoretical implications}\/}.
\newblock \mnras, {\bf 434}, 2380

\bibitem{Bodaghee13}
{Bodaghee} A., {Tomsick} J.A., {Pottschmidt} K. {\em et~al.\/} (2013).
\newblock {\em {Gamma-Ray Observations of the Microquasars Cygnus X-1, Cygnus
  X-3, GRS 1915+105, and GX 339-4 with the Fermi Large Area Telescope}\/}.
\newblock \apj, {\bf 775}, 98

\bibitem{Mariotti10}
{Mariotti} M. (2010).
\newblock {\em {No significant enhancement in the VHE gamma-ray flux of the
  Crab Nebula measured by MAGIC in September 2010}\/}.
\newblock The Astronomer's Telegram, {\bf 2967}, 1

\bibitem{Ong10}
{Ong} R.A. (2010).
\newblock {\em {Search for an Enhanced TeV Gamma-Ray Flux from the Crab Nebula
  with VERITAS}\/}.
\newblock The Astronomer's Telegram, {\bf 2968}, 1

\bibitem{Abramowski14Tb}
{H.~E.~S.~S.~Collaboration}, {Abramowski} A., {Aharonian} F. {\em et~al.\/}
  (2014).
\newblock {\em {H.E.S.S. observations of the Crab during its March 2013 GeV
  gamma-ray flare}\/}.
\newblock \aap, {\bf 562}, L4

\bibitem{Aleksic10T}
{Aleksi{\'c}} J., {Antonelli} L.A., {Antoranz} P. {\em et~al.\/} (2010).
\newblock {\em {Magic Constraints on {$\gamma$}-ray Emission from Cygnus
  X-3}\/}.
\newblock \apj, {\bf 721}, 843

\bibitem{Ahnen15}
{Ahnen} M.L., {Ansoldi} S., {Antonelli} L.A. {\em et~al.\/} (2015).
\newblock {\em {Very high-energy {$\gamma$}-ray observations of novae and dwarf
  novae with the MAGIC telescopes}\/}.
\newblock \aap, {\bf 582}, A67

\bibitem{Albert07}
{Albert} J., {Aliu} E., {Anderhub} H. {\em et~al.\/} (2007).
\newblock {\em {Very High Energy Gamma-Ray Radiation from the Stellar Mass
  Black Hole Binary Cygnus X-1}\/}.
\newblock \apjl, {\bf 665}, L51

\bibitem{Peng16b}
{Peng} F.K., {Tang} Q.W. \& {Wang} X.Y. (2016).
\newblock {\em {Search for High-energy Gamma-ray Emission from Tidal Disruption
  Events with the Fermi Large Area Telescope}\/}.
\newblock \apj, {\bf 825}, 47

\bibitem{Aliu11T}
{Aliu} E., {Arlen} T., {Aune} T. {\em et~al.\/} (2011).
\newblock {\em {VERITAS Observations of the Unusual Extragalactic Transient
  Swift J164449.3+573451}\/}.
\newblock \apjl, {\bf 738}, L30

\bibitem{Aleksic13T}
{Aleksi{\'c}} J., {Antonelli} L.A., {Antoranz} P. {\em et~al.\/} (2013).
\newblock {\em {Very high energy gamma-ray observation of the peculiar
  transient event Swift J1644+57 with the MAGIC telescopes and AGILE}\/}.
\newblock \aap, {\bf 552}, A112

\bibitem{Abdalla17}
{H.~E.~S.~S.~Collaboration}, {Abdalla} H., {Abramowski} A. {\em et~al.\/}
  (2017).
\newblock {\em {First limits on the very-high energy gamma-ray afterglow
  emission of a fast radio burst. H.E.S.S. observations of FRB 150418}\/}.
\newblock \aap, {\bf 597}, A115

\bibitem{Santander15}
{Santander} M., {VERITAS} f.t. \& {IceCube Collaborations} (2015).
\newblock {\em {Searching for TeV gamma-ray emission associated with IceCube
  high-energy neutrinos using VERITAS}\/}.
\newblock proc. 34th ICRC The Hague, Netherlands

\bibitem{Schuessler15}
{Sch{\"u}ssler} F., {Balzer} A., {Brun} F. {\em et~al.\/} (2015).
\newblock {\em {The H.E.S.S. multi-messenger program}\/}.
\newblock proc. 34th ICRC The Hague, Netherlands

\bibitem{AdrianMartinez16a}
{Adri{\'a}n-Mart{\'{\i}}nez} S., {Ageron} M., {Albert} A. {\em et~al.\/}
  (2016).
\newblock {\em {Optical and X-ray early follow-up of ANTARES neutrino
  alerts}\/}.
\newblock \jcap, {\bf 2}, 062

\bibitem{Aartsen16}
{IceCube Collaboration}, {Aartsen} M.G., {Abraham} K. {\em et~al.\/} (2016).
\newblock {\em {Very High-Energy Gamma-Ray Follow-Up Program Using Neutrino
  Triggers from IceCube}\/}.
\newblock arXiv:1610.01814

\bibitem{KM3NeT}
{\em {Km3NeT} {T}echnical {D}esign {R}eport\/}.
\newblock \url{http://www.km3net.org/TDR/TDRKM3NeT.pdf}

\bibitem{GVD}
{\em {BAIKAL-GVD Scientific-Technical Report}\/}.
\newblock \url{http://baikalweb.jinr.ru/gvd/BAIKAL-GVD\_En.pdf}

\bibitem{Aartsen15b}
{The IceCube-Gen2 Collaboration}, {:}, {Aartsen} M.G. {\em et~al.\/} (2015).
\newblock {\em {IceCube-Gen2 - The Next Generation Neutrino Observatory at the
  South Pole: Contributions to ICRC 2015}\/}.
\newblock arXiv:1510.05228

\bibitem{AdrianMartinez16b}
{ANTARES Collaboration}, {IceCube Collaboration}, {LIGO Scientific
  Collaboration} {\em et~al.\/} (2016).
\newblock {\em {High-energy Neutrino follow-up search of Gravitational Wave
  Event GW150914 with ANTARES and IceCube}\/}.
\newblock arXiv:1602.05411

\bibitem{Perna16}
{Perna} R., {Lazzati} D. \& {Giacomazzo} B. (2016).
\newblock {\em {Short Gamma-Ray Bursts from the Merger of Two Black Holes}\/}.
\newblock \apjl, {\bf 821}, L18

\bibitem{Lehner14}
{Lehner} L. \& {Pretorius} F. (2014).
\newblock {\em {Numerical Relativity and Astrophysics}\/}.
\newblock \araa, {\bf 52}, 661

\bibitem{Kyutoku14}
{Kyutoku} K., {Ioka} K. \& {Shibata} M. (2014).
\newblock {\em {Ultrarelativistic electromagnetic counterpart to binary neutron
  star mergers}\/}.
\newblock \mnras, {\bf 437}, L6

\bibitem{Tanvir13}
{Tanvir} N.R., {Levan} A.J., {Fruchter} A.S. {\em et~al.\/} (2013).
\newblock {\em {A `kilonova' associated with the short-duration {$\gamma$}-ray
  burst GRB 130603B}\/}.
\newblock \nat, {\bf 500}, 547

\bibitem{Godet14}
{Godet} O., {Nasser} G., {Atteia} J.. {\em et~al.\/} (2014).
\newblock {\em {The x-/gamma-ray camera ECLAIRs for the gamma-ray burst mission
  SVOM}\/}.
\newblock In {\em Society of Photo-Optical Instrumentation Engineers (SPIE)
  Conference Series\/}, volume 9144 of {\em Society of Photo-Optical
  Instrumentation Engineers (SPIE) Conference Series\/}

\bibitem{Finnegan11}
{Finnegan} G. \& {for the VERITAS Collaboration} (2011).
\newblock {\em {Orbit Mode observation Technique Developed for VERITAS}\/}.
\newblock proc. Fermi Symposium 2011

\bibitem{Greiner11}
{Greiner} J., {Kr{\"u}hler} T., {Klose} S. {\em et~al.\/} (2011).
\newblock {\em {The nature of ``dark'' gamma-ray bursts}\/}.
\newblock \aap, {\bf 526}, A30

\bibitem{Bagoly09}
{Bagoly} Z., {Bal{\'a}zs} L.G., {Horv{\'a}th} I. {\em et~al.\/} (2008).
\newblock {\em {Different satellites-different GRB redshift distributions?}\/}.
\newblock In Y.F. {Huang}, Z.G. {Dai} \& B.~{Zhang} (editors), {\em American
  Institute of Physics Conference Series\/}, volume 1065 of {\em American
  Institute of Physics Conference Series\/}, pp. 119--122

\bibitem{Loh16}
{Loh} A., {Corbel} S., {Dubus} G. {\em et~al.\/} (2016).
\newblock {\em {High-energy gamma-ray observations of the accreting black hole
  V404 Cygni during its 2015 June outburst}\/}.
\newblock \mnras, {\bf 462}, L111

\bibitem{Abbott16d}
{Abbott} B.P., {Abbott} R., {Abbott} T.D. {\em et~al.\/} (2016).
\newblock {\em {The Rate of Binary Black Hole Mergers Inferred from Advanced
  LIGO Observations Surrounding GW150914}\/}.
\newblock arXiv:1602.03842

\bibitem{Kneiske04}
{Kneiske} T.M., {Bretz} T., {Mannheim} K. {\em et~al.\/} (2004).
\newblock {\em {Implications of cosmological gamma-ray absorption. II.
  Modification of gamma-ray spectra}\/}.
\newblock \aap, {\bf 413}, 807

\bibitem{Gilmore12}
{Gilmore} R.C., {Somerville} R.S., {Primack} J.R. {\em et~al.\/} (2012).
\newblock {\em {Semi-analytic modelling of the extragalactic background light
  and consequences for extragalactic gamma-ray spectra}\/}.
\newblock \mnras, {\bf 422}, 3189

\bibitem{InoueY13}
{Inoue} Y., {Inoue} S., {Kobayashi} M.A.R. {\em et~al.\/} (2013).
\newblock {\em {Extragalactic Background Light from Hierarchical Galaxy
  Formation: Gamma-Ray Attenuation up to the Epoch of Cosmic Reionization and
  the First Stars}\/}.
\newblock \apj, {\bf 768}, 197

\bibitem{Kohri12}
{Kohri} K., {Ohira} Y. \& {Ioka} K. (2012).
\newblock {\em {Gamma-ray flare and absorption in the Crab nebula: lovely
  TeV-PeV astrophysics}\/}.
\newblock \mnras, {\bf 424}, 2249

\bibitem{Abdo09c}
{Fermi LAT Collaboration}, {Abdo} A.A., {Ackermann} M. {\em et~al.\/} (2009).
\newblock {\em {Modulated High-Energy Gamma-Ray Emission from the Microquasar
  Cygnus X-3}\/}.
\newblock Science, {\bf 326}, 1512

\bibitem{Padovani14b}
{Padovani} P. \& {Resconi} E. (2014).
\newblock {\em {Are both BL Lacs and pulsar wind nebulae the astrophysical
  counterparts of IceCube neutrino events?}\/}.
\newblock \mnras, {\bf 443}, 474

\bibitem{Gaisser90}
{Gaisser} T.K. (1990).
\newblock {\em {Cosmic rays and particle physics, (Cambridge University
  Press)}\/}

\bibitem{Antoni2005}
{Antoni} T., {Apel} W.D., {Badea} A.F. {\em et~al.\/} (2005).
\newblock {\em {KASCADE measurements of energy spectra for elemental groups of
  cosmic rays: Results and open problems}\/}.
\newblock Astroparticle Physics, {\bf 24}, 1

\bibitem{Hillas05}
{Hillas} A.M. (2005).
\newblock {\em {TOPICAL REVIEW: Can diffusive shock acceleration in supernova
  remnants account for high-energy galactic cosmic rays?}\/}.
\newblock Journal of Physics G Nuclear Physics, {\bf 31}, 95

\bibitem{Drury83}
{Drury} L.O. (1983).
\newblock {\em {An introduction to the theory of diffusive shock acceleration
  of energetic particles in tenuous plasmas}\/}.
\newblock Reports on Progress in Physics, {\bf 46}, 973

\bibitem{Bell04}
{Bell} A.R. (2004).
\newblock {\em {Turbulent amplification of magnetic field and diffusive shock
  acceleration of cosmic rays}\/}.
\newblock \mnras, {\bf 353}, 550

\bibitem{Drury94}
{Drury} L.O., {Aharonian} F.A. \& {Voelk} H.J. (1994).
\newblock {\em {The gamma-ray visibility of supernova remnants. A test of
  cosmic ray origin}\/}.
\newblock \aap, {\bf 287}, 959

\bibitem{Ackermann13}
{Ackermann} M., {Ajello} M., {Allafort} A. {\em et~al.\/} (2013).
\newblock {\em {Detection of the Characteristic Pion-Decay Signature in
  Supernova Remnants}\/}.
\newblock Science, {\bf 339}, 807

\bibitem{Aharonian13}
{Aharonian} F.A. (2013).
\newblock {\em {Gamma rays from supernova remnants}\/}.
\newblock Astroparticle Physics, {\bf 43}, 71

\bibitem{Bell13b}
{Bell} A.R., {Schure} K.M., {Reville} B. {\em et~al.\/} (2013).
\newblock {\em {Cosmic-ray acceleration and escape from supernova remnants}\/}.
\newblock \mnras, {\bf 431}, 415

\bibitem{Gabici07}
{Gabici} S. \& {Aharonian} F.A. (2007).
\newblock {\em {Searching for Galactic Cosmic-Ray Pevatrons with Multi-TeV
  Gamma Rays and Neutrinos}\/}.
\newblock \apjl, {\bf 665}, L131

\bibitem{Ellison10}
{Ellison} D.C., {Patnaude} D.J., {Slane} P. {\em et~al.\/} (2010).
\newblock {\em {Efficient Cosmic Ray Acceleration, Hydrodynamics, and
  Self-Consistent Thermal X-Ray Emission Applied to Supernova Remnant RX
  J1713.7-3946}\/}.
\newblock \apj, {\bf 712}, 287

\bibitem{Casanova10}
{Casanova} S., {Jones} D.I., {Aharonian} F.A. {\em et~al.\/} (2010).
\newblock {\em {Modeling the Gamma-Ray Emission Produced by Runaway Cosmic Rays
  in the Environment of RX J1713.7-3946}\/}.
\newblock \pasj, {\bf 62}, 1127

\bibitem{Stecker71}
{Stecker} F.W. (1971).
\newblock {\em {Cosmic gamma rays}\/}.
\newblock NASA Special Publication, {\bf 249}

\bibitem{Dermer86}
{Dermer} C.D. (1986).
\newblock {\em {Secondary production of neutral pi-mesons and the diffuse
  galactic gamma radiation}\/}.
\newblock \aap, {\bf 157}, 223

\bibitem{Aharonian08}
{Aharonian} F., {Akhperjanian} A.G., {Barres de Almeida} U. {\em et~al.\/}
  (2008).
\newblock {\em {Energy Spectrum of Cosmic-Ray Electrons at TeV Energies}\/}.
\newblock Physical Review Letters, {\bf 101}, 26, 261104

\bibitem{Fukui03}
{Fukui} Y., {Moriguchi} Y., {Tamura} K. {\em et~al.\/} (2003).
\newblock {\em {Discovery of Interacting Molecular Gas toward the TeV Gamma-Ray
  Peak of the SNR G 347.3--0.5}\/}.
\newblock \pasj, {\bf 55}, L61

\bibitem{Koyama95}
{Koyama} K., {Petre} R., {Gotthelf} E.V. {\em et~al.\/} (1995).
\newblock {\em {Evidence for shock acceleration of high-energy electrons in the
  supernova remnant SN1006}\/}.
\newblock \nat, {\bf 378}, 255

\bibitem{Enomoto02}
{Enomoto} R., {Tanimori} T., {Naito} T. {\em et~al.\/} (2002).
\newblock {\em {The acceleration of cosmic-ray protons in the supernova remnant
  RX J1713.7-3946}\/}.
\newblock \nat, {\bf 416}, 823

\bibitem{Aharonian04}
{Aharonian} F.A., {Akhperjanian} A.G., {Aye} K.M. {\em et~al.\/} (2004).
\newblock {\em {High-energy particle acceleration in the shell of a supernova
  remnant}\/}.
\newblock \nat, {\bf 432}, 75

\bibitem{Aharonian05}
{Aharonian} F., {Akhperjanian} A.G., {Bazer-Bachi} A.R. {\em et~al.\/} (2005).
\newblock {\em {Detection of TeV {$\gamma$}-ray emission from the shell-type
  supernova remnant RX J0852.0-4622 with HESS}\/}.
\newblock \aap, {\bf 437}, L7

\bibitem{Acciari2011}
{Acciari} V.A., {Aliu} E., {Arlen} T. {\em et~al.\/} (2011).
\newblock {\em {Discovery of TeV Gamma-ray Emission from Tycho's Supernova
  Remnant}\/}.
\newblock \apjl, {\bf 730}, L20

\bibitem{Albert2007}
{Albert} J., {Aliu} E., {Anderhub} H. {\em et~al.\/} (2007).
\newblock {\em {Observation of VHE {$\gamma$}-rays from Cassiopeia A with the
  MAGIC telescope}\/}.
\newblock \aap, {\bf 474}, 937

\bibitem{Abdo11a}
{Abdo} A.A., {Ackermann} M., {Ajello} M. {\em et~al.\/} (2011).
\newblock {\em {Observations of the Young Supernova Remnant RX J1713.7-3946
  with the Fermi Large Area Telescope}\/}.
\newblock \apj, {\bf 734}, 28

\bibitem{Fukui13}
{Fukui} Y. (2013).
\newblock {\em {Molecular and Atomic Gas in the Young TeV {$\gamma$}-Ray SNRs
  RX J1713.7-3946 and RX J0852.0-4622; Evidence for the Hadronic Production of
  {$\gamma$}-Rays}\/}.
\newblock In D.F. {Torres} \& O.~{Reimer} (editors), {\em Cosmic Rays in
  Star-Forming Environments\/}, volume~34 of {\em Advances in Solid State
  Physics\/}, p. 249

\bibitem{Sano14}
{Sano} H., {Fukuda} T., {Yoshiike} S. {\em et~al.\/} (2014).
\newblock {\em {A detailed study of non-thermal X-ray properties and
  interstellar gas toward the $\backslash$gamma-ray supernova remnant RX
  J1713.7-3946}\/}.
\newblock arXiv:1401.7418

\bibitem{Abdalla16a}
{H.~E.~S.~S.~Collaboration}, {Abdalla} H., {Abdalla} H. {\em et~al.\/} (2016).
\newblock {\em {H.E.S.S. observations of RX J1713.7-3946 with improved angular
  and spectral resolution; evidence for gamma-ray emission extending beyond the
  X-ray emitting shell}\/}.
\newblock arXiv:1609.08671

\bibitem{Berezhko09}
{Berezhko} E.G., {P{\"u}hlhofer} G. \& {V{\"o}lk} H.J. (2009).
\newblock {\em {Theory of cosmic ray and {$\gamma$}-ray production in the
  supernova remnant RX J0852.0-4622}\/}.
\newblock \aap, {\bf 505}, 641

\bibitem{Abdalla16b}
{H.~E.~S.~S.~Collaboration}, {Abdalla} H., {Abramowski} A. {\em et~al.\/}
  (2016).
\newblock {\em {Deeper H.E.S.S. Observations of Vela Junior (RX J0852.0-4622):
  Morphology Studies and Resolved Spectroscopy}\/}.
\newblock arXiv:1610.01863

\bibitem{Pedaletti:2013}
{Pedaletti} G., {Torres} D.F., {Gabici} S. {\em et~al.\/} (2013).
\newblock {\em {On the potential of the Cherenkov Telescope Array for the study
  of cosmic-ray diffusion in molecular clouds}\/}.
\newblock \aap, {\bf 550}, A123

\bibitem{Peng16}
{Peng} F.K., {Wang} X.Y., {Liu} R.Y. {\em et~al.\/} (2016).
\newblock {\em {First Detection of GeV Emission from an Ultraluminous Infrared
  Galaxy: Arp 220 as Seen with the Fermi Large Area Telescope}\/}.
\newblock \apjl, {\bf 821}, L20

\bibitem{Griffin16}
{Griffin} R.D., {Dai} X. \& {Thompson} T.A. (2016).
\newblock {\em {Constraining Gamma-Ray Emission from Luminous Infrared Galaxies
  with Fermi-LAT; Tentative Detection of Arp 220}\/}.
\newblock \apjl, {\bf 823}, L17

\bibitem{Kennicutt12}
{Kennicutt} R.C. \& {Evans} N.J. (2012).
\newblock {\em {Star Formation in the Milky Way and Nearby Galaxies}\/}.
\newblock \araa, {\bf 50}, 531

\bibitem{Socrates08}
{Socrates} A., {Davis} S.W. \& {Ramirez-Ruiz} E. (2008).
\newblock {\em {The Eddington Limit in Cosmic Rays: An Explanation for the
  Observed Faintness of Starbursting Galaxies}\/}.
\newblock \apj, {\bf 687}, 202

\bibitem{Jubelgas08}
{Jubelgas} M., {Springel} V., {En{\ss}lin} T. {\em et~al.\/} (2008).
\newblock {\em {Cosmic ray feedback in hydrodynamical simulations of galaxy
  formation}\/}.
\newblock \aap, {\bf 481}, 33

\bibitem{Ceccarelli11}
{Ceccarelli} C., {Hily-Blant} P., {Montmerle} T. {\em et~al.\/} (2011).
\newblock {\em {Supernova-enhanced Cosmic-Ray Ionization and Induced Chemistry
  in a Molecular Cloud of W51C}\/}.
\newblock \apjl, {\bf 740}, L4

\bibitem{Papadopoulos13}
{Papadopoulos} P.P. \& {Thi} W.F. (2013).
\newblock {\em {The Initial Conditions of Star Formation: Cosmic Rays as the
  Fundamental Regulators}\/}.
\newblock In D.F. {Torres} \& O.~{Reimer} (editors), {\em Cosmic Rays in
  Star-Forming Environments\/}, volume~34 of {\em Advances in Solid State
  Physics\/}, p.~41

\bibitem{Booth13}
{Booth} C.M., {Agertz} O., {Kravtsov} A.V. {\em et~al.\/} (2013).
\newblock {\em {Simulations of Disk Galaxies with Cosmic Ray Driven Galactic
  Winds}\/}.
\newblock \apjl, {\bf 777}, L16

\bibitem{Salem14b}
{Salem} M., {Bryan} G.L. \& {Hummels} C. (2014).
\newblock {\em {Cosmological Simulations of Galaxy Formation with Cosmic
  Rays}\/}.
\newblock \apjl, {\bf 797}, L18

\bibitem{Aharonian02}
{Aharonian} F., {Akhperjanian} A., {Beilicke} M. {\em et~al.\/} (2002).
\newblock {\em {An unidentified TeV source in the vicinity of Cygnus OB2}\/}.
\newblock \aap, {\bf 393}, L37

\bibitem{Abramowski12a}
{Abramowski} A., {Acero} F., {Aharonian} F. {\em et~al.\/} (2012).
\newblock {\em {Discovery of extended VHE {$\gamma$}-ray emission from the
  vicinity of the young massive stellar cluster Westerlund 1}\/}.
\newblock \aap, {\bf 537}, A114

\bibitem{Casse80}
{Casse} M. \& {Paul} J.A. (1980).
\newblock {\em {Local gamma rays and cosmic-ray acceleration by supersonic
  stellar winds}\/}.
\newblock \apj, {\bf 237}, 236

\bibitem{Meier15}
Meier D.S., Walter F., Bolatto A.D. {\em et~al.\/} (2015).
\newblock {\em Alma multi-line imaging of the nearby starburst ngc 253\/}.
\newblock The Astrophysical Journal, {\bf 801}, 1, 63

\bibitem{Kennicutt98}
{Kennicutt} Jr. R.C. (1998).
\newblock {\em {Star Formation in Galaxies Along the Hubble Sequence}\/}.
\newblock \araa, {\bf 36}, 189

\bibitem{Ackermann12}
{Ackermann} M., {Ajello} M., {Allafort} A. {\em et~al.\/} (2012).
\newblock {\em {GeV Observations of Star-forming Galaxies with the Fermi Large
  Area Telescope}\/}.
\newblock \apj, {\bf 755}, 164

\bibitem{Abramowski12b}
{Abramowski} A., {Acero} F., {Aharonian} F. {\em et~al.\/} (2012).
\newblock {\em {Spectral Analysis and Interpretation of the {$\gamma$}-Ray
  Emission from the Starburst Galaxy NGC 253}\/}.
\newblock \apj, {\bf 757}, 158

\bibitem{Reitberger15}
{Reitberger} K., {Reimer} A., {Reimer} O. {\em et~al.\/} (2015).
\newblock {\em {The first full orbit of {$\eta$} Carinae seen by Fermi}\/}.
\newblock \aap, {\bf 577}, A100

\bibitem{Preibisch11}
{Preibisch} T., {Ratzka} T., {Kuderna} B. {\em et~al.\/} (2011).
\newblock {\em {Deep wide-field near-infrared survey of the Carina Nebula}\/}.
\newblock \aap, {\bf 530}, A34

\bibitem{Hamaguchi07}
{Hamaguchi} K., {Petre} R., {Matsumoto} H. {\em et~al.\/} (2007).
\newblock {\em {Suzaku Observation of Diffuse X-Ray Emission from the Carina
  Nebula}\/}.
\newblock \pasj, {\bf 59}, 151

\bibitem{Ezoe09}
{Ezoe} Y., {Hamaguchi} K., {Gruendl} R.A. {\em et~al.\/} (2009).
\newblock {\em {Suzaku and XMM-Newton Observations of Diffuse X-Ray Emission
  from the Eastern Tip Region of the Carina Nebula}\/}.
\newblock \pasj, {\bf 61}, 123

\bibitem{Townsley11}
{Townsley} L.K., {Broos} P.S., {Chu} Y.H. {\em et~al.\/} (2011).
\newblock {\em {The Chandra Carina Complex Project: Deciphering the Enigma of
  Carina's Diffuse X-ray Emission}\/}.
\newblock \apjs, {\bf 194}, 15

\bibitem{Abramowski12g}
{HESS Collaboration}, {Abramowski} A., {Acero} F. {\em et~al.\/} (2012).
\newblock {\em {HESS observations of the Carina nebula and its enigmatic
  colliding wind binary Eta Carinae}\/}.
\newblock \mnras, {\bf 424}, 128

\bibitem{Abramowski11}
{Abramowski} A., {Acero} F., {Aharonian} F. {\em et~al.\/} (2011).
\newblock {\em {Revisiting the Westerlund 2 field with the HESS telescope
  array}\/}.
\newblock \aap, {\bf 525}, A46+

\bibitem{Bartoli14}
{Bartoli} B., {Bernardini} P., {Bi} X.J. {\em et~al.\/} (2014).
\newblock {\em {Identification of the TeV Gamma-Ray Source ARGO J2031+4157 with
  the Cygnus Cocoon}\/}.
\newblock \apj, {\bf 790}, 152

\bibitem{Popkow15}
{Popkow} A. \& {for the VERITAS Collaboration} (2015).
\newblock {\em {The VERITAS Survey of the Cygnus Region of the Galaxy}\/}.
\newblock arXiv:1508.06684

\bibitem{Kothes07}
{Kothes} R. \& {Dougherty} S.M. (2007).
\newblock {\em {The distance and neutral environment of the massive stellar
  cluster Westerlund 1}\/}.
\newblock \aap, {\bf 468}, 993

\bibitem{Ohm13a}
{Ohm} S., {Hinton} J.A. \& {White} R. (2013).
\newblock {\em {{$\gamma$}-ray emission from the Westerlund 1 region}\/}.
\newblock \mnras, {\bf 434}, 2289

\bibitem{Abdo10f}
{Abdo} A.A., {Ackermann} M., {Ajello} M. {\em et~al.\/} (2010).
\newblock {\em {Fermi Large Area Telescope observations of Local Group
  galaxies: detection of M 31 and search for M 33}\/}.
\newblock \aap, {\bf 523}, L2

\bibitem{Dorfi12}
{Dorfi} E.A. \& {Breitschwerdt} D. (2012).
\newblock {\em {Time-dependent galactic winds. I. Structure and evolution of
  galactic outflows accompanied by cosmic ray acceleration}\/}.
\newblock \aap, {\bf 540}, A77

\bibitem{Voelk89}
{V\"olk} H.J., {Klein} U. \& {Wielebinski} R. (1989).
\newblock {\em {M82, the Galaxy, and the dependence of cosmic ray energy
  production on the supernova rate}\/}.
\newblock \aap, {\bf 213}, L12

\bibitem{Acciari09}
{VERITAS Collaboration}, {Acciari} V.A., {Aliu} E. {\em et~al.\/} (2009).
\newblock {\em {A connection between star formation activity and cosmic rays in
  the starburst galaxy M82}\/}.
\newblock \nat, {\bf 462}, 770

\bibitem{Abdo10}
{Abdo} A.A., {Ackermann} M., {Ajello} M. {\em et~al.\/} (2010).
\newblock {\em {Detection of Gamma-Ray Emission from the Starburst Galaxies M82
  and NGC 253 with the Large Area Telescope on Fermi}\/}.
\newblock \apjl, {\bf 709}, L152

\bibitem{Domingo05}
{Domingo-Santamar{\'{\i}}a} E. \& {Torres} D.F. (2005).
\newblock {\em {High energy {$\gamma$}-ray emission from the starburst nucleus
  of NGC 253}\/}.
\newblock \aap, {\bf 444}, 403

\bibitem{Rephaeli10}
{Rephaeli} Y., {Arieli} Y. \& {Persic} M. (2010).
\newblock {\em {High-energy emission from the starburst galaxy NGC 253}\/}.
\newblock \mnras, {\bf 401}, 473

\bibitem{Thompson06}
{Thompson} T.A., {Quataert} E., {Waxman} E. {\em et~al.\/} (2006).
\newblock {\em {Magnetic Fields in Starburst Galaxies and the Origin of the
  FIR-Radio Correlation}\/}.
\newblock \apj, {\bf 645}, 186

\bibitem{Lacki11}
{Lacki} B.C., {Thompson} T.A., {Quataert} E. {\em et~al.\/} (2011).
\newblock {\em {On the GeV and TeV Detections of the Starburst Galaxies M82 and
  NGC 253}\/}.
\newblock \apj, {\bf 734}, 107

\bibitem{Murphy12}
{Murphy} E.J., {Porter} T.A., {Moskalenko} I.V. {\em et~al.\/} (2012).
\newblock {\em {Characterizing Cosmic-Ray Propagation in Massive Star-forming
  Regions: The Case of 30 Doradus and the Large Magellanic Cloud}\/}.
\newblock \apj, {\bf 750}, 126

\bibitem{Persic08}
{Persic} M., {Rephaeli} Y. \& {Arieli} Y. (2008).
\newblock {\em {Very-high-energy emission from M 82}\/}.
\newblock \aap, {\bf 486}, 143

\bibitem{Torres12}
{Torres} D.F., {Cillis} A., {Lacki} B. {\em et~al.\/} (2012).
\newblock {\em {Building up the spectrum of cosmic rays in star-forming
  regions}\/}.
\newblock \mnras, {\bf 423}, 822

\bibitem{Mannheim12}
{Mannheim} K., {Els{\"a}sser} D. \& {Tibolla} O. (2012).
\newblock {\em {Gamma-rays from pulsar wind nebulae in starburst galaxies}\/}.
\newblock Astroparticle Physics, {\bf 35}, 797

\bibitem{Ohm13b}
{Ohm} S. \& {Hinton} J.A. (2013).
\newblock {\em {Non-thermal emission from pulsar-wind nebulae in starburst
  galaxies}\/}.
\newblock \mnras, {\bf 429}, L70

\bibitem{Sanders96}
{Sanders} D.B. \& {Mirabel} I.F. (1996).
\newblock {\em {Luminous Infrared Galaxies}\/}.
\newblock \araa, {\bf 34}, 749

\bibitem{Pavlidou02}
{Pavlidou} V. \& {Fields} B.D. (2002).
\newblock {\em {The Guaranteed Gamma-Ray Background}\/}.
\newblock \apjl, {\bf 575}, L5

\bibitem{Smith98}
{Smith} H.E., {Lonsdale} C.J., {Lonsdale} C.J. {\em et~al.\/} (1998).
\newblock {\em {A Starburst Revealed---Luminous Radio Supernovae in the Nuclei
  of ARP 220}\/}.
\newblock \apjl, {\bf 493}, L17

\bibitem{Torres04}
{Torres} D.F. (2004).
\newblock {\em {Theoretical Modeling of the Diffuse Emission of Gamma Rays from
  Extreme Regions of Star Formation: The Case of ARP 220}\/}.
\newblock \apj, {\bf 617}, 966

\bibitem{Torres05}
{Torres} D.F. \& {Domingo-Santamar{\'{\i}}a} E. (2005).
\newblock {\em {Some Comments on the High Energy Emission from Regions of Star
  Formation Beyond the Galaxy}\/}.
\newblock Modern Physics Letters A, {\bf 20}, 2827

\bibitem{Albert07a}
{Albert} J., {Aliu} E., {Anderhub} H. {\em et~al.\/} (2007).
\newblock {\em {First Bounds on the Very High Energy {$\gamma$}-Ray Emission
  from Arp 220}\/}.
\newblock \apj, {\bf 658}, 245

\bibitem{Fleischhack15}
{{Fleischhack}, H. and {for the VERITAS Collaboration}} (2015).
\newblock {\em {Upper limits on the VHE gamma-ray flux from the ULIRG Arp 220
  and other galaxies with VERITAS}\/}.
\newblock proc. 34th ICRC, The Hague, Netherlands

\bibitem{Knodlseder16}
{Kn{\"o}dlseder} J., {Mayer} M., {Deil} C. {\em et~al.\/} (2016).
\newblock {\em {GammaLib and ctools. A software framework for the analysis of
  astronomical gamma-ray data}\/}.
\newblock \aap, {\bf 593}, A1

\bibitem{Inoue11}
{Inoue} Y. (2011).
\newblock {\em {High Energy Gamma-ray Absorption and Cascade Emission in Nearby
  Starburst Galaxies}\/}.
\newblock \apj, {\bf 728}, 11

\bibitem{Armstrong16}
Armstrong T., Brown A.M., Chadwick P.M. {\em et~al.\/} (2017).
\newblock {\em {DBSCAN re-applied to Pass 8 Fermi-LAT data above 100 GeV}\/}.
\newblock AIP Conf. Proc., {\bf 1792}, 1, 070001

\bibitem{Zech2013}
Zech A., Cerruti M. \& for~the CTA~consortium (2013).
\newblock {\em {Signatures of relativistic protons in CTA blazar spectra}\/}.
\newblock Proc. of the 33rd ICRC, Rio de Janeiro, Brazil.
\newblock Astro-ph/1307.2232

\bibitem{Cerruti2015}
{Cerruti} M., {Zech} A., {Boisson} C. {\em et~al.\/} (2015).
\newblock {\em {A hadronic origin for ultra-high-frequency-peaked BL Lac
  objects}\/}.
\newblock \mnras, {\bf 448}, 910

\bibitem{Zech2016}
{Zech} A., {Cerruti} M. \& {Mazin} D. (2017).
\newblock {\em {Expected signatures from hadronic emission processes in the TeV
  spectra of BL Lacertae objects}\/}.
\newblock \aap, {\bf 602}, A25

\bibitem{Sol2013igmf}
{Sol} H., {Zech} A., {Boisson} C. {\em et~al.\/} (2013).
\newblock {\em {Prospect on intergalactic magnetic field measurements with
  gamma-ray instruments}\/}.
\newblock In A.G. {Kosovichev}, E.~{de Gouveia Dal Pino} \& Y.~{Yan} (editors),
  {\em Solar and Astrophysical Dynamos and Magnetic Activity\/}, volume 294 of
  {\em IAU Symposium\/}, pp. 459--470

\bibitem{Hardcastle11}
{Hardcastle} M.J. \& {Croston} J.H. (2011).
\newblock {\em {Modelling TeV {$\gamma$}-ray emission from the kiloparsec-scale
  jets of Centaurus A and M87}\/}.
\newblock \mnras, {\bf 415}, 133

\bibitem{Franceschini2008}
{Franceschini} A., {Rodighiero} G. \& {Vaccari} M. (2008).
\newblock {\em {Extragalactic optical-infrared background radiation, its time
  evolution and the cosmic photon-photon opacity}\/}.
\newblock \aap, {\bf 487}, 837

\bibitem{Abramowski2013a}
{H.E.S.S.~Collaboration}, {Abramowski} A., {Acero} F. {\em et~al.\/} (2013).
\newblock {\em {Measurement of the extragalactic background light imprint on
  the spectra of the brightest blazars observed with H.E.S.S.}\/}.
\newblock \aap, {\bf 550}, A4

\bibitem{Ackermann2012}
Ackermann M. {\em et~al.\/} (2012).
\newblock {\em {The Imprint of the Extragalactic Background Light in the
  Gamma-Ray Spectra of Blazars}\/}.
\newblock Science, {\bf 338}, 1190

\bibitem{Blandford1977}
{Blandford} R.D. \& {Znajek} R.L. (1977).
\newblock {\em {Electromagnetic extraction of energy from Kerr black holes}\/}.
\newblock \mnras, {\bf 179}, 433

\bibitem{Ghisellini2014}
{Ghisellini} G., {Tavecchio} F., {Maraschi} L. {\em et~al.\/} (2014).
\newblock {\em {The power of relativistic jets is larger than the luminosity of
  their accretion disks}\/}.
\newblock \nat, {\bf 515}, 376

\bibitem{Ghisellini2010}
{Ghisellini} G., {Tavecchio} F., {Foschini} L. {\em et~al.\/} (2010).
\newblock {\em {General physical properties of bright Fermi blazars}\/}.
\newblock \mnras, {\bf 402}, 497

\bibitem{Boettcher2013}
{B{\"o}ttcher} M., {Reimer} A., {Sweeney} K. {\em et~al.\/} (2013).
\newblock {\em {Leptonic and Hadronic Modeling of Fermi-detected Blazars}\/}.
\newblock \apj, {\bf 768}, 54

\bibitem{Cerruti2013}
{Cerruti} M., {Dermer} C.D., {Lott} B. {\em et~al.\/} (2013).
\newblock {\em {Gamma-Ray Blazars near Equipartition and the Origin of the GeV
  Spectral Break in 3C 454.3}\/}.
\newblock \apjl, {\bf 771}, L4

\bibitem{Dermer2014}
{Dermer} C.D., {Cerruti} M., {Lott} B. {\em et~al.\/} (2014).
\newblock {\em {Equipartition Gamma-Ray Blazars and the Location of the
  Gamma-Ray Emission Site in 3C 279}\/}.
\newblock \apj, {\bf 782}, 82

\bibitem{Poutanen2010}
Poutanen J. \& Stern B. (2010).
\newblock {\em {GeV breaks in blazars as a result of gamma-ray absorption
  within the broad-line region}\/}.
\newblock ApJ Lett, {\bf 717}, L118

\bibitem{Senturk2013}
Senturk G.D., Errando M., Boettcher M. {\em et~al.\/} (2013).
\newblock {\em Gamma-ray observational properties of tev-detected blazars\/}.
\newblock ApJ, {\bf 764}, 119

\bibitem{Brown2013}
{Brown} A.M. (2013).
\newblock {\em {Locating the {$\gamma$}-ray emission region of the flat
  spectrum radio quasar PKS 1510-089}\/}.
\newblock \mnras, {\bf 431}, 824

\bibitem{Abeysekara2015}
{Abeysekara} A.U., {Archambault} S., {Archer} A. {\em et~al.\/} (2015).
\newblock {\em {Gamma-Rays from the Quasar PKS 1441+25: Story of an Escape}\/}.
\newblock \apjl, {\bf 815}, L22

\bibitem{Lindfors2013}
Lindfors E., Nilsson K., Barres~de Almeida U. {\em et~al.\/} (2013).
\newblock {\em {VHE gamma-ray emission from the FSRQs observed by the MAGIC
  telescopes}\/}.
\newblock eConf C121028.
\newblock Proc. of the 2012 Fermi Symposium - eConf C121028, astro-ph/1303.2102

\bibitem{Abdo10b}
{Abdo} A.A., {Ackermann} M., {Ajello} M. {\em et~al.\/} (2010).
\newblock {\em {Spectral Properties of Bright Fermi-Detected Blazars in the
  Gamma-Ray Band}\/}.
\newblock \apj, {\bf 710}, 1271

\bibitem{Costamante2001}
{Costamante} L., {Ghisellini} G., {Giommi} P. {\em et~al.\/} (2001).
\newblock {\em {Extreme synchrotron BL Lac objects. Stretching the blazar
  sequence}\/}.
\newblock \aap, {\bf 371}, 512

\bibitem{Bonnoli2015}
{Bonnoli} G., {Tavecchio} F., {Ghisellini} G. {\em et~al.\/} (2015).
\newblock {\em {An emerging population of BL Lacs with extreme properties:
  towards a class of EBL and cosmic magnetic field probes?}\/}.
\newblock \mnras, {\bf 451}, 611

\bibitem{Katarzynski2006}
{Katarzy{\'n}ski} K., {Ghisellini} G., {Tavecchio} F. {\em et~al.\/} (2006).
\newblock {\em {Hard TeV spectra of blazars and the constraints to the infrared
  intergalactic background}\/}.
\newblock \mnras, {\bf 368}, L52

\bibitem{Lefa2011}
{Lefa} E., {Rieger} F.M. \& {Aharonian} F. (2011).
\newblock {\em {Formation of Very Hard Gamma-Ray Spectra of Blazars in Leptonic
  Models}\/}.
\newblock \apj, {\bf 740}, 64

\bibitem{Asano2014}
{Asano} K., {Takahara} F., {Kusunose} M. {\em et~al.\/} (2014).
\newblock {\em {Time-dependent Models for Blazar Emission with the Second-order
  Fermi Acceleration}\/}.
\newblock \apj, {\bf 780}, 64

\bibitem{Murase2012}
Murase K., Dermer C.D., Takami H. {\em et~al.\/} (2012).
\newblock {\em {Blazars as Ultra-high-energy Cosmic-ray Sources: Implications
  for TeV Gamma-Ray Observations}\/}.
\newblock ApJ, {\bf 749}, 63

\bibitem{Biteau2012}
{Biteau} J. \& {Giebels} B. (2012).
\newblock {\em {The minijets-in-a-jet statistical model and the rms-flux
  correlation}\/}.
\newblock \aap, {\bf 548}, A123

\bibitem{McHardy2011}
McHardy I. (2011).
\newblock {\em The origin of high energy variability in blazars\/}.
\newblock PoS(AGN 2011)017.
\newblock Proc. of the ``AGN Physics in the CTA Era" workshop, Toulouse, France

\bibitem{Marscher2014}
Marscher A.P. (2014).
\newblock {\em Turbulent, extreme multi-zone model for simulating flux and
  polarization variability in blazars\/}.
\newblock ApJ, {\bf 780}, 87

\bibitem{deGouveia2010}
{de Gouveia Dal Pino} E.M., {Piovezan} P.P. \& {Kadowaki} L.H.S. (2010).
\newblock {\em {The role of magnetic reconnection on jet/accretion disk
  systems}\/}.
\newblock \aap, {\bf 518}, A5

\bibitem{Giannios2013}
{Giannios} D. (2013).
\newblock {\em {Reconnection-driven plasmoids in blazars: fast flares on a slow
  envelope}\/}.
\newblock \mnras, {\bf 431}, 355

\bibitem{Kadowaki15}
{Kadowaki} L.H.S., {de Gouveia Dal Pino} E.M. \& {Singh} C.B. (2015).
\newblock {\em {The Role of Fast Magnetic Reconnection on the Radio and
  Gamma-ray Emission from the Nuclear Regions of Microquasars and Low
  Luminosity AGNs}\/}.
\newblock \apj, {\bf 802}, 113

\bibitem{Singh15}
{Singh} C.B., {de Gouveia Dal Pino} E.M. \& {Kadowaki} L.H.S. (2015).
\newblock {\em {On the Role of Fast Magnetic Reconnection in Accreting Black
  Hole Sources}\/}.
\newblock \apjl, {\bf 799}, L20

\bibitem{Khiali15}
{Khiali} B., {de Gouveia Dal Pino} E.M. \& {Sol} H. (2015).
\newblock {\em {Particle Acceleration and gamma-ray emission due to magnetic
  reconnection around the core region of radio galaxies}\/}.
\newblock arXiv:1504.07592

\bibitem{Osmanov2010}
{Osmanov} Z. (2010).
\newblock {\em On the simultaneous generation of high energy emission and
  submillimeter/infrared radiation from active galactic nuclei\/}.
\newblock \apj, {\bf 721}, 318

\bibitem{Levinson2011}
Levinson A. \& Rieger F.M. (2011).
\newblock {\em {Variable TeV emission as a manifestation of jet formation in
  M87?}\/}.
\newblock ApJ, {\bf 730}, 123

\bibitem{Aleksic14a}
{Aleksi{\'c}} J., {Antonelli} L.A., {Antoranz} P. {\em et~al.\/} (2014).
\newblock {\em {Rapid and multiband variability of the TeV bright active
  nucleus of the galaxy IC 310}\/}.
\newblock \aap, {\bf 563}, A91

\bibitem{Khiali15b}
{Khiali} B., {de Gouveia Dal Pino} E.M. \& {del Valle} M.V. (2015).
\newblock {\em {A magnetic reconnection model for explaining the
  multiwavelength emission of the microquasars Cyg X-1 and Cyg X-3}\/}.
\newblock \mnras, {\bf 449}, 34

\bibitem{Abdo10c}
{Abdo} A.A., {Ackermann} M., {Ajello} M. {\em et~al.\/} (2010).
\newblock {\em {Fermi Gamma-Ray Imaging of a Radio Galaxy}\/}.
\newblock Science, {\bf 328}, 725

\bibitem{Ackermann2016}
{Ackermann} M., {Ajello} M., {Baldini} L. {\em et~al.\/} (2016).
\newblock {\em {Fermi Large Area Telescope Detection of Extended Gamma-Ray
  Emission from the Radio Galaxy Fornax A}\/}.
\newblock \apj, {\bf 826}, 1

\bibitem{Acciari09b}
{Acciari} V.A., {Aliu} E., {Arlen} T. {\em et~al.\/} (2009).
\newblock {\em {Radio Imaging of the Very-High-Energy {$\gamma$}-Ray Emission
  Region in the Central Engine of a Radio Galaxy}\/}.
\newblock Science, {\bf 325}, 444

\bibitem{Cheung07}
{Cheung} C.C., {Harris} D.E. \& {Stawarz} L. (2007).
\newblock {\em {Superluminal Radio Features in the M87 Jet and the Site of
  Flaring TeV Gamma-Ray Emission}\/}.
\newblock \apjl, {\bf 663}, L65

\bibitem{Abdo10e}
{Abdo} A.A., {Ackermann} M., {Ajello} M. {\em et~al.\/} (2010).
\newblock {\em {Fermi Large Area Telescope View of the Core of the Radio Galaxy
  Centaurus A}\/}.
\newblock \apj, {\bf 719}, 1433

\bibitem{Sahakyan2013}
{Sahakyan} N., {Yang} R., {Aharonian} F.A. {\em et~al.\/} (2013).
\newblock {\em {Evidence for a Second Component in the High-energy Core
  Emission from Centaurus A?}\/}.
\newblock \apjl, {\bf 770}, L6

\bibitem{Brown2016}
{Brown} A.M., {B{\AA}`hm} C., {Graham} J. {\em et~al.\/} (2017).
\newblock {\em {Discovery of a new extragalactic population of energetic
  particles}\/}.
\newblock \prd, {\bf 95}, 6, 063018

\bibitem{Sahu2012}
{Sahu} S., {Zhang} B. \& {Fraija} N. (2012).
\newblock {\em {Hadronic-origin TeV gamma-rays and ultrahigh energy cosmic rays
  from Centaurus A}\/}.
\newblock \prd, {\bf 85}, 4, 043012

\bibitem{Petropoulou2014}
{Petropoulou} M., {Lefa} E., {Dimitrakoudis} S. {\em et~al.\/} (2014).
\newblock {\em {One-zone synchrotron self-Compton model for the core emission
  of Centaurus A revisited}\/}.
\newblock \aap, {\bf 562}, A12

\bibitem{Cerruti2016}
{Cerruti} M., {Zech} A., {Emery} G. {\em et~al.\/} (2016).
\newblock {\em {Hadronic modeling of TeV AGN: gammas and neutrinos}\/}.
\newblock arXiv:1610.00255

\bibitem{Abdo09dd}
{Abdo} A., {Ackermann} M., {Ajello} M. {\em et~al.\/} (2009).
\newblock {\em {Radio-Loud Narrow-Line Seyfert 1 as a New Class of Gamma-Ray
  Active Galactic Nuclei}\/}.
\newblock \apjl, {\bf 707}, L142

\bibitem{DAmmando2012}
{D'Ammando} F., {Orienti} M., {Finke} J. {\em et~al.\/} (2012).
\newblock {\em {SBS 0846+513: a new {$\gamma$}-ray-emitting narrow-line Seyfert
  1 galaxy}\/}.
\newblock \mnras, {\bf 426}, 317

\bibitem{Foschini2011}
Foschini L. {\em et~al.\/} (2011).
\newblock {\em {The July 2010 outburst of the NLS1 PMN J0948+0022}\/}.
\newblock proc. of the 3rd Fermi symposium, Rome, Italy.
\newblock Astro-ph/1110.5649

\bibitem{DAmmando2013}
{D'Ammando} F., {Tosti} G., {Orienti} M. {\em et~al.\/} (2013).
\newblock {\em {Four Years of Fermi LAT Observations of Narrow-Line Seyfert 1
  Galaxies}\/}.
\newblock arXiv:1303.3030

\bibitem{Marconi2008}
Marconi A. {\em et~al.\/} (2008).
\newblock {\em Weighing black holes from zero to high redshift\/}.
\newblock ApJ, {\bf 678}, 693

\bibitem{Calderone2013}
{Calderone} G., {Ghisellini} G., {Colpi} M. {\em et~al.\/} (2013).
\newblock {\em Black hole mass estimate for a sample of radio-loud narrow-line
  seyfert 1 galaxies\/}.
\newblock \mnras, {\bf 431}, 210

\bibitem{Neronov07}
{Neronov} A. \& {Aharonian} F.A. (2007).
\newblock {\em {Production of TeV Gamma Radiation in the Vicinity of the
  Supermassive Black Hole in the Giant Radio Galaxy M87}\/}.
\newblock \apj, {\bf 671}, 85

\bibitem{Rieger2008}
Rieger F.M. \& Aharonian F.A. (2008).
\newblock {\em {Variable VHE gamma-ray emission from non-blazar AGNs}\/}.
\newblock \aap, {\bf 479}, L5

\bibitem{Istomin2009}
Istomin Y.N. \& Sol H. (2009).
\newblock {\em Acceleration of particles in the vicinity of a massive black
  hole\/}.
\newblock Ap\&SS, {\bf 321}, 57

\bibitem{Biteau2015}
{Biteau} J. \& {Williams} D.A. (2015).
\newblock {\em {The Extragalactic Background Light, the Hubble Constant, and
  Anomalies: Conclusions from 20 Years of TeV Gamma-ray Observations}\/}.
\newblock \apj, {\bf 812}, 60

\bibitem{Dominguez2013}
{Dom{\'{\i}}nguez} A. \& {Prada} F. (2013).
\newblock {\em {Measurement of the Expansion Rate of the Universe from
  {$\gamma$}-Ray Attenuation}\/}.
\newblock \apjl, {\bf 771}, L34

\bibitem{Widrow2002}
Widrow L.M. (2002).
\newblock {\em Origin of galactic and extragalactic magnetic fields\/}.
\newblock Reviews of Modern Physics, {\bf 74}, 775

\bibitem{Kulsrud2008}
Kulsrud R.M. \& Zweibel E.G. (2008).
\newblock {\em On the origin of cosmic magnetic fields\/}.
\newblock Rep. Prog. Phys., {\bf 71}, 4, 046901

\bibitem{Kandus2011}
Kandus A., Kunze K.E. \& Tsagas C.G. (2011).
\newblock {\em Primordial magnetogenesis\/}.
\newblock Phys. Rep., {\bf 505}, 1, 1

\bibitem{Widrow2012}
{Widrow} L.M., {Ryu} D., {Schleicher} D.R.G. {\em et~al.\/} (2012).
\newblock {\em The first magnetic fields\/}.
\newblock Space Sci. Rev., {\bf 166}, 37

\bibitem{Ryu2012}
{Ryu} D., {Schleicher} D.R.G., {Treumann} R.A. {\em et~al.\/} (2012).
\newblock {\em {Magnetic Fields in the Large-Scale Structure of the
  Universe}\/}.
\newblock Space Sci. Rev, {\bf 166}, 1

\bibitem{Kim1989}
{Kim} K.T., {Kronberg} P.P., {Giovannini} G. {\em et~al.\/} (1989).
\newblock {\em {Discovery of intergalactic radio emission in the Coma-A1367
  supercluster}\/}.
\newblock \nat, {\bf 341}, 720

\bibitem{Elyiv2009}
{Elyiv} A., {Neronov} A. \& {Semikoz} D.V. (2009).
\newblock {\em {Gamma-ray induced cascades and magnetic fields in the
  intergalactic medium}\/}.
\newblock \prd, {\bf 80}, 2, 023010

\bibitem{Dermer2011}
{Dermer} C.D., {Cavadini} M., {Razzaque} S. {\em et~al.\/} (2011).
\newblock {\em {Time Delay of Cascade Radiation for TeV Blazars and the
  Measurement of the Intergalactic Magnetic Field}\/}.
\newblock \apjl, {\bf 733}, L21

\bibitem{Chen15}
{Chen} W., {Buckley} J.H. \& {Ferrer} F. (2015).
\newblock {\em {Search for GeV {$\gamma$} -Ray Pair Halos Around Low Redshift
  Blazars}\/}.
\newblock Physical Review Letters, {\bf 115}, 21, 211103

\bibitem{Barkov2010}
{Barkov} M.V., {Aharonian} F.A., {Bogovalov} S.V. {\em et~al.\/} (2012).
\newblock {\em {Rapid TeV Variability in Blazars as a Result of Jet-Star
  Interaction}\/}.
\newblock \apj, {\bf 749}, 119

\bibitem{Essey2010}
Essey W. \& Kusenko A. (2010).
\newblock {\em A new interpretation of the gamma-ray observations of distant
  active galactic nuclei\/}.
\newblock Astroparticle Physics, {\bf 33}, 81

\bibitem{Essey2011}
Essey W., Kalashev O., Kusenko A. {\em et~al.\/} (2011).
\newblock {\em Role of line-of-sight cosmic-ray interactions in forming the
  spectra of distant blazars in tev gamma rays and high-energy neutrinos\/}.
\newblock \apj, {\bf 731}, 51

\bibitem{Takami2013}
{Takami} H., {Murase} K. \& {Dermer} C.D. (2013).
\newblock {\em {Disentangling Hadronic and Leptonic Cascade Scenarios from the
  Very-high-energy Gamma-Ray Emission of Distant Hard-spectrum Blazars}\/}.
\newblock \apjl, {\bf 771}, L32

\bibitem{Abdo11c}
{Abdo} A.A., {Ackermann} M., {Ajello} M. {\em et~al.\/} (2011).
\newblock {\em {Fermi Large Area Telescope Observations of Markarian 421: The
  Missing Piece of its Spectral Energy Distribution}\/}.
\newblock \apj, {\bf 736}, 131

\bibitem{Joshi2013}
Joshi J.C. \& Gupta N. (2013).
\newblock {\em {Testing hadronic models of gamma ray production at the core of
  Cen A}\/}.
\newblock Phys. Rev. D, {\bf 87}, 2, 023002

\bibitem{Rieger2009}
Rieger F.M. \& Aharonian F.A. (2009).
\newblock {\em {Centaurus A as TeV gamma-ray and possible UHE cosmic-ray
  source}\/}.
\newblock \aap, {\bf 506}, L41

\bibitem{Yang2012}
Yang R., Sahakyan N., de~Ona~Wilhelmi E. {\em et~al.\/} (2012).
\newblock {\em {Deep observation of the giant radio lobes of Centaurus A with
  the Fermi Large Area Telescope}\/}.
\newblock A\&A, {\bf 542}, 19

\bibitem{deAngelis2007}
{de Angelis} A., {Roncadelli} M. \& {Mansutti} O. (2007).
\newblock {\em {Evidence for a new light spin-zero boson from cosmological
  gamma-ray propagation?}\/}.
\newblock \prd, {\bf 76}, 12, 121301

\bibitem{Simet2008}
Simet M., Hooper D. \& Serpico P.D. (2008).
\newblock {\em Milky way as a kiloparsec-scale axionscope\/}.
\newblock Phys. Rev. D, {\bf 77}, 6, 063001

\bibitem{Sanchez2009}
{S{\'a}nchez-Conde} M.A., {Paneque} D., {Bloom} E. {\em et~al.\/} (2009).
\newblock {\em {Hints of the existence of axionlike particles from the
  gamma-ray spectra of cosmological sources}\/}.
\newblock \prd, {\bf 79}, 12, 123511

\bibitem{deAngelis2009}
{de Angelis} A., {Mansutti} O., {Persic} M. {\em et~al.\/} (2009).
\newblock {\em {Photon propagation and the very high energy {$\gamma$}-ray
  spectra of blazars: how transparent is the Universe?}\/}.
\newblock \mnras, {\bf 394}, L21

\bibitem{deAngelis2011}
{de Angelis} A., {Galanti} G. \& {Roncadelli} M. (2011).
\newblock {\em {Relevance of axionlike particles for very-high-energy
  astrophysics}\/}.
\newblock \prd, {\bf 84}, 10, 105030

\bibitem{Dominguez2011}
{Dom{\'{\i}}nguez} A., {S{\'a}nchez-Conde} M.A. \& {Prada} F. (2011).
\newblock {\em {Axion-like particle imprint in cosmological very-high-energy
  sources}\/}.
\newblock \jcap, {\bf 11}, 020

\bibitem{deAngelis2013}
{De Angelis} A., {Galanti} G. \& {Roncadelli} M. (2013).
\newblock {\em {Transparency of the Universe to gamma-rays}\/}.
\newblock \mnras, {\bf 432}, 3245

\bibitem{Galanti2015}
{Galanti} G., {Roncadelli} M., {De Angelis} A. {\em et~al.\/} (2015).
\newblock {\em {Axion-like particles explain the unphysical redshift-dependence
  of AGN gamma-ray spectra}\/}.
\newblock arXiv:1503.04436

\bibitem{Horns2012}
Horns D. \& Meyer M. (2012).
\newblock {\em {Indications for a pair-production anomaly from the propagation
  of VHE gamma-rays}\/}.
\newblock J. Cosmology Astropart. Phys., {\bf 2}, 33

\bibitem{Horns2013}
{Horns} D. \& {Meyer} M. (2013).
\newblock {\em Pair-production opacity at high and very-high gamma-ray
  energies\/}.
\newblock DESY-PROC-2013-04.
\newblock Astro-ph/1309.3846

\bibitem{Rubtsov2014}
Rubtsov G.I. \& Troitsky S.V. (2014).
\newblock {\em Breaks in gamma-ray spectra of distant blazars and transparency
  of the universe\/}.
\newblock Soviet Journal of Experimental and Theoretical Physics Letters, {\bf
  100}, 355

\bibitem{Sanchez:2013lla}
Sanchez D.A., Fegan S. \& Giebels B. (2013).
\newblock {\em {Evidence for a cosmological effect in gamma-ray spectra of BL
  Lacs}\/}.
\newblock Astron. Astrophys., {\bf 554}, A75

\bibitem{dominguez2015}
{Dom{\'{\i}}nguez} A. \& {Ajello} M. (2015).
\newblock {\em {Spectral Analysis of Fermi-LAT Blazars above 50 GeV}\/}.
\newblock \apjl, {\bf 813}, L34

\bibitem{Horns2012a}
{Horns} D., {Maccione} L., {Meyer} M. {\em et~al.\/} (2012).
\newblock {\em {Hardening of TeV gamma spectrum of active galactic nuclei in
  galaxy clusters by conversions of photons into axionlike particles}\/}.
\newblock \prd, {\bf 86}, 7, 075024

\bibitem{Meyer2014}
Meyer M., Montanino D. \& Conrad J. (2014).
\newblock {\em On detecting oscillations of gamma rays into axion-like
  particles in turbulent and coherent magnetic fields\/}.
\newblock JCAP, {\bf 9}, 3, 003

\bibitem{Tavecchio2014}
Tavecchio F., Roncadelli M. \& Galanti G. (2015).
\newblock {\em Photons to axion-like particles conversion in active galactic
  nuclei\/}.
\newblock Physics Letters B, {\bf 744}, 375

\bibitem{Tavecchio2012}
Tavecchio F., Roncadelli M., Galanti G. {\em et~al.\/} (2012).
\newblock {\em Evidence for an axion-like particle from pks 1222+216?\/}.
\newblock Phys. Rev. D, {\bf 86}, 8, 085036

\bibitem{Meyer2014a}
{Meyer} M. \& {Conrad} J. (2014).
\newblock {\em {Sensitivity of the Cherenkov Telescope Array to the detection
  of axion-like particles at high gamma-ray opacities}\/}.
\newblock \jcap, {\bf 12}, 016

\bibitem{Abramowski2013b}
{Abramowski} A., {Acero} F., {Aharonian} F. {\em et~al.\/} (2013).
\newblock {\em {Constraints on axionlike particles with H.E.S.S. from the
  irregularity of the PKS 2155-304 energy spectrum}\/}.
\newblock \prd, {\bf 88}, 10, 102003

\bibitem{Ajello16}
{Ajello} M., {Albert} A., {Anderson} B. {\em et~al.\/} (2016).
\newblock {\em {Search for Spectral Irregularities due to
  Photon-Axionlike-Particle Oscillations with the Fermi Large Area
  Telescope}\/}.
\newblock Physical Review Letters, {\bf 116}, 16, 161101

\bibitem{Fermi2009}
{Abdo} A.A., {Ackermann} M., {Ajello} M. {\em et~al.\/} (2009).
\newblock {\em {A limit on the variation of the speed of light arising from
  quantum gravity effects}\/}.
\newblock \nat, {\bf 462}, 331

\bibitem{Abramowski11a}
{H.E.S.S.~Collaboration}, {Abramowski} A., {Acero} F. {\em et~al.\/} (2011).
\newblock {\em {Search for Lorentz Invariance breaking with a likelihood fit of
  the PKS 2155-304 flare data taken on MJD 53944}\/}.
\newblock Astroparticle Physics, {\bf 34}, 738

\bibitem{Kifune:1999ex}
Kifune T. (1999).
\newblock {\em {Invariance violation extends the cosmic ray horizon?}\/}.
\newblock \apj, {\bf 518}, L21

\bibitem{Fairbairn2014}
{Fairbairn} M., {Nilsson} A., {Ellis} J. {\em et~al.\/} (2014).
\newblock {\em {The CTA sensitivity to Lorentz-violating effects on the
  gamma-ray horizon}\/}.
\newblock \jcap, {\bf 6}, 005

\bibitem{Dammando2015}
{D'Ammando} F., {Orienti} M., {Larsson} J. {\em et~al.\/} (2015).
\newblock {\em {The first {$\gamma$}-ray detection of the narrow-line Seyfert 1
  FBQS J1644+2619}\/}.
\newblock \mnras, {\bf 452}, 520

\bibitem{Dammando2016}
{D'Ammando} F., {Orienti} M., {Finke} J. {\em et~al.\/} (2016).
\newblock {\em {A Panchromatic View of Relativistic Jets in Narrow-Line Seyfert
  1 Galaxies}\/}.
\newblock Galaxies, {\bf 4}, 11

\bibitem{Magic2008}
{MAGIC Collaboration}, {Albert} J., {Aliu} E. {\em et~al.\/} (2008).
\newblock {\em {Very-High-Energy gamma rays from a Distant Quasar: How
  Transparent Is the Universe?}\/}.
\newblock Science, {\bf 320}, 1752

\bibitem{Ackermann2013}
Ackermann M. {\em et~al.\/} (2013).
\newblock {\em {The First FERMI-LAT Catalog of sources above 10 GeV}\/}.
\newblock ApJS, {\bf 209}, 34

\bibitem{Shaw2013}
{Shaw} M.S., {Romani} R.W., {Cotter} G. {\em et~al.\/} (2013).
\newblock {\em {Spectroscopy of the Largest Ever {$\gamma$}-Ray-selected BL Lac
  Sample}\/}.
\newblock \apj, {\bf 764}, 135

\bibitem{Pita2014a}
Pita A., Goldoni P., Boisson C. {\em et~al.\/} (2014).
\newblock {\em Spectroscopy of high-energy bl lacertae objects with x-shooter
  on the vlt\/}.
\newblock A\&A, {\bf 565}, A12

\bibitem{Abramowski12f}
{H.E.S.S.~Collaboration}, {Abramowski} A., {Acero} F. {\em et~al.\/} (2012).
\newblock {\em {A multiwavelength view of the flaring state of PKS 2155-304 in
  2006}\/}.
\newblock \aap, {\bf 539}, A149

\bibitem{Marscher2008}
{Marscher} A.P., {Jorstad} S.G., {D'Arcangelo} F.D. {\em et~al.\/} (2008).
\newblock {\em {The inner jet of an active galactic nucleus as revealed by a
  radio-to-{$\gamma$}-ray outburst}\/}.
\newblock \nat, {\bf 452}, 966

\bibitem{Marscher2010}
{Marscher} A.P., {Jorstad} S.G., {Larionov} V.M. {\em et~al.\/} (2010).
\newblock {\em {Probing the Inner Jet of the Quasar PKS 1510-089 with
  Multi-Waveband Monitoring During Strong Gamma-Ray Activity}\/}.
\newblock \apjl, {\bf 710}, L126

\bibitem{Blinov2016}
{Blinov} D., {Pavlidou} V., {Papadakis} I. {\em et~al.\/} (2016).
\newblock {\em {RoboPol: do optical polarization rotations occur in all
  blazars?}\/}.
\newblock \mnras, {\bf 462}, 1775

\bibitem{Meyer16}
{Meyer} M., {Conrad} J. \& {Dickinson} H. (2016).
\newblock {\em {Sensitivity of the Cherenkov Telescope Array to the Detection
  of Intergalactic Magnetic Fields}\/}.
\newblock \apj, {\bf 827}, 147

\bibitem{Pinzke10}
{Pinzke} A. \& {Pfrommer} C. (2010).
\newblock {\em {Simulating the {$\gamma$}-ray emission from galaxy clusters: a
  universal cosmic ray spectrum and spatial distribution}\/}.
\newblock \mnras, {\bf 409}, 449

\bibitem{Zandanel14}
{Zandanel} F., {Pfrommer} C. \& {Prada} F. (2014).
\newblock {\em {On the physics of radio haloes in galaxy clusters: scaling
  relations and luminosity functions}\/}.
\newblock \mnras, {\bf 438}, 124

\bibitem{Aleksic16a}
{Ahnen} M.L., {Ansoldi} S., {Antonelli} L.A. {\em et~al.\/} (2016).
\newblock {\em {Deep observation of the NGC 1275 region with MAGIC: search of
  diffuse {$\gamma$}-ray emission from cosmic rays in the Perseus cluster}\/}.
\newblock \aap, {\bf 589}, A33

\bibitem{Brunetti07b}
{Brunetti} G., {Venturi} T., {Dallacasa} D. {\em et~al.\/} (2007).
\newblock {\em {Cosmic Rays and Radio Halos in Galaxy Clusters: New Constraints
  from Radio Observations}\/}.
\newblock ApJl, {\bf 670}, L5

\bibitem{Ackermann14}
{Ackermann} M., {Ajello} M., {Albert} A. {\em et~al.\/} (2014).
\newblock {\em {Search for Cosmic-Ray-induced Gamma-Ray Emission in Galaxy
  Clusters}\/}.
\newblock ApJ, {\bf 787}, 18

\bibitem{Voit05}
{Voit} G.M. (2005).
\newblock {\em {Tracing cosmic evolution with clusters of galaxies}\/}.
\newblock Reviews of Modern Physics, {\bf 77}, 207

\bibitem{Forman03}
{Forman} W., {Churazov} E., {David} L. {\em et~al.\/} (2003).
\newblock {\em {A High Angular Resolution View of Hot Gas in Clusters, Groups,
  and Galaxies}\/}.
\newblock arXiv:0301476

\bibitem{Miniati15b}
{Miniati} F. \& {Beresnyak} A. (2015).
\newblock {\em {Self-similar energetics in large clusters of galaxies}\/}.
\newblock Nature, {\bf 523}, 59

\bibitem{Brunetti14}
{Brunetti} G. \& {Jones} T.W. (2014).
\newblock {\em {Cosmic Rays in Galaxy Clusters and Their Nonthermal
  Emission}\/}.
\newblock International Journal of Modern Physics D, {\bf 23}, 1430007

\bibitem{Feretti12}
{Feretti} L., {Giovannini} G., {Govoni} F. {\em et~al.\/} (2012).
\newblock {\em {Clusters of galaxies: observational properties of the diffuse
  radio emission}\/}.
\newblock \aapr, {\bf 20}, 54

\bibitem{Berezinsky97}
{Berezinsky} V.S., {Blasi} P. \& {Ptuskin} V.S. (1997).
\newblock {\em {Clusters of Galaxies as Storage Room for Cosmic Rays}\/}.
\newblock ApJ, {\bf 487}, 529

\bibitem{Blasi99}
{Blasi} P. \& {Colafrancesco} S. (1999).
\newblock {\em {Cosmic rays, radio halos and nonthermal X-ray emission in
  clusters of galaxies}\/}.
\newblock Astroparticle Physics, {\bf 12}, 169

\bibitem{Pfrommer08}
{Pfrommer} C., {En{\ss}lin} T.A. \& {Springel} V. (2008).
\newblock {\em {Simulating cosmic rays in clusters of galaxies - II. A unified
  scheme for radio haloes and relics with predictions of the {$\gamma$}-ray
  emission}\/}.
\newblock \mnras, {\bf 385}, 1211

\bibitem{Blasi01}
{Blasi} P. (2001).
\newblock {\em {The non-thermal radiation-cluster merger connection}\/}.
\newblock Astroparticle Physics, {\bf 15}, 223

\bibitem{Inoue05}
{Inoue} S., {Aharonian} F.A. \& {Sugiyama} N. (2005).
\newblock {\em {Hard X-Ray and Gamma-Ray Emission Induced by Ultra-High-Energy
  Protons in Cluster Accretion Shocks}\/}.
\newblock ApJl, {\bf 628}, L9

\bibitem{Vannoni11}
{Vannoni} G., {Aharonian} F.A., {Gabici} S. {\em et~al.\/} (2011).
\newblock {\em {Acceleration and radiation of ultra-high energy protons in
  galaxy clusters}\/}.
\newblock A\&A, {\bf 536}, A56

\bibitem{Armengaud06}
{Armengaud} E., {Sigl} G. \& {Miniati} F. (2006).
\newblock {\em {Secondary gamma rays from ultrahigh energy cosmic rays produced
  in magnetized environments}\/}.
\newblock Physics Review D, {\bf 73}, 8, 083008

\bibitem{Kotera09}
{Kotera} K., {Allard} D., {Murase} K. {\em et~al.\/} (2009).
\newblock {\em {Propagation of Ultrahigh Energy Nuclei in Clusters of Galaxies:
  Resulting Composition and Secondary Emissions}\/}.
\newblock ApJ, {\bf 707}, 370

\bibitem{Kelner08}
{Kelner} S.R. \& {Aharonian} F.A. (2008).
\newblock {\em {Energy spectra of gamma rays, electrons, and neutrinos produced
  at interactions of relativistic protons with low energy radiation}\/}.
\newblock Physics Review D, {\bf 78}, 3, 034013

\bibitem{Croston08}
{Croston} J.H., {Pratt} G.W., {B{\"o}hringer} H. {\em et~al.\/} (2008).
\newblock {\em {Galaxy-cluster gas-density distributions of the representative
  XMM-Newton cluster structure survey (REXCESS)}\/}.
\newblock A\&A, {\bf 487}, 431

\bibitem{Brunetti12}
{Brunetti} G., {Blasi} P., {Reimer} O. {\em et~al.\/} (2012).
\newblock {\em {Probing the origin of giant radio haloes through radio and
  {$\gamma$}-ray data: the case of the Coma cluster}\/}.
\newblock \mnras, {\bf 426}, 956

\bibitem{Pinzke16}
{Pinzke} A., {Oh} S.P. \& {Pfrommer} C. (2016).
\newblock {\em {Turbulence and Particle Acceleration in Giant Radio Haloes: the
  Origin of Seed Electrons}\/}.
\newblock arXiv:1611.07533

\bibitem{ZuHone13}
{ZuHone} J.A., {Markevitch} M., {Brunetti} G. {\em et~al.\/} (2013).
\newblock {\em {Turbulence and Radio Mini-halos in the Sloshing Cores of Galaxy
  Clusters}\/}.
\newblock ApJ, {\bf 762}, 78

\bibitem{Jacob16}
{Jacob} S. \& {Pfrommer} C. (2016).
\newblock {\em {Cosmic ray heating in cool core clusters II: Self-regulation
  cycle and non-thermal emission}\/}.
\newblock arXiv:1609.06322

\bibitem{Storm12}
{Storm} E.M., {Jeltema} T.E. \& {Profumo} S. (2012).
\newblock {\em {Gamma Rays from Star Formation in Clusters of Galaxies}\/}.
\newblock \apj, {\bf 755}, 117

\bibitem{Persic12}
{Persic} M. \& {Rephaeli} Y. (2012).
\newblock {\em {Cosmic rays in star-forming galaxies}\/}.
\newblock Journal of Physics Conference Series, {\bf 355}, 1, 012038

\bibitem{Aleksic10}
{Aleksi{\'c}} J., {Antonelli} L.A., {Antoranz} P. {\em et~al.\/} (2010).
\newblock {\em {MAGIC Gamma-ray Telescope Observation of the Perseus Cluster of
  Galaxies: Implications for Cosmic Rays, Dark Matter, and NGC 1275}\/}.
\newblock \apj, {\bf 710}, 634

\bibitem{Aleksic10a}
--- (2010).
\newblock {\em {Detection of Very High Energy {$\gamma$}-ray Emission from the
  Perseus Cluster Head-Tail Galaxy IC 310 by the MAGIC Telescopes}\/}.
\newblock ApJL, {\bf 723}, L207

\bibitem{Aleksic12}
{Aleksi{\'c}} J., {Alvarez} E.A., {Antonelli} L.A. {\em et~al.\/} (2012).
\newblock {\em {Detection of very-high energy {$\gamma$}-ray emission from NGC
  1275 by the MAGIC telescopes}\/}.
\newblock A\&A, {\bf 539}, L2

\bibitem{Aleksic14}
{Aleksi{\'c}} J., {Ansoldi} S., {Antonelli} L.A. {\em et~al.\/} (2014).
\newblock {\em {Contemporaneous observations of the radio galaxy NGC 1275 from
  radio to very high energy {$\gamma$}-rays}\/}.
\newblock A\&A, {\bf 564}, A5

\bibitem{Wouters13}
{Wouters} D. \& {Brun} P. (2013).
\newblock {\em {Constraints on Axion-like Particles from X-Ray Observations of
  the Hydra Galaxy Cluster}\/}.
\newblock \apj, {\bf 772}, 44

\bibitem{Reimer03}
{Reimer} O., {Pohl} M., {Sreekumar} P. {\em et~al.\/} (2003).
\newblock {\em {EGRET Upper Limits on the High-Energy Gamma-Ray Emission of
  Galaxy Clusters}\/}.
\newblock ApJ, {\bf 588}, 155

\bibitem{Ackermann10a}
{Ackermann} M., {Ajello} M., {Allafort} A. {\em et~al.\/} (2010).
\newblock {\em {Constraints on dark matter annihilation in clusters of galaxies
  with the Fermi large area telescope}\/}.
\newblock JCAP, {\bf 5}, 025

\bibitem{Ackermann10b}
--- (2010).
\newblock {\em {GeV Gamma-ray Flux Upper Limits from Clusters of Galaxies}\/}.
\newblock ApJL, {\bf 717}, L71

\bibitem{Jeltema11}
{Jeltema} T.E. \& {Profumo} S. (2011).
\newblock {\em {Implications of Fermi Observations For Hadronic Models of Radio
  Halos in Clusters of Galaxies}\/}.
\newblock ApJ, {\bf 728}, 53

\bibitem{Han12}
{Han} J., {Frenk} C.S., {Eke} V.R. {\em et~al.\/} (2012).
\newblock {\em {Constraining extended gamma-ray emission from galaxy
  clusters}\/}.
\newblock MNRAS, {\bf 427}, 1651

\bibitem{Ando12}
{Ando} S. \& {Nagai} D. (2012).
\newblock {\em {Fermi-LAT constraints on dark matter annihilation cross section
  from observations of the Fornax cluster}\/}.
\newblock JCAP, {\bf 7}, 017

\bibitem{Huber13}
{Huber} B., {Tchernin} C., {Eckert} D. {\em et~al.\/} (2013).
\newblock {\em {Probing the cosmic-ray content of galaxy clusters by stacking
  Fermi-LAT count maps}\/}.
\newblock A\&A, {\bf 560}, A64

\bibitem{Zandanel14b}
{Zandanel} F. \& {Ando} S. (2014).
\newblock {\em {Constraints on diffuse gamma-ray emission from structure
  formation processes in the Coma cluster}\/}.
\newblock \mnras, {\bf 440}, 663

\bibitem{Prokhorov14}
{Prokhorov} D.A. \& {Churazov} E.M. (2014).
\newblock {\em {Counting gamma rays in the directions of galaxy clusters}\/}.
\newblock A\&A, {\bf 567}, A93

\bibitem{Vazza14}
{Vazza} F. \& {Br{\"u}ggen} M. (2014).
\newblock {\em {Do radio relics challenge diffusive shock acceleration?}\/}.
\newblock MNRAS, {\bf 437}, 2291

\bibitem{Griffin14}
{Griffin} R.D., {Dai} X. \& {Kochanek} C.S. (2014).
\newblock {\em {New Limits on Gamma-Ray Emission from Galaxy Clusters}\/}.
\newblock ApJL, {\bf 795}, L21

\bibitem{Selig15}
{Selig} M., {Vacca} V., {Oppermann} N. {\em et~al.\/} (2015).
\newblock {\em {The denoised, deconvolved, and decomposed Fermi {$\gamma$}-ray
  sky. An application of the D$^{3}$PO algorithm}\/}.
\newblock A\&A, {\bf 581}, A126

\bibitem{Vazza15}
{Vazza} F., {Eckert} D., {Br{\"u}ggen} M. {\em et~al.\/} (2015).
\newblock {\em {Electron and proton acceleration efficiency by merger shocks in
  galaxy clusters}\/}.
\newblock MNRAS, {\bf 451}, 2198

\bibitem{Ackermann15a}
{Ackermann} M., {Ajello} M., {Albert} A. {\em et~al.\/} (2016).
\newblock {\em {Search for Gamma-Ray Emission from the Coma Cluster with Six
  Years of Fermi-LAT Data}\/}.
\newblock \apj, {\bf 819}, 149

\bibitem{Ackermann15b}
--- (2015).
\newblock {\em {Search for Extended Gamma-Ray Emission from the Virgo Galaxy
  Cluster with FERMI-LAT}\/}.
\newblock ApJ, {\bf 812}, 159

\bibitem{Perkins06}
{Perkins} J.S., {Badran} H.M., {Blaylock} G. {\em et~al.\/} (2006).
\newblock {\em {TeV Gamma-Ray Observations of the Perseus and Abell 2029 Galaxy
  Clusters}\/}.
\newblock ApJ, {\bf 644}, 148

\bibitem{Perkins08}
{Perkins} J.S. (2008).
\newblock {\em {VERITAS Observations of the Coma Cluster of Galaxies}\/}.
\newblock In F.A. {Aharonian}, W.~{Hofmann} \& F.~{Rieger} (editors), {\em
  American Institute of Physics Conference Series\/}, volume 1085 of {\em
  American Institute of Physics Conference Series\/}, pp. 569--572

\bibitem{Aharonian09z1}
{Aharonian} F., {Akhperjanian} A.G., {Anton} G. {\em et~al.\/} (2009).
\newblock {\em {Very high energy gamma-ray observations of the galaxy clusters
  Abell 496 and Abell 85 with HESS}\/}.
\newblock A\&A, {\bf 495}, 27

\bibitem{Domainko09}
{Domainko} W., {Nedbal} D., {Hinton} J.A. {\em et~al.\/} (2009).
\newblock {\em {New Results from H.E.S.S. Observations of Galaxy Clusters}\/}.
\newblock International Journal of Modern Physics D, {\bf 18}, 1627

\bibitem{Galante09}
{Galante} N. \& {for the VERITAS Collaboration} (2009).
\newblock {\em {Observation of Radio Galaxies and Clusters of Galaxies with
  VERITAS}\/}.
\newblock arXiv:0907.5000

\bibitem{Kiuchi09}
{Kiuchi} R., {Mori} M., {Bicknell} G.V. {\em et~al.\/} (2009).
\newblock {\em {CANGAROO-III Search for TeV Gamma Rays from Two Clusters of
  Galaxies}\/}.
\newblock ApJ, {\bf 704}, 240

\bibitem{Acciari09z}
{Acciari} V.A., {Aliu} E., {Arlen} T. {\em et~al.\/} (2009).
\newblock {\em {VERITAS Upper Limit on the Very High Energy Emission from the
  Radio Galaxy NGC 1275}\/}.
\newblock ApJL, {\bf 706}, L275

\bibitem{Aleksic12b}
{Aleksi{\'c}} J., {Alvarez} E.A., {Antonelli} L.A. {\em et~al.\/} (2012).
\newblock {\em {Constraining cosmic rays and magnetic fields in the Perseus
  galaxy cluster with TeV observations by the MAGIC telescopes}\/}.
\newblock A\&A, {\bf 541}, A99

\bibitem{Arlen12}
{Arlen} T., {Aune} T., {Beilicke} M. {\em et~al.\/} (2012).
\newblock {\em {Constraints on Cosmic Rays, Magnetic Fields, and Dark Matter
  from Gamma-Ray Observations of the Coma Cluster of Galaxies with VERITAS and
  Fermi}\/}.
\newblock ApJ, {\bf 757}, 123

\bibitem{Abramowski12z}
{Abramowski} A., {Acero} F., {Aharonian} F. {\em et~al.\/} (2012).
\newblock {\em {Constraints on the gamma-ray emission from the cluster-scale
  AGN outburst in the Hydra A galaxy cluster}\/}.
\newblock A\&A, {\bf 545}, A103

\bibitem{Ando08}
{Ando} S. \& {Nagai} D. (2008).
\newblock {\em {Gamma-ray probe of cosmic ray pressure in galaxy clusters and
  cosmological implications}\/}.
\newblock MNRAS, {\bf 385}, 2243

\bibitem{Churazov03}
{Churazov} E., {Forman} W., {Jones} C. {\em et~al.\/} (2003).
\newblock {\em {XMM-Newton Observations of the Perseus Cluster. I. The
  Temperature and Surface Brightness Structure}\/}.
\newblock ApJ, {\bf 590}, 225

\bibitem{Pedlar90}
{Pedlar} A., {Ghataure} H.S., {Davies} R.D. {\em et~al.\/} (1990).
\newblock {\em {The Radio Structure of NGC1275}\/}.
\newblock \mnras, {\bf 246}, 477

\bibitem{Gitti02}
{Gitti} M., {Brunetti} G. \& {Setti} G. (2002).
\newblock {\em {Modeling the interaction between ICM and relativistic plasma in
  cooling flows: The case of the Perseus cluster}\/}.
\newblock A\&A, {\bf 386}, 456

\bibitem{Charles2016}
{Charles} E., {S{\'a}nchez-Conde} M., {Anderson} B. {\em et~al.\/} (2016).
\newblock {\em {Sensitivity projections for dark matter searches with the Fermi
  large area telescope}\/}.
\newblock \physrep, {\bf 636}, 1

\bibitem{Reiprich02}
{Reiprich} T.H. \& {B{\"o}hringer} H. (2002).
\newblock {\em {The Mass Function of an X-Ray Flux-limited Sample of Galaxy
  Clusters}\/}.
\newblock ApJ, {\bf 567}, 716

\bibitem{Vazza16}
{Vazza} F., {Br{\"u}ggen} M., {Wittor} D. {\em et~al.\/} (2016).
\newblock {\em {Constraining the efficiency of cosmic ray acceleration by
  cluster shocks}\/}.
\newblock \mnras, {\bf 459}, 70

\bibitem{Kushnir09}
{Kushnir} D. \& {Waxman} E. (2009).
\newblock {\em {Nonthermal emission from clusters of galaxies}\/}.
\newblock JCAP, {\bf 8}, 002

\bibitem{Bonafede10}
{Bonafede} A., {Feretti} L., {Murgia} M. {\em et~al.\/} (2010).
\newblock {\em {The Coma cluster magnetic field from Faraday rotation
  measures}\/}.
\newblock A\&A, {\bf 513}, A30+

\bibitem{Bonafede13}
{Bonafede} A., {Vazza} F., {Br{\"u}ggen} M. {\em et~al.\/} (2013).
\newblock {\em {Measurements and simulation of Faraday rotation across the Coma
  radio relic}\/}.
\newblock \mnras, {\bf 433}, 3208

\bibitem{Rottgering12}
{R{\"o}ttgering} H., {Afonso} J., {Barthel} P. {\em et~al.\/} (2011).
\newblock {\em {LOFAR and APERTIF Surveys of the Radio Sky: Probing Shocks and
  Magnetic Fields in Galaxy Clusters}\/}.
\newblock Journal of Astrophysics and Astronomy, {\bf 32}, 557

\bibitem{Govoni13}
{Govoni} F., {Murgia} M., {Xu} H. {\em et~al.\/} (2013).
\newblock {\em {Polarization of cluster radio halos with upcoming radio
  interferometers}\/}.
\newblock A\&A, {\bf 554}, A102

\bibitem{Bonafede15}
{Bonafede} A., {Vazza} F., {Br{\"u}ggen} M. {\em et~al.\/} (2015).
\newblock {\em {Unravelling the origin of large-scale magnetic fields in galaxy
  clusters and beyond through Faraday Rotation Measures with the SKA}\/}.
\newblock Advancing Astrophysics with the Square Kilometre Array (AASKA14), 95

\bibitem{Ensslin11}
{En{\ss}lin} T., {Pfrommer} C., {Miniati} F. {\em et~al.\/} (2011).
\newblock {\em {Cosmic ray transport in galaxy clusters: implications for radio
  halos, gamma-ray signatures, and cool core heating}\/}.
\newblock A\&A, {\bf 527}, A99+

\bibitem{Wiener13}
{Wiener} J., {Oh} S.P. \& {Guo} F. (2013).
\newblock {\em {Cosmic ray streaming in clusters of galaxies}\/}.
\newblock \mnras, {\bf 434}, 2209

\bibitem{Govoni04}
{Govoni} F. \& {Feretti} L. (2004).
\newblock {\em {Magnetic Fields in Clusters of Galaxies}\/}.
\newblock International Journal of Modern Physics D, {\bf 13}, 1549

\bibitem{Clarke04}
{Clarke} T.E. (2004).
\newblock {\em {Faraday Rotation Observations of Magnetic Fields in Galaxy
  Clusters}\/}.
\newblock Journal of Korean Astronomical Society, {\bf 37}, 337

\bibitem{Ensslin06}
{En{\ss}lin} T.A. \& {Vogt} C. (2006).
\newblock {\em {Magnetic turbulence in cool cores of galaxy clusters}\/}.
\newblock \aap, {\bf 453}, 447

\bibitem{Kuchar11}
{Kuchar} P. \& {En{\ss}lin} T.A. (2011).
\newblock {\em {Magnetic power spectra from Faraday rotation maps. REALMAF and
  its use on Hydra A}\/}.
\newblock \aap, {\bf 529}, A13+

\bibitem{Dominguez11b}
{Dom{\'{\i}}nguez} A., {Primack} J.R., {Rosario} D.J. {\em et~al.\/} (2011).
\newblock {\em {Extragalactic background light inferred from AEGIS
  galaxy-SED-type fractions}\/}.
\newblock MNRAS, {\bf 410}, 2556

\bibitem{Sijbring93}
{Sijbring} L.G. (1993).
\newblock {\em A Radio Continuum and HI Line Study of the Perseus Cluster\/}.
\newblock Ph.D. thesis, Groningen University

\bibitem{Juliusson72}
Juliusson E., Meyer P. \& M\"{u}ller D. (1972).
\newblock {\em {Composition of Cosmic-Ray Nuclei at High Energies}\/}.
\newblock Physical Review Letters, {\bf 29}, 7, 445

\bibitem{Garcia-Munoz75}
Garcia-Munoz M., Mason G.M. \& Simpson J.A. (1975).
\newblock {\em {The isotopic composition of galactic cosmic-ray lithium,
  beryllium, and boron}\/}.
\newblock The Astrophysical Journal, {\bf 201}, L145

\bibitem{Apel13}
 (2013).
\newblock {\em {KASCADE-Grande} measurements of energy spectra for elemental
  groups of cosmic rays\/}.
\newblock Astroparticle Physics, {\bf 47}, 54

\bibitem{Kieda01}
Kieda D., Swordy S. \& Wakely S. (2001).
\newblock {\em {A high resolution method for measuring cosmic ray composition
  beyond 10 TeV}\/}.
\newblock Astroparticle Physics, {\bf 15}, 3, 287

\bibitem{Aharonian07dc}
Aharonian F., Akhperjanian A., Bazer-Bachi A. {\em et~al.\/} (2007).
\newblock {\em {First ground-based measurement of atmospheric Cherenkov light
  from cosmic rays}\/}.
\newblock Physical Review D, {\bf 75}, 4, 042004

\bibitem{Wissel10}
Wissel S.A. (2010).
\newblock {\em {Observations of direct Cerenkov light in ground-based
  telescopes and the flux of iron nuclei at TeV energies}\/}.
\newblock ProQuest Dissertations And Theses; Thesis (Ph.D.)--The University of
  Chicago

\bibitem{Aharonian95}
{Aharonian} F.A., {Atoyan} A.M. \& {Voelk} H.J. (1995).
\newblock {\em {High energy electrons and positrons in cosmic rays as an
  indicator of the existence of a nearby cosmic tevatron}\/}.
\newblock \aap, {\bf 294}, L41

\bibitem{Kobayashi04}
{Kobayashi} T., {Komori} Y., {Yoshida} K. {\em et~al.\/} (2004).
\newblock {\em {The Most Likely Sources of High-Energy Cosmic-Ray Electrons in
  Supernova Remnants}\/}.
\newblock \apj, {\bf 601}, 340

\bibitem{Abdo09e}
{Abdo} A.A., {Ackermann} M., {Ajello} M. {\em et~al.\/} (2009).
\newblock {\em {Measurement of the Cosmic Ray e$^{+}$+e$^{-}$ Spectrum from
  20GeV to 1TeV with the Fermi Large Area Telescope}\/}.
\newblock Physical Review Letters, {\bf 102}, 18, 181101

\bibitem{Adriani11}
{Adriani} O., {Barbarino} G.C., {Bazilevskaya} G.A. {\em et~al.\/} (2011).
\newblock {\em {Cosmic-Ray Electron Flux Measured by the PAMELA Experiment
  between 1 and 625 GeV}\/}.
\newblock Physical Review Letters, {\bf 106}, 20, 201101

\bibitem{Aguilar14}
{Aguilar} M., {Aisa} D., {Alvino} A. {\em et~al.\/} (2014).
\newblock {\em {Electron and Positron Fluxes in Primary Cosmic Rays Measured
  with the Alpha Magnetic Spectrometer on the International Space Station}\/}.
\newblock Physical Review Letters, {\bf 113}, 12, 121102

\bibitem{Abdollahi17}
Abdollahi S., Ackermann M., Ajello M. {\em et~al.\/} (2017).
\newblock {\em Cosmic-ray electron-positron spectrum from 7 gev to 2 tev with
  the fermi large area telescope\/}.
\newblock Phys. Rev. D, {\bf 95}, 082007

\bibitem{Malyshev09}
{Malyshev} D., {Cholis} I. \& {Gelfand} J. (2009).
\newblock {\em {Pulsars versus dark matter interpretation of ATIC/PAMELA}\/}.
\newblock \prd, {\bf 80}, 6, 063005

\bibitem{Adriani09}
{Adriani} O., {Barbarino} G.C., {Bazilevskaya} G.A. {\em et~al.\/} (2009).
\newblock {\em {An anomalous positron abundance in cosmic rays with energies
  1.5-100GeV}\/}.
\newblock \nat, {\bf 458}, 607

\bibitem{Aguilar13}
{Aguilar} M., {Alberti} G., {Alpat} B. {\em et~al.\/} (2013).
\newblock {\em {First Result from the Alpha Magnetic Spectrometer on the
  International Space Station: Precision Measurement of the Positron Fraction
  in Primary Cosmic Rays of 0.5-350 GeV}\/}.
\newblock Physical Review Letters, {\bf 110}, 14, 141102

\bibitem{Borlatridon11}
{Borla Tridon} D. (2011).
\newblock {\em {Measurement of the cosmic electron spectrum with the MAGIC
  telescopes}\/}.
\newblock International Cosmic Ray Conference, {\bf 6}, 47

\bibitem{Staszak15}
{Staszak} D. \& {for the VERITAS Collaboration} (2015).
\newblock {\em {A Cosmic-ray Electron Spectrum with VERITAS}\/}.
\newblock proc. 34th ICRC The Hague, Netherlands

\bibitem{Parsons11}
{Parsons} R.D. (2011).
\newblock {\em Towards a Measurement of the Cosmic Ray Electron Spectrum at the
  Highest Energies, using the Next-Generation Cherenkov Array CTA\/}.
\newblock Ph.D. thesis, University of Leeds

\bibitem{Gaug14}
{Gaug} M., {Berge} D., {Daniel} M. {\em et~al.\/} (2014).
\newblock {\em {Calibration strategies for the Cherenkov Telescope Array}\/}.
\newblock In {\em Observatory Operations: Strategies, Processes, and Systems
  V\/}, volume 9149 of {\em \procspie\/}, p. 914919

\bibitem{Parsons16}
{Parsons} R.D., {Hinton} J.A. \& {Schoorlemmer} H. (2016).
\newblock {\em {Calibration of the Cherenkov telescope array using cosmic ray
  electrons}\/}.
\newblock Astroparticle Physics, {\bf 84}, 23

\bibitem{dEnterria11}
{d'Enterria} D., {Engel} R., {Pierog} T. {\em et~al.\/} (2011).
\newblock {\em {Constraints from the first LHC data on hadronic event
  generators for ultra-high energy cosmic-ray physics}\/}.
\newblock Astroparticle Physics, {\bf 35}, 98

\bibitem{Marrocchesi15}
{Marrocchesi} P.S. (2015).
\newblock {\em {CALET: a high energy astroparticle physics experiment on the
  ISS}\/}.
\newblock arXiv:1512.08059

\bibitem{HanburyBrown74}
{Brown} R.H. (1974).
\newblock {\em {The intensity interferometer: Its application to astronomy,
  (Halsted Press)}\/}

\bibitem{Tuthill14}
Tuthill P.G. (2014).
\newblock {\em The narrabri stellar intensity interferometer: a 50th birthday
  tribute\/}.
\newblock volume 9146, pp. 91460C--91460C--7

\bibitem{LeBohec06}
{Le Bohec} S. \& {Holder} J. (2006).
\newblock {\em {Optical Intensity Interferometry with Atmospheric Cerenkov
  Telescope Arrays}\/}.
\newblock \apj, {\bf 649}, 399

\bibitem{Dravins08}
{Dravins} D. \& {LeBohec} S. (2008).
\newblock {\em {Toward a diffraction-limited square-kilometer optical
  telescope: digital revival of intensity interferometry}\/}.
\newblock In {\em Society of Photo-Optical Instrumentation Engineers (SPIE)
  Conference Series\/}, volume 6986 of {\em Society of Photo-Optical
  Instrumentation Engineers (SPIE) Conference Series\/}

\bibitem{Nunez12}
{Nu{\~n}ez} P.D., {Holmes} R., {Kieda} D. {\em et~al.\/} (2012).
\newblock {\em {High angular resolution imaging with stellar intensity
  interferometry using air Cherenkov telescope arrays}\/}.
\newblock \mnras, {\bf 419}, 172

\bibitem{Dravins14}
{Dravins} D. \& {Lagadec} T. (2014).
\newblock {\em {Stellar intensity interferometry over kilometer baselines:
  laboratory simulation of observations with the Cherenkov Telescope Array}\/}.
\newblock In {\em Optical and Infrared Interferometry IV\/}, volume 9146 of
  {\em \procspie\/}, p. 91460Z

\bibitem{Lacki14}
{Lacki} B.C. (2014).
\newblock {\em {On the use of Cherenkov Telescopes for outer Solar system body
  occultations}\/}.
\newblock \mnras, {\bf 445}, 1858

\bibitem{Hanna09}
{Hanna} D.S., {Ball} J., {Covault} C.E. {\em et~al.\/} (2009).
\newblock {\em {OSETI with STACEE: A Search for Nanosecond Optical Transients
  from Nearby Stars}\/}.
\newblock Astrobiology, {\bf 9}, 345

\bibitem{Abeysekara2016}
{Abeysekara} A.U., {Archambault} S., {Archer} A. {\em et~al.\/} (2016).
\newblock {\em {A Search for Brief Optical Flashes Associated with the SETI
  Target KIC 8462852}\/}.
\newblock \apjl, {\bf 818}, L33

\bibitem{corsika}
{Heck} D., {Knapp} J., {Capdevielle} J. {\em et~al.\/}.
\newblock {\em Corsika a monte-carlo code to simulate extensive air showers\/}.
\newblock Report FZKA 6019 (1998), Forschungszentrum Karls\-ruhe;
  \url{https://web.ikp.kit.edu/corsika/physics_description/corsika_phys.pdf}

\bibitem{Bernlohr08}
{Bernl{\"o}hr} K. (2008).
\newblock {\em {Simulation of imaging atmospheric Cherenkov telescopes with
  CORSIKA and sim\_ telarray}\/}.
\newblock Astroparticle Physics, {\bf 30}, 149

\bibitem{Hassan15}
{Hassan} T., {Arrabito} L., {Bernl{\"o}r} K. {\em et~al.\/} (2015).
\newblock {\em {Second large-scale Monte Carlo study for the Cherenkov
  Telescope Array}\/}.
\newblock arXiv:1508.06075

\end{thebibliography}
\label{BIB}
\addcontentsline{toc}{chapter}{References}	
\markboth{References}{}

\normalsize
\glsaddall
\label{GLOSS}
\printglossaries
\markboth{Glossary}{}

\end{document}